\documentclass[11pt]{article}
\pdfoutput=1
\usepackage{axodraw2}
\usepackage{amssymb,amsmath,amsbsy}
\usepackage{graphicx}
\usepackage{dsfont}
\usepackage{mathrsfs}
\usepackage{hvdashln}
\usepackage{cite}
\usepackage[titletoc,title]{appendix}
\usepackage[hypertexnames=false,pdfencoding=auto]{hyperref}
\usepackage[all]{hypcap}        
\def\signofmetric{0}

\definecolor{Red}{cmyk}{0,1,1,0}
\definecolor{BrickRed}{cmyk}{0,0.89,0.94,0.28}
\definecolor{Blue}{cmyk}{1,1,0,0}
\definecolor{Green}{cmyk}{1,0,1,0}

\if0\signofmetric
\def\BDpos{}
\def\BDneg{-}
\def\BDposs{\phm}
\def\BDneg{-}
\def\BDnegg{-}
\def\BDplus{+}
\def\BDminus{-}
\fi

\if1\signofmetric
\def\BDpos{-}
\def\BDposs{-}
\def\BDneg{}
\def\BDnegg{\phm}
\def\BDplus{-}
\def\BDminus{+}
\fi

\if2\signofmetric
\def\BDpos{{\color{Red}\oplus}}
\def\BDposs{{\color{Red}\oplus}}
\def\BDneg{{\color{Red}\ominus}}
\def\BDnegg{{\color{Red}\ominus}}
\def\BDplus{{\color{Blue}\oplus}}
\def\BDminus{{\color{Blue}\ominus}}
\fi

\if3\signofmetric
\def\BDpos{{\color{Red}\ominus}}
\def\BDposs{{\color{Red}\ominus}}
\def\BDneg{{\color{Red}\oplus}}
\def\BDnegg{{\color{Red}\oplus}}
\def\BDplus{{\color{Blue}\ominus}}
\def\BDminus{{\color{Blue}\oplus}}
\fi

\makeatletter%
\let\@@footnote\footnote
  \renewcommand\footnote[2][]{\def\@tempa{#1}%
    \ifx\@tempa\@empty
      \@@footnote{#2}%
    \else
      \@@footnote[#1]{#2}%
    \fi
    \futurelet\@tempa\@next@footnote}
  \def\@next@footnote{\ifx\@tempa\footnote
      \textsuperscript{,}%
    \fi} 
\makeatother

\makeatletter
\g@addto@macro\bfseries{\boldmath}
\makeatother	

\DeclareTextCommand{\texttilde}{PU}{\83\003}
\DeclareTextCommand{\textoverline}{PU}{\83\004}
\DeclareTextCommand{\textscalar}{PU}{\83\325}
\DeclareTextCommand{\textstarr}{PU}{\9040\360}
\DeclareTextCommand{\textnu}{PU}{\83\275}
\DeclareTextCommand{\texttau}{PU}{\83\304}
\DeclareTextCommand{\textgamma}{PU}{\83\263}

\renewcommand{\theequation}{\arabic{section}.\arabic{equation}}
\renewcommand{\thefigure}{\arabic{section}.\arabic{figure}}
\newlength{\captsize}           \let\captsize=\footnotesize
\newlength{\captwidth}          
\newlength{\beforetableskip}    
\newcommand{\capt}[1]{\begin{minipage}{\captwidth}
              \let\normalsize=\captsize
              \caption[0]{#1}
              \end{minipage}\\ \vspace{\beforetableskip}}

\makeatletter
      \long\def\@makecaption#1#2{\vskip 10 \p@
      \setbox\@tempboxa\hbox{\textbf{#1:} #2}
      \ifdim \wd\@tempboxa >\hsize
            \textbf{#1:} #2\par                 
      \else
         \hbox to \hsize{\box\@tempboxa\hfil}
      \fi}
\makeatother

\DeclareMathOperator{\Tr}{Tr}
\renewcommand{\Re}{\thinspace{\rm Re\thinspace}}
\renewcommand{\Im}{\thinspace{\rm Im\thinspace}}

\def\beq{\begin{equation}}
\def\eeq{\end{equation}}
\newenvironment{Eqnarray}%
     {\arraycolsep 0.14em\begin{eqnarray}}{\end{eqnarray}}
\def\beqa{\begin{Eqnarray}}
\def\eeqa{\end{Eqnarray}}
\def\bea{\begin{Eqnarray*}}
\def\eea{\end{Eqnarray*}}
\def\ifmath#1{\relax\ifmmode #1\else $#1$\fi}
\def\half{\ifmath{{\textstyle{\frac{1}{2}}}}}
\def\halftheta{\frac{\theta}{2}}
\def\third{\ifmath{{\textstyle{\frac{1}{3}}}}}
\def\quarter{\ifmath{{\textstyle{\frac{1}{4}}}}}
\def\refs#1#2{refs.~\cite{#1,#2}}
\def\Ref#1{ref.~\cite{#1}}
\def\Rref#1{Ref.~\cite{#1}}
\def\app#1{Appendix~#1}
\def\eq#1{eq.~(\ref{#1})}
\def\Eq#1{Eq.~(\ref{#1})}
\def\Eqs#1#2{Eqs.~(\ref{#1}) and (\ref{#2})}
\def\eqs#1#2{eqs.~(\ref{#1}) and (\ref{#2})}

\def\eqor#1#2{eqs.~(\ref{#1}) or (\ref{#2})}
\def\eqss#1#2#3{eqs.~(\ref{#1}), (\ref{#2}) and (\ref{#3})}
\def\Eqss#1#2#3{Eqs.~(\ref{#1}), (\ref{#2}) and (\ref{#3})}
\def\eqst#1#2{eqs.~(\ref{#1})--(\ref{#2})}
\def\Eqst#1#2{Eqs.~(\ref{#1})--(\ref{#2})}
\def\sec#1{Section~\ref{#1}}
\def\secs#1#2{Sections~\ref{#1} and \ref{#2}}
\def\secst#1#2{Sections~\ref{#1}--\ref{#2}}
\def\Fig#1{Fig.~\ref{#1}}
\def\fig#1{Fig.~\ref{#1}}
\def\figs#1#2{Figs.~\ref{#1} and \ref{#2}}
\def\figst#1#2{Figs.~\ref{#1}--\ref{#2}}
\def\mathbold#1{\boldsymbol{#1}}

\def\bra#1{\left\langle #1\right|}
\def\ket#1{\left| #1\right\rangle}

\def\vev#1{\left\langle #1\right\rangle}
\def\ls#1{\ifmath{_{\lower1.5pt\hbox{$\scriptstyle #1$}}}}
\def\lsub#1{\ifmath{_{\lower2.5pt\hbox{$\scriptstyle #1$}}}}
\def\rsup#1{^{\raise 1pt\hbox{$\scriptstyle#1$}}}
\def\rsuper#1{^{\raise 2pt\hbox{$\scriptstyle#1$}}}
\def\rrsup#1{^{\raise 4pt\hbox{$\scriptstyle#1$}}}
\def\lsup#1{^{\lower 6pt\hbox{$\scriptstyle#1$}}}
\def\l2sup#1{^{\lower 4pt\hbox{$\scriptstyle#1$}}}
\def\llsup#1{^{\lower 2pt\hbox{$\scriptstyle#1$}}}
\def\lllsup#1{^{\lower 1pt\hbox{$\scriptstyle#1$}}}

\def\LH{$(\half,0)$}
\def\RH{$(0,\half)$}
\def\nicefrac#1#2{\hbox{${\frac{#1}{#2}}$}}

\def\cw{c_W}

\def\ha{A^0}

\def\phm{\phantom{-}}
\def\phs{\phantom{*}}

\let\tf=\textfrac
\def\ip#1#2{\langle #1|#2\rangle}
\def\T{{\mathsf T}}

\def\ubar{\bar u}
\def\vbar{\bar v}
\def\nubar{\bar \nu}
\def\fbar{\bar f}
\def\dbar{\bar d}
\def\bbar{\bar b}
\def\tbar{\bar t}
\def\Psibar{\overline{\Psi}}
\def\Fbar{\overline{F}}
\def\Ebar{\bar{e}}
\def\Ubar{\bar{u}}
\def\Dbar{\bar{d}}
\def\Doverline{\overline D}
\def\Omegabar{\overline{\Omega}}
\def\Sigmabar{\overline{\Sigma}}
\def\Sbar{\overline{S}}
\def\mbar{\overline{m}}
\def\Mbar{\overline{M}}
\def\Jbar{\overline J}
\def\iso{\mathchoice{\cong}{\cong}{\isoS}{\cong}}
\def\isoS{\vbox{\baselineskip 0pt  \lineskip 0.5pt
    \ialign{$ \mathsurround=0pt  \scriptstyle \hfil ## \hfil $\crcr
        \sim \crcr = \crcr}}}
\def\bold#1{\setbox0=\hbox{$#1$}%
     \kern-.025em\copy0\kern-\wd0
     \kern.05em\copy0\kern-\wd0
     \kern-.025em\raise.0433em\box0 }
\def\hhalf{\scriptscriptstyle{\frac12}}

\def\centeron#1#2{{\setbox0=\hbox{#1}\setbox1=\hbox{#2}\ifdim
\wd1>\wd0\kern.5\wd1\kern-.5\wd0\fi
\copy0\kern-.5\wd0\kern-.5\wd1\copy1\ifdim\wd0>\wd1
\kern.5\wd0\kern-.5\wd1\fi}}
\def\ltap{\;\centeron{\raise.35ex\hbox{$<$}}{\lower.65ex\hbox{$\sim$}}\;}
\def\gtap{\;\centeron{\raise.35ex\hbox{$>$}}{\lower.65ex\hbox{$\sim$}}\;}
\def\gsim{\mathrel{\gtap}}
\def\lsim{\mathrel{\ltap}}
\def\metric{g}

\def\reversed#1{{#1}_r}

\def\N1{\widetilde N_1}
\def\Ni{\widetilde N_i}
\def\Nj{\widetilde N_j}
\def\Nk{\widetilde N_k}
\def\Nl{\widetilde N_\ell}
\def\GG{\widetilde G}
\def\stilde{\widetilde}

\def\Ci{{\widetilde C_{i}}}
\def\Cj{{\widetilde C_{j}}}
\def\Ciplus{\widetilde C^+_{i}}
\def\Cjplus{\widetilde C^+_{j}}
\def\Ciminus{\widetilde C^-_{i}}

\def\ra{\rightarrow}
\def\newcdot{\kern.06em{\cdot}\kern.06em}
\def\hsm{h_{\rm SM}}

\def\lam{\lambda}
\def\kap{\kappa}
\def\eps{\epsilon}

\def\slashchar#1{\setbox0=\hbox{$#1$}           
   \dimen0=\wd0                                 
   \setbox1=\hbox{/} \dimen1=\wd1               
   \ifdim\dimen0>\dimen1                        
      \rlap{\hbox to \dimen0{\hfil/\hfil}}      
      #1                                        
   \else                                        
      \rlap{\hbox to \dimen1{\hfil$#1$\hfil}}   
      /                                         
   \fi}                                        %

\def\Sslash{\not{\hbox{\kern-3pt $\mathscr{S}$}}}

\def\singleandthirdspaced{\baselineskip=\normalbaselineskip\multiply
    \baselineskip by 130\divide\baselineskip by 100}

\newcommand{\newc}{\newcommand}
\newc{\sigmabar}{\overline\sigma}
\newc{\MSbar}{\overline{{\rm MS}}}
\newc{\DRbar}{\overline{{\rm DR}}}

\begin{document}

\setcounter{footnote}{0}
\setcounter{page}{1}
\setcounter{section}{0}
\setcounter{subsection}{0}
\setcounter{subsubsection}{0}
\setcounter{tocdepth}{3}
\interfootnotelinepenalty=10000

\renewcommand{\topfraction}{.8}
\renewcommand{\bottomfraction}{.8}
\renewcommand{\textfraction}{.2}
\renewcommand{\floatpagefraction}{.51}

\setcounter{totalnumber}{5}
\singleandthirdspaced
\thispagestyle{empty}

\begin{flushright}
BN-TH-2008-12\\[-1mm]
SCIPP-08/08\\[-1mm]
arXiv:0812.1594v6 [hep-ph]\\[-1mm]
\end{flushright}
\vspace*{1cm}

\begin{center}
{\LARGE \bf Two-component spinor techniques and Feynman rules }\\[0.3cm]
{\LARGE \bf  for quantum field theory and supersymmetry}\\[1.7cm]
\end{center}

\vspace{-0.5in}

\begin{center}
{\sc Herbi K.~Dreiner$^1$, Howard E.~Haber$^2$ and Stephen P.~Martin$^3$}

\vspace{.1in}
$^1${\em Bethe Center for Theoretical Physics and
Physikalisches Institut der Universit\"at Bonn,
Nu\ss allee 12, 53115 Bonn, Germany}

$^2${\em Santa Cruz Institute for Particle Physics, University of
California, Santa Cruz CA 95064}

$^3${\em Department of Physics, Northern Illinois University, DeKalb IL
60115}

\end{center}

\vspace{0.2in}

\begin{abstract}
Two-component spinors are the basic ingredients for describing fermions
in quantum field theory in $3+1$ spacetime dimensions. We develop
and review the techniques of the two-component spinor formalism and
provide a complete set of Feynman rules for fermions using two-component
spinor notation. These rules are suitable for practical calculations of
cross-sections, decay rates, and radiative corrections in the Standard
Model and its extensions, including supersymmetry, and many explicit
examples are provided. The unified treatment presented in this review
applies to massless Weyl fermions and massive Dirac and Majorana
fermions. We exhibit the relation between the two-component spinor
formalism and the more traditional four-component spinor formalism, and
indicate their connections to the spinor helicity method and techniques
for the computation of helicity amplitudes.

\end{abstract}
\clearpage
\tableofcontents
\clearpage

\section{\texorpdfstring{Introduction}{Introduction}}
\label{sec:intro}
\setcounter{equation}{0}
\setcounter{footnote}{0}
\setcounter{figure}{0}
\setcounter{table}{0}

A crucial feature of the Standard Model of particle physics is the
chiral nature of fermion quantum numbers and interactions. According
to the modern understanding of the electroweak interactions, the
fundamental degrees of freedom for quarks and leptons are
two-component Weyl-van der Waerden fermions~\cite{vdWaerden1},
i.e. two-component Lorentz spinors that transform as irreducible
representations under the gauge group SU(2)$_L \times$U(1$)_Y$.
Furthermore, within the context of supersymmetric
field theories, two-component spinors enter naturally, due to the
spinorial nature of the symmetry generators themselves,
and the holomorphic structure of the superpotential.  Despite
this, most pedagogical treatments and practical calculations in
high-energy physics continue to use the four-component Dirac spinor notation,
which combines distinct irreducible representations of the Lorentz symmetry
algebra.  Parity-conserving theories such as QED and QCD are
well-suited to the four-component fermion methods.  There is also a
certain perceived advantage to familiarity. However, as we progress to
phenomena at and above the scale of electroweak symmetry breaking, it
seems increasingly natural to employ two-component fermion notation,
in harmony with the irreducible transformation properties dictated by
the physics.

One occasionally encounters the misconception that two-component
fermion notations are somehow inherently ill-suited or unwieldy for
practical use.  Perhaps this is due in part to a lack of examples of
calculations using two-component language in the pedagogical
literature. In this review, we seek to dispel this idea by presenting
Feynman rules for fermions using two-component spinor notation,
intended for practical calculations of cross-sections, decays, and
radiative corrections.  This formalism employs a unified framework
that applies equally well to Dirac fermions~\cite{Dirac} 
such as the Standard Model
quarks and charged leptons, and to Majorana fermions~\cite{Majorana} 
such as the light
neutrinos
of the seesaw extension of the Standard
Model~\cite{seesaw,Valle-Schechter1} or the neutralinos of the minimal
supersymmetric extension of the Standard Model (MSSM)
\cite{Nilles:1983ge,HaberKane,primer,Chung:2003fi,haberpdg}.

Spinors were introduced by E.~Cartan in 1913 as
projective representations of the rotation group
\cite{Cartan,Cartan2}, and
entered into physics via the Dirac equation in 1928 \cite{Dirac}. In
the same year, H.~Weyl discussed the representations of the Lorentz
group~\cite{WeylHistory},
including the two-component spinor representations, in terms of
stereographic projective coordinates~\cite{Weyl-book}. The extension
of the tensor calculus (or tensor analysis) to spinor calculus (or spinor
analysis) was given by B.L.~van~der~Waerden \cite{vdWaerden1}, upon the
instigation of P.~Ehrenfest. It is in this paper that
van~der~Waerden (not Weyl as often claimed in the literature) first
introduced the notation of dotted and undotted indices for the
irreducible (\half,0) and (0,\half) representations of the Lorentz
group. Both Weyl~\cite{WeylDirac}
and van der Waerden independently considered the
decomposition of the Dirac equation into two coupled differential
equations for two-component spinors.  In the 1930s, more pedagogical
presentations of two-component spinors were given in refs.~\cite{Uhlenbeck,
vdWaerden2,vdWaerden3}.  In particular,
\Ref{Uhlenbeck} was the first paper in English to employ the
dotted and undotted index notation.  \Rref{vdWaerden2} also presents the
first two-component spinor analysis for general relativity.
In the early 1950s, comprehensive reviews of two-component spinor
techniques were published in English by Bade and Jehle~\cite{Bade-Jehle}
and in German by Cap~\cite{cap}. Shortly thereafter, Bergmann reintroduced
two-component spinors into the formalism of general
relativity~\cite{Bergmann}, which was followed by significant developments
by Penrose~\cite{PenroseGR}.\footnote{For typographical reasons,
Penrose replaced the dotted indices with primed indices, a notation
still employed by most general relativists today.} 
Two-component spinor techniques in curved space
are reviewed in \refs{Pirani}{curved}, with an extensive bibliography
given in \Ref{penrosebib}.  A recent mathematical treatment of
two-component spinors and their geometry can be found in \Ref{Canarutto}.
Two-component spinors also play a central role in the covariant
formulation of relativistic wave equations~\cite{PJ}. 

The formalism of two-component spinors
has also been discussed in many textbooks on relativistic quantum
mechanics, quantum field theory, elementary
particle physics, group theoretical methods
in physics, general relativity, and supersymmetry.
For a guide to the non-supersymmetric literature, see
for example, refs.~\cite{Weyl-book,Corson,Umezawa,Rzewuski,
Roman,Gelfand,Aharoni,Barut-book,Davis,LifshitzI,Cbook,MTW,Novozhilov,LifshitzII,Barut,
Scheck,Penrose,Wald,tung,Rao,Stewart,naber,Ryder,Ticciati,Siegel,Hladik,CarmeliSpin,
CarmeliGroup,PCT,Pokorski,Sexl,CarmeliGR,Akhiezer,Blagojevic,ODonnell,Moroi,stephani,Saller,PK,Srednicki,Banks,Lang}.
Among the early books, we would like to draw attention to
\Ref{Corson}, which has an extensive discussion of two-component
spinor methods.
Scheck \cite{Scheck} includes a short discussion of
the field theory of two-component spinors, including the propagator.
A more extensive field theoretic treatment, including Feynman rules and
applications, is given by Ticciati
\cite{Ticciati}. 
A modern textbook on quantum field theory by Srednicki~\cite{Srednicki}
includes a comprehensive treatment of two-component fermions and
their quantization.  Most textbooks and introductory reviews of
supersymmetry~\cite{Nilles:1983ge,HaberKane,primer,Chung:2003fi,
WessBagger,Gates,Ross,Sohnius,Srivastava,Piguet,MullerKirsten,Derendinger,
West,Bailin,Buchbinder,Lykken,Soni,Galperin,bilalsusy,
FigueroaO'Farrill:2001tr,Mohapatra,Drees,Aitchison,Binetruy,Terning,MDine}
include a discussion of two-component
spinors on some level, with a treatment of
dotted and undotted indices and a collection of identities
involving two-component spinors and
the sigma matrices. Particularly extensive and useful sets
of identities can be found in refs.~\cite{WessBagger,Srivastava,
MullerKirsten,Bailin,FigueroaO'Farrill:2001tr,Drees}.
Finally,
some mathematically sophisticated textbook treatments of spinors
can be found in refs.~\cite{Trautman,Benn,Hurley}.

The standard technique for
computing scattering cross-sections with initial and final state
fermions involves squaring the quantum $S$-matrix
amplitude, summing over the spin
states and then computing the traces of products of
gamma matrices (in the four-component spinor formalism), or products of
sigma matrices (in the two-component spinor formalism).  We employ
this latter technique throughout this paper (with a translation
to the four-component formalism provided in an appendix).
However, the
computational effort rises rapidly as the number of
interfering diagrams increases.
The standard techniques typically become impractical with four or
more particles in the final state. One approach to make such extensive
calculations manageable is the helicity amplitude technique. Here the
scattering process is decomposed into the scattering of helicity
eigenstates.  Then the individual amplitudes are computed analytically
in terms of Lorentz scalar invariants, i.e. a complex
number that can be readily computed. It is then a simple numerical
task to sum all the contributing amplitudes and
compute the square of the complex magnitude of the resulting sum.
Such methods were first explored in
refs.~\cite{Bjorken:1966kh, Henry:1967jm, Fearing:1972pt,Eeg:1979wq},
using four-component spinors (see also refs.~\cite{Passarino,
Kleiss:1984dp, Kleiss:1985yh,Ballestrero:1994jn,xzc}).
Spinor techniques in the helicity formalism were also developed in
ref.~\cite{chalmers}.  In fact, the natural spinor formalism for the helicity
amplitude techniques makes use of the two-component Weyl-van der Waerden
spinors, which we discuss in detail in this review. They were
implemented in the helicity amplitude technique in
refs.~\cite{Farrar:1983wk,Kersch:1985rn, Hagiwara:1985yu,
Berends:1987cv,Giele,Dittmaier:1993jj,Dittmaier:1998nn}.  
Recently, the
two-component formalism has been implemented in a computer program for
the numerical computation of amplitudes and cross-sections for event
generators multi-particle processes~\cite{hahncomp}.

This review is outlined as follows. In Section~2, we present our
conventions and notation (with some additional discussion of our
conventions in Appendix~A).  We also establish numerous identities
involving sigma matrices, epsilon symbols and two-component spinors. 
In Section~3, we derive the basic properties of
the quantized two-component fermion fields.  For a generic collection of
$N$ two-component
fermion fields with identical conserved quantum numbers, the
corresponding mass matrix is an $N\times N$ complex symmetric matrix.
To identify the corresponding mass eigenstates, one must perform
a fermion-mass diagonalization that differs from the usual unitary
similarity transformation of an hermitian matrix
that is employed for a collection of scalar fields.
In Section~4, we derive the
Feynman rules for two-component spinors and describe how to write down
amplitudes in our formalism.  We demonstrate how to employ the
two-component formalism for both tree-level and loop-level processes.
In Section~5, we establish a naming convention for
fermion and antifermion particle states and the corresponding fields.
This is important as it provides an unambiguous procedure for obtaining
the amplitudes for a given physical process,
and for comparing these computations in the two-component and
four-component spinor formalisms. In
Section~6 we provide an extensive number of examples of computations using the
two-component spinor formalism. This is the central part of our review.

We have relegated many details to a set of twelve appendices.
In Appendix A, we summarize our metric and sigma matrix
conventions and indicate
how to translate between conventions with opposite metric signature.
With our definition of the sigma matrices, one can
switch easily between the two conventions by computing one overall sign
factor.  In Appendix B,
we provide a comprehensive list of sigma matrix identities,
and indicate which of these
identities can be generalized to $d\neq 4$ dimensions required for
loop computations that employ
dimensional regularization.   Explicit forms for
the two-component spinor wave functions are given in Appendix C (where
we exhibit two of the most common phase conventions employed in the
literature).  The mathematics of fermion mass diagonalization is
discussed in Appendix D.  In contrast to the unitary similarity
transformation of the scalar squared-mass matrix, fermion mass
diagonalization involves the Takagi diagonalization~\cite{takagi} of a
complex symmetric matrix (for neutral fermions) or the singular
value decomposition of a complex matrix (for charged fermions).
In Appendix~E, we review some of the basic facts of Lie groups
and Lie algebras needed in the treatment of gauge theories.
The two-component fermion propagators (derived in Section~4
using canonical field theory techniques) can also be obtained
by path integral methods, as exhibited in Appendix~F.

As most textbooks on quantum field theory and elementary particle
physics employ the four-component spinor formalism for
fermions, we provide
in Appendix G a dictionary that allows one to translate between
the two-component and four-component spinor techniques.
We use the two-component spinor methods developed in this review
to establish a generalization of the standard four-component spinor
Feynman rules that incorporate Majorana fermions in a
natural way.  In Appendix H, we develop a method for computing
helicity amplitudes in terms of Lorentz-invariant scalar quantities.
This method, which makes use of the
Bouchiat-Michel formulae~\cite{bouchmich}
(originally established in the four-component spinor formalism) is developed
in the language of two-component spinors.  However, these methods
are somewhat limited in scope and must be generalized in the case of
multi-particle final states.  This was accomplished by
Hagiwara and Zeppenfeld (HZ) based on a two-component
spinor treatment~\cite{Hagiwara:1985yu}.  In Appendix~I,
we provide a translation between
the HZ formalism and the two-component spinor formalism of this
review.  We also demonstrate that the spinor helicity method
that is now commonly used in obtaining compact expressions for
helicity amplitudes of multi-particle processed has a very
simple development within the two-component spinor formalism.
Finally, the two-component spinor Feynman rules for the
Standard Model, the seesaw-extended Standard Model (which incorporates
massive neutrinos),
the minimal supersymmetric extension of the Standard Model
(MSSM), and the R-parity-violating
extension of the MSSM are given in Appendices J, K and~L.

\section{\texorpdfstring{Essential conventions, notations and two-component spinor 
identities}{Essential conventions, notations and two-component spinor identities}}
\label{sec:notations}
\setcounter{equation}{0}
\setcounter{figure}{0}
\setcounter{table}{0}

We begin with a discussion of necessary conventions. The metric tensor
is taken to be:\footnote{\label{metricsign}%
The published version of this paper employs the
$(+,-,-,-)$  Minkowski space
metric. An otherwise identical version, using the
$(-,+,+,+)$ metric favored by one of the authors~(SPM),
may be found at
\href{http://www.niu.edu/spmartin/spinors/}{http://www.niu.edu/spmartin/spinors/}.
It can also be constructed by changing a single macro at the beginning
of the \LaTeX\ source file~\cite{source},
in an obvious way. You can tell which version you are
presently reading from \eq{signofmetric}. See 
\app{A} 
for further details and rules for translating between metric conventions.}
\beq
\label{signofmetric}
g_{\mu\nu}=
g^{\mu\nu}={\rm diag}(\BDplus 1 , \BDminus 1, \BDminus 1, \BDminus 1)\, ,
\eeq
where $\mu, \nu= 0,1,2,3$ are spacetime vector indices.
Contravariant four-vectors (e.g. positions and momenta) are defined with
raised indices, and covariant four-vectors (e.g. derivatives) with
lowered indices:
\beqa
x^\mu &=& (t\,;\,\mathbold{\vec x})\,,\\
p^\mu &=& (E\,;\,\mathbold{\vec p})\,,\\
\partial_\mu \equiv\frac{\partial}{\partial x^\mu}
&=& (\partial/\partial t\,;\,\mathbold{\vec \nabla})\,,
\eeqa
in units with $c=1$.  The totally antisymmetric pseudo-tensor
$\eps^{\mu\nu\rho\sigma}$ is defined such that
\beq
\eps^{0123}=-\eps\ls{0123}=+1\,.
\eeq
More details on our conventions can be found in \app{A}.

The irreducible building blocks for spin-1/2 fermions are fields that
transform either under the left-handed $(\half,0)$ or the right-handed
$(0,\half)$ representation of the Lorentz group.  Hermitian
conjugation interchanges these two representations.  A
Majorana fermion field can be constructed from either representation;
this is the spin-1/2 analogue of a real scalar field.  A Dirac
fermion field
combines two equal mass two-component fields into a reducible
representation of the form $(\half,0)\oplus(0,\half)$; this is the
spin-1/2 analogue of a complex scalar field.  It is also possible to use
four-component notation to describe a Majorana fermion by imposing a
reality condition on the spinor in order to reduce the number of
degrees of freedom in half. Details of this construction are given in
Appendix G.1.
However, in this review, we shall focus
primarily on two-component spinor notation for all fermions.  In the
following, $(\half,0)$ spinors carry undotted indices
$\alpha,\beta,\ldots = 1,2$, and $(0,\half)$ spinors carry dotted
indices $\dot{\alpha},\dot{\beta},\ldots = 1,2$.

We first provide a brief introduction to the
Lorentz group and its two-dimensional spinor
representations.
Under an active Lorentz transformation, a contravariant four-vector
$x^\mu$ transforms as
\beq \label{ltx}
x^\mu \rightarrow x^{\prime\,\mu} =  {\Lambda^\mu}{}_\nu x^\nu\,,
\eeq
where $\Lambda\in$\,SO(3,1) satisfies
$\Lambda^\mu{}_\nu g_{\mu\rho}\Lambda^\rho
{}_\lambda=g_{\nu\lambda}$.  It then follows that
the transformation of the corresponding covariant four-vector
$x_\mu\equiv g_{\mu\nu}x^\nu$ satisfies:
\beq \label{ltxcov}
x_\nu =  x^\prime_\mu{\Lambda^\mu}{}_\nu\,.
\eeq
The most general proper orthochronous
Lorentz transformation (which is continuously connected to the
identity), corresponding to a rotation by an angle $\theta$ about an
axis $\mathbold{\widehat n}$ [$\mathbold{\vec\theta}\equiv\theta
\mathbold{\widehat n}$] and a boost vector $\mathbold{\vec\zeta}
\equiv \mathbold{\hat v}\tanh^{-1}\beta$ [where $\mathbold{\hat{v}}
\equiv \mathbold{\vec{v}}/|\mathbold{\vec{v}}|$ and $\beta\equiv |
\mathbold{\vec v}|$], is a $4\times 4$ matrix given by:
\begin{equation} \label{lambda44}
\Lambda=\exp\left(\BDneg\half i\theta^{\rho\sigma}
\mathcal{S}_{\rho\sigma}\right)
=\exp\left(-i\mathbold{{\vec\theta}\newcdot}\boldsymbol{\vec{\mathcal{S}}}
-i\mathbold{{\vec\zeta}\newcdot}\boldsymbol{\vec{\mathcal{K}}}\right)\,,
\end{equation}
where $\theta^{\rho\sigma}\!=\!-\theta^{\sigma\rho}$ and $\mathcal{S}_{\rho\sigma}\!=\!-\mathcal{S}_{\sigma\rho}$.  In particular,  
$\theta^i \equiv \BDpos\half\eps^{ijk} \theta_{jk}$,
$\zeta^i \equiv \BDpos\theta^{i0}=\BDneg\theta^{0i}$,
$\mathcal{S}^i \equiv \half\eps^{ijk}\mathcal{S}_{jk}$, 
$\mathcal{K}^i \equiv \mathcal{S}^{0i}= -\mathcal{S}^{i0}$, 
and
\begin{equation} \label{explicitsmunu}
(\mathcal{S}_{\rho\sigma})^\mu{}_\nu= \BDpos i(\delta_\rho^\mu\,g_{\sigma\nu}-\delta_\sigma^\mu
\,g_{\rho\nu})\,.
\end{equation}
Here, the indices $i,j,k=1,2,3$ and $\epsilon^{123}=+1$.

Thus, an infinitesimal
orthochronous Lorentz transformation is given by
\beq
\Lambda^\mu{}_\nu\simeq \delta^\mu_\nu\BDminus\half i\theta^{\rho\sigma}(\mathcal{S}_{\rho\sigma})^\mu{}_\nu 
=\delta^\mu_\nu+\half\theta^{\rho\sigma}(\delta_\rho^\mu\,g_{\sigma\nu}-\delta_\sigma^\mu
\,g_{\rho\nu})= \delta^\mu_\nu+\half(\theta^\mu{}_\nu-\theta_\nu{}^\mu)\,.
\eeq
Since $\theta^\mu{}_\nu=g_{\alpha\nu}\theta^{\mu\alpha}=-g_{\alpha\nu}\theta^{\alpha\mu}=-\theta_\nu{}^\mu$, it follows that
\beq \label{infinitesimalLT}
\Lambda^\mu{}_\nu\simeq \delta^\mu_\nu+\theta^\mu{}_\nu\,.
\eeq
Moreover, the
infinitesimal boost parameter is $\mathbold{\vec{\zeta}}\equiv
\mathbold{\hat v}\tanh^{-1}\beta\simeq \beta\mathbold{\hat v}
\equiv\mathbold{\vec{\beta}}$, since $\beta\ll 1$ for an infinitesimal
boost.  Hence, the actions of the infinitesimal boosts and rotations
on the spacetime coordinates are
\beqa
\mbox{Rotations:}&\quad&
\left \{ \begin{array}{ll}
\boldsymbol{\vec x} \rightarrow
&\boldsymbol{\vec x}' \simeq \boldsymbol{\vec x}
+ (\boldsymbol{\vec \theta} \times \boldsymbol{\vec x})\,,
\\[5pt]
t \rightarrow & t' \simeq t\,,
\end{array}
\right.
\\[9pt]
\mbox{Boosts:}&\quad&
\left \{ \begin{array}{ll}
\boldsymbol{\vec x} \rightarrow
&\boldsymbol{\vec x}' \simeq\boldsymbol{\vec x}
+ \boldsymbol{\vec \beta}\, t\,,
\\[5pt]
t \rightarrow & t' \simeq t + \boldsymbol{\vec{\beta}\newcdot\vec{x}}\,,
\end{array}
\right.
\eeqa
with exactly analogous transformations for any contravariant four-vector.

With respect to the Lorentz transformation $\Lambda$, a general
$n$-component field $\Phi$ transforms according to a representation
$R$ of the Lorentz group
as $\Phi(x^\mu)\to
\Phi^\prime(x^{\prime\,\mu})=M_R(\Lambda)\,\Phi(x^\mu)$,
where $M_R(\Lambda)$ is the corresponding
(finite) $d_R$-dimensional matrix representation.
Equivalently, the functional form of the transformed field
$\Phi$ obeys
\beq
\Phi'(x^{\mu}) = M_R(\Lambda)
\Phi( [\Lambda^{-1}]^\mu{}_\nu x^\nu)\,,
\label{eq:generalLT}
\eeq
after using \eq{ltx}.
For proper orthochronous Lorentz transformations,
\beq
\label{generallorentzmatrix}
M_R=\exp \left (\BDneg\half i\theta_{\mu\nu} J^{\mu\nu}\right ) \simeq
\mathds{1}_{d_R\times d_R}-i\mathbold{{\vec\theta}\newcdot}\boldsymbol{\vec J}
-i\mathbold{{\vec\zeta}\newcdot}\boldsymbol{\vec K}\,,
\eeq
where $\mathds{1}_{d_R\times d_R}$ is the $d_R\times d_R$ identity matrix and
$\theta_{\mu\nu}$ parameterizes the
Lorentz transformation $\Lambda$ [\eq{lambda44}].
The six independent components of the matrix-valued antisymmetric
tensor $J^{\mu\nu}$ are the $d_R$-dimensional generators of the Lorentz group  
and satisfy the commutation relations:
\beq \label{JJ}
[J^{\mu\nu}\,,\,J^{\lambda\kappa}]=\BDpos i(g^{\mu\kappa}\,J^{\nu\lambda} +
g^{\nu\lambda}\,J^{\mu\kappa} - g^{\mu\lambda}\,J^{\nu\kappa} -
g^{\nu\kappa}\,J^{\mu\lambda})\,.
\eeq
We identify
$\boldsymbol{\vec J}$ and $\boldsymbol{\vec K}$ as the generators of
rotations parameterized by $\boldsymbol{\vec\theta}$
and boosts
parameterized by $\boldsymbol{\vec\zeta}$, respectively, where
\beq \label{jkdef}
J^i \equiv\half \eps^{ijk} J_{jk}\,,\qquad\qquad K^i \equiv J^{0i}\,.
\eeq

Here we focus on the simplest non-trivial irreducible representations
of the Lorentz algebra.  These are the two-dimensional (inequivalent)
representations: $(\half,0)$ and $(0,\half)$.  In the $(\half,0)$
representation, $\boldsymbol{\vec J}=\mathbold{\vec\sigma}/2$ and
$\boldsymbol{\vec K}=-i\mathbold{\vec\sigma}/2$ in
eq.~(\ref{generallorentzmatrix}), which yields
\beq
\label{lorentzmatrix}
M_{(\half,0)}\equiv M\simeq
\mathds{1}_{2\times 2}
-i\mathbold{{\vec\theta}\newcdot}\boldsymbol{\vec \sigma}/2
-\mathbold{{\vec\zeta}\newcdot}\boldsymbol{\vec \sigma}/2\,,
\eeq
where $\boldsymbol{\vec\sigma}\equiv(\sigma^1\,,\,\sigma^2\,,\,\sigma^3)$
are the Pauli matrices [cf.~\eq{pauli}].
By definition $M$ carries undotted spinor indices, as indicated by
$M_\alpha{}^\beta$.
A two-component $(\half,0)$ spinor is denoted by
$\psi_\alpha$ and transforms as $\psi_\alpha\to
M_\alpha{}^\beta\psi_\beta$,
omitting the coordinate arguments of the fields, which are as in
eq.~(\ref{eq:generalLT}).  In our conventions for
the location of the spinor indices, we sum implicitly over a repeated index
pair in which one index is lowered and one index is raised.

In the $(0,\half)$ representation, $\boldsymbol{\vec J}=-\mathbold
{\vec\sigma}{}^*/2$ and $\boldsymbol{\vec K}=-i\mathbold{\vec\sigma}
{}^*/2$ in eq.~(\ref{generallorentzmatrix}), so that its
representation matrix is $M^\ast$, the complex conjugate of
eq.~(\ref{lorentzmatrix}).  By definition, the indices carried by
$M^\ast$ are dotted, as indicated by $(M^*)_{\dot{\alpha}}{}^{\dot
{\beta}}$.  A two-component $(0,\half)$ spinor is denoted by
$\psi^\dagger_{\dot{\alpha}}$ and transforms as
$\psi^\dagger_{\dot{\alpha}}\to
(M^*)_{\dot{\alpha}}{}^{\dot{\beta}}\psi^\dagger_{\dot{\beta}}$, again
suppressing the coordinate arguments of the fields, which are as in
eq.~(\ref{eq:generalLT}).  We distinguish between the
undotted and dotted spinor index types because they cannot be directly
contracted with each other to form a Lorentz invariant
quantity.

It follows that the $(\half,0)$ and $(0,\half)$ representations are
related by hermitian conjugation. That is, if $\psi_\alpha$
is a $(\half,0)$ fermion, then $(\psi_\alpha)^\dagger$
transforms as a $(0,\half)$ fermion.  This means that we can, and
will, describe all fermion degrees of freedom using only fields
defined as left-handed $(\half ,0)$ fermions $\psi_\alpha$, and their
conjugates. In combining spinors to make Lorentz tensors
[as in \eq{xisbareta}], it is
useful to regard $\psi^\dagger_{\dot{\alpha}}$ as a row vector, and
$\psi_\alpha$ as a column vector, with:\footnote{In the
early literature that employed the van der Waerden spinor index notation
(surveyed in \sec{sec:intro}),
no dagger was used in conjunction with the dotted index.
The advantage to attaching the dagger to the dotted spinor field
is that it permits the development of a spinor-index-free notation
for Lorentz-covariant
spinor products [see \eqst{suppressionrule}{xisogeta} and the
accompanying text].}\footnote{Other conventions for the dotted spinor
are possible.  For example, in \Ref{MullerKirsten}, the hermitian
conjugate of $\psi_\alpha$ yields a spinor with a \textit{raised} dotted
index, i.e. $\psi^{\dagger\,\dot\alpha}\equiv (\psi_\alpha)^\dagger$,
which the authors rewrite as $\psi^{\dagger\,\dot\alpha}=
\sigmabar^{\dot\alpha\beta}(\psi_\beta)^\dagger$ in an attempt to maintain
the same index structure on both sides of the equation.  Here,
$\sigmabar^{0}\equiv\mathds{1}_{2\times 2}$ [defined
in \eq{pauli}] appears as a formal device.  This latter convention
leads to a number of complications; e.g., $(M_\alpha{}^\beta)^*\neq
(M^*)_{\dot\alpha}{}^{\dot\beta}$, etc. (see \Ref{MullerKirsten}
for further details).  Although this alternative convention seems
self-consistent, we have adopted the more convenient
\eqs{eq:defbardagger}{eq:defbardagger2} in this review.}
\beq
\psi^\dagger_{\dot{\alpha}}\equiv (\psi_\alpha)^\dagger .
\label{eq:defbardagger}
\eeq
The Lorentz transformation property of
$\psi^\dagger_{\dot{\alpha}}$ then follows from
$(\psi_\alpha)^\dagger \to(\psi_\beta)
^\dagger (M^\dagger)^{\dot{\beta}}{}_{\dot{\alpha}}$
[with coordinate arguments of the fields again suppressed], where $(M^
\dagger)^{\dot{\beta}}{}_{\dot{\alpha}}=(M^*)_{\dot{\alpha}}{}^{\dot
{\beta}}$ reflects the definition of the hermitian adjoint matrix as
the complex conjugate transpose of the matrix.
Again the coordinate
arguments of the fields have been suppressed, and are as in
eq.~(\ref{eq:generalLT}).

In this review, we shall employ the dotted-index notation in
association with the dagger to denote hermitian
conjugation, as specified in \eq{eq:defbardagger}. This is the notation
for hermitian conjugation of spinors found in most field theory textbooks
(e.g., see refs.~\cite{Srednicki,Terning,Peskin:1995ev}).  However,
it should be noted that many references in the supersymmetry
literature (e.g., see refs.~\cite{WessBagger,Gates,Ross,
Sohnius,Srivastava,Piguet,MullerKirsten,Derendinger,West,
Bailin,Buchbinder,Lykken,Soni,Galperin,
bilalsusy,FigueroaO'Farrill:2001tr,Mohapatra,Drees,Binetruy,Aitchison})
employ the bar notation made popular by Wess and Bagger \cite{WessBagger}
where $\overline\psi_{\dot{\alpha}}\equiv\psi^\dagger_{\dot{\alpha}}
\equiv(\psi_\alpha)^\dagger$.

Spinors labeled with one undotted or one dotted index are sometimes called 
spinors of rank one [or more precisely, spinors of rank $(1,0)$ or
$(0,1)$, respectively].  One can also define spinors of higher rank, 
which possess more than one spinor index, with
Lorentz transformation properties that depend on the number of undotted and
dotted spinor 
indices\cite{Uhlenbeck,Bade-Jehle,cap,Pirani,PJ,Corson,Umezawa,Rzewuski,Roman,Gelfand,Aharoni,Barut-book,Davis,LifshitzI,Cbook,MTW,Novozhilov,LifshitzII,Siegel,CarmeliSpin,ODonnell,Buchbinder,Massa}.  
In particular, for a spinor of rank $(m,n)$ denoted by
$S_{\alpha_1\alpha_2\cdots\alpha_m\dot{\beta}_1\dot{\beta}_2\cdots
\dot{\beta}_n}$, each lowered undotted
$\alpha$-index transforms separately according to 
${M_{\alpha_i'}}^{\alpha_i}$ in 
eq.~(\ref{lorentzmatrix}) and each lowered dotted $\dot\beta$-index
transforms according to ${(M^*)_{\dot\beta_i'}}^{\dot\beta_i}$.

There are two additional spin-1/2 irreducible representations of the
Lorentz group, $(M^{-1})^{\T}$ and $(M^{-1})^\dagger$, but these are
equivalent representations to the $(\half,0)$ and the $(0,\half)$
representations, respectively.  The spinors that transform under these
representations have raised spinor indices,
$\psi^\alpha$ and $\psi^{\dagger\dot{\alpha}}$, with transformation
laws $\psi^\alpha\to[(M^{-1})^{\T}]^\alpha{}_\beta\psi^\beta$ and
$\psi^{\dagger\dot{\alpha}}\to[(M^{-1})^\dagger]^{\dot{\alpha}}{}_{\dot{\beta}}
\psi^{\dagger\dot{\beta}}$, respectively
(with coordinate arguments of the fields again suppressed).  
It is convenient to rewrite the 
transformation law for the undotted spinor 
as $\psi^\alpha\to\psi^\beta (M^{-1})_\beta{}\rsup\alpha$.
In combining spinors to make Lorentz tensors [as in \eq{xisogeta}], it is
useful to regard $\psi^\alpha$ as a row vector, and
$\psi^{\dagger\,\dot\alpha}$ as a column vector, with:
\beq \label{eq:defbardagger2}
\psi^{\dagger\,\dot\alpha}\equiv(\psi^\alpha)^\dagger\,.
\eeq
The Lorentz transformation property of
$\psi^{\dagger\,\dot{\alpha}}$ then follows from
$(\psi^\alpha)^\dagger \to [(M^{-1})^\dagger]^{\dot{\alpha}}{}_{\dot{\beta}}
(\psi^\beta)^\dagger$. 

The spinor indices are raised and lowered with the two-index antisymmetric
epsilon symbol with non-zero components,\footnote{%
For related earlier work on the epsilon~symbol and its properties,
see refs.~\cite{vdWaerden2,Uhlenbeck,Bade-Jehle,epsilon-rest}.
Various subsets of the subsequent identities in this
section involving commuting and anticommuting two-component spinors,
as well as the $\eps$ symbol and the sigma matrices appear in
many books and reviews
(e.g., see refs.~\cite{vdWaerden3,WessBagger,Gates,Ross,
Srivastava,Sohnius,Piguet,MullerKirsten,Derendinger,West,
Bailin,Buchbinder,Lykken,Soni,Galperin,bilalsusy,FigueroaO'Farrill:2001tr,
Drees,Binetruy,Aitchison,Terning}) and in papers
(e.g., see refs.~\cite{Farrar:1983wk,
Kersch:1985rn,Hagiwara:1985yu,Berends:1987cv,Giele,Dittmaier:1993jj,
Dittmaier:1998nn}).  
}   
\beq \label{epssign}
\eps^{12} = - \eps^{21} = \eps_{21} =
-\epsilon_{12} = 1\,,
\eeq
and the same set of sign conventions for the
corresponding dotted spinor indices.  In particular, we formally define
$\epsilon^{\dot\alpha\dot\beta}\equiv(\epsilon^{\alpha\beta})^*$ and
$\epsilon_{\dot\alpha\dot\beta}\equiv(\epsilon_{\alpha\beta})^*$.
Viewed as a $2\times 2$ matrix, the epsilon symbol with lowered
undotted [dotted] indices is the inverse of the epsilon symbol with
raised undotted [dotted] indices.
Thus, consistent with \eqs{eq:defbardagger}{eq:defbardagger2},
one can write:\footnote{In the general relativity 
literature (see e.g., 
refs.~\cite{PenroseGR,Pirani,MTW,Penrose,Wald,Stewart,CarmeliSpin,ODonnell,stephani,PK}), 
the more common convention for the epsilon symbol (also adopted 
in refs.~\cite{Bade-Jehle,Corson,Rzewuski,Cbook,Rao,naber,Hurley,Dittmaier:1998nn}) 
is $\epsilon^{\alpha\beta}=
\epsilon_{\alpha\beta}$ with $\epsilon^{12}=-\epsilon^{21}=1$, and
similarly for the epsilon symbol with dotted spinor indices.
In this 
convention, one writes
$\psi^\alpha=\epsilon^{\alpha\beta}\psi_\beta$ as above,
but in contrast to \eq{epsalphabeta}, 
$\psi_\alpha=\psi^\beta\epsilon_{\beta\alpha}$,
and similarly for the corresponding equations with dotted spinor indices.  
That is, in raising [lowering] an
index of a spinor quantity, \textit{adjacent} spinor indices are
summed over when multiplied on the left [right] by the
appropriate epsilon symbol.
The various identities
involving the epsilon symbols given in this review
must then be modified by a minus sign for
every epsilon symbol with lowered spinor indices.   
There are some benefits for this alternative convention; e.g.,
the minus signs appearing in \eq{extrasign2} are absent.
However, one must keep track of other minus signs that arise because
$\epsilon_{\alpha\beta}$ is the \textit{negative} of the inverse of 
$\epsilon^{\alpha\beta}$.  
In this review, we have adopted the convention of \eq{epssign},
which is consistent with most of the supersymmetry literature.%
\label{fnepsconv}}\footnote{In \refs{Siegel}{Gates}, 
one finds yet another convention 
in which the spinor indices are
raised and lowered by a two-index antisymmetric quantity,
$C_{\alpha\beta}=-C^{\alpha\beta}=C_{\dot\alpha\dot\beta}
=-C^{\dot\alpha\dot\beta}=\left(\begin{smallmatrix} 0 & -i\\
i & \phm 0\end{smallmatrix}\right)$, which play the role of the
epsilon symbols.  As in footnote \ref{fnepsconv},
$C_{\alpha\beta}$ is the negative inverse of $C^{\alpha\beta}$ in
which case $\psi^\alpha=C^{\alpha\beta}\psi_\beta$ whereas
$\psi_\alpha=\psi^\beta C_{\beta\alpha}$, and similarly for the
corresponding equations 
with dotted spinor indices.  However, in this convention
where $C$ is pure imaginary,
if $\psi^{\dagger\,\dot\alpha}\equiv(\psi^\alpha)^\dagger$ as in
\eq{eq:defbardagger2}, then
$\psi^\dagger_{\dot\alpha}=
-(\psi_\alpha)^\dagger$ in contrast to \eq{eq:defbardagger}.
We choose not to pursue the alternative epsilon symbol conventions
of footnotes~\ref{fnepsconv} or \ref{fnsiegel} in this review.
\label{fnsiegel}}
\beq \label{epsalphabeta}
\psi_\alpha =
\epsilon_{\alpha\beta} \psi^\beta\,, \qquad
\psi^\alpha =\epsilon^{\alpha\beta} \psi_\beta\,, \qquad
\psi^\dagger_{\dot{\alpha}} = \epsilon_{\dot{\alpha}\dot{\beta}}
\psi^{\dagger\dot{\beta}}\,, \qquad
\psi^{\dagger\dot{\alpha}} = \epsilon^{\dot{\alpha}\dot{\beta}}
\psi^\dagger_{\dot{\beta}}\,,
\eeq
which respects Lorentz covariance 
due to the properties of $M$ given in \eqs{covar1}{covar2}. 
The epsilon~symbols $\eps^{\alpha\beta}$ ($\eps_{\alpha\beta}$) and
$\eps^{\dot{\alpha}\dot{\beta}}$ ($\eps_{\dot{\alpha}\dot{\beta}}$),
first introduced in this context in \Ref{vdWaerden1}, are also called the
\textit{spinor metric tensors}, as they
raise (lower) the undotted and dotted spinor indices, 
respectively. Note that in raising or lowering an
index of a spinor quantity, \textit{adjacent} spinor indices are
summed over when multiplied on the \textit{left} by the
appropriate epsilon symbol.

The epsilon symbols can also be used to raise or lower undotted or dotted
indices of spinors of higher rank.  For example, for an object with two
undotted indices it is natural to define
\beqa
A^{\gamma\delta} = \epsilon^{\gamma\alpha}\epsilon^{\delta\beta}
A_{\alpha\beta}\,,\qquad\qquad\qquad
A_{\gamma\delta}= \epsilon_{\gamma\alpha}\epsilon_{\delta\beta}
A^{\alpha\beta}\,.\label{extrasign1}
\eeqa
In the special
case that $A^{\alpha\beta} = \psi^\alpha \chi^\beta$ is a product of
rank-one spinors, 
eq.~(\ref{extrasign1}) is not just natural but necessary, 
as it follows directly from \eq{epsalphabeta}.  However,
in other cases there can be a different sign associated (by convention)
with raising and lowering spinor indices, because of the antisymmetry of
the epsilon symbols (in contrast to the symmetry of the spacetime metric
used to raise and lower spacetime indices). This sign convention can be
defined independently for distinct higher-rank spinors (even in the
case where the 
higher-rank spinors possess the same index structure). Indeed, as a
consequence of our epsilon symbol conventions 
of \eq{epssign}, the epsilon symbols themselves satisfy:
\beqa
&&\epsilon^{\gamma\delta}=-\epsilon^{\gamma\alpha}\epsilon^{\delta\beta}
\epsilon_{\alpha\beta}\,,\qquad\qquad\qquad
\epsilon_{\gamma\delta}=-\epsilon_{\gamma\alpha}\epsilon_{\delta\beta}
\epsilon^{\alpha\beta}\,,\label{extrasign2}
\eeqa
in contrast to \eq{extrasign1}.
The above results (and similar ones with dotted indices) show that some
care is required~\cite{Aharoni}, since the extra overall minus signs of
eq.~(\ref{extrasign2}) in comparison to eq.~(\ref{extrasign1}) might 
otherwise have been unexpected 
[e.g., see \eqs{psipsi}{psidpsid} below].\footnote{It would be 
perhaps more transparent to simply
replace the symbol $\epsilon_{\alpha\beta}$ with 
$\epsilon^{-1}_{\alpha\beta}$, in which case $\epsilon^{\alpha\beta}$
is used to raise spinor indices and $\epsilon^{-1}_{\alpha\beta}$ is
used to lower spinor indices (cf.~\Ref{Novozhilov}).  
Although this convention avoids an apparent conflict
between \eqs{extrasign1}{extrasign2}, it doubles the number of
distinct epsilon symbols.  We shall not adopt such an approach
in this review.\label{fntransparent}}
This reflects an awkwardness imposed by the 
epsilon symbol conventions of \eq{epssign}, 
rather than an inconsistency.
Practitioners of spinor algebra in the conventions used
in this review should be wary of this sign issue
when using the epsilon symbols to explicitly raise or lower two or more spinor
indices of higher-rank spinors.\footnote{In 
the alternative convention mentioned in
footnote~\ref{fnepsconv}, this particular awkwardness is absent; 
the minus signs in the analogue of eq.~(\ref{extrasign2}) do not occur, 
in which case the rules
for raising and lowering the spinor indices in
\eqs{extrasign1}{extrasign2} are identical. More generally, in the
convention of footnote~\ref{fnepsconv}, the indices of all higher-rank
spinors can be raised [lowered] 
via multiplication on the left [right] by the appropriate
epsilon symbol, including the epsilon
symbols themselves, with no extra signs.}  
Fortunately, such manipulations are quite rare in practical calculations.

We also introduce the two-index symmetric Kronecker delta symbol,
\beq \label{deltaK}
\delta^{1}_1=\delta^2_2=1\,,\qquad\qquad \delta^1_2=\delta_2^1=0\,,
\eeq
and $\delta_{\dot\alpha}^{\dot\beta}\equiv(\delta_\alpha^\beta)^*$.
\Eq{deltaK} implies that the numerical values of the
undotted and dotted Kronecker delta symbols coincide. 
The epsilon symbols with undotted and with dotted indices
respectively satisfy:
\beq \label{epsdelta} 
\epsilon_{\alpha\beta} \epsilon^{\gamma\delta} =
-\delta_\alpha^\gamma \delta_\beta^\delta +\delta_\alpha^\delta
\delta_\beta^\gamma ,\qquad\qquad \epsilon_{\dot{\alpha}\dot{\beta}}
\epsilon^{\dot{\gamma}\dot{\delta}} =
-\delta_{\dot{\alpha}}^{\dot{\gamma}}\delta_{\dot{\beta}}^{\dot{\delta}}
+\delta_{\dot{\alpha}}^{\dot{\delta}}\delta_{\dot{\beta}}^{\dot{\gamma}}
\,,
\eeq
from which it follows that:
\beqa
&&\eps_{\alpha\beta}
\eps^{\beta\gamma} = \eps^{\gamma\beta}\eps_{\beta\alpha} =
\delta_\alpha^\gamma ,\qquad\qquad\qquad\quad\,\,
\epsilon_{\dot{\alpha}\dot{\beta}}
\eps^{\dot{\beta}\dot{\gamma}} =
\eps^{\dot{\gamma}\dot{\beta}}\eps_{\dot{\beta}\dot{\alpha}} =
\delta_{\dot{\alpha}}^{\dot{\gamma}}\,,\\
&&\eps_{\alpha\beta}\eps_{\gamma\delta}+\eps_{\alpha\gamma}\eps_{\delta\beta}
+\eps_{\alpha\delta}\eps_{\beta\gamma}=0\,, \qquad\qquad
\eps_{\dot{\alpha}\dot{\beta}}\eps_{\dot{\gamma}\dot{\delta}}
+\eps_{\dot{\alpha}\dot{\gamma}}\eps_{\dot{\delta}\dot{\beta}}
+\eps_{\dot{\alpha}\dot{\delta}}\eps_{\dot{\beta}\dot{\gamma}}=0\,.
\label{schouten}
\eeqa
In the literature, \eq{schouten} is often referred to as
the Schouten identities.\footnote{%
The Schouten identities also follow from the observation that a 
rank-four spinor must vanish if it is antisymmetric
with respect to more than two undotted 
or dotted two-component spinor indices.}

To construct Lorentz invariant Lagrangians and observables, one needs
to first combine products of spinors to make objects that transform as
Lorentz tensors.  In particular, Lorentz vectors are obtained by
introducing the sigma matrices $\sigma^\mu_{\alpha{\dot{\beta}}}$ and
$\sigmabar^{\mu\,\dot{\alpha}\beta}$ defined by \cite{Weyl-book,
vdWaerden1,vdWaerden2,vdWaerden3}
\beqa
&&\sigma^0 = \sigmabar^0 = \begin{pmatrix} 1\quad &0\\[3pt]
0\quad & 1\end{pmatrix}\,,\qquad
\>\>\>\>\>\>\>\>
\sigma^1 = -\sigmabar^1 = \begin{pmatrix}
0\quad &1\\[3pt] 1\quad &0\end{pmatrix}\,,
\nonumber\\[6pt]
&&
\sigma^2 = -\sigmabar^2 = \begin{pmatrix} 0\quad &\!\!-i
\\[3pt] i\quad &\!\!\phm 0\end{pmatrix}\,,\qquad
\>\>\>
\sigma^3 = -\sigmabar^3 =
\begin{pmatrix} 1\quad &\!\!\phm 0\\[3pt] 0\quad
&\!\!-1\end{pmatrix}
\,.
\label{pauli}
\eeqa
The sigma matrices are hermitian, and have been defined 
above with an upper (contravariant)
index.  We denote the 
$2\times 2$ identity matrix by $\mathds{1}_{2\times 2}$ and 
the three-vector of Pauli matrices by
$\boldsymbol{\vec\sigma}\equiv (\sigma^1\,,\,\sigma^2\,,\,
\sigma^3)$.  Hence, \eq{pauli} is equivalent to:
\beq
\sigma^\mu=(\mathds{1}_{2\times 2}\,;
\,\mathbold{\vec\sigma})
\,,\qquad\qquad\qquad
\sigmabar^\mu=(\mathds{1}_{2\times  2}\,;\,-\mathbold{\vec\sigma})\,.
\eeq
We also define the corresponding quantities with 
lower (covariant) indices:
\beq
\sigma_\mu=g_{\mu\nu}\sigma^\nu=
(\BDpos \mathds{1}_{2\times 2}\,;\,
\BDneg \mathbold{\vec\sigma})
\,,\qquad\qquad
\sigmabar_\mu=g_{\mu\nu}\sigmabar^\nu=
(\BDpos \mathds{1}_{2\times 2}\,;\,
 \BDpos \mathbold{\vec\sigma})\,.
\eeq
The relations between $\sigma^\mu$ and $\sigmabar^\mu$ are
\beqa
\sigma^\mu_{\alpha{\dot{\alpha}}} = \epsilon_{\alpha\beta}
\epsilon_{\dot{\alpha}\dot{\beta}} \sigmabar^{\mu\,\dot{\beta}\beta}
\,, \qquad\qquad\qquad
\sigmabar^{\mu\,\dot{\alpha}\alpha} = \epsilon^{\alpha\beta}
\epsilon^{\dot{\alpha}\dot{\beta}} \sigma^{\mu}_{\beta\dot{\beta}}\,,
\label{sigsig1}
\\
\epsilon^{\alpha\beta} \sigma^\mu_{\beta\dot{\alpha}} =
\epsilon_{\dot{\alpha}\dot{\beta}}\sigmabar^{\mu\dot{\beta}\alpha}
\,, \qquad\qquad\qquad
\epsilon^{\dot{\alpha}\dot{\beta}} \sigma^\mu_{\alpha\dot{\beta}} =
\epsilon_{{\alpha}{\beta}}\sigmabar^{\mu\dot{\alpha}\beta}
\> .\label{sigsig2}
\eeqa

Consider a spinor of rank $(n,n)$ denoted by
$S_{\alpha_1\alpha_2\ldots\alpha_n\dot\beta_1\dot\beta_2\ldots\dot\beta_n}$.
The object obtained by multiplying $S$ by
$\sigmabar^{\mu_1\,\dot{\beta}_1\alpha_1}
\cdots\sigmabar^{\mu_n\,\dot{\beta}_n\alpha_n}$ has the transformation
properties of an $n$th rank contravariant 
Lorentz tensor~\cite{Umezawa,Gelfand,Massa}.  For example,
there is a one-to-one
correspondence between each bi-spinor $V_{\alpha\dot{\beta}}$ and the
associated Lorentz four-vector 
$V^\mu$~\cite{vdWaerden1,vdWaerden2,Bade-Jehle,cap,Corson,Umezawa},\footnote{In the
general relativity literature~\cite{Penrose,Stewart,ODonnell,Hurley}, 
the more common normalization is
$V^\mu\equiv \frac{1}{\sqrt{2}}\sigmabar^{\mu\dot\beta\alpha}
V_{\alpha\dot{\beta}}$, which yields
$V_{\alpha\dot{\beta}}=\BDpos \frac{1}{\sqrt{2}}
V^\mu\sigma_{\mu\alpha\dot{\beta}}$.  In this context, the  
$\frac{1}{\sqrt{2}}\sigma^\mu_{\alpha\dot{\beta}}$ 
are often called the Infeld-van der Waerden
symbols.} 
\beq \label{spinorvector}
V^\mu\equiv \half\sigmabar^{\mu\dot\beta\alpha}
V_{\alpha\dot{\beta}}
\,,\qquad\qquad
 V_{\alpha\dot{\beta}}=\BDpos 
V^\mu\sigma_{\mu\alpha\dot{\beta}}\,.
\eeq
In particular, if $V^\mu$ is a real four-vector then $V_{\alpha\dot\beta}$ is 
hermitian (and vice versa).  To clarify this last remark, 
consider the bi-spinor $V_{\alpha\dot\beta}$ regarded as a $2\times 2$
matrix.  Then,\footnote{As stressed in \Ref{Bade-Jehle}, taking 
the transpose of $V_{\alpha\dot\beta}$ interchanges its rows and columns 
without altering the fact that the first spinor index 
is undotted and the second spinor index is dotted.  Moreover, it is often
useful to further
simplify the notation by \textit{defining} $V\ls{\dot\alpha\beta}\equiv
(V_{\alpha\dot\beta})^*$.  In this notation, an hermitian
bi-spinor satisfies 
$V_{\alpha\dot\beta}=V\ls{\dot\beta\alpha}$.}
\beq \label{Vmu}
(V^{\T})_{\alpha\dot\beta}\equiv V_{\beta\dot\alpha}\,,
\qquad\quad
(V^{*})\ls{\alpha\dot\beta}\equiv(V_{\alpha\dot\beta})^*\,,
\qquad\quad 
(V^{\dagger})_{\alpha\dot\beta}\equiv (V\ls{\beta\dot\alpha})^*\,.
\eeq
An hermitian bi-spinor satisfies $V=V^\dagger$, or equivalently
$V_{\alpha\dot\beta}=(V_{\beta\dot\alpha})^*$.  

Rank-two spinors (with two undotted or with two dotted indices) can
also be interpreted as $2\times 2$ matrices.  
In the case of the rank-two spinor
$W_\alpha{}^\beta$, it is convenient to define:
\beq
(W^{\T})^\beta{}_\alpha\equiv W_\alpha{}^\beta\,,
\qquad\quad
(W^*)_{\dot\alpha}{}^{\dot\beta}\equiv (W_\alpha{}^\beta)^*\,,
\qquad\quad
(W^\dagger)^{\dot\beta}{}_{\dot\alpha}\equiv (W_\alpha{}^\beta)^*=
(W^*)_{\dot\alpha}{}^{\dot\beta}\,.
\eeq
Note that the matrix transposition of $W_\alpha{}^\beta$ interchanges 
the rows and columns of $W$ 
without altering the relative heights of the $\alpha$
and $\beta$ indices.  Similar results hold for $W_{\alpha\beta}$ and
$W^{\alpha\beta}$ by either lowering or raising the relevant spinor indices
with the appropriate epsilon symbol.

When constructing Lorentz tensors from fermion fields, the heights of
spinor indices must be consistent in the sense that lowered
indices must only be contracted with raised indices.
As a convention, \textit{descending} contracted undotted indices and 
\textit{ascending} contracted dotted indices, 
\beq
{}^\alpha{}_\alpha\qquad\qquad {\rm and} \qquad\qquad
{}_{\dot{\alpha}}{}^{\dot{\alpha}}\> ,
\label{suppressionrule}
\eeq
can be suppressed. In all spinor products given in this paper,
contracted indices always have heights that conform to
eq.~(\ref{suppressionrule}).  For example, in an index-free notation,
we define:
\beqa
\xi\eta &\equiv & \xi^\alpha\eta_\alpha ,\label{xieta}
\\
\xi^\dagger \eta^\dagger &\equiv &
\xi^\dagger_{\dot\alpha} \eta^{\dagger \dot \alpha}
,\label{xidetad}
\\
\xi^\dagger\sigmabar^\mu\eta &\equiv &  \xi^\dagger_{\dot{\alpha}}
\sigmabar^{\mu\dot{\alpha}\beta}\eta_\beta ,\label{xisbareta}
\\
\xi\sigma^\mu \eta^\dagger &\equiv & \xi^{{\alpha}}
\sigma^{\mu}_{\alpha \dot \beta} \eta^{\dagger\dot \beta} . \label{xisogeta}
\eeqa
All the spinor-index-contracted products above have natural interpretations as
products of matrices and vectors by regarding $\eta_\alpha$ 
and $\eta^{\dagger\dot\alpha}$ as 
column vectors and $\xi^\dagger_{\dot\alpha}$ and $\xi^\alpha$ 
as row vectors of the two-dimensional spinor space.  However, 
the reader is cautioned that 
in the index-free notation
(with undotted and dotted indices suppressed), 
the undaggered and daggered spinors cannot be uniquely
identified as column or row vectors until their locations within the
spinor product are specified.  Nevertheless, the proper identifications
are straightforward, as any spinor on the left end of a spinor product
can be identified as a row vector and any spinor on the right end of a
spinor product can be identified as a column vector.

For an \textit{anticommuting} two-component spinor $\psi$, the product 
$\psi^\alpha\psi^\beta$ is antisymmetric with respect to the
interchange of the spinor indices $\alpha$ and $\beta$.  Hence, this product
of spinors
must be proportional to $\epsilon^{\alpha\beta}$.  Similar conclusions
hold for the corresponding spinor products with raised undotted indices
and with lowered and raised dotted indices, respectively.  Thus,
\beqa
\psi^\alpha\psi^\beta&=&-\half\epsilon^{\alpha\beta}\psi\psi\,,
\qquad\qquad
\psi_\alpha\psi_\beta=\half\epsilon_{\alpha\beta}\psi\psi\,,\label{psipsi}\\
\psi^{\dagger\,\dot\alpha}\psi^{\dagger\,\dot\beta}&=&
\half\epsilon^{\dot\alpha\dot\beta}\psi^\dagger\psi^\dagger\,,
\qquad\qquad
\psi^\dagger_{\dot\alpha}\psi^\dagger_{\dot\beta}
=-\half\epsilon_{\dot\alpha\dot\beta}\psi^\dagger\psi^\dagger\,,
\label{psidpsid}
\eeqa
where $\psi\psi\equiv\psi^\alpha\psi_\alpha$ and
$\psi^\dagger\psi^\dagger\equiv
\psi^\dagger_{\dot\alpha}\psi^{\dagger\,\dot\alpha}$
as in \eqs{xieta}{xidetad}.
Note that the minus signs above can be understood to be a consequence of
the extra minus sign that arises when the 
indices of the epsilon symbol are lowered or raised
[cf.~\eqs{extrasign1}{extrasign2}].

The behavior of the spinor products under hermitian
conjugation (for quantum field operators) or complex conjugation (for
classical fields) is as follows:
\beqa
&&(\xi\eta)^\dagger=\eta^\dagger \xi^\dagger\,,
\label{eq:conbil} 
\\
&&(\xi\sigma^\mu\eta^\dagger)^\dagger=\eta\sigma^\mu\xi^\dagger ,
\label{eq:conbilsig}
\\
&&(\xi^\dagger \sigmabar^\mu \eta)^\dagger = \eta^\dagger \sigmabar^\mu \xi ,
\label{eq:conbilsigbar}
\\
&&(\xi\sigma^\mu\sigmabar^\nu\eta)^\dagger
=
\eta^\dagger \sigmabar^\nu\sigma^\mu \xi^\dagger\,,
\eeqa
where we have used the hermiticity properties, $(\sigma^\mu)^\dagger=
\sigma^\mu$ and $(\sigmabar^\mu)^\dagger=
\sigmabar^{\mu}$.
More generally,
\beq \label{eq:conbilgen}
(\xi \Sigma \eta)^\dagger = \eta^\dagger \reversed{\Sigma} \xi^\dagger\,,
\qquad\quad
(\xi \Sigma \eta^\dagger)^\dagger = \eta \reversed{\Sigma} \xi^\dagger\,,
\qquad\quad (\xi^\dagger  \Sigma \eta)^\dagger = \eta^\dagger \reversed{\Sigma}
\xi\,,
\eeq
where in each case $\Sigma$ stands for any sequence of alternating
$\sigma$ and $\sigmabar$ matrices, and $\reversed{\Sigma}$
is obtained from $\Sigma$ by reversing the order of all of the
$\sigma$ and $\sigmabar$ matrices, since the sigma matrices are hermitian.
\Eqst{eq:conbil}{eq:conbilgen}
are applicable both to anticommuting and to
commuting spinors.

The properties of the two-component spinor fields under the discrete
C, P and T transformations are elucidated in \refs{LifshitzII}{duret}.  The
corresponding behaviors of the spinor products under C, P and T are
easily obtained (and are left as an exercise for the reader).

The following identities can be used to systematically simplify
expressions involving products of $\sigma$ and $\sigmabar$
matrices:\footnote{Since the Kronecker delta symbol is symmetric under
the interchange of its two indices, naively there is nothing gained in writing
$\delta_\alpha{}^\beta$ and  $\delta^{\dot\beta}{}_{\dot\alpha}$, with
the spinor indices staggered as shown,
instead of $\delta_\alpha^\beta$ and
$\delta^{\dot\beta}_{\dot\alpha}$, respectively.  
Nevertheless, we often prefer to employ the
former rather than the latter as it provides some insight into the spinor index
structure of the equation.  For example, in \eq{eq:ssbarsym},
$\alpha$ labels the row and $\beta$ labels the column of the
product of sigma matrices.  Neither $\sigma^\mu\sigmabar^\nu$ nor
$\sigma^\nu\sigmabar^\mu$ is symmetric under the interchange of the
(suppressed) spinor indices
(although the sum of the two
is symmetric).  By writing $\delta_\alpha{}^\beta$ on the right-hand
side of \eq{eq:ssbarsym}, 
one formally maintains the index structure of each of the separate terms
of the equation.}
\beqa
&&
\sigma^\mu_{\alpha\dot{\alpha}} \sigmabar_\mu^{\dot{\beta}\beta}
= \BDpos 2
\delta_{\alpha}{}^{\beta} \delta^{\dot{\beta}}{}_{\dot{\alpha}}\,,
\label{mainfierz}
\\
&&
\sigma^\mu_{\alpha\dot{\alpha}} \sigma_{\mu\beta\dot{\beta}}
= \BDpos 2
\epsilon_{\alpha\beta} \epsilon_{\dot{\alpha}\dot{\beta}}\,,
\label{mainfierz2}
\\
&&
\sigmabar^{\mu\dot{\alpha}\alpha} \sigmabar_\mu^{\dot{\beta}\beta}
= \BDpos 2
\epsilon^{\alpha\beta} \epsilon^{\dot{\alpha}\dot{\beta}}\,,
\label{mainfierz3}
\\
&&
{[\sigma^\mu\sigmabar^\nu + \sigma^\nu \sigmabar^\mu ]_\alpha}^\beta
= \BDpos 2\metric^{\mu\nu} \delta_{\alpha}{}^{\beta}\,,
\label{eq:ssbarsym}
\\
&&[\sigmabar^\mu\sigma^\nu + \sigmabar^\nu \sigma^\mu
]^{\dot{\alpha}}{}_{\dot{\beta}}
= \BDpos 2\metric^{\mu\nu} \delta^{\dot{\alpha}}{}_{\dot{\beta}}\,,
\label{eq:sbarssym}
\eeqa
In the literature, one sometimes sees
\eqs{mainfierz2}{mainfierz3} rewritten using the identity
$\epsilon_{ab}\epsilon_{cd}=\delta_{ac}\delta_{bd}-\delta_{ad}
\delta_{bc}$.  However, as
this latter result does not formally respect
covariance with respect to the dotted and undotted indices,
we shall not make use of it here.   Products of three or more sigma matrices can be reduced to sums of terms involving at most two sigma matrices by employing the identities,
\beqa
&& \sigma^\mu \sigmabar^\nu \sigma^\rho =
\BDpos \metric^{\mu\nu} \sigma^\rho
\BDminus \metric^{\mu\rho} \sigma^\nu
\BDplus \metric^{\nu\rho} \sigma^\mu
\BDplus i \epsilon^{\mu\nu\rho\kappa} \sigma_\kappa\,,
\label{eq:simplifyssbars}
\\
&& \sigmabar^\mu \sigma^\nu \sigmabar^\rho =
\BDpos \metric^{\mu\nu} \sigmabar^\rho
\BDminus \metric^{\mu\rho} \sigmabar^\nu
\BDplus \metric^{\nu\rho} \sigmabar^\mu
\BDminus i \epsilon^{\mu\nu\rho\kappa} \sigmabar_\kappa\,.
\label{eq:simplifysbarssbar}
\eeqa

Computations of cross-sections and decay rates generally require
traces of alternating products of $\sigma$ and $\sigmabar$ matrices
(e.g., see \Ref{Kersch:1985rn}):
\beqa
&&{\rm Tr}[\sigma^\mu \sigmabar^\nu ] =
{\rm Tr}[\sigmabar^\mu \sigma^\nu ] = \BDpos 2 \metric^{\mu\nu} \,,
\label{trssbar}
\\
&&{\rm Tr}[\sigma^\mu \sigmabar^\nu \sigma^\rho \sigmabar^\kappa ] =
2 \left ( \metric^{\mu\nu} \metric^{\rho\kappa} - \metric^{\mu\rho}
\metric^{\nu\kappa} + \metric^{\mu\kappa} \metric^{\nu\rho} + i
\epsilon^{\mu\nu\rho\kappa} \right )\,, \qquad\phantom{xx}
\label{trssbarssbar}
\\
&&{\rm Tr}[\sigmabar^\mu \sigma^\nu \sigmabar^\rho \sigma^\kappa ] =
2 \left ( \metric^{\mu\nu} \metric^{\rho\kappa} - \metric^{\mu\rho}
\metric^{\nu\kappa} + \metric^{\mu\kappa} \metric^{\nu\rho} - i
\epsilon^{\mu\nu\rho\kappa} \right )\,.
\label{trsbarssbars}
\eeqa
Traces involving a larger even number of $\sigma$ and $\sigmabar$
matrices can be systematically obtained from
\eqst{trssbar}{trsbarssbars} by repeated use of
\eqs{eq:ssbarsym}{eq:sbarssym} and the cyclic property of the trace.
Traces involving an odd number of $\sigma$ and $\sigmabar$ matrices
cannot arise, since there is no way to connect the spinor indices
consistently.

In addition to manipulating expressions containing anticommuting
fermion quantum fields, we often must deal with products of {\it commuting}
spinor wave functions that arise when evaluating the Feynman rules.
In the following expressions we denote the generic spinor by $z_i$.
In the various identities listed below, an extra minus sign arises
when interchanging the order of two anticommuting fermion
fields of a given spinor index height.
It is convenient to introduce the notation:
\beq
(-1)^A\equiv\left\{\begin{array}{ll} +1\,, & \textrm{commuting~spinors,}\\[6pt]
-1\,, & \textrm{anticommuting~spinors}.\end{array}\right.
\eeq
The following identities hold for the $z_i$:
\beqa
&&z_1 z_2 = -(-1)^A z_2 z_1\,, \label{zonetwo}
\\
&&z^\dagger_1 z^\dagger_2 = -(-1)^A z^\dagger_2 z^\dagger_1\,,
\label{barzonetwo}
\\
&&z_1 \sigma^\mu z^\dagger_2 = (-1)^A z^\dagger_2 \sigmabar^\mu z_1\,,
\label{europeanvacation} \\
&&z_1 \sigma^\mu \sigmabar^\nu z_2 =
-(-1)^A z_2 \sigma^\nu \sigmabar^\mu z_1\,,
 \label{zsmunuz}\\
&&z^\dagger_1 \sigmabar^\mu \sigma^\nu z^\dagger_2 =
-(-1)^A z^\dagger_2 \sigmabar^\nu \sigma^\mu z^\dagger_1\,,
 \label{zsbarmunuz}\\
&&z^\dagger_1 \sigmabar^\mu \sigma^\rho \sigmabar^\nu z_2=(-1)^A
z_2 \sigma^\nu \sigmabar^\rho \sigma^\mu z^\dagger_1\,,\label{zsssmunuz}
\eeqa
and so on.\footnote{In particular,
if $z$ is a \textit{commuting} spinor, then 
$zz=z^\dagger z^\dagger=0$, as emphasized in \refs{Uhlenbeck}{Bade-Jehle}.}
The hermiticity properties of the spinor products given
in \eqst{eq:conbil}{eq:conbilgen} hold for both commuting
and anticommuting spinors, with no additional sign factor.

Two-component spinor products can often be simplified by using
Fierz identities.
Due to the antisymmetry of the suppressed two-index
epsilon symbol 
[or equivalently, using the Schouten identities given in \eq{schouten}],
the following identities are obtained:
\beqa
(z_1 z_2)(z_3 z_4) &=& -(z_1 z_3) (z_4 z_2) - (z_1 z_4)(z_2 z_3)\,,
\label{eq:twocompfierzone}
\\
(z^\dagger_1 z^\dagger_2)(z^\dagger_3 z^\dagger_4) &=&
- (z^\dagger_1 z^\dagger_3) (z^\dagger_4 z^\dagger_2)
- (z^\dagger_1 z^\dagger_4) (z^\dagger_2 z^\dagger_3)\,,
\label{eq:twocompfierztwo}
\eeqa
where we have used \eqs{zonetwo}{barzonetwo} to eliminate
any residual factors of $(-1)^A$. Similarly,
\eqst{mainfierz}{mainfierz3} can be used to derive
additional Fierz identities,
\beqa
(z_1 \sigma^\mu z^\dagger_2)(z^\dagger_3 \sigmabar_\mu z_4)
&=& \BDneg 2 (z_1 z_4) (z^\dagger_2 z^\dagger_3)\,,
\label{twocompfierza} \\
(z^\dagger_1 \sigmabar^\mu z_2)(z^\dagger_3 \sigmabar_\mu z_4)
&=&\BDpos 2 (z^\dagger_1 z^\dagger_3) (z_4 z_2)\,,
\label{twocompfierzb}
\\[5pt]
(z_1 \sigma^\mu z^\dagger_2)(z_3 \sigma_\mu z^\dagger_4)
&=& \BDpos 2 (z_1 z_3) (z^\dagger_4 z^\dagger_2)\,.
\label{twocompfierzc}
\eeqa
Having eliminated all factors of $(-1)^A$, 
\eqst{eq:twocompfierzone}{twocompfierzc}
hold for both commuting and anticommuting spinors.

{}From the sigma matrices, one can construct the antisymmetrized products:\footnote{The
reader is cautioned that $\sigma^{\mu\nu}$ and $\sigmabar^{\mu\nu}$ are sometimes
defined in the literature without the factor of $i$ 
in \eqs{sigmamunu}{sigmabarmunu} (as in \Ref{Bailin}), or with an overall factor
of $\half i$ (as in \Ref{Sohnius}) 
instead of $\frac{1}{4} i$.}
\beqa
(\sigma^{\mu\nu})_\alpha{}^\beta
&\equiv&
\frac{i}{4} \left(\sigma^\mu{}_{\!\!\!\!\alpha\dot{\gamma}}
\sigmabar^{\nu\dot{\gamma}\beta}-\sigma^\nu{}_{\!\!\!\!\alpha\dot{\gamma}}
\sigmabar^{\mu\dot{\gamma}\beta}\right)\,,
\label{sigmamunu}
\\[5pt]
(\sigmabar^{\mu\nu})^{\dot{\alpha}}{}_{\dot{\beta}}
&\equiv&
\frac{i}{4}\left(\sigmabar^\mu{}^{\dot{\alpha}\gamma}
\sigma^\nu{}_{\!\!\!\!\gamma\dot{\beta}}-\sigmabar^\nu{}^{\dot{\alpha}\gamma}
\sigma^\mu{}_{\!\!\!\!\gamma\dot{\beta}}\right)\,.
\label{sigmabarmunu}
\eeqa
Equivalently, we can use \eqs{eq:ssbarsym}{eq:sbarssym} to write:
\beqa
(\sigma^\mu\sigmabar^\nu)_\alpha{}^\beta &=&  \BDpos
g^{\mu\nu}\delta_\alpha{}^\beta
-2i(\sigma^{\mu\nu})_\alpha{}^\beta\,,\label{smunualt}\\
(\sigmabar^\mu\sigma^\nu)^{\dot{\alpha}}{}_{\dot{\beta}}
&=&  \BDpos g^{\mu\nu}\delta^{\dot{\alpha}}{}_{\dot{\beta}}
-2i(\sigmabar^{\mu\nu})^{\dot{\alpha}}{}_{\dot{\beta}}\,.\label{sbarmunualt}
\eeqa
The components of $\sigma^{\mu\nu}$ and $\sigmabar^{\mu\nu}$
are easily evaluated:
\beq
\sigma^{ij} = \sigmabar^{ij} = \half\eps^{ijk}\sigma^k\,,
\qquad\quad
\sigma^{i0} = -\sigma^{0i} = -\sigmabar^{i0} =
   \sigmabar^{0i} = \half i\sigma^i \,.
\eeq
The matrices $\sigma^{\mu\nu}$ and $ \sigmabar^{\mu\nu}$
satisfy self-duality relations,
\beq
\sigma^{\mu\nu}=
-\half i\epsilon^{\mu\nu\rho\kappa}\sigma_{\rho\kappa}\,,
\qquad\qquad \sigmabar^{\mu\nu}=
\half i\epsilon^{\mu\nu\rho\kappa}\sigmabar_{\rho\kappa}\,.
\label{eq:selfduality}
\eeq
The self-duality relations can be used to obtain the following
two identities:
\beqa
\metric^{\kappa\rho}\sigma^{\mu\nu}
-\metric^{\nu\rho}\sigma^{\mu\kappa}
+\metric^{\mu\rho}\sigma^{\nu\kappa}
-i\epsilon^{\mu\nu\kappa}{}_\lambda\sigma^{\lambda\rho}&=&0\,,
\label{sigmunueps1}\\
\metric^{\kappa\rho}\sigmabar^{\mu\nu}
-\metric^{\nu\rho}\sigmabar^{\mu\kappa}
+\metric^{\mu\rho}\sigmabar^{\nu\kappa}
+ i\epsilon^{\mu\nu\kappa}{}_\lambda\sigmabar^{\lambda\rho}&=&0
\,.\label{sigmunueps2}
\eeqa

A number of useful properties and identities involving
$\sigma^{\mu\nu}$ and $\sigmabar^{\mu\nu}$ can be derived.
For example, \eq{epsdelta} implies that: 
\beqa
(\sigma^{\mu\nu}){}_\alpha{}^\beta&=&
\epsilon_{\alpha\tau}\epsilon^{\beta\gamma}
(\sigma^{\mu\nu}){}_\gamma{}^\tau\,,\qquad\qquad
(\sigmabar^{\mu\nu}){}^{\dot{\alpha}}{}_{\dot{\beta}}}=
\epsilon^{\dot{\alpha}\dot{\tau}}\epsilon_{\dot{\beta}\dot{\gamma}}
(\sigmabar^{\mu\nu}){}^{\dot{\gamma}}{}_{\dot{\tau}\,,
\label{sigmunurel1}
\\
\epsilon^{\tau\alpha}(\sigma^{\mu\nu}){}_\alpha{}^\beta&=&
\epsilon^{\beta\gamma}(\sigma^{\mu\nu}){}_\gamma{}^\tau\,,\qquad\qquad
\epsilon_{\dot{\tau}\dot{\alpha}}
(\sigmabar^{\mu\nu}){}^{\dot{\alpha}}{}_{\dot{\beta}}=
\epsilon_{\dot{\beta}\dot{\gamma}}
(\sigmabar^{\mu\nu}){}^{\dot{\gamma}}{}_{\dot{\tau}}\,,
\label{sigmunurel2}\\
\epsilon_{\gamma\beta}(\sigma^{\mu\nu})_\alpha{}^\beta &=&
\epsilon_{\alpha\tau}(\sigma^{\mu\nu})_\gamma{}^\tau\,,\qquad\qquad
\epsilon^{\dot{\gamma}\dot{\beta}}
(\sigmabar^{\mu\nu}){}^{\dot{\alpha}}{}_{\dot{\beta}}=
\epsilon^{\dot{\alpha}\dot{\tau}}
(\sigmabar^{\mu\nu}){}^{\dot{\gamma}}{}_{\dot{\tau}}\,.
\label{sigmunurel3}
\eeqa
Using \eqst{mainfierz}{eq:simplifysbarssbar}, the following
identities can be obtained:
\beqa
&&\hspace{-0.4in}
(\sigma^{\mu\nu})_\alpha{}\rsup{\beta} 
(\sigma_{\mu\nu})\ls{\gamma}{}\rsup{\tau}=
\epsilon_{\alpha\gamma}\,\epsilon^{\beta\tau}
+\delta_\alpha{}\rsup{\tau}\delta\ls{\gamma}{}\rsup{\beta}=
2\delta_\alpha{}\rsup{\tau}\delta\ls{\gamma}{}\rsup{\beta}-
\delta_\alpha{}\rsup{\beta}\delta\ls{\gamma}{}\rsup{\tau}\,,\label{ssfierza}\\
&&\hspace{-0.4in}
(\sigmabar^{\mu\nu})^{\dot{\alpha}}{}_{\dot{\beta}} 
(\sigmabar_{\mu\nu})^{\dot{\gamma}}{}_{\dot{\tau}}=
\epsilon^{\dot{\alpha}\dot{\gamma}}\,\epsilon_{\dot{\beta}\dot{\tau}}
+\delta^{\dot{\alpha}}{}_{\dot{\tau}}\,\delta^{\dot{\gamma}}{}_{\dot{\beta}}=
2\delta^{\dot{\alpha}}{}_{\dot{\tau}}\,\delta^{\dot{\gamma}}{}_{\dot{\beta}}-
\delta^{\dot{\alpha}}{}_{\dot{\beta}}\,\delta^{\dot{\gamma}}{}_{\dot{\tau}}\,,
\label{ssfierzb}\\
&&\hspace{-0.4in}
(\sigma^{\mu\nu})_\alpha{}^\beta
(\sigmabar_{\mu\nu})^{\dot{\gamma}}{}_{\dot{\tau}}=0\,,
\label{ssfierzc}\\
&&\hspace{-0.4in} \sigma^{\mu\nu}\sigma^\rho = \BDpos\nicefrac{1}{2}i\left(
 \metric^{\nu\rho}\sigma^\mu
-\metric^{\mu\rho}\sigma^\nu
+ i\epsilon^{\mu\nu\rho\kappa}\sigma_\kappa\right)\,,
\label{sigmunuid1}
\\
&&\hspace{-0.4in} \sigmabar^{\mu\nu}\sigmabar^\rho = \BDpos\nicefrac{1}{2}i
\left(\metric^{\nu\rho}\sigmabar^\mu
-\metric^{\mu\rho}\sigmabar^\nu
-i\epsilon^{\mu\nu\rho\kappa}\sigmabar_\kappa\right)\,,
\label{sigmunuid2}
\\
&&\hspace{-0.4in} \sigmabar^\mu\sigma^{\nu\rho}=\BDpos\nicefrac{1}{2}i\left(
\metric^{\mu\nu}\sigmabar^\rho
-\metric^{\mu\rho}\sigmabar^\nu
-i\epsilon^{\mu\nu\rho\kappa}\sigmabar_\kappa\right)\,,
\label{sigmunuid3}
\\
&&\hspace{-0.4in} \sigma^\mu\sigmabar^{\nu\rho}=\BDpos\nicefrac{1}{2}i\left(
\metric^{\mu\nu}\sigma^\rho
- \metric^{\mu\rho}\sigma^\nu
+ i\epsilon^{\mu\nu\rho\kappa}\sigma_\kappa\right)\,,
\label{sigmunuid4} \\
&&\hspace{-0.4in} 
\sigma^{\mu\nu}\sigma^{\rho\kappa}=-\quarter\left(\metric^{\nu\rho}
\metric^{\mu\kappa}-\metric^{\mu\rho}\metric^{\nu\kappa}
+i\epsilon^{\mu\nu\rho\kappa}\right)
\BDplus \nicefrac{1}{2}i\left(\metric^{\nu\rho}\sigma^{\mu\kappa}
+\metric^{\mu\kappa}\sigma^{\nu\rho}-\metric^{\mu\rho}\sigma^{\nu\kappa}
-\metric^{\nu\kappa}\sigma^{\mu\rho}\right)\,,
\label{sigmunuid5} \\
&&\hspace{-0.4in} 
\sigmabar^{\mu\nu}\sigmabar^{\rho\kappa}=-\quarter\left(\metric^{\nu\rho}
\metric^{\mu\kappa}-\metric^{\mu\rho}\metric^{\nu\kappa}
-i\epsilon^{\mu\nu\rho\kappa}\right)
\BDplus \nicefrac{1}{2}i\left(\metric^{\nu\rho}\sigmabar^{\mu\kappa}
+\metric^{\mu\kappa}\sigmabar^{\nu\rho}-\metric^{\mu\rho}\sigmabar^{\nu\kappa}
-\metric^{\nu\kappa}\sigmabar^{\mu\rho}\right)\,.\label{sigmunuid6}
\eeqa
\Eqs{sigmunuid5}{sigmunuid6} and the antisymmetry of
$\sigma^{\mu\nu}$ and $\sigmabar^{\mu\nu}$ yield
the following trace formulae:
\beqa
{\rm Tr}~\sigma^{\mu\nu}&=&{\rm Tr}~\sigmabar^{\mu\nu}=0\,,\\
{\rm Tr}[\sigma^{\mu\nu}\sigma^{\rho\kappa}]&=&\half
\left[g^{\mu\rho}g^{\nu\kappa}
-g^{\mu\kappa}g^{\nu\rho}-i\epsilon^{\mu\nu\rho\kappa}\right]\,,\\
{\rm Tr}[\sigmabar^{\mu\nu}\sigmabar^{\rho\kappa}]&=&
\half\left[g^{\mu\rho}g^{\nu\kappa}
-g^{\mu\kappa}g^{\nu\rho}+i\epsilon^{\mu\nu\rho\kappa}\right]\,.
\eeqa

The properties of spinor products involving $\sigma^{\mu\nu}$ and
$\sigmabar^{\mu\nu}$ are easily derived.  Under hermitian
conjugation (for quantum field operators) or complex conjugation (for
classical fields),
\beq
(\xi\sigma^{\mu\nu}\eta)^\dagger=\eta^\dagger\sigmabar^{\mu\nu}\xi^\dagger\,,
\eeq
due to the hermiticity relation, 
$(\sigma^{\mu\nu})^\dagger=\sigmabar^{\mu\nu}$.
Next, we use \eqs{zsmunuz}{zsbarmunuz} to obtain:
\beqa
z_1\sigma^{\mu\nu}z_2=(-1)^A z_2\sigma^{\mu\nu}z_1\,,\label{zsigmunuz1}\\
z_1^\dagger\sigmabar^{\mu\nu}z_2^\dagger
=(-1)^A z_2^\dagger\sigmabar^{\mu\nu}z_1^\dagger\,.\label{zsigmunuz2}
\eeqa
One can also derive additional Fierz identities, which follow from 
\eqst{ssfierza}{ssfierzc},
\beqa
(z_1\sigma^{\mu\nu}z_2)(z_3\sigma_{\mu\nu}z_4)&=&-2(z_1 z_4)(z_2 z_3)
-(z_1 z_2)(z_3 z_4)\,,\label{sigsigfierza}\\
(z_1^\dagger\sigmabar^{\mu\nu}z_2^\dagger)
(z_3^\dagger\sigmabar_{\mu\nu}z_4^\dagger)&=&
-2(z_1^\dagger z_4^\dagger)(z_2^\dagger z_3^\dagger)
-(z_1^\dagger z_2^\dagger)(z_3^\dagger z_4^\dagger)\,,\label{sigsigfierzb}\\
(z_1\sigma^{\mu\nu}z_2)(z_3^\dagger\sigmabar_{\mu\nu}z_4^\dagger)&=&
0\,,\label{sigsigfierzc}
\eeqa
where we have again used \eqs{zonetwo}{barzonetwo} to eliminate
any residual factors of $(-1)^A$.  
Thus, \eqst{sigsigfierza}{sigsigfierzc}
hold for both commuting and anticommuting spinors.
A more comprehensive list of sigma matrix identities and their
associated Fierz identities are given in \app{B.1} (see also 
Appendix B of \Ref{Bailin}).

The $\sigma^{\mu\nu}$ and $\sigmabar^{\mu\nu}$ 
satisfy the commutation relations of 
the $J^{\mu\nu}$ [cf.~\eq{JJ}], and thus can be identified as 
the generators of the Lorentz group in the $(\half,0)$ and
$(0,\half)$ representations, respectively.  That is, for
the $(\half,0)$ representation
with a lowered undotted index (e.g. $\psi_\alpha$),
$J^{\mu\nu}=\sigma^{\mu\nu}$, while for the $(0,\half)$
representation with a raised dotted index (e.g.
$\psi^{\dagger\dot{\alpha}}$), $J^{\mu\nu}=\sigmabar^{\mu\nu}$.
In particular, the infinitesimal forms for
the $4\times 4$ Lorentz transformation matrix
$\Lambda$ [cf.~\eq{infinitesimalLT}] and the corresponding matrices $M$ and $(M^{-1})^\dagger$
that transform the $(\half,0)$ and
$(0,\half)$ spinors, respectively, are given by:
\beqa
\Lambda^\mu{}_\nu &\simeq&
\delta^\mu_\nu + \theta^\mu{}_\nu
\,,
\label{Linf}\\
M& \simeq& \mathds{1}_{2\times 2}
\BDminus \half i\theta_{\mu\nu}\sigma^{\mu\nu}\,, \label{Minf} \\
(M^{-1})^\dagger &\simeq & \mathds{1}_{2\times 2}
\BDminus \half i \theta_{\mu\nu}\sigmabar^{\mu\nu}\,.\label{MDinf}
\eeqa
The inverses of these
quantities are obtained (to first order in $\theta$) by replacing
$\theta\to -\theta$ in the above formulae.  Using 
eqs.~(\ref{sigmunurel1}), (\ref{Minf}) and (\ref{MDinf}), it follows that:
\beqa
(M^{-1})_\gamma{}^\tau &=& \epsilon^{\tau\alpha}M_\alpha{}^\beta
\epsilon\ls{\beta \gamma}\,,\label{covar1}\\
(M^{-1\,\dagger})^{\dot\gamma}{}_{\dot\tau} &=& \epsilon_{\dot\tau\dot\alpha}\,
(M^\dagger)^{\dot\alpha}{}_{\dot\beta}\,\epsilon^{\dot\beta\dot\gamma}\,.
\label{covar2}
\eeqa
These results can be used to
demonstrate the covariance (with respect to Lorentz transformations)
of the spinor index raising and lowering
properties of the epsilon symbols defined in \eq{epsalphabeta}.
The infinitesimal forms given by \eqst{Linf}{MDinf} can also be used 
[with the assistance of \eqst{sigmunuid1}{sigmunuid3}]
to establish the following two results:
\beqa
M^\dagger\sigmabar^\mu M
&=&\Lambda^\mu{}_\nu\,\sigmabar^\nu\,,\label{lorinv1}\\
M^{-1}\sigma^\mu (M^{-1})^\dagger&=&\Lambda^\mu{}_\nu\,\sigma^\nu\,.
\label{lorinv2}
\eeqa
Using the Lorentz transformation properties of the undotted and dotted
two-component spinor fields, 
\eqs{lorinv1}{lorinv2} can be used, respectively, to prove that
the spinor products 
$\xi^\dagger\sigmabar^\mu\eta$ and $\xi\sigma^\mu\eta^\dagger$
transform as Lorentz four-vectors.

As an example, consider a pure boost from the rest frame to
a frame where $p^\mu=(E_{\boldsymbol{p}}\,,\,\boldsymbol{\vec p})$,
which corresponds to $\theta_{ij}=0$ and
$\zeta^i = \BDpos \theta^{i0} = \BDneg \theta^{0i}$.
We assume that the mass-shell condition is satisfied, i.e. $p^0=
E_{\boldsymbol{\vec p}}\equiv(|\boldsymbol{\vec p}|^2+m^2)^{1/2}$.
The matrices $M_\alpha{}^\beta$
and $[(M^{-1})^\dagger]^{\dot\alpha}{}_{\dot\beta}$ that govern
the Lorentz transformations of spinor fields with a lowered undotted
index and spinor fields with a raised dotted index, respectively, are
given by:
\begin{equation} \label{spinboost}
\exp\left(\BDneg\half i\theta_{\mu\nu} J^{\mu\nu}\right)=
{\everymath{\displaystyle}
\begin{cases}
M=\exp\left(-\half\mathbold{\vec{\zeta}\cdot\vec{\sigma}}\right)=
\sqrt{\frac{\BDpos p\newcdot\sigma}{m}}\,,
 & \text{\quad for $(\half,0)$}\,, \\[12pt]
(M^{-1})^\dagger=\exp\left(\half\mathbold{\vec{\zeta}\cdot\vec{\sigma}}\right)=
\sqrt{\frac{\BDpos p\newcdot\sigmabar}{m}}\,,
 & \text{\quad for $(0,\half)$} \,,
\end{cases}
}
\end{equation}
where
\beqa
\sqrt{\BDpos p\newcdot\sigma}&
\equiv &
\frac{(E_{\boldsymbol{p}}+m)\,\mathds{1}_{2\times 2}
-\boldsymbol{\vec\sigma\newcdot\vec p}}
{\sqrt{2(E_{\boldsymbol{p}}+m)}}
\,, \label{sqpsigma} \\[6pt]
\sqrt{\BDpos p\newcdot\sigmabar}& \equiv &
\frac{(E_{\boldsymbol{p}}+m)\,\mathds{1}_{2\times 2}
+\boldsymbol{\vec\sigma\newcdot\vec p}}
{\sqrt{2(E_{\boldsymbol{p}}+m)}}
\,.
\label{sqpsigmabar}
\eeqa
These matrix square roots are defined to be the unique non-negative
definite hermitian matrices (i.e., with non-negative eigenvalues)
whose squares are equal
to the non-negative definite hermitian matrices
$\BDpos p\newcdot\sigma$ and $\BDpos p\newcdot\sigmabar$,
respectively.\footnote{Note that $\BDpos p\newcdot\sigma$
and $\BDpos p\newcdot\sigmabar$ are non-negative
matrices due to the implicit mass-shell condition
satisfied by $p^\mu$.}

According to \eq{spinboost},
the spinor index structure of $\sqrt{\BDpos p\newcdot\sigma}$ and
$\sqrt{\BDpos p\newcdot\sigmabar}$
corresponds to that of $M_\alpha{}^\beta$ and
$[(M^{-1})^\dagger]^{\dot\alpha}{}_{\dot\beta}$, respectively.  In
this case, we can rewrite \eqs{sqpsigma}{sqpsigmabar} as:
\beqa
\bigl[\sqrt{\BDpos p\newcdot\sigma}\,\bigr]_\alpha{}^\beta
\equiv
\bigl[\sqrt{\BDpos p\newcdot\sigma\,\sigmabar^0}\,\bigr]_\alpha{}^\beta
&=& \frac{\BDpos (p\newcdot\sigma_{\alpha\dot{\alpha}})
\sigmabar^{0\,\dot{\alpha}\beta} + m\delta_\alpha^\beta}
{\sqrt{2(E_{\boldsymbol{p}}+m)}}
\,, \label{sqpsigma1} \\[6pt]
\bigl[\sqrt{\BDpos p\newcdot\sigmabar}\,\bigr]^{\dot{\alpha}}{}_{\dot{\beta}}
\equiv
\bigl[\sqrt{\BDpos p\newcdot\sigmabar\,\sigma^0}\,
\bigr]^{\dot{\alpha}}{}_{\dot{\beta}}
& =&
\frac{\BDpos (p\newcdot\sigmabar^{\dot{\alpha}\alpha})
\sigma^0_{\alpha\dot{\beta}} + m\delta^{\dot{\alpha}}_{\dot{\beta}}}
{\sqrt{2(E_{\boldsymbol{p}}+m)}}
\,,
\label{sqpsigmabar1}
\eeqa
 since $\sigma^0=\sigmabar^0=\mathds{1}_{2\times 2}$.
Using \eqs {eq:simplifyssbars}{eq:simplifysbarssbar}, one can
easily verify that:
\beqa \label{squaring1}
\bigl[\sqrt{\BDpos p\newcdot\sigma}\,\bigr]_\alpha{}^\gamma
\bigl[\sqrt{\BDpos p\newcdot\sigma}\,\bigr]_\gamma{}^\beta
&=& (\BDpos p\newcdot\sigma\,\sigmabar^0)_\alpha{}^\beta\,,\\[5pt]
\bigl[\sqrt{\BDpos p\newcdot\sigmabar}\,\bigr]^{\dot{\alpha}}{}_{\dot{\gamma}}
\bigl[\sqrt{\BDpos p\newcdot\sigmabar}\,\bigr]^{\dot{\gamma}}{}_{\dot{\beta}}
&=& (\BDpos p\newcdot\sigmabar\,\sigma^0)^{\dot{\alpha}}{}_{\dot{\beta}}\,,
\eeqa
where implicit factors of $\sigmabar^0$ and $\sigma^0$ inside the square roots
of \eq{squaring1} have been suppressed.

Due to the fact that $\BDpos p\newcdot\sigma$ and
$\BDpos p\newcdot\sigmabar$
are hermitian, we could have defined their
hermitian matrix square
roots by the hermitian conjugate of \eq{spinboost}.  In this case,
the spinor index structure of $\sqrt{\BDpos p\newcdot\sigma}$ and
$\sqrt{\BDpos p\newcdot\sigmabar}$ would
correspond to that of $[(M^\dagger]^{\dot\alpha}{}_{\dot\beta}$ and
$[M^{-1}]_\alpha{}^\beta$, respectively.  That is, instead of
\eqs{sqpsigma1}{sqpsigmabar1}, we would now
rewrite \eqs{sqpsigma}{sqpsigmabar} in the following form:
\beqa
\bigl[\sqrt{\BDpos p\newcdot\sigma}\,\bigr]^{\dot{\alpha}}{}_{\dot{\beta}}
\equiv
\bigl[\sqrt{\BDpos \sigmabar^0\, p\newcdot\sigma}\,
\bigr]^{\dot{\alpha}}{}_{\dot{\beta}}
&=& \frac{\BDpos \sigmabar^{0\,\dot{\alpha}\beta}
(p\newcdot\sigma_{\beta\dot{\beta}}) + m\delta^{\dot{\alpha}}_{\dot{\beta}}}
{\sqrt{2(E_{\boldsymbol{p}}+m)}}
\,, \label{sqpsigma2} \\[6pt]
\bigl[\sqrt{\BDpos p\newcdot\sigmabar}\,\bigr]_\alpha{}^\beta
\equiv \bigl[\sqrt{\BDpos \sigma^0\, p\newcdot\sigmabar}\,\bigr]_\alpha{}^\beta
& =&
\frac{\BDpos \sigma^0_{\alpha\dot{\beta}}
(p\newcdot\sigmabar^{\dot{\beta}\beta})
 + m\delta_\alpha^\beta}
{\sqrt{2(E_{\boldsymbol{p}}+m)}}
\,.
\label{sqpsigmabar2}
\eeqa
Using \eqs {eq:simplifyssbars}{eq:simplifysbarssbar}, one can
again confirm that:
\beq \label{squaring2}
\bigl[\sqrt{\BDpos p\newcdot\sigma}\,\bigr]^{\dot{\alpha}}{}_{\dot{\gamma}}
\bigl[\sqrt{\BDpos p\newcdot\sigma}\,\bigr]^{\dot{\gamma}}{}_{\dot{\beta}}
= (\BDpos\sigmabar^0\,p\newcdot\sigma)^{\dot{\alpha}}{}_{\dot{\beta}}
\,,\qquad\quad
\bigl[\sqrt{\BDpos p\newcdot\sigmabar}\,\bigr]_\alpha{}^\gamma
\bigl[\sqrt{\BDpos p\newcdot\sigmabar}\,\bigr]_\gamma{}^\beta
= (\BDpos \sigma^0\,p\newcdot\sigmabar)_\alpha{}^\beta\,,
\eeq
where implicit factors of $\sigmabar^0$ and $\sigma^0$ inside the square roots
of \eq{squaring2} have been suppressed.

The proper choice of the spinor index structure for
 $\sqrt{\BDpos p\newcdot\sigma}$ and
$\sqrt{\BDpos p\newcdot\sigmabar}$ can always be
determined for any covariant expression.
That is, if we employ the spinor index-free notation
(and suppress the factors of $\sigma^0$ and $\sigmabar^0$), it will always be
clear from the context which spinor index structure for
$\sqrt{\BDpos p\newcdot\sigma}$ and
$\sqrt{\BDpos p\newcdot\sigmabar}$ is implicit.

As an example that will prove valuable later on,
consider an arbitrary four-vector $S^\mu$, defined in a reference frame
where $p^\mu=(E\,;\,{\boldsymbol{\vec p}})$, whose rest frame value
is $S_R^\mu$, i.e.
\beq \label{lorentzboost}
S^\mu=\Lambda^\mu{}_\nu S_R^\nu\,,\qquad {\rm with} \qquad
\Lambda=\begin{pmatrix} E/m & \quad p^j/m \\[5pt]
p^i/m & \quad \delta_{ij}+\displaystyle\frac{p^i p^j}{m(E+m)}
\end{pmatrix}\,.
\eeq
Then, using \eqss{ltxcov}{lorinv2}{spinboost}, it follows
that:
\beq \label{ssr12}
\sqrt{\BDpos p\newcdot\sigma}\,S\newcdot\sigmabar\,
\sqrt{\BDpos p\newcdot\sigma}
= mS_R\newcdot
\sigmabar\,,\qquad\quad
\sqrt{\BDpos p\newcdot\sigmabar}\,S\newcdot\sigma\,
\sqrt{\BDpos p\newcdot\sigmabar}
= mS_R\newcdot\sigma\,.
\eeq
The spinor index structures of \eq{ssr12} are easily established:
\beqa
\bigl[\sqrt{\BDpos p\newcdot\sigma}\,\bigr]^{\dot{\beta}}{}_{\dot{\gamma}}
\,S\newcdot\sigmabar^{\dot{\gamma}\alpha}
\bigl[\sqrt{\BDpos p\newcdot\sigma}\,\bigr]_\alpha{}^\beta &=&
mS_R\newcdot\sigmabar^{\dot{\beta}\beta}\,,\label{ssrr1} \\
\bigl[\sqrt{\BDpos p\newcdot\sigmabar}\,\bigr]_\beta{}^\gamma
\,S\newcdot\sigma_{\gamma\dot{\alpha}}
\bigl[\sqrt{\BDpos p\newcdot\sigmabar}\,\bigr]^{\dot{\alpha}}{}_{\dot{\beta}}
&=& mS_R\newcdot\sigma_{\beta\dot{\beta}}\,. \label{ssrr2}
\eeqa
Using \eqst{sqpsigma1}{lorentzboost}
and (\ref{eq:simplifyssbars})--(\ref{eq:simplifysbarssbar}), one can directly
verify the above results.

The two-component spinor formalism established in this section
will be applied to the quantum field theory of fermions in Minkowski
space of one time and three space dimensions in
this review.  We also direct the reader's attention to Appendices
G.1 and G.2,
which provide details of the correspondence between the two-component
and four-component spinor notation.

For certain applications, the spinor formalism
in four-dimensional Minkowski space is not sufficient.  For example,
in order to obtain instanton
solutions~\cite{instantons,instanton2,instanton3},
it is necessary
to formulate quantum field theory in Euclidean space.  One also
needs the Euclidean space formalism for a rigorous definition of the
path integral~\cite{zinn,fujikawa}.
The Green functions derived from the Euclidean path
integral can be related to the Green functions of the Minkowski space
theory by a Wick rotation~\cite{wick}.  In addition,
to evaluate the loop-corrected
Green functions of the theory, it is often most convenient to
apply a regularization scheme that involves dimensional continuation
away from $d=4$ spacetime dimensions~\cite{dimreg}. Thus, we also need to
generalize the spinor results of this section to $d\neq 4$.

The treatment of fermions in Euclidean space is
subtle~\cite{Osterwalder:1973kn,mehta,waldron,wet}.  Here, we
focus briefly on the mathematics of fermions in $d=4$ Euclidean
dimensions, where the relevant spacetime symmetry group is
SO(4) rather than SO(3,1).
The two-dimensional representations of
SO(3,1)$\iso$SL(2,$\mathbb{C})$, denoted
in this section by $(\half,0)$ and $(0,\half)$, respectively, are
complex representations that are related by hermitian conjugation.  In
contrast, the two-dimensional representations of
SO(4)$\iso$SU(2)$\times$SU(2),
also denoted by $(\half,0)$ and $(0,\half)$, respectively,%
\footnote{These SO(4) representations transform as a doublet under
one of the SU(2) groups and as a singlet under the other
SU(2) group.}
are independent pseudo-real representations, i.e.~\textit{not}
related by hermitian conjugation.
A two-component spinor notation can be formulated for fields
that transform respectively
under the $(\half,0)$ and $(0,\half)$ representations
of SO(4).  Details can be found in refs.~\cite{instanton2,sherry,LN}.

In Feynman diagram calculations, one can adopt the standard procedure
for the Wick rotation in order to evaluate the loop integrals in Euclidean
space.  We shall employ the standard Euclidean metric $\delta^{\mu\nu}$
in computing scalar products of four-vectors.
Moreover, one can define Euclidean sigma matrices,
$\sigma_E^\mu=(-i\boldsymbol{\vec\sigma}\,,\,\sigma_E^4)$ and
$\sigmabar_E^\mu=(i\boldsymbol{\vec\sigma}\,,\,\sigmabar_E^{\,4})$,
where $\sigma_E^4=\sigmabar_E^{\,4}\equiv \mathds{1}_{2\times 2}$.
In this convention, the Wick-rotated versions of
\eqst{eq:ssbarsym}{trsbarssbars} are preserved [after making the replacements
$g^{\mu\nu}\to \BDpos \delta^{\mu\nu}$ and
$i\epsilon^{ijk0}\to \BDpos \epsilon^{ijk4}$,
with $\epsilon^{1234}=\epsilon_{1234}=+1$].\footnote{In
practical computations of one-loop matrix elements, one can carry
out all the sigma matrix algebra in
Minkowski space \textit{before} Wick-rotating
to Euclidean space in order to perform the loop integrals.}
Further details
of our Euclidean space conventions are provided at the end of
Appendix~A.

The generalization of the spinor results of this section to $d\neq 4$,
useful for dimensional continuation regularization schemes, is
discussed in \app{B.2}.  In particular, the 
identities of \app{B.1} used to derive Fierz identities 
[cf.~\eqst{eq:twocompfierzone}{twocompfierzc} and 
(\ref{sigsigfierza})--(\ref{sigsigfierzc})]
and any identities involving the four-dimensional Levi-Civita
\protect{$\epsilon$}-tensor are not valid unless
\protect{$\mu$} is a Lorentz vector index
in exactly four dimensions. In our treatment of two-component spinor identities in $d\neq 4$
dimensions given in \app{B.2}, we take the Lorentz vector indices
to formally run over $d$ values, whereas the undotted and dotted
spinor indices continue to take on two possible values.  This is
sufficient when used as a regularization procedure for
divergent integrals that arise in loop computations.  However
in generic $d$-dimensional field theories
where $d$ is an integer greater than 4,
the two-component spinor formalism of this review is no longer applicable.
Suitable methods for treating spinors in
diverse spacetime dimensions and signatures~\cite{brauer,Trautman,scherk,vanN,coq,kugo,sohniusapp,vanN2,wetterich,finkelstein,Benn,thacker,andrade,tanii,vanpro}
are briefly presented in \app{G.3}.

\section{\texorpdfstring{Properties of fermion fields}{Properties of fermion fields}}
\label{sec:externalfermions}

In this review, we refer to spin-1/2 particles as
Majorana or Dirac fermions depending
on the nature of the global symmetry\footnote{A subgroup of the global
symmetry group may be gauged (and hence promoted to a local
symmetry).  Degrees of freedom not associated with the gauged subgroup
are typically referred to as flavor degrees of freedom.}
that governs the fermion Lagrangian and dictates the form of the fermion mass
terms.
A \textit{Majorana fermion} is a two-component massive field that 
is completely neutral (i.e.~a singlet with respect to the
symmetry group) or transforms
as a non-trivial real representation of the symmetry group 
(cf.~footnote~\ref{fnmajorana}).  A \textit{Dirac fermion}
consists of a pair of two-component massive fields that are 
oppositely charged  
with respect to a conserved O(2) symmetry.  As shown in 
\sec{subsec:generalmass}, Dirac fermions arise when a multiplet of 
two-component fermions transforms as a complex or pseudo-real
representation of the symmetry group.\footnote{Majorana and Dirac
fermions can also be described in terms of four-component Majorana
and Dirac spinor fields, as in \app{G}.  However, keep in mind that 
the terms \textit{Majorana spinor} and \textit{Dirac spinor} are defined
strictly in the context of the four-component spinor formalism as in \app{G.1},
or in the more general context of a $d$-dimensional spacetime as
in \app{G.3}.}

The case of a massless fermion is
special, as the absence of mass terms leads to an enhanced global
symmetry group. 
Each physical spin-1/2 zero-mass eigenstate 
is fundamentally a two-component spinor.
Thus, following the standard nomenclature
used for massless neutrinos, 
it is common to employ the term \textit{massless Weyl fermion} 
to describe any massless spin-1/2 particle.\footnote{%
Two-component fermions are often called Weyl fermions, 
due to their association with the two-dimensional spinor
representations of the Lorentz group introduced by Weyl in
\refs{Weyl-book}{WeylDirac}.  It is now common practice to 
define a Weyl spinor as the left or right-handed projection of a 
four-component spinor [as in \eq{plprdefs}].  Of course, there is a
one-to-one correspondence between these two definitions.}

\subsection{The two-component fermion field
and spinor wave functions
\label{subsec:singleWeyl}}
\renewcommand{\theequation}{\arabic{section}.\arabic{subsection}.\arabic{equation}}
\renewcommand{\thefigure}{\arabic{section}.\arabic{subsection}.\arabic{figure}}
\renewcommand{\thetable}{\arabic{section}.\arabic{subsection}.\arabic{table}}
\setcounter{equation}{0}
\setcounter{figure}{0}
\setcounter{table}{0}

We begin by describing the properties of a free neutral massive
anticommuting spin-1/2 field, denoted $\xi_\alpha(x)$, which
transforms as $(\half,0)$ under the Lorentz group.  The field $\xi_
\alpha$ therefore describes a Majorana fermion~\cite{Majorana}. The free-field
Lagrangian density is \cite{Uhlenbeck}:
\beq \label{lagsingleMajorana} \mathscr{L}=
i\xi^\dagger
\sigmabar^\mu\partial_\mu\xi - \half m (\xi \xi + \xi^\dagger \xi^\dagger )\,.
\eeq
On-shell, $\xi$ satisfies the free-field Dirac equation
\cite{Dirac,Weyl-book,vdWaerden1,Case,MajReview},
\beq \label{dirac}
i \sigmabar^{\mu\dot{\alpha}\beta} \partial_\mu \xi_\beta
=m\xi^{\dagger\dot{\alpha}}\,.
\eeq
Consequently after quantization, $\xi_\alpha$ can be expanded in a
Fourier series \cite{Case}:
\beq
\label{modes}
\xi_\alpha(x)=\sum_s\int\,\frac{d^3 \boldsymbol{\vec p}}
{(2\pi)^{3/2}(2E_{\boldsymbol p})^{1/2}}
\left[x_\alpha(\boldsymbol{\vec p},s)a(\boldsymbol{\vec
p},s)e^{\BDneg ip\newcdot x}+y_\alpha(\boldsymbol{\vec p},s)a^\dagger(
\boldsymbol{\vec p},s)e^{\BDpos ip\newcdot x}\right]\,,
\eeq
where $E_{\boldsymbol p}\equiv (|{\boldsymbol{\vec p}}|^2+m^2)^{1/2}$,
and the creation and annihilation operators $a^\dagger$ and $a$
satisfy anticommutation relations:
\beq
\{
a ({\boldsymbol{\vec p},s}),\,
a^\dagger({\boldsymbol{\vec p}^{\,\prime},s^\prime})
\}
=\delta^3(\boldsymbol{\vec p}
-\boldsymbol{\vec p}^{\,\prime})
\delta_{ss'}\,,
\eeq
and all other anticommutators vanish.
It follows that
\beq
\label{antimodes}
\xi^\dagger_{\dot \alpha}(x) \equiv
(\xi_\alpha)^\dagger
=\sum_s\int\,\frac{d^3 \boldsymbol{\vec p}}
{(2\pi)^{3/2}(2E_{\boldsymbol p})^{1/2}}
\left[x^\dagger_{\dot\alpha}(\boldsymbol{\vec p},s)a^\dagger
(\boldsymbol{\vec p},s)e^{\BDpos ip\newcdot x}
+y^\dagger_{\dot \alpha}(\boldsymbol{\vec  p},s)a(
\boldsymbol{\vec p},s)e^{\BDneg ip\newcdot x}\right]
\, .
\eeq
We employ covariant
normalization of the one-particle states, i.e., we act with one
creation operator on the vacuum with the following convention
\beq
\ket{\boldsymbol{\vec p},s}\equiv (2\pi)^{3/2}(2E_{\boldsymbol p})^{1/2}
a^\dagger({\boldsymbol{\vec p}},s)\ket{0}\,,
\label{eq:Majoranastate}
\eeq
so that $\vev{\boldsymbol{\vec p},s|\boldsymbol{\vec p}^{\,\prime},s'}
= (2\pi)^{3}(2E_{\boldsymbol p})\delta^3(\boldsymbol{\vec p}
-\boldsymbol{\vec p}^{\,\prime})\delta_{ss'}$.
Therefore,
\beqa
\bra{0}\xi_\alpha(x)\ket{\boldsymbol{\vec p},s} &=&
x_\alpha({\boldsymbol{\vec p}},s)e^{\BDneg ip\newcdot x}\,,
\qquad\quad\,\,
\bra{0}
\xi^\dagger_{\dot{\alpha}}(x)\ket{\boldsymbol{\vec p},s}=
y^\dagger_{\dot{\alpha}}({\boldsymbol{\vec p}},s)e^{\BDneg ip\newcdot x}\,,
\label{instate}\\
\bra{{\boldsymbol{\vec p}},s}\xi_\alpha(x)\ket{0} &=&
y_\alpha({\boldsymbol{\vec p}},s)e^{\BDpos ip\newcdot x}\,,
\qquad\qquad
\bra{{\boldsymbol{\vec p}},s}\xi^\dagger_{\dot{\alpha}}(x)\ket{0}=
x^\dagger_{\dot{\alpha}}({\boldsymbol{\vec p}},s)e^{\BDpos ip\newcdot x}
\,.
\label{outstate}
\eeqa
It should be emphasized that $\xi_\alpha(x)$ is an anticommuting
spinor field, whereas $x_\alpha$ and $y_\alpha$ are {\it commuting}
two-component spinor wave functions.  The anticommuting properties of
the fields are carried by the creation and annihilation operators.

Applying \eq{dirac} to \eq{modes}, we find that the $x_\alpha$ and
$y_\alpha$ satisfy momentum space Dirac equations.  These conditions
can be written down in a number of
equivalent ways:
\beqa
&& (p\newcdot\sigmabar)^{\dot{\alpha}\beta} x_{\beta} =
\BDpos m y^{\dagger\dot{\alpha}} \, \,, \qquad\qquad\qquad \,\,
(p\newcdot\sigma)_{\alpha\dot{\beta}} y^{\dagger\dot{\beta}} =
\BDpos m x_\alpha
\> \,,
\label{onshellone}
\\
&& (p\newcdot\sigma)_{\alpha\dot{\beta}}
x^{\dagger\dot{\beta}} = \BDneg m y_\alpha \, \,, \qquad\qquad\qquad\!\!
(p\newcdot\sigmabar)^{\dot{\alpha}\beta} y_{\beta} =
\BDneg m x^{\dagger\dot{\alpha}} \> \,, \label{onshelltwo}
\\
&& x^{\alpha}
(p\newcdot\sigma)_{\alpha\dot{\beta}} = \BDneg m y^\dagger_{\dot{\beta}} \> \,,
\qquad\qquad\qquad\!\!  y^\dagger_{\dot{\alpha}}
(p\newcdot\sigmabar)^{\dot{\alpha}\beta} = \BDneg mx^{\beta} \> \,,
\label{onshellthree}
\\
&& x^\dagger_{\dot{\alpha}}
(p\newcdot\sigmabar)^{\dot{\alpha}\beta} = \BDpos my^{\beta} \> \,,
\qquad\qquad\qquad\,\, y^{\alpha} (p\newcdot\sigma)_{\alpha\dot{\beta}} =
\BDpos m x^\dagger_{\dot{\beta}} \> \, .
\label{onshellfour}
\eeqa
Using the identities
$[(p\newcdot\sigma)(p\newcdot\sigmabar)]_\alpha{}^\beta
=\BDpos p^2\,\delta_\alpha{}^\beta$ and
$[(p\newcdot\sigmabar)(p\newcdot\sigma)]^{\dot{\alpha}}{}_{\dot{\beta}}=
\BDpos p^2\,\delta^{\dot{\alpha}}{}_{\dot{\beta}}$, one can
check that both $x_\alpha$ and $y_\alpha$ must satisfy the
mass-shell condition, $p^2= \BDpos m^2$ (or equivalently,
$p^0=E_{\boldsymbol{p}}$).  We will later see that
\eqst{onshellone}{onshellfour}
are often useful for simplifying matrix elements.

The quantum number $s$ labels the spin or helicity of the spin-1/2
fermion.  We shall examine two approaches for constructing the
spin-1/2 states.  In the first approach, we consider the particle in
its rest frame and quantize the spin along a fixed axis specified by
the unit vector ${\boldsymbol{\hat s}}\equiv(\sin\theta\cos\phi\,,\,
\sin\theta\sin\phi\,,\,\cos\theta)$ with polar angle $\theta$ and
azimuthal angle $\phi$
with respect to a fixed $z$-axis.\footnote{In the literature, it is
a common practice to choose ${\boldsymbol{\hat s}}=\boldsymbol{\hat z}$.
However in order to be somewhat more general, we shall not assume this
convention here.}
The corresponding spin states will be called fixed-axis spin states.
The relevant basis of
two-component spinors $\chi\ls{s}$ are eigenstates
of $\half{\boldsymbol{\vec\sigma\newcdot\hat s}}$, i.e.,
\beq
\half {\boldsymbol{\vec\sigma\newcdot\hat s}}\,\chi\ls{s} =
s\chi\ls{s}\,,
  \qquad s = \pm\half\,.\label{chiess}
\eeq
Explicit forms for the two-component spinors $\chi\ls{s}$ and their
properties are given in \app{C}.

The fixed-axis spin states described above are not very convenient for
particles in relativistic motion.  Moreover, these states cannot be
employed for massless particles since no rest frame exists.
Thus, a second approach is to consider helicity states and the corresponding
basis of two-component helicity spinors $\chi\ls{\lambda}$ that are
eigenstates of $\half{\boldsymbol{\vec\sigma\newcdot\hat p}}$, i.e.,
\beq
\half {\boldsymbol{\vec\sigma\newcdot\hat p}}\,\chi\ls\lambda = \lambda\chi\ls
\lambda,     \qquad \lambda = \pm\half\,.\label{chilambda}
\eeq
Here ${\boldsymbol{\hat p}}$ is the unit
vector in the direction of the three-momentum,
with polar angle $\theta$ and azimuthal angle $\phi$
with respect to a fixed $z$-axis.  That is,
the two-component helicity spinors can be obtained from the fixed-axis
spinors by replacing ${\boldsymbol{\hat s}}$ by
${\boldsymbol{\hat p}}$ and identifying $\theta$ and $\phi$ as the
polar and azimuthal angles of ${\boldsymbol{\hat p}}$.

For fermions of mass $m\neq 0$, one can define a spin
four-vector $S^\mu$, which reduces to $(0; {\boldsymbol{\hat s}})$
in the rest frame.  For fixed-axis spin states, the unit three-vector
${\boldsymbol{\hat s}}$ corresponds to the axis of spin quantization.
In an arbitrary reference frame, the spin four-vector
satisfies $S\newcdot p=0$ and $S\newcdot S= \BDneg 1$.
Boosting from the rest frame to a frame where $p^\mu=(E\,,
\,{\boldsymbol{\vec{p}}})$ [using \eq{lorentzboost}] yields
\beq \label{fixedsvect}
S^\mu=\left(\frac{\boldsymbol{\vec{p}\newcdot\hat{s}}}{m}\,;\,
{\boldsymbol{\hat{s}}}+\frac{{\boldsymbol{(\vec{p}\newcdot\hat{s})\,
\vec{p}}}}{m(E+m)}\right)\,.
\eeq
If necessary, we shall write $S^\mu({\boldsymbol{\hat s}})$
to emphasize the dependence of $S^\mu$ on $\boldsymbol{\hat s}$.

The spin four-vector for helicity states is defined by taking
${\boldsymbol{\hat s}}={\boldsymbol{\hat p}}$.  \Eq{fixedsvect} then
reduces~to
\beq
S^\mu = \frac{1}{m} \left(|{\boldsymbol{\vec p}}|\,;\,
E{\boldsymbol{\hat p}}
\right)\,.\label{spinvec}
\eeq
In the non-relativistic limit, the spin four-vector for helicity
states is $S^\mu \simeq (0\,;\,{\boldsymbol{\hat p}})$, as
expected.\footnote{\label{fnone}%
Strictly speaking, ${\boldsymbol{\hat p}}$ is not
defined in the rest frame.  In practice, helicity states are defined
in some moving frame with momentum ${\boldsymbol{\vec p}}$.  The rest
frame is achieved by boosting in the direction of $-{\boldsymbol{\vec p}}$.}
In the high energy limit ($E\gg m$), $S^\mu =
p^\mu/m+{\mathcal{O}}(m/E)$.  For a massless fermion, the spin
four-vector does not exist (as there is no rest frame).  Nevertheless,
one can obtain consistent results by working with massive helicity
states and taking the $m\to 0$ limit at the end of the computation.
In this case, one can simply use $S^\mu = p^\mu/m+{\mathcal{O}}(m/E)$;
in practical computations the final result will be well-defined in the
zero mass limit.  In contrast, for massive fermions at rest, the
helicity state does not exist without reference to some particular
boost direction as noted in footnote \ref{fnone}.

Using \eq{ssr12} with $S_R^\mu=(0\,;\,{\boldsymbol{\hat s}})$,
two important formulae are obtained:
\beq \label{psp12}
\sqrt{\BDpos p\newcdot\sigma}\,S\newcdot\sigmabar\,
\sqrt{\BDpos p\newcdot\sigma} = \BDpos m\,
{\boldsymbol{\vec\sigma\newcdot\hat s}}\,, \qquad\quad
\sqrt{\BDpos p\newcdot\sigmabar}\,S\newcdot\sigma\,
\sqrt{\BDpos p\newcdot\sigmabar} = \BDneg m\,
{\boldsymbol{\vec\sigma\newcdot\hat s}}\,.
\eeq
These results can also be derived directly by employing
the explicit form for the spin vector $S^\mu$ [\eq{fixedsvect}]
and the results of \eqs{sqpsigma}{sqpsigmabar}.

The two-component spinor wave functions $x$ and $y$ can now be given
explicitly in terms of the $\chi\ls{s}$ defined in \eq{twocomp}.
First, we note that \eq{onshellone} when evaluated in the rest frame
yields $x_1=y^{\dagger 1}$ and $x_2=y^{\dagger 2}$.  That is, as
column vectors,
$x_\alpha(\boldsymbol{\vec p}=0) =
y^{\dagger\dot\alpha}(\boldsymbol{\vec p}=0)$ can be expressed in general as
some linear combination of the $\chi\ls{s}$ ($s=\pm\half$).  Hence, we
may choose $x_\alpha(\boldsymbol{\vec p}=0,s) =
y^{\dagger\dot\alpha}(\boldsymbol{\vec p}=0,s)=\sqrt{m}\chi_s$, where the
factor of $\sqrt{m}$ reflects the standard relativistic normalization
of the rest frame spin states.  These wave functions can be boosted to
an arbitrary frame using \eq{spinboost}.  The resulting undotted
spinor wave functions are given by:\footnote{Explicit forms for two-component spinor wave 
functions have been exhibited
a number of times in the literature.  For example, see
\refs{Kersch:1985rn}{Hagiwara:1985yu} and \app{I.1}.}
\beqa
x_\alpha(\boldsymbol{\vec p},s)
&=&\sqrt{\BDpos p\newcdot\sigma}\,\chi\ls{s}\,,
\qquad\qquad\quad\,\,
x^\alpha(\boldsymbol{\vec p},s)
=-2s\chi^\dagger\ls{-s}\sqrt{\BDpos p\newcdot\sigmabar}\,,
\label{explicitxa} \\
y_\alpha(\boldsymbol{\vec p},s)&=&2s
\sqrt{\BDpos p\newcdot\sigma}\,\chi\ls{-s}\,,\qquad\quad\,\,\,\,
y^\alpha(\boldsymbol{\vec p},s)=
\chi^\dagger\ls{s}\sqrt{\BDpos p\newcdot\sigmabar}\,,
\label{explicitya}
\eeqa
and the dotted spinor wave functions are given by
\beqa
x^{\dagger\dot{\alpha}}(\boldsymbol{\vec p},s)&=&
-2s\sqrt{\BDpos p\newcdot\sigmabar}\,\chi\ls{-s}\,,\qquad\qquad
x^\dagger_{\dot{\alpha}}(\boldsymbol{\vec p},s)
=\chi^\dagger\ls{s}\sqrt{\BDpos p\newcdot\sigma}\,,\\
\label{explicitxb}
y^{\dagger\dot{\alpha}}(\boldsymbol{\vec p},s)&=&
\sqrt{\BDpos p\newcdot\sigmabar}\,\chi\ls{s}\,,\qquad\qquad\qquad\,\,
y^\dagger_{\dot{\alpha}}(\boldsymbol{\vec p},s)
=2s\chi^\dagger\ls{-s}\sqrt{\BDpos p\newcdot\sigma}\,,
\label{explicityb}
\eeqa
where $\sqrt{\BDpos p\newcdot\sigma}$ and
$\sqrt{\BDpos p\newcdot\sigmabar}$ are defined either by
\eqs{sqpsigma1}{sqpsigmabar1} or by \eqs{sqpsigma2}{sqpsigmabar2},
respectively (as mandated by the spinor index 
structure).
Note that \eqst{explicitxa}{explicityb} imply that the $x$ and $y$
spinors are related:
\beq \label{xyrelation}
y({\boldsymbol{\vec p}},s)=2s x({\boldsymbol{\vec p}},-s)\,,\qquad\qquad
y^\dagger({\boldsymbol{\vec p}},s)=2s x^\dagger({\boldsymbol{\vec p}},-s)\,.
\eeq

The phase choices in \eqst{explicitxa}{explicityb} are consistent with
those employed for four-component spinor wave functions [see \app{G}].
We again emphasize that in \eqst{explicitxa}{explicityb}, one may
either choose $\chi\ls{s}$ to be an eigenstate of $\boldsymbol{\vec
\sigma\newcdot\hat s}$, where the spin is measured in the rest frame
along the quantization axis $\boldsymbol{\hat s}$, or choose $\chi\ls
{s}$ to be an eigenstate of $\boldsymbol{\vec\sigma\newcdot\hat p}$
(in this case we shall write $s=\lambda$), which yields the helicity spinor
wave functions.

The following equations can now be derived:
\beqa
&&
(S\newcdot\sigmabar)^{\dot{\alpha}\beta}
x_{\beta}({\boldsymbol{\vec{p}}},s) =
\BDpos 2s y^{\dagger\dot{\alpha}}({\boldsymbol{\vec{p}}},s)
\, \,, \qquad\qquad \,\,
(S\newcdot\sigma)_{\alpha\dot{\beta}}
y^{\dagger\dot{\beta}}({\boldsymbol{\vec{p}}},s)
= \BDneg 2s x_\alpha({\boldsymbol{\vec{p}}},s)
\> \,, \label{spinone}
\\ &&
(S\newcdot\sigma)_{\alpha\dot{\beta}}
x^{\dagger\dot{\beta}}({\boldsymbol{\vec{p}}},s)
= \BDneg 2s y_\alpha({\boldsymbol{\vec{p}}},s)
\, \,, \qquad\qquad\!\!
(S\newcdot\sigmabar)^{\dot{\alpha}\beta}
y_{\beta}({\boldsymbol{\vec{p}}},s)
= \BDpos 2s x^{\dagger\dot{\alpha}}({\boldsymbol{\vec{p}}},s)
\> \,, \label{spintwo}
\\ &&
x^{\alpha}({\boldsymbol{\vec{p}}},s)
(S\newcdot\sigma)_{\alpha\dot{\beta}}
= \BDneg 2s
y^\dagger_{\dot{\beta}}({\boldsymbol{\vec{p}}},s)
\> \,, \qquad\qquad\!\!
y^\dagger_{\dot{\alpha}}({\boldsymbol{\vec{p}}},s)
(S\newcdot\sigmabar)^{\dot{\alpha}\beta} =
\BDpos 2s
x^{\beta}({\boldsymbol{\vec{p}}},s)
\> \,, \label{spinthree}
\\   &&
x^\dagger_{\dot{\alpha}}({\boldsymbol{\vec{p}}},s)
(S\newcdot\sigmabar)^{\dot{\alpha}\beta}
= \BDpos 2s y^{\beta}({\boldsymbol{\vec{p}}},s)
\> \,, \qquad\qquad\,\,
y^{\alpha}({\boldsymbol{\vec{p}}},s)
(S\newcdot\sigma)_{\alpha\dot{\beta}}
= \BDneg 2s x^\dagger_{\dot{\beta}}({\boldsymbol{\vec{p}}},s)
\> \,. \label{spinfour}
\eeqa
For example, using \eq{psp12} and the definitions above for
$x_\alpha({\boldsymbol{\vec{p}}},s)$ and
$y^{\dagger\dot{\alpha}}({\boldsymbol{\vec{p}}},s)$, we find
(suppressing spinor indices),
\beq \label{psigmaderiv}
\sqrt{\BDpos p\newcdot\sigma}\,S\newcdot\sigmabar\,x({\boldsymbol{\vec{p}}},s)
=
\sqrt{\BDpos p\newcdot\sigma}
\,S\newcdot\sigmabar\,\sqrt{\BDpos p\newcdot\sigma}\,\chi\ls{s}
= \BDpos m{\boldsymbol{\vec\sigma\newcdot\hat{s}}}\,\chi\ls{s}
= \BDpos 2sm\,\chi\ls{s}\,.
\eeq
Multiplying both sides of \eq{psigmaderiv}
by $\sqrt{\BDpos p\newcdot\sigmabar}$ and noting that
$\sqrt{\BDpos p\newcdot\sigmabar}\sqrt{\BDpos p\newcdot\sigma}=m$, we end up with
\beq
S\newcdot\sigmabar\,x({\boldsymbol{\vec{p}}},s)=
\BDpos 2s\sqrt{\BDpos p\newcdot\sigmabar}\,\chi\ls{s}
= \BDpos 2s y^\dagger({\boldsymbol{\vec{p}}},s)\,.
\eeq
All the results of \eqst{spinone}{spinfour} can be derived in this manner.

The consistency of \eqst{spinone}{spinfour} can also be checked as follows.
First, each of these equations yields
\beq
(S\newcdot\sigma)_{\alpha{\dot{\alpha}}}
(S\newcdot\sigmabar)^{\dot{\alpha}\beta}=-\delta_\alpha^\beta\,,\qquad\qquad
(S\newcdot\sigmabar)^{\dot{\alpha}\alpha}
(S\newcdot\sigma)_{\alpha\dot{\beta}}=-\delta^{\dot{\alpha}}_{\dot{\beta}}\,,
\eeq
after noting that $4s^2=1$ (for $s=\pm\half$).
{}From \eqs{eq:ssbarsym}{eq:sbarssym}
it follows that $S\newcdot S = \BDneg 1$, as required.  Second, if
one applies
\beqa
(p\newcdot \sigma\, S\newcdot\sigmabar+S\newcdot\sigma
\,p\newcdot\sigmabar)_\alpha{}^\beta &=&
\BDpos 2p\newcdot S\,\delta_\alpha{}^\beta\,,\\
(p\newcdot \sigmabar\, S\newcdot\sigma+S\newcdot\sigmabar
\,p\newcdot\sigma)^{\dot{\alpha}}{}_{\dot{\beta}} &=&
\BDpos 2p\newcdot S\,\delta^{\dot{\alpha}}{}_{\dot{\beta}}\,,
\eeqa
to \eqst{onshellone}{onshellfour} and \eqst{spinone}{spinfour}, it follows
that $p\newcdot S =0$.

It is useful to combine the results of
\eqst{onshellone}{onshellfour} and \eqst{spinone}{spinfour} as follows:
\beqa
&&(p^\mu-2s mS^\mu)\sigmabar_\mu^{\dot{\alpha}\beta}
x_\beta({\boldsymbol{\vec{p}}},s)=0\,, \qquad\qquad
(p_\mu-2s mS_\mu)\sigma^\mu_{\alpha\dot{\beta}}
x^{\dagger\dot{\beta}}({\boldsymbol{\vec{p}}},s)=0\,,
\label{combineone}
\\
&&(p^\mu+2s mS^\mu)\sigmabar_\mu^{\dot{\alpha}\beta}
y_\beta({\boldsymbol{\vec{p}}},s)=0\,, \qquad\qquad
(p_\mu+2s mS_\mu)\sigma^\mu_{\alpha\dot{\beta}}
y^{\dagger\dot{\beta}}({\boldsymbol{\vec{p}}},s)=0\,,
\label{combinetwo}
\\
&&
x^\alpha({\boldsymbol{\vec{p}}},s)
\sigma^\mu_{\alpha\dot{\beta}}
(p_\mu-2s mS_\mu)
=0\,,\qquad\qquad
x^\dagger_{\dot{\alpha}}({\boldsymbol{\vec{p}}},s)
\sigmabar_\mu^{\dot{\alpha}\beta}
(p^\mu-2s mS^\mu)
=0\,,
\label{combinethree}
\\
&&y^\alpha({\boldsymbol{\vec{p}}},s)
\sigma^\mu_{\alpha\dot{\beta}}
(p_\mu+2s mS_\mu)
=0\,,\qquad\qquad
y^\dagger_{\dot{\alpha}}({\boldsymbol{\vec{p}}},s)
\sigmabar_\mu^{\dot{\alpha}\beta}
(p^\mu+2s mS^\mu)
=0\,.
\label{combinefour}
\eeqa
\Eqst{explicitxa}{combinefour}
also apply to the helicity wave functions
$x(\boldsymbol{\vec p},\lambda)$ and $y(\boldsymbol{\vec p},\lambda)$
simply by replacing $s$ with $\lambda$ and
$S^\mu({\boldsymbol{\hat{s}}})$ [\eq{fixedsvect}]
with $S^\mu({\boldsymbol{\hat{p}}})$
[\eq{spinvec}].

The above results are applicable only for massive fermions (where the
spin four-vector $S^\mu$ exists).  We may treat the case of massless
fermions directly by employing helicity spinors in
\eqst{explicitxa}{explicityb}.  Putting $E=|\boldsymbol{\vec{p}}|$
and $m=0$, we easily obtain:
\beqa
x_\alpha(\boldsymbol{\vec p},\lambda)
&=&\sqrt{2E}\,\,(\half-\lambda)\,\chi\ls\lambda\,,
\qquad\qquad\quad\,\,
x^\alpha(\boldsymbol{\vec p},\lambda)
=\sqrt{2E}\,\,(\half-\lambda)\chi^\dagger\ls{-\lambda}\,,
\label{helexplicitxa} \\
y_\alpha(\boldsymbol{\vec p},\lambda)&=&
\sqrt{2E}\,\,(\half+\lambda)
\,\chi\ls{-\lambda}\,,\qquad\qquad\,\,\,\,\,
y^\alpha(\boldsymbol{\vec p},\lambda)=\sqrt{2E}\,\,(\half+\lambda)
\chi^\dagger\ls\lambda\,,
\label{helexplicitya}
\eeqa
or equivalently,
\beqa
x^{\dagger\dot{\alpha}}(\boldsymbol{\vec p},\lambda)&=&\sqrt{2E}\,\,
(\half-\lambda)\,\chi\ls{-\lambda}\,,\qquad\qquad\,\,
x^\dagger_{\dot{\alpha}}(\boldsymbol{\vec p},\lambda)
=\sqrt{2E}\,\,(\half-\lambda)\chi^\dagger\ls\lambda\,,\\
\label{helexplicitxb}
y^{\dagger\dot{\alpha}}(\boldsymbol{\vec p},\lambda)&=&
\sqrt{2E}\,\,(\half+\lambda)\,\chi\ls{\lambda}\,,\qquad\qquad\quad
y^\dagger_{\dot{\alpha}}(\boldsymbol{\vec p},\lambda)
=\sqrt{2E}\,\,(\half+\lambda)\chi^\dagger\ls{-\lambda}\,.
\label{helexplicityb}
\eeqa
It follows that:
\beqa \label{masslesshel}
\left(\half+\lambda\right)
x({\boldsymbol{\vec p}},\lambda)=0\,,&&\qquad\qquad
\left(\half+\lambda\right)
x^\dagger({\boldsymbol{\vec p}},\lambda)=0\,,\label{masslesshela}\\
\left(\half-\lambda\right)
y({\boldsymbol{\vec p}},\lambda)=0\,,&&\qquad\qquad
\left(\half-\lambda\right)
y^\dagger({\boldsymbol{\vec p}},\lambda)=0\,.\label{masslesshelb}
\eeqa
The significance of \eqs{masslesshela}{masslesshelb}
is clear; for massless fermions, only
one helicity component of $x$ and $y$ is non-zero.  Applying this
result to neutrinos, we find that massless neutrinos are left-handed
($\lambda=-1/2$), while antineutrinos are right-handed
($\lambda=+1/2$).

\Eqs{masslesshela}{masslesshelb}
can also be derived by carefully taking the $m\to 0$ limit
of \eqs{combineone}{combinetwo}
applied to the helicity wave functions
$x(\boldsymbol{\vec p},\lambda)$ and $y(\boldsymbol{\vec p},\lambda)$
[i.e., replacing $s$ with $\lambda$].  We then replace $mS^\mu$ with
$p^\mu$, which is the leading term in the limit of $E\gg m$.
Using the results of \eqs{onshellone}{onshelltwo} and dividing out by
an overall factor of $m$
(before finally taking the $m\to 0$ limit)
reproduces \eqs{masslesshela}{masslesshelb}.

Having defined explicit forms for the two-component spinor wave
functions, we can now write down the spin projection matrices.
Noting that
$\half(1+2s\,\boldsymbol{\vec{\sigma}\newcdot\hat{s}})\chi\ls{s'}
=\half(1+4ss')\chi_{s'}=\delta_{ss'}\chi_{s'}$ (since
$s$,~$s'=\pm\half$), one can write:
\beq \label{chischis}
\chi\ls{s}\chi\ls{s}^\dagger =\half
\left(1+2s\,\boldsymbol{\vec{\sigma}\newcdot\hat{s}} \right)
\sum_{s'}\chi\ls{s'}\chi\ls{s'}^\dagger
=\half
\left(1+2s\,\boldsymbol{\vec{\sigma}\newcdot\hat{s}} \right)\,,
\eeq
where at the second step, we have employed
the completeness relation given in \eq{completeness}.  Making use of
\eq{psp12} for ${\boldsymbol{\vec{\sigma}\newcdot\hat{s}}}$,
it follows that
\beq
\chi\ls{s}\chi\ls{s}^\dagger
=\half\left(1
\BDplus \frac{2s}{m}\sqrt{\BDpos p\newcdot\sigma}\,S\newcdot\sigmabar\,
\sqrt{\BDpos p\newcdot\sigma}\right)\,.
\eeq
Hence, with both spinor indices in the lowered position,
\beqa
x(\boldsymbol{\vec p},s) x^\dagger(\boldsymbol{\vec p},s)&=&
\sqrt{\BDpos p\newcdot\sigma}\,\chi\ls{s}\chi^\dagger\ls{s}\,
\sqrt{\BDpos p\newcdot\sigma} \nonumber \\
&=& \half\sqrt{\BDpos p\newcdot\sigma}\left[1
\BDplus
\frac{2s}{m}\sqrt{\BDpos p\newcdot\sigma}
S\newcdot\sigmabar\sqrt{\BDpos p\newcdot\sigma}\right]
\sqrt{\BDpos p\newcdot\sigma}
\nonumber \\
&=&\half\left[\BDpos p\newcdot\sigma \BDplus \frac{2s}{m}p\newcdot\sigma S\newcdot\sigmabar
p\newcdot\sigma\right] \nonumber \\
&=& \half\left[\BDpos p\newcdot\sigma \BDminus 2smS\newcdot\sigma\right]\,.
\eeqa
In the final step above, we simplified the product of three
dot products by noting that $p\newcdot S=0$ implies that $S\newcdot
\sigmabar\; p\newcdot\sigma =-p\newcdot\sigmabar\;S\newcdot\sigma$.
The other spin projection formulae for massive fermions can be
similarly derived.  The complete set of such formulae is given below:%
\footnote{Similar formulae for the products of two-component spinor
wave functions are given in ref.~\cite{Kersch:1985rn}.}
\beqa
&&x_\alpha({\boldsymbol{\vec p}},s) x^\dagger_{\dot{\beta}}
({\boldsymbol{\vec p}},s) = \half (\BDpos p_\mu \BDminus 2smS_\mu)
\sigma^\mu_{\alpha\dot{\beta}}
\,,
\label{xxdagmassive}
\\
&&y^{\dagger\dot{\alpha}}({\boldsymbol{\vec p}},s) y^\beta
({\boldsymbol{\vec p}},s) = \half (\BDpos p^\mu \BDplus 2sm S^\mu)
\sigmabar_\mu^{\dot{\alpha}{\beta}}
\,,
\\
&&x_\alpha({\boldsymbol{\vec p}},s)
y^\beta({\boldsymbol{\vec p}},s)
 = \half \left ( m \delta_\alpha{}^\beta - 2s
[S\newcdot \sigma \, p \newcdot \sigmabar]_\alpha{}^\beta \right )
\,,
\\
&&y^{\dagger\dot{\alpha}}
({\boldsymbol{\vec p}},s)
x^\dagger_{\dot{\beta}}({\boldsymbol{\vec p}},s)
 = \half \left ( m \delta^{\dot{\alpha}}{}_{\dot{\beta}} + 2s
[S\newcdot \sigmabar \, p \newcdot \sigma]^{\dot{\alpha}}{}_{\dot{\beta}}
\right )
\,.
\label{ydagxdagmassive}
\eeqa
By taking the hermitian conjugate of the above results, one obtains an
equivalent set of formulae,
\beqa
&&x^{\dagger\dot{\alpha}}({\boldsymbol{\vec p}},s)
x^\beta ({\boldsymbol{\vec p}},s)
= \half (\BDpos p^\mu \BDminus 2sm S^\mu)
\sigmabar_\mu^{\dot{\alpha}{\beta}}
\,,
\\
&&y_\alpha({\boldsymbol{\vec p}},s)
y^\dagger_{\dot{\beta}}
({\boldsymbol{\vec p}},s) = \half (\BDpos p_\mu \BDplus 2smS_\mu)
\sigma^\mu_{\alpha\dot{\beta}}
\,,
\\
&&y_\alpha({\boldsymbol{\vec p}},s)
x^\beta({\boldsymbol{\vec p}},s)
 = -\half \left (m \delta_\alpha{}^\beta + 2s
[S\newcdot \sigma \, p \newcdot \sigmabar]_\alpha{}^\beta \right )
\,,
\\
&&x^{\dagger\dot{\alpha}}({\boldsymbol{\vec p}},s)
y^\dagger_{\dot{\beta}}({\boldsymbol{\vec p}},s)
 = -\half \left ( m \delta^{\dot{\alpha}}{}_{\dot{\beta}}-2s
[S\newcdot \sigmabar \, p \newcdot \sigma]^{\dot{\alpha}}{}_{\dot{\beta}}
\right )
\,.
\label{eq:spinprojend}
\eeqa

For the case of massless spin-1/2 fermions, we must use helicity
spinor wave functions.  The corresponding massless projection
operators can be obtained directly from the explicit forms for the
two-component spinor wave functions given in
\eqst{helexplicitxa}{helexplicityb}:
\beqa
&& x_\alpha({\boldsymbol{\vec p}},\lambda) x^\dagger_{\dot{\beta}}
({\boldsymbol{\vec p}},\lambda) =
(\BDpos \half \BDminus \lambda)
p \newcdot \sigma_{\alpha \dot{\beta}}
\,,
\qquad \quad \quad\!\!\!\!\!
x^{\dagger\dot{\alpha}}({\boldsymbol{\vec p}},\lambda) x^\beta
({\boldsymbol{\vec p}},\lambda) =
(\BDpos \half \BDminus \lambda)
p \newcdot \sigmabar^{\dot{\alpha}\beta}
\,,
\label{xxdagmassless}
\\
&&
y^{\dagger\dot{\alpha}}({\boldsymbol{\vec p}},\lambda) y^\beta
({\boldsymbol{\vec p}},\lambda) =
(\BDpos \half \BDplus \lambda)
p \newcdot \sigmabar^{\dot{\alpha}\beta}
\,,
\qquad \quad \quad\,\!\!\!\!\!
y_\alpha({\boldsymbol{\vec p}},\lambda) y^\dagger_{\dot{\beta}}
({\boldsymbol{\vec p}},\lambda) =
(\BDpos \half \BDplus \lambda)
p \newcdot \sigma_{\alpha \dot{\beta}}
\,,
\label{yydagmassless}
\\
&&
x_\alpha({\boldsymbol{\vec p}},\lambda) y^\beta({\boldsymbol{\vec p}},\lambda)
= 0
\,,
\qquad \qquad \qquad\,\,\,\quad \quad
y_\alpha({\boldsymbol{\vec p}},\lambda) x^\beta({\boldsymbol{\vec p}},\lambda)
= 0
\,,
\\
&&
y^{\dagger\dot{\alpha}}({\boldsymbol{\vec p}},\lambda)
x^\dagger_{\dot{\beta}}({\boldsymbol{\vec p}},\lambda)= 0
\,,
\qquad \qquad \qquad\quad \,\,\,\quad
x^{\dagger\dot{\alpha}}({\boldsymbol{\vec p}},\lambda)
y^\dagger_{\dot{\beta}}({\boldsymbol{\vec p}},\lambda)= 0
\,.\phantom{xxx}
\label{ydagxdagmassless}
\eeqa
As a check, one can verify that the above results follow from
\eqst{xxdagmassive}{eq:spinprojend}, by replacing $s$ with
$\lambda$, setting $mS^\mu=p^\mu$, and taking
the $m\to 0$ limit at the end of the computation.

Having listed the projection operators for definite spin projection or
helicity, we may now sum over spins to derive the spin sum identities.
These arise when computing squared matrix elements for unpolarized
scattering and decay.  There are only four basic identities, but for
convenience we list each of them with the two-index height
permutations that can occur in squared amplitudes by following the
rules given in this paper.  The results can be derived by inspection
of the spin projection operators, since summing over $s=\pm\half$
simply removes all terms linear in the spin four-vector $S^\mu$.
\beqa
\sum_s x_\alpha({\boldsymbol{\vec p}},s)
x^\dagger_{\dot{\beta}}({\boldsymbol{\vec p}},s) =
\BDpos p\newcdot\sigma_{\alpha\dot{\beta}}
\,,
\qquad\qquad\qquad
\sum_s x^{\dagger\dot{\alpha}}({\boldsymbol{\vec p}},s)
 x^{\beta}({\boldsymbol{\vec p}},s) =
\BDpos p\newcdot\sigmabar^{\dot{\alpha}\beta}
\,,
\label{xxdagsummed}
\\
 \sum_s y^{\dagger\dot{\alpha}}({\boldsymbol{\vec p}},s)
y^{\beta}({\boldsymbol{\vec p}},s)
= \BDpos p \newcdot \sigmabar^{\dot{\alpha}\beta}
\,,
\qquad\qquad\qquad
\sum_s y_\alpha({\boldsymbol{\vec p}},s)
y^\dagger_{\dot{\beta}}({\boldsymbol{\vec p}},s) =
\BDpos p \newcdot \sigma_{\alpha\dot{\beta}}
\,,
\label{yydagsummed}
\\
\sum_s x_\alpha({\boldsymbol{\vec p}},s)
y^\beta({\boldsymbol{\vec p}},s) = m \delta_\alpha{}^\beta
\,,
\qquad\qquad\qquad
\sum_s y_\alpha({\boldsymbol{\vec p}},s)
x^\beta({\boldsymbol{\vec p}},s) = -m \delta_\alpha{}^\beta
\,,
\label{yxsummed}
\\
\sum_s y^{\dagger\dot{\alpha}}({\boldsymbol{\vec p}},s)
x^\dagger_{\dot{\beta}}({\boldsymbol{\vec p}},s) = m
\delta^{\dot{\alpha}}{}_{\dot{\beta}}
\,,
\qquad\qquad\qquad
\sum_s x^{\dagger\dot{\alpha}}({\boldsymbol{\vec p}},s)
y^\dagger_{\dot{\beta}}({\boldsymbol{\vec p}},s) = - m
\delta^{\dot{\alpha}}{}_{\dot{\beta}}
\,.
\label{ydagxdagsummed}
\eeqa
These results are applicable both to spin sums and helicity sums, and
hold for both massive and massless spin-1/2 fermions.

One can generalize the above massive and massless
projection operators by considering products of two-component spinor wave
functions, where the spin or helicity of each spinor can be
different.  These are the Bouchiat-Michel formulae~\cite{bouchmich},
which are derived in \app{H.3}.

\subsection{Fermion mass diagonalization in a
general theory}
\label{subsec:generalmass}
\renewcommand{\theequation}{\arabic{section}.\arabic{subsection}.\arabic{equation}}
\renewcommand{\thefigure}{\arabic{section}.\arabic{subsection}.\arabic{figure}}
\renewcommand{\thetable}{\arabic{section}.\arabic{subsection}.\arabic{table}}
\setcounter{equation}{0}
\setcounter{figure}{0}
\setcounter{table}{0}

Consider a collection of free anticommuting two-component spin-1/2
fields, $\hat\xi_{\alpha i}(x)$, which transform as $(\half,0)$ fields
under the Lorentz group.  Here, $\alpha$ is the spinor index, and $i$
labels the distinct fields of the collection.  The free-field
Lagrangian is given by (e.g., see ref.~\cite{Valle-Schechter1}):
\beq \label{Ldiag}
\mathscr{L} =
i{\hat\xi}^{\dagger i}\sigmabar^\mu\partial_\mu\hat\xi_i
- \half M^{ij}\hat\xi_i\hat\xi_j
- \half M_{ij}{\hat\xi}^{\dagger i}{\hat\xi}^{\dagger j}\,,
\eeq
where
\beq
M_{ij}\equiv (M^{ij})^*.
\label{eq:complexindexconvention}
\eeq
Note that $M$ is a complex symmetric matrix, since the
product of anticommuting two-component fields satisfies
$\hat\xi_i\hat\xi_j=\hat\xi_j\hat\xi_i$
[with the spinor contraction rule according to \eq{suppressionrule}].

In \eq{Ldiag}, we have employed the U($N$)-covariant tensor
calculus~\cite{tung,Cv} for ``flavor-tensors'' labeled by the
flavor indices $i$ and $j$. Each left-handed $(\half,0)$ fermion
always has an index with the opposite height of the corresponding
right-handed $(0,\half)$ fermion.  Raised indices can only be
contracted with lowered indices and vice versa.  Flipping the heights
of all flavor indices of an object corresponds to complex conjugation,
as in \eq{eq:complexindexconvention}.  In particular, we generalize
\eq{eq:defbardagger} as follows:\footnote{\label{fnglob}%
In the case at hand, we
have more specifically chosen all of the left-handed fermions to have
lowered flavor indices, which implies that all of the right-handed
fermions have raised flavor indices. However, in cases where a subset
of left-handed fermions transform according to some representation $R$
of a (global) symmetry and a different subset of left-handed
fermions transform according to the conjugate representation $R^*$, it
is often more convenient to employ a raised flavor index for the
latter subset of left-handed fields.}
\beq
\psi^{\dagger\,i}_{\dot{\alpha}}\equiv (\psi_{\alpha i})^\dagger\,.
\eeq
If $M=0$, then the free-field Lagrangian is invariant under a global U($N$)
symmetry.  That is, for a unitary matrix $U$, with matrix elements
$U_i{}^j$, and its hermitian conjugate defined by:
\beq \label{Udagger}
(U^\dagger)_i{}^j=(U_j{}^i)^*\equiv U^j{}_i\,,
\eeq
with $U_i{}^k (U^\dagger)_k{}^j=\delta_i^j$,
the massless free-field Lagrangian is invariant
under the transformations:
\beq \label{unsym}
\hat\xi_i\longrightarrow  U_i{}^j\hat\xi_j\,,\qquad\qquad \hat\xi^{\dagger i}
\longrightarrow U^i{}_j\hat\xi^{\dagger j}\,.
\eeq
For $M\neq 0$, \eq{Ldiag} remains formally invariant under the global
U($N$)-symmetry if $M$ acts as a spurion field~\cite{spurion}
with the appropriate
tensorial transformation law, $M^{ij}\longrightarrow U^i{}_kU^j{}_\ell
M^{k\ell}$.

Expressions consisting of flavor-vectors and second-rank
flavor-tensors have natural interpretations as products of vectors
and matrices.  As a result, the flavor indices can be suppressed,
and the resulting expressions can be written in an index-free
matrix notation.  To accomplish this, one must first assign a
particular flavor index structure to the matrices that will appear
in the index-free expression.  For example, given the second-rank
flavor-tensors introduced above, we define the matrix elements of
$M$ to be $M^{ij}$ and the matrix elements of $U$ to be $U_i{}^j$.
Note that $(U^\dagger)_i{}^j$ has the same flavor-index structure as
$U$.\footnote{The reader should not be tempted
to substitute $U^\dagger$ for $U$ in \eq{Udagger}, as the resulting
flavor-index structure for $U$ and $U^\dagger$ would then disagree with the
original flavor-index assignments.}

As a simple example, in an index-free notation \eq{unsym} reads:
$\hat\xi\longrightarrow
U\hat\xi$ and $\hat\xi^\dagger\longrightarrow\hat\xi^\dagger
U^\dagger$.  A slightly more complicated example is exhibited below:
\beq
U^i{}_k M^{k\ell}=(U^\dagger)_k{}^i M^{k\ell}=(U^*M)^{i\ell}\,,
\eeq
where we have used $(U^\dagger)^{\T}=U^*$ in obtaining the final
result.  That is, in matrix
notation with suppressed indices, $U^i{}_k M^{k\ell}$ corresponds to
the matrix $U^*M$.  Thus, in an index-free notation,
the tensorial transformation law for the spurion field $M$ is
given by $M\longrightarrow U^*MU^\dagger$.

We can diagonalize the mass matrix $M$ and rewrite the
Lagrangian in terms of
mass eigenstates $\xi_{\alpha i}$ and
(real non-negative) masses $m_i$.  To do this,
we introduce a unitary matrix $\Omega$,
\beq \label{Omegadef}
\hat\xi_i= \Omega_i{}^k\xi_k\,,
\eeq
and demand that $M^{ij}\Omega_i{}^k\Omega_j{}^\ell=m_k\delta^{k\ell}$
(no sum over $k$), where the $m_k$ are real and non-negative.
Equivalently, in matrix notation with suppressed
indices, $\hat\xi=\Omega\xi$ and\footnote{\label{fsing}%
In general, the $m_i$ are \textit{not} the eigenvalues of $M$.  Rather,
they are the \textit{singular values} of the matrix $M$, which are
defined to be the non-negative square roots of the eigenvalues of
$M^\dagger M$.  See \app{D} for further details.}
\beq
\label{takagi}
\Omega^{\T} M\, \Omega = {\boldsymbol m} = {\rm diag}(m_1,m_2,\ldots).
\eeq
This is the Takagi diagonalization~\cite{takagi,horn} of
an arbitrary complex symmetric matrix, which is discussed in more
detail in \app{D.2}. To compute the values of the diagonal elements
of~${\boldsymbol m}$, note that
\beq \label{diagmm}
\Omega^\dagger M^\dagger M\Omega = {\boldsymbol
m}^2\,.
\eeq
Indeed $M^\dagger M$ is hermitian and thus it can be diagonalized
by a unitary matrix. Hence, the elements of the diagonal matrix
$\boldsymbol{m}$ are the non-negative square roots of the
corresponding eigenvalues of $M^\dagger M$.  However, in cases where
$M^\dagger M$ has degenerate eigenvalues, \eq{diagmm}
\textit{cannot} be employed to determine the unitary matrix
$\Omega$ that satisfies \eq{takagi}. A more general technique for
determining $\Omega$ that works in all cases is given in
\app{D.2}.

In terms of the mass eigenstates,
\beq \label{lagMajorana}
\mathscr{L}=
i\xi^{\dagger i}\sigmabar^\mu\partial_\mu\xi_i- \half m_{i}(\xi_i\xi_i+
\xi^{\dagger i} \xi^{\dagger i})\,,
\eeq
where the sum over $i$ is implicit.  If the $m_i\neq 0$ 
are non-degenerate, then
the corresponding field $\xi_i$ describes a neutral Majorana fermion
consisting of two on-shell real degrees of freedom.  
The case of mass degeneracies will be treated explicitly below.
If $m_i=0$, then we shall denote the corresponding field $\xi_i$
as a massless Weyl fermion~\cite{WeylDirac}.

Each $\xi_{\alpha i}$  can now be expanded in a Fourier series, exactly
as in eq.~(\ref{modes}):
\beq \label{twocompmodes}
\xi_{\alpha i}(x)=
\sum_s\int\,\frac{d^3 \boldsymbol{\vec p}}
{(2\pi)^{3/2}(2E_{\boldsymbol{p}i})^{1/2}}
\left[x_{\alpha}(\boldsymbol{\vec p},s)a_i(\boldsymbol{\vec
p},s)e^{\BDneg ip\newcdot x}+y_{\alpha}(\boldsymbol{\vec p},s)a_i^\dagger(
\boldsymbol{\vec p},s)e^{\BDpos ip\newcdot x}\right]\,,
\eeq
where $E_{\boldsymbol{p}i}\equiv (|{\boldsymbol{\vec p}}|^2+m_i^2)^{1/2}$,
and the creation and annihilation operators, $a_i^\dagger$ and $a_i$
satisfy anticommutation relations:
\beq \label{etacr}
\{a_i(\boldsymbol{\vec p},s),a_j^\dagger(\boldsymbol{\vec
p}^{\,\prime},s^\prime)\}=\delta^3(\boldsymbol{\vec p}
-\boldsymbol{\vec p}^{\,\prime})
\delta_{ss'}\delta_{ij}\,.
\eeq
We employ covariant normalization of the one-particle states,
i.e., we act with one creation operator on the vacuum with the
following convention
\beq
\label{xistate}
\ket{\boldsymbol{\vec p},i,s}\equiv (2\pi)^{3/2}(2E_{\boldsymbol{p}i})^{1/2}
a_i^\dagger({\boldsymbol{\vec p}},s)\ket{0}\,,
\eeq
so that $\vev{\boldsymbol{\vec p},i,s\,|\,
\boldsymbol{\vec p}^{\,\,\prime},j,s'} =
(2\pi)^{3}(2E_{\boldsymbol{p}i})\delta^3(\boldsymbol{\vec p}
-\boldsymbol{\vec p}^{\,\,\prime}) \delta_{ij} \delta_{ss'}$.

In the case of two mass-degenerate massive fermion fields, 
$m_1=m_2\neq 0$,
\eq{lagMajorana} possesses a global internal O(2)
flavor symmetry,
$\xi_i\to \mathcal{O}_i{}^j\xi_j$ ($i=1,2$),
where $\mathcal{O}^{\T}\mathcal{O}=
\mathds{1}_{2\times 2}$.  Corresponding to this symmetry is a
conserved hermitian Noether current:
\beq \label{Noether}
J^\mu=i(\xi^{\dagger 1}\sigmabar^\mu\xi_2
- \xi^{\dagger 2}\sigmabar^\mu\xi_1)\,,
\eeq
with a corresponding conserved charge, $Q=\int J^0 d^3 x$.  In the
$\xi_1$--$\xi_2$ basis, the Noether current is off-diagonal.  However, it is
convenient to define a new basis of fields:
\beq \label{chietadirac}
\chi\equiv\frac{1}{\sqrt{2}}(\xi_1+i\xi_2)\,,\qquad\quad
\eta\equiv\frac{1}{\sqrt{2}}(\xi_1-i\xi_2)\,.
\eeq
With respect to the $\chi$--$\eta$ basis, the Noether current is
diagonal:
\beq
J^\mu =  \chi^\dagger\sigmabar^\mu\chi
- \eta^\dagger\sigmabar^\mu\eta\,.
\eeq
That is, the fermions $\chi$ and $\eta$ are eigenstates of the charge
operator $Q$ with corresponding eigenvalues $\pm 1$.  In terms of
the fermion fields of definite charge, the free-field fermion
Lagrangian [\eq{lagMajorana} with $i=1,2$ and $m_1=m_2\equiv m$] is
given by~\cite{Uhlenbeck}:\footnote{Although the
fermion mass matrix is not diagonal in the
$\chi$--$\eta$ basis, this is not an obstacle to the subsequent
analysis, as one only
needs a diagonal {\em squared}-mass matrix, $M^\dagger M$, to ensure that the
denominators of propagators are diagonal.
\Eq{chietadirac}
provides the explicit Takagi diagonalization of the
Dirac fermion matrix $\left(\begin{smallmatrix} 0 & 1 \\ 1 & 0
\end{smallmatrix}\right)$.
See \app{D.3} for the mathematical interpretation of this
special case.}
\beq
\label{lagDirac}
\mathscr{L}= i\chi^\dagger\sigmabar^\mu\partial_\mu\chi +
  i\eta^\dagger\sigmabar^\mu\partial_\mu\eta-m(\chi\eta
  +\chi^\dagger\eta^\dagger)\,.
\eeq
On-shell, $\chi$ and $\eta$ satisfy the
free-field Dirac equations:
\beq
i\,\sigmabar^\mu\partial_\mu\chi-m\eta^\dagger=0\,,\qquad\quad
i\,\sigmabar^\mu\partial_\mu\eta-m\chi^\dagger=0\,.
\label{diraceq1}
\eeq
In the $\chi$--$\eta$ basis,
the global internal SO(2) symmetry
(which is continuously connected to the identity) is
realized as the U(1) symmetry $\chi\to e^{i\theta}\chi$ and
$\eta\to e^{-i\theta}\eta$, where $\theta$ is the rotation angle that
defines the SO(2) rotation matrix.

Together, $\chi$ and $\eta^\dagger$ constitute a single Dirac fermion.
We can then write:
\beqa
 \chi_\alpha(x) &=&\sum_s
\int\,\frac{d^3 \boldsymbol{\vec p}}
{(2\pi)^{3/2}(2E_{\boldsymbol p})^{1/2}}
\left[x_\alpha(\boldsymbol{\vec p},s)a(\boldsymbol{\vec
p},s)e^{\BDneg ip\newcdot x}+y_\alpha(\boldsymbol{\vec p},s)b^\dagger(
\boldsymbol{\vec p},s)e^{\BDpos ip\newcdot x}\right]\,, \label{chiexpansion}
\\[5pt]
 \eta_\alpha(x) &=&\sum_s
\int\,\frac{d^3 \boldsymbol{\vec p}}
{(2\pi)^{3/2}(2E_{\boldsymbol p})^{1/2}}
\left[x_\alpha(\boldsymbol{\vec p},s)b(\boldsymbol{\vec
p},s)e^{\BDneg ip\newcdot x}+y_\alpha(\boldsymbol{\vec p},s)a^\dagger(
\boldsymbol{\vec p},s)e^{\BDpos ip\newcdot x}\right]\,, \label{etaexpansion}
\eeqa
where $E_{\boldsymbol p}\equiv (|{\boldsymbol{\vec p}}|^2+m^2)^{1/2}$,
the creation and annihilation operators, $a^\dagger$, $b^\dagger$, $a$
and $b$ satisfy anticommutation relations:
\beq
\{a(\boldsymbol{\vec p},s),a^\dagger(\boldsymbol{\vec p}^{\,\prime},s^\prime)\}
=
\{b(\boldsymbol{\vec p},s),b^\dagger
(\boldsymbol{\vec p}^{\,\prime},s^\prime)\}=\delta^3(\boldsymbol{\vec p}
-\boldsymbol{\vec p}^{\,\prime})
\delta_{s,s'}
\,,
\eeq
and all other anticommutators vanish.  We now must distinguish between
two types of one-particle states, which we can call fermion $(F)$ and
antifermion $(\overline{F})$:
\beq
\ket{\boldsymbol{\vec p},s;F}\equiv  (2\pi)^{3/2}(2E_{\boldsymbol
p})^{1/2}
a^\dagger({\boldsymbol{\vec p}},s)\ket{0}\,, \qquad\quad
\ket{\boldsymbol{\vec p},s;\overline{F}}
\equiv  (2\pi)^{3/2}(2E_{\boldsymbol p})^{1/2}
b^\dagger({\boldsymbol{\vec p}},s)\ket{0}\,. \label{chietastate}
\eeq
Note that both $\eta(x)$ and $\chi^\dagger(x)$ can create
$\ket{\boldsymbol{\vec p},s;F}$ from the vacuum, while $\eta^\dagger(x)$
and $\chi(x)$ can create $\ket{\boldsymbol{\vec p},s;\overline{F}}$.  The
one-particle wave functions are given by:
\beqa
\bra{0}\chi_\alpha(x)\ket{\boldsymbol{\vec p},s;F} &=&
x_\alpha({\boldsymbol{\vec p}},s)e^{\BDneg ip\newcdot x}\,,
\qquad\quad
\bra{0}\eta^\dagger_{\dot{\alpha}}(x)\ket{\boldsymbol{\vec p},s;F}=
y^\dagger_{\dot{\alpha}}({\boldsymbol{\vec p}},s)e^{\BDneg ip\newcdot x}\,,
\label{chainstate}\\
\bra{F;{\boldsymbol{\vec p}},s}\eta_\alpha(x)\ket{0} &=&
y_\alpha({\boldsymbol{\vec p}},s)e^{\BDpos ip\newcdot
x}\,,
\qquad\quad
\bra{F;{\boldsymbol{\vec p}},s}\chi^\dagger_{\dot{\alpha}}(x)\ket{0}=
x^\dagger_{\dot{\alpha}}({\boldsymbol{\vec p}},s)e^{\BDpos ip\newcdot x}\,,
\\
\label{chaoutstate}
\bra{0}\eta_\alpha(x)\ket{\boldsymbol{\vec p},s;\overline{F}} &=&
x_\alpha({\boldsymbol{\vec p}},s)e^{\BDneg ip\newcdot x}\,,
\qquad\quad
\bra{0}\chi^\dagger_{\dot{\alpha}}(x)\ket{\boldsymbol{\vec p},s;\overline{F}}=
y^\dagger_{\dot{\alpha}}({\boldsymbol{\vec p}},s)e^{\BDneg ip\newcdot x}\,,
\label{chbinstate}\\
\bra{\overline{F};{\boldsymbol{\vec p}},s}\chi_\alpha(x)\ket{0} &=&
y_\alpha({\boldsymbol{\vec p}},s)e^{\BDpos ip\newcdot
x}\,,
\qquad\quad
\bra{\overline{F};{\boldsymbol{\vec p}},s}\eta^\dagger_{\dot{\alpha}}(x)\ket{0}=
x^\dagger_{\dot{\alpha}}({\boldsymbol{\vec p}},s)e^{\BDpos ip\newcdot x}\,,
\label{chboutstate}
\eeqa
and the eight other single-particle matrix elements vanish.

More generally, consider a collection of free anticommuting
charged Dirac fermions, which can be represented by
pairs of two-component fields
$\hat\chi_{\alpha i}(x)$, $\hat\eta_{\alpha }^i(x)$.
These fields transform in (possibly
reducible) representations of
the unbroken symmetry group that are conjugates of each other.
This accounts for the opposite flavor index heights of $\hat\chi_i$
and~$\hat\eta^i$ [cf.~footnote~\ref{fnglob}].  The free-field Lagrangian
is given by
\beq
\label{lagDiracmixed}
\mathscr{L}=
  i{\hat\chi}^{\dagger i}\sigmabar^\mu\partial_\mu\hat\chi_i
+
  i{\hat\eta}^\dagger_{i}\sigmabar^\mu\partial_\mu\hat\eta^i
-M^i{}_j \hat\chi_i\hat\eta^j
-M_i{}^j  {\hat\chi}^{\dagger i}\hat\eta^\dagger_{ j}\,,
\eeq
where $M$ is an arbitrary complex matrix with matrix elements
$M^i{}_j$, and
\beq \label{Mdagger}
M_i{}^j\equiv (M^i{}_j)^*\,.
\eeq
If $M=0$, then the free-field Lagrangian is invariant under a global
U($N$)$\times$U($N$)
symmetry.  That is, for a pair of unitary matrices $U_L$ and $U_R$,
with matrix elements given respectively by $(U_L)_i{}^j$
and $(U_R)^i{}_j$, and the corresponding hermitian
conjugates defined by:
\beq \label{ulurdagger}
(U^\dagger_{L})_j{}^i=[(U_{L})_i{}^j]^*\equiv (U_{L})^i{}_j\,,\qquad\quad
(U^\dagger_{R})^j{}_i=[(U_{R})^i{}_j]^*\equiv (U_{R})_i{}^j\,,
\eeq
the massless free-field
Lagrangian is invariant under the transformations:
\beq
\hat\chi\ls{i}\longrightarrow  (U_L)_i{}^j\hat\chi\ls{j}\,,\qquad
\hat\chi^{\dagger i}\longrightarrow (U_L)^i{}_j
\hat\chi^{\dagger j}\,,\qquad
\hat\eta^i\longrightarrow  (U_R)^i{}_j\hat\eta^j\,,\qquad
\hat\eta^\dagger_i\longrightarrow (U_R)_i{}^j\hat\eta^\dagger_j\,.
\label{unun}
\eeq
For $M\neq 0$, \eq{lagDiracmixed} remains formally
invariant under the U($N$)$\times$U($N$)
symmetry if $M$ acts as a spurion
field~\cite{spurion}
with the appropriate tensorial transformation law,
$M^i{}_j\to (U_L)^i{}_k(U_R)_j{}^\ell M^k{}_\ell$
(or equivalently, in an index-free matrix
notation with suppressed flavor indices,
$M\longrightarrow U_L^{*} MU_R^\dagger$).

In order to diagonalize the mass
matrix, we introduce the mass eigenstates $\chi_i$ and
$\eta^i$ and unitary matrices $L$ and $R$, with matrix elements
given respectively by $L_i{}^k$ and $R^i{}_k$, such that
\beq
\label{lrdef}
\hat\chi_i=L_i{}^k\chi_k\,,\qquad\qquad
\hat\eta^i=R^i{}_k\eta^k\,,
\eeq
and demand that
$M^i{}_j L_i{}^k R^j{}_\ell = m_k\delta^k_\ell$ (no sum over $k$),
where the $m_k$ are real and non-negative.
Equivalently, in matrix notation with suppressed indices,
$\hat\chi=L\chi\,,\,\hat\eta=R\eta$ and
\beq
\label{svd}
L^{\T} M R= {\boldsymbol{m}}={\rm diag}(m_1,m_2,\ldots),
\eeq
with the $m_i$ real and non-negative (cf.~footnote~\ref{fsing}).
The singular value
decomposition of linear algebra, discussed more fully in \app{D.1},
states that for any complex matrix $M$, unitary matrices $L$ and $R$
exist such that \eq{svd} is satisfied.  It then follows that:
\beqa
L^{\T}(MM^\dagger)L^* \,=\,
R^\dagger(M^\dagger M) R \,=\, {\boldsymbol{m}}^2 .
\eeqa
That is, since $MM^\dagger$
and $M^\dagger M$ are both hermitian,
they can be diagonalized by unitary matrices.
The diagonal elements of $\boldsymbol{m}$ are therefore
the non-negative square roots of the
corresponding eigenvalues of $MM^\dagger$ (or equivalently, $M^\dagger M$).
In terms of the
mass eigenstates,
\beq
\label{lagDiracdiag}
\mathscr{L}= i{\chi}^{\dagger i}\sigmabar^\mu\partial_\mu\chi_i+
  i{\eta}^\dagger_i \sigmabar^\mu \partial_\mu\eta^i
- m_i(\chi_i\eta^i + \chi^{\dagger i} \eta^\dagger_i)\,.
\eeq
The mass matrix now consists of $2\times 2$ blocks
$\bigl(\begin{smallmatrix}0 & m_i \\ m_i & 0\end{smallmatrix}\bigr)$
along the diagonal.  More importantly, the squared-mass matrix is
diagonal with doubly degenerate entries $m_i^2$ that will appear in
the denominators of the propagators of the theory.
For $m_i\neq 0$, each $\chi_i$--$\eta^i$ pair describes a charged
Dirac fermion consisting of four on-shell real degrees of freedom.%
\footnote{Of course, one could
always choose instead to treat the Dirac fermions in a
non-charge-eigenstate basis with a fully
diagonalized mass matrix, as in \eq{lagMajorana}.
Inverting \eq{chietadirac} for each Dirac fermion yields
$\xi_{2i-1} = (\chi_i + \eta^i)/\sqrt{2}$ and $\xi_{2i} = i
(\eta_i - \chi^i)/\sqrt{2}$.
However, it is rarely, if ever,
convenient to do so; practical calculations only require that the
squared-mass matrix $M^\dagger M$
is diagonal, and it is of course more convenient to
employ fields that carry well-defined charges.}
In addition, \eq{lagDiracdiag} yields
an even number of massless Weyl fermions.%

Given an arbitrary collection of two-component left-handed $(\half,0)$
fermions, the distinction between Majorana and Dirac fermions
depends on whether the Lagrangian is invariant under
a global (or local) continuous
symmetry group $G$, and the corresponding multiplet structure of the fermion
fields~\cite{Napoly}.
If no such continuous symmetry exist, then the fermion
mass eigenstates will consist of Majorana fermions.  If the
Lagrangian is invariant under a symmetry group~$G$, then the
collection of two-component fermions will break up into a sum of
multiplets that transform irreducibly under $G$.
As described in \app{E}, a representation $R$ can be either
a real, pseudo-real, or complex representation of $G$.  If a multiplet
transforms under a real
representation of $G$, then the corresponding fermion mass eigenstates
are Majorana fermions.\footnote{This is a slight
generalization of the more restrictive
definition that requires Majorana fermions to transform trivially
under the group $G$.
Gluinos, which transform under the (real) adjoint representation of
the color SU(3) group, are Majorana fermions according to our more
general definition. \label{fnmajorana}}
If a multiplet transforms under a complex
representation of $G$, then the corresponding fermion mass eigenstates
are Dirac fermions.  In particular [as noted above \eq{lagDiracmixed}], if the
$\chi\ls{i}$ transform under the representation $R$, then the
$\eta^i$ transform under the conjugate representation $R^*$.

The case where a multiplet
of two-component left-handed fermions transform under a pseudo-real
representation of $G$ has not been explicitly treated above.
The simplest example of this kind is a model
of $2n$ multiplets (or ``flavors'')
of two-component SU(2)-doublet\footnote{The doublet
representation of SU(2) is pseudo-real.}
fermions, $\hat{\psi}_{ia}$
(where $i=1,2,\ldots,2n$ labels the flavor index and
$a$ labels the SU(2) doublet index).
The free-field Lagrangian is given
by:
\beq \label{LagPseudoreal}
\mathscr{L}=i\hat{\psi}^{\dagger ia}\sigmabar^\mu\partial_\mu\hat{\psi}_{ia}
-\half \left(M^{ij}\epsilon^{ab}\hat{\psi}_{ia}\hat{\psi}_{jb}+{\rm h.c.}
\right)\,,
\eeq
where $\epsilon^{ab}$
is the antisymmetric SU(2)-invariant tensor,
defined such that $\epsilon^{12}=-\epsilon^{21}=+1$.
As $\epsilon^{ab}\hat{\psi}_{ia}\hat{\psi}_{jb}$ is antisymmetric
under the interchange of flavor indices
$i$ and $j$, it follows that $M$ is a
complex antisymmetric matrix.  To identify the fermion
mass eigenstates
$\psi_{ja}$,
we introduce a unitary matrix $U$
(with matrix elements $U_i{}^j$)
such that
$\hat{\psi}_{ia}=U_i{}^j\psi_{ja}$
and demand that:
\beq \label{pseudonormal}
U^{\T} M U = \boldsymbol{N}
\equiv {\rm diag} \left\{\begin{pmatrix} \phm 0 & m_1 \\ -m_1 & 0
\end{pmatrix}\,,\,\begin{pmatrix} \phm 0 & m_2 \\ -m_2 & 0
\end{pmatrix}\,,\,\cdots \,,\begin{pmatrix} \phm 0 & m_n \\ -m_n & 0
\end{pmatrix}\right\}\,,
\eeq
where $\boldsymbol{N}$ is written in block-diagonal form consisting of
$2\times 2$ matrix blocks appearing along the diagonal,
and the $m_j$ are real and non-negative. \Eq{pseudonormal}
corresponds to the reduction of a complex antisymmetric matrix
to its real normal form~\cite{zum}, which is discussed in
more detail in \app{D.4}.  In order to compute the $m_k$, we first
note that
\beq \label{doublemass}
U^\dagger M^\dagger M U={\rm diag}(m_1^2\,,\,m_1^2\,,\,m_2^2\,,\,m_2^2\,,\,
\ldots\,,\,m_n^2\,,\,m_n^2)\,.
\eeq
Hence, the $m_j$ are the non-negative square roots of the
corresponding eigenvalues of $M^\dagger M$.  Since
the dimension of the doublet representation of SU(2) provides an
additional degeneracy factor of 2, \eq{doublemass} implies that
the mass spectrum consists of $2n$ pairs of mass-degenerate
two-component fermions, which are equivalent to $2n$ Dirac
fermions. In particular,
\beq \label{pseudoDirac}
\mathscr{L}=\sum_{i=1}^{2n}i\psi^{\dagger
  ia}\sigmabar^\mu\partial_\mu\psi_{ia}
-\sum_{i=1}^n\left(
m_i \epsilon^{ab}\psi_{2i-1,\,a}\psi_{2i,\,b}+{\rm h.c.}
\right)\,.
\eeq

In the general case of a pseudo-real representation $R$ (of
dimension $d_R$), the SU(2)-invariant
$\epsilon$-tensor is replaced by a more general $d_R\times d_R$
unitary antisymmetric matrix, $C$ [defined in \eq{unitaryC}].
Thus, the analysis above can be repeated
virtually unchanged.  By defining
\beq
\qquad\qquad\quad
\chi_{ia}\equiv \psi_{2i-1,\,a}\,,\qquad \eta^{ia}\equiv
C^{ab}\psi_{2i,\,b}\,,\qquad i=1,2,\ldots,n\,;\quad a=1,2,\ldots,d_R\,,
\eeq
with an implicit sum over the repeated index $b$,
the resulting Lagrangian given by
\beq \label{lagDiracpseudo}
\mathscr{L}= \sum_{i=1}^n
i{\chi}^{\dagger ia}\sigmabar^\mu\partial_\mu\chi_{ia}+
 i{\eta}^\dagger_{ia} \sigmabar^\mu \partial_\mu\eta^{ia}
-m_i\left(\chi_{ia}\eta^{ia}+\chi^{\dagger ia}\eta^\dagger_{ia}\right)\,,
\eeq
describes a free field theory of $nd_R$ Dirac fermions
[cf.~\eq{lagDiracdiag}].  Therefore, if a multiplet
of two-component left-handed fermions transforms under a pseudo-real
representation of $G$, then the corresponding fermion mass eigenstates
are Dirac fermions~\cite{Napoly}.
If \eq{LagPseudoreal}
contains an odd number of pseudo-real fermion
multiplets, then
the (antisymmetric) mass matrix $M$ is odd-dimensional and thus has
an odd number of zero eigenvalues
[according to \eq{antinormal}].  But as $d_R$ must be even,
it follows that the pseudo-real fermion multiplet
contains an even number of massless
Weyl fermions.

In conclusion, the mass diagonalization procedure of an arbitrary 
field theory of fermions 
yields (in general) a set of massless Weyl fermions, a set of massive 
neutral Majorana fermions [as
in \eq{lagMajorana}], and a set of massive charged Dirac fermions
[as in \eq{lagDiracdiag}].  
The Feynman rules for these mass eigenstate
two-component fermion fields are given in
\sec{sec:externalfermionrules}.

For completeness, we review the squared-mass matrix diagonalization
procedure for scalar fields.  First, consider a collection of free
commuting real spin-0 fields, $\hat\varphi_i(x)$, where the flavor
index $i$ labels the distinct scalar fields of the collection.
The free-field Lagrangian is given by%
\footnote{Since the scalar fields
are real, there is no need to distinguish between
raised and lowered flavor indices.}
\beq \label{Lagrscalars}
\mathscr{L}=
\BDpos \half \partial_\mu\hat\varphi_i\partial^\mu\hat\varphi_i
-\half M^2_{ij}\hat\varphi_i\hat\varphi_j\,,
\eeq
where $M^2$ is a real symmetric matrix.  We diagonalize the
scalar squared-mass matrix by introducing mass eigenstates $\varphi_i$
and the orthogonal matrix $Q$ such that
$\hat\varphi_i=Q_{ij}\varphi_j$,
with $M^2_{ij} Q_{ik}Q_{j\ell}=m^2_k\delta_{k\ell}$
(no sum over $k$).
In matrix form,
\beq \label{qtrans}
Q^{\T}M^2Q=\boldsymbol{m}^2={\rm diag}(m_1^2,m_2^2,\ldots)\,,
\eeq
where the squared-mass eigenvalues $m_k^2$ are real.\footnote{\label{fntwo}%
If the vacuum corresponds to
a local minimum (or flat direction)
of the scalar potential, then the squared-mass eigenvalues of
$M^2$ are real \textit{and} non-negative.}
This is the standard diagonalization problem for a real symmetric
matrix.

Next, consider a collection of free
commuting complex spin-0 fields, $\hat\Phi_i(x)$.
For complex fields, we follow the
conventions for flavor indices enunciated below \eq{eq:complexindexconvention}
[e.g. $\hat\Phi^i=(\hat\Phi_i)^\dagger$].
The free-field Lagrangian is given by
\beq \label{Lagcscalars}
\mathscr{L}=
\BDpos \partial_\mu\hat\Phi^i\partial^\mu\hat\Phi_i
-(M^2)^i{}_j\hat\Phi_i\hat\Phi^j\,,
\eeq
where $M^2$ is an hermitian matrix [i.e.,
$(M^2)^i{}_j=(M^2)_j{}^i$ in the notation
of \eq{ulurdagger}].

We diagonalize the
scalar squared-mass matrix by introducing mass eigenstates $\Phi_i$
and the unitary matrix $W$ such that
$\hat\Phi_i=W_i{}^k\Phi_k$ (and $\hat\Phi^i=W^i{}_k\Phi^k$),
with $(M^2)^i{}_j W_i{}^k W^j{}_\ell=m^2_k\delta^k_\ell$
(no sum over $k$).  In matrix form,
\beq \label{wtrans}
W^\dagger M^2 W=\boldsymbol{m}^2={\rm diag}(m_1^2,m_2^2,\ldots)\,.
\eeq
where the squared-mass eigenvalues
$m_k^2$ are real (cf.~footnote~\ref{fntwo}).
This is the standard diagonalization problem for an hermitian
matrix.

\section{\texorpdfstring{Feynman rules with two-component spinors}{Feynman rules with two-component spinors}}
\label{sec:externalfermionrules}
\renewcommand{\theequation}{\arabic{section}.\arabic{equation}}
\renewcommand{\thefigure}{\arabic{section}.\arabic{figure}}
\renewcommand{\thetable}{\arabic{section}.\arabic{table}}
\setcounter{equation}{0}
\setcounter{figure}{0}
\setcounter{table}{0}

In order to systematically perform perturbative
calculations using two-component spinors, we present the basic
Feynman rules. The Feynman rules for the Standard Model (and its
seesaw extension) and the MSSM (including possible R-parity-violating
interactions) are given in
Appendices J, K and L. Feynman rules for two-component spinors have also been
treated in
refs.~\cite{Ticciati,Berends:1987cv,Dittmaier:1998nn}.

\subsection{External fermion and boson rules\label{subsec:externallines}}
\label{externalrules}
\renewcommand{\theequation}{\arabic{section}.\arabic{subsection}.\arabic{equation}}
\renewcommand{\thefigure}{\arabic{section}.\arabic{subsection}.\arabic{figure}}
\renewcommand{\thetable}{\arabic{section}.\arabic{subsection}.\arabic{table}}
\setcounter{equation}{0}
\setcounter{figure}{0}
\setcounter{table}{0}

Consider a general theory, for which we may assume that the
mass matrix for fermions has been diagonalized as discussed in
\sec{subsec:generalmass}.
The rules for assigning two-component external state spinors
are then as follows:
\beqa
\bullet &&\,\,\textrm{For an initial state (incoming)
left-handed \LH\ fermion:} \quad x \nonumber \\
\bullet &&\,\,\textrm{For an initial state (incoming)
right-handed \RH\ fermion:} \quad y^\dagger \nonumber \\
\bullet &&\,\,\textrm{For a final state (outgoing) left-handed
\LH\ fermion:} \quad  x^\dagger\nonumber \\
\bullet &&\,\,\textrm{For a final state (outgoing) right-handed
\RH\ fermion:}\quad y\nonumber
\eeqa
where we have suppressed the momentum and spin arguments of the
spinor wave functions.  These rules are summarized in the mnemonic diagram
of \fig{fig:mnemonic}.

\begin{figure}[ht!]
\vspace{.5cm}
\begin{picture}(350,150)(-200,-75)
\thicklines
\GCirc(0,0){21}{0.8}
\put(-60.5,42){$x$}
\ArrowLine(-57,57)(-14.3,14.3)
\ArrowLine(14.3,14.3)(57,57)
\put(51.5,40.5){$x^\dagger$}
\ArrowLine(-14.3,-14.3)(-57,-57)
\put(-60.5,-42){$y^\dagger$}
\ArrowLine(57,-57)(14.3,-14.3)
\put(51.5,-42){$y$}
\put(-37,70){$L$~~\LH\ fermion}
\put(-37,-77){$R$~~\RH\ fermion}
\put (-138,0){Initial State}
\put (82,0){Final State}
\end{picture}
\vspace{.4cm}
\caption[0]{\label{fig:mnemonic} The external wave function spinors
should be assigned as indicated here, for initial state and final state
left-handed \protect{$(\half,0)$} and right-handed
\protect{$(0,\half)$} fermions.}
\end{figure}
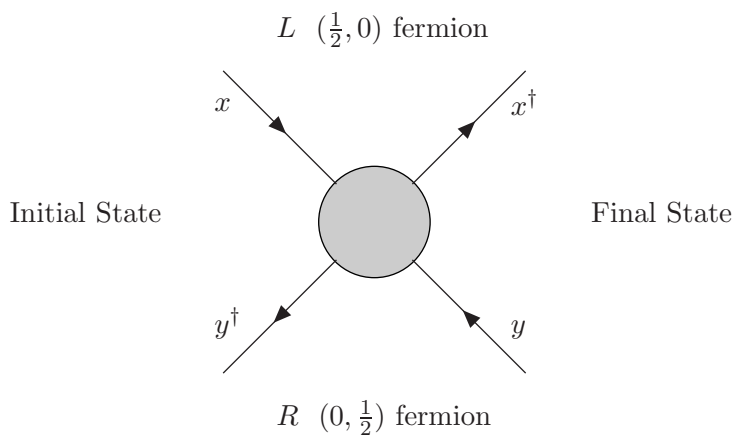

In general, the two-component external state fermion wave
functions are
distinguished by their Lorentz group transformation properties,
rather than by their particle or antiparticle status
as in four-component Feynman rules. This helps to explain why
two-component notation is especially convenient for (i) theories with
Majorana
particles, in which there is no fundamental distinction between particles
and antiparticles, and (ii) theories like the Standard Model and MSSM in
which the left and right-handed fermions transform under different
representations of the gauge group
and (iii) problems with polarized particle beams.

In contrast to four-component Feynman rules (given in \app{G.5}), the
direction of the arrow does {\it not} correspond to the flow of
charge or fermion number.
The two-component Feynman rules for external fermion lines simply correspond to
the formulae for the one-particle wave functions exhibited in
\eqs{instate}{outstate} [with the convention that
$\ket{\boldsymbol{\vec p},s}$ is an initial state fermion and
$\bra{\boldsymbol{\vec p},s}$ is a final state fermion].
In particular, the arrows indicate the spinor index structure, with
fields of undotted indices flowing \textit{into} any vertex and
fields of dotted indices flowing \textit{out} of any vertex.

The rules above apply to any mass eigenstate two-component
fermion external wave functions. It is noteworthy that the same rules
apply for the two-component fermions governed by the Lagrangians of
\eq{lagMajorana} [Majorana] and eqs.~(\ref{lagDiracdiag})
or (\ref{lagDiracpseudo}) [Dirac].

The corresponding rules for external boson lines are well-known (see, e.g
\Ref{Peskin:1995ev}).
\beqa
\bullet &&\,\,\textrm{For an initial state (incoming) or
final state (outgoing) spin-0 boson}:
\qquad\qquad\quad 1\nonumber \\
\hspace{-2in}\bullet &&\,\,\textrm{For an initial state (incoming)
spin-1 boson of momentum}~\boldsymbol{\vec k}~\textrm{and
helicity}~\lambda:
\quad \quad\!\! \varepsilon^\mu(\boldsymbol{\vec k}\,,\,\lambda) \nonumber \\
\bullet &&\,\,\textrm{For a final state (outgoing)
spin-1 boson of momentum}~\boldsymbol{\vec k}~\textrm{and
helicity}~\lambda:
\quad \qquad\varepsilon^{\mu}(\boldsymbol{\vec k}\,,\,\lambda)^\ast
\nonumber
\eeqa
The explicit form of the helicity $\pm 1$
(massless or massive) spin-1 polarization
vector $\varepsilon^\mu$ is given in \eq{varepsk}.
The helicity zero massive spin-1 polarization vector is given
in \eq{massivespinone}.

\subsection{Propagators}
\label{subsec:fermionprops}
\renewcommand{\theequation}{\arabic{section}.\arabic{subsection}.\arabic{equation}}
\renewcommand{\thefigure}{\arabic{section}.\arabic{subsection}.\arabic{figure}}
\renewcommand{\thetable}{\arabic{section}.\arabic{subsection}.\arabic{table}}
\setcounter{equation}{0}
\setcounter{figure}{0}
\setcounter{table}{0}

Next we turn to the subject of fermion propagators for two-component
fermions. A derivation of the two-component fermion propagators using
path integral techniques is given in \app{F}.  Here, we will follow
the more elementary approach typically given in an initial textbook
treatment of quantum field theory.

Fermion propagators are the Fourier transforms of the free-field vacuum
expectation values of time-ordered products of two fermion fields.
They are obtained by inserting the free-field expansion of the
two-component fermion field and evaluating the spin sums using the
formulae given in \eqs{xxdagsummed}{ydagxdagsummed}.  For the case of
a single neutral two-component fermion field $\xi(x)$ of mass $m$,
\eqs{twocompmodes}{etacr} yield
\cite{Case,Ticciati,Valle-Schechter2,Berends:1987cv,Dittmaier:1993jj,
Dittmaier:1998nn}:
\beqa
\bra{0}T\xi_\alpha(x)\xi^\dagger_{\dot{\beta}}(y)\ket{0}_{\rm FT} &=&
\frac{\BDpos i}{p^2 \BDminus m^2 \BDplus i\epsilon}\sum_s\,
x_\alpha({\boldsymbol{\vec p}},s)
x^\dagger_{\dot{\beta}}({\boldsymbol{\vec p}},s)
=\frac{ i}{p^2 \BDminus m^2 \BDplus i\epsilon}
\,p\newcdot\sigma_{\alpha\dot{\beta}}\,,
\phantom{x} \label{ft1}
\\[5pt]
\bra{0}T
\xi^{\dagger\dot{\alpha}}(x)\xi^\beta(y)\ket{0}_{\rm FT} &=&
\frac{\BDpos i}{p^2 \BDminus m^2 \BDplus i\epsilon}\sum_s\,
y^{\dagger\dot{\alpha}}({\boldsymbol{\vec p}},s)
y^\beta({\boldsymbol{\vec p}},s)
=\frac{i}{p^2 \BDminus m^2 \BDplus i\epsilon}
\,p\newcdot\sigmabar^{\dot{\alpha}\beta}\,,
\phantom{x} \label{ft2}
\\[5pt]
\bra{0}T
\xi^{\dagger\dot{\alpha}}(x)
\xi^\dagger_{\dot{\beta}}(y)\ket{0}_{\rm FT} &=&
\frac{\BDpos i}{p^2 \BDminus m^2 \BDplus i\epsilon}\sum_s\,
y^{\dagger\dot{\alpha}}({\boldsymbol{\vec p}},s)
x^\dagger_{\dot{\beta}}({\boldsymbol{\vec p}},s)
=\frac{\BDpos i}{p^2 \BDminus m^2 \BDplus i\epsilon}
\,m\delta^{\dot{\alpha}}{}_{\dot{\beta}}\,,
\phantom{x} \label{ft3}
\\[5pt]
\bra{0}T\xi_\alpha(x)\xi^\beta(y)\ket{0}_{\rm FT} &=&
\frac{\BDpos i}{p^2 \BDminus m^2 \BDplus i\epsilon}\sum_s\,
x_\alpha({\boldsymbol{\vec p}},s)
y^\beta({\boldsymbol{\vec p}},s)
=\frac{\BDpos i}{p^2 \BDminus m^2 \BDplus i\epsilon}
\,m\delta_\alpha{}^\beta\,,
\phantom{x} \label{ft4}
\eeqa
where FT indicates the Fourier transform from position to momentum
space.\footnote{\label{footnotefnft}%
The Fourier transform of a translationally invariant
function $f(x,y)\equiv f(x-y)$ is given by
$$
f(x,y)=\int\frac{d^4 p}{(2\pi)^4}\,\widehat f(p)\,e^{\BDneg ip\newcdot (x-y)}
\,,\qquad {\rm where} \qquad
\widehat f(p) = \int d^4 x\, f(x) e^{\BDpos ip\newcdot x}\,.
$$
In the notation of the text above,
$f(x,y)\ls{\rm FT}\equiv\widehat f(p)$.}
These results have a clear diagrammatic representation, as shown in
\fig{fig:neutprop}.
\begin{figure}[t!]
\centerline{
\begin{picture}(300,60)(-135,-26)
\thicklines
\LongArrow(-110,25)(-70,25)
\ArrowLine(-130,15)(-50,15)
\put(-170,10){(a)}
\put(10,10){(b)}
\put(-90,30){$p$}
\put(-56,1){$\alpha$}
\put(-130,4){$\dot{\beta}$}
\LongArrow(70,25)(110,25)
\ArrowLine(130,15)(50,15)
\put(90,30){$p$}
\put(50,3){$\beta$}
\put(120,4){$\dot{\alpha}$}
\put(-115,-20){$\displaystyle
 \frac{ip\newcdot\sigma_{\alpha\dot{\beta}}}{p^2 \BDminus m^2}
\qquad\qquad\qquad\qquad\qquad\qquad\quad\>
\frac{ip\newcdot\sigmabar^{\dot{\alpha}\beta}}{p^2 \BDminus m^2}
$
}
\end{picture}
}
\vspace{.3cm}
\centerline{
\begin{picture}(300,65)(-135,-26)
\thicklines
\ArrowLine(-130,15)(-90,15)
\ArrowLine(-50,15)(-90,15)
\put(-170,10){(c)}
\put(10,10){(d)}
\put(-130,4){$\dot{\beta}$}
\put(-56,1){$\dot{\alpha}$}
\ArrowLine(90,15)(130,15)
\ArrowLine(90,15)(50,15)
\put(120,4){$\alpha$}
\put(50,3){$\beta$}
\put(-115,-20){$\displaystyle
\frac{\BDpos im}{p^2 \BDminus m^2} \delta^{\dot{\alpha}}{}_{\dot{\beta}}
\qquad\qquad\qquad\qquad\qquad\quad\>
\frac{\BDpos im}{p^2 \BDminus m^2} \delta_{{\alpha}}{}^{{\beta}}
$}
\end{picture}
}
\caption[0]{\label{fig:neutprop} Feynman rules for propagator lines of
a neutral two-component fermion with mass $m$.
(For simplicity, the $\BDplus i\epsilon$ terms in the denominators
are omitted in all propagator rules.)}
\end{figure}
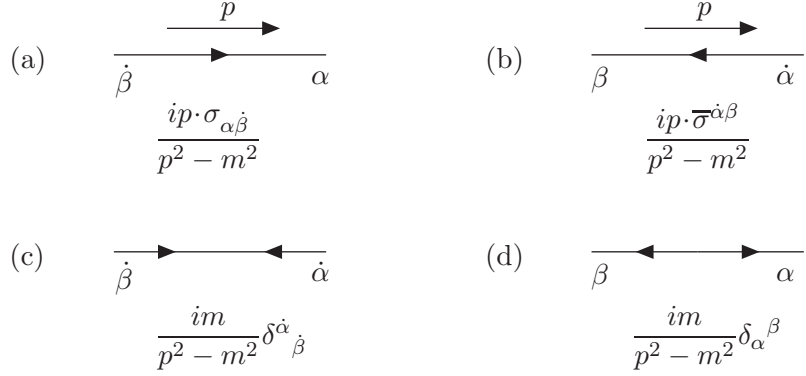
Note that the direction of the momentum flow $p^\mu$ here is
determined by the creation operator that appears in the evaluation of
the free-field propagator.  Arrows on fermion lines always run away
from dotted indices at a vertex and toward undotted indices at a
vertex.

There are clearly two types of fermion propagators.
The first type preserves the direction of arrows,
so it has one dotted and one undotted index.  For this type of propagator,
it is convenient to establish a convention where $p^\mu$ in the
diagram is defined to be the momentum flowing in the direction of the
arrow on the fermion propagator.  With this convention, the two
rules above for propagators of the first type can be summarized by
one rule, as shown in \fig{fig:neutproprev}.
Here the choice of the $\sigma$ or the $\sigmabar$ version of the rule is
uniquely determined by the height of the indices on the vertex
to which the propagator is connected.\footnote{The second form
of the rule in \fig{fig:neutproprev}
arises when one flips diagram~(b) of \fig{fig:neutprop}
around by a 180$^\circ$ rotation (about an axis perpendicular to the
plane of the diagram), and then relabels $p\to -p$,
$\dot\alpha\to\dot\beta$ and $\beta\to\alpha$.}
These heights should always be chosen so that they are contracted
as in eq.~(\ref{suppressionrule}).  It should be noted that
in diagrams~(a) and (b) of \fig{fig:neutprop} as drawn, the
indices on the $\sigma$ and $\sigmabar$ read from right to left.
In particular, the Feynman rules for the propagator can be employed with
the spinor indices suppressed provided that the arrow-preserving
propagator lines are traversed in the
direction parallel [antiparallel] to the arrowed line segment for the
$\sigmabar$ [$\sigma$] version of the rule, respectively.

\begin{figure}[t!]
\centerline{
\begin{picture}(300,38)(-50,-3)
\thicklines
\LongArrow(-20,25)(20,25)
\ArrowLine(-40,15)(40,15)
\put(-40,1){$\dot{\beta}$}
\put(33,6){$\alpha$}
\put(0,30){$p$}
\put(92,16){$\displaystyle
 \frac{ip\newcdot \sigma_{\alpha\dot{\beta}}}{p^2 \BDminus m^2}$
\quad{\underline{or}}\quad
$\displaystyle
 \frac{-ip\newcdot \sigmabar^{\dot{\beta}\alpha}}{p^2 \BDminus m^2}$
}
\end{picture}
}
\caption[0]{\label{fig:neutproprev} This rule summarizes the results
of both Figs.~\ref{fig:neutprop}(a) and (b) for a neutral two-component
fermion with mass $m$.}
\end{figure}

The second type of propagator shown in diagrams (c) and (d) of
\fig{fig:neutprop} does not preserve
the direction of arrows, and corresponds to an
odd number of mass insertions. The indices on
$\delta^{\dot{\alpha}}{}_{\dot{\beta}}$ and $\delta_{\alpha}{}^{\beta}$
are staggered as shown to
indicate that $\dot\alpha$ and $\alpha$ are to be contracted with
expressions to the left, while $\dot\beta$ and $\beta$ are to
be contracted with expressions to the right, in accord with
eq.~(\ref{suppressionrule}).\footnote{As in \fig{fig:neutproprev},
alternative and equivalent
versions of the rules corresponding to diagrams (c) and
(d) of \fig{fig:neutprop}
can be given for which the indices on the Kronecker
deltas are staggered as $\delta^{\dot\beta}{}_{\dot{\alpha}}$
and $\delta_\beta{}^\alpha$.  These versions correspond to
flipping the two respective diagrams by 180$^\circ$ and relabeling
the indices $\dot\alpha\to\dot\beta$ and $\beta\to\alpha$.}

Starting with massless fermion propagators, one can also derive the
massive fermion propagators by employing mass insertions as
interaction vertices, as shown in \fig{fig:massinsertion}.
By summing up an infinite chain of such mass insertions between massless
fermion propagators, one can reproduce the massive
fermion propagators of both types.

\begin{figure}[ht!]
\centerline{
\begin{picture}(350,60)(-180,-20)
\thicklines
\ArrowLine(-126,15)(-90,15)
\ArrowLine(-54,15)(-90,15)
\put(-126,19){$\beta$}
\put(-59,19){$\alpha$}
\ArrowLine(90,15)(126,15)
\ArrowLine(90,15)(54,15)
\put(-97,11.5){\Large{$\boldsymbol\times$}}
\put(83,11.5){\Large{$\boldsymbol\times$}}
\put(121,19){$\dot{\alpha}$}
\put(54,19){$\dot{\beta}$}
\put(-107,-14){$\displaystyle
-i m
\delta_{\alpha}{}^{\beta}
$}
\put(73,-14){$\displaystyle
-i m
\delta^{\dot{\alpha}}{}_{\dot{\beta}}
$}
\end{picture}
}
\caption[0]{\label{fig:massinsertion} Fermion mass insertions
(indicated by the crosses) can
be treated as a type of interaction vertex, using the Feynman rules shown
here.}
\end{figure}

The above results for the propagator of a Majorana fermion can be
generalized to a multiplet of mass eigenstate
Majorana fermions, $\xi_{\alpha a}(x)$ [such as a color octet of gluinos],
which transforms as a real representation $R$ of a
(gauge or flavor) group~$G$
(where $a=1,2,\ldots, d_R$ for a representation of dimension $d_R$).
In this case, the Feynman graphs given in
\figst{fig:neutprop}{fig:massinsertion}
are modified simply by specifying a group index $a$ and $b$ at either
end of the propagator line.
The corresponding Feynman rules then include an additional
Kronecker delta factor in the group indices.  In particular,
if we associate the index $a$ with the spinor indices $\alpha$,
$\dot{\alpha}$ and the index $b$ with the spinor indices $\beta$,
$\dot{\beta}$, then the rules exhibited in \fig{fig:neutprop}(a) and (b)
would include the following Kronecker delta factors:
\beq
(a)~~\delta_a^b\,,\phantom{ m^{cd} \delta^b_d = m_a\delta^{ab}}
\qquad  \qquad\qquad  \qquad
 (b)~~\delta^a_b\,,\phantom{ m_{cd} \delta_b^d = m_a \delta_{ab}}
\eeq
and the factors of $m$ in the rules exhibited in \fig{fig:neutprop}(c)
and (d) would be replaced by
\beq
(c)~~\delta_c^a m^{cd} \delta^b_d = m_a\delta^{ab} \,,
\qquad\qquad  \qquad  \qquad
(d)~~\delta_a^c m_{cd} \delta_b^d = m_a \delta_{ab} \,,
\eeq
(with no sum over the repeated index $a$),
where $m^{cd}$ and $m_{cd}\equiv m^{cd}$ are diagonal matrices
with real non-negative diagonal elements $m_c$.  Here, we have introduced the
separate symbol $m_{cd}$ in order to maintain the convention that two
repeated group indices are summed when one index is raised and one
index is lowered.  Of course,  if the Lagrangian is invariant under
the symmetry group $G$, then a multiplet of
Majorana fermions corresponding to an irreducible representation $R$
has a common mass $m=m_a$.

It is convenient to treat separately the
case of charged massive fermions.  Consider a
charged Dirac fermion of mass $m$, which is described by a pair of
two-component fields $\chi(x)$ and $\eta(x)$ [cf.~\eq{lagDirac}].
Using the free-field expansions
[\eqs{chiexpansion}{etaexpansion}] and the spin sums
[\eqst{xxdagsummed}{ydagxdagsummed}], the two-component free-field
propagators are obtained:
\beqa
\bra{0}T\chi_{\alpha}(x)\chi^\dagger_{\dot{\beta}}(y)\ket{0}_{\rm FT}
=\bra{0}T\eta_{\alpha}(x)\eta^\dagger_{\dot{\beta}}(y)\ket{0}_{\rm FT}&=&
\frac{i}{p^2 \BDminus m^2}
\,p\newcdot\sigma_{\alpha\dot{\beta}}\,,\label{chprop1} \\[5pt]
\bra{0}T\chi^{\dagger\dot{\alpha}}(x)\chi^{\beta}(y)\ket{0}_{\rm FT}
=\bra{0}T\eta^{\dagger\dot{\alpha}}(x)\eta^{\beta}(y)\ket{0}_{\rm FT} &=&
\frac{i}{p^2 \BDminus m^2}
\,p\newcdot\sigmabar^{\dot{\alpha}\beta}\,, \label{chprop2} \\[5pt]
\bra{0}T\chi_{\alpha}(x)\eta^\beta(y)\ket{0}_{\rm FT}
=\bra{0}T\eta_{\alpha}(x)\chi^\beta(y)\ket{0}_{\rm FT}&=&
\frac{\BDpos i}{p^2 \BDminus m^2}
\,m\,\delta_\alpha{}^\beta\,,\label{chprop3}  \\[5pt]
\bra{0}T \chi^{\dagger\dot{\alpha}}(x)
\eta^\dagger_{\dot{\beta}}(y)\ket{0}_{\rm FT}
=\bra{0}T
\eta^{\dagger\dot{\alpha}}(x)\chi^\dagger_{\dot{\beta}}(y)\ket{0}_{\rm FT}
&=& \frac{\BDpos i}{p^2 \BDminus m^2}
\,m\,\delta^{\dot{\alpha}}{}_{\dot{\beta}}\,. \label{chprop4}
\eeqa
For all other combinations of fermion bilinears, the corresponding
two-point functions vanish.
These results again have a simple diagrammatic representation, as shown
in \fig{fig:Diracpropagators}.
\begin{figure}[t!]
\centerline{
\begin{picture}(350,65)(-150,-26)
\thicklines
\ArrowLine(-130,15)(-50,15)
\LongArrow(-110,25)(-70,25)
\put(-180,10){(a)}
\put(50,10){(b)}
\put(-140,15){$\chi$}
\put(-45,15){$\chi$}
\put(185,15){$\eta$}
\put(90,15){$\eta$}
\put(-90,30){$p$}
\put(-56,4){$\alpha$}
\put(-130,1){$\dot{\beta}$}
\ArrowLine(100,15)(180,15)
\LongArrow(120,25)(160,25)
\put(100,1){$\dot{\beta}$}
\put(170,4){$\alpha$}
\put(140,30){$p$}
\put(-155,-26){$\displaystyle
 \frac{ip\newcdot \sigma_{\alpha\dot{\beta}}}{p^2 \BDminus m^2}
\qquad{\rm \underline{or}}\qquad
 \frac{-ip\newcdot \sigmabar^{\dot{\beta}\alpha}}{p^2 \BDminus m^2}
$}
\put(80,-26){$\displaystyle
 \frac{ip\newcdot \sigma_{\alpha\dot{\beta}}}{p^2 \BDminus m^2}
\qquad{\rm \underline{or}}\qquad
 \frac{-ip\newcdot \sigmabar^{\dot{\beta}\alpha}}{p^2 \BDminus m^2}
$}
\end{picture}
}

\vspace{.57cm}
\centerline{
\begin{picture}(350,65)(-150,-26)
\thicklines
\ArrowLine(-130,15)(-90,15)
\ArrowLine(-50,15)(-90,15)
\put(-140,15){$\chi$}
\put(-45,15){$\eta$}
\put(185,15){$\eta$}
\put(90,15){$\chi$}
\put(-180,10){(c)}
\put(50,10){(d)}
\put(-130,1){$\dot{\beta}$}
\put(-56,4){$\dot{\alpha}$}
\ArrowLine(140,15)(180,15)
\ArrowLine(140,15)(100,15)
\put(170,4){$\alpha$}
\put(100,2){$\beta$}
\put(-115,-20){$\displaystyle
\frac{\BDpos im}{p^2 \BDminus m^2} \delta^{\dot{\alpha}}{}_{\dot{\beta}}
\,\,\,\qquad\qquad\qquad\qquad\qquad\qquad\qquad\quad
\frac{\BDpos im}{p^2 \BDminus m^2} \delta_{{\alpha}}{}^{{\beta}}
$}
\end{picture}
}
\caption{\label{fig:Diracpropagators}
Feynman rules for propagator lines of a pair of charged two-component
fermions with a Dirac mass $m$.  As in \fig{fig:neutproprev},
the direction of the momentum is taken to flow from the dotted to
the undotted index in diagrams (a) and (b).}
\end{figure}
Note that for Dirac fermions, the propagators with
opposing arrows
(proportional to a mass) necessarily change the identity ($\chi$ or
$\eta$) of the
two-component fermion, while the single-arrow propagators are
diagonal in the fields.
In processes involving such a charged fermion, one must of course
distinguish between the $\chi$ and $\eta$ fields.

The above results for the propagator of a Dirac fermion can be
generalized to a multiplet of mass eigenstate Dirac
fermions, $\chi\ls{\alpha i}$, $\eta\ls{\alpha}^i$,
which transform under a (gauge or flavor) group~$G$.  In this case,
the Feynman graphs given in \fig{fig:Diracpropagators}
are modified simply by specifying a group index $i$ and $j$ at either
end of the propagator line.  The
corresponding Feynman rules then include an additional
Kronecker delta factor in the group indices.
In particular, if we associate the group index $i$
with the spinor indices $\alpha$, $\dot{\alpha}$ and the index $j$
with the spinor indices $\beta$, $\dot{\beta}$, then the rules
exhibited in \fig{fig:Diracpropagators}(a) and (b) would include the
following Kronecker delta factors:
\beq
(a)~~\delta_i^j \phantom{ m_\ell{}^n \delta^j_n = m_i\delta_i^j}
\qquad\qquad\qquad\qquad \,
(b)~~\delta_j^i\,,
\phantom{ m^\ell{}_n \delta_j^n = m_i \delta^i_j}
\eeq
and the factors of $m$ in the rules exhibited in
\fig{fig:Diracpropagators}(c) and (d) would be replaced by
\beq
(c)~~\delta_i^\ell m_\ell{}^n \delta^j_n = m_i\delta_i^j \,,
\qquad\qquad  \qquad  \qquad
(d)~~\delta^i_\ell m^\ell{}_n \delta_j^n = m_i \delta^i_j \,,
\eeq
where $m^\ell{}_n$ and $m_\ell{}^n\equiv m^\ell{}_n$ are diagonal matrices
with real non-negative diagonal elements $m_\ell$, and there is no
sum over the repeated index $i$.  (Here, we have introduced the
separate symbol $m_\ell{}^n$ in order to maintain the convention that two
repeated group indices are summed when one index is raised and one
index is lowered.)  As before, if the Lagrangian is invariant under
the symmetry group $G$,
then an irreducible multiplet of Dirac fermions has a common mass $m=m_i$.

For completeness, we exhibit in \fig{fig:bosonprops} the
Feynman rules for the propagators of the
(neutral or charged) scalar boson and gauge
boson in the $R_\xi$ gauge, with gauge parameter~$\xi$~\cite{chengli}.
\vskip 0.1in
\begin{figure}[hb!]
\begin{center}
\begin{picture}(320,11)(0,5)
\thicklines
\DashLine(10,10)(90,10)6
\Text(125,10)[l]{$\displaystyle \frac{\BDpos i}{p^2 \BDminus m^2}$}
\end{picture}
\end{center}
\begin{center}
\begin{picture}(320,25)(0,3)
\thicklines
\Photon(10,10)(90,10){3}{6}
\Text(16,1)[c]{$\mu,a$}
\Text(84,1)[c]{$\nu,b$}
\Text(125,8.5)[l]{$\displaystyle \frac{-i}{p^2 \BDminus m^2}
\left [
\metric^{\mu\nu} - (1-\xi) \frac{p^\mu p^\nu}{p^2 \BDminus \xi m^2}
\right ]\,\delta^{ab}
$}
\end{picture}
\end{center}
\caption{\label{fig:bosonprops}
Feynman rules for the (neutral or charged) scalar and gauge boson propagators,
in the $R_\xi$ gauge,
where $p^\mu$ is the propagating four-momentum.  In the gauge
boson propagator,
$\xi=1$ defines the 't~Hooft-Feynman gauge, $\xi=0$ defines the Landau gauge,
and $\xi\to\infty$ defines the unitary gauge.  For
the propagation of a non-abelian gauge
boson, one must also specify the adjoint representation indices $a$, $b$.
}
\end{figure}

\subsection{Fermion interactions with bosons}
\label{subsec:fermioninteractions}
\renewcommand{\theequation}{\arabic{section}.\arabic{subsection}.\arabic{equation}}
\renewcommand{\thefigure}{\arabic{section}.\arabic{subsection}.\arabic{figure}}
\renewcommand{\thetable}{\arabic{section}.\arabic{subsection}.\arabic{table}}
\setcounter{equation}{0}
\setcounter{figure}{0}
\setcounter{table}{0}

We next discuss the interaction vertices for fermions with bosons.
Renormalizable Lorentz-invariant interactions involving fermions
must consist of bilinears in the fermion fields, which transform as a Lorentz
scalar or vector, coupled to the appropriate bosonic scalar or vector
field to make an overall Lorentz scalar quantity.

Let us write all of the two-component left-handed
$(\half, 0)$ fermions of the theory
as $\hat\psi_i$, where $i$ runs over all of the gauge group representation
and flavor degrees of freedom.  In general, the
$(\half, 0)$-fermion fields $\hat\psi_i$ consist
of Majorana fermions $\hat \xi_i$, and
Dirac fermion pairs $\hat \chi_i$ and $\hat \eta^i$ after mass
terms (both explicit and coming from spontaneous symmetry breaking) are
taken into account.  Likewise, consider a multiplet of scalar fields
$\hat\phi_I$, where $I$ runs over all of the gauge group representation
and flavor degrees of freedom.  In general, the scalar fields
$\hat\phi_I$ consist of real scalar fields
$\hat\varphi_I$ and pairs of complex scalar fields
$\hat\Phi_I$ and $\hat\Phi^{I}\equiv(\hat\Phi_I)^\dagger$.
In matrix form,
\beq
\hat \psi \equiv \begin{pmatrix}\hat\xi \\ \hat\chi \\ \hat\eta
\end{pmatrix}\,,\qquad\qquad
\hat \phi \equiv \begin{pmatrix}\hat\varphi^{\phs} \\
\hat\Phi^{\phs} \\ \hat \Phi^\dagger \end{pmatrix}\,.
\eeq
By dividing up the fermions into Majorana and Dirac fermions and the
spin-zero fields into real and complex scalars, we are
assuming implicitly that some of the indices $I$ and $i$ correspond to
states of a definite (global) U(1)-charge (denoted in the following
by $q_I$ and $q_i$, respectively).

The most general set of Yukawa interactions of the scalar fields
with a pair of fermion fields is then given by:
\beq
\mathscr{L}_{\rm int} = -\half \hat Y^{Ijk} \hat\phi_I\hat\psi_j\hat\psi_k
-\half \hat Y_{Ijk}\hat\phi^{I} {\hat\psi}^{\dagger j} {\hat\psi}^{\dagger k}
\,,
\label{eq:lintY}
\eeq
where $\hat Y_{Ijk} = (\hat Y^{Ijk})^*$ is symmetric under the interchange of $j$ and $k$.
We have suppressed the spinor indices; the
product of two-component spinors is always performed according to the
index convention indicated in \eq{suppressionrule}.
The Yukawa Lagrangian [\eq{eq:lintY}] must be invariant under:
\beq
\hat\xi_i\to\hat\xi_i\,,\!\!\qquad
\hat\chi_i\to e^{iq_i\theta}\hat\chi_i\,,
\!\!\qquad
\hat\eta^i\to e^{-iq_i\theta}\hat\eta^i\,,\!\!\qquad
\hat\varphi_i\to\hat\varphi_i\,,
\!\!\qquad
\hat\Phi_I\to e^{iq_I\theta}\hat\Phi_I\,,\!\!\qquad
\hat\Phi^I\to e^{-iq_I\theta}\hat\Phi^I\,,
\eeq
where the $q_i$ are the U(1)-charges of the corresponding Dirac fermions
and the $q_I$ are the U(1)-charges of the corresponding complex scalars.
Consequently, the form of the $\hat Y^{Ijk}$ is constrained:

\beq \label{qcon}
\hat Y^{Ijk}=0\,,\quad {\rm unless}\quad  q_I+q_j+q_k=0\,.
\eeq
Of course, any other conserved symmetries will impose additional selection
rules on the Yukawa couplings $\hat Y^{Ijk}$.

The hatted fields are the interaction eigenstate fields.
However, in general the mass eigenstates can be different, as discussed in
\sec{subsec:generalmass}.
The computation of matrix elements for physical processes
is more conveniently done in terms of the propagating
mass eigenstate fields.   The mass eigenstate
basis $\psi$ is related to the interaction eigenstate basis $\hat \psi$ by
a unitary rotation ${U_i}^j$ on the flavor indices.
In matrix form:
\beq
\label{massrotate}
\hat \psi \equiv \begin{pmatrix}\hat\xi\ls{i} \\[3pt] \hat\chi\ls{i} \\[3pt]
\hat\eta^i
\end{pmatrix}= U \psi
\equiv \begin{pmatrix}\Omega_i{}^j & 0 & 0 \\[3pt]
                0 & L_i{}^j & 0 \\[3pt]
                0 & 0 & R^i{}_j\end{pmatrix}
\begin{pmatrix}\xi\ls{j} \\[3pt] \chi\ls{j} \\[3pt] \eta^j\end{pmatrix} \,,
\eeq
where $\Omega$, $L$, and $R$ are constructed as described previously in
\sec{subsec:generalmass} [see \eqs{takagi}{svd}].
Likewise, the mass eigenstate
basis $\phi$ is related to the interaction eigenstate basis $\hat \phi$ by
a unitary rotation ${V_I}^J$ on the flavor indices.
In matrix form,
\beq
\label{mass2rotate}
\hat \phi \equiv \begin{pmatrix}\hat\varphi\ls{I}
\\[3pt] \hat\Phi\ls{I} \\[3pt] \hat\Phi^I
\end{pmatrix}= V \phi
\equiv \begin{pmatrix}Q_I{}^J & 0 & 0\\[3pt]
                0 & W_I{}^J & 0 \\[3pt] 0 & 0 & W^I{}_J\end{pmatrix}
\begin{pmatrix}\varphi\ls{J} \\[3pt] \Phi\ls{J} \\[3pt]
\Phi^J \end{pmatrix} \,,
\eeq
where $ W^I{}_J=(W_I{}^J)^*$, and
$Q$ and $W$ are constructed according to \eqs{qtrans}{wtrans}.

\begin{figure}[t!]
\begin{center}
\begin{picture}(200,64)(30,20)
\DashArrowLine(10,40)(60,40)5
\ArrowLine(100,70)(60,40)
\ArrowLine(100,10)(60,40)
\Text(18,30)[]{$I$}
\Text(75,10)[]{$k, \beta$}
\Text(75,70)[]{$j, \alpha$}
\Text(128,40)[l]{$-i Y^{Ijk}\delta_\alpha{}^\beta \quad
{\rm or} \quad -i Y^{Ijk}\delta_\beta{}^\alpha$}
\Text(-40,40)[]{(a)}
\end{picture}
\end{center}
\begin{center}
\begin{picture}(200,75)(30,20)
\DashArrowLine(60,40)(10,40)5
\ArrowLine(60,40)(100,70)
\ArrowLine(60,40)(100,10)
\Text(18,30)[]{$I$}
\Text(75,10)[]{$k, \dot\beta$}
\Text(75,70)[]{$j, \dot\alpha$}
\Text(128,40)[l]{$-i Y_{Ijk}\delta^{\dot\alpha}{}_{\dot\beta}
\quad {\rm or} \quad
-i Y_{Ijk}\delta^{\dot\beta}{}_{\dot\alpha}$}
\Text(-40,40)[]{(b)}
\end{picture}
\end{center}
\caption{\label{fig:Yukawavertexrules}
  {Feynman rules for Yukawa couplings of scalars to two-component
    fermions in a general field theory.  The choice of which rule to
    use depends on how the vertex connects to the rest of the
    amplitude. When indices are suppressed, the spinor index part is
    always just proportional to the identity matrix.}}
\end{figure}
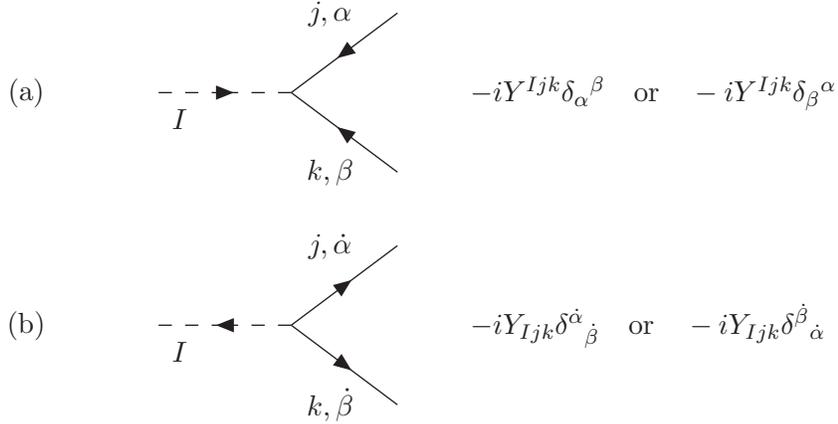

Thus, we may rewrite \eq{eq:lintY} in terms of mass eigenstate fields:
\beq
\mathscr{L}_{\rm int} = -\half Y^{Ijk} \phi_I\psi_j\psi_k
-\half Y_{Ijk} \phi^{I} {\psi}^{\dagger j} {\psi}^{\dagger k}
\,,
\label{eq:lintYmass}
\eeq
where
\beq
Y^{Ijk}=Y^{Ikj}\equiv V_J{}^I U_m{}^j U_n{}^k \hat Y^{Jmn}\,.
\eeq
Note that \eq{qcon} implies that $Y^{Ijk}=0$ unless $q_I+q_j+q_k=0$.
The corresponding Feynman rules
that arise from the Yukawa interaction Lagrangian
are shown in
Fig.~\ref{fig:Yukawavertexrules}.  If the scalar
$\phi_I$ is complex, then one can associate an arrow with the flow
of analyticity, which would point into the vertex in (a)
and would point out of the vertex in (b).
That is, the arrow on the scalar line keeps track of the height of the
scalar flavor index entering or leaving
the vertex.

In \fig{fig:Yukawavertexrules},
two versions are given for each Feynman
rule.  The choice of which rule to use is
dictated by the height of the indices on the fermion lines that connect
to the vertex.  These heights should always be chosen so that they are
contracted as in eq.~(\ref{suppressionrule}).  However,
when all spinor indices are suppressed, the scalar-fermion-fermion rules will
have an identical appearance for both cases, since they are
just proportional to the identity matrix of the $2\times 2$ spinor
space.

To provide a more concrete example of the above results, consider
a real neutral scalar field $\phi$ and a (possibly) complex charged scalar
field $\Phi$ (with U(1)-charge $q\ls{\Phi}$)
that interact with a multiplet of
Majorana fermions $\xi_i$ and Dirac fermion pairs
$\chi_j$ and $\eta^j$ (with U(1)-charges $q_j$ and $-q_j$, respectively).
We assume that all fields are given in the mass eigenstate basis.
The Yukawa interaction Lagrangian is given by:
\beqa \label{concrete}
\mathscr{L}_{\rm int}&=&-\half(\lambda^{ij}\xi_i\xi_j+\lambda_{ij}
\xi^{\dagger i}\xi^{\dagger j})\phi-\kappa^i{}_j\chi_i\eta^j\Phi
+\kappa_i{}^j\chi^{\dagger i}\eta^\dagger_j\Phi^\dagger \nonumber \\
&& -[(\kappa_1)^i{}_j\xi_i \eta^j
+(\kappa_2)_{ij}\xi^{\dagger i} \chi^{\dagger j}]\Phi
-[(\kappa_2)^{ij}\xi_i\chi_j
+(\kappa_1)_i{}^j\xi^{\dagger i} \eta^{\dagger}_j]\Phi^\dagger\,,
\eeqa
where $\lambda$ is a complex symmetric matrix, and $\kappa$,
$\kappa_1$ and $\kappa_2$ are complex matrices such that
$\kappa^i{}_j=0$ unless $q\ls{\Phi}=q_j-q_i$ and
$(\kappa_1)^i{}_j=(\kappa_2)_{ij}=0$ unless $q\ls{\Phi}=q_j$
[flavor index conventions are specified in
\eqs{eq:complexindexconvention}{Mdagger}].
The corresponding Feynman
rules of \fig{fig:Yukawavertexrules}(a) are obtained by identifying
$Y^{Iij}=\lambda^{ij}$, $\kappa^i{}_j$, $(\kappa_1)^i{}_j$
and $(\kappa_2)^{ij}$ for the undotted fermion vertices
$\phi\xi_i\xi_j$, $\Phi\chi\ls{i}\eta^j$, $\Phi\xi_i\eta^j$ and
$\Phi^\dagger \xi_i\chi_j$, respectively.\footnote{For
the $\Phi^\dagger\xi_i\chi_j$ vertex, we should reverse
the direction of the arrow
on the scalar line in \fig{fig:Yukawavertexrules}(a)
[and likewise for the corresponding hermitian-conjugated vertex of
\fig{fig:Yukawavertexrules}(b)], in which case
all arrows on the charged scalar and fermion lines would represent the
direction of flow of the conserved U(1)-charge.}
The corresponding Feynman rules of \fig{fig:Yukawavertexrules}(b)
for the dotted fermion vertices are governed by the complex-conjugated
Yukawa couplings, $Y_{Ijk}\equiv (Y^{Ijk})^*$.

The renormalizable interactions of vector bosons with
fermions and scalars arise from gauge interactions.  These interaction
terms of the Lagrangian derive from the respective kinetic energy terms of
the fermions and scalars when the derivative is promoted to the covariant
derivative:
\beq
(D_\mu)_i{}^j\equiv
\delta_i{}^j\partial_\mu
\BDplus ig_a A_\mu^a({\boldsymbol{T^a}})_i{}^j\,,
\eeq
where the index $a$ labels the real (interaction eigenstate) vector bosons
$A_a^\mu$ and is summed over.
The index $a$ runs over the
adjoint representation of the gauge group,\footnote{Since the adjoint
representation is a real representation, the height of the adjoint
index $a$ is not significant.  The choice of a subscript or
superscript adjoint index is based solely on typographical considerations.}
and the $(\boldsymbol{T^a})_i{}^j$
are hermitian representation matrices
of the generators of the Lie algebra of the gauge group acting on the
left-handed fermions (for further details, see \app{E}).
For a $U(1)$ gauge
group, the $\boldsymbol{T^a}$ are replaced by real numbers
corresponding to the U(1) charges of the left-handed $(\half,0)$
fermions.  There is a separate coupling
$g_a$ for each simple group or U(1) factor of the
gauge group $G$.\footnote{That is, the generators $\boldsymbol{T^a}$
separate out into distinct classes, each of which is associated with
a simple group or one of the U(1) factors contained in the direct
product that defines $G$.  In particular, $g_a=g_b$ if $\boldsymbol{T^a}$ and
$\boldsymbol{T^b}$ are in the same class.
If $G$ is simple, then $g_a=g$ for all $a$.}

In the gauge-interaction basis for the
left-handed $(\half, 0)$ two-component fermions the corresponding interaction
Lagrangian is given by
\beqa
\mathscr{L}_{\rm int} =
\BDneg g_a A_a^{\mu} {\hat\psi}^{\dagger i}\,
\sigmabar_\mu ({\boldsymbol T}^a)_i{}^j \hat\psi_j \,.
\label{eq:lintG}
\eeqa
In the case of spontaneously broken gauge theories, one must
diagonalize the vector boson squared-mass matrix.  The form of
\eq{eq:lintG} still applies where $A_\mu^a$ are gauge boson fields of
definite mass, although in this case for a fixed value of $a$, $g_a
{\boldsymbol T}^a$ [which multiplies $A_\mu^a$ in \eq{eq:lintG}] is
some linear combination of the original $g_a {\boldsymbol T}^a$ of the
unbroken theory.
That is, the hermitian matrix gauge field $(A_\mu)_i{}^j\equiv
A_\mu^a (\boldsymbol{T^a})_i{}^j$ appearing in \eq{eq:lintG} can always
be re-expressed in terms of the
\textit{physical} mass eigenstate gauge boson fields.

If an unbroken U(1) (global or local)
symmetry exists, then the physical gauge bosons will
be eigenstates of the
conserved U(1)-charge.\footnote{\label{vectormass}
In terms of the physical gauge boson fields, $A_\mu^a\boldsymbol{T^a}$
consists of a sum over real neutral gauge fields
multiplied by hermitian generators, and
complex charged gauge fields
multiplied by non-hermitian generators.
For example, in the electroweak Standard
Model, $G$=SU(2)$\times$U(1) with gauge bosons and
generators $W_\mu^a$ and ${\boldsymbol T}^a=\half\tau^a$ for SU(2)
and $B_\mu$ and $\boldsymbol{Y}$ for U(1),
where the $\tau^a$ are the usual Pauli matrices.
After diagonalizing the gauge boson squared-mass matrix\cite{chengli}:
\beq \label{fnew}
gW_\mu^a \boldsymbol{T^a}+ g' B_\mu \boldsymbol{Y}=
\frac{g}{\sqrt{2}}(W_\mu^+\boldsymbol{T^+}
+W_\mu^-\boldsymbol{T^-})+
\frac{g}{\cos\theta_W}\left(\boldsymbol{T^3}-\boldsymbol{Q}
\sin^2\theta_W\right)Z_\mu+e\boldsymbol{Q}A_\mu\,,
\eeq
where
$\boldsymbol{Q}=\boldsymbol{T^3}+\boldsymbol{Y}$ is the generator
of the unbroken U(1)$_{\rm EM}$,
$\boldsymbol{T^\pm}\equiv \boldsymbol{T^1}\pm i\boldsymbol{T^2}$,
and $e=g\sin\theta_W=g'\cos\theta_W$.
The massive gauge boson charge-eigenstate fields
of the broken theory
consist of a charged massive gauge boson pair,
$W^\pm\equiv (W^1\mp iW^2)/\sqrt{2}$, a neutral massive gauge boson,
$Z\equiv W^3\cos\theta_W-B\sin\theta_W$, and the massless photon,
$A\equiv W^3\sin\theta_W+B\cos\theta_W$.}
If the U(1) symmetry group is orthogonal to the
gauge group under which the $A^\mu_a$
transform, then all the gauge bosons are neutral with respect to
the U(1)-charge.  For example, in the case of
the interaction of a gluon with a pair of Majorana
fermion gluinos, the gluon is a gauge
boson that transforms under
the SU(3) color group, which is orthogonal to
the conserved U(1)$_{\rm EM}$.
That is, gluinos are color octet, electrically neutral fermions.
In contrast, in the case of the interaction of
a $Z^0$ with pair of Majorana neutralinos, U(1)$_{\rm EM}$ is not
orthogonal to the electroweak  SU(2)$\times$U(1)
gauge group. Nevertheless, the $Z^0$-gauge boson
interactions of the neutralinos
are allowed as they conserve electric charge.

To obtain the desired Feynman rule, we
rewrite \eq{eq:lintG} in terms of mass eigenstate fermion fields.
The resulting interaction Lagrangian can be rewritten as
\beqa
\mathscr{L}_{\rm int} =
\BDneg A_a^{\mu} \psi^{\dagger i}\,
\sigmabar_\mu (G^a)_i{}^j \psi_j \,,
\label{eq:lintG2}
\eeqa
where the $A_a^\mu$ are the mass eigenstate gauge fields
(of definite U(1)-charge, if relevant), and
\beq \label{gadef}
(G^a)_i{}^j= g_a U^k{}_i(\boldsymbol{T^a})_k{}^m U_m{}^j\,,
\eeq
or in matrix form, $G^a=g_a U^\dagger \boldsymbol{T^a} U$ (no sum over~$a$).
For values of $a$ corresponding to the neutral gauge fields, the $G^a$
are hermitian matrices.  The corresponding Feynman rule is shown in
\fig{fig:Gaugevertexrules}.

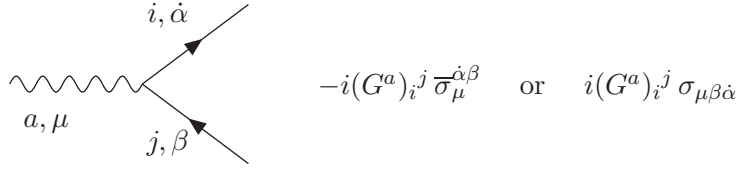
\begin{figure}[tb!]
\begin{center}
\begin{picture}(200,55)(30,20)
\Photon(60,40)(10,40){3}{5}
\ArrowLine(60,40)(100,70)
\ArrowLine(100,10)(60,40)
\Text(24,25)[]{$a, \mu$}
\Text(70,18)[]{$j, \beta$}
\Text(70,66)[]{$i, \dot\alpha$}
\Text(123,40)[l]{
$\BDneg i (G^a)_i{}^{j}\, \sigmabar_\mu^{\dot{\alpha}\beta}$
\quad or \quad
$\BDpos i (G^a)_i{}^{j}\, \sigma_{\mu\beta\dot{\alpha}}$}
\end{picture}
\end{center}
\caption{\label{fig:Gaugevertexrules}
The Feynman rules for two-component fermion interactions
with gauge bosons.   The choice of which rule to
use depends on how the vertex connects to the rest of the
amplitude.  The $G^a$ are defined in \eq{gadef}.
The index $a$ runs over both neutral
and charged (mass eigenstate) gauge bosons, consistent with
charge conservation at the vertex.
}
\end{figure}

The above treatment of the gauge interactions of (two-component) fermions
is general.  Nevertheless, it is useful to
consider separately
three cases where the gauge bosons couple
to a pair of Majorana fermions, a pair of Dirac fermions, and
a fermion pair consisting of one Majorana and one Dirac fermion.

First, consider the gauge interactions of neutral Majorana fermions.
The Majorana fermions consist of left-handed $(\half,0)$
interaction eigenstate fermions $\hat\xi_i$
that transform under a real representation of the gauge group.
After converting from the
interaction eigenstates $\hat\xi_i$ to the mass eigenstates $\xi_i$
using \eq{Omegadef}, the Lagrangian for the gauge
interactions of Majorana fermions is given by:
\beq
\mathscr{L}_{\rm int} =
\BDneg A^a_{\mu} \xi^{\dagger i}\,
\sigmabar^\mu (G^a)_i{}^j \xi_j \,,
\label{eq:lintG2Maj}
\eeq
where the $A_\mu^a$ are \textit{neutral} (real) mass eigenstate gauge
fields, and
\beq \label{gadefMaj}
(G^a)_i{}^j= g_a \Omega^k{}_i(\boldsymbol{T^a})_k{}^m \Omega_m{}^j\,,
\eeq
or in matrix form, $G^a=g_a \Omega^\dagger \boldsymbol{T^a} \Omega$
(no sum over~$a$).  Note that the $G^a$ are hermitian matrices.
The corresponding Feynman rule takes the same form as the generalized
rule shown in \fig{fig:Gaugevertexrules}, with $a$ restricted to
values corresponding to the neutral mass eigenstate gauge
bosons.

Next, consider the gauge interactions
of charged Dirac fermions.
The Dirac fermions consist of
pairs of left-handed $(\half,0)$
interaction eigenstate fermions $\hat\chi_i$ and $\hat\eta^i$
that transform as
conjugate representations
of the gauge group (hence the opposite flavor index heights).
The fermion mass matrix couples $\chi$ and $\eta$ type fields
as in eq.~(\ref{lagDiracmixed}).  In the coupling to the
interaction eigenstate gauge fields,
if the $({\boldsymbol T}^a)_i{}^j$ are matrix elements of the hermitian
representation matrices
of the generators acting on the $\hat\chi_i$, then the $\hat\eta^i$
transform in the complex conjugate representation with
the corresponding generator
matrices $-({\boldsymbol T}^a)^*
= -({\boldsymbol T}^a)^{\T}$, i.e. with matrix elements
$-({\boldsymbol T}^a)_j{}^i$. Hence,
the Lagrangian for the gauge interactions
of Dirac fermions can be written in the form:
\beq \label{gaugedirac}
\mathscr{L}_{\rm int} =
\BDneg g_a A_a^{\mu} {\hat\chi}^{\dagger i}\, \sigmabar_\mu
({\boldsymbol T}^a)_i{}^j \hat\chi_j
\BDplus g_a A_a^{\mu} {\hat\eta}^\dagger_{i}\, \sigmabar_\mu
({\boldsymbol T}^a)_j{}^i \hat\eta^j \,.
\eeq
We now rewrite \eq{gaugedirac}
in terms of mass eigenstate fermion fields
using eq.~(\ref{lrdef}), and express the hermitian matrix gauge
field $A^\mu\equiv A^\mu_a\boldsymbol{T^a}$ in terms of
mass eigenstate gauge fields (of definite U(1)-charge, if relevant).
The resulting interaction Lagrangian is then given by:
\beq \label{massgaugedirac}
\mathscr{L}_{\rm int} =
\BDneg A_a^{\mu}\left[ {\chi}^{\dagger i}\, \sigmabar_\mu
(G_L^a)_i{}^j \chi_j
-{\eta}^\dagger_{i}\, \sigmabar_\mu
(G_R^a)_j{}^i \eta^j\right] \,,
\eeq
where $A^\mu_a G^a_L$ and $A^\mu_a G^a_R$ are hermitian matrix-valued
gauge fields, with:
\beqa
(G_L^a)_i{}^j &=& g_a L^k{}_i(\boldsymbol{T^a})_k{}^m L_m{}^j\,,
\label{gachidef}\\
(G_R^a)_j{}^i &=& g_a R^m{}_j(\boldsymbol{T^a})_m{}^k R_k{}^i\,.
\label{gaetadef}
\eeqa
In matrix form, \eqs{gachidef}{gaetadef} read:
$G_L^a=g_a L^\dagger \boldsymbol{T^a} L$ and
$G_R^a=g_a R^\dagger \boldsymbol{T^a} R$ (no sum over~$a$).
For values of $a$ corresponding to the neutral gauge fields,
$G^a_L$ and $G^a_R$ are hermitian matrices.
The corresponding Feynman
rules for the gauge interactions of Dirac
fermions are shown  in \fig{fig:GaugevertexDirac}.  Note that
$\chi_i$ with its arrow pointing out of the vertex and $\eta^i$
with its arrow pointing into the vertex represent the same
Dirac fermion.
\begin{figure}[tb!]
\begin{center}
\begin{picture}(200,55)(36,20)
\Photon(60,40)(10,40){3}{5}
\ArrowLine(60,40)(100,70)
\ArrowLine(100,10)(60,40)
\Text(24,25)[]{$a, \mu$}
\Text(70,18)[]{$\beta$}
\Text(70,66)[]{$\dot\alpha$}
\Text(136,40)[l]{
$\BDneg i (G_L^a)_i{}^{j}\, \sigmabar_\mu^{\dot{\alpha}\beta}$
\quad or \quad
$\BDpos i (G_L^a)_i{}^{j}\, \sigma_{\mu\beta\dot{\alpha}}$}
\Text(108,72)[]{$\chi_i$}
\Text(108,12)[]{$\chi_j$}
\Text(-40,40)[]{(a)}
\end{picture}
\end{center}
\begin{center}
\begin{picture}(200,75)(36,20)
\Photon(60,40)(10,40){3}{5}
\ArrowLine(100,70)(60,40)
\ArrowLine(60,40)(100,10)
\Text(24,25)[]{$a, \mu$}
\Text(70,18)[]{$\dot\beta$}
\Text(70,66)[]{$\alpha$}
\Text(136,40)[l]{
$\BDpos i (G_R^a)_i{}^{j}\, \sigmabar_\mu^{\dot{\beta}\alpha}$
\quad or \quad
$\BDneg i (G_R^a)_i{}^{j}\, \sigma_{\mu\alpha\dot{\beta}}$}
\Text(108,72)[]{$\eta^i$}
\Text(108,12)[]{$\eta^j$}
\Text(-40,40)[]{(b)}
\end{picture}
\end{center}
\caption{\label{fig:GaugevertexDirac}
The Feynman rules for the interaction of a gauge boson
and a pair of Dirac fermions (each formed by $\chi$ and $\eta$ of the
appropriate flavor index).  The fermion lines are labeled by
the corresponding two-component left-handed $(\half,0)$ fermion fields.
The matrices $G_L^a$ and $G_R^a$ depend on the group generators
for the representation carried by the $\chi_i$ according to
\eqs{gachidef}{gaetadef}.  The index $a$ runs over both neutral
and charged (mass eigenstate) gauge bosons, consistent with
charge conservation at the vertex.
}
\end{figure}
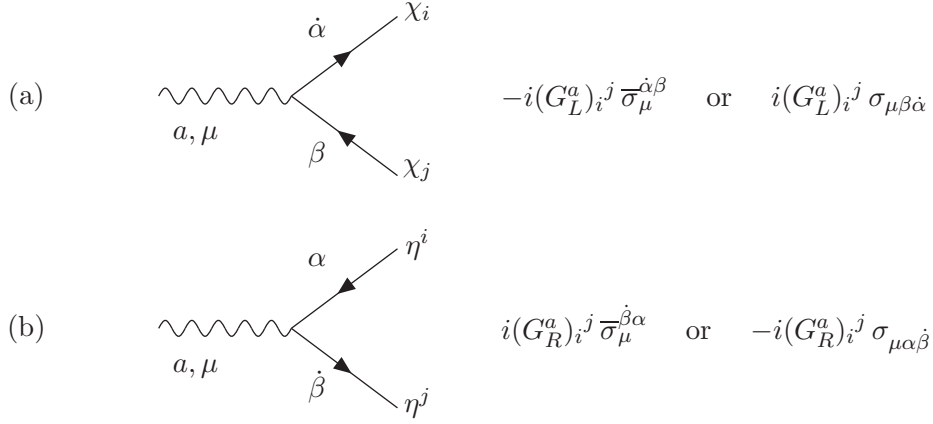

Finally, consider the interaction of a charged
vector boson $W$ (with U(1)-charge $q\ls{W}$) with a fermion pair consisting of
one Majorana and one Dirac fermion.  As before, we denote the
Majorana fermion by $\xi_i$ and the Dirac fermion
pair by $\chi_j$ and $\eta^j$ (with U(1)-charges $q_j$ and $-q_j$,
respectively).  All fields are assumed to be in
the mass eigenstate basis.  The interaction Lagrangian is given
by:\footnote{The sign in front of $G_2$ is
conventionally chosen to match the
sign of the term proportional to $G_R^a$ in \eq{massgaugedirac}.}
\beq  \label{lintwplus}
\mathscr{L}_{\rm int} =
\BDneg W_\mu[(G_1)_j{}^i\chi^{\dagger j}\sigmabar^\mu\xi_i
-(G_2)_{ij}\xi^{\dagger i}\sigmabar^\mu \eta^j]
\BDminus W_\mu^\dagger[(G_1)^j{}_i\,\xi^{\dagger i}\sigmabar^\mu\chi_j
-(G_2)^{ij}\eta^{\dagger}_j\sigmabar^\mu\xi_i]\,,
\eeq
where $G_1$ and $G_2$ are arbitrary complex matrices,
with $(G_1)^i{}_j\equiv [(G_1)_i{}^j]^*$ and
$(G_2)^{ij}\equiv [(G_2)_{ij}]^*$, such that
$(G_1)_j{}^i=(G_2)_{ij}=0$ unless $q\ls{W}=q_j$.
The interactions of \eq{lintwplus} yield the
Feynman rules exhibited in \fig{fig:GaugevertexMajDirac}.
Note that rules (c) and (d) are the complex conjugates of
rules (a) and (b), respectively, corresponding to a reversal
of the flow of the U(1)-charge through the interaction vertex.

\begin{figure}[t!]
\begin{center}
\begin{picture}(200,68)(0,0)
\Photon(60,40)(10,40){3}{5}
\LongArrow(20,50)(50,50)
\ArrowLine(60,40)(100,70)
\ArrowLine(100,10)(60,40)
\Text(30,25)[]{$\mu$}
\Text(70,20)[]{$\alpha$}
\Text(70,67)[]{$\dot{\beta}$}
\Text(135,60)[l]{$\phantom{\BDpos}
\BDneg i(G_1)_j{}^i\sigmabar_\mu^{\dot{\beta}\alpha}$}
\Text(170,40)[]{or}
\Text(135,20)[l]{$\phantom{\BDneg}
\BDpos i(G_1)_j{}^i\sigma_{\mu\alpha\dot{\beta}}$}
\Text(110,70)[]{$\chi_j$}
\Text(110,10)[]{$\xi_i$}
\Text(0,70)[]{(a)}
\end{picture}
\hspace{1.4cm}
\begin{picture}(200,68)(0,0)
\Photon(60,40)(10,40){3}{5}
\LongArrow(20,50)(50,50)
\ArrowLine(100,70)(60,40)
\ArrowLine(60,40)(100,10)
\Text(30,25)[]{$\mu$}
\Text(70,20)[]{$\dot{\alpha}$}
\Text(70,67)[]{$\beta$}
\Text(135,60)[l]{$\phantom{\BDneg}
\BDpos i(G_2)_{ij}\sigmabar_\mu^{\dot{\alpha}\beta}$}
\Text(170,40)[]{or}
\Text(135,20)[l]{$\phantom{\BDpos}
\BDneg i(G_2)_{ij}\sigma_{\mu\beta\dot{\alpha}}$}
\Text(110,70)[]{$\eta^j$}
\Text(110,10)[]{$\xi_i$}
\Text(0,70)[]{(b)}
\end{picture}
\end{center}
\vspace{0.2cm}
\begin{center}
\begin{picture}(200,58)(0,10)
\Photon(60,40)(10,40){3}{5}
\LongArrow(50,50)(20,50)
\ArrowLine(100,70)(60,40)
\ArrowLine(60,40)(100,10)
\Text(30,25)[]{$\mu$}
\Text(70,20)[]{$\dot{\alpha}$}
\Text(70,67)[]{$\beta$}
\Text(135,60)[l]{$\phantom{\BDpos}
\BDneg (G_1)^j{}_i\sigmabar_\mu^{\dot{\alpha}\beta}$}
\Text(170,40)[]{or}
\Text(135,20)[l]{$\phantom{\BDneg}
\BDpos (G_1)^j{}_i\sigma_{\mu\beta\dot{\alpha}}$}
\Text(110,70)[]{$\chi_j$}
\Text(110,10)[]{$\xi_i$}
\Text(0,70)[]{(c)}
\end{picture}
\hspace{1.4cm}
\begin{picture}(200,58)(0,10)
\Photon(60,40)(10,40){3}{5}
\LongArrow(50,50)(20,50)
\ArrowLine(60,40)(100,70)
\ArrowLine(100,10)(60,40)
\Text(30,25)[]{$\mu$}
\Text(70,20)[]{$\alpha$}
\Text(70,67)[]{$\dot{\beta}$}
\Text(135,60)[l]{$\phantom{\BDneg}
\BDpos i(G_2)^{ij}\sigmabar_\mu^{\dot{\beta}\alpha}$}
\Text(170,40)[]{or}
\Text(135,20)[l]{$\phantom{\BDpos}
\BDneg i(G_2)^{ij}\sigma_{\mu\alpha\dot{\beta}}$}
\Text(110,70)[]{$\eta^j$}
\Text(110,10)[]{$\xi_i$}
\Text(0,70)[]{(d)}
\end{picture}
\end{center}
\caption{\label{fig:GaugevertexMajDirac}
The Feynman rules for the interactions of a
charged vector boson (with U(1)-charge $q\ls{W}$)
with a fermion pair consisting of
one Majorana fermion $\xi_i$ and one Dirac fermion formed by
$\chi_j$ and $\eta^j$ (with corresponding U(1)-charges $q_j$ and
$-q_j$).   The fermion lines are labeled by
the corresponding two-component left-handed $(\half,0)$ fermion fields.
The matrix couplings $G_1$
and $G_2$ are defined in \eq{lintwplus}.  Note that
$(G_1)_j{}^i=(G_2)_{ij}=0$ unless $q\ls{W}=q_j$.  The arrows
indicate the direction of flow of the U(1)-charges of the fermion
and boson fields.
}
\end{figure}
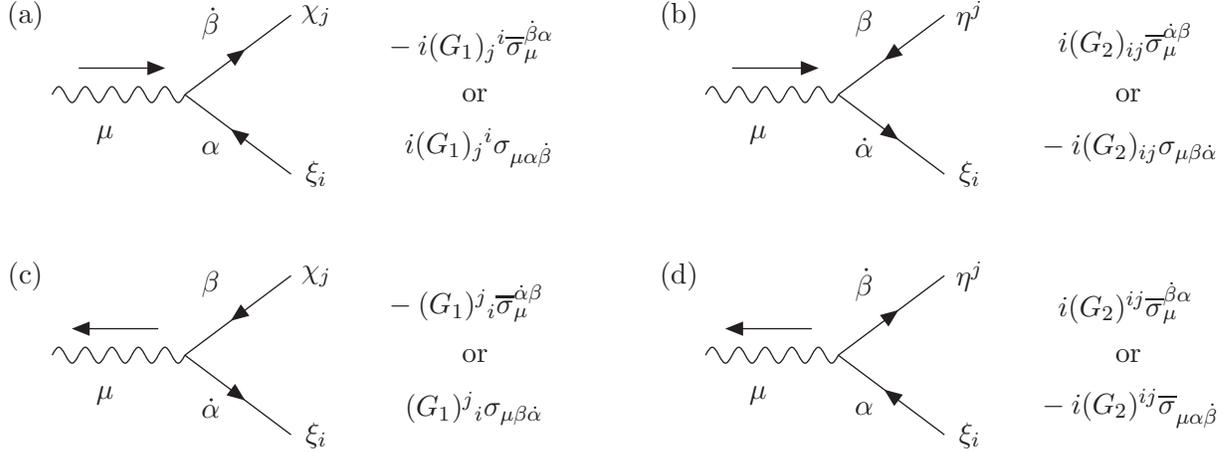

In \figst{fig:Gaugevertexrules}{fig:GaugevertexMajDirac},
two versions are given for each of the boson-fermion-fermion Feynman
rules. The correct version to use depends in a unique way on the
heights of indices used to connect each fermion line to the rest of
the diagram. For example, the way of writing the
vector-fermion-fermion interaction rule depends on whether we used
$\psi^{\dagger i}\sigmabar^\mu\psi_j$, or its equivalent form
$-\psi_j\sigma^\mu \psi^{\dagger i}$, in \eq{eq:lintG}. Note the different
heights of the undotted and dotted spinor indices that adorn
$\sigma^\mu$ and $\sigmabar^\mu$.  The choice of which rule to use is
thus dictated by the height of the indices on the lines that connect
to the vertex.  These heights should always be chosen so that they are
contracted as in eq.~(\ref{suppressionrule}).

The application of the rules of this subsection will be exhibited in
\sec{subsec:simpleapps}.
Many additional examples involving Standard
Model and MSSM processes can be found in \sec{sec:examples}.


\subsection{General structure and rules for Feynman graphs}
\label{subsec:genstructure}
\renewcommand{\theequation}{\arabic{section}.\arabic{subsection}.\arabic{equation}}
\renewcommand{\thefigure}{\arabic{section}.\arabic{subsection}.\arabic{figure}}
\renewcommand{\thetable}{\arabic{section}.\arabic{subsection}.\arabic{table}}
\setcounter{equation}{0}
\setcounter{figure}{0}
\setcounter{table}{0}

When computing an amplitude for a given process, all possible diagrams
should be drawn that conform with the rules given
in \secst{subsec:externallines}{subsec:fermioninteractions}
for external wave functions, propagators, and interactions,
respectively.  Starting from any external
wave function spinor (or from any vertex on a fermion loop), factors
corresponding to each propagator and vertex should be written down from
left to right, following the line until it ends at another external state
wave function (or at the original point on the fermion loop).  If one starts
a fermion line at an $x$ or $y$ external state spinor, it should have a
raised undotted index in accord with eq.~(\ref{suppressionrule}).  Or, if
one starts with an $x^\dagger$ or $y^\dagger$, it should have a lowered dotted
spinor index. Then, all spinor indices should always be contracted as in
eq.~(\ref{suppressionrule}). If one ends with an $x$ or $y$ external state
spinor, it will have a lowered undotted index, while if one ends with an
$x^\dagger$ or $y^\dagger$ spinor, it will have a raised dotted index. For
arrow-preserving fermion propagators and gauge vertices, the preceding
determines whether the $\sigma$ or $\sigmabar$ rule should be used.

With only a little practice, one can write down amplitudes
immediately with all spinor indices suppressed.
In particular, the following must be satisfied:
\beq \label{rule1}
\parbox[t]{5.5in}{$\bullet$
For any scattering matrix amplitude, factors of $\sigma$ and
$\sigmabar$ must alternate. If one or more factors of $\sigma$ and/or
$\sigmabar$ are present, then $x$ and $y$ must be followed [preceded] by a
$\sigma$ [$\sigmabar$], and $x^\dagger$ and $y^\dagger$ must be followed
[preceded] by a $\sigmabar$ [$\sigma$].
}
\eeq

These requirements automatically dictate whether the $\sigma$ or
$\sigmabar$ version of the rule for arrow-preserving fermion
propagators and gauge vertices are employed in any tree-level
Feynman diagram.  In loop diagrams, we must add one further
requirement that governs the order of the $\sigma$ and $\sigmabar$
factors as one traverses around the loop.
\beq \label{rule2}
\parbox[t]{5.5in}{
$\bullet$ Arrow-preserving propagator lines must be traversed in a
direction parallel [anti\-parallel] to the arrowed line segment for the
$\sigmabar$ [$\sigma$] version of the propagator rule.\footnotemark}\\[8pt]
\eeq

\footnotetext{This rule is simply a
consequence of the order of the spinor indices in
\fig{fig:neutproprev}, as noted in \sec{subsec:fermionprops}.}

For fermion lines that are not closed loops, this last requirement is
realized automatically provided that the requirements of \eq{rule1} are
satisfied. However, for closed fermion loops, one must use the correct
fermion propagator corresponding to the direction around the loop one has
chosen to follow in writing down the spinor trace with suppressed indices.
For example, having employed a $\sigma$ [$\sigmabar$] rule at one
vertex attached to the loop, one must then traverse the loop
from that vertex point in a direction
parallel [antiparallel] to the arrow-preserving propagator lines in
the loop.
Indeed, this rule is crucial for obtaining the correct sign for the
triangle anomaly calculation in \sec{sec:anomaly}.

Symmetry factors for identical particles are implemented in the usual way.
Fermi-Dirac statistics are implemented by the following rules:
\begin{itemize}
\item[$\bullet$] Each closed fermion loop gets a factor of $-1$.
\item[$\bullet$] A relative minus sign is imposed between terms
contributing to a given amplitude
whenever the ordering of external state
spinors (written left-to-right in a formula) differs by an odd permutation.
\end{itemize}

\noindent
Amplitudes generated according to these rules will contain objects
of the form:
\beq
a = z_1 \Sigma z_2
\label{preccspinorbilinears}
\eeq
where $z_1$ and $z_2$ are each commuting external spinor wave functions
$x$, $x^\dagger$, $y$, or $y^\dagger$, and $\Sigma$ is a sequence of
alternating $\sigma$ and $\sigmabar$ matrices. The
complex conjugate of this quantity is obtained by applying the
results of \eqst{eq:conbil}{eq:conbilgen}, and is
given by\footnote{For Lorentz-scalar
quantities of the form given by \eq{preccspinorbilinears},
there is no distinction between complex conjugation and
hermitian conjugation.}
\beq
a^* = z^\dagger_2 {\reversed{\Sigma}} z^\dagger_1
\label{ccspinorbilinears}
\eeq
where ${\reversed{\Sigma}}$ is obtained from $\Sigma$ by reversing the
order of all the $\sigma$ and $\sigmabar$ matrices, and using the same
rule for suppressed spinor indices.
[Notice that this rule for taking
complex conjugates has the same form as for anticommuting spinors;
cf.~eqs.~(\ref{eq:conbil})--(\ref{eq:conbilgen}).]
We emphasize that in principle, it does not matter in what direction a
diagram is traversed while applying the rules.  However,
for each diagram one must
include a sign that depends on the ordering of the
external fermions. This sign can be fixed by first choosing some canonical
ordering of the external fermions. Then for any graph that contributes to
the process of interest, the corresponding sign is positive (negative) if
the ordering of external fermions is an even (odd) permutation with
respect to the canonical ordering. If one chooses a different canonical
ordering, then the resulting amplitude changes by an overall phase (is
unchanged) if this ordering is an odd (even) permutation of the original
canonical ordering.\footnote{For a process with exactly two external
fermions, it
is convenient to apply the Feynman rules by starting from the same fermion
external state in all diagrams. That way, all terms in the amplitude
have the same canonical ordering of fermions and there are no additional
minus signs between diagrams. However, if there are four or more external
fermions, it often happens that there is no way to choose the same
ordering of external state spinors for all graphs when the amplitude is
written down. Then the relative signs between different graphs must be
chosen according to the relative sign of the permutation of the
corresponding external fermion spinors. This guarantees that the total
amplitude is antisymmetric under the interchange of any pair of external
fermions.}
This is consistent with the fact that the $S$-matrix element is only
defined up to an overall sign, which is not physically
observable.\footnote{The $S$-matrix element is related
to the invariant matrix element $\mathcal{M}_{fi}$ by
$
S_{fi}=\delta_{fi}+(2\pi)^4\delta^{(4)}(p_f-p_i)\,i\mathcal{M}_{fi}\,,
$
where $p_f$ ($p_i$) is the total four-momentum of the final (initial)
state.  If $f\neq i$ (i.e.~the final and initial states are distinct),
then $\delta_{fi}=0$ in which case the invariant matrix element is only
defined up to an overall (unphysical) sign.
However, if $f=i$, the most convenient choice for the
canonical ordering of external fermions is the one that yields
$\vev{f|i}=\delta_{fi}$ (with no extra minus sign), which then
fixes the absolute sign of the invariant matrix element.
\label{fncanonical}}

Note that different graphs contributing to the same process will often
have different external state wave function spinors, with different arrow
directions, for the same external fermion.  Furthermore, there are no
arbitrary choices to be made for arrow directions, as there are in some
four-component Feynman rules for Majorana fermions
(as discussed in \app{G}.)
Instead, one must add together
{\em all} Feynman graphs that obey the rules.

\subsection{Basic examples of writing down diagrams and amplitudes}
\label{subsec:simpleapps}
\renewcommand{\theequation}{\arabic{section}.\arabic{subsection}.\arabic{equation}}
\renewcommand{\thefigure}{\arabic{section}.\arabic{subsection}.\arabic{figure}}
\renewcommand{\thetable}{\arabic{section}.\arabic{subsection}.\arabic{table}}
\setcounter{equation}{0}
\setcounter{figure}{0}
\setcounter{table}{0}

Some simple examples
will help clarify the rules of \sec{subsec:genstructure}.
In the tree-level Feynman graphs of this subsection,
we label all two-component fermion lines by their corresponding
left-handed $(\half,0)$ fields.  (We shall propose a slightly
different labeling convention in \sec{sec:nomenclature}.)
A larger number of examples, drawn
from practical calculations, are given in \sec{sec:examples}.

\subsubsection{Scalar boson decay to fermion pairs}

Let us first consider a theory with a
multiplet of
uncharged, massive $(\half,0)$ fermions $\xi_i$, and a real scalar $\phi$,
with interaction
\beqa
\mathscr{L}_{\rm int} =
-\half \left ( \lambda^{ij} \xi_i\xi_j +
\lambda_{ij}  \xi^{\dagger i} \xi^{\dagger j} \right ) \phi\,,
\label{eq:phixixiinteraction}
\eeqa
where $\lambda_{ij}\equiv (\lambda^{ij})^*$
and $\lambda^{ij}=\lambda^{ji}$.
Consider the decay
$\phi\to\xi_i(\boldsymbol{\vec p}_1,s_1)\xi_j(\boldsymbol{\vec p_2},s_2)$
[for a fixed choice of $i$ and $j$], where by
$\xi_i(\boldsymbol{\vec p},s)$ we mean the
one particle state given by \eq{xistate}.
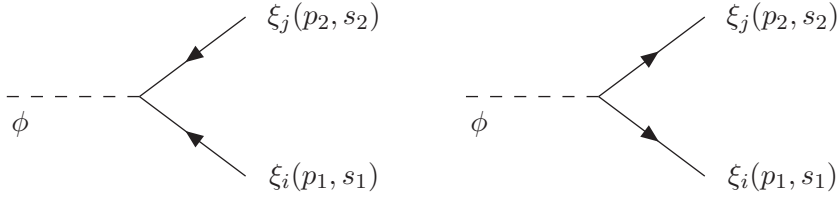
\begin{figure}[ht!]
\begin{center}
\vspace{0.15in}
\begin{picture}(170,57)(0,12)
\DashLine(10,40)(60,40)5
\ArrowLine(100,70)(60,40)
\ArrowLine(100,10)(60,40)
\Text(15,30)[]{$\phi$}
\Text(130,10)[]{$\xi_i(p_1,s_1)$}
\Text(130,70)[]{$\xi_j(p_2,s_2)$}
\end{picture}
\begin{picture}(170,57)(0,12)
\DashLine(60,40)(10,40)5
\ArrowLine(60,40)(100,70)
\ArrowLine(60,40)(100,10)
\Text(15,30)[]{$\phi$}
\Text(130,10)[]{$\xi_i(p_1,s_1)$}
\Text(130,70)[]{$\xi_j(p_2,s_2)$}
\end{picture}
\end{center}
\caption{\label{fig:phixixidecay}
{The two tree-level Feynman diagrams contributing to the decay of a
neutral scalar into a pair of Majorana fermions.}}
\end{figure}

Two diagrams contribute to this process,
as shown in \fig{fig:phixixidecay}.
The matrix element is:
\beqa
i\mathcal{M} &=&
y(\boldsymbol{\vec p}_1,s_1)^\alpha
(-i \lambda^{ij} {\delta_\alpha}^\beta )
y(\boldsymbol{\vec p}_2,s_2)_\beta
+ x^\dagger(\boldsymbol{\vec p}_1,s_1)_{\dot\alpha}
(-i \lambda_{ij} {\delta^{\dot\alpha}}_{\dot\beta})
x^\dagger(\boldsymbol{\vec p}_2,s_2)^{\dot\beta}
\nonumber \\[6pt]
&=& - i \lambda^{ij} y(\boldsymbol{\vec p}_1,s_1)y(\boldsymbol{\vec p}_2,s_2)
- i \lambda_{ij} x^\dagger(\boldsymbol{\vec p}_1,s_1)
x^\dagger(\boldsymbol{\vec p}_2,s_2)\,.
\label{scalardecay}
\eeqa
The second line could be written down directly by recalling that the
sum over suppressed spinor indices is taken according to
\eq{suppressionrule}.  Note that if we reverse the ordering for the
external fermions, the overall sign of the amplitude changes
sign. This is easily checked, since for the commuting spinor
wave functions ($x$ and $y$), the spinor products in
\eq{scalardecay} change sign when the order is reversed [see
\eqs{zonetwo}{barzonetwo}].  This overall sign is not
significant and depends on the order used in constructing the two
particle state.  One could even make the choice
of starting the first diagram from fermion $1$, and the second diagram
from fermion $2$:
\beq
i\mathcal{M}= -i \lambda^{ij} y(\boldsymbol{\vec p}_1,s_1)
y(\boldsymbol{\vec p}_2,s_2) - (-1) i\lambda_{ij}
x^\dagger(\boldsymbol{\vec p}_2,s_2)
x^\dagger(\boldsymbol{\vec p}_1,s_1)\, .
\label{scalardecaytwo}
\eeq
Here, the first term establishes the canonical ordering of fermions
(12), and the contribution from the second diagram therefore includes
the relative minus sign in parentheses.
Indeed, \eqs{scalardecay}{scalardecaytwo} are
equal.  In the computation of the total decay rate for the case
of $i=j$, one must
multiply the integral over the total phase space by $1/2$ to account
for the identical particles.

Next, we consider a theory of a massive neutral scalar boson that
couples to a multiplet of Dirac fermions.  We denote the corresponding
two-component fields by $\chi_i$ and $\eta^i$.  For simplicity, we take
all the U(1)-charges of the $\chi_i$ to be equal (and opposite
to the charges of the $\eta^i$).
The corresponding U(1)-invariant interaction is:
\beq
\mathscr{L}_{\rm int}=-(\kappa^i{}_j\chi_i\eta^j+\kappa_i{}^j
\chi^{\dagger i} \eta_j^\dagger)\phi\,,
\eeq
where $\kappa_i{}^j=(\kappa^i{}_j)^*$.  Consider the decay
$\phi\to f_i(\boldsymbol{\vec p}_1,s_1)\overline{f}\llsup{\,j}
(\boldsymbol{\vec p_2},s_2)$ [for a fixed choice of $i$ and~$j$],
where by $f(\boldsymbol{\vec p},s)$ and $\overline{f}(\boldsymbol{\vec p},s)$
we mean the one particle states given by \eq{chietastate}.
Two diagrams contribute to this process, as shown in
\fig{fig:phichietadecay}.
\begin{figure}[t!]
\begin{center}
\vspace{0.15in}
\begin{picture}(170,57)(0,12)
\DashLine(10,40)(60,40)5
\ArrowLine(100,70)(60,40)
\ArrowLine(100,10)(60,40)
\Text(15,30)[]{$\phi$}
\Text(130,10)[]{$\eta^i(p_1,s_1)$}
\Text(130,70)[]{$\chi_j(p_2,s_2)$}
\end{picture}
\begin{picture}(170,57)(0,12)
\DashLine(60,40)(10,40)5
\ArrowLine(60,40)(100,70)
\ArrowLine(60,40)(100,10)
\Text(15,30)[]{$\phi$}
\Text(130,10)[]{$\chi_i(p_1,s_1)$}
\Text(130,70)[]{$\eta^j(p_2,s_2)$}
\end{picture}
\end{center}
\caption{\label{fig:phichietadecay}
The two tree-level Feynman diagrams contributing to the decay of a
neutral scalar into a pair of Dirac fermions.
The $\chi_i$--$\eta^i$ and  $\chi_j$--$\eta^j$ pairs,
each with oppositely directed arrows,
comprise Dirac fermion states with flavor indices $i$ and $j$,
respectively.
}
\end{figure}
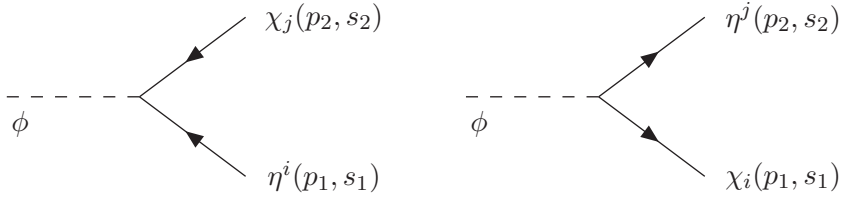
Note that the outgoing fermion lines are
distinguished by their U(1)-charges.
The matrix element is then given by
\beq
i\mathcal{M}
= - i \kappa^j{}_i y(\boldsymbol{\vec p}_1,s_1)y(\boldsymbol{\vec p}_2,s_2)
- i \kappa_i{}^j x^\dagger(\boldsymbol{\vec p}_1,s_1)
x^\dagger(\boldsymbol{\vec p}_2,s_2)\,.
\label{scalardiracdecay}
\eeq

The matrix element for
$\phi\to f_i(\boldsymbol{\vec p}_1,s_1)\overline{f}\llsup{\,j}
(\boldsymbol{\vec p_2},s_2)$ is identical to that of
$\phi\to\xi_i(\boldsymbol{\vec p}_1,s_1)\xi_j(\boldsymbol{\vec p_2},s_2)$
after replacing $\lambda^{ij}$ with $\kappa^i{}_j$.
However for fixed $i=j$, the rate for scalar boson
decay to $f_i\overline{f}\llsup{\,i}$
is twice that of $\xi_i\xi_i$ due to the final state identical particles in
the latter case, as noted above.
One also arrives at the same conclusion if one treats a single Dirac
fermion as a pair of mass-degenerate two-component fields $\xi_1$ and
$\xi_2$ [cf.~\eq{chietadirac}].  Due to the U(1)-symmetry, the
scalar Yukawa interactions are diagonal in the $\xi_1$--$\xi_2$
basis, so the rate for scalar decay into the Dirac fermion pair is
equal to the incoherent sum of the rate for decay into $\xi_1\xi_1$
and $\xi_2\xi_2$.

\subsubsection{Fermion pair annihilation into a scalar boson}

It is also instructive to consider the corresponding $2\to 1$
scattering (annihilation) processes
$\xi(\boldsymbol{\vec p}_1,s_1)\xi(\boldsymbol{\vec
p_2},s_2)\to \phi$ and
$f(\boldsymbol{\vec p}_1,s_1)\overline{f}(\boldsymbol{\vec
p_2},s_2)\to\phi$, respectively.  The corresponding amplitudes are
given by \eqs{scalardecay}{scalardiracdecay} with $y\to x$ and
$x^\dagger \to y^\dagger$ (for simplicity, we neglect flavor).
In the computation of the cross-sections, there
is no extra factor required to account for the case of identical
particles in the initial state.  That is, the cross-section
for $f(\boldsymbol{\vec p}_1,s_1)\overline{f}(\boldsymbol{\vec
p_2},s_2)\to\phi$ is equal to the cross-section for
$\xi(\boldsymbol{\vec p}_1,s_1)\xi(\boldsymbol{\vec
p_2},s_2)\to \phi$ after replacing $\lambda$ with $\kappa$.

This may at first seem puzzling given that a Dirac fermion can be
represented by a pair of mass-degenerate two-component fields $\chi_1$
and $\chi_2$.  But, recall the standard procedure for the calculation
of decay rates and cross-sections in field theory---\textit{average
over unobserved degrees of freedom of the initial state and sum over
unobserved degrees of freedom of the final state}.  This mantra is
well-known for dealing with spin and color degrees of freedom, but it
is also applicable to degrees of freedom
associated with global internal symmetries.
Thus, the cross-section for the annihilation of a Dirac
fermion pair into a neutral scalar boson can be obtained by computing
the \textit{average} of the cross-sections for
$\xi_1(\boldsymbol{\vec p}_1,s_1)\xi_1(\boldsymbol{\vec
p_2},s_2)\to \phi$ and
$\xi_2(\boldsymbol{\vec p}_1,s_1)\xi_2(\boldsymbol{\vec
p_2},s_2)\to \phi$.  Since the
annihilation cross-sections for $\xi_1\xi_1$ and $\xi_2\xi_2$
are equal, we confirm the annihilation cross-section
for the Dirac fermion pair obtained above in the $\chi$--$\eta$ basis.
Since the latter is conceptually simpler, subsequent
computations involving Dirac fermions will be performed in the
$\chi$--$\eta$ basis.

The annihilation rate of fermions enters in the analysis of the event
flux due to the annihilation of dark matter in the halo of our galaxy.
Let us compare the
rates in the case that the dark matter is either a Majorana or a Dirac
fermion.  Suppose the annihilation involves two fermions whose number
densities are $n_1$ and $n_2$ respectively.  Then the
observer on Earth who integrates along the line of sight to the
annihilation events that are detected sees a flux of events
proportional to~\cite{Bergstrom:1997fj}
\beq \label{flux}
\frac{dN_{\rm events}}{dA\, dt}\sim
\int n_1 n_2 \vev{\sigma_{\rm ann} v_{\rm rel}} d\ell\,,
\eeq
where $v_{\rm rel}$ is the relative velocity of the annihilating
initial state particles, $\sigma_{\rm ann}$ is the annihilation
cross-section and $\vev{\cdots}$ refers to a
thermal average~\cite{Gondolo:1990dk} over the
velocity distribution of dark matter particles in the halo.
We now compare the case of the annihilation of a single
species of Majorana particles and the annihilation of a Dirac
fermion-antifermion pair (assumed to have the same mass and couplings).
We assume that the number density of Dirac fermions and antifermions
and the corresponding number density of Majorana fermions are all the
same (and denoted by $n$).
Above, we showed that $\sigma_{\rm ann}$ is the same for the annihilation
of a single species of Majorana and Dirac fermions.
For the Dirac case, $n_1 n_2 = n^2$.  For the Majorana case, because
the Majorana fermions are identical particles, given $N$ initial state
fermions in a volume $V$, there are $N(N-1)/2$ possible scatterings.
In the thermodynamic limit where $N$, $V\to \infty$ at fixed $n\equiv
N/V$, we conclude that $n_1 n_2 =\half n^2$ for a single species of
annihilating Majorana fermions.\footnote{The factor of 1/2, which has
been erroneously omitted in many papers in the literature, was
correctly employed and explained in footnote 1 of \Ref{ullio}.}
Hence the event flux rate for the
annihilation of a Dirac fermion-antifermion pair is double that of a
single species of Majorana fermions.\footnote{This is also consistent
with the interpretation of a Dirac fermion as a pair of mass-degenerate
Majorana fermions.}
The extra factor of $1/2$ can also be understood by noting that in
the case of annihilating dark matter particles (in the large $N$
limit), all possible scattering axes occur and are implicitly
integrated over.  But, integrating over $4\pi$ steradians double
counts the annihilation of identical particles (in the same way it
does in the computation of the decay rate of a scalar into identical
fermions discussed above).  Hence, one must include a factor of $\half$
in this case by replacing $n_1 n_2=n^2$ by $\half n^2$ in \eq{flux}.

The relic abundance of primordial dark matter particles in the universe
is inversely proportional to $\vev{\sigma_{\rm ann} v_{\rm rel}}$~\cite{kolb}.
By similar arguments to the ones just presented, it follows that
the relic abundance of a single species of Majorana fermions
would be twice that of a single species of Dirac fermions.

\subsubsection{Vector boson decay into fermion pairs}

Consider next the decay of a massive neutral vector boson $A_\mu$ into a
pair of Majorana fermions, $A_\mu\to\xi_i(\boldsymbol{\vec p}_1,s_1)\xi_j(
\boldsymbol{\vec p}_2,s_2)$, following from the interaction,
\beq
\mathscr{L}_{\rm int} = \BDneg G_i{}^j A^\mu \xi^{\dagger i}
\sigmabar_\mu \xi_j\,,
\label{eq:Axixiinteraction}
\eeq
where $G$ is an hermitian coupling matrix.  The two diagrams shown in
\fig{fig:AtoMajoranadecay} contribute.
\begin{figure}[ht!]
\vspace{1cm}
\begin{picture}(400,58)(0,0)
\Photon(110,40)(60,40){3}{5}
\ArrowLine(110,40)(150,70)
\ArrowLine(150,10)(110,40)
\Text(50,40)[]{$A_\mu$}
\Text(175,10)[]{$\xi_j(p_2,s_2)$}
\Text(175,70)[]{$\xi_i(p_1,s_1)$}
\Photon(310,40)(260,40){3}{5}
\ArrowLine(350,70)(310,40)
\ArrowLine(310,40)(350,10)
\Text(250,40)[]{$A_\mu$}
\Text(375,10)[]{$\xi_j(p_2,s_2)$}
\Text(375,70)[]{$\xi_i(p_1,s_1)$}
\end{picture}
\caption[0]{\label{fig:AtoMajoranadecay}
The two tree-level Feynman diagrams contributing to the decay of a massive
neutral vector boson $A_\mu$ into a pair of Majorana fermions.}
\end{figure}
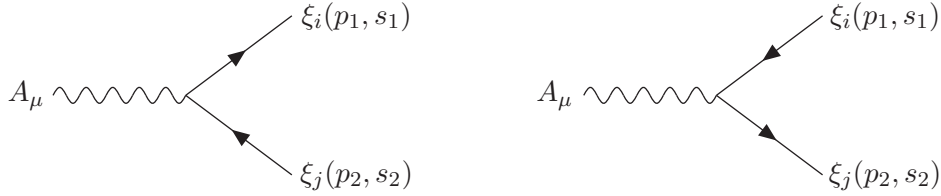

We start from the fermion with momentum $p_1$ and spin vector $s_1$ and
end at the fermion with momentum $p_2$ and spin vector $s_2$,
using the rules of \fig{fig:Gaugevertexrules}.
The resulting amplitude for the decay is
\beq
\label{vectordecay}
i\mathcal{M}=
\varepsilon^\mu
\left [ \BDneg i G_i{}^j
x^\dagger(\boldsymbol{\vec p}_1,s_1)\sigmabar_\mu
y(\boldsymbol{\vec p}_2,s_2)
\BDplus i G_j{}^i y(\boldsymbol{\vec p}_1,s_1)\sigma_\mu
x^\dagger(\boldsymbol{\vec p}_2,s_2)
\right ]\,,
\eeq
where $\varepsilon^\mu$ is the vector boson polarization vector.
We have used
the $\sigmabar$-version of the vector-fermion-fermion rule
[see \fig{fig:Gaugevertexrules}] for the first
diagram of \fig{fig:AtoMajoranadecay} and the $\sigma$-version for the second
diagram of \fig{fig:AtoMajoranadecay},
as dictated by the implicit spinor indices, which we have
suppressed.  However, we could have chosen to evaluate the second
diagram of \fig{fig:AtoMajoranadecay} using the
$\sigmabar$-version of the vector-fermion-fermion rule by starting
from the fermion with momentum~$p_2$ and spin vector $s_2$.  In that case,
the term $\BDpos i G_j{}^i y(\boldsymbol{\vec p}_1,s_1)\sigma_\mu
x^\dagger (\boldsymbol{\vec p}_2,s_2)$ in \eq{vectordecay}
is replaced by
\beq \label{vectordecaypiece}
(-1)[\BDneg i G_j{}^i x^\dagger (\boldsymbol{\vec p}_2,s_2)\sigmabar_\mu
y(\boldsymbol{\vec p}_1,s_1)]\,.
\eeq
In \eq{vectordecaypiece}, the factor of $\BDneg iG_j{}^i$
arises from the use of the
$\sigmabar$-version of the vector-fermion-fermion rule, and the
overall factor of $-1$ appears because the order of the
fermion wave functions has been reversed; i.e. $(21)$ is an
odd permutation of $(12)$.  This is in accord with the ordering rule
stated at the end of \sec{subsec:genstructure}.  Thus, the resulting
amplitude for the decay of the vector boson into the pair of Majorana
fermions now takes the form:
\beq \label{vectordecayalt}
i\mathcal{M}=
\varepsilon^\mu
\left  [\BDneg i G_i{}^j
x^\dagger(\boldsymbol{\vec p}_1,s_1)\sigmabar_\mu y(\boldsymbol{\vec p}_2,s_2)
\BDplus i G_j{}^i x^\dagger(\boldsymbol{\vec p}_2,s_2)\sigmabar_\mu
y(\boldsymbol{\vec p}_1,s_1)\right]
\,,
\eeq
which coincides with \eq{vectordecay} after
using $y\sigma^\mu x^\dagger= x^\dagger\sigmabar^\mu y$
[cf.~\eq{europeanvacation} with commuting spinors].
\Eq{vectordecayalt} explicitly exhibits the property
that the amplitude is antisymmetric under the
interchange of the two external identical fermions.
Again, the absolute sign of the total amplitude is not significant
and depends on the choice of ordering of the outgoing states.

Next, we consider the decay of a massive neutral vector boson into a pair
of Dirac fermions.  Each Dirac fermion is described by the
two-component fields $\chi\ls{i}$ and $\eta^i$, which possess equal
and opposite U(1)-charges, respectively.  The corresponding interaction
Lagrangian is given by:
\beq
\mathscr{L}_{\rm int} =
\BDneg A^\mu [ (G_L)_i{}^j\, \chi^{\dagger i}\sigmabar_\mu \chi_j
- (G_R)_j{}^i\,  \eta_i^\dagger\sigmabar_\mu \eta^j\,] \,,
\label{eq:Achietainteractions}
\eeq
where $G_L$ and $G_R$ are hermitian.
There are two contributing
graphs, as shown in \fig{fig:AtoDiracdecay}.
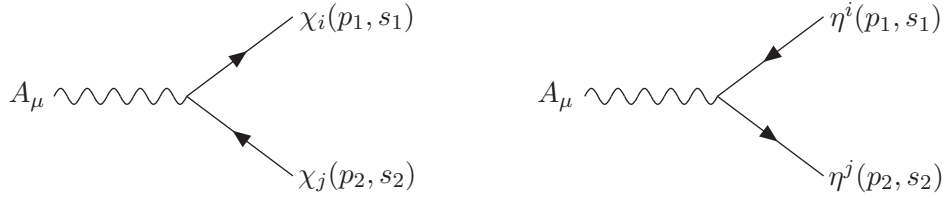
\begin{figure}[ht!]
\vspace{.5cm}
\begin{picture}(400,55)(0,8)
\Photon(110,40)(60,40){3}{5}
\ArrowLine(110,40)(150,70)
\ArrowLine(150,10)(110,40)
\Text(50,40)[]{$A_\mu$}
\Text(175,10)[]{$\chi_j(p_2,s_2)$}
\Text(175,70)[]{$\chi_i(p_1,s_1)$}
\Photon(310,40)(260,40){3}{5}
\ArrowLine(350,70)(310,40)
\ArrowLine(310,40)(350,10)
\Text(250,40)[]{$A_\mu$}
\Text(375,10)[]{$\eta^j(p_2,s_2)$}
\Text(375,70)[]{$\eta^i(p_1,s_1)$}
\end{picture}
\caption[0]{\label{fig:AtoDiracdecay}
The two tree-level Feynman diagrams contributing to the decay of a massive
neutral vector boson $A_\mu$ into a pair of Dirac fermions.
The $\chi_i$--$\eta^i$ and  $\chi_j$--$\eta^j$ pairs,
each with oppositely directed arrows,
comprise Dirac fermion states with flavor indices $i$ and $j$,
respectively.}

\end{figure}

To evaluate the amplitude, we start
with the fermion of momentum $p_1$ and spin vector $s_1$, and
end at the fermion with momentum $p_2$ and spin vector~$s_2$.
Note that the outgoing $\chi_i$ with the arrow pointing outward from
the vertex and the outgoing $\eta^i$ with the arrow pointing inward to
the vertex both correspond to the same outgoing Dirac fermion.
The amplitude for the decay is given by:
\beqa \label{vectordecay2}
i\mathcal{M}&=&
\varepsilon^\mu
\left [\BDneg i (G_L)_i{}^j
x^\dagger(\boldsymbol{\vec p}_1,s_1)\sigmabar_\mu y(\boldsymbol{\vec p}_2,s_2)
\BDminus i (G_R)_i{}^j y(\boldsymbol{\vec p}_1,s_1)\sigma_\mu
x^\dagger(\boldsymbol{\vec p}_2,s_2)\right]\nonumber \\[5pt]
&=&
\varepsilon^\mu
\left [\BDneg i (G_L)_i{}^j
x^\dagger(\boldsymbol{\vec p}_1,s_1)\sigmabar_\mu y(\boldsymbol{\vec p}_2,s_2)
\BDminus i (G_R)_i{}^j x^\dagger(\boldsymbol{\vec p}_2,s_2)\sigmabar_\mu
y(\boldsymbol{\vec p}_1,s_1)\right]\,.
\eeqa
As in the case of the decay to a pair of Majorana fermions, we have
exhibited a second form for the amplitude in \eq{vectordecay2} in which
the  $\sigmabar$-version of the vertex Feynman rule has
been employed in both diagrams.  Of course, the resulting
amplitude must be the same in each method (up to a
possible overall sign of the total
amplitude that is not determined).

The computation of the amplitude for the decay of a charged
vector boson to a fermion pair consisting of one Majorana
fermion and one Dirac fermion, due to the interactions given in
\eq{lintwplus}, is straightforward and will not be
given explicitly here.

\subsubsection{Two-body scattering of a boson and a neutral fermion}

The next level of complexity consists of diagrams that involve fermion
propagators.  In the examples that follow in this and in the next
subsection, we shall ignore the flavor index and consider
scattering processes that involve a single flavor of Majorana or
Dirac fermion.
For our first example of this type, consider the
tree-level matrix element for the scattering of a neutral scalar and a
two-component neutral massive fermion ($\phi\xi\to\phi\xi$), with the
interaction Lagrangian given above in
eq.~(\ref{eq:phixixiinteraction}).  Using the corresponding Feynman
rules, there are eight contributing diagrams.  Four are depicted in
\fig{phixscattering}; there are another four diagrams (not shown)
where the initial and final state scalars are crossed (i.e., the
initial state scalar is attached to the same vertex as the final state
fermion).
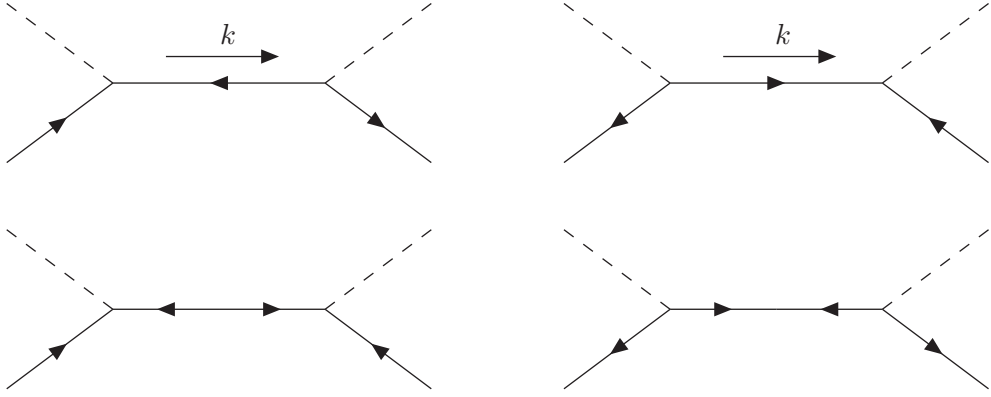
\begin{figure}[t!]
\centerline{
\begin{picture}(330,85)(-135,-26)
\thicklines
\ArrowLine(-60,15)(-140,15)
\DashLine(-140,15)(-180,45)5
\ArrowLine(-180,-15)(-140,15)
\DashLine(-60,15)(-20,45)5
\ArrowLine(-60,15)(-20,-15)
\LongArrow(-120,25)(-80,25)
\put(-100,30){$k$}
\ArrowLine(70,15)(150,15)
\DashLine(70,15)(30,45)5
\ArrowLine(70,15)(30,-15)
\DashLine(150,15)(190,45)5
\ArrowLine(190,-15)(150,15)
\LongArrow(90,25)(130,25)
\put(110,30){$k$}
\end{picture}
}
\vspace{0.2cm}
\centerline{
\begin{picture}(330,68)(-135,-15)
\thicklines
\ArrowLine(-100,15)(-140,15)
\ArrowLine(-100,15)(-60,15)
\DashLine(-140,15)(-180,45)5
\ArrowLine(-180,-15)(-140,15)
\DashLine(-60,15)(-20,45)5
\ArrowLine(-20,-15)(-60,15)
\ArrowLine(70,15)(110,15)
\ArrowLine(150,15)(110,15)
\DashLine(70,15)(30,45)5
\ArrowLine(70,15)(30,-15)
\DashLine(150,15)(190,45)5
\ArrowLine(150,15)(190,-15)
\end{picture}
}
\caption{\label{phixscattering}
Tree-level Feynman diagrams contributing to the elastic
scattering of a neutral scalar and a Majorana fermion.
There are four more diagrams, obtained from these by crossing the
initial and final scalar lines.}
\end{figure}

We shall write down the amplitudes for the four diagrams
shown in \fig{phixscattering},
starting with
the final state fermion
line and moving toward the initial state fermion line.  Then,
\beqa \label{pxscatter}
\!\!\!\!\!\!\!\!
i\mathcal{M} &=&
\frac{i}{k^2 \BDminus m^2_\xi}
\biggl\{
(-i \lambda)(-i\lambda^*)
\left [x^\dagger(\boldsymbol{\vec p}_2,s_2)\,
\sigmabar\newcdot k\,
x(\boldsymbol{\vec p}_1,s_1)+y(\boldsymbol{\vec p}_2,s_2)\,
\sigma\newcdot k\,
y^\dagger(\boldsymbol{\vec p}_1,s_1)\right ] \nonumber \\[4pt]
&&\>\> \BDplus  \,m_\xi\left[(-i \lambda)^2
y(\boldsymbol{\vec p}_2,s_2)x(\boldsymbol{\vec p}_1,s_1)+(-i \lambda^*)^2
x^\dagger(\boldsymbol{\vec p}_2,s_2)
y^\dagger(\boldsymbol{\vec p}_1,s_1)
\right]\biggr\}+{\rm (crossed)}\,,
\eeqa
where $k^\mu$ is the sum of the two incoming (or outgoing) four-momenta,
$(p_1,s_1)$ are the momentum and
spin four-vectors of the incoming fermion, and $(p_2,s_2)$ are those
of the outgoing fermion. The notation ``(crossed)'' refers to
the contribution to the amplitude from diagrams
which have the initial and final scalars interchanged.
Note that we could
have evaluated the diagrams above by starting with the initial vertex
and moving toward the final vertex.  It is easy to check that the
resulting amplitude is the negative of the one obtained in
\eq{pxscatter}; the overall sign change simply corresponds to swapping
the order of the two fermions and has no physical consequence.  The
overall minus sign is a consequence of
\eqst{zonetwo}{europeanvacation} and the minus sign difference between
the two ways of evaluating the propagator that preserves the arrow
direction.

Next, we compute the tree-level matrix
element for the scattering of a neutral vector boson and a neutral
massive two-component
fermion $\xi$ with the interaction Lagrangian of
eq.~(\ref{eq:Axixiinteraction}).  Again there are eight diagrams: the four
diagrams depicted in \fig{gamxscattering}
plus another four (not shown)
where the initial and final state vector
bosons are crossed.
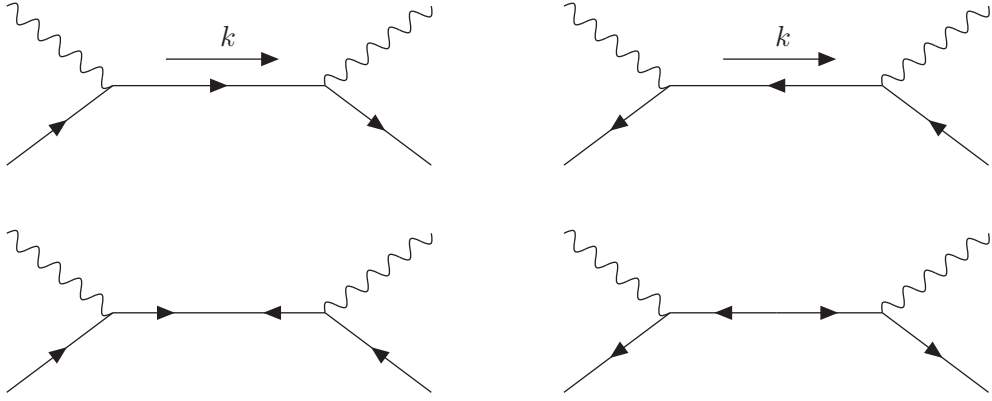
\begin{figure}[t!]
\centerline{
\begin{picture}(330,75)(-135,-26)
\thicklines
\ArrowLine(-140,15)(-60,15)
\Photon(-140,15)(-180,45){3}{5}
\ArrowLine(-180,-15)(-140,15)
\Photon(-60,15)(-20,45){3}{5}
\ArrowLine(-60,15)(-20,-15)
\LongArrow(-120,25)(-80,25)
\put(-100,30){$k$}
\ArrowLine(150,15)(70,15)
\Photon(70,15)(30,45){3}{5}
\ArrowLine(70,15)(30,-15)
\Photon(150,15)(190,45){3}{5}
\ArrowLine(190,-15)(150,15)
\LongArrow(90,25)(130,25)
\put(110,30){$k$}
\end{picture}
}
\centerline{
\begin{picture}(330,75)(-135,-16)
\thicklines
\ArrowLine(-140,15)(-100,15)
\ArrowLine(-60,15)(-100,15)
\Photon(-140,15)(-180,45){3}{5}
\ArrowLine(-180,-15)(-140,15)
\Photon(-60,15)(-20,45){3}{5}
\ArrowLine(-20,-15)(-60,15)
\ArrowLine(110,15)(70,15)
\ArrowLine(110,15)(150,15)
\Photon(70,15)(30,45){3}{5}
\ArrowLine(70,15)(30,-15)
\Photon(150,15)(190,45){3}{5}
\ArrowLine(150,15)(190,-15)
\end{picture}
}
\caption{\label{gamxscattering}
Tree-level Feynman diagrams contributing to the elastic
scattering of a neutral vector boson and a Majorana fermion.
There are four more diagrams, obtained from these by crossing the
initial and
final scalar lines.}
\end{figure}

Starting with the final state fermion line and moving toward the initial
state, we obtain
\beqa
\label{gxscatter}
i\mathcal{M} &=& \frac{i}{k^2 \BDminus m^2_\xi}
\biggl\{
(\BDneg iG)^2 x^\dagger(\boldsymbol{\vec p}_2,s_2)
\,\sigmabar\newcdot\varepsilon\ls{2}^*\,
\sigma\newcdot k\,
\sigmabar\newcdot\varepsilon\ls{1}\, x(\boldsymbol{\vec p}_1,s_1)
+ (\BDpos iG)^2 y(\boldsymbol{\vec p}_2,s_2)\,
\sigma\newcdot\varepsilon\ls{2}^*\,
\sigmabar\newcdot k
\,\sigma\newcdot\varepsilon\ls{1} y^\dagger(\boldsymbol{\vec p}_1,s_1)
\nonumber \\[4pt]
&&\> \BDplus  (\BDneg iG)(\BDpos iG)
m_\xi
\left[
y(\boldsymbol{\vec p}_2,s_2)\,\sigma\newcdot\varepsilon\ls{2}^*\,
\sigmabar\newcdot\varepsilon\ls{1}\, x(\boldsymbol{\vec p}_1,s_1)+
x^\dagger(\boldsymbol{\vec p}_2,s_2)\,\sigmabar\newcdot\varepsilon\ls{2}^*\,
\sigma\newcdot\varepsilon\ls{1}\,y^\dagger(\boldsymbol{\vec p}_1,s_1)
\right]\biggr\}
\nonumber \\[5pt]
&&
+{\rm (crossed)}\,,
\eeqa
where $\varepsilon\ls{1}$ and $\varepsilon\ls{2}$
are the initial and final vector boson
polarization four-vectors, respectively.
As before,
$k^\mu$ is the sum of the two incoming (or outgoing) four-momenta,
$(p_1,s_1)$ and $(p_2,s_2)$ are the momentum and spin four-vectors
of the incoming and outgoing fermions, respectively, and
``(crossed)'' indicates the terms from diagrams in
which the initial and final vector bosons are interchanged.
Alternatively, if one starts with an initial state fermion
and moves toward the final state,
the resulting amplitude is the negative of the one obtained
in \eq{gxscatter}, as expected.

The computation of the amplitude for the
scattering of a charged scalar or vector boson and a Majorana fermion is
straightforward and will not be given explicitly here.

\subsubsection{Two-body scattering of a boson and a charged fermion}

We first consider the scattering of a Dirac fermion
with a neutral scalar.  We denote the Dirac mass of the fermion by $m_D$.
The left-handed fields $\chi$ and $\eta$ have opposite charges
(which we take to be $Q=+1$ and $-1$ respectively), and
interact with
the scalar $\phi$ according to
\beq
\mathscr{L}_{\rm int} = -\phi
[\kappa \chi \eta + \kappa^* \chi^\dagger \eta^\dagger]\,,
\eeq
where $\kappa$ is a coupling parameter. Then, for the elastic
scattering of the $Q=+1$ fermion and a scalar, the diagrams of
\fig{phifscattering} contribute at tree level plus another four
diagrams (not shown) where the initial and final state scalars are
crossed.  Now, these diagrams match precisely those of
\fig{phixscattering}.  Thus, applying the Feynman rules yields the
same matrix element, \eq{pxscatter}, previously obtained for the
scattering of a neutral scalar and neutral two-component fermion, with
the replacement of $\lambda$ with $\kappa$ and $m_\xi$ with $m_D$.

\begin{figure}[t!]
\centerline{
\begin{picture}(300,55)(-135,-5)
\thicklines
\ArrowLine(-60,15)(-140,15)
\DashLine(-140,15)(-180,45)5
\ArrowLine(-180,-15)(-140,15)
\DashLine(-60,15)(-20,45)5
\ArrowLine(-60,15)(-20,-15)
\LongArrow(-120,25)(-80,25)
\put(-100,30){$k$}
\put(-170,-15){$\chi$}
\put(-100,3){$\eta$}
\put(-40,-15){$\chi$}
\ArrowLine(70,15)(150,15)
\DashLine(70,15)(30,45)5
\ArrowLine(70,15)(30,-15)
\DashLine(150,15)(190,45)5
\ArrowLine(190,-15)(150,15)
\LongArrow(90,25)(130,25)
\put(110,30){$k$}
\put(40,-15){$\eta$}
\put(105,3){$\chi$}
\put(175,-15){$\eta$}
\end{picture}
}
\vspace{1.5cm}
\centerline{
\begin{picture}(300,55)(-135,-26)
\thicklines
\ArrowLine(-100,15)(-140,15)
\ArrowLine(-100,15)(-60,15)
\DashLine(-140,15)(-180,45)5
\ArrowLine(-180,-15)(-140,15)
\DashLine(-60,15)(-20,45)5
\ArrowLine(-20,-15)(-60,15)
\put(-170,-15){$\chi$}
\put(-125,3){$\eta$}
\put(-85,3){$\chi$}
\put(-35,-15){$\eta$}
\ArrowLine(70,15)(110,15)
\ArrowLine(150,15)(110,15)
\DashLine(70,15)(30,45)5
\ArrowLine(70,15)(30,-15)
\DashLine(150,15)(190,45)5
\ArrowLine(150,15)(190,-15)
\put(40,-15){$\eta$}
\put(85,3){$\chi$}
\put(130,3){$\eta$}
\put(170,-15){$\chi$}
\end{picture}
}
\caption{\label{phifscattering}
Tree-level Feynman diagrams contributing to the elastic
scattering of a neutral scalar and a charged fermion. There are four more
diagrams, obtained from these by crossing the initial and
final scalar lines.}
\end{figure}
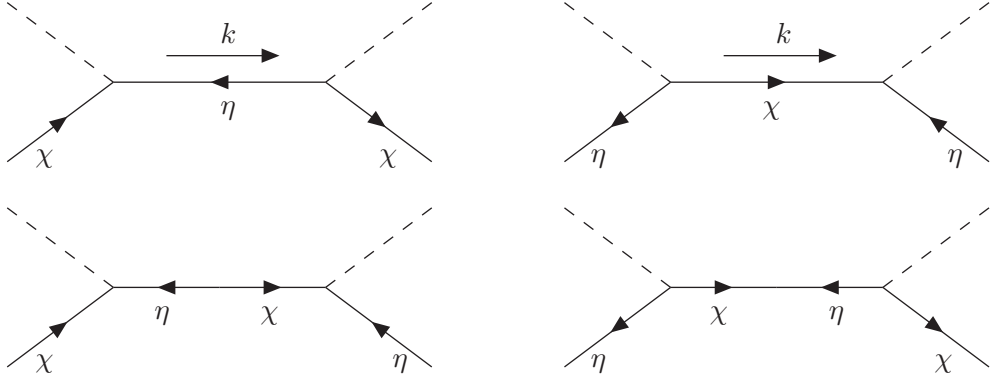

We next examine the scattering of a Dirac fermion and a charged
scalar, where both the scalar and fermion have the same absolute value
of the charge.
As above, we denote the charged $Q=\pm 1$ fermion by the pair of
two-component fermions $\chi$ and $\eta$ and the
(intermediate state) neutral two-component
fermion by $\xi$.  The charged $Q=\pm 1$ scalar is represented by the
complex scalar field $\Phi$ and its hermitian conjugate.
The interaction Lagrangian takes the form:
\beq
\mathscr{L}_{\rm int} = -\Phi
[ \kappa_1 \eta \xi+\kappa_2^* \chi^\dagger \xi^\dagger ]
-\Phi^\dagger [\kappa_2 \chi \xi + \kappa_1^* \eta^\dagger \xi^\dagger]\,.
\eeq
Consider the
scattering of an initial boson-fermion state into its
charge-conjugated final state via the exchange of a neutral fermion.
The relevant diagrams are shown in \fig{cphifscattering}
plus the corresponding diagrams with the initial and final scalars
crossed.  We define the four-momentum~$k$ to be the sum of the
two initial state four-momenta as shown in  \fig{cphifscattering}.
The derivation of the amplitude is similar to the ones given previously,
and we end up with
\beqa \label{chsfscatter}
i\mathcal{M} &=& \frac{-i}{k^2 \BDminus m^2_\xi}
\biggl\{\kappa_1^*\kappa_2[
x^\dagger(\boldsymbol{\vec p}_2,s_2)\,\sigmabar\newcdot k\,
x(\boldsymbol{\vec p}_1,s_1)+
y(\boldsymbol{\vec p}_2,s_2)\,\sigma\newcdot k\,
y^\dagger(\boldsymbol{\vec p}_1,s_1)] \nonumber \\[4pt]
&&\quad \BDplus \,m_\xi\left[\kappa_2^2
y(\boldsymbol{\vec p}_2,s_2)x(\boldsymbol{\vec p}_1,s_1)+(\kappa_1^*)^2
x^\dagger(\boldsymbol{\vec p}_2,s_2)
y^\dagger(\boldsymbol{\vec p}_1,s_1)
\right]\biggr\}+{\rm (crossed)}\,.
\eeqa

\begin{figure}[t!]
\centerline{
\begin{picture}(300,53)(-135,-5)
\thicklines
\ArrowLine(-60,15)(-140,15)
\DashArrowLine(-140,15)(-180,45)5
\ArrowLine(-180,-15)(-140,15)
\DashArrowLine(-60,15)(-20,45)5
\ArrowLine(-60,15)(-20,-15)
\LongArrow(-120,25)(-80,25)
\put(-100,30){$k$}
\put(-170,-15){$\chi$}
\put(-103,2){$\xi$}
\put(-40,-15){$\eta$}
\ArrowLine(70,15)(150,15)
\DashArrowLine(70,15)(30,45)5
\ArrowLine(70,15)(30,-15)
\DashArrowLine(150,15)(190,45)5
\ArrowLine(190,-15)(150,15)
\LongArrow(90,25)(130,25)
\put(110,30){$k$}
\put(40,-15){$\eta$}
\put(107,2){$\xi$}
\put(175,-15){$\chi$}
\end{picture}
}
\vspace{1.5cm}
\centerline{
\begin{picture}(300,65)(-135,-26)
\thicklines
\ArrowLine(-100,15)(-140,15)
\ArrowLine(-100,15)(-60,15)
\DashArrowLine(-140,15)(-180,45)5
\ArrowLine(-180,-15)(-140,15)
\DashArrowLine(-60,15)(-20,45)5
\ArrowLine(-20,-15)(-60,15)
\put(-170,-15){$\chi$}
\put(-101,3){$\xi$}
\put(-35,-15){$\chi$}
\ArrowLine(70,15)(110,15)
\ArrowLine(150,15)(110,15)
\DashArrowLine(70,15)(30,45)5
\ArrowLine(70,15)(30,-15)
\DashArrowLine(150,15)(190,45)5
\ArrowLine(150,15)(190,-15)
\put(40,-15){$\eta$}
\put(108,3){$\xi$}
\put(170,-15){$\eta$}
\end{picture}
}
\caption{\label{cphifscattering}
Tree-level Feynman diagrams contributing to the
scattering of an initial charged scalar and a charged fermion into its
charge-conjugated final state.  The unlabeled intermediate state is a
neutral fermion.  There are four more
diagrams, obtained from these by crossing the initial and
final scalar lines.}
\end{figure}
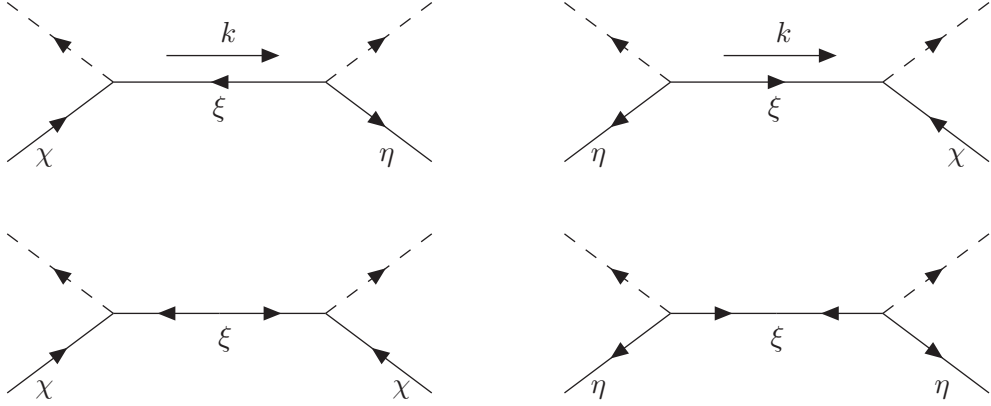
\begin{figure}[t!]
\centerline{
\begin{picture}(300,57)(-135,-5)
\thicklines
\ArrowLine(-140,15)(-60,15)
\Photon(-140,15)(-180,45){3}{5}
\ArrowLine(-180,-15)(-140,15)
\Photon(-60,15)(-20,45){3}{5}
\ArrowLine(-60,15)(-20,-15)
\LongArrow(-120,25)(-80,25)
\put(-100,30){$k$}
\put(-170,-15){$\chi$}
\put(-105,3){$\chi$}
\put(-40,-15){$\chi$}
\ArrowLine(150,15)(70,15)
\Photon(70,15)(30,45){3}{5}
\ArrowLine(70,15)(30,-15)
\Photon(150,15)(190,45){3}{5}
\ArrowLine(190,-15)(150,15)
\LongArrow(90,25)(130,25)
\put(110,30){$k$}
\put(45,-15){$\eta$}
\put(110,3){$\eta$}
\put(175,-15){$\eta$}
\end{picture}
}
\vspace{1.5cm}
\centerline{
\begin{picture}(300,65)(-135,-26)
\thicklines
\ArrowLine(-140,15)(-100,15)
\ArrowLine(-60,15)(-100,15)
\Photon(-140,15)(-180,45){3}{5}
\ArrowLine(-180,-15)(-140,15)
\Photon(-60,15)(-20,45){3}{5}
\ArrowLine(-20,-15)(-60,15)
\put(-170,-15){$\chi$}
\put(-125,3){$\chi$}
\put(-85,3){$\eta$}
\put(-35,-15){$\eta$}
\ArrowLine(110,15)(70,15)
\ArrowLine(110,15)(150,15)
\Photon(70,15)(30,45){3}{5}
\ArrowLine(70,15)(30,-15)
\Photon(150,15)(190,45){3}{5}
\ArrowLine(150,15)(190,-15)
\put(40,-15){$\eta$}
\put(85,3){$\eta$}
\put(130,3){$\chi$}
\put(170,-15){$\chi$}
\end{picture}
}
\caption{\label{gamfscattering}
Tree-level Feynman diagrams contributing to the elastic
scattering of a neutral vector boson and a Dirac fermion.
There are four more diagrams, obtained from these by crossing the initial
and final vector lines.}
\end{figure}
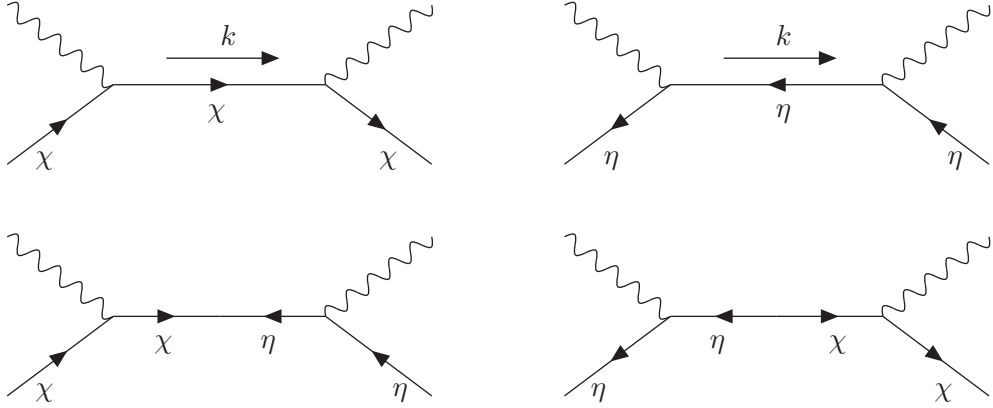

The scattering of a charged fermion
and a neutral spin-1 vector boson can be
similarly treated.  For example, consider the
amplitude for the elastic scattering of a charged
fermion of mass $m_D$ and a neutral vector boson.  Again taking the
interactions as given in eq.~(\ref{eq:Achietainteractions}),
the relevant
diagrams are those shown in \fig{gamfscattering}, plus four
diagrams
(not shown) obtained from these by crossing the initial and final
state vector bosons.
Applying the Feynman rules of \fig{fig:GaugevertexDirac},
one obtains the following matrix element:
\beqa \label{gfscatter}
&&i\mathcal{M} = \frac{-i}{k^2 \BDminus m^2_D}\biggl\{
G_L^2  x^\dagger(
\boldsymbol{\vec p}_2,s_2)\,
\sigmabar\newcdot\varepsilon\ls{2}^*\,\sigma\newcdot k\,
\sigmabar\newcdot\varepsilon\ls{1}\, x(\boldsymbol{\vec p}_1,s_1)
+G_R^2 y(\boldsymbol{\vec p}_2,s_2)\,\sigma\newcdot\varepsilon\ls{2}^*\,
\sigmabar\newcdot k
\,\sigma\newcdot\varepsilon\ls{1} y^\dagger(\boldsymbol{\vec p}_1,s_1)
\phantom{xx}
\nonumber \\[6pt]
&&\qquad\quad\BDplus m_D G_L G_R \left[
y(\boldsymbol{\vec p}_2,s_2)\,\sigma\newcdot\varepsilon\ls{2}^*\,
\sigmabar\newcdot\varepsilon\ls{1}\, x(\boldsymbol{\vec p}_1,s_1)+
x^\dagger(\boldsymbol{\vec p}_2,s_2)\,\sigmabar\newcdot\varepsilon\ls{2}^*\,
\sigma\newcdot\varepsilon\ls{1}\,
y^\dagger(\boldsymbol{\vec p}_1,s_1)
\right]\biggr\} +{\rm (crossed)}\,,\nonumber \\
&&\phantom{line}
\eeqa
and the assignments of momenta and spins are as before.

The computation of the amplitude for the scattering of a charged
fermion and a charged vector boson is straightforward and will not be
given explicitly here.

\subsubsection{Two-body fermion--fermion scattering}
\label{subsec:ff}

Finally, let us work out an example with four external state fermions.
Consider the case of elastic scattering of two identical Majorana
fermions due to scalar exchange, governed by the interaction of
eq.~(\ref{eq:phixixiinteraction}).  The  diagrams for
scattering initial fermions labeled $1,2$ into final state fermions
labeled $3,4$ are shown in \fig{ffscatt}.
\vskip 0.1in

\begin{figure}[ht!]
\begin{center}
\begin{picture}(90,47)(0,0)
\ArrowLine(0,44)(25,22)
\ArrowLine(0,0)(25,22)
\ArrowLine(90,44)(65,22)
\ArrowLine(90,0)(65,22)
\DashLine(25,22)(65,22){5}
\Text(1,34)[c]{$1$}
\Text(1,10)[c]{$2$}
\Text(89,34)[c]{$3$}
\Text(89,10)[c]{$4$}
\end{picture}
\hspace{0.75cm}
\begin{picture}(90,47)(0,0)
\ArrowLine(25,22)(0,44)
\ArrowLine(25,22)(0,0)
\ArrowLine(65,22)(90,44)
\ArrowLine(65,22)(90,0)
\DashLine(25,22)(65,22){5}
\Text(1,34)[c]{$1$}
\Text(1,10)[c]{$2$}
\Text(89,34)[c]{$3$}
\Text(89,10)[c]{$4$}
\end{picture}
\hspace{0.75cm}
\begin{picture}(90,47)(0,0)
\ArrowLine(0,44)(25,22)
\ArrowLine(0,0)(25,22)
\ArrowLine(65,22)(90,44)
\ArrowLine(65,22)(90,0)
\DashLine(25,22)(65,22){5}
\Text(1,34)[c]{$1$}
\Text(1,10)[c]{$2$}
\Text(89,34)[c]{$3$}
\Text(89,10)[c]{$4$}
\end{picture}
\hspace{0.75cm}
\begin{picture}(90,47)(0,0)
\ArrowLine(25,22)(0,44)
\ArrowLine(25,22)(0,0)
\ArrowLine(90,44)(65,22)
\ArrowLine(90,0)(65,22)
\DashLine(25,22)(65,22){5}
\Text(1,34)[c]{$1$}
\Text(1,10)[c]{$2$}
\Text(89,34)[c]{$3$}
\Text(89,10)[c]{$4$}
\end{picture}
\hspace{0.75cm}

\vspace{0.38cm}

\begin{picture}(90,50)(0,0)
\ArrowLine(0,40)(45,40)
\ArrowLine(0,0)(45,0)
\ArrowLine(90,40)(45,40)
\ArrowLine(90,0)(45,0)
\DashLine(45,40)(45,0){5}
\Text(8,48)[c]{$1$}
\Text(8,8)[c]{$2$}
\Text(82,48)[c]{$3$}
\Text(82,8)[c]{$4$}
\end{picture}
\hspace{0.75cm}
\begin{picture}(90,50)(0,0)
\ArrowLine(45,40)(0,40)
\ArrowLine(45,0)(0,0)
\ArrowLine(45,40)(90,40)
\ArrowLine(45,0)(90,0)
\DashLine(45,40)(45,0){5}
\Text(8,48)[c]{$1$}
\Text(8,8)[c]{$2$}
\Text(82,48)[c]{$3$}
\Text(82,8)[c]{$4$}
\end{picture}
\hspace{0.75cm}
\begin{picture}(90,50)(0,0)
\ArrowLine(45,40)(0,40)
\ArrowLine(0,0)(45,0)
\ArrowLine(45,40)(90,40)
\ArrowLine(90,0)(45,0)
\DashLine(45,40)(45,0){5}
\Text(8,48)[c]{$1$}
\Text(8,8)[c]{$2$}
\Text(82,48)[c]{$3$}
\Text(82,8)[c]{$4$}
\end{picture}
\hspace{0.75cm}
\begin{picture}(90,50)(0,0)
\ArrowLine(0,40)(45,40)
\ArrowLine(45,0)(0,0)
\ArrowLine(90,40)(45,40)
\ArrowLine(45,0)(90,0)
\DashLine(45,40)(45,0){5}
\Text(8,48)[c]{$1$}
\Text(8,8)[c]{$2$}
\Text(82,48)[c]{$3$}
\Text(82,8)[c]{$4$}
\end{picture}
\end{center}

\vspace{0.37cm}

\begin{center}
\begin{picture}(90,50)(0,0)
\ArrowLine(0,40)(45,40)
\ArrowLine(0,0)(45,0)
\Line(90,0)(67.5,20)
\ArrowLine(67.5,20)(45,40)
\ArrowLine(65.25,18)(45,0)
\Line(90,40)(69.75,22)
\DashLine(45,40)(45,0){5}
\Text(8,48)[c]{$1$}
\Text(8,8)[c]{$2$}
\Text(75,38)[c]{$3$}
\Text(75,5)[c]{$4$}
\end{picture}
\hspace{0.75cm}
\begin{picture}(90,50)(0,0)
\ArrowLine(45,40)(0,40)
\ArrowLine(45,0)(0,0)
\Line(90,0)(67.5,20)
\ArrowLine(45,40)(67.5,20)
\ArrowLine(45,0)(65.25,18)
\Line(90,40)(69.75,22)
\DashLine(45,40)(45,0){5}
\Text(8,48)[c]{$1$}
\Text(8,8)[c]{$2$}
\Text(75,38)[c]{$3$}
\Text(75,5)[c]{$4$}
\end{picture}
\hspace{0.75cm}
\begin{picture}(90,50)(0,0)
\ArrowLine(45,40)(0,40)
\ArrowLine(0,0)(45,0)
\Line(90,0)(67.5,20)
\ArrowLine(45,40)(67.5,20)
\ArrowLine(65.25,18)(45,0)
\Line(90,40)(69.75,22)
\DashLine(45,40)(45,0){5}
\Text(8,48)[c]{$1$}
\Text(8,8)[c]{$2$}
\Text(75,38)[c]{$3$}
\Text(75,5)[c]{$4$}
\end{picture}
\hspace{0.75cm}
\begin{picture}(90,50)(0,0)
\ArrowLine(0,40)(45,40)
\ArrowLine(45,0)(0,0)
\Line(90,0)(67.5,20)
\ArrowLine(67.5,20)(45,40)
\ArrowLine(45,0)(65.25,18)
\Line(90,40)(69.75,22)
\DashLine(45,40)(45,0){5}
\Text(8,48)[c]{$1$}
\Text(8,8)[c]{$2$}
\Text(75,38)[c]{$3$}
\Text(75,5)[c]{$4$}
\end{picture}
\end{center}
\caption{\label{ffscatt}
Tree-level Feynman diagrams contributing to the elastic
scattering of identical
Majorana fermions via scalar exchange in the $s$-channel
(top row), $t$-channel (middle row), and
$u$-channel (bottom row).}
\end{figure}
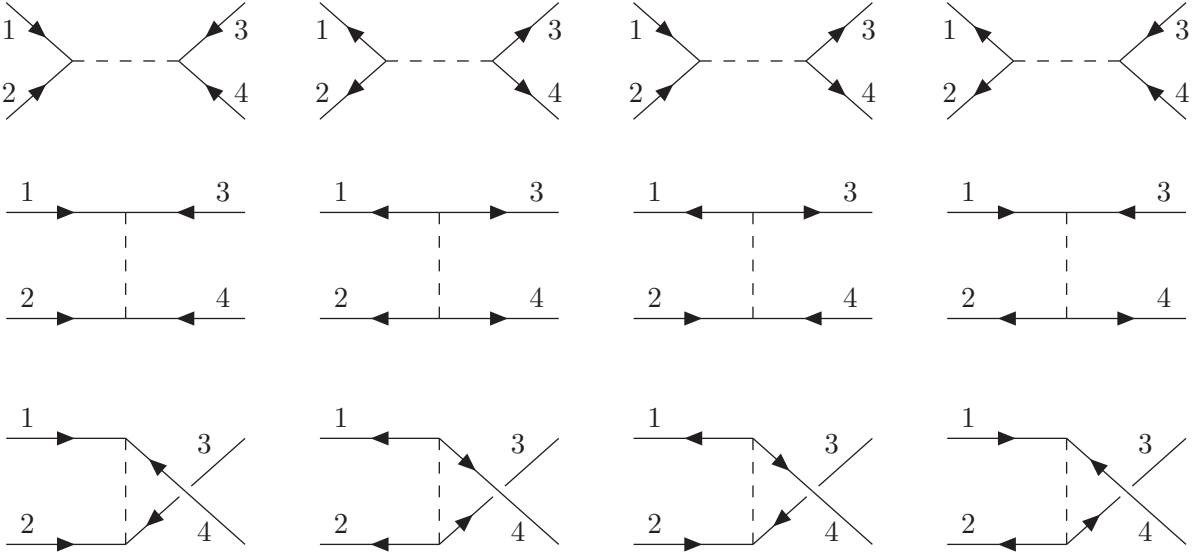

The resulting invariant matrix element is:
\beqa
i\mathcal{M} &=& (-1)\frac{- i}{s - m_\phi^2}\left\{\lambda^2 (x_1 x_2)(y_3 y_4)
+(\lambda^\ast)^2(y^\dagger_1 y^\dagger_2) (x^\dagger_3 x^\dagger_4)
+|\lambda|^2\left[(x_1 x_2)(x^\dagger_3 x^\dagger_4)
+(y^\dagger_1 y^\dagger_2)(y_3 y_4)
\right]
\right\}\nonumber \\[5pt]
&&\quad +\frac{- i}{t - m_\phi^2}\left\{\lambda^2
(y_3 x_1)(y_4 x_2)+(\lambda^\ast)^2(x^\dagger_3 y^\dagger_1)
(x^\dagger_4 y^\dagger_2)
+|\lambda|^2\left[(x^\dagger_3 y^\dagger_1)(y_4 x_2)
+ (y_3 x_1)(x^\dagger_4 y^\dagger_2)
\right]
\right\}\nonumber \\[5pt]
&&\quad +(-1)
\frac{- i}{u - m_\phi^2}\left\{\lambda^2 (y_4 x_1)(y_3 x_2)
+(\lambda^\ast)^2(x^\dagger_4 y^\dagger_1)
(x^\dagger_3 y^\dagger_2) \right.\nonumber \\
&&\hspace{2in}\left.
+|\lambda|^2\left[(x^\dagger_4 y^\dagger_1)(y_3 x_2)
+ (y_4 x_1)(x^\dagger_3 y^\dagger_2)
\right]\right\}\,,
\label{eq:ffscatttwo}
\eeqa
where $x_i\equiv x(\boldsymbol{\vec p}_i,s_i)$,
$y_i\equiv y(\boldsymbol{\vec p}_i,s_i)$,
$m_\phi$ is the mass of the exchanged scalar, $s=\BDpos (p_1+p_2)^2$,
$t = \BDpos (p_1-p_3)^2$
and $u = \BDpos (p_1-p_4)^2$. We have chosen the canonical ordering
of external fermions to be $3142$ (corresponding
to the $t$-channel contribution).  For elastic scattering,
this choice of canonical ordering
guarantees that if no scattering occurs then
the $S$-matrix is just equal to the unit operator with no extraneous
minus sign (cf.~footnote~\ref{fncanonical}).   The
relative minus signs  between the $t$-channel diagram
and the $s$ and $u$-channel
diagrams [shown in parentheses in \eq{eq:ffscatttwo}]
are obtained by observing that both $1234$ and $4132$ are both odd
permutations of $3142$.  Note that we would have obtained the same
relative signs for
the $u$-channel diagrams had we crossed the initial state fermion lines
instead of the final state fermion lines.

\Eq{eq:ffscatttwo} can be factorized with respect to the scalar line:
\beqa
i\mathcal{M} &=& \frac{i}{s - m_\phi^2}
(\lambda x_1 x_2 + \lambda^* y^\dagger_1 y^\dagger_2)
(\lambda y_3 y_4 +\lambda^\ast x^\dagger_3 x^\dagger_4)
+\frac{-i}{t - m_\phi^2}
(\lambda y_3 x_1 + \lambda^* x^\dagger_3 y^\dagger_1)
(\lambda y_4 x_2 + \lambda^* x^\dagger_4 y^\dagger_2)
\nonumber \\[5pt] &&
+\frac{i}{u - m_\phi^2}
(\lambda y_4 x_1 + \lambda^* x^\dagger_4 y^\dagger_1)
(\lambda y_3 x_2 + \lambda^* x^\dagger_3 y^\dagger_2)\,.
\label{eq:ffscatttwofact}
\eeqa
This is
a common feature of Feynman graphs with a virtual boson. This
example also illustrates that in contrast to the four-component
fermion formalism, the two-component fermion
Feynman rules typically yield many more diagrams, but
the contribution of each of the diagrams is correspondingly simpler.

\subsubsection{Non-relativistic potential due to scalar or
pseudoscalar exchange}
\label{subsec:potential}

Consider two distinguishable fermions, and a scalar-fermion-fermion
Yukawa interaction given by \eq{concrete}.  We can compute
the force law that the fermions experience due to exchange
of a spinless boson.
That is, we shall derive the Yukawa potential as a function of the
separation distance of the two fermions in the static limit.

To carry out this computation, we compute the invariant matrix
element for two-body fermion-fermion elastic scattering in the
non-relativistic limit.  The relevant diagrams are shown in \fig{ffscatt}.
As our two fermions are distinguishable, only the $t$-channel graphs
(shown in the middle row of \fig{ffscatt}) are relevant.  As a result,
the matrix element for the elastic scattering of two Majorana fermions
is given by the $t$-channel
contribution of \eq{eq:ffscatttwofact},
\beq
i\mathcal{M} =
\frac{i}{m_\phi^2-t}
(\lambda y_3 x_1 + \lambda^* x^\dagger_3 y^\dagger_1)
(\lambda y_4 x_2 + \lambda^* x^\dagger_4 y^\dagger_2)\,.
\label{eq:tchannelffscatt}
\eeq
The choice of the overall sign
is fixed by the canonical ordering of the external
fermions.\footnote{As noted in \sec{subsec:ff},
the canonical ordering of the external fermions in
two-body elastic scattering
is determined by the requirement that
$\vev{f|i}=+1$ for $f=i$ (cf.~footnote~\ref{fncanonical}).}
Although the two fermions
are distinguishable, we have assumed for simplicity that
their (complex) Yukawa coupling strengths are the same and given by
$\lambda$.  For the scattering of two distinguishable Dirac fermions,
the resulting expression for the scattering amplitude is identical to
\eq{eq:tchannelffscatt}, with $\lambda$ replaced by the appropriate
complex Yukawa coupling $\kappa$.

We denote the masses of the distinguishable fermions by $m_1$ and $m_2$.
In the non-relativistic limit, $p_1\simeq (m_1\,;\,\boldsymbol{\vec p_1})$
and  $p_3\simeq (m_1\,;\,\boldsymbol{\vec p_3})$, so that
\beq
m_\phi^2-t\simeq |\boldsymbol{\vec p_1}-\boldsymbol{\vec p_3}|^2
+m_\phi^2\equiv |\boldsymbol{\vec q}|^2+m_\phi^2\,,
\eeq
where
\beq
\boldsymbol{\vec q}\equiv\boldsymbol{\vec p_3}-\boldsymbol{\vec p_1}
=\boldsymbol{\vec p_2}-\boldsymbol{\vec p_4}
\eeq
is the momentum-transfer three-vector.  Two separate cases will be considered.

In the first case, $\lambda$ is a real coupling.
This corresponds to the exchange of a $J^{PC}=0^{++}$ scalar.
Using the non-relativistic forms of \eqs{nrxy}{nrxyb} for the
spinor bilinears, it is only necessary to keep the leading term.  We
then find:
\beq
i\mathcal{M} =
\frac{4i|\lambda|^2m_1 m_2}{|\boldsymbol{\vec q}|^2+m_\phi^2}
\delta_{s_1 s_3}\delta_{s_2 s_4}\,,
\eeq
in agreement with eq.~(4.123) of \Ref{Peskin:1995ev}.

In the second case, $\lambda$ is purely imaginary, and we will write
$\lambda=i|\lambda|$ (the overall sign is not significant).
This corresponds to the exchange of a $J^{PC}=0^{-+}$ pseudoscalar.
Again, we use  the non-relativistic forms of \eqs{nrxy}{nrxyb} for the
spinor bilinears.  However, in this case the leading term cancels
and we must retain the $\mathcal{O}(|\boldsymbol{\vec p}|/m)$ terms
appearing in the non-relativistic limit of the spinor bilinears.
In this case, we find
\beq \label{pseudoexch}
i\mathcal{M} = \frac{i|\lambda|^2}{|\boldsymbol{\vec q}|^2+m_\phi^2}
(\boldsymbol{\vec q\newcdot\hat s^a}\tau^a_{s_3 s_1})\,
(\boldsymbol{\vec q\newcdot\hat s^b}\tau^b_{s_4 s_2})\,.
\eeq
We choose the spin quantization axis to lie along the $z$-direction.
That is, according to \eq{saexplicit}, we choose
\beq \label{hatsaexplicit}
({\boldsymbol{{\hat s}^1}}\,,\,{\boldsymbol{{\hat s}^2}}\,,\,
{\boldsymbol{{\hat s}^3}})=
(\boldsymbol{{\hat x}}\,,\,
\boldsymbol{{\hat y}}\,,\,\boldsymbol{{\hat z}})\,,
\eeq
in which case one can rewrite \eq{pseudoexch} in the more traditional
way,
\beq \label{pseudoexch2}
i\mathcal{M} = \frac{i|\lambda|^2}{|\boldsymbol{\vec q}|^2+m_\phi^2}
(\boldsymbol{\vec q\newcdot\vec\sigma}_{s_3 s_1})\,
(\boldsymbol{\vec q\newcdot\vec\sigma}_{s_4 s_2})\,,
\eeq
where $\boldsymbol{\vec\sigma}\equiv \boldsymbol{\hat x}\tau^1
+\boldsymbol{\hat y}\tau^2+\boldsymbol{\hat z}\tau^3$
are the usual spin-1/2 Pauli matrices.\footnote{%
The subscripted spin labels on $\boldsymbol{\vec\sigma}$
should be interpreted in the same way as outlined in footnote~\ref{fntau}.}.  
Thus, pseudoscalar exchange yields a spin-dependent
force law.

The non-relativistic potential that arises from the $t$-channel
scalar or pseudoscalar exchange is obtained by comparing the
relativistic scattering amplitude $\mathcal{M}$
with the Born approximation for scattering
off a potential $V(\boldsymbol{\vec x})$ in non-relativistic quantum
mechanics.  Taking into account the difference
between the conventions for the normalization of relativistic and
non-relativistic single-particle states, one finds
that the static potential is given by~\cite{Maggiore}
\beq \label{nrpot}
V(\boldsymbol{\vec x})=-\frac{1}{4m_1 m_2}\int \frac{d^3
  q}{(2\pi)^3}\,\mathcal{M}(\boldsymbol{\vec q})
e^{i\boldsymbol{\vec q\newcdot\vec x}}\,,
\eeq
in a convention where the invariant amplitude is defined as in
footnote~\ref{fncanonical}.
Inserting the scattering amplitude for scalar (S) exchange, one obtains
the well-known attractive spin-independent Yukawa potential
\beq
V(\boldsymbol{\vec x})\ls{\rm S}
=-\frac{|\lambda|^2}{4\pi r}\,e^{-m_\phi r}\,\delta_{s_1 s_3}
\delta_{s_2 s_4}\,,
\eeq
where $r\equiv |\boldsymbol{\vec x}|$.
For the case of pseudoscalar (PS) exchange, one can easily evaluate
the integral in \eq{nrpot} by writing
$q_j q_k e^{i\boldsymbol{\vec q\newcdot\vec x}}=-\nabla_j\nabla_k
e^{i\boldsymbol{\vec q\newcdot\vec x}}$.  The end result is~\cite{sakurai}:
\beqa
V(\boldsymbol{\vec x})\ls{\rm PS}&=&
\frac{|\lambda|^2}{16\pi m_1 m_2}
(\boldsymbol{\vec\sigma}_{s_3 s_1}\boldsymbol{\newcdot\vec\nabla})
(\boldsymbol{\vec\sigma}_{s_4 s_2}\boldsymbol{\newcdot\vec\nabla})
\,\frac{e^{-m_\phi r}}{r}
\nonumber \\[8pt]
&=&\frac{|\lambda|^2 m_\phi^2}{16\pi m_1 m_2}
\Biggl\{\left[
-\frac{4\pi}{3m_\phi^2}\delta^{(3)}(\boldsymbol{\vec x})
+\frac{e^{-m_\phi r}}{r}
\right]\boldsymbol{\vec\sigma}_{s_3 s_1}
\boldsymbol{\newcdot\vec\sigma}_{s_4 s_2}
\nonumber \\[6pt]
&&\qquad\qquad\,\, +\left[\frac{1}{(m_\phi r)^2}+\frac{1}{(m_\phi r)}
+\frac{1}{3}\right]\left[\frac{3\,
(\boldsymbol{\vec\sigma}_{s_3 s_1}\boldsymbol{\newcdot
\vec x})(\boldsymbol{\vec\sigma}_{s_4 s_2}\boldsymbol{\newcdot \vec x})}{r^2}-
\boldsymbol{\vec\sigma}_{s_3 s_1}
\boldsymbol{\newcdot\vec\sigma}_{s_4 s_2}\right]
\frac{e^{-m_\phi r}}{r}\Biggr\}\,,\nonumber \\
\phantom{line}
\eeqa
where we have used~\cite{frahm}:
\beq
\nabla_i\nabla_j\left(\frac{1}{r}\right)=-\frac{4\pi}{3}\delta_{ij}\,
\delta^{(3)}(\boldsymbol{\vec x})+\frac{3x_i x_j-r^2\delta_{ij}}{r^5}\,.
\eeq

\subsection{Self-energy functions and pole masses for two-component fermions}
\label{subsec:selfenergies}
\renewcommand{\theequation}{\arabic{section}.\arabic{subsection}.\arabic{equation}}
\renewcommand{\thefigure}{\arabic{section}.\arabic{subsection}.\arabic{figure}}
\renewcommand{\thetable}{\arabic{section}.\arabic{subsection}.\arabic{table}}
\setcounter{equation}{0}
\setcounter{figure}{0}
\setcounter{table}{0}

In this section, we discuss the self-energy functions for fermions in
two-component notation, taking into account the possibilities of
loop-induced mixing and absorptive parts corresponding to decays to
intermediate states. This formalism is useful in the computation of
loop-corrected physical pole masses.

Consider a theory with left-handed fermion degrees of freedom $\hat\psi_i$
labeled by an index $i=1,2,\ldots,N$. Associated with each $\hat\psi_i$
is a right-handed fermion ${\hat\psi}^{\dagger i}$, where the flavor labels are
treated as described below \eq{eq:complexindexconvention}. The theory is
assumed to contain arbitrary interactions, which we will not need to refer
to explicitly. As discussed in \sec{subsec:generalmass}, we diagonalize
the fermion mass matrix and identify the fermion mass eigenstates $\psi_i$
as indicated in \eq{massrotate}.  In general, the mass eigenstates consist
of Majorana fermions $\xi_k$ ($k=1,\ldots N-2n$) and Dirac fermion
pairs $\chi_\ell$ and $\eta_\ell$ ($\ell=1,\ldots, n$).\footnote{In order
to have a unified description, we shall take the flavor index of all
left-handed fields (including $\eta_k$) in the lowered position 
(in contrast to the convention adopted in
\secs{subsec:generalmass}{subsec:fermioninteractions})
when considering a collection of two-component fermion fields that
contains both Majorana and Dirac fermions.} With respect to
this basis, the symmetric $N\times N$ tree-level fermion mass matrix,
${\boldsymbol m}^{ij}$, is made up of diagonal elements $m_k$ and $2\times
2$ blocks $\bigl(\begin{smallmatrix}0 & m_\ell \\ m_\ell &
0\end{smallmatrix}\bigr)$ along the diagonal, where the $m_k$ and $m_\ell$
are real and non-negative.  Since $\boldsymbol{m}^{ij}$ is real, the
height of the flavor indices is not significant.  Nevertheless, it is
useful to define ${\boldsymbol \mbar}_{ij} \equiv {{\boldsymbol
m}}^{ij}$ in order to maintain the convention that two repeated flavor
indices are summed when one index is raised and the other is
lowered.\footnote{We will soon be suppressing the indices, so 
it is convenient to employ the bar on ${\boldsymbol \mbar}_{ij}$ 
to indicate its lowered index structure.} Note 
that ${\boldsymbol \mbar}_{ik}
{\boldsymbol m}^{kj} ={\boldsymbol m}^{ik} {\boldsymbol \mbar}_{kj} =
m^2_i \delta_i^j$ is a diagonal matrix.

The full, loop-corrected Feynman propagators with four-momentum $p^\mu$
are defined by the Fourier transforms of vacuum expectation values of
time-ordered products of bilinears of the fully interacting two-component
fermion fields [cf.~footnote \ref{footnotefnft}].
Following \eqst{ft1}{ft4}, we define:
\beqa
\label{ftfull1}
\bra{0}T\psi_{\alpha i}(x)\psi^{\dagger j}_{\dot{\beta}}(y)\ket{0}_{\rm FT} &=&
\BDpos ip\newcdot\sigma_{\alpha\dot{\beta}}\,
{\boldsymbol C}_i{}^j(s)\,,
\\[5pt]
\bra{0}T\psi^{\dagger \dot{\alpha} i}(x)\psi_j^\beta(y)\ket{0}_{\rm FT} &=&
\BDpos ip\newcdot\sigmabar^{\dot{\alpha}\beta}\,
({\boldsymbol C}^{\T})^{\,i}{}_j(s)\,,
\label{ftfull2}
\\[5pt]
\bra{0}T \psi^{\dagger\dot{\alpha} i}(x)\psi^{\dagger j}_{\dot{\beta}}(y)
\ket{0}_{\rm FT} &=&
i\delta^{\dot{\alpha}}{}_{\dot{\beta}}\,{\boldsymbol D}^{ij}(s)\,,
\label{ftfull3}
\\[5pt]
\bra{0}T\psi_{\alpha i}(x)\psi_j^\beta(y)\ket{0}_{\rm FT} &=&
i\delta_\alpha{}^\beta\,{\boldsymbol \Doverline}_{ij}(s)\,,
\label{ftfull4}
\eeqa
where $s \equiv \BDpos p^2$ and
\beq \label{ctij}
({\mathbold{C}}^{\T})^i{}_j\equiv {\mathbold{C}}_j{}^{\,i}\,.
\eeq
One can derive \eq{ftfull2} from \eq{ftfull1}
by first writing
\beq \label{ctid}
\psi^{\dagger \dot\alpha i}(x)\psi_j^{\beta}(y)=
-\epsilon^{\beta\alpha}\,\epsilon^{\dot{\alpha}\dot{\beta}}
\psi_{\alpha j}(y)\psi^{\dagger i}_{\dot\beta}(x)\,,
\eeq
where the minus sign arises due to the anticommutativity of the fields,
and then using \eq{sigsig1}; the interchange of $x$ and $y$ (after FT)
simply changes $p^\mu$ to $-p^\mu$.

In general, ${\boldsymbol D}$ and ${\boldsymbol \Doverline}$ are complex
symmetric matrices, and ${\boldsymbol \Doverline}={\boldsymbol D}^\star$.
The matrix ${\boldsymbol C}$ satisfies the hermiticity condition
$[{\boldsymbol C}^{\T}]^\star= {\boldsymbol C}$.  Here, we have introduced
the star symbol to mean that a quantity $Q^\star$ is obtained from $Q$ by
taking the complex conjugate of all Lagrangian parameters appearing in its
calculation, but not taking the complex conjugates of Euclideanized loop
integral functions, whose imaginary (absorptive) parts correspond to
fermion decay widths to multi-particle intermediate states.  That is, the
dispersive part of $\boldsymbol{C}$ is hermitian and the absorptive part
of $\boldsymbol{C}$ is anti-hermitian.

\begin{figure}[b!]
\begin{center}
\begin{picture}(80,70)(0,8)
\ArrowLine(28,40)(0,40)
\ArrowLine(80,40)(52,40)
\GBox(28,28)(52,52){0.85}
\Text(76,49)[]{$\dot\beta$}
\Text(6,48)[]{$\alpha$}
\Text(74,32)[]{$j$}
\Text(6,32)[]{$i$}
\Text(40,72)[c]{$p$}
\LongArrow(56,64)(24,64)
\Text(40,8)[c]{$\BDpos ip\newcdot\sigma_{\alpha\dot\beta}
\,{\boldsymbol C}_{i}{}^j$}
\end{picture}
\hspace{1.1cm}
\begin{picture}(80,70)(0,8)
\ArrowLine(0,40)(28,40)
\ArrowLine(52,40)(80,40)
\GBox(28,28)(52,52){0.85}
\Text(6,49)[]{$\dot\alpha$}
\Text(74,49)[]{$\beta$}
\Text(6,32)[]{$i$}
\Text(74,32)[]{$j$}
\Text(40,72)[c]{$p$}
\LongArrow(56,64)(24,64)
\Text(48,8)[c]{$\BDpos ip\newcdot\sigmabar^{\dot{\alpha}\beta}
\,({\boldsymbol C}^{\T})^{i}{}_{j}$}
\end{picture}
\hspace{1.1cm}
\begin{picture}(80,70)(0,8)
\ArrowLine(0,40)(28,40)
\ArrowLine(80,40)(52,40)
\GBox(28,28)(52,52){0.85}
\Text(6,48)[]{$\dot\alpha$}
\Text(76,49)[]{$\dot\beta$}
\Text(6,32)[]{$i$}
\Text(74,32)[]{$j$}
\Text(40,8)[c]{$i\delta^{\dot\alpha}{}_{\dot\beta}\, {{\boldsymbol D}}^{ij}$}
\end{picture}
\hspace{1.1cm}
\begin{picture}(80,70)(0,8)
\ArrowLine(28,40)(0,40)
\ArrowLine(52,40)(80,40)
\GBox(28,28)(52,52){0.85}
\Text(6,48)[]{$\alpha$}
\Text(74,48)[]{$\beta$}
\Text(6,32)[]{$i$}
\Text(74,32)[]{$j$}
\Text(40,8)[c]{$i\delta_\alpha{}^\beta \,{\boldsymbol \Doverline}_{ij}$}
\end{picture}
\end{center}
\caption{The full, loop-corrected propagators for two-component fermions
are associated with functions ${\boldsymbol C}(p^2)_{i}{}^j$ [and its
matrix transpose],
${\boldsymbol D}(p^2)^{ij}$, and
${\boldsymbol \Doverline}(p^2)_{ij}$, as shown.
The shaded boxes represent the sum of all connected
Feynman diagrams, with external legs included.
The four-momentum $p$ flows from right to left.
}
\label{fig:fullprops}
\end{figure}
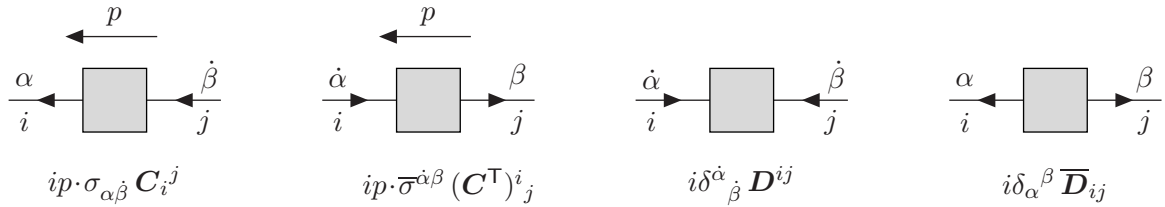

The diagrammatic representations of the full propagators are displayed in
\fig{fig:fullprops}, where ${\boldsymbol C}_{i}{}^j$, ${\boldsymbol
D}^{ij}$, and ${\boldsymbol \Doverline}_{ij}$ defined above are each
$N\times N$ matrix functions.
Note that the second diagram of \fig{fig:fullprops}, when flipped
by $180^\circ$ about the axis that bisects the diagram,
is equivalent to the first diagram of
\fig{fig:fullprops} (with $p\to -p$, $\alpha\to\beta$,
$\dot{\beta}\to\dot{\alpha}$ and $i\leftrightarrow j$).  In analogy with
\fig{fig:neutproprev}, one could replace the first two diagrammatic rules
of \fig{fig:fullprops} with a single rule shown in \fig{fullpropchoices},
where we have used \eq{ctij} to rewrite the second version of the rule in
terms of ${\boldsymbol C}^{\T}$.  Indeed, by using the $\sigmabar$-version
of the rule shown in \fig{fullpropchoices}
and flipping the corresponding diagram by $180^\circ$ as
described above, one reproduces the rule of the second diagram of
\fig{fig:fullprops}.

\begin{figure}[t]
\begin{center}
\thicklines
\begin{picture}(100,47)(20,25)
\ArrowLine(-32,40)(-80,40)
\ArrowLine(40,40)(-8,40)
\GBox(-32,28)(-8,52){0.85}
\Text(14,52)[]{$\dot\beta$}
\Text(-54,52)[]{$\alpha$}
\Text(14,30)[]{$j$}
\Text(-54,30)[]{$i$}
\Text(-20,72)[c]{$p$}
\LongArrow(0,64)(-40,64)
\Text(150,40)[]{$\BDpos ip\newcdot\sigma_{\alpha\dot\beta}
\,{\boldsymbol C}_{i}{}^j$
\quad \underline{or} \quad $\BDneg ip\newcdot\sigmabar^{\dot{\beta}\alpha}
\,({\boldsymbol C}^{\T})^{j}{}_{i}$}
\end{picture}
\end{center}
\caption{The first two diagrammatic rules of \fig{fig:fullprops} can
be summarized by a single diagram. Here, the choice of the $\sigma$
or $\sigmabar$ version of the rule is uniquely determined by the
height of the spinor indices on the vertex to which the
full loop-corrected propagator is connected (cf.~\fig{fig:neutproprev}
and the accompanying text).
\label{fullpropchoices}}
\end{figure}

In what follows, we prefer to keep the first two rules of
\fig{fig:fullprops} as separate entities.  This will permit us to
conveniently assemble the four diagrams of \fig{fig:fullprops} into a
$2\times 2$ block matrix of two-component propagators
[cf.~\eq{twofourprop}]. In addition, by choosing the momentum flow in the
two-component propagators from right to left, the left-to-right orderings
of the spinor labels of the diagrams coincide with the ordering of spinor
indices that appear in the corresponding algebraic representations.
Thus, we can multiply diagrams together and interpret them as the product
of the respective algebraic quantities taken from left to right in the
normal fashion.

Given the tree-level propagators of
\fig{fig:neutprop}, the full propagator functions
are given by:
\beqa
{\boldsymbol C}_{i}{}^j &=& \delta_i{}^j/(s - m_i^2) + \ldots
\label{treeprop1} \\
{\boldsymbol D}^{ij} &=&   {\boldsymbol m}^{ij}/(s - m_i^2)
+ \ldots
\label{treeprop2}\\
{\boldsymbol \Doverline}_{ij} &=&
{\boldsymbol \mbar}_{ij}/(s -  m_i^2)
+ \ldots \,,\label{treeprop3}
\eeqa
with no sum on $i$ in each case.  They are functions of the external
momentum invariant $s$ and the masses and couplings of the theory.
Inserting the leading terms [\eqst{treeprop1}{treeprop3}] into
\fig{fig:fullprops} and organizing the result in a $2\times 2$ block
matrix of two-component propagators reproduces the usual four-component
fermion tree-level propagator given in \eq{twofourprop}.

The computation of the full propagators can be organized, as usual in
quantum field theory, in terms of one-particle irreducible (1PI)
self-energy functions.  These are formally defined to be the sum of
Feynman diagrams to all orders in perturbation theory (with
the corresponding tree-level graph \textit{excluded})
that contribute to the 1PI
two-point Green function.  Diagrammatically, the 1PI self-energy functions
are defined in \fig{fig:selfenergies}.
\begin{figure}[tb!]
\begin{center}
\begin{picture}(80,70)(0,8)
\Text(40,72)[c]{$p$}
\LongArrow(56,64)(24,64)
\ArrowLine(28,40)(0,40)
\ArrowLine(80,40)(52,40)
\GCirc(40,40){12}{0.85}
\Text(74,48)[]{$\beta$}
\Text(6,48)[]{$\dot\alpha$}
\Text(6,32)[]{$i$}
\Text(74,32)[]{$j$}
\Text(40,8)[c]{$\BDneg i p\newcdot\sigmabar^{\dot{\alpha}\beta}
{\boldsymbol \Xi}_i{}^j$}
\end{picture}
\hspace{1.1cm}
\begin{picture}(80,70)(0,8)
\Text(40,72)[c]{$p$}
\LongArrow(56,64)(24,64)
\ArrowLine(0,40)(28,40)
\ArrowLine(52,40)(80,40)
\GCirc(40,40){12}{0.85}
\Text(74,48)[]{$\dot\beta$}
\Text(6,48)[]{$\alpha$}
\Text(6,32)[]{$i$}
\Text(74,32)[]{$j$}
\Text(40,8)[c]{$\BDneg i p\newcdot\sigma_{\alpha\dot\beta}
({\boldsymbol\Xi}^{\T})^i{}_j$}
\end{picture}
\hspace{1.1cm}
\begin{picture}(80,70)(0,8)
\ArrowLine(0,40)(28,40)
\ArrowLine(80,40)(52,40)
\GCirc(40,40){12}{0.85}
\Text(6,48)[]{$\alpha$}
\Text(74,48)[]{$\beta$}
\Text(6,32)[]{$i$}
\Text(74,32)[]{$j$}
\Text(40,8)[c]{$-i\delta_\alpha{}^\beta {\boldsymbol \Omega}^{ij}$}
\end{picture}
\hspace{1.1cm}
\begin{picture}(80,70)(0,8)
\ArrowLine(28,40)(0,40)
\ArrowLine(52,40)(80,40)
\GCirc(40,40){12}{0.85}
\Text(6,48)[]{$\dot\alpha$}
\Text(74,48)[]{$\dot\beta$}
\Text(6,32)[]{$i$}
\Text(74,32)[]{$j$}
\Text(40,8)[c]{$-i\delta^{\dot\alpha}{}_{\dot\beta}
{\boldsymbol \Omegabar}_{ij}$}
\end{picture}
\end{center}
\caption{The self-energy functions for two-component fermions are
associated with functions ${\boldsymbol \Xi}(s)_i{}^j$ [and its
matrix transpose], ${\boldsymbol \Omega}(s)^{ij}$, and $
{\boldsymbol \Omegabar}(s)_{ij}$, as shown.  The shaded circles represent
the sum of all one-particle irreducible, connected Feynman diagrams, and
the external legs are amputated. The four-momentum $p$ flows from right to
left. \label{fig:selfenergies} }
\end{figure}
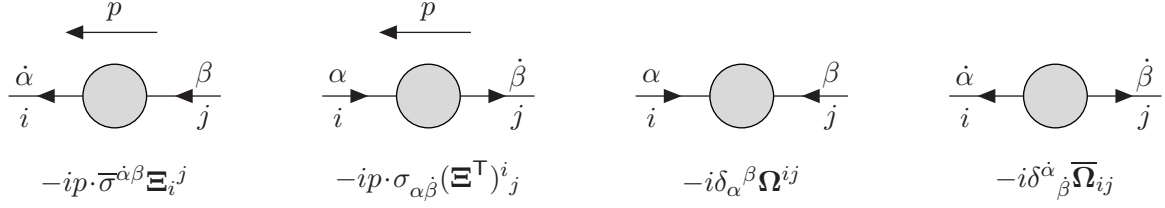
As in the case of the full loop-corrected propagators,  $\boldsymbol{\Omega}$ and
$\boldsymbol{\Omegabar}$ are complex symmetric matrices and
the self-energy function satisfy hermiticity conditions,
$[{\boldsymbol{\Xi}}^{\T}]^\star={\boldsymbol \Xi}$ and
${\boldsymbol{\Omegabar}}={\boldsymbol \Omega}^\star$, where
the star symbol was defined in the paragraph following \eq{ctid} and
$({\boldsymbol{\Xi}}^{\T})^i{}_j \equiv {\boldsymbol \Xi}_j{}^i$.

We illustrate the computation of the full propagator by considering
first the following diagrammatic identity
(with momentum $p$ flowing from right to left):
\beqa \label{digidentity}
&&\mbox{
\begin{picture}(215,120)(135,-60)
\ArrowLine(48,40)(20,40)
\ArrowLine(100,40)(72,40)
\GBox(48,28)(72,52){0.85}
\Text(94,49)[]{$\dot\beta$}
\Text(26,48)[]{$\alpha$}
\Text(94,32)[]{$j$}
\Text(26,32)[]{$i$}
\Text(120,40)[]{$=$}
\Text(212,49)[]{$\dot\beta$}
\Text(146,48)[]{$\alpha$}
\Text(212,32)[]{$j$}
\Text(146,32)[]{$i$}
\Text(162,0)[]{$\dot\gamma$}
\Text(126,0)[]{$\alpha$}
\Text(162,-18)[]{$k$}
\Text(126,-18)[]{$i$}
\Text(205,2)[]{$\delta$}
\Text(205,-18)[]{$\ell$}
\Text(258,1)[]{$\dot\beta$}
\Text(258,-18)[]{$j$}
\Text(354,0)[]{$\dot\gamma$}
\Text(318,0)[]{$\alpha$}
\Text(354,-18)[]{$k$}
\Text(318,-18)[]{$i$}
\Text(397,2)[]{$\dot\delta$}
\Text(397,-18)[]{$\ell$}
\Text(450,1)[]{$\dot\beta$}
\Text(450,-18)[]{$j$}
\Text(162,-50)[]{$\gamma$}
\Text(126,-50)[]{$\alpha$}
\Text(162,-68)[]{$k$}
\Text(126,-68)[]{$i$}
\Text(205,-48)[]{$\delta$}
\Text(205,-68)[]{$\ell$}
\Text(258,-49)[]{$\dot\beta$}
\Text(258,-68)[]{$j$}
\Text(354,-50)[]{$\gamma$}
\Text(318,-50)[]{$\alpha$}
\Text(354,-68)[]{$k$}
\Text(318,-68)[]{$i$}
\Text(397,-48)[]{$\dot\delta$}
\Text(397,-68)[]{$\ell$}
\Text(450,-49)[]{$\dot\beta$}
\Text(450,-68)[]{$j$}
\Text(120,40)[]{$=$}
\ArrowLine(218,40)(140,40)
\Text(100,-10)[]{$+$}
\ArrowLine(168,-10)(120,-10)
\GCirc(180,-10){12}{0.85}
\ArrowLine(220,-10)(192,-10)
\GBox(220,-22)(244,2){0.85}
\ArrowLine(273,-10)(244,-10)
\Text(292,-10)[]{$+$}
\ArrowLine(360,-10)(312,-10)
\GCirc(372,-10){12}{0.85}
\ArrowLine(384,-10)(412,-10)
\GBox(412,-22)(436,2){0.85}
\ArrowLine(464,-10)(436,-10)
\Text(100,-60)[]{$+$}
\ArrowLine(144,-60)(120,-60)
\ArrowLine(144,-60)(168,-60)
\GCirc(180,-60){12}{0.85}
\ArrowLine(220,-60)(192,-60)
\GBox(220,-72)(244,-48){0.85}
\ArrowLine(273,-60)(244,-60)
\Text(292,-60)[]{$+$}
\ArrowLine(336,-60)(312,-60)
\ArrowLine(336,-60)(360,-60)
\GCirc(372,-60){12}{0.85}
\ArrowLine(384,-60)(412,-60)
\GBox(412,-72)(436,-48){0.85}
\ArrowLine(464,-60)(436,-60)
\end{picture}
}
\nonumber \\
&&\phantom{line} \nonumber \\
&&\phantom{line}
\eeqa

\noindent
Similar diagrammatic
identities can be constructed for the three other full loop-corrected
propagators of \fig{fig:fullprops}.  The resulting four equations can
be neatly summarized by:
\beq \label{fttsf}
F=T+TSF\,,
\eeq
where $F$ is the matrix of full loop-corrected propagators, $T$ is the
matrix of tree-level propagators and $S$ is the matrix of self-energy
functions.  Expressing \eq{fttsf} in terms of diagrams,
\vspace{0.1in}
\beqa \label{matdiag}
\!\!\!
\begin{pmatrix}
\mbox{\begin{picture}(35,21)(0,10.5)
\SetScale{0.7}
\ArrowLine(17.5,25)(0,25)
\ArrowLine(32.5,25)(50,25)
\GBox(17.5,17.5)(32.5,32.5){0.85}
\end{picture}}
&\,\,\,
\mbox{\begin{picture}(35,21)(0,10.5)
\SetScale{0.7}
\ArrowLine(17.5,25)(0,25)
\ArrowLine(50,25)(32.5,25)
\GBox(17.5,17.5)(32.5,32.5){0.85}
\end{picture}}
\\
\mbox{\begin{picture}(35,21)(0,10.5)
\SetScale{0.7}
\ArrowLine(0,25)(17.5,25)
\ArrowLine(32.5,25)(50,25)
\GBox(17.5,17.5)(32.5,32.5){0.85}
\end{picture}}
&\,\,\,
\mbox{\begin{picture}(35,21)(0,10.5)
\SetScale{0.7}
\ArrowLine(0,25)(17.5,25)
\ArrowLine(50,25)(32.5,25)
\GBox(17.5,17.5)(32.5,32.5){0.85}
\end{picture}}
\end{pmatrix}
&=&
\begin{pmatrix}
\mbox{\begin{picture}(17.5,21)(0,10.5)
\SetScale{0.7}
\ArrowLine(15,25)(0,25)
\ArrowLine(15,25)(25,25)
\end{picture}}
&\quad
\mbox{\begin{picture}(17.5,21)(0,10.5)
\SetScale{0.7}
\ArrowLine(30,25)(0,25)
\end{picture}}\,\,
\\
\mbox{\begin{picture}(17.5,21)(0,10.5)
\SetScale{0.7}
\ArrowLine(0,25)(25,25)
\end{picture}}
&\quad
\mbox{\begin{picture}(17.5,21)(0,10.5)
\SetScale{0.7}
\ArrowLine(0,25)(15,25)
\ArrowLine(30,25)(15,25)
\end{picture}}\,\,
\end{pmatrix}
\left[
\begin{pmatrix}
\mbox{\begin{picture}(9.8,21)(0,0)
\Text(4.9,10.5)[c]{$\boldsymbol 1$}
\end{picture}}
&\,\,\,\,\,
\mbox{\begin{picture}(9.8,21)(0,0)
\Text(4.9,10.5)[c]{$\boldsymbol 0$}
\end{picture}}
\\
\mbox{\begin{picture}(9.8,21)(0,0)
\Text(4.9,10.5)[c]{$\boldsymbol 0$}
\end{picture}}
&\,\,\,\,\,
\mbox{\begin{picture}(9.8,21)(0,0)
\Text(4.9,10.5)[c]{$\boldsymbol 1$}
\end{picture}}
\end{pmatrix}
+
\begin{pmatrix}
\mbox{\begin{picture}(35,21)(0,10.5)
\SetScale{0.7}
\ArrowLine(0,25)(17.5,25)
\ArrowLine(50,25)(32.5,25)
\GCirc(25,25){7.5}{0.85}
\end{picture}}
&\,\,\,
\mbox{\begin{picture}(35,21)(0,10.5)
\SetScale{0.7}
\ArrowLine(0,25)(17.5,25)
\ArrowLine(32.5,25)(50,25)
\GCirc(25,25){7.5}{0.85}
\end{picture}}
\\
\mbox{\begin{picture}(35,21)(0,10.5)
\SetScale{0.7}
\ArrowLine(17.5,25)(0,25)
\ArrowLine(50,25)(32.5,25)
\GCirc(25,25){7.5}{0.85}
\end{picture}}
&\,\,\,
\mbox{\begin{picture}(35,21)(0,10.5)
\SetScale{0.7}
\ArrowLine(17.5,25)(0,25)
\ArrowLine(32.5,25)(50,25)
\GCirc(25,25){7.5}{0.85}
\end{picture}}
\end{pmatrix}
\begin{pmatrix}
\mbox{\begin{picture}(35,21)(0,10.5)
\SetScale{0.7}
\ArrowLine(17.5,25)(0,25)
\ArrowLine(32.5,25)(50,25)
\GBox(17.5,17.5)(32.5,32.5){0.85}
\end{picture}}
&\,\,\,
\mbox{\begin{picture}(35,21)(0,10.5)
\SetScale{0.7}
\ArrowLine(17.5,25)(0,25)
\ArrowLine(50,25)(32.5,25)
\GBox(17.5,17.5)(32.5,32.5){0.85}
\end{picture}}
\\
\mbox{\begin{picture}(35,21)(0,10.5)
\SetScale{0.7}
\ArrowLine(0,25)(17.5,25)
\ArrowLine(32.5,25)(50,25)
\GBox(17.5,17.5)(32.5,32.5){0.85}
\end{picture}}
&\,\,\,
\mbox{\begin{picture}(35,21)(0,10.5)
\SetScale{0.7}
\ArrowLine(0,25)(17.5,25)
\ArrowLine(50,25)(32.5,25)
\GBox(17.5,17.5)(32.5,32.5){0.85}
\end{picture}}
\end{pmatrix}
\right]
\nonumber \\
&&\phantom{line}
\eeqa
which, when expanded out, yields \eq{digidentity} and the corresponding
identities for the three other full loop-corrected
propagators of \fig{fig:fullprops}.
Note that
we have chosen the labeling and momentum flow in
\figs{fig:fullprops}{fig:selfenergies} such that the spinor and flavor
labels of the diagrams appear in the appropriate left-to-right order to
permit the interpretation of \eq{matdiag} as a matrix equation.
To solve for $F$,\footnote{\label{fniter}%
Alternatively, one can solve \eq{matdiag} by iteration
and summing the resulting geometric series.  This yields:
\beqa
\label{geomsum} F&=& T+TS(T+TS(T+TS(\cdots)))
=T+TST+TSTST+\ldots
=T[1+ST+(ST)^2+\ldots]
\nonumber \\
&=&T[1-ST]^{-1}=(T^{-1})^{-1}[1-ST]^{-1}=[(1-ST)T^{-1}]^{-1}
=[T^{-1}-S]^{-1}\,,\nonumber
\eeqa
which is equivalent to \eq{fullproppict}.}
we multiply \eq{fttsf}
on the left by $T^{-1}$ and on the right by $F^{-1}$ to obtain
$T^{-1}=F^{-1}+S$. Thus, $F=[T^{-1}-S]^{-1}$.  In pictures:
\beq \label{fullproppict}
\begin{pmatrix}
\mbox{\begin{picture}(35,21)(0,10.5)
\SetScale{0.7}
\ArrowLine(17.5,25)(0,25)
\ArrowLine(32.5,25)(50,25)
\GBox(17.5,17.5)(32.5,32.5){0.85}
\end{picture}}
&\quad
\mbox{\begin{picture}(35,21)(0,10.5)
\SetScale{0.7}
\ArrowLine(17.5,25)(0,25)
\ArrowLine(50,25)(32.5,25)
\GBox(17.5,17.5)(32.5,32.5){0.85}
\end{picture}}
\\
\mbox{\begin{picture}(35,21)(0,10.5)
\SetScale{0.7}
\ArrowLine(0,25)(17.5,25)
\ArrowLine(32.5,25)(50,25)
\GBox(17.5,17.5)(32.5,32.5){0.85}
\end{picture}}
&\quad
\mbox{\begin{picture}(35,21)(0,10.5)
\SetScale{0.7}
\ArrowLine(0,25)(17.5,25)
\ArrowLine(50,25)(32.5,25)
\GBox(17.5,17.5)(32.5,32.5){0.85}
\end{picture}}
\end{pmatrix}
=
\left [
\begin{pmatrix}
\mbox{\begin{picture}(17.5,21)(0,10.5)
\SetScale{0.7}
\ArrowLine(15,25)(0,25)
\ArrowLine(15,25)(30,25)
\end{picture}}
&\quad
\mbox{\begin{picture}(17.5,21)(0,10.5)
\SetScale{0.7}
\ArrowLine(25,25)(0,25)
\end{picture}}
\,\,\\
\mbox{\begin{picture}(17.5,21)(0,10.5)
\SetScale{0.7}
\ArrowLine(0,25)(25,25)
\end{picture}}
&\quad
\mbox{\begin{picture}(17.5,21)(0,10.5)
\SetScale{0.7}
\ArrowLine(0,25)(15,25)
\ArrowLine(30,25)(15,25)
\end{picture}}
\end{pmatrix}^{-1}
-
\begin{pmatrix}
\mbox{\begin{picture}(35,21)(0,10.5)
\SetScale{0.7}
\ArrowLine(0,25)(17.5,25)
\ArrowLine(50,25)(32.5,25)
\GCirc(25,25){7.5}{0.85}
\end{picture}}
&\quad
\mbox{\begin{picture}(35,21)(0,10.5)
\SetScale{0.7}
\ArrowLine(0,25)(17.5,25)
\ArrowLine(32.5,25)(50,25)
\GCirc(25,25){7.5}{0.85}
\end{picture}}
\\
\mbox{\begin{picture}(35,21)(0,10.5)
\SetScale{0.7}
\ArrowLine(17.5,25)(0,25)
\ArrowLine(50,25)(32.5,25)
\GCirc(25,25){7.5}{0.85}
\end{picture}}
&\quad
\mbox{\begin{picture}(35,21)(0,10.5)
\SetScale{0.7}
\ArrowLine(17.5,25)(0,25)
\ArrowLine(32.5,25)(50,25)
\GCirc(25,25){7.5}{0.85}
\end{picture}}
\end{pmatrix}
\right ]^{-1}.
\eeq

We evaluate the tree-level propagator matrix and its inverse using
\eqst{treeprop1}{treeprop3}, keeping in mind that the direction of
momentum flow is from right to left:
\beqa
&& \begin{pmatrix}
\mbox{\begin{picture}(17.5,21)(0,10.5)
\SetScale{0.7}
\ArrowLine(15,25)(0,25)
\ArrowLine(15,25)(30,25)
\end{picture}}
&\quad
\mbox{\begin{picture}(17.5,21)(0,10.5)
\SetScale{0.7}
\ArrowLine(25,25)(0,25)
\end{picture}}
\,\,\\
\mbox{\begin{picture}(17.5,21)(0,10.5)
\SetScale{0.7}
\ArrowLine(0,25)(25,25)
\end{picture}}
&\quad
\mbox{\begin{picture}(17.5,21)(0,10.5)
\SetScale{0.7}
\ArrowLine(0,25)(15,25)
\ArrowLine(30,25)(15,25)
\end{picture}}
\end{pmatrix}
\,\,=\,\,
\frac{1}{s - m_i^2}\begin{pmatrix}
i\,{\boldsymbol{\mbar}}_{ij}\,\delta_{\alpha}{}^{\beta}
& \quad
\BDpos i p\newcdot\sigma_{\alpha\dot{\beta}}\,\delta_i{}^j  \\[8pt]
\BDpos i p\newcdot\sigmabar^{\dot{\alpha}\beta}\,\delta^i{}_j &
\quad
i\,{\boldsymbol{m}}^{ij}\,\delta^{\dot{\alpha}}{}_{\dot\beta}
\end{pmatrix}
\,,\\[10pt]
&& \begin{pmatrix}
\mbox{\begin{picture}(17.5,21)(0,10.5)
\SetScale{0.7}
\ArrowLine(15,25)(0,25)
\ArrowLine(15,25)(30,25)
\end{picture}}
&\quad
\mbox{\begin{picture}(17.5,21)(0,10.5)
\SetScale{0.7}
\ArrowLine(25,25)(0,25)
\end{picture}}
\,\,\label{matrixprops}\\
\mbox{\begin{picture}(17.5,21)(0,10.5)
\SetScale{0.7}
\ArrowLine(0,25)(25,25)
\end{picture}}
&\quad
\mbox{\begin{picture}(17.5,21)(0,10.5)
\SetScale{0.7}
\ArrowLine(0,25)(15,25)
\ArrowLine(30,25)(15,25)
\end{picture}}
\end{pmatrix}^{-1}
=
\begin{pmatrix}
i\,{\boldsymbol{m}}^{ij}\,\delta_\alpha{}^\beta
& \quad
 \BDneg ip\newcdot\sigma_{\alpha\dot{\beta}}\,\delta^i{}_j
\\[8pt]
 \BDneg ip\newcdot\sigmabar^{\dot{\alpha}\beta}\,\delta_i{}^j & \quad
i\,{\boldsymbol{\mbar}}_{ij}\,\delta^{\dot\alpha}{}_{\dot\beta}
\end{pmatrix}
\,,\label{inverseprop}
\eeqa
where we follow the index structure defined in
\figs{fig:fullprops}{fig:selfenergies}. Inserting \eq{inverseprop} into
\eq{fullproppict}, one obtains a $4N\times 4N$ matrix equation for the
full propagator functions:
\beqa \label{matfullprop}
\begin{pmatrix}
  i {\boldsymbol \Doverline} & \quad
 \BDpos i p \newcdot \sigma \, {\boldsymbol C}
\\[8pt]
\quad
 \BDpos i p\newcdot \sigmabar \, {\boldsymbol C}^{\T} & \quad
i{{\boldsymbol D}}
\end{pmatrix}
=
\begin{pmatrix}
 i ({\boldsymbol m} + {\boldsymbol \Omega})
 & \quad
 \BDneg i p\newcdot \sigma \,
({\boldsymbol 1} - {\boldsymbol \Xi}^{\,\T})
\\[8pt]
 \BDneg i p\newcdot \sigmabar\,
({\boldsymbol 1} - {\boldsymbol \Xi})
 & \quad
i ({\boldsymbol \mbar}+{\boldsymbol \Omegabar})\end{pmatrix}^{-1}\,,
\eeqa
where $\boldsymbol{1}$ is the $N\times N$ identity matrix.  The right
hand side of \eq {matfullprop} can be evaluated by employing the
following identity for the inverse of a block-partitioned
matrix \cite{matrixref}:
\beq \label{invid}
\begin{pmatrix} P & Q \\ R & S
\end{pmatrix}^{-1}=\begin{pmatrix} \,(P-QS^{-1}R)^{-1} & \,\,
(R-SQ^{-1}P)^{-1} \\ \,(Q-PR^{-1}S)^{-1} & \,\,
(S-RP^{-1}Q)^{-1}\end{pmatrix}\,,
\eeq
under the assumption that all inverses appearing in \eq{invid} exist.
Applying this result to \eq{matfullprop}, we obtain
\beqa
{\boldsymbol{C}}^{-1}&=& s ({\boldsymbol{1}}-{\boldsymbol{\Xi}})-
({\boldsymbol{\mbar}}+{\boldsymbol{\Omegabar }})
({\boldsymbol{1}}-{\boldsymbol{\Xi^{\T}}})^{-1}
({\boldsymbol{m}}+{\boldsymbol{\Omega}})\,, \label{cinv}
\\
{\boldsymbol{D}}^{-1}&=& s ({\boldsymbol{1}}-{\boldsymbol{\Xi}})
({\boldsymbol{m}}+{\boldsymbol{\Omega}})^{-1}
({\boldsymbol{1}}-{\boldsymbol{\Xi^{\T}}})-
({\boldsymbol{\mbar}}+{\boldsymbol{\Omegabar }})\,,\label{dinv} \\
{\boldsymbol{\Doverline}}^{-1}&=&
s ({\boldsymbol{1}}-{\boldsymbol{\Xi^{\T}}})
({\boldsymbol{\mbar}}+{\boldsymbol{\Omegabar }})^{-1}
({\boldsymbol{1}}-{\boldsymbol{\Xi}})-
({\boldsymbol{m}}+{\boldsymbol{\Omega}})\,.\label{dbarinv}
\eeqa
Note that \eq{dbarinv} is consistent with \eq{dinv} as $\boldsymbol{\Xi^\star}
=\boldsymbol{\Xi^{\T}}$.

The pole mass can be found most easily by considering the rest frame of the
(off-shell) fermion, in which the space components of $p^\mu$ vanish.
This reduces the spinor index dependence to a triviality.
Setting $p^\mu=(\sqrt{s}\,;\,{\boldsymbol 0})$, we search for values of $s$
where the inverse of the full propagator has a zero eigenvalue.
This is equivalent to setting the determinant of the
inverse of the full propagator to zero.  Here we shall use the
well-known formula for the determinant of a block-partitioned
matrix~\cite{matrixref}:
\beq
{\rm det}~\begin{pmatrix} P & \quad Q\\[5pt] R & \quad S\end{pmatrix}
= {\rm det}~P\,\,{\rm det}~(S-RP^{-1}Q )\,.
\eeq
The end result is that the poles of the full propagator
(which are in general complex),
\beq
s_{{\rm pole},j} \equiv M^2_j - i \Gamma_j M_j,
\eeq
are formally the solutions to the non-linear equation\footnote{The
determinant of the inverse of the full propagator [the inverse of
\eq{matfullprop}] is equal to \eq{eq:polemass} multiplied by
${\rm det}~[-({\boldsymbol 1} - {\boldsymbol \Xi}) ({\boldsymbol
1} - {\boldsymbol \Xi}^{\T})]$.  We assume that the latter does not
vanish.  This must be true perturbatively since the eigenvalues of
$\boldsymbol{\Xi}$ are one-loop (or higher) quantities, which one
assumes cannot be as large as $1$.}
\beqa
{\rm det}~\bigl[
s {\boldsymbol 1} -
({\boldsymbol 1} - {\boldsymbol \Xi}^{\T})^{-1}
({\boldsymbol m} + {\boldsymbol \Omega})
({\boldsymbol 1} - {\boldsymbol \Xi})^{-1}
({\boldsymbol \mbar} + {{\boldsymbol \Omegabar}})
\bigr] = 0 \, .
\label{eq:polemass}
\eeqa

Some care is required in using \eq{eq:polemass}, since the pole
squared mass always has a {\em non-positive} imaginary part, while the loop
integrals used to find the self-energy functions are complex functions of
a real variable $s$ that is given an infinitesimal {\em positive}
imaginary part. Therefore, \eq{eq:polemass} should be solved iteratively
by first expanding the self-energy function matrices ${\boldsymbol
\Xi}$, ${\boldsymbol \Omega}$ and ${\boldsymbol \Omegabar}$ in a
series in $s$ about either $m^2_j + i \epsilon$ or $M^2_j + i
\epsilon$. The complex quantities $s_{{\rm pole},j}$, which can be 
identified as the complex pole squared masses, are
renormalization group and gauge invariant physical observables.  Examples
are given in \secs{subsec:toppole}{subsec:gluinopole}.

The results of this section can be applied to an arbitrary collection
of fermions (both Majorana or Dirac).  However, it is convenient to
treat separately the case where all fermions are Dirac fermions
(consisting of pairs of two-component fields $\chi_i$ and $\eta^i$).
As discussed in \sec{subsec:generalmass}, the Dirac fermion
mass eigenstates are defined in \eq{lrdef} and are determined by the
singular value decomposition of the Dirac fermion mass matrix.
With respect to the mass basis, we denote the diagonal Dirac fermion
mass matrix by $\boldsymbol{M}^{ij}$.  The elements of this matrix are
real and non-negative.  Nevertheless, it will be convenient as before
to define ${\boldsymbol \Mbar}_{ij}\equiv \boldsymbol{M}^{ij}$ to
maintain covariance when manipulating tensors with flavor indices.

At tree level, there are four propagators for each pair of $\chi$ and
$\eta$ fields as shown in \fig{fig:Diracpropagators}.  The
corresponding full, loop-corrected propagators are
shown in \fig{fig:fullDiracprops}.
\begin{figure}[htb!]
\begin{center}
\begin{picture}(80,70)(0,8)
\ArrowLine(28,40)(0,40)
\ArrowLine(80,40)(52,40)
\GBox(28,28)(52,52){0.85}
\Text(-8,38)[]{$\chi$}
\Text(88,38)[]{$\chi$}
\Text(76,49)[]{$\dot\beta$}
\Text(6,48)[]{$\alpha$}
\Text(74,32)[]{$j$}
\Text(6,32)[]{$i$}
\Text(40,72)[c]{$p$}
\LongArrow(56,64)(24,64)
\Text(40,8)[c]{$\BDpos ip\newcdot\sigma_{\alpha\dot\beta}
\,\boldsymbol{S_R}_{\,i}{}^j$}
\end{picture}
\hspace{1.1cm}
\begin{picture}(80,70)(0,8)
\Text(-8,38)[]{$\eta$}
\Text(88,38)[]{$\eta$}
\ArrowLine(0,40)(28,40)
\ArrowLine(52,40)(80,40)
\GBox(28,28)(52,52){0.85}
\Text(6,49)[]{$\dot\alpha$}
\Text(74,49)[]{$\beta$}
\Text(6,32)[]{$i$}
\Text(74,32)[]{$j$}
\Text(40,72)[c]{$p$}
\LongArrow(56,64)(24,64)
\Text(48,8)[c]{$\BDpos ip\newcdot\sigmabar^{\dot{\alpha}\beta}
\,(\boldsymbol{S_L^{\T}})^{i}{}_{j}$}
\end{picture}
\hspace{1.1cm}
\begin{picture}(80,70)(0,8)
\Text(-8,38)[]{$\eta$}
\Text(88,38)[]{$\chi$}
\ArrowLine(0,40)(28,40)
\ArrowLine(80,40)(52,40)
\GBox(28,28)(52,52){0.85}
\Text(6,48)[]{$\dot\alpha$}
\Text(76,49)[]{$\dot\beta$}
\Text(6,32)[]{$i$}
\Text(74,32)[]{$j$}
\Text(40,8)[c]{$i\delta^{\dot\alpha}{}_{\dot\beta}\, {\boldsymbol{S_D}}^{ij}$}
\end{picture}
\hspace{1.1cm}
\begin{picture}(80,70)(0,8)
\Text(-8,38)[]{$\chi$}
\Text(88,38)[]{$\eta$}
\ArrowLine(28,40)(0,40)
\ArrowLine(52,40)(80,40)
\GBox(28,28)(52,52){0.85}
\Text(6,48)[]{$\alpha$}
\Text(74,48)[]{$\beta$}
\Text(6,32)[]{$i$}
\Text(74,32)[]{$j$}
\Text(40,8)[c]{$i\delta_\alpha{}^\beta \,
(\boldsymbol{\Sbar_D\llsup{\, \T}})_{ij}$}
\end{picture}
\end{center}
\caption{The full, loop-corrected propagators for Dirac fermions,
represented by pairs of two-component
(oppositely charged) fermion fields $\chi_i$ and
$\eta_i$,
are associated with functions $\boldsymbol{S_R}(s)_{i}{}^j$,
$\boldsymbol{S_L^{\T}}(s)^{i}{}_j$,
$\boldsymbol{S_D}(s)^{ij}$, and
$\boldsymbol{\Sbar_D\llsup{\,\T}}(s)_{ij}$, as shown.
The shaded boxes represent the sum of all connected
Feynman diagrams, with external legs included.
The four-momentum $p$ and the charge of $\chi$ flow from right to left.
}
\label{fig:fullDiracprops}
\end{figure}
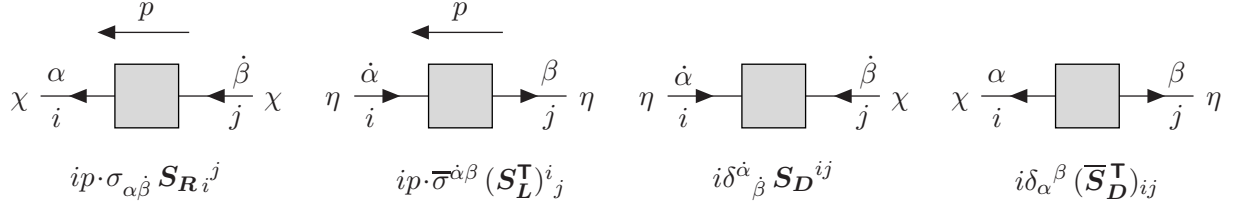
The naming and sign conventions
employed for the full, loop-corrected Dirac fermion
propagator functions in \fig{fig:fullDiracprops}
derives from the corresponding functions used
in the more traditional four-component treatment presented in
\app{G} [cf.~\eq{essex}].

In general, the complex matrices $\boldsymbol{S_R}$ and
$\boldsymbol{S_L}$ satisfy hermiticity conditions
$[\boldsymbol{S_R^{\T}}]^\star=\boldsymbol{S_R}$ and
$[\boldsymbol{S_L^{\T}}]^\star=\boldsymbol{S_L}$, whereas the complex
matrices $\boldsymbol{S_D}$ and $\boldsymbol{\Sbar_D}$ are
related by $\boldsymbol{\Sbar_D}=\boldsymbol{S_D^{\,\star}}$,
where the star symbol is defined in the paragraph below \eq{ctid}.  In
contrast to the general case of an arbitrary collection of fermions
treated earlier, $\boldsymbol{S_R}$ and
$\boldsymbol{S_L}$ are unrelated and $\boldsymbol S_D$ is
a complex matrix (not necessarily symmetric).

Instead of working in a $\chi$--$\eta$ basis for the
two-component Dirac fermion fields, one can Takagi-diagonalize the
fermion mass matrix.  In the new $\psi$-basis,
the loop-corrected propagators of \fig{fig:fullprops} are
applicable.  It is easy to check that the number of independent
functions is the same in both methods for treating Dirac fermions.
In particular, the loop-corrected propagator functions in
the $\psi$-basis are given in terms of the corresponding functions
in the $\chi$--$\eta$ basis by:\footnote{\label{fnpropnotation}%
The simple forms
of $\boldsymbol{C}$ in \eq{cddbar}
and $\boldsymbol{\Xi}$ in \eq{xisigsigbar} motivate our
definitions of $\boldsymbol{S_L}$ and $\boldsymbol{\Sigma_R}$
with the transpose
as indicated in \figs{fig:fullDiracprops}{fig:diracselfenergies},
respectively.}
\beq \label{cddbar}
\boldsymbol{C}=\begin{pmatrix} {\boldsymbol S_R}\,\, & 0 \\ 0\,\, &
{\boldsymbol S_L}
\end{pmatrix}
\,,\qquad\quad
{\boldsymbol D}=\begin{pmatrix}  0\,\, & {\boldsymbol S_D^{\T}}
\\ {\boldsymbol S_D} \,\,& 0
\end{pmatrix}\,,
\qquad\quad
{\boldsymbol \Doverline}
=\begin{pmatrix}  0\,\, & \boldsymbol{\Sbar_D^{\T}}
\\ \boldsymbol{\Sbar_D} \,\,& 0
\end{pmatrix}\,.
\eeq

We similarly introduce the 1PI self-energy matrix functions
for the Dirac fermions in the $\chi$--$\eta$ basis,
where the corresponding self-energy
functions are defined in \fig{fig:diracselfenergies}.
As before, the naming and sign conventions
employed for the Dirac fermion self-energy
functions above derives from the corresponding functions used
in the more traditional four-component treatment of \app{G}
[cf.~\eq{sigmaex}].
\begin{figure}[tbh]
\begin{center}
\begin{picture}(80,70)(0,8)
\Text(40,72)[c]{$p$}
\LongArrow(56,64)(24,64)
\ArrowLine(28,40)(0,40)
\ArrowLine(80,40)(52,40)
\Text(-8,38)[]{$\chi$}
\Text(88,38)[]{$\chi$}
\GCirc(40,40){12}{0.85}
\Text(74,48)[]{$\beta$}
\Text(6,48)[]{$\dot\alpha$}
\Text(6,32)[]{$i$}
\Text(74,32)[]{$j$}
\Text(40,8)[c]{$\BDneg i p\newcdot\sigmabar^{\dot{\alpha}\beta}
\boldsymbol{\Sigma_L}_{\,i}{}^j$}
\end{picture}
\hspace{1.1cm}
\begin{picture}(80,70)(0,8)
\Text(-8,38)[]{$\eta$}
\Text(88,38)[]{$\eta$}
\Text(40,72)[c]{$p$}
\LongArrow(56,64)(24,64)
\ArrowLine(0,40)(28,40)
\ArrowLine(52,40)(80,40)
\GCirc(40,40){12}{0.85}
\Text(74,48)[]{$\dot\beta$}
\Text(6,48)[]{$\alpha$}
\Text(6,32)[]{$i$}
\Text(74,32)[]{$j$}
\Text(40,8)[c]{$\BDneg i p\newcdot
\sigma_{\alpha\dot\beta}(\boldsymbol{\Sigma_R^{\T}})^i{}_j$}
\end{picture}
\hspace{1.1cm}
\begin{picture}(80,70)(0,8)
\Text(-8,38)[]{$\eta$}
\Text(88,38)[]{$\chi$}
\ArrowLine(0,40)(28,40)
\ArrowLine(80,40)(52,40)
\GCirc(40,40){12}{0.85}
\Text(6,48)[]{$\alpha$}
\Text(74,48)[]{$\beta$}
\Text(6,32)[]{$i$}
\Text(74,32)[]{$j$}
\Text(40,8)[c]{$-i\delta_\alpha{}^\beta \boldsymbol{\Sigma_D}^{ij}$}
\end{picture}
\hspace{1.1cm}
\begin{picture}(80,70)(0,8)
\Text(-8,38)[]{$\chi$}
\Text(88,38)[]{$\eta$}
\ArrowLine(28,40)(0,40)
\ArrowLine(52,40)(80,40)
\GCirc(40,40){12}{0.85}
\Text(6,48)[]{$\dot\alpha$}
\Text(74,48)[]{$\dot\beta$}
\Text(6,32)[]{$i$}
\Text(74,32)[]{$j$}
\Text(40,8)[c]{$-i\delta^{\dot\alpha}{}_{\dot\beta}
(\boldsymbol{\Sigmabar_D\llsup{\,\T}})_{ij}$}
\end{picture}
\end{center}
\caption{The self-energy functions for two-component Dirac fermions,
  represented by pairs of two-component (oppositely charged) fermion
  fields $\chi_i$ and $\eta_i$, are associated with functions
  $\boldsymbol{\Sigma_L}(s)_{i}{}^j$, $\boldsymbol{
    \Sigma_R^{\T}}(s)^{i}{}_j$, $\boldsymbol{\Sigma_D}(s)^{ij}$, and
  $\boldsymbol{\Sigmabar_D\llsup{\,\T}}(s)_{ij}$, as shown.  The
  shaded circles represent the sum of all one-particle irreducible,
  connected Feynman diagrams, and the external legs are amputated.
  The four-momentum $p$ flows from right to left.
\label{fig:diracselfenergies}
}
\end{figure}
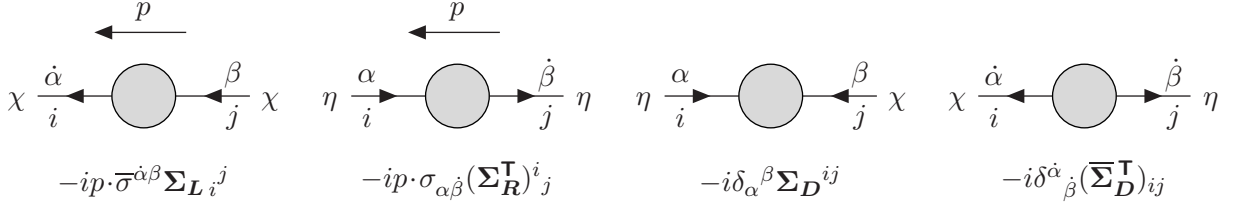

Once again, the complex matrices $\boldsymbol{\Sigma_L}$ and
$\boldsymbol{\Sigma_R}$ satisfy
hermiticity conditions $[\boldsymbol{\Sigma_L^{\T}}]^\star
=\boldsymbol{\Sigma_L}$
and $[\boldsymbol{\Sigma_R^{\T}}]^\star=\boldsymbol{\Sigma_R}$,
whereas the complex
matrices $\boldsymbol{\Sigma_D}$ and
$\boldsymbol{\Sigmabar_D}$ are related
by $\boldsymbol{\Sigmabar_D}=\boldsymbol{\Sigma_D^\star}$,
where the star symbol is defined in the paragraph
below \eq{ctid}.  Likewise, $\boldsymbol{\Sigma_L}$ and
$\boldsymbol{\Sigma_R}$ are unrelated and
$\boldsymbol{\Sigma_D}$ is a complex matrix (not necessarily
symmetric). The self-energy functions in
the $\psi$-basis are given in terms of the corresponding functions
in the $\chi$--$\eta$ basis by:$^{\ref{fnpropnotation}}$
\beq \label{xisigsigbar}
\boldsymbol{\Xi}=\begin{pmatrix} {\boldsymbol \Sigma_L}\,\, & 0 \\ 0\,\, &
{\boldsymbol \Sigma_R}
\end{pmatrix}
\,,\qquad\quad
\boldsymbol{\Omega}=\begin{pmatrix}  0\,\, & {\boldsymbol\Sigma_D^\T}
\\ {\boldsymbol\Sigma_D} \,\,& 0
\end{pmatrix}
\,,\qquad\quad
{\boldsymbol{\Omegabar}}=
\begin{pmatrix}  0\,\, & \boldsymbol{\Sigmabar_D^\T}
\\ \boldsymbol{\Sigmabar_D} \,\,& 0
\end{pmatrix}
\,.
\eeq

In the case of Dirac fermions fields,
\eq{fullproppict} still holds in the $\chi$--$\eta$ basis,
which yields:
\beqa \label{matfullDiracprop}
\begin{pmatrix}
i \boldsymbol{\Sbar_D\llsup{\,\T}}
& \quad
\BDpos i p \newcdot \sigma \, {\boldsymbol S_R}
\\[8pt]
\BDpos i p\newcdot \sigmabar \, \boldsymbol{S_L^{\T}}  & \quad
i{{\boldsymbol S_D}}
\end{pmatrix}
=
\begin{pmatrix}
i ({\boldsymbol M} + {\boldsymbol \Sigma_D})
& \quad
 \BDneg i p\newcdot \sigma \,({\boldsymbol 1} - {\boldsymbol \Sigma_R^{\T}})
\\[8pt]
 \BDneg i p\newcdot \sigmabar\, ({\boldsymbol 1}
- {\boldsymbol \Sigma_L} ) & \quad
i ({\boldsymbol \Mbar}+\boldsymbol{\Sigmabar_D\llsup{\,\T}})
\end{pmatrix}^{-1}\,.
\eeqa
Using \eq{invid}, it follows that:
\beqa
{\boldsymbol{S\ls{L}}}^{\!\!-1}
&=& s ({\boldsymbol{1}}-{\boldsymbol{\Sigma_R}})-
({\boldsymbol{\Mbar}}+{\boldsymbol{\Sigmabar_D}})
({\boldsymbol{1}}-{\boldsymbol{\Sigma_L^{\T}}})^{-1}
({\boldsymbol{M}}+{\boldsymbol{\Sigma_D^{\T}}})\,,
\label{sinv1}
\\
{\boldsymbol{S\ls{R}}}^{\!\!-1}&=&
s ({\boldsymbol{1}}-{\boldsymbol{\Sigma_L}})-
({\boldsymbol{\Mbar}}+{\boldsymbol{\Sigmabar_D^{\T}}})
({\boldsymbol{1}}-{\boldsymbol{\Sigma_R^{\T}}})^{-1}
({\boldsymbol{M}}+{\boldsymbol{\Sigma_D}})\,,
\label{sinv2}
\\
{\boldsymbol{S\ls{D}}}^{\!\!-1}
&=& s ({\boldsymbol{1}}-{\boldsymbol{\Sigma_L}})
({\boldsymbol{M}}+{\boldsymbol{\Sigma_D}})^{-1}
({\boldsymbol{1}}-{\boldsymbol{\Sigma_R^{\T}}})-
({\boldsymbol{\Mbar}}+{\boldsymbol{\Sigmabar_D^{\T}}})\,,
\label{sinv3}
\\
{\boldsymbol{\Sbar\ls{D}}}^{\!\!-1}&=&
s ({\boldsymbol{1}}-{\boldsymbol{\Sigma_L^{\T}}})
({\boldsymbol{\Mbar}}+{\boldsymbol{\Sigmabar_D}})^{-1}
({\boldsymbol{1}}-{\boldsymbol{\Sigma_R}})
-({\boldsymbol{M}}+{\boldsymbol{\Sigma_D^{\T}}})\,.\label{sinv4}
\eeqa
Note that \eq{sinv4} is consistent with \eq{sinv3} as
$\boldsymbol{\Sigma_{L,R}^\star}
=\boldsymbol{\Sigma_{L,R}^{\T}}$.

The pole mass is now easily computed
using the technique previously outlined.  In particular, \eq{eq:polemass}
becomes:
\beqa
{\rm det}~\bigl[
s {\boldsymbol 1} -
({\boldsymbol 1} - {\boldsymbol \Sigma_R^{\T}})^{-1}
({\boldsymbol M} + {\boldsymbol \Sigma_D})
({\boldsymbol 1} - {\boldsymbol \Sigma_L})^{-1}
({\boldsymbol \Mbar} + \boldsymbol{\Sigmabar_D\llsup{\,\T}})
\bigr] = 0 \,,
\label{eq:diracpolemass}
\eeqa
which determines the complex pole squared masses,
$s_{\rm pole}$, of the corresponding Dirac fermions.
Again, the self-energy functions should be expanded in a series in $s$ about
a point with an infinitesimal positive imaginary part.

Finally, we examine the special case of a parity-conserving vectorlike
theory of Dirac fermions (such as QED or QCD).  In this case, the
following relations hold among the loop-corrected propagator functions
and self-energy functions, respectively:\footnote{These relations
are derived using four-component spinor methods in \app{G}
[cf.~\eqs{app:vectorlike1}{app:vectorlike2}].}
\beqa
&& \boldsymbol{S_R}_i{}^j=(\boldsymbol{S_L^{\T}})^i{}_j\,,
\qquad\qquad \boldsymbol{S_D}^{ij}=
(\boldsymbol{\Sbar\llsup{\,\T}_D})_{ij}\,,\label{vectorlike1}\\
&& \boldsymbol{\Sigma_L}_i{}^j=(\boldsymbol{\Sigma_R^{\T}})^i{}_j\,,
\qquad\qquad \boldsymbol{\Sigma_D}^{ij}
=(\boldsymbol{\Sigmabar_D\llsup{\,\T}})_{ij}
\,.\label{vectorlike2}
\eeqa
By imposing \eq{vectorlike2} on \eqst{sinv1}{sinv4} and recalling that
$\boldsymbol{\Mbar}_{ij}=\boldsymbol{M}^{ij}$, it is
straightforward to verify that \eq{vectorlike1} is satisfied.

\section{\texorpdfstring{Conventions for fermion and antifermion names and fields}{Conventions for fermion and antifermion names and fields}}
\label{sec:nomenclature}
\renewcommand{\theequation}{\arabic{section}.\arabic{equation}}
\renewcommand{\thefigure}{\arabic{section}.\arabic{figure}}
\renewcommand{\thetable}{\arabic{section}.\arabic{table}}
\setcounter{equation}{0}
\setcounter{figure}{0}
\setcounter{table}{0}

In this section, we establish conventions for labeling Feynman diagrams
that contain two-component
fermion fields of the Standard Model (SM) and its minimal
supersymmetric extension (MSSM).
In the case of Majorana fermions, there is a one-to-one correspondence
between the particle names and the undaggered $(\half,0)$
[left-handed] fields.  In contrast, for Dirac
fermions there are always two distinct two-component fields that
correspond to each particle name. For a quark or lepton generically
denoted by $f$, we employ the two-component
undaggered $(\half,0)$ [left-handed] fields $f$ and $\bar f$
(where the bar is part of the field name and does \textit{not}
refer to complex conjugation of any kind).  This is illustrated in
Table~\ref{tab:nomenclature}, which lists the
SM and MSSM fermion particle names
together with the corresponding two-component fields.
For each particle, we list the two-component field with the same quantum
numbers, i.e., the field that contains the annihilation operator for that
one-particle state (which creates the one-particle state when acting to the
\textit{left} on the vacuum $\langle 0|$).

\renewcommand{\arraystretch}{1.55}
\begin{table}[t!]
\caption{Fermion and antifermion names and two-component fields in the
Standard Model and the MSSM.  In the listing of two-component fields,
the first is an undaggered $(\half,0)$ [left-handed] field and the
second is a daggered $(0,\half)$ [right-handed] field.
The bars on the two-component (antifermion)
fields are part of their names, and do not
denote some form of complex conjugation.
(In this table, neutrinos are considered to be exactly massless
and the left-handed antineutrino $\bar\nu$ is absent from the spectrum.)
\label{tab:nomenclature}}
\begin{center}
\begin{tabular}{|c|c|}
\hline
Fermion name & Two-component fields  
\\
\hline\hline
$\ell^-$ (lepton) & $\ell\> , \> {\bar\ell}^\dagger$
\\  \hline
$\ell^+$ (antilepton) & $\bar\ell \> ,{\ell}^\dagger \> $
\\  \hline
$\nu$ (neutrino) & $\nu\> , \> {\rm -}$
\\  \hline
$\nubar$ (antineutrino) & ${\rm -} \> , \> \nu^\dagger$
\\  \hline
$q$ (quark) & $q\> , \>{\bar q}^\dagger$
\\  \hline
$\bar q$ (antiquark) & $\bar q\> , \> {q}^\dagger$
\\  \hline
$f$ (quark or lepton) & $f\> , \>{\bar f}^\dagger$
\\  \hline
$\fbar$ (antiquark or antilepton) & $\bar f\> , \> {f}^\dagger$
\\  \hline
$\stilde N_i$ (neutralino) & $\chi^0_i\> ,
        \> {\chi^0_i}^\dagger$
\\ \hline
$\stilde C_i^+$ (chargino) & $\chi^+_i\> ,
        \> {\chi^-_i}^\dagger$
\\ \hline
$\stilde C_i^-$ (anti-chargino) & $\chi^-_i\> ,
        \> {\chi^+_i}^\dagger$
\\ \hline
$\stilde g$ (gluino) & $\> \stilde g\> , \> {\stilde g}^\dagger$
\\ \hline
\end{tabular}
\end{center}
\end{table}

There is an option of labeling fermion lines in Feynman diagrams
by particle names or by field names;
each choice has advantages and
disadvantages.\footnote{Unfortunately, the notation for
fermion names
can be ambiguous because some of the symbols used also appear as names
for one of the two-component fermion fields.  In practice, it should
be clear from the context which set of names are being employed.}
In all of the examples that follow, we have chosen
to eliminate the possibility of ambiguity as follows. We always label
fermion lines with two-component fields (rather than particle
names), and adopt the following
conventions:

$\bullet$ In the Feynman rules for interaction vertices,
the external lines are always labeled by the undaggered
$(\half,0)$ [left-handed]
field, regardless of whether the corresponding arrow is
pointed in or out of the vertex.
Two-component fermion lines with arrows
pointing away from the vertex
correspond to dotted indices, and two-component
fermion lines with arrows pointing
toward the vertex  always correspond to
undotted indices.
This also applies to
Feynman diagrams where
the roles of the initial state and the final state are ambiguous
(such as self-energy diagrams).

$\bullet$ Internal fermion lines in Feynman diagrams are also always
labeled by the undaggered $(\half,0)$
[left-handed] field(s). Internal fermion lines containing
a propagator with opposing arrows can carry two labels (e.g., see
\fig{phifscattering}).

$\bullet$ Initial state external fermion lines
(which always have physical
three-momenta pointing into the vertex) in Feynman diagrams for complete
processes are labeled by
the corresponding
undaggered ($\half,0)$ [left-handed]
field if the arrow is into the vertex, and by the
daggered $(0,\half)$ [right-handed] field if the arrow is away from the vertex.

$\bullet$ Final state external fermion lines (which always have physical
three-momenta pointing out of the vertex)
in Feynman diagrams
for complete processes are labeled by the corresponding
daggered $(0,\half)$
[right-handed] field if the arrow is into the vertex, and by the
undaggered $(\half,0)$ [left-handed] field if the arrow is away from the vertex.
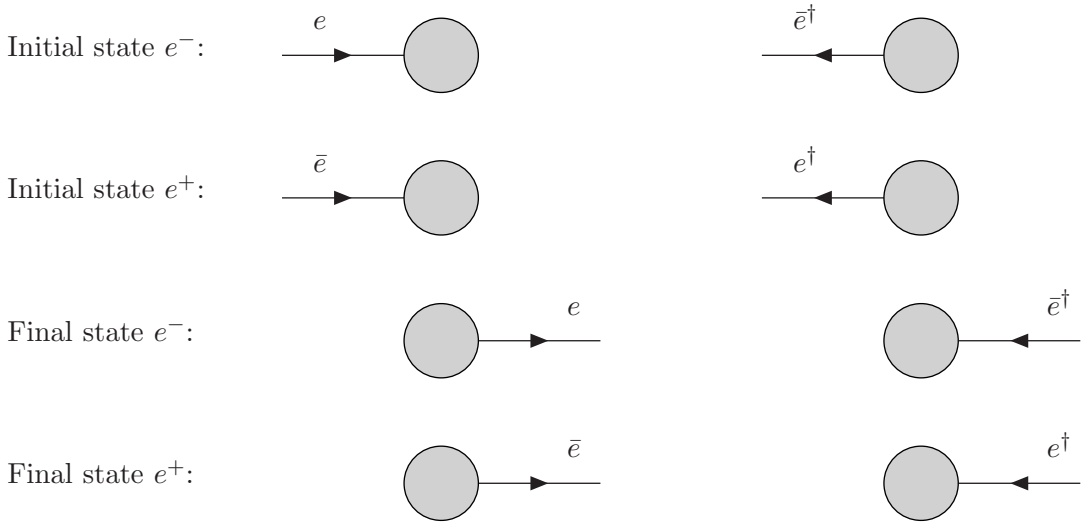
\begin{figure}[tbp!]
\begin{center}
\begin{picture}(80,60)(0,0)
\put(0,20){Initial state $e^-$:}
\end{picture}
\begin{picture}(160,40)(0,0)
\ArrowLine(20,20)(66,20)
\GCirc(80,20){14}{0.82}
\put(32,30){$e$}
\end{picture}
\hspace{0.5cm}
\begin{picture}(160,40)(0,0)
\ArrowLine(66,20)(20,20)
\GCirc(80,20){14}{0.82}
\put(32,30){${\bar e}^\dagger$}
\end{picture}
\end{center}
\begin{center}
\begin{picture}(80,41)(0,0)
\put(0,20){Initial state $e^+$:}
\end{picture}
\begin{picture}(160,41)(0,0)
\ArrowLine(20,20)(66,20)
\GCirc(80,20){14}{0.82}
\put(32,30){$\bar e$}
\end{picture}
\hspace{0.5cm}
\begin{picture}(160,41)(0,0)
\ArrowLine(66,20)(20,20)
\GCirc(80,20){14}{0.82}
\put(32,30){${e}^\dagger$}
\end{picture}
\end{center}
\begin{center}
\begin{picture}(80,41)(0,0)
\put(0,20){Final state $e^-$:}
\end{picture}
\begin{picture}(160,41)(0,0)
\ArrowLine(94,20)(140,20)
\GCirc(80,20){14}{0.82}
\put(128,30){$e$}
\end{picture}
\hspace{0.5cm}
\begin{picture}(160,41)(0,0)
\ArrowLine(140,20)(94,20)
\GCirc(80,20){14}{0.82}
\put(128,30){${\bar e}^\dagger$}
\end{picture}
\end{center}
\begin{center}
\begin{picture}(80,41)(0,0)
\put(0,20){Final state $e^+$:}
\end{picture}
\begin{picture}(160,41)(0,0)
\ArrowLine(94,20)(140,20)
\GCirc(80,20){14}{0.82}
\put(128,30){$\bar e$}
\end{picture}
\hspace{0.5cm}
\begin{picture}(160,41)(0,0)
\ArrowLine(140,20)(94,20)
\GCirc(80,20){14}{0.82}
\put(128,30){${e}^\dagger$}
\end{picture}
\end{center}
\caption{The
two-component field labeling conventions for external Dirac fermion lines in a
Feynman diagram for a physical process.
The top row corresponds to an initial state electron,
the second row to an initial state positron, the
third row to a final state electron, and the fourth row to a final state
positron.
The labels above each line are the two-component field names.
The corresponding conventions for a massless neutrino are obtained by
deleting the diagrams with $\bar e$ or ${\bar e}^\dagger$, and
changing
$e$ and ${e}^\dagger$ to $\nu$ and $\nu^\dagger$, respectively.}
\label{labelconvention}
\end{figure}

The rules for labeling external Dirac fermions
are summarized in \fig{labelconvention}.
These labeling conventions differ slightly from the ones employed
in \sec{subsec:simpleapps}, where \textit{all} internal and
external initial state
and final state fermion lines were labeled by the corresponding
\textit{undaggered} $(\half,0)$ left-handed fields.  In this latter
convention, the conserved quantities
(charges, lepton numbers, baryon numbers, etc.)~of the labeled fields
follow the direction of the arrow that adorns
the corresponding fermion line in the diagram.
In contrast, in the convention of \fig{labelconvention},
the field labels used for external fermion lines
always correspond to the physical particle, and the corresponding
conserved quantities of the labeled fields
follow the direction of the particle
three-momentum.  As an example, for either initial or final states,
the two-component fields $e$ and ${\bar e}^\dagger$ both
represent a negatively charged electron, conventionally denoted by
$e^-$, whereas both $\bar e$ and~${e}^\dagger$ represent a positively
charged positron, conventionally denoted by $e^+$
(cf.~Table~\ref{tab:nomenclature}).

The application of our labeling conventions to processes involving
Majorana fermions is completely straightforward.  For example,
the conventions for employing the neutralino states as external particles are
summarized in \fig{labelconventionN}.
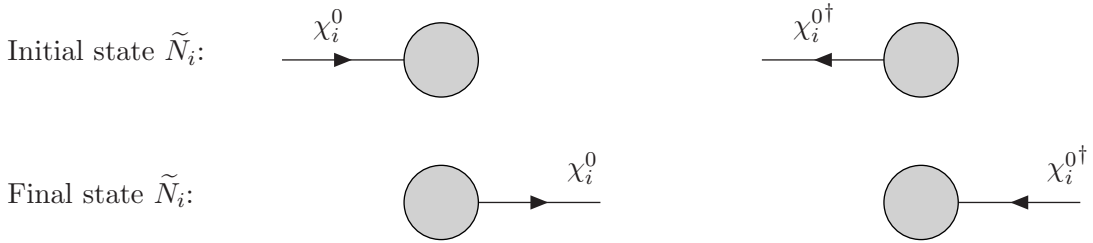
\begin{figure}[tbp!]
\begin{center}
\begin{picture}(80,60)(0,0)
\put(0,20){Initial state $\stilde N_i$:}
\end{picture}
\begin{picture}(160,40)(0,0)
\ArrowLine(20,20)(66,20)
\GCirc(80,20){14}{0.82}
\put(32,30){$\chi_i^0$}
\end{picture}
\hspace{0.5cm}
\begin{picture}(160,40)(0,0)
\ArrowLine(66,20)(20,20)
\GCirc(80,20){14}{0.82}
\put(32,30){${\chi_i^0}^\dagger$}
\end{picture}
\end{center}
\begin{center}
\begin{picture}(80,41)(0,0)
\put(0,20){Final state $\stilde N_i$:}
\end{picture}
\begin{picture}(160,41)(0,0)
\ArrowLine(94,20)(140,20)
\GCirc(80,20){14}{0.82}
\put(128,30){$\chi_i^0$}
\end{picture}
\hspace{0.5cm}
\begin{picture}(160,41)(0,0)
\ArrowLine(140,20)(94,20)
\GCirc(80,20){14}{0.82}
\put(128,30){${\chi_i^0}^\dagger$}
\end{picture}
\end{center}
\caption{The
two-component field labeling conventions for external
Majorana fermion lines in a
Feynman diagram for a physical process.
The top row corresponds to an initial state neutralino,
and the second row to a final state neutralino.
The labels above each line are the two-component field names.
(The neutralino is its own antiparticle.)}
\label{labelconventionN}
\end{figure}
\begin{figure}[tbp!]
\vspace{0.1in}
\begin{center}
\begin{picture}(200,68)(0,0)
\Photon(40,40)(-10,40){3}{5}
\ArrowLine(40,40)(80,70)
\ArrowLine(80,10)(40,40)
\Text(10,25)[]{$\gamma$}
\Text(50,65)[]{$\dot{\alpha}$}
\Text(50,20)[]{$\beta$}
\Text(140,40)[l]{$
\BDpos ie \sigmabar_\mu^{\dot{\alpha}\beta}$
\quad \rm{or}\quad $\BDneg ie\sigma_{\mu\beta\dot{\alpha}}$}
\Text(90,70)[]{$e$}
\Text(90,10)[]{$e$}
\put(-50,40){(a)}
\end{picture}
\end{center}
\vspace{0.1in}
\begin{center}
\begin{picture}(200,68)(0,0)
\Photon(40,40)(-10,40){3}{5}
\ArrowLine(40,40)(80,70)
\ArrowLine(80,10)(40,40)
\Text(10,25)[]{$\gamma$}
\Text(50,20)[]{$\beta$}
\Text(50,65)[]{$\dot{\alpha}$}
\Text(140,40)[l]{$\BDneg ie\sigmabar_\mu^{\dot{\alpha}\beta}$
\quad \rm{or}\quad $\BDpos ie\sigma_{\mu\beta\dot{\alpha}}$}
\Text(92,70)[]{$\bar e$}
\Text(92,10)[]{$\bar e$}
\put(-50,40){(b)}
\end{picture}
\end{center}
\caption{\label{eorebar} The two-component Feynman rules for the QED
vertex.  Following the conventions outlined in
\sec{sec:nomenclature},
we label these rules with the $(\half,0)$ [left-handed] fields $e$
and $\bar e$, which comprise the Dirac electron.  Note that $Q_e=-1$,
and the electromagnetic coupling constant $e$ (not to be confused with the
two-component electron field that is denoted by the same letter)
is conventionally defined such that $e>0$ [cf.~\fig{SMintvertices}].}
\end{figure}
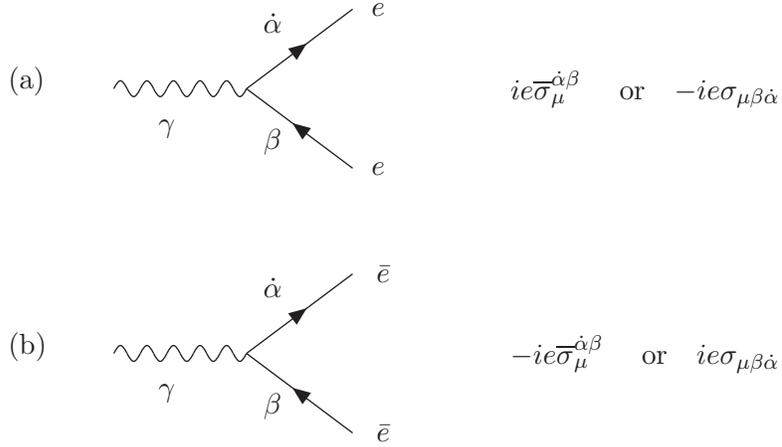

As a simple example, consider Bhabha scattering ($e^-e^+\to e^- e^+$)
\cite{bhabha}. We require 
the two-component Feynman rules for the QED coupling of
electrons and positrons to the photon, which are exhibited in
\fig{eorebar}.
Consider the $s$-channel tree-level Feynman
diagrams that contribute to the invariant amplitude for $e^-e^+\to e^- e^+$.
If we were to label the external fermion lines
according to the corresponding particle names (which does \textit{not}
conform to the conventions introduced above), the result is shown in
\fig{fig:Bhabhaparticlelabels}. One can find the identity of the
external two-component fermion fields by carefully observing the
direction of the arrow of each fermion line.  For contrast, the same
diagrams, relabeled with two-component fields following the
conventions established in this section
(cf.~\fig{labelconvention}), are shown in \fig{fig:Bhabhalabels}.
An explicit computation of the invariant amplitude is given
in \sec{subsec:Bhabha}.
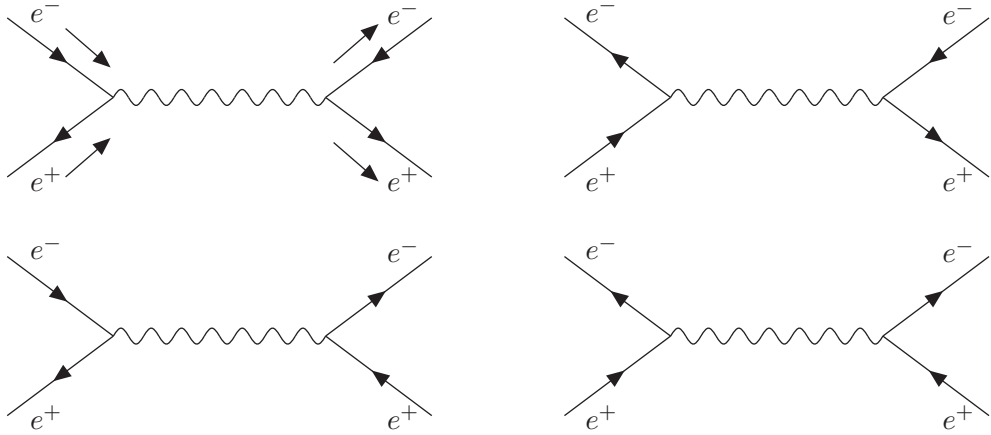
\begin{figure}[tbp!]
\centerline{
\begin{picture}(300,185)(-135,-25)
\thicklines
\Photon(-140,105)(-60,105){3}{7}
\ArrowLine(-180,135)(-140,105)
\ArrowLine(-140,105)(-180,75)
\ArrowLine(-20,135)(-60,105)
\ArrowLine(-60,105)(-20,75)
\put(-172,134){$e^-$}
\put(-172,70){$e^+$}
\put(-37,133){$e^-$}
\put(-37,70){$e^+$}
\LongArrow(-158,131)(-143,118)
\LongArrow(-158,75)(-143,88)
\LongArrow(-57,118)(-42,131)
\LongArrow(-57,88)(-42,75)
\Photon(70,105)(150,105){3}{7}
\ArrowLine(70,105)(30,135)
\ArrowLine(30,75)(70,105)
\ArrowLine(190,135)(150,105)
\ArrowLine(150,105)(190,75)
\put(38,134){$e^-$}
\put(38,70){$e^+$}
\put(173,133){$e^-$}
\put(173,70){$e^+$}
\Photon(-140,15)(-60,15){3}{7}
\ArrowLine(-180,45)(-140,15)
\ArrowLine(-140,15)(-180,-15)
\ArrowLine(-60,15)(-20,45)
\ArrowLine(-20,-15)(-60,15)
\put(-172,44){$e^-$}
\put(-172,-20){$e^+$}
\put(-37,43){$e^-$}
\put(-37,-20){$e^+$}
\Photon(70,15)(150,15){3}{7}
\ArrowLine(70,15)(30,45)
\ArrowLine(30,-15)(70,15)
\ArrowLine(150,15)(190,45)
\ArrowLine(190,-15)(150,15)
\put(38,44){$e^-$}
\put(38,-20){$e^+$}
\put(173,43){$e^-$}
\put(173,-20){$e^+$}
\end{picture}
}
\caption{Tree-level $s$-channel Feynman diagrams for $e^- e^+\to e^- e^+$,
with the external lines labeled according to the particle names.
The initial state is on the left, and the final state is
on the right. Thus, the physical momentum flow
of the external particles,
as well as the flow of the
labeled charges,
are indicated by the arrows adjacent to
the corresponding fermion lines in the upper left diagram.}
\label{fig:Bhabhaparticlelabels}
\end{figure}
\begin{figure}[htbp!]
\vspace{0.2in}
\centerline{
\begin{picture}(300,165)(-135,-25)
\thicklines
\Photon(-140,105)(-60,105){3}{7}
\ArrowLine(-180,135)(-140,105)
\ArrowLine(-140,105)(-180,75)
\ArrowLine(-20,135)(-60,105)
\ArrowLine(-60,105)(-20,75)
\put(-172,134){$e$}
\put(-172,70){$e^\dagger$}
\put(-37,133){${\bar e}^\dagger$}
\put(-37,70){$\bar e$}
\Photon(70,105)(150,105){3}{7}
\ArrowLine(70,105)(30,135)
\ArrowLine(30,75)(70,105)
\ArrowLine(190,135)(150,105)
\ArrowLine(150,105)(190,75)
\put(38,134){${\bar e}^\dagger$}
\put(38,70){$\bar e$}
\put(173,133){${\bar e}^\dagger$}
\put(173,70){$\bar e$}
\Photon(-140,15)(-60,15){3}{7}
\ArrowLine(-180,45)(-140,15)
\ArrowLine(-140,15)(-180,-15)
\ArrowLine(-60,15)(-20,45)
\ArrowLine(-20,-15)(-60,15)
\put(-172,44){$e$}
\put(-172,-20){$e^\dagger$}
\put(-37,43){$e$}
\put(-37,-20){$e^\dagger$}
\Photon(70,15)(150,15){3}{7}
\ArrowLine(70,15)(30,45)
\ArrowLine(30,-15)(70,15)
\ArrowLine(150,15)(190,45)
\ArrowLine(190,-15)(150,15)
\put(38,44){${\bar e}^\dagger$}
\put(38,-20){$\bar e$}
\put(173,43){$e$}
\put(173,-20){$e^\dagger$}
\end{picture}
}
\caption{Tree-level $s$-channel Feynman diagrams for $e^+ e^-\to e^+ e^-$.
These diagrams are the same as in \fig{fig:Bhabhaparticlelabels},
but with the external lines
relabeled by the two-component fermion fields according to the
conventions of \fig{labelconvention}.}
\label{fig:Bhabhalabels}
\end{figure}

\section{\texorpdfstring{Practical examples from the Standard Model and its SUSY extension}{Practical examples from the Standard Model and its SUSY extension}}
\label{sec:examples}

\setcounter{equation}{0}
\setcounter{figure}{0}
\setcounter{table}{0}

In this section we will present some examples to illustrate the use of
the rules presented in this paper. These examples are chosen from the
Standard Model \cite{Weinberg:1967tq} and its supersymmetric (SUSY) extension
\cite{Nilles:1983ge,HaberKane,primer,Chung:2003fi,haberpdg}, in
order to provide an
unambiguous point of reference.  In all cases, the fermion lines in
Feynman diagrams are labeled by two-component field names, rather than
the particle names, as explained in \sec{sec:nomenclature}.

\renewcommand{\theequation}{\arabic{section}.\arabic{subsection}.\arabic{equation}}
\renewcommand{\thefigure}{\arabic{section}.\arabic{subsection}.\arabic{figure}}
\renewcommand{\thetable}{\arabic{section}.\arabic{subsection}.\arabic{table}}

\subsection{Top quark decay: \texorpdfstring{$t\ra b W^+$}{t\textrightarrow bW\textplussuperior}}
\label{tdecay}
\setcounter{equation}{0}
\setcounter{figure}{0}
\setcounter{table}{0}

We begin by calculating the decay width of a top quark into a bottom
quark and $W^+$ vector boson.  For simplicity, we treat this as a
one-generation problem and ignore Cabibbo-Kobayashi-Maskawa (CKM)
\cite{Kobayashi:1973fv} mixing among the three quark generations [see
\eq{ckm-matrix} and the surrounding text].  Let the four-momenta and
helicities of these particle be $(p_t,\lam_t)$, $(k_b,\lam_b)$ and
$(k\ls{W},\lam_ W)$, respectively. Then $p_t^2 = \BDpos m_t^2$, $k_b^2
= \BDpos m_b^2$ and $k\ls{W}^2 = \BDpos m\ls{W}^2$ and
\beqa
2 p_t \newcdot k\ls{W} &=& \BDpos m_t^2 \BDminus m_b^2 \BDplus
m_W^2\,,\\ 2 p_t \newcdot k_b &=& \BDpos m_t^2 \BDplus m_b^2 \BDminus
m_W^2\,,\\ 2 k\ls{W} \newcdot k_b &=& \BDpos m_t^2 \BDminus m_b^2
\BDminus m_W^2 \,.
\eeqa
Because only left-handed top quarks couple to the $W$ boson, the only
Feynman diagram for $t\ra b W^+$ is the one shown in
\fig{fig:topdecay}.
\begin{figure}[t!]
\begin{picture}(400,80)(0,0)
\ArrowLine(155,40)(210,40)
\Photon(210,40)(250,70){3}{5}
\ArrowLine(210,40)(250,10)
\Text(132,40)[]{$t(p_t,\lambda_t)$}
\Text(290,70)[]{$W^+(k_W,\lambda_W)$}
\Text(280,10)[]{$b(k_b,\lambda_b)$}
\end{picture}
\caption{The Feynman diagram for $t\ra b W^+$ at tree level.}
\label{fig:topdecay}
\end{figure}
The corresponding amplitude can be read off of the Feynman rule of
\fig{SMintvertices} in \app{J}. Here the initial state top quark is
a two-component field $t$ going into the vertex and the final state
bottom quark is created by a two-component field $b^\dagger$. Therefore
the amplitude is given by:
\beq
i {\cal M} = \BDneg i \frac{g}{\sqrt{2}}  \varepsilon_\mu^*
x^\dagger_b \sigmabar^\mu x_t\,,
\eeq
where $\varepsilon_\mu^* \equiv \varepsilon_\mu (k_W,\lambda_W)^*$ is
the polarization vector of the $W^+$, and $x^\dagger_b \equiv  x^\dagger(
\boldsymbol {\vec k}_b,\lambda_b)$ and
$x_t \equiv x(\boldsymbol{\vec p}_t,\lambda_t)$
are the external state wave function factors for the
bottom and top quark. Squaring this amplitude
using \eq{eq:conbilsigbar} yields:
\beq
|{\cal M}|^2 = \frac{g^2}{2}
\varepsilon_\mu^* \varepsilon_\nu
(x^\dagger_b \sigmabar^\mu x_t)\,
(x^\dagger_t \sigmabar^\nu x_b) \, .
\eeq
Next, we can
average over the top quark spin polarizations using
\eq{xxdagsummed}:
\beq
\frac{1}{2} \sum_{\lambda_t} |{\cal M}|^2
=
\frac{g^2}{4} \varepsilon_\mu^* \varepsilon_\nu  x^\dagger_b
\sigmabar^\mu \,p_t \newcdot \sigma\, \sigmabar^\nu x_b\, .
\eeq
Summing over the bottom quark spin polarizations in the same way
yields a trace over spinor indices:
\beqa
\frac{1}{2} \sum_{\lambda_t,\lambda_b} |{\cal M}|^2
&=&
\frac{g^2}{4} \varepsilon_\mu^* \varepsilon_\nu \,
{\rm Tr}[\sigmabar^\mu p_t \newcdot \sigma \,\sigmabar^\nu k_b \newcdot
\sigma] \nonumber
\\ &=&
\frac{g^2}{2} \varepsilon_\mu^*
\varepsilon_\nu \left ( p_t^\mu k_b^\nu + k_b^\mu p_t^\nu -
\metric^{\mu\nu} p_t \newcdot k_b
- i \epsilon^{\mu\rho\nu\kappa} p_{t\rho} k_{b\kappa}
\right ) \, ,
\eeqa
where we have used eq.~(\ref{trsbarssbars}).  Finally we can sum over
the $W^+$ polarizations according to:
\beq
\sum_{\lambda_W}
\varepsilon_\mu^* \varepsilon_\nu = \BDneg \metric_{\mu\nu}
+ (k\ls{W})_\mu (k\ls{W})_\nu/m\ls{W}^2\,.
\eeq
The end result is:
\beqa
\frac{1}{2} \sum_{\mbox{spins}} |{\cal M}|^2 &=&
\frac{g^2}{2} \left [ \BDpos p_t \newcdot k_b +
2 (p_t \newcdot k\ls{W})(k_b \newcdot k\ls{W})/m\ls{W}^2 \right ]
\,.
\eeqa
After performing the phase space integration, one obtains:
\beqa
&& \hspace{-0.15in}
\Gamma (t \ra b W^+) = \frac{1}{16 \pi m_t^3}
\lambda^{1/2} (m_t^2 ,m\ls{W}^2, m_b^2 )
\left ( \frac{1}{2} \sum_{\mbox{spins}} |{\cal M}|^2 \right )
\nonumber \\[4pt]
&&
= \frac{g^2}{64\pi m_W^2 m_t^3}
\lambda^{1/2}(m_t^2,m\ls{W}^2, m_b^2)
\left [
(m_t^2 + 2m_W^2)(m_t^2 - m_W^2)+ m_b^2 (m_W^2 - 2 m_t^2) + m_b^4
\right]
,\phantom{xxxx}
\label{eq:gammat}
\eeqa
where the kinematical triangle
function $\lambda^{1/2}$ is defined by~\cite{BK}:
\beq
\lambda(x,y,z) \equiv x^2 + y^2 + z^2 - 2xy-2xz-2yz .
\label{eq:deftrianglefunction}
\eeq
In the approximation $m_b \ll m\ls{W}, m_t$, one ends up with
the well-known result \cite{BP}
\beq
\Gamma
(t \ra b W^+) = \frac{g^2 m_t}{64 \pi}
\left ( 2 + \frac{m_t^2}{m\ls{W}^2} \right ) \left
( 1 - \frac{m\ls{W}^2}{m_t^2} \right )^2\,,
\eeq
which exhibits the Nambu-Goldstone enhancement factor $(m_t^2/m\ls{W}
^2)$ for the longitudinal $W$ contribution compared to the two
transverse $W$ contributions \cite{BP}.

\subsection{\texorpdfstring{$Z^0$}{Z\textzerosuperior} vector boson decay:
\texorpdfstring{$Z^0\ra f \fbar$}{Z\textzerosuperior\textrightarrow f{f\textoverline}}}
\label{zff}
\setcounter{equation}{0}
\setcounter{figure}{0}
\setcounter{table}{0}

Consider the partial decay width of the $Z^0$ boson into a Standard
Model fermion-antifermion pair.  As in the generic example of
\fig{fig:AtoDiracdecay}, there are two contributing Feynman
diagrams, shown in \fig{fig:Zdecay}.
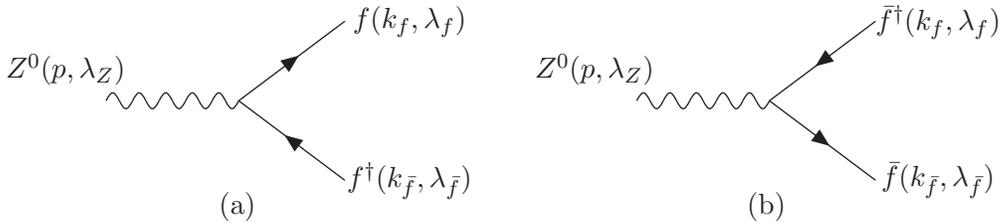
\begin{figure}[b!]
\begin{picture}(400,80)(0,0)
\Photon(110,40)(60,40){3}{5}
\ArrowLine(110,40)(150,70)
\ArrowLine(150,10)(110,40)
\Text(45,52)[]{$Z^0(p,\lambda_Z)$}
\Text(175,70)[]{$f(k_f,\lambda_f)$}
\Text(175,10)[]{$f^\dagger(k_{\fbar},\lambda_{\fbar})$}
\Photon(310,40)(260,40){3}{5}
\ArrowLine(350,70)(310,40)
\ArrowLine(310,40)(350,10)
\Text(245,52)[]{$Z^0(p,\lambda_Z)$}
\Text(375,70)[]{${\bar f}^\dagger(k_f,\lambda_f)$}
\Text(375,10)[]{$\bar f(k_{\fbar},\lambda_{\fbar})$}
\Text(110,0)[]{(a)}
\Text(310,0)[]{(b)}
\end{picture}
\caption{The Feynman diagrams for $Z^0$ decay into a
fermion-antifermion pair.  Fermion lines are labeled according to the
two-component fermion field labeling convention established in
\sec{sec:nomenclature}.}
\label{fig:Zdecay}
\end{figure}
In diagram (a), the fermion particle $f$
in the final state is created by
a two-component field $f$ in the Feynman rule, and the
antifermion particle $\fbar$ by a two-component field $f^\dagger$.  In
diagram (b), the fermion particle $f$ in the final state is
created by a
two-component field $\bar f$, and the antifermion particle $\fbar$
by a two-component field ${\bar f}^\dagger$.
Denote the initial $Z^0$ four-momentum and helicity ($p$, $\lam_Z$)
and the final state fermion ($f$) and antifermion ($\fbar$)
momentum and helicities
($k_f,\lambda_f$) and ($k_{\fbar}, \lambda_{\fbar}$),
respectively.  Then, $k_f^2 = k_{\fbar}^2
= \BDpos m_f^2$ and $p^2 =
\BDpos m\ls{Z}^2$, and
\beqa
&& k_f \newcdot  k_{\fbar}
= \BDpos \frac{1}{2} m\ls{Z}^2 \BDminus m_f^2
\,,
\label{Zffkintwo}\\
&& p \newcdot k_f = p\newcdot k_{\fbar}  = \BDpos\half m\ls{Z}^2
\> . \label{Zffkinthree}
\eeqa
According to the rules of \fig{SMintvertices}, the matrix elements for
the two Feynman graphs are:
\beqa
i {\cal M}_{a} &\,=\,& \BDneg i \frac{g}{c\ls{W}}(T_3^f - s\ls{W}^2 Q_f) \,
\varepsilon_\mu  x^\dagger_f \sigmabar^\mu y_{\fbar}\,, \label{zffamp1}
\\
i {\cal M}_{b} &\,=\,&
\BDpos i g \frac{s\ls{W}^2}{c\ls{W}} Q_f\, \varepsilon_\mu
y_f \sigma^\mu  x^\dagger_{\fbar}\,, \label{zffamp2}
\eeqa
where $x_i\equiv x(\boldsymbol{\vec k}_i,\lambda_i)$ and
$y_i\equiv y(\mathbold{\vec k_i},\lambda_i)$,
for $i=f,\fbar$, and
$\varepsilon_\mu\equiv\varepsilon_\mu(p,\lambda_Z)$.

Using the Bouchiat-Michel formulae developed in \app{H.3}, one
can explicitly evaluate $\mathcal{M}_a$ and $\mathcal{M}_b$ as a
function of the final state fermion helicities.  The result of this
computation is given in \eqs{zffhel1}{zffhel2}.   If the final state
helicities are not measured, then it is simpler to square the
amplitude and sum over the final state spins.

It is convenient to define:
\beq
a_f \equiv T_3^f - Q_f s\ls{W}^2 \,, \qquad\qquad
b_f \equiv  -Q_f s\ls{W}^2\,.
\eeq
Then the squared matrix element for the decay
is, using eqs.~(\ref{eq:conbilsig}) and (\ref{eq:conbilsigbar}),
\beq
|{\cal M}|^2 =\frac{g^2}{c^2\ls{W}}
\varepsilon_\mu\varepsilon^*_\nu
\left ( a_f  x^\dagger_f \sigmabar^\mu y_{\fbar} +
b_f y_f \sigma^\mu  x^\dagger_{\fbar} \right )
\left ( a_f  y^\dagger_{\fbar} \sigmabar^\nu x_f +
b_f x_{\fbar} \sigma^\nu  y^\dagger_f \right ) .\>\phantom{x}
\eeq
Summing over the antifermion helicity using
\eqst{xxdagsummed}{ydagxdagsummed} gives:
\beqa
\sum_{\lambda_{\fbar}}|{\cal M}|^2 &=& \frac{g^2}{c^2\ls{W}}
\varepsilon_\mu \varepsilon^*_\nu
\Bigl (
a_f^2  x^\dagger_f \sigmabar^\mu k_{\fbar} \newcdot \sigma
\sigmabar^\nu
x_f+
b_f^2 y_f \sigma^\mu k_{\fbar} \newcdot \sigmabar
\sigma^\nu  y^\dagger_f
\cr && \qquad\qquad
- m_f a_f b_f  x^\dagger_f \sigmabar^\mu \sigma^\nu  y^\dagger_f
- m_f a_f b_f y_f \sigma^\mu\sigmabar^\nu x_f
\Bigr ) \, .
\eeqa
Next, we sum over the fermion helicity:
\beqa
\sum_{\lambda_f, \lambda_{\fbar}} |{\cal M}|^2
&=& \frac{g^2}{c^2\ls{W}}
\varepsilon_\mu \varepsilon^*_\nu
\Bigl (
a_f^2 {\rm Tr}[\sigmabar^\mu k_{\fbar} \newcdot \sigma \sigmabar^\nu
k_f\newcdot\sigma ]
+
b_f^2  {\rm Tr}[\sigma^\mu k_{\fbar} \newcdot \sigmabar \sigma^\nu
k_f\newcdot\sigmabar ]
\cr
&& \qquad\qquad
- m^2_f a_f b_f {\rm Tr}[\sigmabar^\mu \sigma^\nu ]
- m^2_f a_f b_f {\rm Tr}[\sigma^\mu\sigmabar^\nu ]
\Bigr )\, .
\eeqa
Averaging over the $Z^0$ polarization using
\beq
\frac{1}{3} \sum_{\lambda_Z}
\varepsilon_\mu \varepsilon_{\nu}^* = \frac{1}{3} \left (
\BDneg \metric_{\mu\nu} + \frac{p_\mu p_\nu}{m\ls{Z}^2}
\right )\,,
\eeq
and applying \eqst{trssbar}{trsbarssbars}, one gets:
\beqa
\frac{1}{3} \sum_{\rm spins} |{\cal M}|^2 &=& \frac{g^2}{3 c^2\ls{W}}
\left [
(a_f^2 + b_f^2) \left (
\BDpos 2 k_f \newcdot k_{\fbar}
+ 4 \, k_f \newcdot p \, k_{\fbar} \newcdot p/m\ls{Z}^2
\right ) + 12 a_f b_f m_f^2 \right
] \nonumber
\\
 &=&
\frac{2 g^2}{3 c^2\ls{W}}
\left [
(a_f^2 + b_f^2) (m\ls{Z}^2 - m_f^2 ) + 6 a_f b_f m_f^2 \right
]\, ,
\eeqa
where we have used \eqs{Zffkintwo}{Zffkinthree}. After the
standard phase space integration, we obtain the well-known result for
the partial width of the $Z^0$:
\beqa
\Gamma(Z^0 \ra f \fbar) &=& \frac{N_c^f}{16 \pi m\ls{Z}} \left (
1 - \frac{4m_f^2}{m\ls{Z}^2}\right )^{1/2}\, \left (
\frac{1}{3} \sum_{\rm spins} |{\cal M}|^2 \right )
\nonumber \\[5pt] &=&
\frac{N_c^f g^2 m\ls{Z}}{24 \pi c^2\ls{W}}
\left (
1 - \frac{4m_f^2}{m\ls{Z}^2}\right )^{1/2}
\left [
(a_f^2 + b_f^2) \left(1 - \frac{m_f^2}{m\ls{Z}^2}\right)
+ 6 a_f b_f \frac{m_f^2}{m\ls{Z}^2}
\right ]\, .
\eeqa
Here we have also included a factor of $N_c^f$ (equal to $1$ for
leptons and $3$ for quarks) for the sum over colors.
Since the $Z^0$ is a color singlet,
the color factor is simply equal to the dimension of the
color representation of the outgoing fermions.

\subsection{Bhabha scattering: \texorpdfstring{$e^- e^+ \ra e^- e^+$}{e\textminussuperior e\textplussuperior \textrightarrow e\textminussuperior e\textplussuperior}} 
\label{subsec:Bhabha}
\setcounter{equation}{0}
\setcounter{figure}{0}
\setcounter{table}{0}

In our next example, we consider the computation of Bhabha scattering
in QED (that is, we consider photon exchange but neglect
$Z^0$-exchange) \cite{bhabha}. Bhabha scattering
has also been computed using two-component spinors in
\cite{Kersch:1985rn}.  We denote the initial state electron and
positron momenta and helicities by ($p_1,\lambda_1$) and ($p_2,\lambda
_2$) and the final state electron and positron momenta and helicities
by ($p_3,\lambda_3$) and ($p_4,\lambda_4$), respectively.
Neglecting the electron mass, we have in terms of the usual Mandelstam
variables $s,t,u$:
\beqa
&&p_1 \newcdot p_2 = p_3 \newcdot p_4 \equiv \BDpos\half s\, ,\\
&&p_1 \newcdot p_3 = p_2 \newcdot p_4 \equiv \BDneg \half t\, ,\\
&&p_1 \newcdot p_4 = p_2 \newcdot p_3 \equiv \BDneg \half u\,,
\eeqa
and $p_i^2=0$ for $i=1,\ldots, 4$.
There are eight distinct Feynman diagrams.
First, there are four $s$-channel diagrams, as shown in
\fig{fig:Bhabhalabels}
with amplitudes that follow from the Feynman rules of \fig{eorebar}
(more generally, see \fig{SMintvertices} in \app{J}):
\beqa
i\mathcal{M}_s &=& \left ( \frac{ \BDneg i\metric^{\mu\nu}}{s} \right )
\Bigl [
(\BDneg ie\, x_1 \sigma_\mu y^\dagger_2)(\BDpos ie\, y_3 \sigma_\nu x^\dagger_4)
+
(\BDneg ie\, y^\dagger_1 \sigmabar_\mu x_2)(\BDpos ie\, y_3 \sigma_\nu  x^\dagger_4)
\phantom{xx}
\nonumber \\ &&
\qquad\qquad +
(\BDneg ie\, x_1 \sigma_\mu y^\dagger_2 )(\BDpos ie\, x^\dagger_3\sigmabar_\nu y_4)
+
(\BDneg ie\, y^\dagger_1 \sigmabar_\mu x_2)(\BDpos ie\, x^\dagger_3\sigmabar_\nu y_4)
\Bigr ]\,,
\label{bhabhaschannel}
\eeqa
where $x_i\equiv x(\mathbold{\vec p_i},\lambda_i)$ and
$y_i\equiv y(\mathbold{\vec p_i},\lambda_i)$, for $i=1,4$.
The photon propagator in Feynman gauge is
$-i \metric^{\mu\nu}/(p_1+ p_2)^2 = \BDneg i \metric^{\mu\nu}/s$.
Here, we have chosen to write the external fermion spinors in the order
$1,2,3,4$. This
dictates in each term the use of either the $\sigmabar$ or
$\sigma$ forms of the Feynman rules of \fig{eorebar}.
One can group the terms of \eq{bhabhaschannel} together more compactly:
\beq
i {\cal M}_s = e^2
\left ( \frac{\BDneg i \metric^{\mu\nu}}{s} \right )
\left (x_1 \sigma_\mu y^\dagger_2 + y^\dagger_1 \sigmabar_\mu x_2 \right )
\left (y_3 \sigma_\nu x^\dagger_4 + x^\dagger_3 \sigmabar_\nu y_4 \right ) .
\eeq

There are also four $t$-channel diagrams, as shown in
\fig{fig:Bhabhatchannel}.
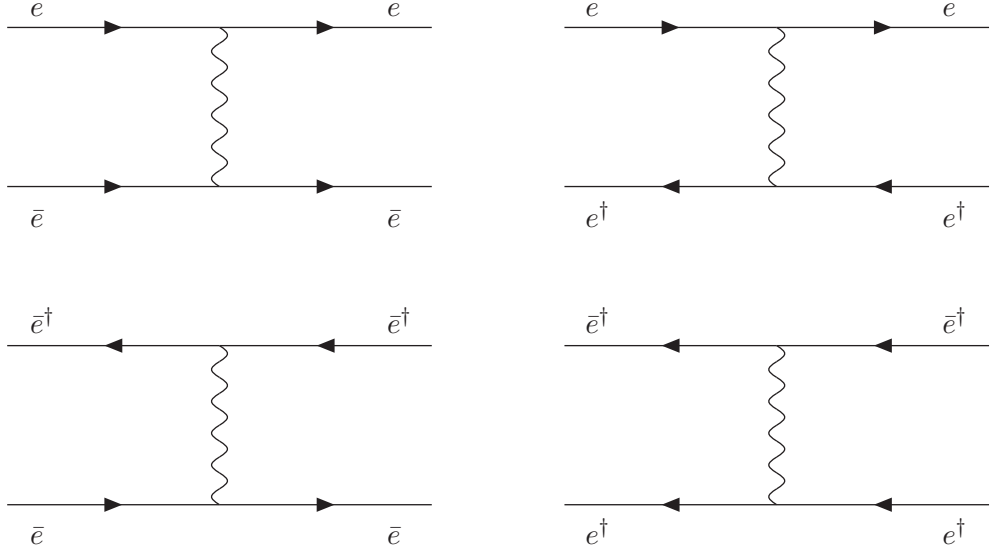
\begin{figure}[t!]
\centerline{
\begin{picture}(300,165)(-135,-55)
\thicklines
\Photon(-100,135)(-100,75){3}{5}
\ArrowLine(-180,75)(-100,75)
\ArrowLine(-180,135)(-100,135)
\ArrowLine(-100,75)(-20,75)
\ArrowLine(-100,135)(-20,135)
\put(-172,60){$\bar e$}
\put(-172,140){$e$}
\put(-37,60){$\bar e$}
\put(-37,140){$e$}
\Photon(110,135)(110,75){3}{5}
\ArrowLine(190,75)(110,75)
\ArrowLine(110,135)(190,135)
\ArrowLine(110,75)(30,75)
\ArrowLine(30,135)(110,135)
\put(38,60){${e}^\dagger$}
\put(38,140){$e$}
\put(173,60){${e}^\dagger$}
\put(173,140){$e$}
\Photon(-100,15)(-100,-45){3}{5}
\ArrowLine(-180,-45)(-100,-45)
\ArrowLine(-100,15)(-180,15)
\ArrowLine(-100,-45)(-20,-45)
\ArrowLine(-20,15)(-100,15)
\put(-172,-60){$\bar e$}
\put(-172,20){${\bar e}^\dagger$}
\put(-37,-60){$\bar e$}
\put(-37,20){${\bar e}^\dagger$}
\Photon(110,15)(110,-45){3}{5}
\ArrowLine(190,-45)(110,-45)
\ArrowLine(110,15)(30,15)
\ArrowLine(110,-45)(30,-45)
\ArrowLine(190,15)(110,15)
\put(38,-60){${e}^\dagger$}
\put(38,20){${\bar e}^\dagger$}
\put(173,-60){${e}^\dagger$}
\put(173,20){${\bar e}^\dagger$}
\end{picture}
}
\caption{Tree-level $t$-channel Feynman diagrams for $e^- e^+\to e^- e^+$,
  with the external lines labeled according to the two-component field
  names.  The momentum flow of the external particles is from left to
  right.}
\label{fig:Bhabhatchannel}
\end{figure}
The corresponding amplitudes for these four diagrams can be written:
\beqa
i \mathcal{M}_t &=& (-1) e^2
\left ( \frac{\BDneg i \metric^{\mu\nu}}{t} \right )
\left(
 x_1 \sigma_\mu x^\dagger_3
+
  y^\dagger_1 \sigmabar_\mu y_3\right ) \left (
 x_2 \sigma_\nu  x^\dagger_4 +
  y^\dagger_2 \sigmabar_\nu y_4 \right )
 . \phantom{xx}
\eeqa
Here, the overall factor of $(-1)$ comes from
Fermi-Dirac statistics, since the
external fermion wave functions are written in an odd permutation
$(1,3,2,4)$ of
the original
order $(1,2,3,4)$ established by the first term in
eq.~(\ref{bhabhaschannel}).

Fierzing each term using \eqst{twocompfierza}{twocompfierzc}, and
using \eqs{zonetwo}{barzonetwo}, the total amplitude can be written
as:
\beqa
\mathcal{M} &=& {\cal M}_s + {\cal M}_t =
         2 e^2 \biggl [
\frac{1}{s}
(x_1 y_3) (y^\dagger_2 x^\dagger_4)
+ \frac{1}{s} (y^\dagger_1 x^\dagger_3)(x_2 y_4)
+\left (\frac{1}{s}+\frac{1}{t}\right )(y^\dagger_1 x^\dagger_4)(x_2 y_3)
\nonumber \\[4pt] &&
+\left (\frac{1}{s}+\frac{1}{t}\right )(x_1 y_4)
(y^\dagger_2 x^\dagger_3)
- \frac{1}{t} (x_1 x_2)(x^\dagger_3 x^\dagger_4)\,
- \frac{1}{t}(y^\dagger_1 y^\dagger_2)(y_3 y_4)
\biggr ]. \phantom{xxx}
\label{bhabhamat}
\eeqa
Squaring this amplitude and summing over spins, all of the cross terms
will vanish in the $m_e \ra 0 $ limit. This is because each
cross term will have an $x$ or an $ x^\dagger$ for some electron or
positron combined with a $y$ or a $y^\dagger$ for the same particle, and
the corresponding spin sum is proportional to $m_e$ [see
\eqs{yxsummed}{ydagxdagsummed}].  Hence, summing over final state
spins and averaging over initial state spins, the end result contains
only
the sum of the squares of the six terms in eq.~(\ref{bhabhamat}):
\beqa
\frac{1}{4}
\sum_{\rm spins} |{\cal M}|^2 &=& e^4\sum_{ \lambda_1,
  \lambda_2, \lambda_3, \lambda_4} \biggl \lbrace \frac{1}{s^2} \left [
  (x_1 y_3)(y^\dagger_3 x^\dagger_1) (y^\dagger_2x^\dagger_4)(x_4 y_2)+
  (y^\dagger_1x^\dagger_3)(x_3 y_1)(x_2 y_4)(y^\dagger_4 x^\dagger_2) \right ]
\nonumber \\[4pt] && \qquad\>\>\> + \left (\frac{1}{s} + \frac{1}{t}
   \right )^2 \left[(y^\dagger_1 x^\dagger_4)(x_4 y_1) {(x_2 y_3 )(
  y^\dagger_3x^\dagger_2)}+ (x_1 y_4)(y^\dagger_4 x^\dagger_1)
  {(y^\dagger_2x^\dagger_3 )(
 x_3y_2)}
\right ] \nonumber \\[5pt] && \qquad\>\>\> +\frac{1}{t^2} \left[(x_1
  x_2)(x^\dagger_2 x^\dagger_1)(x^\dagger_3 x^\dagger_4)(x_4 x_3)+
  (y^\dagger_1
  y^\dagger_2)(y_2 y_1)(y_3 y_4)(y^\dagger_4 y^\dagger_3) \right ] \biggr \rbrace
\, .
\eeqa
Here we have used eq.~(\ref{eq:conbil}) to get the complex square
of the fermion bilinears.
Performing these spin sums using
\eqs{xxdagsummed}{yydagsummed} and using the trace identities
\eq{APPtrssbar}:
\beqa \frac{1}{4} \sum_{\rm spins} |{\cal
  M}|^2 &=& {8}e^4 \biggl [ \frac{p_2 \newcdot p_4\,
 p_1 \newcdot  p_3}{s^2} +\frac{p_1 \newcdot p_2 \,
p_3 \newcdot p_4}{t^2}
+ \left (\frac{1}{s}+\frac{1}{t}\right )^2 p_1 \newcdot
p_4 \, p_2 \newcdot p_3 \biggr ]
\nonumber \\[3pt]
&=& 2e^4 \biggl [ \frac{t^2}{s^2}+\frac{s^2}{t^2} + \left (\frac{u}{s}
    +\frac{u}{t}\right )^2 \biggr ]\, .
\eeqa
Thus, the differential cross-section for Bhabha scattering is given
by:
\beq
\frac{d \sigma}{dt}
   = \frac{1}{16 \pi s^2} \left(\frac{1}{4}\sum_{\rm spins} |{\cal
    M}|^2 \right ) = \frac{2 \pi \alpha^2}{s^2} \biggl [ \frac{t^2}{s^2}
   +\frac{s^2}{t^2} + \left (\frac{u}{s}+\frac{u}{t}\right )^2
\biggr ]\, .\phantom{xxxxxx}
\eeq
This agrees with the result given in problem 5.2 of
ref.~\cite{Peskin:1995ev}.

\subsection{Polarized muon decay}
\setcounter{equation}{0}
\setcounter{figure}{0}
\setcounter{table}{0}

So far we have only treated cases where the initial state fermion
spins are averaged and the final state spins are summed. In the case
of the polarized decay of a particle or polarized scattering we must
project out the appropriate polarization of the particles in the spin
sums. This is achieved by replacing the spin sums given in
\eqst{xxdagsummed}{ydagxdagsummed} by the relevant
polarized spin projections exhibited in \eqst{xxdagmassive}{ydagxdagmassless}.
As an example, we consider the decay of a polarized muon.
Polarized muon decay has also been computed using
two-component spinors in \Ref{Kersch:1985rn}, however with an
effective four-fermion interaction.\footnote{In a related calculation
given in \Ref{Kulyabov}, two component spinor techniques are applied
to the computation of matrix element for $\nu+n\to p+e^-$ using an
effective four-fermion $V-A$ interaction.}
The leading order Feynman
diagram for muon decay
is shown in \fig{fig:muondecay} (and the relevant four-momenta are indicated).

\begin{figure}[ht!]
\begin{center}
\begin{picture}(80,95)(0,10)
\ArrowLine(-80,60)(-30,60)
\ArrowLine(-30,60)(10,90)
\Photon(-30,60)(10,30)4 5 
\ArrowLine(10,30)(50,60)
\ArrowLine(50,00)(10,30)
\Text(-77,74)[]{$\mu\,(p,s)$}
\Text(63,70)[]{$e \,(k_e,\lambda_e)$}
\Text(68,12)[]{$\nu_e^\dagger\,(k_{\nubar_e},\lambda_{\nubar_e})$}
\Text(22,97)[]{$\nu_\mu\,(k_{\nu_\mu},\lambda_{\nu_\mu})$}
\put(-33,30){$W^-$}
\end{picture}
\end{center}
\caption{Feynman diagram for electroweak muon decay.}
\label{fig:muondecay}
\end{figure}
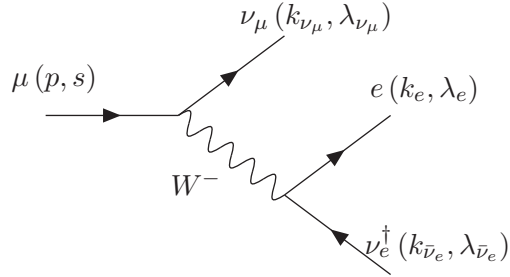
 
In our computation, the mass of the muon 
is denoted by $m_\mu$ and the electron mass is neglected.
The spin of the muon is measured in its rest frame with respect
to a fixed $z$-axis.
Assume that the muon at rest is polarized such that its spin component along
the $\boldsymbol{\hat z}$-direction is $s=+\half$.
Then, the
decay amplitude is given by\footnote{Throughout this subsection $\mu$ and
$\nu$ are particle labels.  Hence, we employ
$\rho$ and $\tau$ as Lorentz vector indices.}
\beq \label{mudecayamp}
i{\cal  M}=\left(\frac{\BDneg ig}{\sqrt{2}}\right)^2
\left( x^\dagger_{\nu_\mu} \sigmabar_\rho x_\mu \right )
 \left( x^\dagger_e\sigmabar_\tau y_{\nubar_e} \right )
\left (\frac{-ig^{\rho\tau}}{D_W}\right )\,,
\eeq
where $D_W= (p - k_{\nu_\mu})^2 \BDminus m_W^2$ is the denominator of the
$W$-boson propagator. In \eq{mudecayamp},
$x_\mu\equiv x(\boldsymbol{\vec p},s=\half)$
for the spin-polarized initial state muon, and
$x^\dagger_{\nu_\mu}\equiv x(\boldsymbol
{\vec k}_{\nu_\mu},\lam_{\nu_\mu})$,
$x^\dagger_e\equiv x^\dagger(\boldsymbol {\vec
k}_e,\lam_e)$, and $y_{\nubar_e}\equiv y(\boldsymbol
{\vec k}_{\nubar_e},\lam_{\nubar_e})$. Squaring the amplitude using
eq.~(\ref{eq:conbilsigbar}), we obtain
\beqa
|{\cal M}|^2&=&\frac{g^4}{4 D_W^2}
\left(x^\dagger_{\nu_\mu}\sigmabar^\rho x_\mu\right)
\left(x^\dagger_\mu\sigmabar^\tau x_{\nu_\mu}\right)
\left(x^\dagger_e\sigmabar_\rho y_{\nubar_e}\right)
\left(y^\dagger_{\nubar_e}\sigmabar_\tau x_e\right)\,.
\eeqa
Summing over the neutrino and electron spins using
\eqst{xxdagsummed}{yydagsummed},
and using
\eq{xxdagmassive} for the muon spin (with $s=\half$) yields:
\beqa
\sum_{\lambda_{\nu_\mu}\lambda_{e}\lambda_{\nubar_e}}|{\cal M}|^2
&=& \frac{g^4}{8 D_W^2}
{\rm Tr}[k_{\nu_\mu}\newcdot\sigma\,\sigmabar^\rho
(p\newcdot\sigma-m_\mu S\newcdot\sigma) \,\sigmabar^\tau]
\,{\rm Tr}[k_e\newcdot\sigma \,\sigmabar_\rho
k_{\nubar_e}\newcdot\sigma \,\sigmabar_\tau]\nonumber \\
&=&
\frac{2 g^4}{D_W^2} \,k_e\newcdot k_{\nu_\mu}\,
k_{\nubar_e}\newcdot (p - m_\mu S)\,,
\eeqa
where $S^\mu$ in an arbitrary frame is given by \eq{fixedsvect}
[with $\boldsymbol{\hat s}=\boldsymbol{\hat z}$].
To obtain the second line we have used the trace identity
eq.~(\ref{trssbarssbar}) twice; note that the resulting
terms linear in
the
antisymmetric tensor do not contribute, but the term quadratic
in the antisymmetric tensor does.

The differential decay amplitude is now given by
\beq
d\Gamma =\frac{1}{2m_\mu} |{\cal M}|^2
\frac{d^3\boldsymbol{\vec k}_e}{(2\pi)^3 2E_e}
\frac{d^3\boldsymbol{\vec k}_{\nubar_e}}{(2\pi)^3 2E_{\nubar_e}}
\frac{d^3\boldsymbol{\vec k}_{\nu_\mu}}{(2\pi)^3 2E_{\nu_\mu}}
(2\pi)^4\delta^4(p-k_e-k_{\nubar_e}-k_{\nu_\mu})\,,
\eeq
where $E_i,$ $i=e,\nubar_e,\nu_\mu$ are the energies of the final state
particles in the muon rest frame.  In the following we shall neglect
both the electron mass and
the momentum in the $W$-propagator compared to the $W$-boson mass, so
$D_W^2 \rightarrow m_W^4$. We can now use the following identity to
integrate over the neutrino momenta \cite{okun}
\beq
\int
\frac{d^3\boldsymbol{\vec k}_{\nubar_e}}{(2\pi)^3 2E_{\nubar_e}}
\frac{d^3\boldsymbol{\vec k}_{\nu_\mu}}{(2\pi)^3 2E_{\nu_\mu}} (2\pi)^4
\delta^4(q-k_{\nubar_e}-k_{\nu_\mu})
k_{\nubar_e}^\rho k_{\nu_\mu}^\tau
= \frac{1}{96\pi} (q^2g^{\rho\tau}+2q^\rho q^\tau)
\,,
\eeq
where $q=p-k_e$. It follows that
\beq
d\Gamma= \frac{g^4}{1536\pi^4m_\mu m_W^4}
\left[q^2 \,k_e\newcdot (p - m_\mu S)
\,+\, 2 q \newcdot k_e \, q \newcdot (p-m_\mu S)
\right ]
\frac{d^3\boldsymbol{\vec k}_e}{E_e}\,.
\eeq
In the muon rest frame,
$k_e=E_e(1;\cos\phi\sin\theta,\sin\phi\sin\theta,\cos\theta)$
and $S=(0;0,0,1)$, so that
$q^2 = \BDpos m_\mu^2 \BDminus 2 E_e m_\mu$ and
$k_e\newcdot (p - m_\mu S) = \BDpos m_\mu E_e (1 + \cos\theta)$ and
$q\newcdot k_e= \BDpos m_\mu E_e $ and
$q\newcdot (p-m_\mu S) = m_\mu (\BDpos m_\mu \BDminus E_e \BDminus E_e
\cos\theta)$. Noting that the maximum energy of the electron is
$m_\mu/2$ (when the neutrino and antineutrino both recoil in the opposite
direction), we obtain
\beqa
\frac{d\Gamma}{d(\cos\theta)}&=&\frac{g^4m_\mu^2}{768\pi^3m_W^4}
\int_0^{m_\mu/2} dE_eE_e^2\left[ 3-\frac{4E_e}{m_\mu}+
\left(1-\frac{4E_e}{m_\mu}\right)\cos\theta\right] \nonumber \\
&=& \frac{g^4m_\mu^5}{3\newcdot 2^{12}\pi^3m_W^4}
\left(1-\third\cos\theta\right)\,,\label{polmu}
\eeqa
in agreement with ref.~\cite{okun}.  Introducing the Fermi constant,
$G_F\equiv \sqrt{2}g^2/(8m_W^2)$, we can rewrite \eq{polmu} as:
\beq
\frac{d\Gamma}{d(\cos\theta)}=\frac{G_F^2 m_\mu^5}{384\pi^3}
\left(1-\third\cos\theta\right)\,.
\eeq
Integrating over $\cos\theta$ reproduces the well-known total muon
decay width,
\beq
\Gamma=\frac{G_F^2 m_\mu^5}{192\pi^3}\,.
\eeq

\subsection{Neutral MSSM Higgs boson decays: \texorpdfstring{$\phi^0
\rightarrow f \fbar$ for $\phi^0=h^0,H^0,A^0$}{\textscalar\textzerosuperior\textrightarrow f{f\textoverline} \nobreakspace for \textscalar\textzerosuperior = h\textzerosuperior, H\textzerosuperior, A\textzerosuperior}}
\setcounter{equation}{0}
\setcounter{figure}{0}
\setcounter{table}{0}

In this subsection, we consider the decays of the neutral Higgs scalar
bosons $\phi^0 = h^0$, $H^0$, and $A^0$ of the MSSM
into Standard Model fermion-antifermion pairs. The relevant
tree-level Feynman diagrams are shown in \fig{fig:hHffbardecay}.
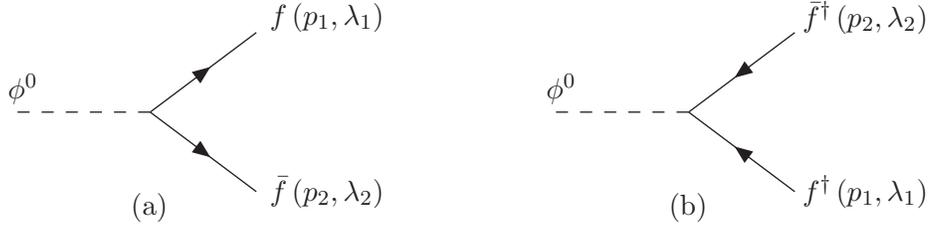
\begin{figure}[tb]
\begin{center}
\begin{picture}(200,80)(0,0)
\DashLine(60,40)(10,40)5
\ArrowLine(60,40)(100,70)
\ArrowLine(60,40)(100,10)
\Text(12,49)[]{$\phi^0$}
\Text(127,10)[]{$\bar f \,(p_2,\lam_2)$}
\Text(127,76)[]{$f \,(p_1,\lam_1)$}
\Text(60,4)[]{(a)}
\end{picture}
\begin{picture}(200,80)(0,0)
\DashLine(10,40)(60,40)5
\ArrowLine(100,70)(60,40)
\ArrowLine(100,10)(60,40)
\Text(12,49)[]{$\phi^0$}
\Text(127,10)[]{$f^\dagger \,(p_1,\lam_1)$}
\Text(127,76)[]{${\bar f}^\dagger \,(p_2,\lam_2)$}
\Text(60,4)[]{(b)}
\end{picture}
\end{center}
\caption{The Feynman diagrams for the decays $\phi^0 \ra f \fbar$,
where $\phi^0 = h^0, H^0, A^0$ are the neutral Higgs scalar bosons
of the MSSM, and $f$ is a Standard Model quark or lepton,
and $\fbar$ is the corresponding antiparticle.
We have labeled the external fermions according to the two-component
field names.}
\label{fig:hHffbardecay}
\end{figure}
The final state fermion is assigned four-momentum $p_1$ and polarization
$\lambda_1$, and the antifermion is assigned four-momentum $p_2$
and polarization $\lambda_2$.
We will first work out the case that $f$ is a charge $-1/3$ quark or
a charged lepton, and later note the simple change needed for
charge $+2/3$ quarks. The Feynman rules of
\fig{nehiggsqq} of \app{K} tell us that
the amplitudes are:
\beqa
i \mathcal{M}_{a} &=&
-\frac{i}{\sqrt{2}}\, Y_f \, k_{d\phi^0}^* \, x^\dagger_1 x^\dagger_2
\, ,
\\
i \mathcal{M}_{b} &=&
-\frac{i}{\sqrt{2}}\, Y_f \, k_{d\phi^0} \, y_1 y_2
\, .
\eeqa
Here $Y_f$ is the Yukawa coupling of the fermion, $k_{d\phi^0}$ is
the Higgs mixing parameter from eq.~(\ref{eq:defkdphi0}), and
the external wave functions are denoted
$x_1 \equiv x(\boldsymbol{\vec p}_1,\lam_ 1)$,
$y_1 \equiv y(\boldsymbol{\vec p}_1,\lam_1)$ for the fermion and
$x_2 \equiv x(\boldsymbol{\vec p}_2,\lam_2)$,
$y_2 \equiv y(\boldsymbol{\vec p}_2,\lam_2)$ for the antifermion.
Squaring the total amplitude
$i {\cal M} = i {\cal M}_a + i {\cal M}_b$ using eq.~(\ref{eq:conbil})
results in:
\beq
|{\cal M}|^2 =
\frac{1}{2} |Y_f|^2 \left [
|k_{d\phi^0}|^2 ( y_1 y_2 \, y^\dagger_2 y^\dagger_1
+ x^\dagger_1 x^\dagger_2\, x_2 x_1)
+ (k^*_{d\phi^0})^2 x^\dagger_1 x^\dagger_2\, y^\dagger_2 y^\dagger_1
+ (k_{d\phi^0})^2 y_1 y_2\, x_2 x_1
\right ] .
\eeq
Summing over the final state antifermion spin using
\eqst{xxdagsummed}{ydagxdagsummed} gives:
\beq
\sum_{\lam_2} |{\cal M}|^2 =
\frac{1}{2} |Y_f|^2 \left [
\BDpos |k_{d\phi^0}|^2 ( y_1 p_2 \newcdot \sigma y^\dagger_1
+ x^\dagger_1 p_2 \newcdot \sigmabar x_1)
- (k^*_{d\phi^0})^2 m_f x^\dagger_1 y^\dagger_1
- (k_{d\phi^0})^2 m_f y_1  x_1
\right ] .
\eeq
Summing over the fermion spins in the same way yields:
\beqa
\sum_{\lam_1,\lam_2} |{\cal M}|^2 &=&
\frac{1}{2} |Y_f|^2 \left \{
|k_{d\phi^0}|^2 (
{\rm Tr}[p_2 \newcdot \sigma p_1 \newcdot \sigmabar]
+ {\rm Tr}[p_2 \newcdot \sigmabar p_1 \newcdot \sigma])
- 2 (k^*_{d\phi^0})^2 m_f^2 - 2 (k_{d\phi^0})^2 m_f^2
\right \} \phantom{xxx.} \nonumber
\\
&=& |Y_f|^2 \left \{ \BDpos 2 |k_{d\phi^0}|^2 p_1 \newcdot p_2
- 2 {\rm Re}[(k_{d\phi^0})^2] m_f^2 \right \} \nonumber
\\
&=& |Y_f|^2 \left \{ |k_{d\phi^0}|^2 (m^2_{\phi^0} - 2 m_f^2)
- 2 {\rm Re}[(k_{d\phi^0})^2] m_f^2 \right \} ,
\eeqa
where we have used the trace identity eq.~(\ref{trssbar}) to obtain
the second equality. The corresponding expression for charge $+2/3$
quarks can be obtained by simply replacing $k_{d\phi^0}$ with
$k_{u\phi^0}$.  The total decay rates now follow from integration over
phase space \cite{sundaresan}
\beq
\Gamma (\phi^0 \rightarrow f \fbar) =
\frac{N_c^f}{16 \pi m_{\phi^0}} \left (1 - 4 m_f^2/m_{\phi^0}^2
\right )^{1/2}
\sum_{\lam_1,\lam_2} |{\cal M}|^2 .
\eeq
The factor of $N_c^f = 3$ for quarks and 1 for leptons comes from the
sum over colors.

Results for special cases are obtained by putting in the relevant
values for the couplings and the mixing parameters from
eqs.~(\ref{eq:defkuphi0}) and (\ref{eq:defkdphi0}). In
particular,
for the CP-even Higgs bosons $h^0$ and $H^0$, $k_{d\phi^0}$ and
$k_{u\phi^0}$ are real, so
one obtains:
\beqa
\Gamma(h^0 \ra b \bbar)
&=& \frac{3}{16 \pi}\,Y_b^2 \, \sin^2\!\alpha \,
m_{h^0}
\left (1 - {4m_b^2/ m_{h^0}^2} \right )^{3/2},
\label{hbbresult}
\\
\Gamma(h^0 \ra c \bar c)
&=& \frac{3}{16 \pi}\,Y_c^2 \, \cos^2\!\alpha \,
m_{h^0}
\left (1 - {4m_c^2/ m_{h^0}^2} \right )^{3/2},
\label{hccresult}
\\
\Gamma(h^0 \ra \tau^+ \tau^-)
&=& \frac{1}{16 \pi}\,Y_\tau^2 \, \sin^2\!\alpha \,
m_{h^0}
\left (1 - {4m_\tau^2/ m_{h^0}^2} \right )^{3/2},
\label{htautauresult}
\\
\Gamma(H^0 \ra t \tbar)
&=& \frac{3}{16 \pi}\,Y_t^2 \, \sin^2\!\alpha \,
m_{H^0}
\left (1 - {4m_t^2/ m_{H^0}^2} \right )^{3/2},
\label{Httresult}
\\
\Gamma(H^0 \ra b \bbar)
&=& \frac{3}{16 \pi}\,Y_b^2 \, \cos^2\alpha \,
m_{H^0}
\left (1 - {4m_b^2/ m_{H^0}^2} \right )^{3/2},
\label{Hbbresult}
\eeqa
etc., which check with the expressions in \app{C} of
ref.~\cite{HHG}.  For the CP-odd Higgs boson $A^0$, the mixing
parameters $k_{uA^0} = i \cos\!\beta_0$ and $k_{dA^0} = i \sin\!\beta_0$
are purely imaginary, so
\beqa \Gamma(A^0 \ra t \tbar) &=& \frac{3}{16\pi}
\,Y_t^2 \, \cos^2\!\beta_0 \, m_{A^0} \left (1 - {4m_t^2/
    m_{A^0}^2} \right )^{1/2},
\label{Attresult}
\\
\Gamma(A^0 \ra b \bbar)
&=& \frac{3}{16 \pi}\,Y_b^2 \, \sin^2\!\beta_0 \,
m_{A^0}
\left (1 - {4m_b^2/ m_{A^0}^2} \right )^{1/2},
\label{Abbresult}
\\
\Gamma(A^0 \ra \tau^+ \tau^-)
&=& \frac{1}{16 \pi}\,Y_\tau^2 \, \sin^2\!\beta_0 \, m_{A^0}
\left (1 - {4m_\tau^2/ m_{A^0}^2} \right )^{1/2} .
\label{Atautauresult}
\eeqa

Note that the differing kinematic factors for the CP-odd Higgs decays
came about because of the different relative sign between the two Feynman
diagrams. For example, in the case of $h^0 \rightarrow b \bbar$,
the matrix element is
\beq
i {\cal M} = \frac{i}{\sqrt{2}} Y_b \sin\!\alpha \,(y_1 y_2
+ x^\dagger_1 x^\dagger_2),
\eeq
while for $A^0 \rightarrow b \bbar$, it is
\beq
i {\cal M} =\frac{1}{\sqrt{2}} Y_b \sin\!\beta_0 \,(y_1 y_2 -
x^\dagger_1 x^\dagger_2).
\eeq
The differing relative sign between $y_1 y_2$ and $x^\dagger_1 x^\dagger_2$
follows from the imaginary pseudoscalar Lagrangian
coupling, which is complex conjugated in the second diagram.


\subsection{Sneutrino decay: \texorpdfstring{$\widetilde \nu_e \ra \Ciplus e^-$}{\textnu\texttilde\texteinferior\textrightarrow C\texttilde\textiinferior\textplussuperior e\textminussuperior}}
\setcounter{equation}{0}
\setcounter{figure}{0}
\setcounter{table}{0}

Next we consider the process of sneutrino decay
$\widetilde \nu_e \ra \Ciplus
e^- $ in the MSSM. Because only the left-handed electron
can couple to the chargino and sneutrino (with the excellent
approximation that the electron Yukawa coupling vanishes), there is just
one Feynman diagram, shown in \fig{fig:snuCedecay}.
\begin{figure}[bt]
\begin{center}
\begin{picture}(200,53)(0,12)
\DashLine(10,30)(60,30)5
\ArrowLine(60,30)(100,60)
\ArrowLine(60,30)(100,0)
\Text(15,38)[]{$\stilde \nu_e$}
\Text(128,2)[]{$e\, (k_e,\lam_e)$}
\Text(133,60)[]{$\chi_i^+ \, (k_{\tilde C},\lam_{\tilde C})$}
\end{picture}
\end{center}
\caption{The Feynman diagram for $ \widetilde \nu_e \ra
\Ciplus e^- $ in the MSSM.
}
\label{fig:snuCedecay}
\end{figure}
The external wave functions of the electron and chargino are denoted
as $x_e\equiv x(\boldsymbol{\vec k}_e,\lam_e),$ and $x_{\tilde C}\equiv
x(\boldsymbol{\vec k}_{\tilde C},\lam_{\tilde C})$, respectively.
{}From the corresponding Feynman rule given in \fig{cqsq} of
\app{K}, the amplitude is:
\beq i{\cal M} = -ig V_{i1}\,x^\dagger_{\tilde C} x^\dagger_e ,
\eeq
where $V_{ij}$ is one of the two matrices used to diagonalize the
chargino masses [cf.~\eq{u-and-v}]. Squaring this
using eq.~(\ref{eq:conbil}) yields:
\beq |{\cal M}|^2 = g^2 |V_{i1}|^2\, (x^\dagger_{\tilde C}x^\dagger_e) (x_e
x_{\tilde C})\> .  \eeq 
Summing over the electron and chargino
spin polarizations using eq.~(\ref{xxdagsummed}) yields
\beq
\sum_{\lambda_e, \lambda_{\tilde C}} |{\cal M}|^2 = g^2 |V_{i1}|^2
{\rm Tr}[k_e \newcdot \sigmabar \, k_{\tilde C} \newcdot \sigma ] =
\BDpos 2 g^2 |V_{i1}|^2\, k_e \newcdot k_{\tilde C} = g^2 |V_{i1}|^2
(m_{\tilde \nu_e}^2 - m_{\tilde C_i}^2) \> ,
\eeq
where we have used $\BDpos 2 k_e\newcdot k_{\tilde C}=m_{\tilde\nu_
  e}^2 - m_{\tilde C_i}^2$, neglecting the electron mass. Therefore,
after integrating over phase space in the standard way, the decay
width is:
\beqa
\Gamma (\stilde \nu_e \ra \Ciplus e^- ) = \frac{1}{16 \pi
  m_{\tilde \nu_e}} \left ( 1 - \frac{m_{\tilde C_i}^2}{m_{\tilde
      \nu_e}^2} \right ) \left[\sum_{\lambda_e, \lambda_{\tilde C}}
  |{\cal M}|^2\right] = \frac{g^2}{16 \pi}|V_{i1}|^2 m_{\tilde \nu_e}
\left (1 - \frac{m_{\tilde C_i}^2}{m_{\tilde \nu_e}^2} \right )^2\!\!,
\eeqa
which agrees with ref.~\cite{DicusNandiTata} and eq.~(3.8) in
ref.~\cite{HaberKane}.

\subsection{Chargino decay: \texorpdfstring{$\Ciplus \ra \widetilde \nu_e e^+$}{C\texttilde\textiinferior\textplussuperior\textrightarrow \textnu\texttilde\texteinferior e\textplussuperior}}
\setcounter{equation}{0}
\setcounter{figure}{0}
\setcounter{table}{0}

Here again, there is just one Feynman diagram (neglecting the
electron mass in the Yukawa coupling) shown in \fig{fig:Csnuedecay}.
The external wave functions for the chargino and the positron are
denoted by $x_{\tilde C}\equiv x(\boldsymbol{\vec p}_{\tilde C},
\lam_{\tilde C})$ and
$y_{e}\equiv y(\boldsymbol{\vec k}_e,\lam_e)$, respectively.
The fermion momenta
and helicities are denoted as in \fig{fig:Csnuedecay}.
As in the previous example, the amplitude
can be directly determined using the Feynman rule given in \fig{cqsq}
in \app{K}:
\begin{figure}[t!]
\begin{center}
\begin{picture}(150,53)(0,12)
\ArrowLine(10,30)(60,30)
\DashLine(60,30)(100,60)5
\ArrowLine(100,0)(60,30)
\Text(13,44)[]{$\chi_i^+\, (p_{\tilde C},\lam_{\tilde C})$}
\Text(128,2)[]{$ e^\dagger\, (k_e,\lam_e)$}
\Text(110,60)[]{$\stilde \nu_e$}
\end{picture}
\end{center}
\caption{The Feynman diagram for $\Ciplus\,\ra \widetilde\nu_e e^+ $ in
the MSSM.}
\label{fig:Csnuedecay}
\end{figure}
\beq
{\cal M} = -ig V^*_{i1}\; x_{\tilde C} \,y_{e} \> .
\eeq
Squaring this using eq.~(\ref{eq:conbil}) yields:
\beq
|{\cal M}|^2 = g^2 |V_{i1}|^2 \, (x_{\tilde C} y_{e})\,
(y^\dagger_{e} x^\dagger_{\tilde C}) \> .
\eeq
Summing over the electron helicity and averaging over the chargino
helicity using eqs.~(\ref{xxdagsummed}) and (\ref{yydagsummed})
we obtain:
\beq
\half  \sum_{\lambda_e, \lambda_{\tilde C}} |{\cal M}|^2
= \half g^2 |V_{i1}|^2 {\rm Tr}[k_e \newcdot \sigma \,
   p_{\tilde C} \newcdot \sigmabar ]
= \BDpos g^2 |V_{i1}|^2 k_e \newcdot p_{\tilde C}
= \frac{g^2}{2} |V_{i1}|^2 (m_{\tilde C_i}^2 -m_{\tilde \nu_e}^2) \> .
\eeq
So the decay width is, neglecting the electron mass:
\beq
\Gamma
(\Ciplus \ra \stilde \nu e^+ )= \frac{1}{16
\pi m_{\tilde C_i}} \left ( 1 - \frac{m_{\tilde \nu_e}^2}{
 m_{\tilde C_i}^2} \right )
 \left ( \frac{1}{2} \sum_{\lam_e, \lam_{\tilde C}}
 |{\cal M}|^2 \right )
=\frac{g^2}{32 \pi}|V_{i1}|^2 m_{\tilde C_i}
 \left (1 - \frac{m_{\tilde \nu_e}^2}{m_{\tilde C_i}^2} \right )^2\!\!,
\eeq
which agrees with ref.~\cite{DicusNandiTata}.


\subsection{Neutralino decays:
\texorpdfstring{$\Ni \ra \phi^0 \Nj$ for $\phi^0 = h^0, H^0, A^0$}{N\texttilde\textiinferior\textrightarrow\textscalar\textzerosuperior N\texttilde\textjinferior \nobreakspace for  \textscalar\textzerosuperior = h\textzerosuperior, H\textzerosuperior, A\textzerosuperior}}
\setcounter{equation}{0}
\setcounter{figure}{0}
\setcounter{table}{0}

Next we consider the decay of a neutralino to a lighter neutralino and
neutral Higgs boson $\phi^0 = h^0$, $H^0$, or $A^0$.
The two tree-level Feynman
graphs are shown in \fig{fig:neut1toneut2h}, where we have also
labeled the momenta and helicities.
We denote the masses for the
neutralinos and the Higgs boson as $m_{\Ni}$, $m_{\Nj},$ and
$m_{\phi^0}$.
Using the Feynman rules of \fig{inohiggsboson}, the amplitudes
are respectively given by
\beqa
i{\cal M}_1 &=& -i Y \,x_iy_j\,, \\
i{\cal M}_2 &=& -i Y^* \, y^\dagger_i x^\dagger_j\,,
\eeqa
where the coupling $Y \equiv Y^{\phi^0\chi^0_i\chi^0_j}$ is defined in
\eq{higgs-gauginos1},
and the external wave functions are
$x_i\equiv x(\boldsymbol {\vec p}_i,\lam_i)$,
$y^\dagger_i\equiv y^\dagger(\boldsymbol {\vec p}_i,\lam_i)$,
$y_j\equiv y(\boldsymbol{\vec k}_j,\lam_j)$, and
$x^\dagger_j\equiv x^\dagger(\boldsymbol{\vec k}_j,\lam_j)$.

Taking the square of the total matrix element using
eq.~(\ref{eq:conbil}) gives: \beqa |{\cal M}|^2 = |Y|^2 (x_i y_j
y^\dagger_j x^\dagger_i +
y^\dagger_i x^\dagger_j x_j y_i) + Y^2 x_i y_j x_j y_i +
Y^{*2} y^\dagger_i x^\dagger_j y^\dagger_j x^\dagger_i .  \eeqa
Summing over the final state neutralino spins using
\eqst{xxdagsummed}{ydagxdagsummed} yields
\beqa
\sum_{\lam_j} |{\cal M}|^2 =
\BDpos |Y|^2 (x_i k_j \newcdot \sigma x^\dagger_i +
 y^\dagger_i k_j \newcdot \sigmabar y_i) - Y^2 m_{\Nj} x_i y_i - Y^{*2}
m_{\Nj}  y^\dagger_i x^\dagger_i .  \eeqa Averaging over the initial state
neutralino spins in the same way gives
\begin{figure}[t!]
\begin{center}
\begin{picture}(150,70)(0,12)
\ArrowLine(-80,40)(-30,40)
\DashLine(-30,40)(10,70)5
\ArrowLine(10,10)(-30,40)
\Text(-77,54)[]{$\chi^0_i\,(p_i,\lam_i)$}
\Text(38,12)[]{${\chi^{0\,\dagger}_j} \,(k_j,\lam_j)$}
\Text(22,73)[]{$\phi^0$}
\ArrowLine(170,40)(120,40)
\DashLine(170,40)(210,70)5
\ArrowLine(170,40)(210,10)
\Text(123,54)[]{${\chi^{0\,\dagger}_i}\, (p_i,\lam_i)$}
\Text(238,12)[]{$\chi^0_j \,(k_j,\lam_j)$}
\Text(222,73)[]{$\phi^0$}
\end{picture}
\end{center}
\caption{The Feynman diagrams for $\stilde N_i \ra \stilde N_j
 \phi^0 $ in the MSSM.}
\label{fig:neut1toneut2h}
\end{figure}
\beqa
\frac{1}{2}
\sum_{\lam_i,\lam_j} |{\cal M}|^2 &=& \frac{1}{2} |Y|^2 ({\rm Tr}[ k_j
\newcdot \sigma p_i \newcdot \sigmabar] + {\rm Tr}[k_j \newcdot
\sigmabar p_i \newcdot \sigma]) + {\rm Re}[Y^2] m_{\Ni} m_{\Nj} {\rm
  Tr}[\mathds{1}_{2\times 2}]\nonumber
\\
&=& \BDpos 2 |Y|^2 p_i\newcdot k_j +2 {\rm Re}[Y^2] m_{\Ni} m_{\Nj}
\nonumber \\
&=& |Y|^2 (m_{\Ni}^2 + m_{\Nj}^2 - m_{\phi^0}^2) + 2 {\rm Re}[Y^2]
m_{\Ni} m_{\Nj} , \eeqa where we have used eq.~(\ref{trssbar}) to
obtain the second equality.  The total decay rate is therefore
\beqa
\Gamma (\Ni \ra \phi^0\Nj) &=& \frac{1}{16 \pi m_{\Ni}^3 }
\lambda^{1/2} (m_{\Ni}^2, m\ls{\phi^0}^2 , m_{\Nj}^2 ) \left ( \frac{1}{2}
    \sum_{\lambda_i,\lambda_j} |{\cal M}|^2 \right ) \nonumber
\\
&=& \frac{ m_{\Ni}}{16\pi} \lambda^{1/2}(1,r_\phi,r_j)
\left[|Y^{\phi^0\chi^0_i\chi^0_j}|^2 (1+r_j-r_\phi ) +2{\rm Re}
  \Bigl[\bigl(Y^{\phi^0\chi^0_i\chi^0_j}\bigr)^2\Bigr] \sqrt{r_j}
\right],\phantom{xxxx} \label{N2phiN}
\eeqa
where the triangle function $\lambda^{1/2}$ is defined in
\eq{eq:deftrianglefunction},
$r_j\equiv {m_{\Nj}^2}/{m_{\Ni}^2}$ and $r_\phi \equiv
{m_{\phi^0}^2}/{m_{\Ni}^2}$.  The results for $\phi^0 = h^0,
H^0, A^0$ can now be obtained by using eqs.~(\ref{eq:defkuphi0}) and
(\ref{eq:defkdphi0}) in eq.~(\ref{higgs-gauginos1}).  In comparing
\eq{N2phiN} with the original calculation in \Ref{Gunion:1987yh},
it is helpful to employ eqs.~(4.51) and (4.53) of \cite{gunhab}. The
results agree.


\subsection{\texorpdfstring{$\Ni \ra Z^0 \Nj$}{N\texttilde\textiinferior\textrightarrow Z\textzerosuperior N\texttilde\textjinferior}}
\setcounter{equation}{0}
\setcounter{figure}{0}
\setcounter{table}{0}

For this two-body decay there are two tree-level Feynman diagrams,
shown in \fig{fig:neut1toneut2Z} with the definitions of the
helicities and the momenta.
Using the Feynman rules of \fig{nnboson},
the two amplitudes are given by\footnote{When comparing with the
  four-component Feynman rule in ref.~\cite{HaberKane} note that ${\cal
    O}_{ij}^{\prime\prime L}=-{\cal O}_{ij}^{\prime\prime R*}$
    [cf.~\eq{eq:defOLpp}].}
\beqa
i{\cal M}_1&=& \BDneg i \frac{g}{c_W} {\cal O}_{ji}^{\prime\prime L}
              x_i \sigma^\mu x^\dagger_j \varepsilon^*_\mu\,,
\\
i{\cal M}_2&=& \BDpos i \frac{g}{c_W} {\cal O}_{ij}^{\prime\prime L}
               y^\dagger_i \sigmabar^\mu y_j \varepsilon^*_\mu\,,
\eeqa
where the external wave functions are
$x_i=x(\boldsymbol{\vec p}_i,\lam_i)$,
$ y^\dagger_i= y^\dagger(\boldsymbol{\vec p}_i,\lam_i)$,
$x^\dagger_j=x^\dagger(\boldsymbol{\vec k}_j,\lam_j)$,
$y_j=y(\boldsymbol{\vec k}_j,\lam_j)$, and
$\varepsilon^*_\mu = \varepsilon_\mu ({\boldsymbol{\vec k}}_Z,\lam_Z)^*$.
Noting that ${\cal O}_{ji}^{\prime\prime L} =
{\cal O}_{ij}^{\prime\prime L *}$ [see eq.~(\ref{eq:defOLpp})], and applying
eqs.~(\ref{eq:conbilsig}) and (\ref{eq:conbilsigbar}), we find that the
squared matrix element is:
\beqa
|{\cal M}|^2 &=& \frac{g^2}{c_W^2} \varepsilon^*_\mu \varepsilon_\nu
\biggl [
|{\cal O}_{ij}^{\prime\prime L}|^2
(x_i \sigma^\mu x^\dagger_j x_j \sigma^\nu x^\dagger_i +
 y^\dagger_i \sigmabar^\mu y_j  y^\dagger_j \sigmabar^\nu y_i)
\nonumber \\ &&
- \left ({\cal O}_{ij}^{\prime\prime L} \right )^2
 y^\dagger_i \sigmabar^\mu y_j x_j \sigma^\nu  x^\dagger_i
- \left ({\cal O}_{ij}^{\prime\prime L*} \right )^2
x_i \sigma^\mu  x^\dagger_j  y^\dagger_j \sigmabar^\nu y_i
\biggr ]\,.
\eeqa
Summing over the final state neutralino spin using
\eqst{xxdagsummed}{ydagxdagsummed} yields:
\beqa
\sum_{\lambda_j} |{\cal M}|^2
&=& \frac{g^2}{c_W^2} \varepsilon^*_\mu \varepsilon_\nu
\biggl [
\BDpos |{\cal O}_{ij}^{\prime\prime L}|^2
(x_i \sigma^\mu k_j \newcdot \sigmabar \sigma^\nu  x^\dagger_i +
 y^\dagger_i \sigmabar^\mu k_j \newcdot \sigma \sigmabar^\nu y_i)
\nonumber \\ &&
+ \left ({\cal O}_{ij}^{\prime\prime L} \right )^2
m_{\Nj}  y^\dagger_i \sigmabar^\mu \sigma^\nu  x^\dagger_i
+ \left ({\cal O}_{ij}^{\prime\prime L*} \right )^2
m_{\Nj} x_i \sigma^\mu \sigmabar^\nu y_i
\biggr ] .
\eeqa
\begin{figure}[t]
\begin{center}
\begin{picture}(150,70)(0,5)
\ArrowLine(-80,40)(-30,40)
\Photon(-30,40)(10,70)54
\ArrowLine(-30,40)(10,10)
\Text(-77,54)[]{$\chi^0_i\,(p_i,\lam_i)$}
\Text(38,12)[]{$\chi^0_j \,(k_j,\lam_j)$}
\Text(33,76)[]{$Z^0\,(k_Z,\lambda_Z)$}
\ArrowLine(170,40)(120,40)
\Photon(170,40)(210,70)54
\ArrowLine(210,10)(170,40)
\Text(123,54)[]{${\chi^{0\,\dagger}_i}\, (p_i,\lam_i)$}
\Text(238,12)[]{${\chi^{0\,\dagger}_j} \,(k_j,\lam_j)$}
\Text(233,76)[]{$Z^0\,(k_Z, \lambda_Z)$}
\end{picture}
\end{center}
\caption{The Feynman diagrams for $\stilde N_i \ra \stilde N_j
 Z^0 $ in the MSSM.}
\label{fig:neut1toneut2Z}
\end{figure}
Averaging over the initial state neutralino spins in the same way gives
\beqa
\frac{1}{2} \sum_{\lambda_i,\lambda_j} |{\cal M}|^2
&=& \frac{g^2}{2 c_W^2} \varepsilon^*_\mu \varepsilon_\nu
\biggl [
|{\cal O}_{ij}^{\prime\prime L}|^2
\Bigl ( {\rm Tr}[ \sigma^\mu k_j \newcdot \sigmabar \sigma^\nu p_i \newcdot
\sigmabar]
+
{\rm Tr}[\sigmabar^\mu k_j \newcdot \sigma \sigmabar^\nu p_i \newcdot
\sigma] \Bigr )
\nonumber \\ &&
\quad
- \left ({\cal O}_{ij}^{\prime\prime L} \right )^2
m_{\Ni} m_{\Nj} {\rm Tr} [\sigmabar^\mu \sigma^\nu ]
- \left ({\cal O}_{ij}^{\prime\prime L*} \right )^2
m_{\Ni} m_{\Nj} {\rm Tr} [\sigma^\mu \sigmabar^\nu ]
\biggr ] \nonumber
\\
&=&
\frac{2 g^2}{c_W^2} \varepsilon^*_\mu \varepsilon_\nu
\biggl \{ |{\cal O}_{ij}^{\prime\prime L}|^2 \left (
k_j^\mu p_i^\nu
+ p_i^\mu k_j^\nu
- p_i \newcdot k_j \metric^{\mu\nu} \right )
\BDminus
{\rm Re}\Bigl [\bigl ({\cal O}_{ij}^{\prime\prime L} \bigr )^2 \Bigr]
m_{\Ni} m_{\Nj} \metric^{\mu\nu}
\biggr \} , \phantom{xxx}
\eeqa
where in the last equality we have applied
\eqst{trssbar}{trsbarssbars}.
Using
\beq
\sum_{\lambda_Z}\varepsilon^{\mu*}\varepsilon^\nu =
\BDneg g^{\mu\nu} + k_Z^\mu k_Z^\nu/{m_Z^2}\,,
\eeq
we obtain
\beqa
\frac{1}{2} \sum_{\lam_i,\lam_j,\lam_Z}|{\cal M}|^2
&=& \frac{2 g^2}{c_W^2}
\biggl \{ |{\cal O}_{ij}^{\prime\prime L}|^2 \left (
\BDpos p_i \newcdot k_j + 2 p_i \newcdot k_Z k_j \newcdot k_Z/m_Z^2 \right
)
+ 3 m_{\Ni} m_{\Nj}
{\rm Re}\bigl [\bigl ({\cal O}_{ij}^{\prime\prime L} \bigr )^2 \bigr]
\biggr \}\,.
\phantom{xxx}
\eeqa
Using
$2k_j\newcdot k_Z = \BDpos m_{\Ni}^2 \BDminus m_{\Nj}^2 \BDminus m_Z^2$,
$2p_i\newcdot k_j = \BDpos m_{\Ni}^2 \BDplus m_{\Nj}^2 \BDminus m_Z^2$,
and
$2p_i\newcdot k_Z = \BDpos m_{\Ni}^2 \BDminus m_{\Nj}^2 \BDplus m_Z^2$,
we obtain the total decay width:
\beqa
&&\Gamma (\Ni \ra Z^0\Nj) = \frac{1}{16 \pi m_{\Ni}^3 }
\lambda^{1/2}\bigl(m_{\Ni}^2 ,m\ls{Z}^2, m_{\Nj}^2\bigr)
\left ( \frac{1}{2} \sum_{\lam_i,\lam_j,\lam_Z} |{\cal M}|^2 \right )
\nonumber \\
&&\quad
= \frac{g^2 m_{\Ni} }{16\pi\cw^2} \lambda^{1/2}(1,r_Z,r_j)
\biggl[|{\cal O}_{ij}^{\prime\prime L}|^2\left(1+r_j-2r_Z+(1-r_j)^2/r_Z
\right)
+6{\rm Re}\bigl[\bigl ({\cal O}_{ij}^{\prime\prime L} \bigr )^2\bigr]
\sqrt{r_j} \biggr],\phantom{xxxxx} \label{chi2chiZ}
\eeqa
where
\beq
r_j\equiv {m_{\Nj}^2}/{m_{\Ni}^2}\,,\qquad\quad
r_Z\equiv {m_{Z}^2}/{m_{\Ni}^2}\,,
\eeq
and the triangle function $\lambda^{1/2}$ is defined in
eq.~(\ref{eq:deftrianglefunction}). The result
obtained in \eq{chi2chiZ} agrees with
the original calculation in \Ref{Gunion:1987yh}.

\subsection{Selectron pair production in electron-electron collisions}
\setcounter{equation}{0}
\setcounter{figure}{0}
\setcounter{table}{0}

\subsubsection{\texorpdfstring{$e^-e^- \ra \stilde e_L^- \stilde e_R^-$}{e\textminussuperior e\textminussuperior \textrightarrow e\texttilde\textscl\textminussuperior e\texttilde\textscr\textminussuperior}} 

Here there are two Feynman graphs (neglecting the electron mass
and Yukawa couplings),
shown in \fig{fig:eeseLseR}.
Note that these two graphs are related by interchange of the
identical initial state electrons. Let the electrons have momenta
$p_1$ and $p_2$ and the selectrons have momenta
$k_{\stilde e_L}$ and $k_{\stilde e_R}$,
so that $p_1^2 = p_2^2 = 0$;
$k_1^2 = \BDpos m_{\tilde e_L}^2$;
$k_2^2 = \BDpos m_{\tilde e_R}^2$;
$s = \BDpos (p_1 + p_2)^2 = \BDpos (k_1 + k_2)^2$;
$t = \BDpos (k_1 - p_1)^2 = \BDpos (k_2 - p_2)^2$;
$u = \BDpos (k_1 - p_2)^2 = \BDpos (k_2 - p_1)^2$.
\begin{figure}[ht!]
\hspace{0.4cm}
\begin{picture}(400,100)(-200,29)
\ArrowLine(-200,105)(-125,105)
\ArrowLine(-125,50)(-125,105)
\ArrowLine(-125,50)(-200,50)
\DashLine(-125,105)(-50,105){5}
\DashLine(-125,50)(-50,50){5}
\put(-201,111){$e\,(p_1,\lam_1)$}
\put(-201,56){${\bar e}^\dagger\,(p_2,\lam_2)$}
\put(-58,111){$\stilde e_L^-\,(k_1)$}
\put(-58,56){$\stilde e_R^- \,(k_2)$}
\put(-119,74){$\chi^0_i$}
\Line(125,105)(87.5,77.5)
\ArrowLine(50,50)(87.5,77.5)
\ArrowLine(125,50.)(125,105)
\Line(125,50)(92.75,73.65)
\ArrowLine(82.25,81.35)(50,105)
\DashLine(125,105)(200,105){5}
\DashLine(125,50)(200,50){5}
\put(50,38){$e \, (p_2,\lam_2)$}
\put(50,111){${\bar e}^\dagger \,(p_1,\lam_1)$}
\put(192,111){$\stilde e_L^-\, (k_1)$}
\put(192,56){$\stilde e_R^-\, (k_2)$}
\put(132,74){$\chi_i^0 $}
\end{picture}
\caption{Feynman diagrams for $e^-e^-\ra{\stilde e_L}^-
{\stilde e_R}^-$.\label{fig:eeseLseR}}
\end{figure}

Using the Feynman rules of \fig{nqsq},
the matrix element for the first graph, for each neutralino
$\stilde N_i$ exchanged in the $t$ channel, is:
\beq
i {\cal M}_t =
\left [ i \frac{g}{\sqrt{2}} \left (N^*_{i2} + \frac{s_W}{c_W}
    N^*_{i1}\right ) \right ]
\left [ - i \sqrt{2} g \frac{s_W}{c_W} N_{i1}
\right ]
\> x_1 \left [ \frac{i (k_1 - p_1) \newcdot \sigma}{(k_1- p_1)^2
    \BDminus m_{\tilde N_i}^2 } \right ]  y^\dagger_2 \, .
\eeq
We employ the notation for the external wave functions $x_i=(
\boldsymbol{\vec p}_i,\lam_i)$, $i=1,2$ and analogously for $y_i, x^\dagger_i,
 y^\dagger_i$. The matrix element for the second
($u$-channel) graph is the same
with the two incoming electrons exchanged, $e_1 \leftrightarrow e_2$:
\beq
i{\cal M}_u = (-1)
\left [ i \frac{g}{\sqrt{2}} \left (N^*_{i2} + \frac{s_W}{c_W}
N^*_{i1}\right ) \right ]
\left [ - i \sqrt{2} g \frac{s_W}{c_W} N_{i1} \right]
    \> x_2 \left [ \frac{
     i (k_1 - p_2) \newcdot \sigma}{(k_1- p_2)^2
    \BDminus m_{\tilde N_i}^2 } \right ]  y^\dagger_1 \> .
\label{opp-order}
\eeq
Note that since we have written the fermion wave function spinors in
the opposite order in ${\cal M}_2$ compared to ${\cal M}_1$, there is
a factor $(-1)$ for Fermi-Dirac statistics.  Alternatively, starting
at the electron with momentum $p_1$ and using the Feynman rules as
above, we can directly write:
\beq
i{\cal M}_u =  \left [ i \frac{g}{\sqrt{2}} \left
    (N^*_{i2} + \frac{s_W}{c_W} N^*_{i1}\right ) \right ]
\left [ - i \sqrt{2} g
  \frac{s_W}{c_W} N_{i1} \right ]
\>  y^\dagger_1
\left [ \frac{- i (k_1 - p_2) \newcdot \sigmabar}
  {(k_1- p_2)^2 \BDminus m_{\tilde N_i}^2 } \right ] x_2 \, .
\label{same-order}
\eeq
This has no Fermi-Dirac factor $(-1)$ because the wave function
spinors are written in the same order as in ${\cal M}_t$. However, now
the Feynman rule for the propagator has an extra minus sign, as can be
seen in \fig{fig:neutproprev}. We can also obtain \eq{same-order}
from \eq{opp-order} by using \eq{europeanvacation}. So we can write for
the total amplitude:
\beq
{\cal M} ={\cal M}_t +{\cal M}_u = x_1 a\newcdot
\sigma  y^\dagger_2 +  y^\dagger_1 b \newcdot \sigmabar x_2 \, ,
\eeq
where
\beqa
a^\mu &\equiv&   \BDpos \frac{g^2s_W}{c_W} (k_1^\mu -
p_1^\mu) \sum_{i=1}^4 N_{i1} (N^*_{i2} + \frac{s_W}{c_W} N^*_{i1} )
\frac{1}{t - m^2_{\tilde N_i}} \> ,
\\
b^\mu &\equiv&
\BDneg \frac{g^2 s_W}{c_W} (k_1^\mu - p_2^\mu)
\sum_{i=1}^4 N_{i1} (N^*_{i2} + \frac{s_W}{c_W} N^*_{i1} ) \frac{1}{u -
  m^2_{\tilde N_i}} \> .
\eeqa
Hence, using eqs.~(\ref{eq:conbilsig}) and (\ref{eq:conbilsigbar}):
\beqa
|{\cal M}|^2 &=&
\left(x_1 a\newcdot \sigma  y^\dagger_2 \right)
\left(y_2 a^* \newcdot \sigma  x^\dagger_1 \right)
+ \left( y^\dagger_1 b \newcdot \sigmabar x_2 \right)
  \left( x^\dagger_2 b^* \newcdot \sigmabar y_1\right)\nonumber \\[5pt]
&&
+ \left(x_1 a \newcdot \sigma  y^\dagger_2\right)
     \left( x^\dagger_2 b^* \newcdot \sigmabar y_1\right)
+ \left( y^\dagger_1 b \newcdot \sigmabar x_2\right)
  \left(y_2 a^* \newcdot \sigma  x^\dagger_1\right)\,.
\eeqa
Averaging over the initial state electron spins using
\eqst{xxdagsummed}{ydagxdagsummed},
the $a,b^*$ and $a^*, b$ cross terms
are proportional to $m_e$ and can thus be
neglected in our approximation.  We get:
\beq
\frac{1}{4} \sum_{\lambda_1, \lambda_2} |{\cal M}|^2 =
\frac{1}{4}{\rm Tr}[a\newcdot \sigma\; p_2 \newcdot \sigmabar\; a^*
\newcdot \sigma\; p_1\newcdot \sigmabar ] +\frac{1}{4} {\rm Tr}[b\newcdot
\sigmabar\; p_2 \newcdot \sigma\; b^* \newcdot \sigmabar\; p_1 \newcdot
\sigma ] \> .
\eeq
These terms can be simplified using the identities:
\beqa
{\rm Tr}[(k_1 - p_1) \newcdot \sigma\; p_2\newcdot
\sigmabar\; (k_1 - p_1) \newcdot \sigma\; p_1 \newcdot \sigmabar]
&=& {\rm
  Tr}[(k_1 - p_2) \newcdot \sigmabar\; p_2 \newcdot\sigma \;
  (k_1 - p_2)
\newcdot \sigmabar\; p_1 \newcdot \sigma]
\phantom{xxx}\nonumber
\\ & = &
tu - m_{\tilde e_L}^2 m_{\tilde e_R}^2 ,
\phantom{x}
\eeqa
which follow from eq.~(\ref{trssbarssbar}) and (\ref{trsbarssbars}),
resulting in:
\beqa
\frac{1}{4} \sum_{\lambda_1, \lambda_2} |{\cal M}|^2 &=&
\frac{g^4 s_W^2}{4
  c_W^2} (tu - m_{\tilde e_L}^2 m_{\tilde e_R}^2) \sum_{i,j = 1}^4
N_{j1}N^*_{i1} (N^*_{j2} + \frac{s_W}{c_W} N^*_{j1}) (N_{i2} +
   \frac{s_W}{c_W} N_{i1})
\phantom{xxx}
\cr &&\qquad\qquad\>\>\>
  \left [
      \frac{1}{(t - m_{\tilde N_i}^2)(t - m_{\tilde N_j}^2) }
    + \frac{1}{(u - m_{\tilde N_i}^2)(u - m_{\tilde N_j}^2) }
  \right ] \, .
\eeqa
To get the differential cross-section $d\sigma/dt$, multiply this by
$1/(16 \pi s^2)$:
\beqa
\frac{d \sigma}{dt} &=& \frac{\pi \alpha^2}{4
  s_W^2 c_W^2} \left ( \frac{tu - m_{\stilde e_L}^2 m_{\stilde e_R}^2}
   {s^2} \right ) \sum_{i,j = 1}^4 N_{j1}N^*_{i1} (N^*_{j2} + \frac{s_W}
   {c_W} N^*_{j1}) (N_{i2} + \frac{s_W}{c_W} N_{i1}) \cr &&
\qquad\qquad\qquad\>\>\>\> \left [ \frac{1}{(t - m_{\tilde N_i}^2)(t -
    m_{\tilde N_j}^2) } + \frac{1}{(u - m_{\tilde N_i}^2)(u -
    m_{\tilde N_j}^2) } \right ] \, .
\eeqa
To compare with the original calculation in \Ref{Keung:1983nq} and
with eq.~E26, p.~244 in ref.~\cite{HaberKane}, note that for a pure
photino exchange, $N_{i1} \ra c_W{\delta_{i1}}$ and $N_{i2} \ra
s_W{\delta_{i1}}$, so that
\beq \frac{1}{4 s_W^2 c_W^2} |N_{i1}|^2 |N_{i2} + \frac{s_W}{c_W}
N_{i1}|^2 \> \ra \> 1\;.
\eeq
Also note that in \Ref{Keung:1983nq} polarized electron beams are
assumed. The result checks.

\subsubsection{\texorpdfstring{$e^-e^- \ra \stilde e_R^- \stilde e_R^-$}{e\textminussuperior e\textminussuperior \textrightarrow e\texttilde\textscr\textminussuperior e\texttilde\textscr\textminussuperior}} 

For this process, there are again two Feynman graphs, which are
related by the exchange of identical electrons in the initial state or
equivalently by exchange of the identical selectrons in the final
state, as shown in \fig{fig:ee2seRseR}.
(We again neglect the electron mass and thus the
higgsino coupling to the electron.)
Let the electrons have momenta $p_1$ and $p_2$ and the selectrons have
momenta $k_1$ and $k_2$, so that $p_1^2 = p_2^2 = 0$; $k_1^2 = k_2^2 =
\BDpos m_{\stilde e_R}^2$; $s = \BDpos (p_1 + p_2)^2$;
$t = \BDpos (k_1 -p_1)^2$; $u = \BDpos (k_1 - p_2)^2$.
\begin{figure}[ht!]
\centerline{
\begin{picture}(400,100)(-200,29)
\ArrowLine(-125,105)(-200,105)
\ArrowLine(-125,105)(-125,77.5)
\ArrowLine(-125,50)(-125,77.5)
\ArrowLine(-125,50)(-200,50)
\DashLine(-125,105)(-50,105)5
\DashLine(-125,50)(-50,50)5
\put(-201,111){${\bar e}^\dagger\,(p_1,\lam_1)$}
\put(-201,56){${\bar e}^\dagger\,(p_2,\lam_2)$}
\put(-58,111){$\stilde e_R^-\,(k_1)$}
\put(-58,56){$\stilde e_R^-\,(k_2)$}
\put(-119,74){$\chi^0_i$}
\Line(125,105)(87.5,77.5)
\ArrowLine(87.5,77.5)(50,50)
\ArrowLine(125,105)(125,77.5)
\ArrowLine(125,50)(125,77.5)
\Line(125,50)(92.75,73.65)
\ArrowLine(82.25,81.35)(50,105)
\DashLine(125,105)(200,105)5
\DashLine(125,50)(200,50)5
\put(50,38){${\bar e}^\dagger\,(p_2,\lam_2)$}
\put(50,111){${\bar e}^\dagger\,(p_1,\lam_1)$}
\put(192,111){$\stilde e_R^-\,(k_1)$}
\put(192,56){$\stilde e_R^-\,(k_2)$}
\put(132,74){$\chi^0_i$}
\end{picture}
}
\caption{\label{fig:ee2seRseR} The two Feynman diagrams for $e^-e^-
\ra{\stilde e_R}^-  {\stilde e_R}^-$ in the limit where $m_e\ra0$.}
\end{figure}

Using the Feynman rules of \fig{nqsq},
the amplitude for the first graph is:
\beq
i{\cal M}_t = \left ( -i \sqrt{2} g \frac{s_W}{c_W} N_{i1} \right )^2
\left [ \frac{\BDpos i\, m_{\stilde N_i}}{(k_1 - p_1)^2
\BDminus m_{\stilde N_i}^2 }
\right ]  y^\dagger_1  y^\dagger_2\,,
\eeq
for each exchanged neutralino.
The amplitude for the second graph is the same, but with the electrons
interchanged:
\beq
i{\cal M}_u = \left ( -i \sqrt{2} g \frac{s_W}{c_W} N_{i1}
\right )^2
\left [ \frac{\BDpos i\, m_{\stilde N_i}}{(k_1 - p_2)^2 \BDminus
m_{\stilde N_i}^2 }
\right ]  y^\dagger_1  y^\dagger_2 \> .
\eeq
Since we have chosen to write the external state wave function spinors
in the same order in ${\cal M}_t$ and ${\cal M}_u$, there is no factor of
$(-1)$ for Fermi-Dirac statistics.
So, applying eq.~(\ref{eq:conbil}), the total amplitude squared is:
\beqa
|{\cal M}|^2 &=& \frac{4 g^4 s_W^4}{c_W^4}  ( y^\dagger_1  y^\dagger_2)
(y_2 y_1) \left | \sum_{i = 1}^4
(N_{i1})^2 m_{\stilde N_i}
\left (
\frac{1}{t - m_{\stilde N_i}^2} + \frac{1}{u - m_{\stilde N_i}^2}
\right )
\right |^2 .
\eeqa
The sum over the electron spins is obtained from
\beq
\sum_{\lam_1,\lam_2} ( y^\dagger_1  y^\dagger_2) (y_2 y_1) =
{\rm Tr}[p_2 \newcdot
\sigmabar p_1 \newcdot \sigma] = \BDpos 2 p_2 \newcdot p_1 = s\> .
\eeq
So, using eq.~(\ref{yydagsummed}),
the spin-averaged differential cross-section is:
\beqa  \label{ererstuff}
\frac{d \sigma}{d t} =\frac{1}{16 \pi s^2}
\left (\frac{1}{4}
\sum_{\lambda_1,\lambda_2}
|{\cal M}|^2 \right )
=
\frac{\pi \alpha^2}{2 c_W^4 s}
\left |
\sum_{i=1}^4
(N_{i1})^2 m_{\stilde N_i}
\left (
\frac{1}{t - m_{\stilde N_i}^2} + \frac{1}{u - m_{\stilde N_i}^2}
\right ) \right |^2
.\qquad\phantom{xxxx}
\eeqa
After integrating over $t$ to obtain the total cross-section, the resulting \pagebreak
expression must be multiplied
by a factor of $1/2$ to account for the 
identical sleptons in the final state (to avoid the double counting of indistinguishable particles).

To compare with \cite{Keung:1983nq} and also with eq.~E27 of
ref.~\cite{HaberKane}, note that for a pure photino exchange, $N_{i1}
\ra c_W\delta_{i1}$, so it checks.

\subsubsection{\texorpdfstring{$e^-e^- \ra \stilde e_L^- \stilde e_L^-$}{e\textminussuperior e\textminussuperior \textrightarrow e\texttilde\textscl\textminussuperior e\texttilde\textscl\textminussuperior}} 

Again, in the limit of vanishing electron mass, there are two Feynman
graphs, which are related by the exchange of identical electrons in
the initial state or equivalently by exchange of the identical
selectrons in the final state. As shown in \fig{fig:ee2seLseL},
they are exactly like the previous example, but with all arrows
reversed.
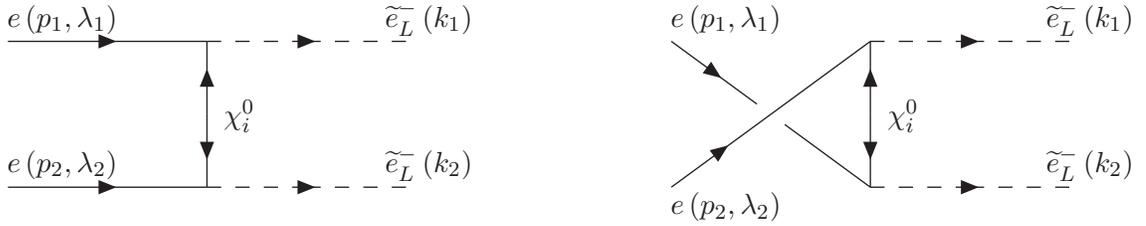
\begin{figure}[ht!]
\centerline{
\begin{picture}(400,100)(-200,29)
\ArrowLine(-200,105)(-125,105)
\ArrowLine(-125,77.5)(-125,105)
\ArrowLine(-125,77.5)(-125,50)
\ArrowLine(-200,50)(-125,50)
\DashArrowLine(-125,105)(-50,105)5
\DashArrowLine(-125,50)(-50,50)5
\put(-201,111){$e\,(p_1,\lam_1)$}
\put(-201,56){$e\,(p_2,\lam_2)$}
\put(-58,111){$\stilde e_L^-\,(k_1)$}
\put(-58,56){$\stilde e_L^-\,(k_2)$}
\put(-119,74){$\chi^0_i$}
\Line(125,105)(87.5,77.5)
\ArrowLine(50,50)(87.5,77.5)
\ArrowLine(125,77.5)(125,105)
\ArrowLine(125,77.5)(125,50)
\Line(125,50)(92.75,73.65)
\ArrowLine(50,105)(82.25,81.35)
\DashArrowLine(125,105)(200,105)5
\DashArrowLine(125,50)(200,50)5
\put(50,40){$e\,(p_2,\lam_2)$}
\put(50,111){$e\,(p_1,\lam_1)$}
\put(192,111){$\stilde e_L^-\,(k_1)$}
\put(192,56){$\stilde e_L^-\,(k_2)$}
\put(132,74){$\chi^0_i$}
\end{picture}
}
\caption{\label{fig:ee2seLseL} The two Feynman diagrams for $e^-e^-
\ra{\stilde e_L}^-  {\stilde e_L}^-$ in the limit of vanishing
electron mass.}
\end{figure}

Using the Feynman rules of \fig{nqsq},
the amplitude for the first graph is:
\beq
{\cal M}_t = \left ( i \frac{g}{\sqrt{2}} [ N_{i2}^* + \frac{s_W}{c_W}
N_{i1}^* ] \right )^2 \left [ \frac{\BDpos i\, m_{\stilde N_i}}{
(p_1 - k_1)^2 \BDminus m_{\stilde N_i}^2 }\right ]  x_1 x_2\,,
\eeq
for each exchanged neutralino. The amplitude for the second graph is
the same, but with $p_1\leftrightarrow p_2$:
\beq
{\cal M}_u = \left ( i \frac{g}{\sqrt{2}} [ N_{i2}^* + \frac{s_W}{c_W}
N_{i1}^* ] \right )^2 \left [ \frac{\BDpos i\, m_{\stilde N_i}}{
(p_2 - k_1)^2 \BDminus m_{\stilde N_i}^2 }\right ]  x_1 x_2\,.
\eeq
Since we have chosen to write the external state wave function spinors
in the same order in ${\cal M}_1$ and ${\cal M}_2$, there is no factor
of $(-1)$ for Fermi-Dirac statistics.  The total amplitude squared is:
\beqa
|{\cal M}|^2 &=& \frac{g^4}{4} (x_1 x_2) ( x^\dagger_2  x^\dagger_1)
\left |
\sum_{i = 1}^4
(N^*_{i2} + \frac{s_W}{c_W} N^*_{i1})^2
m_{\stilde N_i}
\left (
\frac{1}{t - m_{\stilde N_i}^2} + \frac{1}{u - m_{\stilde N_i}^2}
\right )
\right |^2
 \> .
\eeqa
The average over the electron spins follows from
eq.~(\ref{xxdagsummed}):
\beq
\sum_{\lam_1,\lam_2} (x_1 x_2) ( x^\dagger_2  x^\dagger_1)
= {\rm Tr}[p_2 \newcdot \sigma p_1 \newcdot \sigmabar ]
= \BDpos 2 p_2 \newcdot p_1 = s \> .
\eeq
So the spin-averaged differential cross-section is:
\beqa
&&
\frac{d \sigma}{d t} = \frac{1}{16 \pi s^2}
\left (\frac{1}{4}
\sum_{\lambda_1,\lambda_2}
|{\cal M}|^2 \right )
=
\frac{\pi \alpha^2}{16 s_W^4 s}
\left | \sum_{i=1}^4
(N^*_{i2} + \frac{s_W}{c_W} N^*_{i1})^2
m_{\stilde N_i}
\left (\frac{1}{t - m_{\stilde N_i}^2} + \frac{1}{u - m_{\stilde N_i}^2}
\right )
\right |^2 .
 \nonumber \\  
 && \phantom{line}\label{elelstuff}
\eeqa
After integrating over $t$ to obtain the total cross-section, one must multiply the resulting expression
by a factor of $1/2$ to account for the identical sleptons in the final state [as noted below \eq{ererstuff}]. 
To compare with \cite{Keung:1983nq} and also with eq.~(E27) of
ref.~\cite{HaberKane}, note that for a pure photino exchange, $N_{i1}
\ra c_W\delta_{i1}$ and $N_{i2} \ra s_W\delta_{i1}$, so it checks.

\subsection{\texorpdfstring{$e^-e^+ \ra \stilde \nu \stilde\nu^*$}{e\textminussuperior e\textplussuperior\textrightarrow\textnu\texttilde \textnu\texttilde \textstar}}
\setcounter{equation}{0}
\setcounter{figure}{0}
\setcounter{table}{0}

Consider now the pair production of sneutrinos in electron-positron
collisions. There are two graphs featuring the $s$-channel exchange of
the $Z^0$.  We will neglect the electron mass and Yukawa coupling, so
there is only one graph involving the $t$-channel exchange of the
charginos.  These three Feynman diagrams are shown in
\fig{fig:ee2snusnu}, where we have also defined the helicities
and momenta of the particles.
The Mandelstam variables can be expressed in terms of the external
momenta and the sneutrino mass:
\beqa
&& \BDpos 2 p_1 \newcdot p_2 = s\,,\qquad\qquad\qquad\qquad\qquad
\BDpos 2 k_1\newcdot k_2 = s - 2m_{\tilde \nu}^2\,,\label{eq:eesnusnukinA}
\\
&& \BDpos 2p_1 \newcdot k_1 = \BDpos 2p_2\newcdot k_2 =
  m_{\tilde \nu}^2 -t\,,\qquad\quad\,\,\,
\BDpos 2p_1 \newcdot k_2 = \BDpos 2p_2 \newcdot k_1 =
  m_{\tilde \nu}^2 -u\,.
\label{eq:eesnusnukinB}
\eeqa
Using the Feynman rules of \fig{SMintvertices}, the
amplitudes for the two $s$-channel $Z$ boson exchange diagrams
are:\footnote{Because we neglect the electron mass, we may drop the
$Q^\mu Q^\nu$ term of the $Z$ propagator, where $Q\equiv p_1+p_2$ is the
propagating four-momentum in the $s$-channel [cf.~\fig{fig:bosonprops}].
\label{fnqq}}
\beqa
i {\cal M}_1 &=&
\left [\BDneg i \frac{g}{2c_W} (k_1 - k_2)_\mu \right ]
\left [ \frac{\BDneg i g^{\mu\nu}}{D_Z} \right ]
\left [\BDpos i\frac{g}{c_W} (s_W^2 - \half) \right]
x_1 \sigma_\nu  y^\dagger_2\,,
\phantom{xxxx}
\\
i {\cal M}_2 &=&
\left [\BDneg i \frac{g}{2c_W} (k_1 - k_2)_\mu \right ]
\left [ \frac{\BDneg i g^{\mu\nu}}{D_Z} \right ]
\left [\BDpos i\frac{g s_W^2}{c_W}  \right]
 y^\dagger_1 \sigmabar_\nu x_2\,,
\eeqa
where the first factor in each case is the Feynman rule from the $Z$
boson coupling to the sneutrinos (see Fig.~72c, ref.~\cite{HaberKane}),
and $D_Z\equiv s-m_Z^2+i\Gamma_Zm_Z$
is the denominator of the $Z$ boson propagator.\footnote{The explicit
inclusion of the finite decay width in the propagator of an unstable particle
involves subtle issues of gauge invariance and unitarity, particularly
in higher loop computations.  The authors of \Ref{dd} recommend the
complex-mass scheme for perturbative calculations with unstable
particles, first introduced in \Ref{complexmass}.} 
The $t$-channel diagram due to each chargino gives a contribution
\begin{figure}[tb!]
\begin{center}
\begin{picture}(400,100)(-200,30)
\ArrowLine(-200,115)(-152,87.5)
\ArrowLine(-152,87.5)(-200,60)
\Photon(-152,87.5)(-98,87.5)46
\DashArrowLine(-98,87.5)(-50,115)5
\DashArrowLine(-50,60)(-98,87.5)5
\put(-205,120){$e(p_1,\lam_1)$}
\put(-205,51){${e}^\dagger(p_2,\lam_2)$}
\put(-51,120){$\stilde \nu(k_1)$}
\put(-50,48){$\stilde \nu^*(k_2)$}
\put(-129,100){$Z^0$}
\ArrowLine(98,87.5)(50,115)
\ArrowLine(50,60)(98,87.5)
\Photon(98,87.5)(152,87.5)46
\DashArrowLine(152,87.5)(200,115)5
\DashArrowLine(200,60)(152,87.5)5
\put(45,120){${\bar e}^\dagger(p_1,\lam_1)$}
\put(45,49){$\bar e(p_2,\lam_2)$}
\put(199,120){$\stilde \nu(k_1)$}
\put(200,48){$\stilde \nu^*(k_2)$}
\put(121,100){$Z^0$}
\end{picture}
\end{center}
\begin{center}
\begin{picture}(400,80)(-300,42)
\ArrowLine(-200,105)(-125,105)
\ArrowLine(-125,50)(-125,105)
\ArrowLine(-125,50)(-200,50)
\DashArrowLine(-125,105)(-50,105)5
\DashArrowLine(-50,50)(-125,50)5
\put(-201,111){$e\,(p_1,\lam_1)$}
\put(-201,56){${{e}^\dagger}\,(p_2,\lam_2)$}
\put(-58,111){$\stilde \nu\,(k_1)$}
\put(-58,56){$\stilde \nu^*\,(k_2)$}
\put(-144,74){$\chi_i^+$}
\end{picture}
\end{center}
\caption{\label{fig:ee2snusnu} The Feynman diagrams for $e^-e^+
\ra{\stilde\nu}  {\stilde\nu}^*$.
}
\end{figure}%
\beqa
i{\cal M}_3 &=&
\left (-ig V_{i1}^* \right )
\left (-ig V_{i1} \right )
x_1
\left [ \frac{ i (k_1 - p_1) \newcdot \sigma}{(k_1 - p_1)^2
\BDminus m_{\tilde C_i}^2} \right ]
 y^\dagger_2,
\eeqa
using the rules of \fig{cqsq}.
Therefore, the total amplitude can be rewritten as:
\beqa
{\cal M} =
c_1 x_1 (k_1 - k_2)\newcdot \sigma  y^\dagger_2
+ c_2  y^\dagger_1 (k_1 - k_2) \newcdot \sigmabar x_2
+ c_3 x_1 (k_1 - p_1) \newcdot \sigma
 y^\dagger_2\,,
\eeqa
where
\beqa
c_1 &\equiv& \BDpos \frac{g^2(1 - 2s_W^2 )}{4 c_W^2 D_Z},
\quad\qquad
c_2 \equiv \BDneg \frac{g^2 s_W^2}{2 c_W^2 D_Z},
\quad\qquad
c_3 \equiv
\BDpos g^2 \sum_{i=1}^2 \frac{|V_{i1}|^2}{m_{\tilde C_j}^2 - t}\,.
\phantom{xxx}
\eeqa

Squaring the amplitude and summing over the electron and
positron spins, the interference terms involving $c_2$
will vanish in the massless electron limit due to
\eqs{yxsummed}{ydagxdagsummed}. Therefore, we obtain
\beqa
&& \hspace{-0.5in}\sum_{\lambda_1,\lambda_2} |{\cal M}|^2=
\sum_{\lambda_1,\lambda_2} \biggl \{
|c_1|^2 \, x_1 (k_1 - k_2) \newcdot \sigma  y^\dagger_2
        \, y_2 (k_1 - k_2) \newcdot \sigma  x^\dagger_1
+
|c_2|^2 \,  y^\dagger_1 (k_1 - k_2) \newcdot \sigmabar x_2
        \,  x^\dagger_2 (k_1 - k_2) \newcdot \sigmabar y_1
\nonumber \\ &&
\qquad\quad
+ c_3^2 \, x_1 (k_1 - p_1) \newcdot \sigma  y^\dagger_2
        \, y_2 (k_1 - p_1) \newcdot \sigma  x^\dagger_1
+ 2 {\rm Re}[c_1 c_3 \,
        x_1 (k_1 - k_2) \newcdot \sigma  y^\dagger_2
      \, y_2 (k_1 - p_1) \newcdot \sigma  x^\dagger_1]
\biggr \} \nonumber
\\[8pt]
&&=
|c_1|^2 \,{\rm Tr} [(k_1 - k_2) \newcdot \sigma p_2 \newcdot \sigmabar
                  (k_1 - k_2) \newcdot \sigma p_1 \newcdot \sigmabar]
+
|c_2|^2\, {\rm Tr} [(k_1 - k_2) \newcdot \sigmabar p_2 \newcdot \sigma
                  (k_1 - k_2) \newcdot \sigmabar p_1 \newcdot \sigma]
\nonumber \\[6pt] &&
\quad
+ c_3^2 \,
{\rm Tr} [(k_1 - p_1) \newcdot \sigma p_2 \newcdot \sigmabar
          (k_1 - p_1) \newcdot \sigma p_1 \newcdot \sigmabar]
+ 2 {\rm Re}[c_1] c_3 \,
{\rm Tr} [(k_1 - k_2) \newcdot \sigma p_2 \newcdot \sigmabar
          (k_1 - p_1) \newcdot \sigma p_1 \newcdot
          \sigmabar],\nonumber \\
\phantom{line}
\eeqa
where we have used eqs.~(\ref{xxdagsummed}) and (\ref{yydagsummed})
to perform the spin sums.  Applying the trace
identities
\eqs{trssbarssbar}{trsbarssbars} and simplifying the results
using \eqst{eq:eesnusnukinA}{eq:eesnusnukinB}
and $u = 2 m_{\tilde \nu}^2 - s - t$, we get
\beqa
\sum_{\lambda_1,\lambda_2} |{\cal M}|^2
= -[st + (t - m_{\tilde \nu}^2)^2] \left (
4 |c_1|^2 + 4 |c_2|^2 + c_3^2 + 4 {\rm Re}[c_1] c_3\right) .
\eeqa
When $m_{\tilde C_1} = m_{\tilde C_2}$, this agrees with
eqs.~(E46)--(E48) of \Ref{HaberKane}\footnote{There is a typographical
error in eq.~(E48) of \cite{HaberKane}; the right-hand side should be
  multiplied by 1/$\cos^2\theta_W$.} and with \Ref{lackner}.
The differential cross-section follows in the standard way by
averaging over the initial state spins:
\beqa
\frac{d\sigma}{dt} =
\frac{1}{16\pi s^2}
\biggl (
\frac{1}{4} \sum_{\lambda_1,\lambda_2} |{\cal M}|^2
\biggr )\,.
\eeqa
Note that
\beqa
t &=& m_{\tilde \nu}^2 - \half(1 - \beta \cos\theta) s\,,\qquad\qquad
\beta \equiv \left(1 - \frac{4 m_{\tilde \nu}^2}{s}\right)^{1/2},
\eeqa
where $\theta$ is the angle between the initial state electron and the
final state sneutrino in the center-of-momentum frame.  The upper and
lower limits $t_+$ and $t_-$ are obtained by inserting $\cos\theta =
\pm1$ above, respectively.

Performing the integration over $t$ to obtain the total cross-section,
one obtains
\beq
\sigma = \int_{t_-}^{t_+} \frac{d\sigma}{dt} dt = \frac{g^4}{64\pi s}
\biggl ( S_Z + \sum_{i,j = 1}^2 S_{ij} + \sum_{i=1}^2 S_{Zi} \biggr ),
\eeq
where
\beqa
S_Z &=& \frac{ \beta^3}{24 c_W^4} (8 s_W^4 - 4 s_W^2 +
1) \frac{s^2}{|D_Z|^2}
,
\\
S_{ii} &=& |V_{i1}|^4 \left [ (1 - 2 \gamma_i) L_i - 2 \beta \right ]
,
\\
S_{12} = S_{21} &=& |V_{11} V_{12}|^2
\left \{
\frac{(m_{\tilde C_2}^2 + s \gamma_2^2) L_2
-(m_{\tilde C_1}^2 + s \gamma_1^2) L_1}
{m_{\tilde C_2}^2 - m_{\tilde C_1}^2}
-\beta
\right \}
,
\phantom{xxxx}
\\
S_{Zi} &=&
\frac{(2s_W^2 - 1)}{c_W^2} |V_{i1}|^2
\left [
(m_{\tilde C_i}^2 + s \gamma_i^2) L_i + s \beta (\gamma_i - 1/2)
\right ]
\frac{(s-m_Z^2)}{|D_Z|^2}
,
\eeqa
with
\beqa
\gamma_i &\equiv& \frac{m_{\tilde \nu}^2 - m_{\tilde C_i}^2}{s},
\qquad\qquad\quad
L_i \equiv {\rm ln}\biggl (
\frac{m_{\tilde C_i}^2 - t_- }{m_{\tilde C_i}^2- t_+}
\biggr ).
\eeqa
This agrees with eqs.~(E49)-(E52) of ref.~\cite{HaberKane} in the
limit of degenerate charginos, or of a single wino chargino with
$|V_{11}| = 1$ and $V_{12} = 0$. It also agrees with \cite{lackner}.

\subsection{\texorpdfstring{$e^-e^+ \ra \Ni\Nj$}{e\textminussuperior {e\textplussuperior \textrightarrow N\texttilde\textiinferior N\texttilde\textjinferior}}}
\label{eechichi}
\setcounter{equation}{0}
\setcounter{figure}{0}
\setcounter{table}{0}

Next we consider the pair production of neutralinos via
$e^-e^+$ annihilation.
There are four Feynman graphs for
$s$-channel $Z^0$ exchange, shown in \fig{fig:ee2neutneut},
and four for $t$/$u$-channel selectron exchange, shown
in \fig{fig:ee2neutneut2}.
The momenta and polarizations are as labeled in the graphs.
We denote the neutralino masses as
$m_{{\widetilde N}_{i}},m_{{\widetilde N}_{j}}$ and the
selectron masses as $m_{\tilde e_L}$ and $m_{\tilde e_R}$. The
electron mass will again be neglected. The kinematic variables are
then given by
\beqa
s &=&
\BDpos 2p_1\newcdot p_2 = m_{\Ni}^2 + m_{\Nj}^2 \BDplus 2k_i\newcdot k_j ,
\\
t &=& m_{\Ni}^2 \BDminus  2p_1\newcdot k_i =
    m_{\Nj}^2 \BDminus  2p_2\newcdot k_j ,
\\
u &=& m_{\Ni}^2 \BDminus  2p_2\newcdot k_i =
      m_{\Nj}^2 \BDminus  2p_1\newcdot k_j .
\eeqa
\begin{figure}[!t]
\centerline{
\begin{picture}(450,230)(-210,30)
\ArrowLine(-200,225)(-152,197.5)
\ArrowLine(-152,197.5)(-200,170)
\Photon(-152,197.5)(-98,197.5)46
\ArrowLine(-98,197.5)(-50,225)
\ArrowLine(-50,170)(-98,197.5)
\put(-205,230){$e\, (p_1,\lam_1)$}
\put(-205,161){${e}^\dagger\, (p_2,\lam_2)$}
\put(-57,230){$\chi^0_i \, (k_i,\lam_i)$}
\put(-55,157){${\chi^{0\,\dagger}_j} \,(k_j,\lam_j)$}
\put(-129,210){$Z^0$}
\ArrowLine(98,197.5)(50,225)
\ArrowLine(50,170)(98,197.5)
\Photon(98,197.5)(152,197.5)46
\ArrowLine(152,197.5)(200,225)
\ArrowLine(200,170)(152,197.5)
\put(45,230){${\bar e}^\dagger\,(p_1,\lam_1)$}
\put(45,159){$\bar e\,(p_2,\lam_2)$}
\put(193,230){$\chi^0_i\,(k_i,\lam_i)$}
\put(195,157){${\chi^{0\,\dagger}_j}\,(k_j,\lam_j)$}
\put(121,210){$Z^0$}
\ArrowLine(-200,115)(-152,87.5)
\ArrowLine(-152,87.5)(-200,60)
\Photon(-152,87.5)(-98,87.5)46
\ArrowLine(-50,115)(-98,87.5)
\ArrowLine(-98,87.5)(-50,60)
\put(-205,120){$e\, (p_1,\lam_1)$}
\put(-205,51){${e}^\dagger\, (p_2,\lam_2)$}
\put(-57,120){${\chi^{0\,\dagger}_i}\, (k_i,\lam_i)$}
\put(-55,48){$\chi^0_j\, (k_j,\lam_j)$}
\put(-129,100){$Z^0$}
\ArrowLine(98,87.5)(50,115)
\ArrowLine(50,60)(98,87.5)
\Photon(98,87.5)(152,87.5)46
\ArrowLine(200,115)(152,87.5)
\ArrowLine(152,87.5)(200,60)
\put(45,120){${\bar e}^\dagger\,(p_1,\lam_1)$}
\put(45,49){$\bar e\,(p_2,\lam_2)$}
\put(193,120){${\chi^{0\,\dagger}_i}\,(k_i,\lam_i)$}
\put(195,48){$\chi^0_j\,(k_j,\lam_j)$}
\put(121,100){$Z^0$}
\end{picture}
}
\caption{\label{fig:ee2neutneut} The four Feynman diagrams for $e^-e^+
\ra \Ni\Nj$ via $s$-channel $Z^0$ exchange.}
\end{figure}
\begin{figure}[tbp]
\centerline{
\begin{picture}(450,230)(-210,30)
\ArrowLine(-200,225)(-125,225)
\ArrowLine(-125,170)(-200,170)
\DashArrowLine(-125,225)(-125,170)5
\ArrowLine(-50,225)(-125,225)
\ArrowLine(-125,170)(-50,170)
\put(-205,230){$e\, (p_1,\lam_1)$}
\put(-205,161){${e}^\dagger\, (p_2,\lam_2)$}
\put(-57,231){${\chi^{0\,\dagger}_i}\, (k_i,\lam_i)$}
\put(-55,158){${\chi^0_j}\, (k_j,\lam_j)$}
\put(-145,197.5){$\stilde e_L$}
\ArrowLine(125,225)(50,225)
\ArrowLine(50,170)(125,170)
\DashArrowLine(125,170)(125,225)5
\ArrowLine(125,225)(200,225)
\ArrowLine(200,170)(125,170)
\put(45,230){${\bar e}^\dagger\,(p_1,\lam_1)$}
\put(45,159){$\bar e\,(p_2,\lam_2)$}
\put(193,230){$\chi^0_i\,(k_i,\lam_i)$}
\put(195,157){${\chi^{0\,\dagger}_j}\,(k_j,\lam_j)$}
\put(105,197.5){$\stilde e_R^{\,*}$}
\ArrowLine(-200,115)(-125,115)
\ArrowLine(-125,60)(-200,60)
\DashArrowLine(-125,115)(-125,60)5
\Line(-125,115)(-91.25,90.25)
\ArrowLine(-50,60)(-83.75,84.75)
\Line(-87.5,87.5)(-125,60)
\ArrowLine(-87.5,87.5)(-50,115)
\put(-205,120){$e\,(p_1,\lam_1)$}
\put(-205,51){${e}^\dagger\,(p_2,\lam_2)$}
\put(-57,120){$\chi^0_i\,(k_i,\lam_i)$}
\put(-55,46){${\chi^{0\,\dagger}_j}\,(k_j,\lam_j)$}
\put(-145,87.5){$\stilde e_L$}
\ArrowLine(125,115)(50,115)
\ArrowLine(50,60)(125,60)
\DashArrowLine(125,60)(125,115)5
\ArrowLine(166.25,84.75)(200,60)
\Line(125,115)(158.75,90.25)
\ArrowLine(200,115)(162.5,87.5)
\Line(125,60)(162.5,87.5)
\put(45,120){${\bar e}^\dagger\,(p_1,\lam_1)$}
\put(45,49){$\bar e\,(p_2,\lam_2)$}
\put(193,121){${\chi^{0\,\dagger}_i}\,(k_i,\lam_i)$}
\put(195,48){$\chi^0_j\,(k_j,\lam_j)$}
\put(105,87.5){$\stilde e_R^{\,*}$}
\end{picture}
}
\caption{\label{fig:ee2neutneut2} The four Feynman diagrams for $e^-e^+
\ra\Ni\Nj$ via $t$/$u$-channel selectron exchange.}
\end{figure}

By applying the Feynman rules of \figs{SMintvertices}{nnboson},
we obtain for the sum of the $s$-channel diagrams
in \fig{fig:ee2neutneut} [cf.~footnote~\ref{fnqq}],
\beqa
i{\cal M}_Z &=&
\frac{\BDneg i \metric^{\mu\nu}}{D_Z}
\biggl [\BDpos \frac{ig(s_W^2 -\half)}{c_W} x_1 \sigma_\mu  y^\dagger_2
\BDplus  \frac{igs_W^2}{c_W}  y^\dagger_1 \sigmabar_\mu x_2
\biggr ]
\biggl [
\BDpos \frac{ig}{c_W} O_{ij}^{\prime\prime L} x^\dagger_i \sigmabar_\nu y_j
\BDminus  \frac{ig}{c_W} O_{ji}^{\prime\prime L} y_i \sigma_\nu  x^\dagger_j
\biggr ]\,,
\phantom{xxxxx}
\label{eq:eeNNs}
\eeqa
where $O_{ij}^{\prime\prime}$ is given in eq.~(\ref{eq:defOLpp}),
and $D_Z\equiv s-m_Z^2+i\Gamma_Zm_Z$.  The
fermion spinors are denoted by $x_1 \equiv x(\boldsymbol{\vec
  p}_1,\lam_1)$, $ y^\dagger_2 \equiv  y^\dagger(\boldsymbol{\vec p}_2,\lam_2)$,
$ x^\dagger_i \equiv  x^\dagger(\boldsymbol{\vec k}_i,\lam_i)$, $y_j \equiv
y(\boldsymbol{\vec k}_j,\lam_j)$, etc.  Note that we have combined the
matrix elements of the four diagrams by factorizing with respect to
the common boson propagator.  For the four $t$/$u$-channel
diagrams, we obtain, by applying the rules of \fig{nqsq}:
\beqa
i {\cal M}^{(t)}_{\tilde e_L} &=&
(-1)
\biggl [ \frac{i}{t - m_{\tilde e_L}^2} \Bigr ]
\biggl [
\frac{ig}{\sqrt{2}} \Bigl (N_{i2}^* + \frac{s_W}{c_W} N_{i1}^* \Bigr )
\biggr ]
\biggl [
\frac{ig}{\sqrt{2}} \Bigl (N_{j2} + \frac{s_W}{c_W} N_{j1} \Bigr )
\biggr ]
x_1 y_i  y^\dagger_2  x^\dagger_j
,
\label{eq:eeNNtL}
\\
i {\cal M}^{(u)}_{\tilde e_L} &=&
\biggl [ \frac{i}{u - m_{\tilde e_L}^2} \Bigr ]
\biggl [
\frac{ig}{\sqrt{2}} \Bigl (N_{j2}^* + \frac{s_W}{c_W} N_{j1}^* \Bigr )
\biggr ]
\biggl [
\frac{ig}{\sqrt{2}} \Bigl (N_{i2} + \frac{s_W}{c_W} N_{i1} \Bigr )
\biggr ]
x_1 y_j  y^\dagger_2  x^\dagger_i
,
\label{eq:eeNNuL}
\\
i {\cal M}^{(t)}_{\tilde e_R} &=&
(-1)
\biggl [ \frac{i}{t - m_{\tilde e_R}^2} \Bigr ]
\Bigl (
-i\sqrt{2} g \frac{s_W}{c_W} N_{i1}
\Bigr )
\Bigl (
-i\sqrt{2} g \frac{s_W}{c_W} N_{j1}^*
\Bigr )
 y^\dagger_1  x^\dagger_i x_2 y_j
,
\label{eq:eeNNtR}
\\
i {\cal M}^{(u)}_{\tilde e_R} &=&
\biggl [ \frac{i}{u - m_{\tilde e_R}^2} \Bigr ]
\Bigl (
-i\sqrt{2} g \frac{s_W}{c_W} N_{j1}
\Bigr )
\Bigl (
-i\sqrt{2} g \frac{s_W}{c_W} N_{i1}^*
\Bigr )
 y^\dagger_1  x^\dagger_j x_2 y_i .
\label{eq:eeNNuR}
\eeqa
The first factors of $(-1)$ in each of eqs.~(\ref{eq:eeNNtL})
and (\ref{eq:eeNNtR}) are present because the order of the
spinors in each case is an odd permutation of the ordering $(1,2,i,j)$
established by the $s$-channel contribution.
The other contributions have spinors in an even permutation of
that ordering.

The $s$-channel diagram contribution of eq.~(\ref{eq:eeNNs})
can be profitably rearranged
using the Fierz identities of
\eqs{twocompfierza}{twocompfierzb}.
Then, combining the result with the $t$/$u$-channel and $s$-channel
contributions, we have for the total:
\beq
{\cal M} =
c_1 x_1 y_j  y^\dagger_2  x^\dagger_i
+ c_2 x_1 y_i  y^\dagger_2  x^\dagger_j
+ c_3  y^\dagger_1  x^\dagger_i x_2 y_j
+ c_4  y^\dagger_1  x^\dagger_j x_2 y_i ,
\label{eq:eeNNniceform}
\eeq
where
\beqa
c_1 &=& \frac{g^2}{c_W^2} \left [
(1 - 2 s_W^2) O_{ij}^{\prime\prime L}/D_Z
- \half(c_W N_{i2} + s_W N_{i1}) (c_W N_{j2}^* + s_W N_{j1}^*)/
(u - m^2_{\tilde e_L}) \right ],\phantom{xxxx}
\\
c_2 &=& \frac{g^2}{c_W^2} \left [
(2 s_W^2 - 1) O_{ji}^{\prime\prime L}/D_Z
+ \half(c_W N_{i2}^* + s_W N_{i1}^*) (c_W N_{j2} + s_W N_{j1})/
(t - m^2_{\tilde e_L}) \right ],\phantom{xxxxx.}
\\
c_3 &=&
\frac{2 g^2 s_W^2}{c_W^2} \left [
-O_{ij}^{\prime\prime L}/D_Z
+  N_{i1} N_{j1}^*/
(t - m^2_{\tilde e_R}) \right ]
,
\\
c_4 &=&
\frac{2 g^2 s_W^2}{c_W^2} \left [
O_{ji}^{\prime\prime L}/D_Z
-  N_{i1}^* N_{j1}/
(u - m^2_{\tilde e_R}) \right ] .
\eeqa
Squaring the amplitude
and averaging over electron and positron spins, only terms involving
$x_1  x^\dagger_1$ or $y_1  y^\dagger_1$, and
$x_2  x^\dagger_2$ or $y_2  y^\dagger_2$ survive in
the massless electron limit.  Thus,
\beqa
\sum_{\lam_1,\lam_2} |{\cal M}|^2 &=&
\sum_{\lam_1,\lam_2} \biggl (
|c_1|^2  y^\dagger_j  x^\dagger_1 x_1 y_j x_i y_2  y^\dagger_2  x^\dagger_i
+ |c_2|^2  y^\dagger_i  x^\dagger_1 x_1 y_i x_j y_2  y^\dagger_2  x^\dagger_j
\nonumber \\ &&
\qquad
+ |c_3|^2 x_i y_1  y^\dagger_1  x^\dagger_i  y^\dagger_j  x^\dagger_2 x_2 y_j
+ |c_4|^2 x_j y_1  y^\dagger_1  x^\dagger_j  y^\dagger_i  x^\dagger_2 x_2 y_i
\nonumber \\ &&
\qquad
+ 2 {\rm Re}\bigl [c_1 c_2^*  y^\dagger_i  x^\dagger_1 x_1 y_j x_j y_2  y^\dagger_2  x^\dagger_i
\bigr ]
+ 2 {\rm Re}\bigl [c_3 c_4^* x_j y_1  y^\dagger_1  x^\dagger_i  y^\dagger_i  x^\dagger_2 x_2 y_j \bigr ]
\biggr )\phantom{xxxx} \nonumber
\\
&=&
|c_1|^2  y^\dagger_j p_1\newcdot \sigmabar y_j\, x_i p_2\newcdot\sigma  x^\dagger_i
+ |c_2|^2  y^\dagger_i p_1\newcdot \sigmabar y_i\, x_j p_2\newcdot \sigma
 x^\dagger_j
\nonumber \\ &&
\quad
+ |c_3|^2 x_i p_1 \newcdot \sigma  x^\dagger_i\,
   y^\dagger_j p_2 \newcdot \sigmabar y_j
+ |c_4|^2 x_j p_1 \newcdot \sigma  x^\dagger_j\,
   y^\dagger_i p_2 \newcdot \sigmabar y_i
\nonumber \\ &&
\quad
+ 2 {\rm Re}\bigl [c_1 c_2^*  y^\dagger_i p_1 \newcdot \sigmabar y_j\,
                        x_j p_2 \newcdot \sigma  x^\dagger_i \bigr ]
+ 2 {\rm Re}\bigl [c_3 c_4^* x_j p_1 \newcdot \sigma  x^\dagger_i \,
                         y^\dagger_i p_2 \newcdot \sigmabar y_j\bigr ]\,,
\eeqa
after employing the results of \eqst{xxdagsummed}{ydagxdagsummed}.

We now perform the remaining spin sums using
\eqst{xxdagsummed}{ydagxdagsummed} again, obtaining:
\beqa
\sum_{\lam_1,\lam_2,\lam_i,\lam_j} |{\cal M}|^2 &=&
|c_1|^2 {\rm Tr}[p_1 \newcdot \sigmabar k_j \newcdot \sigma]
        {\rm Tr}[p_2 \newcdot \sigma k_i \newcdot \sigmabar]
+|c_2|^2 {\rm Tr}[p_1 \newcdot \sigmabar k_i \newcdot \sigma]
        {\rm Tr}[p_2 \newcdot \sigma k_j \newcdot \sigmabar]
\nonumber \\[-6pt] &&
+|c_3|^2 {\rm Tr}[p_1 \newcdot \sigma k_i \newcdot \sigmabar]
        {\rm Tr}[p_2 \newcdot \sigmabar k_j \newcdot \sigma]
+|c_4|^2 {\rm Tr}[p_1 \newcdot \sigma k_j \newcdot \sigmabar]
        {\rm Tr}[p_2 \newcdot \sigmabar k_i \newcdot \sigma]
\nonumber \\ &&
+ 2 {\rm Re}[c_1 c_2^*] m_{\Ni} m_{\Nj}
    {\rm Tr}[p_2 \newcdot \sigma p_1 \newcdot \sigmabar]
+ 2 {\rm Re}[c_3 c_4^*] m_{\Ni} m_{\Nj}
    {\rm Tr}[p_1 \newcdot \sigma p_2 \newcdot \sigmabar] .
    \phantom{xxxx}
\eeqa
Applying the trace identity of eq.~(\ref{trssbar}) to this yields
\beqa
\sum_{\rm spins} |{\cal M}|^2 &=&
(|c_1|^2 + |c_4|^2) 4 p_1 \newcdot k_j\> p_2 \newcdot k_i
+ (|c_2|^2 + |c_3|^2) 4 p_1 \newcdot k_i\> p_2 \newcdot k_j
\nonumber \\[-6pt] &&
\quad
\BDplus 4 {\rm Re}[c_1 c_2^* + c_3 c_4^*] m_{\Ni} m_{\Nj}
p_1 \newcdot p_2
\nonumber \\
&=&
(|c_1|^2 + |c_4|^2) (u - m_{\Ni}^2)(u - m_{\Nj}^2)
+ (|c_2|^2 + |c_3|^2) (t - m_{\Ni}^2)(t - m_{\Nj}^2)\phantom{xxxx}
\nonumber \\ &&
\quad
+ 2 {\rm Re}[c_1 c_2^* + c_3 c_4^*] m_{\Ni} m_{\Nj} s .
\label{eq:eeNNnicerform}
\eeqa
The differential cross-section then follows:
\beq
\frac{d\sigma}{dt} = \frac{1}{16\pi s^2}
\left (\frac{1}{4} \sum_{\rm spins} |{\cal M}|^2 \right ) .
\eeq
This agrees with the first complete calculation presented in
\Ref{eeNN}. For the case of pure photino pair production, i.e.
$N_{i1}\ra c_W\delta_{i1}$ and $N_{i2} \ra s_W\delta_{i1}$ and for
degenerate selectron masses this also agrees with eq.~(E9) of the
erratum of \cite{HaberKane}. Other earlier calculations with some
simplifications are given in refs.~\cite{eeNNprime,Dawson:1983fw}.

Defining $\cos\theta={\boldsymbol{\hat p_1}}\newcdot{\boldsymbol{\hat
k_i}}$ (the cosine of the angle between the initial state electron and
one of the neutralinos in the center-of-momentum frame), the
Mandelstam variables $t,u$ are given by
\beqa
t &=& \frac{1}{2} \left
[m_{\Ni}^2 + m_{\Nj}^2 -s + \lam^{1/2}(s,m_{\Ni}^2,m_{\Nj}^2) \cos\theta
\right ], \label{eq:t}
\\
u &=& \frac{1}{2}
\left
[m_{\Ni}^2 + m_{\Nj}^2 -s - \lam^{1/2}(s,m_{\Ni}^2,m_{\Nj}^2) \cos\theta
\right ]\,,
\eeqa
where the triangle function $\lam^{1/2}$ is defined in
\eq{eq:deftrianglefunction}.
Taking into account the identical fermions in the final state when $i=j$,
the total cross-section is
\beq
\sigma = \frac{1}{1+\delta_{ij}} \int_{t_-}^{t_+} \frac{d\sigma}{dt} dt\,,
\eeq
where $t_-$ and $t_+$ are obtained by inserting $\cos\theta = \mp1$ in
\eq{eq:t}, respectively.

\subsection{\texorpdfstring{$\N1\N1 \ra f\bar f$}{N\texttilde\textoneinferior N\texttilde\textoneinferior \textrightarrow f{f\textoverline}}}
\label{chichiee}
\setcounter{equation}{0}
\setcounter{figure}{0}
\setcounter{table}{0}

In this section, we compute the annihilation
rate for $\N1\N1 \ra f\bar f$, where $f$ is any
kinematically allowed quark, charged lepton or
neutrino.
The case of $f=e^-$ is the reversed reaction of the
process examined in \sec{eechichi} (with $i=j=1$).
In R-parity-conserving supersymmetric models in which $\N1$ is the
lightest supersymmetric particle (and hence is stable), the $\N1\N1$
annihilation process is relevant for the computation of the neutralino
relic density \cite{Goldberg:1983nd}.
In particular, $\N1\N1 \ra f\bar f$ can be an important contribution to cold
dark matter annihilation \cite{Goldberg:1983nd,Ellisetal,Griest,Griestreview}.
Neutralino dark matter is typically heavier than about 6 GeV
\cite{Bottino:2003iu}; for lighter neutralinos see
\Ref{Dreiner:2009ic}.

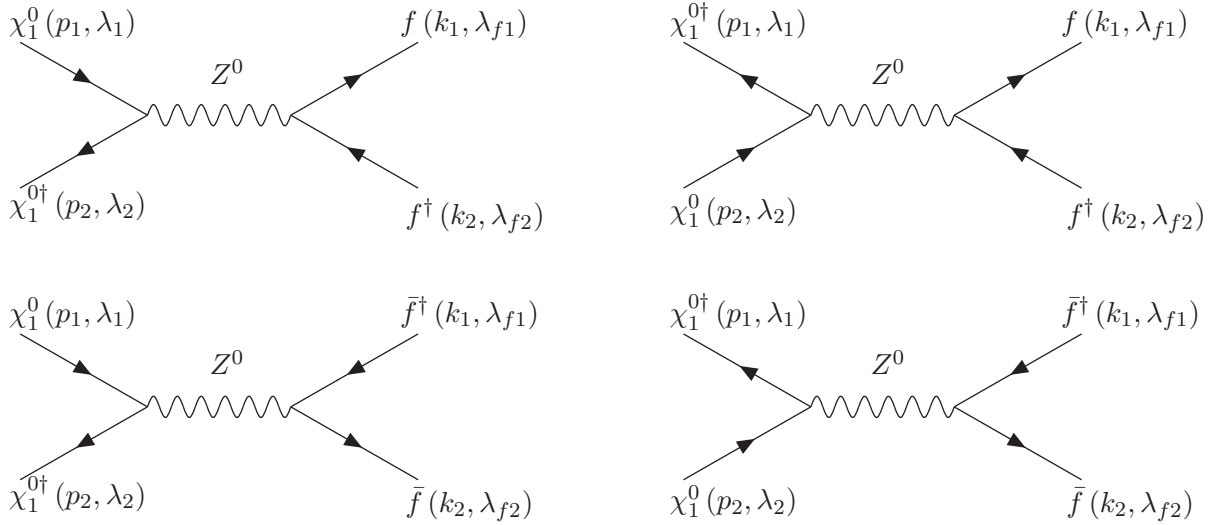
\begin{figure}[!t]
\centerline{
\begin{picture}(450,230)(-210,30)
\ArrowLine(-200,225)(-152,197.5)
\ArrowLine(-152,197.5)(-200,170)
\Photon(-152,197.5)(-98,197.5)46
\ArrowLine(-98,197.5)(-50,225)
\ArrowLine(-50,170)(-98,197.5)
\put(-205,230){$\chi^0_1\, (p_1,\lam_1)$}
\put(-205,161){${\chi}^{0\dagger}_1\, (p_2,\lam_2)$}
\put(-57,230){$f \, (k_1,\lam_{f1})$}
\put(-55,157){${f^{\dagger}} \,(k_2,\lam_{f2})$}
\put(-129,210){$Z^0$}
\ArrowLine(98,197.5)(50,225)
\ArrowLine(50,170)(98,197.5)
\Photon(98,197.5)(152,197.5)46
\ArrowLine(152,197.5)(200,225)
\ArrowLine(200,170)(152,197.5)
\put(45,230){${\chi}^{0\dagger}_1\,(p_1,\lam_1)$}
\put(45,159){$\chi^0_1\,(p_2,\lam_2)$}
\put(193,230){$f\,(k_1,\lam_{f1})$}
\put(195,157){${f^{\dagger}}\,(k_2,\lam_{f2})$}
\put(121,210){$Z^0$}
\ArrowLine(-200,115)(-152,87.5)
\ArrowLine(-152,87.5)(-200,60)
\Photon(-152,87.5)(-98,87.5)46
\ArrowLine(-50,115)(-98,87.5)
\ArrowLine(-98,87.5)(-50,60)
\put(-205,120){$\chi^0_1 \, (p_1,\lam_1)$}
\put(-205,51){${\chi}^{0\dagger}_1\, (p_2,\lam_2)$}
\put(-57,120){${\bar f^\dagger}\, (k_1,\lam_{f1})$}
\put(-55,48){$\bar f\, (k_2,\lam_{f2})$}
\put(-129,100){$Z^0$}
\ArrowLine(98,87.5)(50,115)
\ArrowLine(50,60)(98,87.5)
\Photon(98,87.5)(152,87.5)46
\ArrowLine(200,115)(152,87.5)
\ArrowLine(152,87.5)(200,60)
\put(45,120){${\chi}^{0\dagger}_1\,(p_1,\lam_1)$}
\put(45,49){$\chi^0_1 \,(p_2,\lam_2)$}
\put(193,120){${\bar f^{\dagger}}\,(k_1,\lam_{f1})$}
\put(195,48){$\bar f\,(k_2,\lam_{f2})$}
\put(121,100){$Z^0$}
\end{picture}
}
\caption{\label{fig:neutneut2ee} The four Feynman diagrams for
$\N1\N1\ra f\bar f$ via $s$--channel $Z^0$ exchange, where $f$ is
a quark or lepton.}
\end{figure}

In the computation of the relic density, one computes $v_{\rm rel}\sigma_{\rm ann}$,
where $\sigma_{\rm ann}$ is the $\N1\N1$ annihilation cross-section
and $v_{\rm rel}$ is the relative velocity of the two neutralinos in the
center-of-momentum frame.  The square of the relative velocity is taken to
be its thermal average, $v_{\rm rel}^2\simeq 6k_B T/m_{\N1}$ \cite{Goldberg:1983nd},
which is typically non-relativistic when the temperature
is of order the freeze-out temperature \cite{Griest} (where the neutralino falls out
of thermal equilibrium).  Hence, it is sufficient to compute the
annihilation cross-section for $\N1\N1 \ra f\bar f$ in the non-relativistic limit.

As in \sec{eechichi}, there
are four Feynman graphs for $s$-channel $Z^0$ exchange, shown in
\fig{fig:neutneut2ee}.  In addition, there are $s$-channel neutral Higgs exchange
graphs, shown in \fig{fig:neutneuthiggs}, that yield contributions to the
annihilation amplitude proportional to the fermion mass, $m_f$.\footnote{In regions of parameter
space where $m_{\N1}\simeq\half m_Z$ or $m_{\N1}\simeq\half m_{\phi^0}$
(where $\phi^0=h^0$, $H^0$ or $A^0$), the resonant $2\to 1$ annihilation
$\N1\N1\to Z^0$ or $\N1\N1\to\phi^0$ dominates
the $2\to 2$ annihilation processes considered here.}
Likewise, as in \sec{eechichi}, there are
four Feynman graphs for $t$/$u$-channel $\widetilde f_L$ and $\widetilde f_R$ exchange,
shown in \fig{fig:neutneut2ee2}.  However, because we do not set $m_f$ to
zero, four additional $t$/$u$-channel graphs contribute,
shown in \fig{fig:neutneut2ee3},
that are sensitive to the higgsino components of the neutralino.

\begin{figure}[tb]
\centerline{
\begin{picture}(450,230)(-210,30)
\ArrowLine(-200,225)(-152,197.5)
\ArrowLine(-200,170)(-152,197.5)
\DashLine(-152,197.5)(-98,197.5)5
\ArrowLine(-98,197.5)(-50,225)
\ArrowLine(-98,197.5)(-50,170)
\put(-205,230){$\chi^0_1\, (p_1,\lam_1)$}
\put(-205,161){${\chi}^{0}_1\, (p_2,\lam_2)$}
\put(-57,230){$f \, (k_1,\lam_{f1})$}
\put(-55,157){${\bar{f}} \,(k_2,\lam_{f2})$}
\put(-129,210){$\phi^0$}
\ArrowLine(50,225)(98,197.5)
\ArrowLine(50,170)(98,197.5)
\DashLine(98,197.5)(152,197.5)5
\ArrowLine(200,225)(152,197.5)
\ArrowLine(200,170)(152,197.5)
\put(45,230){${\chi}^{0}_1\,(p_1,\lam_1)$}
\put(45,159){$\chi^0_1\,(p_2,\lam_2)$}
\put(193,230){$\bar{f}^\dagger\,(k_1,\lam_{f1})$}
\put(195,157){${f^{\dagger}}\,(k_2,\lam_{f2})$}
\put(121,210){$\phi^0$}
\ArrowLine(-152,87.5)(-200,115)
\ArrowLine(-152,87.5)(-200,60)
\DashLine(-152,87.5)(-98,87.5)5
\ArrowLine(-98,87.5)(-50,115)
\ArrowLine(-98,87.5)(-50,60)
\put(-205,120){$\chi^{0\dagger}_1 \, (p_1,\lam_1)$}
\put(-205,51){${\chi}^{0\dagger}_1\, (p_2,\lam_2)$}
\put(-57,120){${f}\, (k_1,\lam_{f1})$}
\put(-55,48){$\bar f\, (k_2,\lam_{f2})$}
\put(-129,100){$\phi^0$}
\ArrowLine(98,87.5)(50,115)
\ArrowLine(98,87.5)(50,60)
\DashLine(98,87.5)(152,87.5)5
\ArrowLine(200,115)(152,87.5)
\ArrowLine(200,60)(152,87.5)
\put(45,120){${\chi}^{0\dagger}_1\,(p_1,\lam_1)$}
\put(45,49){$\chi^{0\dagger}_1 \,(p_2,\lam_2)$}
\put(193,120){${\bar f^{\dagger}}\,(k_1,\lam_{f1})$}
\put(195,48){$f^\dagger\,(k_2,\lam_{f2})$}
\put(121,100){$\phi^0$}
\end{picture}
}
\caption{\label{fig:neutneuthiggs} Feynman diagrams for
$\N1\N1\ra f\bar f$ via $s$--channel Higgs exchange.  There are
four diagrams for each possible neutral Higgs state $\phi^0= h^0$,
$H^0$ and $A^0$.}
\end{figure}
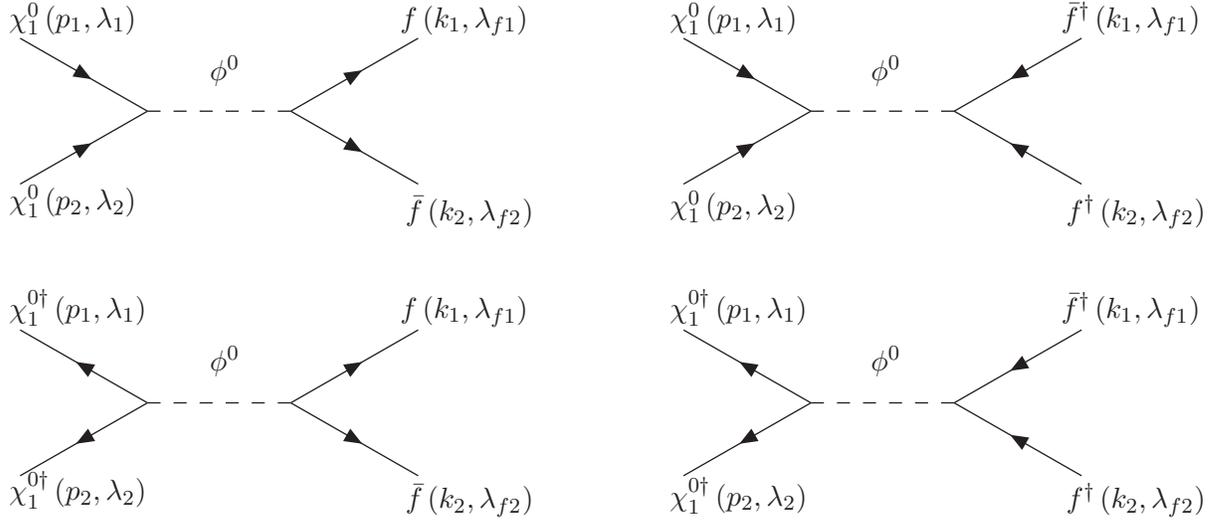

\begin{figure}[tbp]
\centerline{
\begin{picture}(450,230)(-210,30)
%
%
\ArrowLine(-125,225)(-200,225)
\ArrowLine(-200,170)(-125,170)
\DashArrowLine(-125,170)(-125,225)5
\ArrowLine(-125,225)(-50,225)
\ArrowLine(-50,170)(-125,170)
\put(-205,230){$\chi^{0\dagger}_1\, (p_1,\lam_1)$}
\put(-205,159){${\chi}^{0}_1\, (p_2,\lam_2)$}
\put(-57,231){${f}\, (k_1,\lam_{f1})$}
\put(-55,158){${f}^\dagger\, (k_2,\lam_{f2})$}
\put(-145,197.5){$\stilde f_L$}
%
%
\ArrowLine(50,225)(125,225)
\ArrowLine(125,170)(50,170)
\DashArrowLine(125,225)(125,170)5
\ArrowLine(200,225)(125,225)
\ArrowLine(125,170)(200,170)
\put(45,230){${\chi}^{0}_1\,(p_1,\lam_1)$}
\put(45,159){$\chi^{0\dagger}_1\,(p_2,\lam_2)$}
\put(193,230){$\bar f^\dagger\,(k_1,\lam_{f1})$}
\put(195,157){${\bar f}\,(k_2,\lam_{f2})$}
\put(105,197.5){$\stilde f_R^{\,*}$}
%
%
\ArrowLine(-200,115)(-125,115)
\ArrowLine(-125,60)(-200,60)
\DashArrowLine(-125,115)(-125,60)5
\Line(-125,115)(-91.25,90.25)
\ArrowLine(-50,60)(-83.75,84.75)
\Line(-87.5,87.5)(-125,60)
\ArrowLine(-87.5,87.5)(-50,115)
\put(-205,120){$\chi^{0}_1\,(p_1,\lam_1)$}
\put(-205,48){${\chi}^{0\dagger}_1\,(p_2,\lam_2)$}
\put(-57,120){$f\,(k_1,\lam_{f1})$}
\put(-55,46){$f^\dagger\,(k_2,\lam_{f2})$}
\put(-145,87.5){$\stilde f_L$}
%
%
\ArrowLine(125,115)(50,115)
\ArrowLine(50,60)(125,60)
\DashArrowLine(125,60)(125,115)5
\ArrowLine(166.25,84.75)(200,60)
\Line(125,115)(158.75,90.25)
\ArrowLine(200,115)(162.5,87.5)
\Line(125,60)(162.5,87.5)
\put(45,120){${\chi}^{0\dagger}_1\,(p_1,\lam_1)$}
\put(45,49){$\chi^0_1\,(p_2,\lam_2)$}
\put(193,121){${\bar f}^\dagger\,(k_1,\lam_{f1})$}
\put(195,48){$\bar f\,(k_2,\lam_{f2})$}
\put(105,87.5){$\stilde f_R^{\,*}$}
\end{picture}
}
\caption{\label{fig:neutneut2ee2} The four Feynman diagrams for
$\Ni\Nj\ra f\bar f$ via $t$/$u$--channel $\widetilde f_L$ and
$\widetilde f_R$ exchange,
where $\widetilde f_L$ and $\widetilde f_R$ couple to the gaugino components of the
neutralino.}
\end{figure}
\begin{figure}[h!t]
\centerline{
\begin{picture}(450,230)(-210,30)
%
%
\ArrowLine(-200,225)(-125,225)
\ArrowLine(-125,170)(-200,170)
\DashArrowLine(-125,170)(-125,225)5
\ArrowLine(-50,225)(-125,225)
\ArrowLine(-125,170)(-50,170)
\put(-205,230){$\chi^{0}_1\, (p_1,\lam_1)$}
\put(-205,159){${\chi}^{0\dagger}_1\, (p_2,\lam_2)$}
\put(-57,231){${\bar{f}^\dagger}\, (k_1,\lam_{f1})$}
\put(-55,158){$\bar{f}\, (k_2,\lam_{f2})$}
\put(-145,197.5){$\stilde f_L$}
%
%
\ArrowLine(125,225)(50,225)
\ArrowLine(50,170)(125,170)
\DashArrowLine(125,225)(125,170)5
\ArrowLine(125,225)(200,225)
\ArrowLine(200,170)(125,170)
\put(45,230){${\chi}^{0\dagger}_1\,(p_1,\lam_1)$}
\put(45,159){$\chi^{0}_1\,(p_2,\lam_2)$}
\put(193,230){$f\,(k_1,\lam_{f1})$}
\put(195,157){$f^\dagger\,(k_2,\lam_{f2})$}
\put(105,197.5){$\stilde f_R^{\,*}$}
%
%
\ArrowLine(-125,115)(-200,115)
\ArrowLine(-200,60)(-125,60)
\DashArrowLine(-125,115)(-125,60)5
\Line(-125,115)(-91.25,90.25)
\ArrowLine(-83.75,84.75)(-50,60)
\Line(-87.5,87.5)(-125,60)
\ArrowLine(-50,115)(-87.5,87.5)
\put(-205,120){$\chi^{0\dagger}_1\,(p_1,\lam_1)$}
\put(-205,48){${\chi}^{0}_1\,(p_2,\lam_2)$}
\put(-57,120){$\bar{f}^\dagger\,(k_1,\lam_{f1})$}
\put(-55,46){$\bar{f}\,(k_2,\lam_{f2})$}
\put(-145,87.5){$\stilde f_L$}
%
%
\ArrowLine(50,115)(125,115)
\ArrowLine(125,60)(50,60)
\DashArrowLine(125,60)(125,115)5
\ArrowLine(200,60)(166.25,84.75)
\Line(125,115)(158.75,90.25)
\ArrowLine(162.5,87.5)(200,115)
\Line(125,60)(162.5,87.5)
\put(45,120){${\chi}^{0}_1\,(p_1,\lam_1)$}
\put(45,49){$\chi^{0\dagger}_1\,(p_2,\lam_2)$}
\put(193,121){$f\,(k_1,\lam_{f1})$}
\put(195,48){$f^\dagger\,(k_2,\lam_{f2})$}
\put(105,87.5){$\stilde f_R^{\,*}$}
\end{picture}
}
\caption{\label{fig:neutneut2ee3} The four Feynman diagrams for
$\Ni\Nj\ra f\bar f$ via $t$/$u$--channel $\widetilde f_L$ and
$\widetilde f_R$ exchange,
where $\widetilde f_L$ and $\widetilde f_R$ couple to the higgsino components of the
neutralino.}
\end{figure}

The neutralino and the final state fermion four-momenta and polarizations are as
labeled in the Feynman graphs.
In the center-of-momentum (CM) frame, the four-momenta are
\beq
p_1^\mu=(E\,;\,\boldsymbol{\vec{p}})\,,\qquad p_2^\mu=(E\,;\,-\boldsymbol{\vec{p}})\,,
\qquad k_1^\mu=E(1\,;\,\beta\boldsymbol{\hat{k}})\,,\qquad k_2^\mu=E(1\,;\,-\beta
\boldsymbol{\hat{k}})\,,
\eeq
where
\beq \label{betaE}
\beta\equiv\sqrt{1-\frac{m_f^2}{E^2}}\,.
\eeq
In the non-relativistic limit where $|\boldsymbol{\vec{p}}|\ll m_{\N1}$,
\beq \label{Enonrel}
E\simeq m_{\N1}+\displaystyle\frac{|\boldsymbol{\vec{p}}|^2}{2m_{\N1}}\,,
\eeq
and the kinematic invariants are given by
\beqa
s &=& \BDpos (p_1+p_2)^2 = 4E^2=4m_{\N1}^2+4|\boldsymbol{\vec{p}}|^2\,,
\\
t &=& \BDpos (p_1-k_1)^2=m_{\N1}^2+m_f^2 \BDminus 2p_1\newcdot k_1
\simeq -m_{\N1}^2+m_f^2+2\beta m_{\N1}|\boldsymbol{\vec{p}}|\cos\theta-2|\boldsymbol{\vec{p}}|^2\,,
\\
u &=& \BDpos (p_1-k_2)^2=m_{\N1}^2+m_f^2 \BDminus 2p_1\newcdot k_2
\simeq -m_{\N1}^2+m_f^2-2\beta m_{\N1}|\boldsymbol{\vec{p}}|\cos\theta-2|\boldsymbol{\vec{p}}|^2\,,
\eeqa
where $\theta$ is the CM scattering angle.
Subsequently, we shall neglect the subdominant
$\mathcal{O}(|\boldsymbol{\vec{p}}|)$ terms in the $t$ and $u$-channel propagator denominators by
setting $t\simeq u\simeq -m_{\N1}^2+m_f^2$.

By applying the Feynman rules of \figs{SMintvertices}{nnboson}, and using the
unitary gauge for the $Z$-boson propagator,
we obtain for the sum of the $s$-channel $Z$-exchange diagrams
of \fig{fig:neutneut2ee},
\beq
i\mathcal{M}_Z =
\frac{i \left(\BDneg \metric^{\mu\nu} + Q^\mu Q^\nu/m_Z^2\right)}{D_Z}
\left(\frac{-ig}{c_W}\right)^2 O_{11}^{\prime\prime L}
\biggl [
x_1 \sigma_\mu y_2^\dagger -
 y_1^\dagger \sigmabar_\mu  x_2
\biggr ]
\biggl [(T_3^f-s_W^2Q_f) x_{f1}^\dagger \sigmabar_\nu  y\ls{f2}
-  s_W^2 Q_f  y\ls{f1} \sigma_\nu x_{f2}^\dagger
\biggr ]\,,
\label{eq:NNees}
\eeq
where $O_{11}^{\prime\prime\,L}$ is given in eq.~(\ref{eq:defOLpp}),
$D_Z\equiv s-m_Z^2+i\Gamma_Zm_Z$, and $Q\equiv p_1+p_2=k_1+k_2$.
The spinor wave functions are denoted by
$x_1 \equiv x(\boldsymbol{\vec p}_1,\lam_1)$, $ y^\dagger_2 \equiv
y^\dagger(\boldsymbol{\vec p}_2,\lam_2)$, $ x^\dagger_{f1} \equiv
x^\dagger(\boldsymbol{\vec k}_1,\lam_{f1})$, $y\ls{f2} \equiv
y(\boldsymbol{\vec k}_2,\lam_{f2})$, etc.  In obtaining
\eq{eq:NNees}, we have combined
the matrix elements of the four diagrams by factorizing with respect
to the common $Z$-boson propagator.   Note that all four terms
in \eq{eq:NNees} have the same order of spinor wave functions (1,2,$f1$,$f2$).
Thus, no additional relative signs arise (beyond the sign associated with the choice
of the $\sigma$ or $\sigmabar$ version of the vertex Feynman rules).
One can simplify the terms that originate from the $Q^\mu Q^\nu$ part of the
$Z$-boson propagator by writing $Q^\mu=(p_1+p_2)^\mu$ and
$Q^\nu=(k_1+k_2)^\nu$.  Contracting the $\mu$ and $\nu$ indices with the
help of \eqst{onshellone}{onshellfour} yields:
\beqa
\hspace{-0.5in}
(p_1+p_2)^\mu \left[
x_1 \sigma_\mu y_2^\dagger -
 y_1^\dagger \sigmabar_\mu  x_2 \right]&=&
2m_{\N1}\left(\BDpos x_1 x_2 \BDminus y_1^\dagger y_2^\dagger\right)\,,\\
\hspace{-0.5in}
(k_1+k_2)^\nu\left[(T_3^f-s_W^2 Q_f) x_{f1}^\dagger \sigmabar_\nu  y\ls{f2}
-  s_W^2 Q_f y\ls{f1} \sigma_\nu x_{f2}^\dagger\right]
&=&
T_3^f m_f\left(\BDpos y\ls{f1}y\ls{f2}
\BDminus x^\dagger_{f1}x^\dagger_{f2}\right)\,.
\eeqa
Hence, we shall write
\beq
\mathcal{M}_Z\equiv \mathcal{M}^{(1)}_Z+\mathcal{M}^{(2)}_Z\,,
\eeq
where
\beqa
i\mathcal{M}_Z^{(1)}&=&\frac{\BDneg i \metric^{\mu\nu}}{D_Z}\left(\frac{-ig}{c_W}\right)^2
O_{11}^{\prime\prime L}
\biggl [
x_1 \sigma_\mu y_2^\dagger -
 y_1^\dagger \sigmabar_\mu  x_2
\biggr ]
\biggl [(T_3^f-s_W^2 Q_f) x_{f1}^\dagger \sigmabar_\nu  y\ls{f2}
-  s_W^2 Q_f  y\ls{f1} \sigma_\nu x_{f2}^\dagger
\biggr ]\,,\nonumber \\
\phantom{line}
\label{eq:NNees1}\\
\hspace{-0.4in}
i\mathcal{M}_Z^{(2)}&=&
\frac{im_f m_{\N1}}{m_Z^2 D_Z}\left(\frac{-ig}{c_W}\right)^2
O_{11}^{\prime\prime L}(2T_3^f)\left(x_1 x_2-y_1^\dagger y_2^\dagger\right)\left(y\ls{f1}y\ls{f2}
-x^\dagger_{f1}x^\dagger_{f2}\right)\,.\label{eq:NNees2}
\eeqa

Next, we apply the Feynman rules of \figs{nehiggsqq}{inohiggsboson} to obtain
the sum of the four $s$-channel Higgs exchange diagrams (for $\phi^0=h^0$, $H^0$ and $A^0$)
of \fig{fig:neutneuthiggs},
\beq
i\mathcal{M}_H\!=\!\!\sum_{\phi^0=h^0,H^0,A^0}\frac{i}{D_{\phi^0}}\left(\frac{-m_f}{\sqrt{2}\,v_f}
\right)\left[(Y^{\phi^0\chi_1^0\chi_1^0})x_1 x_2+(Y^{\phi^0\chi_1^0\chi_1^0})^*y^\dagger_1 y^\dagger_2\right]
\left[k_{f\phi^0}y\ls{f1}y\ls{f2}+k_{f\phi^0}^* x^\dagger_{f1}x^\dagger_{f2}\right],\label{sHiggs}
\eeq
where
$Y^{\phi^0\chi_1^0\chi_1^0}$ is given by \eq{higgs-gauginos1}, and $D_{\phi^0}\equiv
s-m_{\phi^0}^2+i\Gamma_{\phi^0}m_{\phi^0}$.   In addition, we have introduced
the following notation
\beq \label{vfdef}
k_{f\phi^0}\equiv \begin{cases} k_{d\phi^0}\,, & \text{for}\,\,f=d\,,\,e^- \,,\\
k_{u\phi^0}\,, & \text{for}\,\,f=u\,, \\ 0\,, & \text{for}\,\,f=\nu\,,\end{cases}
\qquad\qquad
v_f\equiv \begin{cases} v_d \,, & \text{for}\,\,f=d\,,\,e^-\,, \\ v_u\,, & \text{for}\,\,f=u\,,\, \nu\,,
\end{cases}
\eeq
where $v_u$, $v_d$ are the neutral Higgs vacuum expectation values [cf.~\eq{vuvddef}] and $k_{u\phi^0}$ and $k_{d\phi^0}$
are defined in \eqs{eq:defkuphi0}{eq:defkdphi0}.
As the order of the spinor wave
functions is $(1,2,f1,f2)$ for all four terms of $\mathcal{M}_H$, no extra minus signs appear.

A good check of the above calculations is to repeat the analysis in the 't
Hooft--Feynman gauge (where the gauge parameter $\xi=1$).
In this gauge, $\mathcal{M}_Z=\mathcal{M}_{Z}^{(1)}$,
since the term proportional to $Q^\mu Q^\nu$ is absent from
the gauge boson propagator.  However, we must now include the diagrams
of \fig{fig:neutneuthiggs} with $\phi^0=G^0$.  In the
't~Hooft--Feynman gauge, $m_{G^0}=m_Z$ and $D_{G^0}=D_Z$.
Moreover, using \eqs{eq:defkuphi0}{eq:defkdphi0},
\beq
\frac{k_{fG^0}}{v_f}=\frac{2iT_f}{v}\,.
\eeq
Hence,
using \eq{sHiggs} with $\phi^0=G^0$,
\beq
i\mathcal{M}_G=\frac{m_f}{\sqrt{2}\,vD_{Z}}(2T_f)
 \,Y^{G^0\chi_1^0\chi_1^0}
\left(x_1 x_2-y^\dagger_1 y^\dagger_2\right)
\left(y\ls{f1}y\ls{f2}-x^\dagger_{f1}x^\dagger_{f2}\right)\,,
\eeq
where we have noted that $iY^{G^0\chi_1^0\chi_1^0}$ is
real.  In particular, using \eq{iy} and recalling that $m_W^2=m_Z^2
c_W^2=\half g^2 v^2$, we confirm that $\mathcal{M}_G=\mathcal{M}_Z^{(2)}$
as expected from gauge invariance.

We next evaluate the $t$/$u$-channel exchange diagrams shown in \figs{fig:neutneut2ee2}{fig:neutneut2ee3}.
We neglect $\widetilde f_L$--$\widetilde f_R$ mixing.  Eight Feynman graphs contribute,
and we denote the total invariant amplitude by:
\beq
\mathcal{M}_{\tilde f}=\sum_{j=1}^2 (\mathcal{M}^{(tj)}_{\tilde f_L}+
\mathcal{M}^{(uj)}_{\tilde f_L}+\mathcal{M}^{(tj)}_{\tilde f_R}+\mathcal{M}^{(uj)}_{\tilde f_R})\,,
\eeq
where $j=1,2$ labels the contributions of \figs{fig:neutneut2ee2}{fig:neutneut2ee3},
respectively, and the other superscripts ($t$ or $u$) and subscripts ($\widetilde f_L$
or $\widetilde f_R$) indicate the exchange channel
and the exchanged particle, respectively.
These matrix elements are evaluated by applying the rules of \fig{nqsq}.

The graphs of \fig{fig:neutneut2ee2} are sensitive to the gaugino components of $\N1$, and yield
\beqa
\hspace{-0.4in}
i \mathcal{M}^{(t1)}_{\tilde f_L} &=&
(-1)\left(-ig\sqrt{2}\right)^2
\left(\frac{i}{t - m_{\tilde f_L}^2} \right)
\left|T_3^f N_{12} + \frac{s_W}{c_W}(Q_f-T_3^f) N_{11}\right|^2
(y_1^\dagger x_{f1}^\dagger)(  x_{2}y\ls{f2})\,,
\label{eq:NNeetL1}
\\
\hspace{-0.4in}
i \mathcal{M}^{(u1)}_{\tilde f_L} &=&
\left(-ig\sqrt{2}\right)^2
\left|T_3^f N_{12} + \frac{s_W}{c_W}(Q_f-T_3^f) N_{11}\right|^2
\left( \frac{i}{u - m_{\tilde f_L}^2} \right)
(x_1 y\ls{f2})(  y_2^\dagger  x_{f1}^\dagger)
\,,
\label{eq:NNeeuL1}
\\
\hspace{-0.4in}
i \mathcal{M}^{(t1)}_{\tilde f_R} &=&
(-1)\left(i\sqrt{2} g\frac{s_W}{c_W}Q_f\right)^2
\left(\frac{i}{t - m_{\tilde f_R}^2} \right)
\left|N_{11}\right|^2
( x_1 y\ls{f1})( y_{2}^\dagger x_{f2}^\dagger)
\,,
\label{eq:NNeetR1}
\\
\hspace{-0.4in}
i \mathcal{M}^{(u1)}_{\tilde f_R} &=&
\left(i\sqrt{2} g\frac{s_W}{c_W}Q_f\right)^2
\left(\frac{i}{u - m_{\tilde f_R}^2}\right)
\left|N_{11}\right|^2
(y_1^\dagger  x_{f2}^\dagger) (x_2 y\ls{f1}) \,.
\label{eq:NNeeuR1}
\eeqa
The explicit factors of $(-1)$ in eqs.~(\ref{eq:NNeetL1}) and
(\ref{eq:NNeetR1}) are present because the order of the spinor wave functions in these
cases is an odd permutation of the ordering $(1,2,f1,f2)$ established
in the computation of the $s$-channel amplitudes.

The graphs of \fig{fig:neutneut2ee3} are sensitive to the higgsino components of $\N1$,
and yield
\beqa
i \mathcal{M}^{(t2)}_{\tilde f_L} &=&
(-1)\left(\frac{-im_f}{v_f}\right)^2
\left(\frac{i}{t - m_{\tilde f_L}^2} \right)
\left|N_{1f}\right|^2
(x_1 y\ls{f1})(  y^\dagger_{2}x^\dagger_{f2})\,,
\label{eq:NNeetL2}
\\
i \mathcal{M}^{(u2)}_{\tilde f_L} &=&
\left(\frac{-im_f}{v_f}\right)^2
\left( \frac{i}{u - m_{\tilde f_L}^2} \right)
\left|N_{1f}\right|^2
(y^\dagger_1 x^\dagger_{f2})(  x_2  y\ls{f1})
\,,
\label{eq:NNeeuL2}
\\
i \mathcal{M}^{(t2)}_{\tilde f_R} &=&
(-1)\left(\frac{-im_f}{v_f}\right)^2
\left(\frac{i}{t - m_{\tilde f_R}^2} \right)
\left|N_{1f}\right|^2
( y^\dagger_1 x^\dagger_{f1})( x_{2} y\ls{f2})
\,,
\label{eq:NNeetR2}
\\
i \mathcal{M}^{(u2)}_{\tilde f_R} &=&
\left(\frac{-im_f}{v_f}\right)^2
\left(\frac{i}{u - m_{\tilde f_R}^2}\right)
\left|N_{1f}\right|^2
(x_1  y\ls{f2}) (y^\dagger_2 x^\dagger_{f1}) \,,
\label{eq:NNeeuR2}
\eeqa
where $v_f$ is defined in \eq{vfdef}, and
\beq
N_{1f}\equiv\begin{cases} N_{13}\,, & \text{for}\,\, f=d\,,\,e^-\,, \\
N_{14}\,, &\text{for} \,\,f=u\,,\\ 0\,, &\text{for}\,\, f=\nu\,.
\end{cases}
\eeq
As before, the explicit factors of $(-1)$ are due to the ordering of the spinor wave functions.

It is convenient to write the total matrix element
for $\N1\N1\to f\bar f$ as the sum of products of separate neutralino and final state fermionic
currents.  The contributions of the $s$-channel diagrams are already in
this form.  The contributions of the $t$-- and $u$--channel diagrams given in
eqs.~(\ref{eq:NNeetL1})--(\ref{eq:NNeeuR2}) can be rearranged using the
Fierz identities of \eqst{twocompfierza}{twocompfierzc},
\beqa
y^\dagger_1 x^\dagger_{f1} x_{2}y\ls{f2} &=&\BDneg\half
(y_1^\dagger\sigmabar^\mu x_2)(x^\dagger_{f1}\sigmabar_\mu y\ls{f2})\,,\\
x_1 y\ls{f2} y_2^\dagger x_{f1}^\dagger &=& \BDneg\half (x_1\sigma^\mu
y_2^\dagger) (x_{f1}^\dagger\sigmabar_\mu y\ls{f2})\,, \\
x_1 y\ls{f1} y^\dagger_{2} x^\dagger_{f2}&=& \BDneg\half (x_1\sigma^\mu
y_2^\dagger)(
y\ls{f1}\sigma_\mu x_{f2}^\dagger)\,,\\
y_1^\dagger x_{f2}^\dagger x_2 y\ls{f1} &=& \BDneg\half
(y_1^\dagger\sigmabar^\mu x_2)(y\ls{f1}\sigma_\mu x^\dagger_{f2})\,.
\eeqa
Combining the result of the $s$, $t$, and $u$--channel contributions,
we have for the total amplitude:
\beqa
&&\!\!\!\!\!\!\mathcal{M} = \frac{m_f m_{\N1}}{m_Z^2}c_0\left(x_1 x_2-y_1^\dagger
y_2^\dagger\right)\left(y\ls{f1}y\ls{f2}-x^\dagger_{f1}x^\dagger_{f2}\right)
\nonumber \\[4pt]
&& \BDplus c_1(y_1^\dagger\sigmabar^\mu x_2)(x^\dagger_{f1}\sigmabar_\mu
y\ls{f2}) \!\BDplus \!
c_2 (x_1 \sigma^\mu y_2^\dagger)(x^\dagger_{f1}\sigmabar_\mu y\ls{f2})
\!\BDplus \!c_3 (x_1 \sigma^\mu y_2^\dagger) (y\ls{f1}\sigma_\mu
x_{f2}^\dagger) \!\BDplus \!
c_4 (y_1^\dagger\sigmabar^\mu x_2)(y\ls{f1}\sigma_\mu x^\dagger_{f2})
\nonumber \\[6pt]
&& +m_f\left[c_5 (x_1 x_2)(y\ls{f1}y\ls{f2})+c_6(x_1 x_2)(x^\dagger_{f1}x^\dagger_{f2})
+c_7(y_1^\dagger y_2^\dagger)(y\ls{f1}y\ls{f2})+c_8(y_1^\dagger y_2^\dagger)
(x^\dagger_{f1}x^\dagger_{f2})\right]\,,
\label{ampl-chichi2ee}
\eeqa
where the coefficients $c_0\,,\,c_1\,,\,\ldots\,,\,c_4$ are given by
\beqa
\hspace{-0.4in}
c_0&=&-g^2\frac{2T_3^f O_{11}^{\prime\prime
      L}}{c_W^2D_Z}\,,\\
\hspace{-0.4in}
c_1 &=& -g^2\left
  [\frac{(T_3^f-s_W^2 Q_f)O_{11}^{\prime\prime
      L}}{c_W^2D_Z}+\frac{|T_3^f
    N_{12} + \frac{s_W}{c_W}(Q_f-T_3^f) N_{11} |^2}{t - m_{\tilde f_L}^2} \right
]-\frac{m^2_f}{2v^2_f}\left(\frac{|N_{1f}|^2}{t - m^2_{\tilde
      f_R}}\right)\,,
\\
\hspace{-0.4in}
c_2 &=& g^2\left [\frac{(T_3^f-s_W^2 Q_f) O_{11}^{\prime\prime L}}{c_W^2
    D_Z}+\frac{|T_3^f N_{12} + \frac{s_W}{c_W}(Q_f-T_3^f) N_{11} |^2}{u -
    m_{\tilde f_L}^2} \right ]+\frac{m^2_f}{2v^2_f}\left(\frac{|N_{1f}|^2}{u - m^2_{\tilde
      f_R}}\right)\,,
\\
\hspace{-0.4in}
c_3 &=& -g^2\frac{s_W^2}{c_W^2}Q_f \left [\frac{O_{11}^{\prime\prime L}}{D_Z}
 +\frac{Q_f|N_{11}|^2}{t - m^2_{\tilde
      f_R}} \right ] -\frac{m^2_f}{2v^2_f}\left(\frac{|N_{1f}|^2}{t - m^2_{\tilde
      f_L}}\right)\,,
\\
\hspace{-0.4in}
c_4 &=& g^2 \frac{s_W^2}{c_W^2}Q_f\left [\frac{O_{11}^{\prime\prime L}}{D_Z} +
  \frac{Q_f|N_{11}|^2}{u - m^2_{\tilde f_R}} \right ]+
  \frac{m^2_f}{2v^2_f}\left(\frac{|N_{1f}|^2}{u - m^2_{\tilde
      f_L}}\right)\,.
\eeqa
The coefficients $c_5,\ldots,c_8$ are obtained from
\eq{sHiggs} and represent the $s$-channel Higgs exchange contributions to the annihilation
matrix element.

In the non--relativistic limit,
$|\boldsymbol{\vec{p}}|\ll m_{\N1}$. Then $t\simeq u\simeq -m_{\N1}^2+m_f^2$, and
we can approximate\footnote{In particular, we assume that
$\widetilde f_L$ and $\widetilde f_R$ are significantly heavier than all other
particles in the annihilation process.  Consequently, we can ignore all
$\mathcal{O}(|\boldsymbol{\vec{p}}|/m\ls{\tilde f_{L,R}})$ terms in
$c_1+c_2$ and $c_3+c_4$.}
$c_1=-c_2$ and $c_3=-c_4$.
Hence, the total amplitude,
\eq{ampl-chichi2ee}, can be written~as
\beqa
\mathcal{M}
&=& \frac{m_f m_{\N1}}{m_Z^2}c_0\left(x_1 x_2-y_1^\dagger
y_2^\dagger\right)\left(y\ls{f1}y\ls{f2}-x^\dagger_{f1}x^\dagger_{f2}\right)
\nonumber \\[4pt]
&&+\left[y_1^\dagger\sigmabar^\mu x_2-x_1 \sigma^\mu
  y_2^\dagger\right] \left[\BDpos c_1(x^\dagger_{f1}\sigmabar_\mu
  y\ls{f2})\BDminus c_3(y_{f1}\sigma_\mu
x^\dagger_{f2})\right]+\mathcal{M}_H \,,
\label{ampl-chichi2ee-nr}
\eeqa
where the $s$-channel Higgs exchange contributions, $\mathcal{M}_H$,
will be neglected for simplicity in the subsequent analysis.
The spin-averaged squared matrix element for $\N1\N1\to f\bar f$ then
takes the following form:
\beqa
\quarter\sum_{s_1,s_2,s\ls{f1},s\ls{f2}} |\mathcal{M}_Z+
\mathcal{M}_{\tilde f}|^2
&=& N_{\mu\nu}\biggl[|c_1|^2 F_1^{\mu\nu}+|c_3|^2 F_2^{\mu\nu}
-2{\rm Re}(c_1 c_3^*)F_{12}^{\mu\nu}\biggr]+
\frac{m_f^2 m_{\N1}^2}{m_Z^4}|c_0|^2 NF
\nonumber \\
&& \BDplus \frac{2 m_f m_{\N1}}{m_Z^2}{\rm Re}[c_0^*(c_1+c_3)]N_\mu
F^\mu\,,\label{mave2}
\eeqa
where
$N_{\mu\nu}$, $N_\mu$ and $N$ are spin-averaged tensor, vector and scalar quantities
that depend on the initial state neutralino kinematics and $F^{\mu\nu}_{1,2,12}$,
$F^\mu$ and $F$ are spin-summed tensor, vector and scalar quantities that depend on the final
state fermion kinematics.  These quantities are easily computed using
the projection operators of \eqst{xxdagsummed}{ydagxdagsummed}
and the standard trace techniques to perform the spin averages and sums.
Explicitly, the spin-averaged neutralino quantities are
\beqa
N&\equiv& \quarter\sum_{s_1,s_2} (x_1 x_2-y_1^\dagger y_2^\dagger)
(x_2^\dagger x_1^\dagger-y_2 y_1)= \BDpos p_1\newcdot p_2 +m_{\N1}^2
=2E^2\,,\label{chi-curr1}\\
N^\mu&\equiv&\quarter\sum_{s_1,s_2}  (y_1^\dagger\sigmabar^{\mu} x_2-x_1
\sigma^{\mu} y_2^\dagger)(x_2^\dagger x_1^\dagger-y_2 y_1)
=-m_{\N1}(p_1+p_2)^\mu=\left\{
\begin{array}{ll}
-2 m_{\N1}E\,, &\qquad\mu=0 \,,\\[2mm]
\quad 0\,,  &\qquad \mu=i \,,\end{array}\right.\nonumber \\
\phantom{line}\label{chi-curr2}
\eeqa
and a symmetric second-rank tensor,
\beqa
N^{\mu\nu}&\equiv&\quarter\sum_{s_1,s_2} (y_1^\dagger\sigmabar^{\mu} x_2-x_1
\sigma^{\mu}
y_2^\dagger)(x_2^\dagger\sigmabar^{\nu} y_1-y_2\sigma^{\nu} x_1^\dagger)
=
p_1^\mu p_2^\nu+p_2^\mu p_1^\nu-g^{\mu\nu}(p_1\newcdot p_2 \BDminus m_{\N1}^2)
\nonumber \\
&&\qquad\qquad\qquad =\left\{\begin{array}{ll}2m_{\N1}^2\,,\quad  & \mu=\nu=0\,, \\[2mm]
\,\,\,0\,, & \mu=0\,,\, \nu=j\,\,\, {\rm or} \,\,\,\mu=i\,,\,\nu=0\,, \\[2mm]
2\left[|\boldsymbol{\vec{p}}|^2\,\delta^{ij}-p^{i}p^{j}\right]\,,\quad &\mu=i,\,\nu=j\,,
\end{array}\right.\label{chi-curr3}
\eeqa
where the final results given in \eqst{chi-curr1}{chi-curr3}
have been evaluated in the CM frame.
Similarly, the spin-summed final state fermion quantities are
\beqa
\hspace{-0.4in}
F&\equiv& \sum_{s\ls{f1},s\ls{f2}}(y\ls{f1}y\ls{f2}-x^\dagger_{f1}x^\dagger_{f2})
(y^\dagger_{f2}y^\dagger_{f1}-x\ls{f2}x\ls{f1})
=4(\BDpos k_1\newcdot k_2+m_f^2)=8E^2\,,\\
\hspace{-0.4in}
F^\mu &\equiv &\sum_{s\ls{f1},s\ls{f2}}
(x^\dagger_{f1}\sigmabar^\mu y\ls{f2})
(y^\dagger_{f2}y^\dagger_{f1}-x\ls{f2}x\ls{f1})
=-\sum_{s\ls{f1},s\ls{f2}}
(y\ls{f1}\sigma^\mu x^\dagger_{f2})
(y^\dagger_{f2}y^\dagger_{f1}-x\ls{f2}x\ls{f1}) \nonumber \\
&& =2m_f(k_1+k_2)^\mu=\left\{
\begin{array}{ll}
4 m_f E\,, &\qquad\mu=0 \,,\\[2mm]
\quad 0\,,  &\qquad \mu=i \,,\end{array}\right.
\eeqa
after evaluating the above quantities in the CM frame,
and
\beqa
\hspace{-0.5in}
F_1^{\mu\nu}&\equiv& \sum_{s\ls{f1},s\ls{f2}} (x_{f1}^\dagger\sigmabar^{\mu} y\ls{f2})
(y_{f2}^\dagger \sigmabar^{\nu} x\ls{f1})=k_{1\rho}k_{2\lambda}\Tr(\sigma^\rho\sigmabar^\mu
\sigma^\lambda\sigmabar^\nu)\,,\\
\hspace{-0.5in}
F_2^{\mu\nu}&\equiv& \sum_{s\ls{f1},s\ls{f2}} (y\ls{f1}\sigma^{\mu} x_{f2}^\dagger)
(x\ls{f2} \sigma^{\nu} y_{f1}^\dagger)=k_{1\rho}k_{2\lambda}\Tr(\sigmabar^\rho\sigma^\mu
\sigmabar^\lambda\sigma^\nu)\,,\\
\hspace{-0.5in}
F_{12}^{\mu\nu}&\equiv& \sum_{s\ls{f1},s\ls{f2}}(y\ls{f1}\sigma^{\mu} x_{f2}^\dagger)
(y_{f2}^\dagger \sigmabar^{\nu} x\ls{f1})=\sum_{s\ls{f1},s\ls{f2}}
 (x_{f1}^\dagger\sigmabar^{\mu} y\ls{f2})(x\ls{f2} \sigma^{\nu} y_{f1}^\dagger)
=-m_f^2\Tr(\sigma^\mu\sigmabar^\nu).
\eeqa
Since $N^{\mu\nu}$ is symmetric, the antisymmetric parts of
$F_1^{\mu\nu}$ and $F_2^{\mu\nu}$ do not contribute in \eq{mave2}.
The symmetric parts of $F_1^{\mu\nu}$ and $F_2^{\mu\nu}$
are equal and given by:
\beqa
[F_1^{\mu\nu}]_{\rm symm}&=&[F_2^{\mu\nu}]_{\rm symm}=2(k_1^{\mu}k_2^{\nu}+k_1^{\nu}
k_2^{\mu}-k_1\cdot k_2 g^{\mu\nu}) \nonumber \\[8pt]
&=&\left\{\begin{array}{ll}
2m_f^2\,,& \mu=\nu=0\,,\\[2mm]
\,\,\,0\,, & \mu=0\,,\, \nu=j\,\,\, {\rm or} \,\,\,\mu=i\,,\,\nu=0\,, \\[2mm]
2m_f^2(2\boldsymbol{\hat{k}}^{i}\boldsymbol{\hat{k}}^{j}-\delta^{ij})
-4E^2(\boldsymbol{\hat{k}}^{i}\boldsymbol{\hat{k}}^{j}-\delta^{ij})\,,\qquad
& \mu=i,\,\nu=j\,,
\end{array}\right.\nonumber \\
\phantom{line}
\eeqa
and
$F_{12}^{\mu\nu}= \BDneg 2m_f^2 g^{\mu\nu}$.
The spin-averaged squared matrix element for $\N1\N1\to f\bar f$ given by
\eq{mave2} can now be fully evaluated, resulting in
\beqa
\nicefrac{1}{4}\sum_{s_1,s_2,s_{f1},s_{f2}}\!\!\!|\mathcal{M}_{Z}+\mathcal{M}_{\tilde f}|^2&=&
4(|c_1|^2+|c_3|^2)\left[m_{\N1}^2 m_f^2+2|\boldsymbol{\vec p}|^2(E^2(1+\cos^2\theta)-m_f^2\cos^2\theta)
\right]\nonumber \\
&&  +8m_f^2\,{\rm Re}(c_1 c_3^*)\left[m_{\N1}^2 -2|\boldsymbol{\vec p}|^2\right] \nonumber\\[4pt]
&&  +\frac{16m_f^2 m_{\N1}^2}{m_Z^4}\,E^2\biggl[E^2|c_0|^2-m_Z^2{\rm Re}[c_0^*(c_1+c_3)]\biggr]\,,
\label{msquaredann}
\eeqa
where $\cos\theta=\boldsymbol{\vec p\newcdot\hat k}/|\boldsymbol{\vec p}|$.
In the non-relativistic limit, we
use \eq{Enonrel} and drop terms of $\mathcal{O}(|\boldsymbol{\vec p}|^4)$.

To compute $v_{\rm rel}\sigma_{\rm ann}$, we make use of the following result for
the differential annihilation cross-section in the CM frame:
\beq \label{dsigdomega}
v_{\rm rel}\left(\frac{d\sigma}{d\Omega}\right)_{\rm CM}=\frac{1}{32\pi^2 s}\left(1-
\frac{4m_f^2}{s}\right)^{1/2}|\mathcal{M}|^2_{\rm ave}\,,
\eeq
where $|\mathcal{M}|^2_{\rm ave}$ is the squared matrix element for the annihilation
process, averaged over initial spins and summed over final spins, and the relative velocity
of the initial state neutralinos in the CM frame is given by
$v_{\rm rel}=4|\boldsymbol{\vec p}|/\sqrt{s}\simeq 2|\boldsymbol{\vec p}|/m_{\N1}$,
after noting that $\sqrt{s}\simeq 2m_{\N1}$ in the non-relativistic limit.
Inserting the squared matrix element obtained above into \eq{dsigdomega} and integrating
over solid angles, we end up with:
\beqa \label{vrelsigann}
v_{\rm rel}\sigma_{\rm ann}&=&\frac{1}{8\pi E^2}\left(1-
\frac{m_f^2}{E^2}\right)^{1/2}\Biggl\{(|c_1|^2+|c_3|^2)\left[m_{\N1}^2 m_f^2+
\frac{2|\boldsymbol{\vec p}|^2}{3}\left(4m_{\N1}^2-m_f^2\right)\right]\nonumber \\[6pt]
&&\qquad
+\frac{4m_f^2 m_{\N1}^2}{m_Z^4}\biggl[m_{\N1}^2(m_{\N1}^2+2|\boldsymbol{\vec{p}}|^2)|c_0|^2
-m_Z^2(m_{\N1}^2+|\boldsymbol{\vec{p}}|^2){\rm Re}[c_0^*(c_1+c_3)]\biggr]\nonumber \\[6pt]
&&\qquad
+2m_f^2\,{\rm Re}(c_1 c_3^*)\left[m_{\N1}^2 -2|\boldsymbol{\vec p}|^2\right]
+\mathcal{O}(|\boldsymbol{\vec p}|^4)\Biggr\}\,,
\eeqa
where the effects of the $s$-channel Higgs boson exchanges have been omitted.

The momentum dependence of \eq{vrelsigann} reflects the famous $p$-wave suppression
of the annihilation cross-section in the $m_f=0$ limit noted in \Ref{Goldberg:1983nd}.\footnote{In
\Ref{Goldberg:1983nd}, the annihilation rate for photinos is computed,
corresponding to $N_{11}=c_W$, $N_{12}=s_W$ and $N_{13}=N_{14}=0$.  In this case,
the $Z$ boson and Higgs boson $s$-channel exchange diagrams are absent.  The result
presented in \Ref{Goldberg:1983nd} should be multiplied by a factor of two
(H.~Goldberg, private communication)---the corrected expression then agrees
with \eq{vrelsigann}.}
In general, the annihilation cross-section in
the non-relativistic limit behaves as $v_{\rm rel}\,\sigma_{\rm ann}\propto
|\boldsymbol{\vec p}|^{2\ell}$.  Applying this result to \eq{vrelsigann}
in the $m_f=0$ limit implies that $\ell=1$.  This is a consequence of the
Majorana nature of the neutralino.  In particular, in the limit of $m_f=0$, the
$f\bar f$ pair is in a $J=1$ angular momentum state.  However, Fermi statistics
dictates that at threshold, a pair of identical Majorana fermions in a $J=1$
state must have relative orbital angular momentum $\ell=1$ (corresponding to
$p$-wave annihilation).  The $s$-wave annihilation (corresponding to
the Majorana fermion pair in a $J=0$ state) is
suppressed by a factor of $m_f^2$, as is evident from \eq{vrelsigann}.

We have checked that \eq{vrelsigann} corresponds to a result first obtained
in \Ref{Ellisetal} (although the latter reference omits
the terms in \eq{vrelsigann} proportional to $c_0$).
However, we emphasize that
this formula neglects the effects of $s$-channel Higgs boson exchanges.
We invite the reader to complete the computation of the annihilation cross-section by
including these terms (along with the effects of interference between the neglected
contributions and the ones computed above).

The annihilation of $\N1\N1$ into heavy quarks ($c$, $b$ and $t$), followed
by the decay of the heavy quarks, can yield
observable signatures suitable for indirect dark matter
detection.  For example, the annihilation of neutralinos in the galaxy provides a possible
source of indirect dark matter detection via the observation of
positrons in cosmic rays~\cite{baltz}.
Neutralino dark matter can also be captured in the 
sun~\cite{Press:1985ug}.
The neutrinos that arise (either directly or indirectly) from the neutralino annihilation in the sun
can be detected on Earth (see, e.g., \Ref{edsjo}).


\subsection{\texorpdfstring{$e^-e^+ \ra \Ciminus \Cjplus$}{e\textminussuperior e\textplussuperior\textrightarrow C\texttilde\textiinferior\textminussuperior C\texttilde\textjinferior\textplussuperior}}
\setcounter{equation}{0}
\setcounter{figure}{0}
\setcounter{table}{0}

Next we consider the pair production of charginos in electron-positron
collisions. The $s$-channel Feynman diagrams are shown in
\fig{fig:ee2charchar}, where we have also introduced the notation
for the fermion momenta and polarizations. The Mandelstam variables
are given by
\beqa
s &=&
\BDpos 2p_1\newcdot p_2 = m_{\Ci}^2 + m_{\Cj}^2 \BDplus 2k_i\newcdot k_j ,
\\
t &=& m_{\Ci}^2 \BDminus  2p_1\newcdot k_i =
    m_{\Cj}^2 \BDminus  2p_2\newcdot k_j ,
\\
u &=& m_{\Ci}^2 \BDminus  2p_2\newcdot k_i =
      m_{\Cj}^2 \BDminus  2p_1\newcdot k_j .
\eeqa
Note that the negatively charged chargino
carries momentum and polarization $(k_i,\lam_i)$, while the
positively charged one carries $(k_j,\lam_j)$.
\begin{figure}[!b]
\centerline{
\begin{picture}(450,230)(-210,30)
\ArrowLine(-200,225)(-152,197.5)
\ArrowLine(-152,197.5)(-200,170)
\Photon(-152,197.5)(-98,197.5)46
\ArrowLine(-98,197.5)(-50,225)
\ArrowLine(-50,170)(-98,197.5)
\put(-205,230){$e\, (p_1,\lam_1)$}
\put(-205,161){${e}^\dagger\, (p_2,\lam_2)$}
\put(-57,230){$\chi^-_i\, (k_i,\lam_i)$}
\put(-55,156){${\chi^{-\,\dagger}_j}\, (k_j,\lam_j)$}
\put(-136.5,210){$\gamma,Z^0$}
\ArrowLine(98,197.5)(50,225)
\ArrowLine(50,170)(98,197.5)
\Photon(98,197.5)(152,197.5)46
\ArrowLine(152,197.5)(200,225)
\ArrowLine(200,170)(152,197.5)
\put(45,230){${\bar e}^\dagger\, (p_1,\lam_1)$}
\put(45,159){$\bar e\, (p_2,\lam_2)$}
\put(193,230){$\chi^-_i\, (k_i,\lam_i)$}
\put(195,156){${\chi^{-\,\dagger}_j} \, (k_j,\lam_j)$}
\put(113.5,210){$\gamma,Z^0$}
\ArrowLine(-200,115)(-152,87.5)
\ArrowLine(-152,87.5)(-200,60)
\Photon(-152,87.5)(-98,87.5)46
\ArrowLine(-50,115)(-98,87.5)
\ArrowLine(-98,87.5)(-50,60)
\put(-205,120){$e\, (p_1,\lam_1)$}
\put(-205,51){${e}^\dagger\, (p_2,\lam_2)$}
\put(-57,120){${\chi_i^{+\,\dagger}}\, (k_i,\lam_i)$}
\put(-55,48){$\chi^+_j\, (k_j,\lam_j)$}
\put(-136.5,100){$\gamma,Z^0$}
\ArrowLine(98,87.5)(50,115)
\ArrowLine(50,60)(98,87.5)
\Photon(98,87.5)(152,87.5)46
\ArrowLine(200,115)(152,87.5)
\ArrowLine(152,87.5)(200,60)
\put(45,120){${\bar e}^\dagger\, (p_1,\lam_1)$}
\put(45,49){$\bar e\, (p_2,\lam_2)$}
\put(193,120){${\chi_i^{+\,\dagger}}\, (k_i,\lam_i)$}
\put(195,48){$\chi_j^+\, (k_j,\lam_j)$}
\put(113.5,100){$\gamma,Z^0$}
\end{picture}
}
\caption{\label{fig:ee2charchar} Feynman diagrams for $e^-e^+
\ra \Ciminus\Cjplus$ via $s$-channel $\gamma$ and $Z^0$ exchange.}
\end{figure}

Using the Feynman rules of \figs{SMintvertices}{nnboson},
the sum of the photon-exchange diagrams is given by:
\beqa
i{\cal M}_{\gamma} = \frac{\BDneg i\metric^{\mu\nu}}{s}
\left (
\BDneg ie \, x_1 \sigma_\mu  y^\dagger_2
\BDminus i e \,  y^\dagger_1 \sigmabar_\mu x_2
\right )
\left (
\BDpos ie\, \delta_{ij} y_i \sigma_\nu  x^\dagger_j
\BDplus ie\, \delta_{ij}  x^\dagger_i \sigmabar_\nu y_j
\right )\,.
\label{eq:eeCCg}
\eeqa
The $Z$-exchange diagrams yields [cf.~footnote~\ref{fnqq}]:
\beq
i {\cal M}_Z=
\frac{\BDneg i \metric^{\mu\nu}}{D_Z}
\Bigl [
\BDpos \frac{ig}{c_W} (s_W^2 - \half) \, x_1 \sigma_\mu  y^\dagger_2
\BDplus \frac{igs_W^2}{c_W} \,  y^\dagger_1 \sigmabar_\mu x_2
\Bigr ]
\Bigl [
\BDneg \frac{ig}{c_W} O_{ji}^{\prime L} \, y_i \sigma_\nu  x^\dagger_j
\BDminus
\frac{ig}{c_W} O_{ji}^{\prime R} \,  x^\dagger_i \sigmabar_\nu y_j
\Bigr ]\,,\phantom{xxxxx}
\label{eq:eeCCZ}
\eeq
where $D_Z\equiv s-m_Z^2+i\Gamma_Zm_Z$.
The $t$-channel Feynman diagram via
sneutrino exchange is shown in \fig{fig:ee2charchar2}.
Applying the rules of \fig{cqsq}, we find:
\beqa
i{\cal M}_{\tilde \nu_e} =
(-1) \frac{i}{t-m^2_{\tilde \nu_e}}
\left ( -i g V_{i1}^* x_1 y_i \right )
\left ( -i g V_{j1}  y^\dagger_2  x^\dagger_j \right ).
\eeqa
The Fermi-Dirac factor $(-1)$ in this equation arises because the spinors
appear an order which is an odd permutation of the order used in all
of the $s$-channel diagram results.

\begin{figure}[t!]
\centerline{
\begin{picture}(400,100)(-300,29)
\ArrowLine(-200,105)(-125,105)
\DashArrowLine(-125,105)(-125,50)5
\ArrowLine(-125,50)(-200,50)
\ArrowLine(-50,105)(-125,105)
\ArrowLine(-125,50)(-50,50)
\put(-201,111){$e\,(p_1,\lam_1)$}
\put(-201,56){${e}^\dagger\,(p_2,\lam_2)$}
\put(-58,111){${\chi^{+\,\dagger}_i}\,(k_i,\lam_i)$}
\put(-58,56){${\chi^{-\,\dagger}_j}\,(k_j,\lam_j)$}
\put(-144,74){$\stilde\nu_e$}
\end{picture}
}
\caption{\label{fig:ee2charchar2} The Feynman diagram for $e^-e^+
  \ra\Ciminus\Cjplus$ via the $t$-channel exchange of a
  sneutrino.}
\end{figure}

One can now apply the Fierz transformation identities
\eqst{twocompfierza}{twocompfierzc} to eqs.~(\ref{eq:eeCCg}) and
(\ref{eq:eeCCZ}) to remove the $\sigma$ and $\sigmabar$ matrices.
The result can be combined with the $t$-channel contribution to
obtain a total matrix element ${\cal M}$ with exactly the
same form as eq.~(\ref{eq:eeNNniceform}), but now with:
\beqa
c_1 &=& 2 \frac{e^2 \delta_{ij}}{s}
- \frac{g^2}{c_W^2 D_Z} (1 - 2 s_W^2) O^{\prime R}_{ji}
,
\\
c_2 &=& \frac{2 e^2 \delta_{ij}}{s}
- \frac{g^2}{c_W^2 D_Z} (1 - 2 s_W^2) O^{\prime L}_{ji}
+ \frac{g^2 V_{i1}^* V_{j1}}{t - m_{\tilde \nu_e}^2}
,
\\
c_3 &=& \frac{2 e^2 \delta_{ij}}{s}
+ \frac{2g^2s_W^2}{c_W^2 D_Z} O^{\prime R}_{ji}\,,\\
c_4 &=& \frac{2 e^2 \delta_{ij}}{s}
+ \frac{2g^2s_W^2}{c_W^2 D_Z} O^{\prime L}_{ji}
.
\eeqa
The rest of this calculation is identical in form to
\eqst{eq:eeNNniceform}{eq:eeNNnicerform}, so that
the result is:
\beqa
\sum_{\rm spins} |{\cal M}|^2 &=&
(|c_1|^2 + |c_4|^2) (u - m_{\Ci}^2)(u - m_{\Cj}^2)
+ (|c_2|^2 + |c_3|^2) (t - m_{\Ci}^2)(t - m_{\Cj}^2)\phantom{xxxx}
\nonumber \\ &&
\quad
+ 2 {\rm Re}[c_1 c_2^* + c_3 c_4^*] m_{\Ci} m_{\Cj} s \,.
\label{eq:eeCCnicerform}
\eeqa
The differential cross-section then follows:
\beq
\frac{d\sigma}{dt} = \frac{1}{16\pi s^2}
\left (\frac{1}{4} \sum_{\rm spins} |{\cal M}|^2 \right ) .
\eeq
As in the previous subsection, we define
$\cos\theta = {\boldsymbol{\hat p_1}}\newcdot {\boldsymbol{\hat k_i}}$
(where $\theta$ is the angle between the initial state
electron and $\tilde C_i^-$ in the center-of-momentum frame).  The
Mandelstam variables $t,u$ are given by
\beqa
t &=& \frac{1}{2} \left
[m_{\Ci}^2 + m_{\Cj}^2 -s + \lam^{1/2}(s,m_{\Ci}^2,m_{\Cj}^2) \cos\theta
\right ], \label{eeCC-t}
\\
u &=& \frac{1}{2}
\left
[m_{\Ci}^2 + m_{\Cj}^2 -s - \lam^{1/2}(s,m_{\Ci}^2,m_{\Cj}^2) \cos\theta
\right ] .
\eeqa
The total cross-section can now be computed as
\beq
\sigma = \int_{t_-}^{t_+} \frac{d\sigma}{dt} dt\,,
\eeq
where $t_-$ and $t_+$ are obtained with $\cos\theta = -1$ and $+1$ in
eq.~(\ref{eeCC-t}), respectively. Our results agree with the original
first complete calculation in \Ref{Bartl:1985fk}. Earlier work with
simplifying assumptions is given in \Ref{CC-early}. An extended
calculation for the production of polarized charginos is given in
\cite{Choi:1998ut}.


\subsection{\texorpdfstring{$u\dbar \ra \Ciplus \Nj$}{u{d\textoverline} \textrightarrow C\texttilde\textiinferior\textplussuperior N\texttilde\textjinferior}}
\setcounter{equation}{0}
\setcounter{figure}{0}
\setcounter{table}{0}

Next we consider the associated production of a chargino and a neutralino
in quark, antiquark collisions. The leading order Feynman diagrams are shown
in \fig{fig:ud2charneut}, where we have also
defined the momenta and the helicities. The corresponding
Mandelstam variables are
\beqa
s &=&
\BDpos 2p_1\newcdot p_2 = m_{\Ci}^2 + m_{\Nj}^2 \BDplus 2k_i\newcdot k_j ,
\\
t &=& m_{\Ci}^2 \BDminus  2p_1\newcdot k_i =
    m_{\Nj}^2 \BDminus  2p_2\newcdot k_j ,
\\
u &=& m_{\Ci}^2 \BDminus  2p_2\newcdot k_i =
      m_{\Nj}^2 \BDminus  2p_1\newcdot k_j .
\eeqa

The matrix elements for the $s$-channel diagrams are obtained by
applying the Feynman rules of \figs{SMintvertices}{ccboson}:
\beqa
i{\cal M}_s = \frac{\BDneg i \metric^{\mu\nu}}{s - m_W^2}
\left (\BDpos \frac{ig}{\sqrt{2}} x_1 \sigma_\mu  y^\dagger_2 \right )
\left (\BDpos ig O_{ji}^{L*}  x^\dagger_i \sigmabar_\nu y_j
\BDplus i g O^{R*}_{ji} y_i \sigma_\nu  x^\dagger_j \right ) .
\eeqa
The external spinors are denoted by
$x_1\equiv x(\boldsymbol{\vec p}_1,\lam_1)$,
$ y^\dagger_2\equiv  y^\dagger(\boldsymbol{\vec p}_2,\lam_2)$,
$ x^\dagger_i\equiv  x^\dagger(\boldsymbol{\vec k}_i,\lam_i)$,
$y_j\equiv y(\boldsymbol{\vec k}_j,\lam_j)$, etc.
The matrix elements for the $t$ and $u$ channel graphs follow from the
rules of \figs{cqsq}{nqsq}:
\beqa
i{\cal M}_t &=& (-1)
\frac{i}{t - m_{\tilde d_L}^2}
\left (-ig U_{i1}^* \right )
\Bigl (
\frac{ig}{\sqrt{2}}\bigl [N_{j2} - \frac{s_W}{3c_W} N_{j1} \bigr ]
\Bigr )
x_1 y_i  y^\dagger_2  x^\dagger_j\,,
\label{eq:eeCNt}
\\
i{\cal M}_u &=&
\frac{i}{u - m_{\tilde u_L}^2}
\left (-ig V_{i1} \right )
\Bigl (
\frac{ig}{\sqrt{2}}\bigl [-N_{j2}^* - \frac{s_W}{3c_W} N_{j1}^* \bigr ]
\Bigr )
x_1 y_j  y^\dagger_2  x^\dagger_i\,.
\eeqa
The first factor of $(-1)$ in eq.~(\ref{eq:eeCNt}) is required because
the order of the spinors $(1,i,2,j)$ is in an odd permutation of
the order $(1,2,i,j)$ used in the $s$-channel and
$u$-channel results.

\begin{figure}[!t]
\centerline{
\begin{picture}(450,230)(-210,30)
\ArrowLine(-200,225)(-152,197.5)
\ArrowLine(-152,197.5)(-200,170)
\Photon(-152,197.5)(-98,197.5)46
\ArrowLine(-98,197.5)(-50,225)
\ArrowLine(-50,170)(-98,197.5)
\put(-205,230){$u \, (p_1,\lam_1)$}
\put(-205,158){${d}^\dagger\, (p_2,\lam_2)$}
\put(-57,230){${\chi_i^+}\, (k_i,\lam_i)$}
\put(-55,158){${\chi^{0\,\dagger}_j}\, (k_j,\lam_j)$}
\put(-129,210){$W^+$}
\ArrowLine(50,225)(98,197.5)
\ArrowLine(98,197.5)(50,170)
\Photon(98,197.5)(152,197.5)46
\ArrowLine(200,225)(152,197.5)
\ArrowLine(152,197.5)(200,170)
\put(45,230){$u\, (p_1,\lam_1)$}
\put(45,159){${d}^\dagger\, (p_2,\lam_2)$}
\put(193,230){${\chi_i^{-\,\dagger}}\,(k_i,\lam_i)$}
\put(195,158){${\chi^0_j}\,(k_j,\lam_j)$}
\put(121,210){$W^+$}
\ArrowLine(-200,115)(-125,115)
\ArrowLine(-125,60)(-200,60)
\DashArrowLine(-125,115)(-125,60)5
\ArrowLine(-125,60)(-50,60)
\ArrowLine(-50,115)(-125,115)
\put(-205,120){$u\, (p_1,\lam_1)$}
\put(-205,49){${d}^\dagger\, (p_2,\lam_2)$}
\put(-57,120){${\chi^{-\,\dagger}_i}\, (k_i,\lam_i)$}
\put(-55,48){${\chi^0_j}\, (k_j,\lam_j)$}
\put(-145,87.5){$\stilde d_L$}
\ArrowLine(50,115)(125,115)
\ArrowLine(125,60)(50,60)
\DashArrowLine(125,115)(125,60)5
\ArrowLine(200,60)(166.25,84.75)
\Line(125,115)(158.75,90.25)
\ArrowLine(162.5,87.5)(200,115)
\Line(125,60)(162.5,87.5)
\put(45,120){$u\, (p_1,\lam_1)$}
\put(45,49){${d}^\dagger\, (p_2,\lam_2)$}
\put(193,120){${\chi^+_i}\, (k_i,\lam_i)$}
\put(195,46){${\chi^{0\,\dagger}_j}\, (k_j,\lam_j)$}
\put(105,87.5){$\stilde u_L$}
\end{picture}
}
\caption{\label{fig:ud2charneut} The four tree-level Feynman diagrams
for $u\dbar\ra\Ciplus\Nj$.}
\end{figure}

Now we can use the Fierz relations eqs.~(\ref{twocompfierza}) and
(\ref{twocompfierzc}) to rewrite the $s$-channel amplitude in a form
without $\sigma$ or $\sigmabar$ matrices. Combining the result with
the $t$-channel and $u$-channel contributions yields a total ${\cal M}$
with exactly the same form as eq.~(\ref{eq:eeNNniceform}), but now with
\beqa
c_1 &=& -\sqrt{2} g^2 \left [\frac{O_{ji}^{L*}}{s-m_W^2} +
\left(\frac{1}{2} N_{j2}^* + \frac{s_W}{6c_W} N_{j1}^*\right)\frac{V_{i1}}{u- m_{\tilde
u_L}}
\right]
,
\\
c_2 &=& -\sqrt{2} g^2 \left [ \frac{O_{ji}^{R*}}{s-m_W^2} +
\left(\frac{1}{2} N_{j2}^* - \frac{s_W}{6c_W} N_{j1}^*\right)\frac{U_{i1}^*}{t- m_{\tilde
d_L}}
\right]
,
\\
c_3 &=& c_4 = 0.
\eeqa
The rest of this calculation is identical in form to
that of \eqst{eq:eeNNniceform}{eq:eeNNnicerform}, leading to:
\beq
\sum_{\rm spins} |{\cal M}|^2 =
|c_1|^2 (u - m_{\Ci}^2)(u - m_{\Nj}^2)
+ |c_2|^2 (t - m_{\Ci}^2)(t - m_{\Nj}^2)
+ 2 {\rm Re}[c_1 c_2^*] m_{\Ci} m_{\Nj} s .
\eeq
{}From this, one can obtain:
\beqa
\frac{d\sigma}{dt} = \frac{1}{16\pi s^2}
\left (\frac{1}{3\cdot4} \sum_{\rm spins} |{\cal M}|^2 \right),
\label{qqCN}
\eeqa
where we have included a factor of $1/3$ from the color average for
the incoming quarks.  As in the previous two subsections,
\eq{qqCN} can be expressed in terms of the
angle between the $u$ quark and the chargino in the center-of-momentum
frame, using
\beqa
t &=& \frac{1}{2} \left
[m_{\Ci}^2 + m_{\Nj}^2 -s + \lam^{1/2}(s,m_{\Ci}^2,m_{\Nj}^2) \cos\theta
\right ],
\\
u &=& \frac{1}{2}
\left
[m_{\Ci}^2 + m_{\Nj}^2 -s - \lam^{1/2}(s,m_{\Ci}^2,m_{\Nj}^2) \cos\theta
\right ] .
\eeqa
This process occurs in proton-antiproton and proton-proton collisions,
where $\sqrt{s}$ is not fixed, and the angle $\theta$ is different
than the lab frame angle. The observable cross-section depends crucially
on experimental cuts. Our result in \eq{qqCN} agrees with the
complete computation in \Ref{Beenakker:1999xh}. Earlier calculations
in special supersymmetric scenarios, e.g.~with photino mass
eigenstates, are given in refs.~\cite{Dawson:1983fw,Baer:1986vf}.


\subsection{\texorpdfstring{$\Ni \ra \Nj \Nk \Nl$}{N\texttilde\textiinferior \textrightarrow N\texttilde\textjinferior N\texttilde\textkinferior N\texttilde\textlinferior}}
\setcounter{equation}{0}
\setcounter{figure}{0}
\setcounter{table}{0}

Next we consider the decay of a neutralino $\Ni$ to three lighter
neutralinos: $\Nj,\Nk,\Nl$.  To the best of our knowledge,
this process has not been computed in the literature.
This decay is not likely to be
phenomenologically relevant, because a variety of two-body decay modes
will always be available. Furthermore, the calculation itself is quite
complicated because of the large number of Feynman diagrams
involved. Therefore, we consider this only as a matter-of-principle
example of a process with four external state Majorana fermions, and
will restrict ourselves to writing down the contributing matrix
element amplitudes.

At tree level, the decay can proceed via a virtual
$Z^0$ boson; the Feynman graphs are shown in
\fig{fig:neut1toneut2neut3neut4}.
\begin{figure}[!b]
\begin{center}
\begin{picture}(150,210)(0,-10)
\ArrowLine(-100,160)(-50,160)
\ArrowLine(-50,160)(-10,190)
\Photon(-50,160)(-10,130)56
\ArrowLine(-10,130)(30,160)
\ArrowLine(30,100)(-10,130)
\Text(-97,174)[]{$\chi^0_i
$}
\Text(43,165)[]{$\chi^0_k
$}
\Text(43,105)[]{${\chi^{0\,\dagger}_\ell}
$}
\Text(2,197)[]{$\chi^0_j
$}
\put(-45,130){$Z^0$}
\ArrowLine(120,160)(170,160)
\ArrowLine(170,160)(210,190)
\Photon(210,130)(170,160)56
\ArrowLine(250,160)(210,130)
\ArrowLine(210,130)(250,100)
\Text(123,174)[]{$\chi^0_i
$}
\Text(263,165)[]{${\chi^{0\,\dagger}_k}
$}
\Text(263,105)[]{$\chi^0_\ell
$}
\Text(222,197)[]{$\chi^0_j
$}
\put(175,130){$Z^0$}
\ArrowLine(-50,40)(-100,40)
\ArrowLine(-10,70)(-50,40)
\Photon(-50,40)(-10,10)56
\ArrowLine(-10,10)(30,40)
\ArrowLine(30,-20)(-10,10)
\Text(-97,54)[]{${\chi^{0\,\dagger}_i}
$}
\Text(43,45)[]{$\chi^0_k
$}
\Text(43,-15)[]{${\chi^{0\,\dagger}_\ell}
$}
\Text(2,77)[]{${\chi^{0\,\dagger}_j}
$}
\put(-45,10){$Z^0$}
\ArrowLine(170,40)(120,40)
\ArrowLine(210,70)(170,40)
\Photon(210,10)(170,40)56
\ArrowLine(250,40)(210,10)
\ArrowLine(210,10)(250,-20)
\Text(123,54)[]{${\chi^{0\,\dagger}_i}
$}
\Text(263,45)[]{${\chi^{0\,\dagger}_k}
$}
\Text(263,-15)[]{$\chi^0_\ell
$}
\Text(222,77)[]{${\chi^{0\,\dagger}_j}
$}
\put(175,10){$Z^0$}
\end{picture}
\end{center}
\caption{Four Feynman diagrams for $\Ni\ra\Nj\Nk\Nl$ in the MSSM via
  $Z^0$ exchange. There are four more where $\Nj\leftrightarrow\Nk$
  and another four where $\Nj\leftrightarrow\Nl$.}
\label{fig:neut1toneut2neut3neut4}
\end{figure}
In addition, it can proceed via the exchange of any of the
neutral scalar Higgs bosons of the MSSM, $\phi^0 = h^0,H^0,A^0$, as shown in
\fig{fig:neut1toneut2neut3neut4-2}.
\begin{figure}[!ht]
\begin{center}
\begin{picture}(150,210)(0,-10)
\ArrowLine(-100,160)(-50,160)
\ArrowLine(-10,190)(-50,160)
\DashLine(-50,160)(-10,130)5
\ArrowLine(30,160)(-10,130)
\ArrowLine(30,100)(-10,130)
\Text(-97,174)[]{$\chi^0_i
$}
\Text(43,165)[]{${\chi^{0\,\dagger}_k}
$}
\Text(43,105)[]{${\chi^{0\,\dagger}_\ell}
$}
\Text(2,197)[]{${\chi^{0\,\dagger}_j}
$}
\put(-75,130){$h^0,H^0,A^0$}
\ArrowLine(120,160)(170,160)
\ArrowLine(210,190)(170,160)
\DashLine(210,130)(170,160)5
\ArrowLine(210,130)(250,160)
\ArrowLine(210,130)(250,100)
\Text(123,174)[]{$\chi^0_i
$}
\Text(263,165)[]{$\chi^0_k
$}
\Text(263,105)[]{$\chi^0_\ell
$}
\Text(222,197)[]{${\chi^{0\,\dagger}_j}
$}
\put(145,130){$h^0,H^0,A^0$}
\ArrowLine(-50,40)(-100,40)
\ArrowLine(-50,40)(-10,70)
\DashLine(-50,40)(-10,10)5
\ArrowLine(30,40)(-10,10)
\ArrowLine(30,-20)(-10,10)
\Text(-97,54)[]{${\chi^{0\,\dagger}_i}
$}
\Text(43,45)[]{${\chi^{0\,\dagger}_k}
$}
\Text(43,-16)[]{${\chi^{0\,\dagger}_\ell}
$}
\Text(2,77)[]{$\chi^0_j
$}
\put(-75,10){$h^0,H^0,A^0$}
\ArrowLine(170,40)(120,40)
\ArrowLine(170,40)(210,70)
\DashLine(210,10)(170,40)5
\ArrowLine(210,10)(250,40)
\ArrowLine(210,10)(250,-20)
\Text(123,54)[]{${\chi^{0\,\dagger}_i}
$}
\Text(263,45)[]{$\chi^0_k
$}
\Text(263,-16)[]{$\chi^0_\ell
$}
\Text(222,77)[]{$\chi^0_j
$}
\put(145,10){$h^0,H^0,A^0$}
\end{picture}
\end{center}
\caption{Four Feynman diagrams for $\Ni \ra \Nj\Nk\Nl$ in the MSSM
  via $\phi^0 = h^0,\,H^0,\,A^0$ exchange. There are four more where $\Nj
  \leftrightarrow\Nk$ and another four where $\Nj \leftrightarrow\Nl$.}
\label{fig:neut1toneut2neut3neut4-2}
\end{figure}
Since any of the final
state neutralinos can directly couple to the initial state neutralino
there are two more diagrams for each one shown in
\figs{fig:neut1toneut2neut3neut4}{fig:neut1toneut2neut3neut4-2},
for a total of 48 tree-level
diagrams (counting each intermediate Higgs boson state as distinct).
In all cases, the four-momenta of the neutralinos $\Ni$, $\Nj$, $\Nk$,
$\stilde N_\ell$ are denoted $p_i$, $k_j$, $k_k$, $k_\ell$ respectively.

We obtain the sum of the four
diagrams in \fig{fig:neut1toneut2neut3neut4}
by implementing the rules of \fig{nnboson},
and using the 't Hooft-Feynman gauge:
\beqa \label{nnnn1}
i{\cal M}_Z^{(1)} = \frac{-i g^2/c_W^2}{(p_i-k_j)^2\BDminus m_Z^2}
\Bigl (
O_{ji}^{\prime\prime L} x_i \sigma_\mu  x^\dagger_j -
O_{ij}^{\prime\prime L}  y^\dagger_i \sigmabar_\mu y_j \Bigr )
\Bigl (
O_{k\ell}^{\prime\prime L}  x^\dagger_k \sigmabar^\mu y_\ell -
O_{\ell k}^{\prime\prime L} y_k \sigma^\mu  x^\dagger_\ell \Bigr )\,.
\phantom{xxx}
\eeqa
The external wave functions are
$x_i\equiv x(\boldsymbol{\vec p}_i,\lam_i)$,
$x_{j,k,\ell}\equiv x(\boldsymbol{\vec k}_{j,k,\ell},\lam_
{j,k,\ell})$, and analogously for $ x^\dagger_{i,j,k,\ell}$,
and $y_{i,j,k,\ell}$ and
$ y^\dagger_{i,j, k,\ell}$.
Note that we have factorized the sum of the four diagrams, taking advantage of
the common virtual boson line propagator.  By a judicious use of the
$\sigma$ or $\sigmabar$ version of the vertex rule, we have ensured
that the order of the four spinor wave functions is the same for each
of the four diagrams.  Hence, no additional relative minus signs are required.

The contributions from the diagrams related to these by permutations
can now be obtained from the appropriate substitutions
$(j\leftrightarrow k)$ and $(j\leftrightarrow \ell)$:
\beqa
i{\cal M}_Z^{(2)} &=& (-1)
\frac{-i g^2/c_W^2}{(p_i-k_k)^2\BDminus m_Z^2}
\Bigl (
O_{ki}^{\prime\prime L} x_i \sigma_\mu  x^\dagger_k -
O_{ik}^{\prime\prime L}  y^\dagger_i \sigmabar_\mu y_k \Bigr )
\Bigl (
O_{j\ell }^{\prime\prime L}  x^\dagger_j \sigmabar^\mu y_\ell -
O_{\ell j}^{\prime\prime L} y_j \sigma^\mu  x^\dagger_\ell \Bigr ),
\phantom{xxxxx}
\\
i{\cal M}_Z^{(3)} &=& (-1)
\frac{-i g^2/c_W^2}{(p_i-k_\ell)^2\BDminus m_Z^2}
\Bigl (
O_{\ell i}^{\prime\prime L} x_i \sigma_\mu  x^\dagger_\ell -
O_{i\ell}^{\prime\prime L}  y^\dagger_i \sigmabar_\mu y_\ell \Bigr )
\Bigl (
O_{kj}^{\prime\prime L}  x^\dagger_k \sigmabar^\mu y_j -
O_{jk}^{\prime\prime L} y_k \sigma^\mu  x^\dagger_j \Bigr ).
\eeqa
The first factors of $(-1)$ in $i{\cal M}_Z^{(2)}$ and
 $i{\cal M}_Z^{(3)}$ are present because
the order of the spinors in each case
appear in an odd permutation of the canonical
order set by $i{\cal M}_Z^{(1)}$. Note that if we were to
proceed to a computation of the decay rate, the very first step
would be to apply the Fierz relations of
\eqst{twocompfierza}{twocompfierzc}
to eliminate all of the $\sigma$ and $\sigmabar$ matrices
in the above amplitudes.

The diagrams in \fig{fig:neut1toneut2neut3neut4-2} combine
to give a contribution:
\beqa
i {\cal M}_{\phi^0}^{(1)} =
\frac{\BDneg i}{(p_i- k_j)^2 \BDminus m_{\phi^0}^2}
(Y^{ij} x_i y_j + Y_{ij}  y^\dagger_i  x^\dagger_j)
(Y^{k\ell} y_k y_\ell + Y_{k\ell}  x^\dagger_k  x^\dagger_\ell)\,,
\eeqa
where we have
used the Feynman rules of \fig{inohiggsboson}, and
adopted the shorthand notation
$Y^{ij} = (Y_{ij})^* = Y^{\phi^0\chi^0_i\chi^0_j}$.
Again we have factored the amplitude using the common
virtual boson propagator. As in the $Z$-exchange diagrams,
the other contributions can be obtained by the appropriate
substitutions:
\beqa
i {\cal M}_{\phi^0}^{(2)} &=& (-1)
\frac{\BDneg i}{(p_i- k_k)^2 \BDminus m_{\phi^0}^2}
(Y^{ik} x_i y_k + Y_{ik}  y^\dagger_i  x^\dagger_k)
(Y^{j\ell} y_j y_\ell + Y_{j\ell}  x^\dagger_j  x^\dagger_\ell)\,,
\\
i {\cal M}_{\phi^0}^{(3)} &=& (-1)
\frac{\BDneg i}{(p_i- k_\ell)^2 \BDminus m_{\phi^0}^2}
(Y^{i\ell} x_i y_\ell + Y_{i\ell}  y^\dagger_i  x^\dagger_\ell)
(Y^{kj} y_k y_j + Y_{kj}  x^\dagger_k  x^\dagger_j)\,.
\eeqa
The first factors of $(-1)$ in $i {\cal M}_{\phi^0}^{(2)}$
and $i {\cal M}_{\phi^0}^{(3)}$ are needed because the spinors
in each case are in an odd permutation of the canonical order
established earlier.

The total matrix element is obtained by adding all the contributing
diagrams:
\beq
{\cal M} = \sum_{n=1}^3 {\cal M}_Z^{(n)} +
\sum_{\phi^0 
}\, \sum_{n=1}^3 {\cal M}_{\phi^0}^{(n)}\,.
\eeq
Squaring the matrix element, dividing by $2M_{\tilde N_i}$,
and integrating over phase space yields the total decay rate.
Note that final states differing by the interchange of
identical particles must be considered as a single state, counted
once~\cite{Novozhilov}.
Given an $N$-body final state made up of $\nu_r$
particles of type $r$ (where $r\leq N$), we define a statistical factor $S$,
\beq
S=\prod_r \nu_r!\,,\qquad {\rm where}\qquad  \sum_r\nu_r=N\,.
\eeq
Then, in computing the total decay rate, the integration over the total
phase space must be divided by $S$ to avoid over-counting.
In the present example, $N=3$ with $S=2$ [or $S=6$] in the case of
two [or three] identical neutralinos in the final state, respectively


\subsection{Three-body slepton decays:
\texorpdfstring{$\widetilde \ell_R^- \ra \ell^- \tau^\pm\, \widetilde\tau_1^\mp$
for $\ell = e,\mu$}{\textell\texttilde\textscr\textminussuperior\textrightarrow\textell\textminussuperior\texttau\textplussuperior\texttau\texttilde\textoneinferior\textminussuperior,
\textell\textminussuperior\texttau\textminussuperior\texttau\texttilde\textoneinferior\textplussuperior
\nobreakspace for \textell=e,\textmu}}
\setcounter{equation}{0}
\setcounter{figure}{0}
\setcounter{table}{0}

We next consider the three-body decays of sleptons
through a virtual neutralino. The usual assumption in supersymmetric
phenomenology is that these decays will have a very small branching
fraction, because a two-body decay to a lighter neutralino and lepton
is always open.  However, in Gauge Mediated Supersymmetry Breaking
models with a non-minimal messenger sector, the sleptons can be
lighter than the lightest neutralino \cite{Dimopoulos:1996vz,
Giudice:1998bp}.  In that case, the mostly R-type smuon and
selectron, $\stilde \mu_R$ and $\tilde e_R$, will decay by $\widetilde
\ell_R^- \ra \ell^- \tau^\pm\widetilde\tau _1^\mp $.  The lightest
stau mass eigenstate, $\stilde\tau_1^\pm$, is a mixture of the weak
eigenstates $\stilde\tau_L^\pm$ and $\stilde
\tau_R^\pm$, as described in \app{K.4}:
\beq
\stilde\tau_1^-
=
R_{\tilde \tau_1}^* \stilde \tau_R^-
+
L_{\tilde \tau_1}^* \stilde\tau_L^-
,
\eeq
and $\stilde \tau_1^+ = (\stilde \tau_1^-)^*$,
while the $\stilde \mu_R$ and $\tilde e_R$ are taken to be
unmixed.

First consider the decay
$\widetilde \ell_R^- \ra \ell^- \tau^+ \widetilde\tau_1^- $,
which proceed by the diagrams in the top row of
\fig{fig:smuRtomutaustau}.
\begin{figure}[tbp]
\begin{center}
\begin{picture}(250,102)(120,-20)
\DashLine(170,40)(120,40)5
\ArrowLine(210,70)(170,40)
\ArrowLine(210,10)(170,40)
\ArrowLine(210,10)(250,40)
\DashLine(210,10)(250,-20)5
\Text(123,54)[]{$\stilde \ell_R^-\, (p)$}
\Text(263,50)[]{$\bar\tau \,(k_2,\lam_2)$}
\Text(263,-8)[]{$\stilde\tau_1^-\,(k_3)$}
\Text(225,77)[]{${\bar\ell}^\dagger \, (k_1,\lam_1)$}
\put(172,10){$\chi^0_j$}
\end{picture}
\begin{picture}(150,102)(-100,-20)
\DashLine(-50,40)(-100,40)5
\ArrowLine(-10,70)(-50,40)
\ArrowLine(-30,25)(-50,40)
\ArrowLine(-30,25)(-10,10)
\ArrowLine(30,40)(-10,10)
\DashLine(30,-20)(-10,10)5
\Text(-97,54)[]{$\stilde \ell_R^-\, (p)$}
\Text(43,50)[]{${\tau}^\dagger \,(k_2,\lam_2)$}
\Text(43,-8)[]{$\stilde\tau_1^-(k_3)$}
\Text(2,77)[]{${\bar\ell}^\dagger (k_1,\lam_1)$}
\put(-47,10){$\chi_j^0$}
\end{picture}
\end{center}
\begin{center}
\begin{picture}(250,110)(-100,108)
\DashLine(-100,160)(-50,160)5
\ArrowLine(-10,190)(-50,160)
\ArrowLine(-10,130)(-50,160)
\ArrowLine(-10,130)(30,160)
\DashLine(30,100)(-10,130)5
\Text(-97,174)[]{$\stilde \ell_R^-\, (p)$}
\Text(43,170)[]{$\tau \,(k_2,\lam_2)$}
\Text(43,112)[]{$\stilde\tau_1^+\, (k_3)$}
\Text(2,197)[]{${\bar\ell}^\dagger\,(k_1,\lam_1)$}
\put(-47,130){$\chi^0_j$}
\end{picture}
\begin{picture}(150,110)(120,108)
\DashLine(120,160)(170,160)5
\ArrowLine(210,190)(170,160)
\ArrowLine(190,145)(210,130)
\ArrowLine(190,145)(170,160)
\ArrowLine(250,160)(210,130)
\DashLine(210,130)(250,100)5
\Text(123,174)[]{$\stilde \ell_R^-\, (p)$}
\Text(263,170)[]{${\bar\tau}^\dagger \,(k_2,\lam_2)$}
\Text(263,112)[]{$\stilde\tau_1^+\, (k_3)$}
\Text(222,197)[]{${\bar\ell}^\dagger\,(k_1,\lam_1)$}
\put(172,130){$\chi^0_j$}
\end{picture}
\end{center}
\caption{Feynman diagrams for the three-body slepton decays
$\stilde\ell_R^- \ra \ell^- \tau^+ \stilde\tau_1^-$
(top row) and
$\stilde\ell_R^- \ra \ell^- \tau^- \stilde\tau_1^+$
(bottom row)
in the MSSM.}
\label{fig:smuRtomutaustau}
\end{figure}
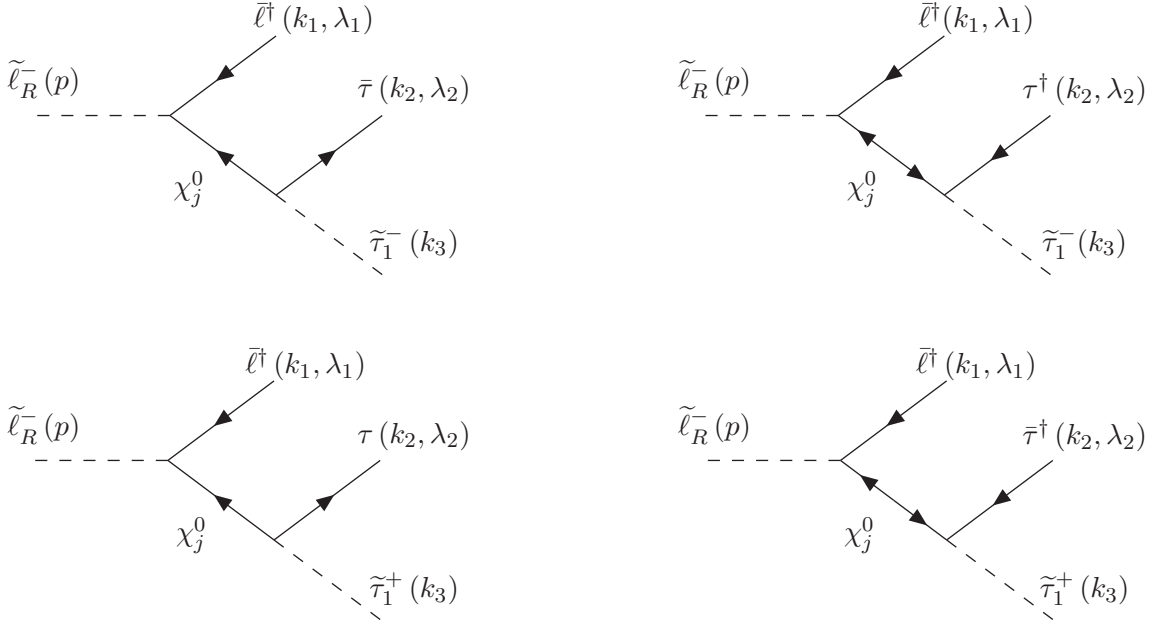
The momenta and polarizations of the particles are also indicated
on the diagram.
Using the Feynman rules of \fig{nqsqmixed}, we find that
the amplitudes
of these two diagrams, for each neutralino $\tilde N_j$
exchanged, are:
\beqa
i{\cal M}_1 &=& (-i a_j^{\tilde \ell *})(-i a_j^{\tilde \tau })
\,
y_1 \biggl [
\frac{- i (p - k_1) \newcdot \sigma}{
(p-k_1)^2 \BDminus m^2_{\tilde N_j}}
\biggr ]  x^\dagger_2 ,
\label{eq:threebodysleptonone}
\\
i{\cal M}_2 &=& (-i a_j^{\tilde \ell *})(-i b_j^{\tilde \tau})
\,
y_1 \Bigl [
\frac{\BDpos i m_{\tilde N_j}}{
(p-k_1)^2 \BDminus m^2_{\tilde N_j}}
\Bigr ] y_2 .
\label{eq:threebodysleptontwo}
\eeqa
where
\beqa
a_j^{\tilde \ell} &=& \sqrt{2} g' N_{j1},
\\
a_j^{\tilde \tau} &=&
Y_\tau  N_{j3} L_{\tilde \tau_1}^*
+ \sqrt{2} g' N_{j1} R_{\tilde \tau_1}^*
,
\\
b_j^{\tilde \tau} &=&
Y_\tau N_{j3}^*  R_{\tilde \tau_1}^*
- \frac{1}{\sqrt{2}} (g N_{j2}^* + g' N_{j1}^*) L_{\tilde \tau_1}^*
.
\eeqa
The spinor wave function factors are
$y_1 = y (\boldsymbol{\vec k}_1, \lam_1)$,
$y_2 = y (\boldsymbol{\vec k}_2, \lam_2)$, and
$ x^\dagger_2 =  x^\dagger (\boldsymbol{\vec k}_2, \lam_2)$.

In the following, we will use the kinematic variables
\beqa
z_\ell
&\equiv& \BDpos 2 p\newcdot k_1/m_{\tilde \ell_R}^2
= 2 E_\ell/m_{\tilde \ell_R},\qquad  z_\tau
\equiv \BDpos 2 p\newcdot k_2/m_{\tilde \ell_R}^2
= 2 E_\tau/m_{\tilde \ell_R} ,
\\
r_{\tilde N_j} &\equiv& m_{\tilde N_j}/m_{\tilde \ell_R},
\qquad\qquad\qquad\quad\;\;\;
r_{\tilde \tau} \equiv m_{\tilde \tau_1}/m_{\tilde \ell_R} ,
\\
r_{\tau} &\equiv& m_{\tau}/m_{\tilde \ell_R} ,
\qquad\qquad \qquad\qquad\;\,
r_{\ell} \equiv m_{\ell}/m_{\tilde \ell_R} .
\eeqa
The total amplitude then can be written as
\beqa
{\cal M} = \sum_{j=1}^4 \left [c_j y_1 (p-k_1) \newcdot \sigma  x^\dagger_2
+ d_j y_1 y_2 \right ]\,,
\eeqa
where
\beqa
c_j &=& \BDneg a_j^{\tilde \ell *} a_j^{\tilde \tau}/
[m_{\ell_R}^2 (r_{\tilde N_j}^2 - 1 + z_\ell)],
\\
d_j &=& a_j^{\tilde \ell *} b_j^{\tilde \tau} m_{\tilde N_j}/
[m_{\ell_R}^2 (r_{\tilde N_j}^2 - 1 + z_\ell)].
\eeqa
We consistently
neglect the electron and muon masses and
Yukawa couplings (so $r_\ell = 0$) in the matrix elements,
but not below in the kinematic integration over phase space, where the
muon mass can be important.

Using eqs.~(\ref{eq:conbil}) and (\ref{eq:conbilsig}), we find
\beqa
|{\cal M}|^2 =\sum_{j,k}
&& \Bigl [
c_j c_k^*\,
y_1 (p-k_1) \newcdot\sigma  x^\dagger_2 \,
x_2 (p - k_1) \newcdot\sigma  y^\dagger_1
+ d_j d_k^* y_1 y_2\,  y^\dagger_2  y^\dagger_1
\nonumber \\ &&
+ c_j d_k^* y_1 (p-k_1) \newcdot\sigma  x^\dagger_2 \,  y^\dagger_2  y^\dagger_1
+ c_j^* d_k x_2 (p - k_1) \newcdot\sigma  y^\dagger_1\,y_1 y_2
\Bigr ] .
\eeqa
Summing over the lepton spins using
\eqst{xxdagsummed}{ydagxdagsummed} gives
\beqa
\sum_{\lam_1,\lam_2} |{\cal M}|^2  = \sum_{j,k}
&& \Bigl [
c_j c_k^* {\rm Tr} [
(p-k_1) \newcdot \sigma k_2 \newcdot \sigmabar
(p-k_1) \newcdot \sigma k_1 \newcdot \sigmabar ]
+ d_j d_k^* {\rm Tr} [ k_2 \newcdot \sigma k_1 \newcdot \sigmabar ]
\nonumber \\ &&
\BDminus c_j d_k^* m_\tau {\rm Tr}[ (p-k_1) \newcdot \sigma k_1 \newcdot \sigmabar]
\BDminus c_j^* d_k m_\tau{\rm Tr}[ (p-k_1) \newcdot \sigma k_1 \newcdot
\sigmabar ]
\Bigr]
.
\eeqa
Taking the traces using eqs.~(\ref{trssbar}) and (\ref{trssbarssbar})
yields
\beqa
\sum_{\rm spins} |{\cal M}|^2  = \sum_{j,k}
&& \Bigl \{
c_j c_k^* [4 k_1 \newcdot (p-k_1) k_2 \newcdot (p-k_1)
-2 k_1 \newcdot k_2 (p-k_1)^2]
\BDplus 2 d_j d_k^* k_1 \newcdot k_2
\nonumber \\ &&
- 4 {\rm Re}[c_j d_k^*] m_\tau k_1 \newcdot (p-k_1)
\Bigr \}
\nonumber \\
=
\sum_{j,k}
&& \Bigl \{
c_j c_k^* m_{\tilde \ell_R}^4 [
(1 - z_\ell)(1- z_\tau) - r_{\tilde \tau}^2 + r_{\tau}^2 ]
\nonumber \\ &&
+ d_j d_k^* m_{\tilde \ell_R}^2
(z_\ell + z_\tau -1 + r_{\tilde \tau}^2 - r_{\tau}^2)
\BDminus 2  {\rm Re}[c_j d_k^*] m_\tau m_{\tilde \ell_R}^2 z_\ell
\Bigr \}\,.
\eeqa
The differential decay rate for
$\widetilde \ell_R^- \ra \ell^- \tau^+ \widetilde\tau_1^- $
then follows:
\beqa
\frac{d^2\Gamma}{dz_\ell dz_\tau} =
\frac{m_{\tilde\ell_R}}{256\pi^3}
\biggl (\sum_{\rm spins} |{\cal M}|^2  \biggr )\,.
\eeqa
The total decay rate in that channel can be found by integrating over $z_\ell$,
$z_\tau$, with the limits (see for example ref.~\cite{BP}):
\beqa
&&
2 r_\ell < z_\ell < 1 + r_\ell^2 - (r_\tau + r_{\tilde \tau})^2
,\\
&& (z_\tau)_{\rm min}<z_\tau<(z_\tau)_{\rm max}\,,
\eeqa
where
\beq
(z_{\tau})_{\rm{min}\,,\,\rm{max}}=
\frac{1}{2(1-z_\ell + r_\ell^2)}
\Bigl [ (2 - z_\ell)(1 + r_\ell^2 + r_\tau^2 - r_{\tilde \tau}^2 - z_\ell)
\mp (z_\ell^2 - 4 r_\ell^2)^{1/2}
\lambda^{1/2}(1 + r_\ell^2 - z_\ell, r_\tau^2, r_{\tilde \tau}^2 ) \Bigr ]
\,,
\eeq
and the triangle function $\lam^{1/2}$ is defined in
\eq{eq:deftrianglefunction}.

Now we turn to the competing decay
$\stilde\ell_R^- \ra \ell^- \tau^- \stilde\tau_1^+$, with diagrams
appearing in the bottom row of
\fig{fig:smuRtomutaustau}.
By appealing again to the Feynman rules of \fig{cqsqmixed},
we find that the amplitude has exactly the same form as in
eqs.~(\ref{eq:threebodysleptonone}) and
(\ref{eq:threebodysleptontwo}), except now
with $a_j^{\tilde \tau} \leftrightarrow b_j^{\tilde \tau *}$.
Therefore, the entire previous calculation goes through precisely
as before, but now with
\beqa
c_j &=& \frac{\BDneg a_j^{\tilde \ell *} b_j^{\tilde \tau *}}
{m_{\ell_R}^2 (r_{\tilde N_j}^2 - 1 + z_\ell)},
\\
d_j &=& \frac{a_j^{\tilde \ell *} a_j^{\tilde \tau *} m_{\tilde N_j}}
{m_{\ell_R}^2 (r_{\tilde N_j}^2 - 1 + z_\ell)}.
\eeqa
The differential decay widths found above can be integrated to find
the total decay widths. The results agree with ref.~\cite{AKM},
except that the signs of the coefficient $c^{(3)}_{ij}$ and
$c^{(4)}_{ij}$ in the published version of that
paper are incorrect; the arXiv eprint version has been corrected.
(Also, the notations for the sfermion mixing angle are different in
that paper.)
If $m_{\tilde \ell_R} - m_{\tilde
\tau_1} - m_{\tau}$ is not too large, the resulting decays can have a
macroscopic length in a detector, and the ratio of the two decay modes
can provide an interesting probe of the supersymmetric Lagrangian.

\subsection{Neutralino decay to photon and
Goldstino: \texorpdfstring{$\stilde N_i \ra \gamma \widetilde G$}{N\texttilde\textiinferior\textrightarrow \textgamma G\texttilde}}
\setcounter{equation}{0}
\setcounter{figure}{0}
\setcounter{table}{0}

The Goldstino $\stilde G$ is a massless Weyl fermion that couples to the
neutralino and photon fields according to the
non-renormalizable Lagrangian term \cite{goldstinoint}:
\beq
\mathscr{L}\, =\, \BDneg \frac{a_i}{2}
(
\chi_i^0
\sigma^\mu
\sigmabar^\rho
\sigma^\nu
\partial_\mu {\stilde G}^\dagger
) \,
( \partial_\nu A_\rho - \partial_\rho A_\nu ) + {\rm h.c.}
\eeq
Here $\chi_i^0$ is the left-handed two-component
fermion field that corresponds to the neutralino $\stilde N_i$ particle,
$\stilde G$ is the two-component fermion field corresponding to the
(nearly) massless Goldstino,
and the effective coupling is
\beq
a_i \equiv \frac{1}{\sqrt{2} \langle F \rangle}
(N_{i1}^* \cos\theta_W + N_{i2}^* \sin\theta_W ) ,
\eeq
where $N_{ij}$ the mixing matrix for the neutralinos
[see eq.~(\ref{eq:neutmix})],
and $\langle F \rangle$ is the $F$-term expectation value
associated with
supersymmetry breaking.
Therefore $\stilde N_i$ can decay to $\gamma$ plus $\stilde G$
through the diagrams shown in \fig{fig:neut1togammaG},
with amplitudes:
\beqa
i{\cal M}_{1} &=& \BDpos i \frac{a_i}{2} \>
x_{\tilde N}
k_{\tilde G} \newcdot \sigma
\left (
\varepsilon^* \newcdot \sigmabar \, k_\gamma \newcdot \sigma
-
k_\gamma \newcdot \sigmabar \,\varepsilon^* \newcdot \sigma
\right )  x^\dagger_{\tilde G}
\, ,
\\
i{\cal M}_{2} &=& \BDneg i \frac{a_i^*}{2} \,
 y^\dagger_{\tilde N}
k_{\tilde G} \newcdot \sigmabar
\left (
\varepsilon^* \newcdot \sigma \,k_\gamma \newcdot \sigmabar
-
k_\gamma \newcdot \sigma \,\varepsilon^* \newcdot \sigmabar
\right )
y_{\tilde G}
\, .
\eeqa
Here $x_{\tilde N} \equiv x(\boldsymbol{\vec p},\lam_{\tilde N})$,
$ y^\dagger_{\tilde N} \equiv  y^\dagger (\boldsymbol{\vec p},\lam_{\tilde N})$,
and
$ x^\dagger_{\tilde G} \equiv
 x^\dagger (\boldsymbol{\vec k}_{\tilde G},\lam_{\tilde G})$,
$y_{\tilde G} \equiv
y (\boldsymbol{\vec k}_{\tilde G},\lam_{\tilde G})$,
and $\varepsilon^* =
\varepsilon^* (\boldsymbol{\vec k}_{\gamma},\lam_{\gamma})$
are the external wave function factors for the neutralino, Goldstino,
and photon, respectively.
Using the on-shell condition $k_\gamma \newcdot \varepsilon^* = 0$,
we have
$
k_\gamma \newcdot \sigma \varepsilon^* \newcdot\sigmabar
= -\varepsilon^*\newcdot\sigma k_\gamma\newcdot\sigmabar
$ and
$
k_\gamma \newcdot \sigmabar \varepsilon^* \newcdot\sigma
 =
-\varepsilon^* \newcdot\sigmabar k_\gamma\newcdot\sigma
$ from
eqs.~(\ref{eq:ssbarsym})
and (\ref{eq:sbarssym}).
So we can rewrite the total amplitude as
\beq
{\cal M} = {\cal M}_1 + {\cal M}_2 =
x_{\tilde N} A  x^\dagger_{\tilde G} +  y^\dagger_{\tilde N} B y_{\tilde G}\,,
\eeq
where
\beqa
A &=&
\BDpos a_i \,k_{\tilde G}\newcdot\sigma\,  \varepsilon^*\newcdot\sigmabar\,
    k_\gamma \newcdot \sigma,
\\
B &=&
\BDneg
a_i^* \, k_{\tilde G}\newcdot\sigmabar\,  \varepsilon^*\newcdot\sigma\,
    k_\gamma \newcdot \sigmabar .
\eeqa
The complex square of the matrix element is therefore
\beqa
|{\cal M}|^2 &=&
x_{\tilde N} A  x^\dagger_{\tilde G} x_{\tilde G} \hat{A}  x^\dagger_{\tilde N}
+  y^\dagger_{\tilde N} B y_{\tilde G}  y^\dagger_{\tilde G} \hat{B} y_{\tilde N}
+ x_{\tilde N} A  x^\dagger_{\tilde G}  y^\dagger_{\tilde G} \hat{B} y_{\tilde N}
+  y^\dagger_{\tilde N} B y_{\tilde G} x_{\tilde G} \hat{A}  x^\dagger_{\tilde N}
,
\phantom{xxx}
\eeqa
where $\hat{A}$ and $\hat{B}$ are obtained from $A$ and $B$
by reversing the order of the $\sigma$ and $\sigmabar$ matrices
and taking the complex conjugates of $a_i$ and $\varepsilon$
[cf.~\eq{ccspinorbilinears} and the associated text].

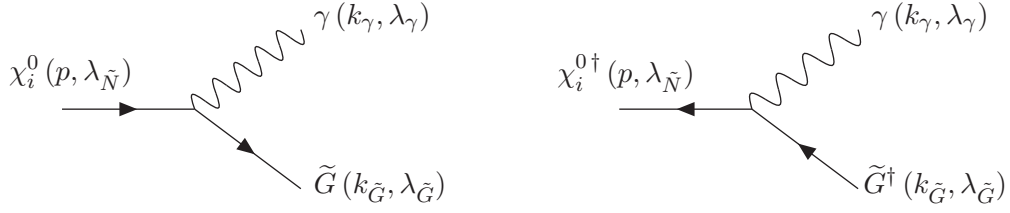
\begin{figure}[t!]
\begin{center}
\begin{picture}(330,67)(-80,10)
\ArrowLine(-90,40)(-40,40)
\Photon(-40,40)(0,70){5}{5}
\ArrowLine(-40,40)(0,10)
\Text(-87,54)[]{$\chi^0_i\,(p,\lam_{\tilde N})$}
\Text(30,12)[]{$\stilde G \,(k_{\tilde G},\lam_{\tilde G})$}
\Text(27,75)[]{$\gamma\, (k_\gamma,\lambda_\gamma)$}
\ArrowLine(170,40)(120,40)
\Photon(170,40)(210,70)54
\ArrowLine(210,10)(170,40)
\Text(123,54)[]{${\chi^{0\,\dagger}_i}\, (p,\lam_{\tilde N})$}
\Text(240,12)[]{${\stilde G}^\dagger \,(k_{\tilde G},\lam_{\tilde G})$}
\Text(238,75)[]{$\gamma\, (k_\gamma,\lambda_\gamma)$}
\end{picture}
\end{center}
\caption{The two Feynman diagrams for $\stilde N_i \ra \gamma\stilde G$
in supersymmetric models with a light Goldstino.}
\label{fig:neut1togammaG}
\end{figure}

Summing over the Goldstino spins using \eqst{xxdagsummed}{ydagxdagsummed}
now yields:
\beqa
\sum_{\lambda_{\tilde G}}
|{\cal M}|^2 &=&
\BDpos
x_{\tilde N} A k_{\tilde G} \newcdot \sigmabar \hat{A}  x^\dagger_{\tilde N}
\BDplus
 y^\dagger_{\tilde N} B  k_{\tilde G} \newcdot \sigma \hat{B} y_{\tilde N}.
\eeqa
(The $A,\hat B$ and $\hat A,B$ cross terms vanish because
of $m_{\tilde G} = 0$.)
Averaging over the neutralino spins using
eqs.~(\ref{xxdagsummed}) and (\ref{yydagsummed}), we find
\beqa
\frac{1}{2} \sum_{\lambda_{\tilde N}, \lambda_{\tilde G}}
|{\cal M}|^2 &=&
\frac{1}{2}
{\rm Tr}[A k_{\tilde G} \newcdot \sigmabar \hat{A} p \newcdot \sigmabar ]
+ \frac{1}{2}
{\rm Tr}[B  k_{\tilde G} \newcdot \sigma \hat{B} p \newcdot \sigma]
\nonumber \\
&=&
\frac{1}{2} |a_i|^2 {\rm Tr} [
\varepsilon^* \newcdot \sigmabar\,
k_\gamma \newcdot \sigma\,
k_{\tilde G} \newcdot \sigmabar\,
k_\gamma \newcdot \sigma\,
\varepsilon \newcdot \sigmabar\,
k_{\tilde G} \newcdot \sigma\,
p \newcdot \sigmabar \,
k_{\tilde G} \newcdot \sigma]
+ (\sigma \leftrightarrow \sigmabar) .\phantom{xxx}
\eeqa
We now use
\beqa
k_\gamma \newcdot \sigma\,
k_{\tilde G}  \newcdot \sigmabar\,
k_\gamma \newcdot \sigma\,
&=&
\BDpos 2 k_{\tilde G} \newcdot k_\gamma\,
k_\gamma \newcdot \sigma,
\\
k_{\tilde G} \newcdot \sigma\,
p \newcdot \sigmabar\,
k_{\tilde G} \newcdot \sigma\,
&= &
\BDpos 2 k_{\tilde G} \newcdot p\,
k_{\tilde G} \newcdot \sigma ,
\eeqa
which follow from eq.~(\ref{eq:simplifyssbars}), and the corresponding
identities with $\sigma \leftrightarrow \sigmabar$,
to obtain:
\beqa
\frac{1}{2} \sum_{\lambda_{\tilde N}, \lambda_{\tilde G}}
|{\cal M}|^2 &=&
2 |a_i|^2 (k_{\tilde G} \newcdot k_\gamma)
(k_{\tilde G} \newcdot p)
{\rm Tr} [
\varepsilon^* \newcdot \sigmabar\,
k_\gamma \newcdot \sigma\,
\varepsilon \newcdot \sigmabar\,
k_{\tilde G} \newcdot \sigma]
+ (\sigma \leftrightarrow \sigmabar) .
\eeqa
Applying the photon spin-sum identity
\beq \label{photonspinsum}
\sum_{\lambda_\gamma} \varepsilon^\mu \varepsilon^{\nu *} =
\BDneg \metric^{\mu\nu}\,,
\eeq
and the trace identities eq.~(\ref{trssbarssbar}) and (\ref{trsbarssbars}),
we get
\beqa
\frac{1}{2} \sum_{\lambda_\gamma, \lambda_{\tilde N}, \lambda_{\tilde G}}
|{\cal M}|^2 &=&
\BDpos 16 |a_i|^2 (k_{\tilde G} \newcdot k_\gamma)^2
(k_{\tilde G} \newcdot p)
= 2 |a_i|^2 m_{\tilde N_i}^6\,.
\eeqa
So, the decay rate is \cite{NGgamma,Dimopoulos:1996vz}:
\beq
\Gamma(\stilde N_i \ra \gamma \GG ) =
\frac{1}{16 \pi m_{\stilde N_i}} \left ( \frac{1}{2}
\sum_{\lambda_\gamma, \lambda_{\tilde N}, \lambda_{\tilde G}}
|{\cal M}|^2\right )
= |N_{i1} \cos\theta_W + N_{i2}\sin\theta_W|^2
\frac{m_{\stilde N_i}^5}{16 \pi |\langle F \rangle|^2}
.
\eeq


\subsection{Gluino pair production from gluon fusion:
\texorpdfstring{$gg \ra \stilde g \stilde g$}{gg\textrightarrow g\texttilde g\texttilde}}
\setcounter{equation}{0}
\setcounter{figure}{0}
\setcounter{table}{0}

In this subsection we will compute the cross-section for the process
$gg \rightarrow \stilde g \stilde g$. The relevant Feynman diagrams
are shown in \fig{fig:gg2ginogino}.
The initial state gluons have $SU(3)_c$ adjoint representation
indices $a$ and $b$, with momenta $p_1$ and $p_2$ and
polarization vectors
$\varepsilon_1^\mu = \varepsilon^\mu(\boldsymbol{\vec p}_1,\lam_1)$ and
$\varepsilon_2^\mu =
\varepsilon^\mu(\boldsymbol{\vec p}_2,\lam_2)$, respectively.
The final state gluinos carry adjoint representation
indices $c$ and $d$,
with
momenta $k_1$ and $k_2$ and wave function spinors
$ x^\dagger_1 =  x^\dagger (\boldsymbol{\vec k}_1,\lam_1')$ or
$y_1 = y (\boldsymbol{\vec k}_1,\lam_1')$
and
$ x^\dagger_2 =  x^\dagger (\boldsymbol{\vec k}_2,\lam_2')$ or
$y_2 = y (\boldsymbol{\vec k}_2,\lam_2')$, respectively.
\begin{figure}[tp]
\begin{center}
\begin{picture}(450,88)(-210,157)
\Gluon(-200,225)(-152,197.5){4}{5}
\Gluon(-200,170)(-152,197.5){4}{5}
\Gluon(-152,197.5)(-98,197.5){4}{5}
\ArrowLine(-98,197.5)(-50,225)
\ArrowLine(-50,170)(-98,197.5)
\put(-205,233){$g_a \,(p_1,\lam_{1})$}
\put(-205,158){$g_b \,(p_2,\lam_{2})$}
\put(-57,230){$\stilde g_c \,(k_1,\lam_1')$}
\put(-55,158){$\stilde g_d^\dagger \,(k_2,\lam_2')$}
\put(-129,210){$g_e$}
\Gluon(50,225)(98,197.5){4}{5}
\Gluon(50,170)(98,197.5){4}{5}
\Gluon(98,197.5)(152,197.5){4}{5}
\ArrowLine(200,225)(152,197.5)
\ArrowLine(152,197.5)(200,170)
\put(45,234){$g_a$}
\put(45,159){$g_b$}
\put(193,230){$\stilde g_c^\dagger$}
\put(195,158){$\stilde g_d$}
\put(121,210){$g_e$}
\end{picture}
\end{center}
\begin{center}
\begin{picture}(450,112)(-210,155)
\Gluon(-200,225)(-125,225){4}{7}
\Gluon(-200,170)(-125,170){4}{7}
\ArrowLine(-125,170)(-125,225)
\ArrowLine(-125,225)(-50,225)
\ArrowLine(-50,170)(-125,170)
\put(-205,235){$g_a$}
\put(-205,155){$g_b$}
\put(-57,233){$\stilde g_c$}
\put(-55,158){$\stilde g_d^\dagger$}
\put(-140,197.5){$\stilde g_e$}
\Gluon(50,225)(125,225){4}{7}
\Gluon(50,170)(125,170){4}{7}
\ArrowLine(200,225)(125,225)
\ArrowLine(125,225)(125,170)
\ArrowLine(125,170)(200,170)
\put(45,235){$g_a$}
\put(45,155){$g_b$}
\put(193,230){$\stilde g_c^\dagger$}
\put(195,158){$\stilde g_d$}
\put(110,197.5){$\stilde g_e$}
\end{picture}
\end{center}
\begin{center}
\begin{picture}(450,111)(-210,154)
\Gluon(-200,225)(-125,225){4}{7}
\Gluon(-200,170)(-125,170){4}{7}
\ArrowLine(-125,197.5)(-125,225)
\ArrowLine(-125,197.5)(-125,170)
\ArrowLine(-125,225)(-50,225)
\ArrowLine(-125,170)(-50,170)
\put(-205,235){$g_a$}
\put(-205,155){$g_b$}
\put(-57,230){$\stilde g_c$}
\put(-55,158){$\stilde g_d$}
\put(-140,197.5){$\stilde g_e$}
\Gluon(50,225)(125,225){4}{7}
\Gluon(50,170)(125,170){4}{7}
\ArrowLine(125,170)(125,197.5)
\ArrowLine(125,225)(125,197.5)
\ArrowLine(200,225)(125,225)
\ArrowLine(200,170)(125,170)
\put(45,235){$g_a$}
\put(45,155){$g_b$}
\put(193,230){$\stilde g_c^\dagger$}
\put(195,158){${\stilde g_d^\dagger}$}
\put(110,197.5){$\stilde g_e$}
\end{picture}
\end{center}
\begin{center}
\begin{picture}(450,112)(-210,43)
\Gluon(-200,115)(-125,115){4}{7}
\Gluon(-200,60)(-125,60){4}{7}
\ArrowLine(-125,115)(-125,60)
\Line(-125,115)(-91.25,90.25)
\ArrowLine(-50,60)(-83.75,84.75)
\Line(-87.5,87.5)(-125,60)
\ArrowLine(-87.5,87.5)(-50,115)
\put(-205,125){$g_a$}
\put(-205,46){$g_b$}
\put(-57,120){${\stilde g_c}$}
\put(-55,48){${\stilde g^\dagger_d}$}
\put(-140,87.5){$\stilde g_e$}
\Gluon(50,115)(125,115){4}{7}
\Gluon(50,60)(125,60){4}{7}
\ArrowLine(125,60)(125,115)
\ArrowLine(166.25,84.75)(200,60)
\Line(125,115)(158.75,90.25)
\ArrowLine(200,115)(162.5,87.5)
\Line(125,60)(162.5,87.5)
\put(45,125){$g_a$}
\put(45,46){$g_b$}
\put(193,120){${\stilde g^\dagger_c}$}
\put(195,48){${\stilde g_d}$}
\put(110,87.5){$\stilde g_e$}
\end{picture}
\end{center}
\begin{center}
\begin{picture}(450,111)(-210,43)
\Gluon(-200,115)(-125,115){4}{7}
\Gluon(-200,60)(-125,60){4}{7}
\ArrowLine(-125,87.5)(-125,60)
\ArrowLine(-125,87.5)(-125,115)
\Line(-125,115)(-91.25,90.25)
\ArrowLine(-83.75,84.75)(-50,60)
\Line(-87.5,87.5)(-125,60)
\ArrowLine(-87.5,87.5)(-50,115)
\put(-205,125){$g_a$}
\put(-205,46){$g_b$}
\put(-57,120){${\stilde g_c}$}
\put(-55,48){${\stilde g_d}$}
\put(-140,87.5){$\stilde g_e$}
\Gluon(50,115)(125,115){4}{7}
\Gluon(50,60)(125,60){4}{7}
\ArrowLine(125,60)(125,87.5)
\ArrowLine(125,115)(125,87.5)
\ArrowLine(200,60)(166.25,84.75)
\Line(125,115)(158.75,90.25)
\ArrowLine(200,115)(162.5,87.5)
\Line(125,60)(162.5,87.5)
\put(45,125){$g_a$}
\put(45,46){$g_b$}
\put(193,120){${\stilde g^\dagger_c}$}
\put(195,48){${\stilde g^\dagger_d}$}
\put(110,87.5){$\stilde g_e$}
\end{picture}
\end{center}
\caption{\label{fig:gg2ginogino} The ten Feynman diagrams for $gg
\ra\stilde g\stilde g$. The momentum and spin polarization assignments
are indicated on the first diagram.}
\end{figure}
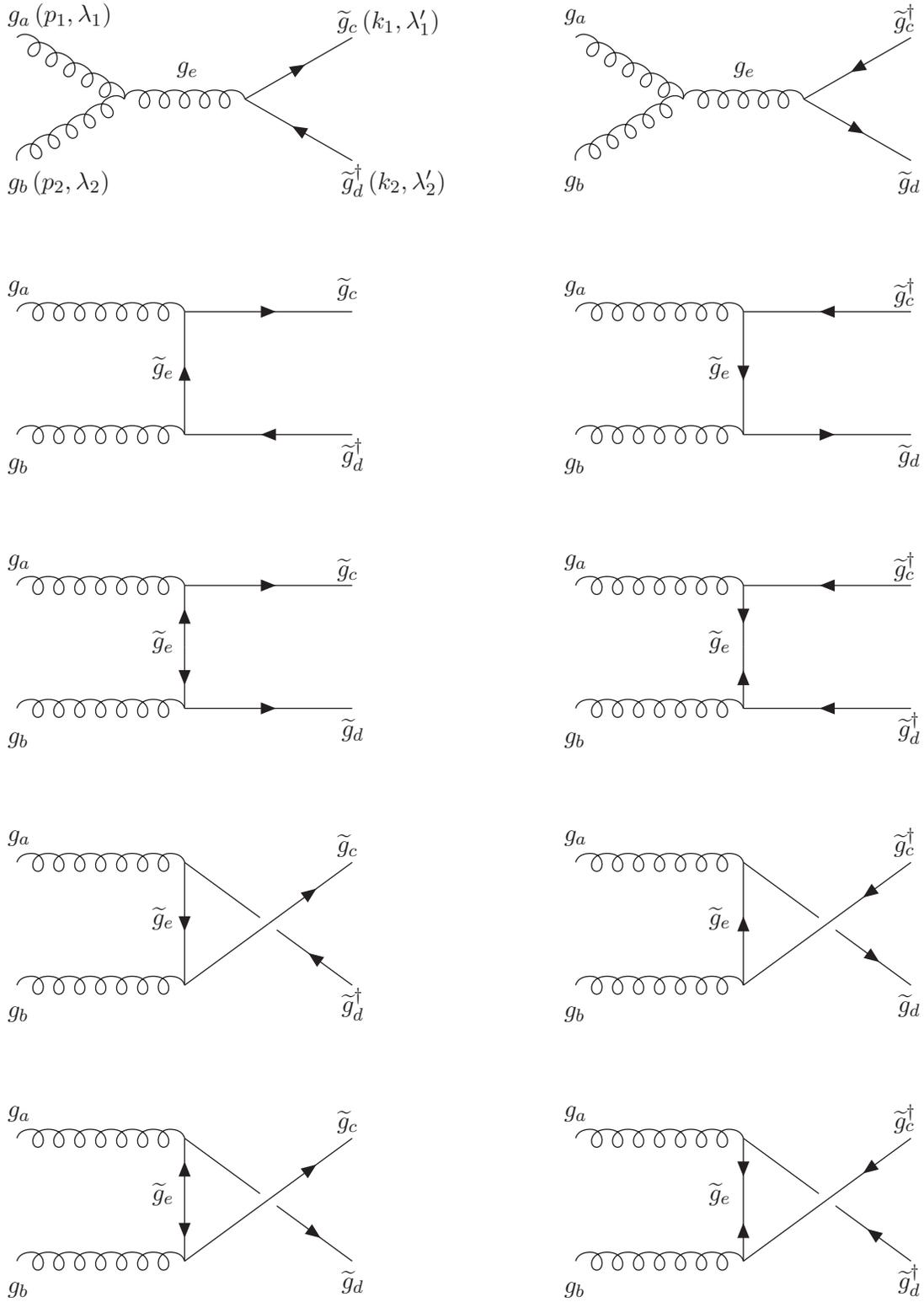

The Feynman rules for the gluino couplings in the supersymmetric
extension of QCD are given in \fig{fig:SUSYQCDgluonrules}.
For the two $s$-channel amplitudes, we obtain:
\beqa
i{\cal M}_{s} &=&
\left (
-g_s f^{abe} [ \metric_{\mu\nu} (p_1 - p_2)_\rho
+ \metric_{\nu\rho} (p_1 + 2 p_2)_\mu
- \metric_{\mu\rho} (2p_1 + p_2)_\nu ]
\right )
\left ( \frac{\BDneg i \metric^{\rho\kappa}}{s} \right )
\varepsilon_1^\mu \varepsilon_2^\nu
\nonumber \\ &&
\times\,
\left [
(\BDneg g_s f^{cde})\,
 x^\dagger_1 \sigmabar_\kappa y_2
\,+\, (\BDpos g_s f^{dce})\,
y_1 \sigma_\kappa  x^\dagger_2
\right ]
.
\eeqa
The first factor is the Feynman rule for the three-gluon
interaction of standard QCD, and the second factor is the gluon
propagator. The next four ($t$-channel) diagrams have a total
amplitude:
\beqa
i {\cal M}_t &=&
\bigl (\BDneg g_s f^{cea} \varepsilon_1^\mu \bigr )
\bigl (\BDneg g_s f^{edb}\varepsilon_2^\nu \bigr )
\,
 x^\dagger_1 \sigmabar_\mu \left [
\frac{i (k_1 - p_1) \newcdot \sigma}{(k_1 - p_1)^2 \BDminus
m_{\tilde g}^2}
\right ] \sigmabar_\nu y_2
\nonumber \\ &&
+
\bigl (\BDpos g_s f^{eca} \varepsilon_1^\mu \bigr )
\bigl (\BDpos g_s f^{deb}\varepsilon_2^\nu \bigr )
\,
y_1 \sigma_\mu \left [
\frac{i (k_1 - p_1) \newcdot \sigmabar}{(k_1 - p_1)^2 \BDminus
m_{\tilde g}^2}
\right ] \sigma_\nu  x^\dagger_2
\nonumber \\ &&
+
\bigl (\BDneg g_s f^{cea} \varepsilon_1^\mu \bigr )
\bigl (\BDpos g_s f^{deb}\varepsilon_2^\nu \bigr )
\,
 x^\dagger_1 \sigmabar_\mu \left [
\frac{\BDpos i m_{\tilde g}}{(k_1 - p_1)^2 \BDminus
m_{\tilde g}^2}
\right ] \sigma_\nu  x^\dagger_2
\nonumber \\ &&
+
\bigl (\BDpos g_s f^{eca} \varepsilon_1^\mu \bigr )
\bigl (\BDneg g_s f^{edb}\varepsilon_2^\nu \bigr )
\,
y_1 \sigma_\mu \left [
\frac{\BDpos i m_{\tilde g}}{(k_1 - p_1)^2 \BDminus
m_{\tilde g}^2}
\right ] \sigmabar_\nu y_2 .
\label{eq:ggggt}
\eeqa
Finally, the $u$-channel Feynman diagrams result in:
\beqa
i {\cal M}_u &=&
\bigl (\BDneg g_s f^{eda} \varepsilon_1^\mu \bigr )
\bigl (\BDneg g_s f^{ceb} \varepsilon_2^\nu \bigr )
\,
 x^\dagger_1 \sigmabar_\nu \left [
\frac{i (k_1 - p_2) \newcdot \sigma}{(k_1 - p_2)^2 \BDminus
m_{\tilde g}^2}
\right ] \sigmabar_\mu y_2
\nonumber \\ &&
+
\bigl (\BDpos g_s f^{dea}\varepsilon_1^\mu \bigr )
\bigl (\BDpos g_s f^{ecb} \varepsilon_2^\nu \bigr )
\,
y_1 \sigma_\nu \left [
\frac{i (k_1 - p_2) \newcdot \sigmabar}{(k_1 - p_2)^2 \BDminus
m_{\tilde g}^2}
\right ] \sigma_\mu  x^\dagger_2
\nonumber \\ &&
+
\bigl (\BDpos g_s f^{dea}\varepsilon_1^\mu \bigr )
\bigl (\BDneg g_s f^{ceb} \varepsilon_2^\nu \bigr )
\,
 x^\dagger_1 \sigmabar_\nu \left [
\frac{\BDpos i m_{\tilde g}}{(k_1 - p_2)^2 \BDminus
m_{\tilde g}^2}
\right ] \sigma_\mu  x^\dagger_2
\nonumber \\ &&
+
\bigl (\BDneg g_s f^{eda}\varepsilon_1^\mu \bigr )
\bigl (\BDpos g_s f^{ecb} \varepsilon_2^\nu \bigr )
\,
y_1 \sigma_\nu \left [
\frac{\BDpos i m_{\tilde g}}{(k_1 - p_2)^2 \BDminus
m_{\tilde g}^2}
\right ] \sigmabar_\mu y_2 .
\label{eq:ggggu}
\eeqa

We choose to work with {\em real} transverse polarization vectors
$\varepsilon_1$, $\varepsilon_2$.
These vectors must both be orthogonal to the
initial state collision axis in the center-of-momentum frame.
Hence,
\beqa
\varepsilon_1 \newcdot \varepsilon_1 &=&
\varepsilon_2 \newcdot \varepsilon_2 = \BDneg 1\,,
\\[-3pt]
\varepsilon_1 \newcdot p_1 &=&
\varepsilon_2 \newcdot p_1 =
\varepsilon_1 \newcdot p_2 =
\varepsilon_2 \newcdot p_2 = 0,
\\[-3pt]
\varepsilon_1 \newcdot k_2 &=& -\varepsilon_1 \newcdot k_1,
\\[-3pt]
\varepsilon_2 \newcdot k_2 &=& -\varepsilon_2 \newcdot k_1 ,
\eeqa
for each choice of $\lambda_1, \lambda_2$.
The sums over gluon polarizations will be performed using
[cf.~\eq{epsthree}]:
\beqa
\sum_{\lambda_1} \varepsilon_1^\mu \varepsilon_1^\nu
=
\sum_{\lambda_2} \varepsilon_2^\mu \varepsilon_2^\nu
=
\BDneg \metric^{\mu\nu} + \frac{2\left (p_1^\mu p_2^\nu
+ p_2^\mu p_1^\nu \right )}{s} .
\label{eq:ggggsumpol}
\eeqa
Note that in QCD processes with two or more external gluons, the term
$2\left (p_1^\mu p_2^\nu+ p_2^\mu p_1^\nu \right )/s$
in \eq{eq:ggggsumpol} cannot in general be dropped~\cite{cutler}.
This is to be contrasted to the photon polarization sum
[cf.~\eq{photonspinsum}], where this latter term
can always be neglected (due to a Ward identity of
quantum electrodynamics).

Before taking the complex square of the amplitude,  it is convenient to
rewrite the last two terms in each of
eqs.~(\ref{eq:ggggt}) and (\ref{eq:ggggu}) by using the identities
[see eq.~(\ref{onshellfour})]:
\beqa
m_{\tilde g}  x^\dagger_1  = \BDpos y_1 (k_1\newcdot \sigma)\,,
\qquad\qquad
m_{\tilde g} y_1  = \BDpos  x^\dagger_1 (k_1  \newcdot\sigmabar)\,.
\eeqa
Using eqs.~(\ref{eq:simplifyssbars}) and (\ref{eq:simplifysbarssbar}),
the resulting total matrix element is then
reduced to
a sum of terms that each contain exactly one $\sigma$ or $\sigmabar$
matrix.
We define convenient factors:
\beqa
G_s &\equiv& g_s^2 f^{abe} f^{cde}/s,
\\
G_t &\equiv& g_s^2 f^{ace} f^{bde}/(t - m_{\tilde g}^2),
\\
G_u &\equiv& g_s^2 f^{ade} f^{bce}/(u - m_{\tilde g}^2) .
\eeqa
where the usual Mandelstam variables are:
\beqa
s &=& \BDpos (p_1 + p_2)^2 = \BDpos (k_1 + k_2)^2
,
\\
t &=& \BDpos (k_1 - p_1)^2 = \BDpos (k_2 - p_2)^2
,
\\
u &=& \BDpos (k_1 - p_2)^2 = \BDpos (k_2 - p_1)^2.
\eeqa
Then the total amplitude is  (noting that the gluon polarizations
$\varepsilon_1, \varepsilon_2$ were
chosen real):
\beqa
{\cal M} = {\cal M}_s + {\cal M}_t + {\cal M}_u
=  x^\dagger_1 a \newcdot \sigmabar y_2 +
            y_1 a^* \newcdot \sigma  x^\dagger_2,
\eeqa
where
\beqa
a^\mu &\equiv&
- (G_t + G_s) \varepsilon_1 \newcdot \varepsilon_2 \,p_1^\mu
- (G_u - G_s) \varepsilon_1 \newcdot \varepsilon_2 \,p_2^\mu
- 2 G_t k_1\newcdot \varepsilon_1 \,\varepsilon_2^\mu
- 2 G_u k_1\newcdot \varepsilon_2 \,\varepsilon_1^\mu
\nonumber \\ &&
- i \epsilon^{\mu\nu\rho\kappa}
\varepsilon_{1\nu} \varepsilon_{2\rho} (G_t p_1 - G_u p_2)_\kappa .
\label{eq:ggggdefa}
\eeqa

Squaring the amplitude using eqs.~(\ref{eq:conbilsig})
and (\ref{eq:conbilsigbar}), we get:
\beq
|{\cal M}|^2 =
 x^\dagger_1 a \newcdot \sigmabar y_2
 y^\dagger_2 a^* \newcdot \sigmabar x_1
+
y_1 a^* \newcdot \sigma  x^\dagger_2
x_2 a \newcdot \sigma  y^\dagger_1
+
 x^\dagger_1 a \newcdot \sigmabar y_2
x_2 a \newcdot \sigma  y^\dagger_1
+
y_1 a^* \newcdot \sigma  x^\dagger_2
 y^\dagger_2 a^* \newcdot \sigmabar x_1 .
\phantom{xxxx}
\eeq
Summing over the gluino spins using
\eqst{xxdagsummed}{ydagxdagsummed}, we find:
\beqa
\sum_{\lambda'_1,\lambda'_2} |{\cal M}|^2 &=&
{\rm Tr}[
a \newcdot \sigmabar k_2 \newcdot \sigma
a^* \newcdot \sigmabar k_1 \newcdot \sigma]
+
{\rm Tr}[
a^* \newcdot \sigma k_2 \newcdot \sigmabar
a \newcdot \sigma k_1 \newcdot \sigmabar]
\nonumber \\[-9pt] &&
- m^2_{\tilde g} {\rm Tr}[a \newcdot \sigmabar a \newcdot \sigma]
- m^2_{\tilde g} {\rm Tr}[a^* \newcdot \sigma a^* \newcdot \sigmabar ].
\eeqa
Performing the traces with \eqst{trssbar}{trsbarssbars} then yields:
\beqa
\sum_{\lambda'_1,\lambda'_2} |{\cal M}|^2
=
8\, {\rm Re}[a \newcdot k_1 a^* \newcdot k_2]
- 4 a \newcdot a^* \, k_1 \newcdot k_2
- 4 i \epsilon^{\mu\nu\rho\kappa} k_{1\mu} k_{2\nu} a_\rho a_\kappa^*
\BDminus 4 m_{\tilde g}^2 {\rm Re}[a^2]
.
\eeqa
Inserting the explicit form for $a^\mu$ [\eq{eq:ggggdefa}] 
into the above result, we obtain:
\beqa
\sum_{\lambda'_1,\lambda'_2} |{\cal M}|^2 &=&
2 (t - m_{\tilde g}^2)(u - m_{\tilde g}^2)
[(G_t + G_u)^2 + 4 (G_s + G_t) (G_s - G_u)
(\varepsilon_1 \newcdot \varepsilon_2)^2]
\nonumber \\[-9pt] &&
\BDplus
16 (G_t + G_u) [G_s (t-u) + G_t (t-m_{\tilde g}^2)
+ G_u (u-m_{\tilde g}^2)]
(\varepsilon_1 \newcdot \varepsilon_2)
(k_1 \newcdot \varepsilon_1)
(k_1 \newcdot \varepsilon_2)
\phantom{xx}
\nonumber \\ &&
-32 (G_t + G_u)^2
(k_1 \newcdot \varepsilon_1)^2
(k_1 \newcdot \varepsilon_2)^2 .
\eeqa
The sums over gluon polarizations can be done using
eq.~(\ref{eq:ggggsumpol}), which implies:
\beqa
&&
\sum_{\lambda_1,\lambda_2} 1 = 4,
\qquad\qquad\quad
\sum_{\lambda_1,\lambda_2} (\varepsilon_1 \newcdot \varepsilon_2)^2 = 2,
\\
&&\sum_{\lambda_1,\lambda_2}
(\varepsilon_1 \newcdot \varepsilon_2)
(k_1 \newcdot \varepsilon_1)
(k_1 \newcdot \varepsilon_2) =
\BDpos m_{\tilde g}^2 \BDminus (t - m_{\tilde g}^2)(u - m_{\tilde g}^2)/s
,
\\
&&\sum_{\lambda_1,\lambda_2}
(k_1 \newcdot \varepsilon_1)^2
(k_1 \newcdot \varepsilon_2)^2 =
\left (
m_{\tilde g}^2 - (t - m_{\tilde g}^2)(u - m_{\tilde g}^2)/s
\right )^2 .
\eeqa
Summing over colors using $f^{abe} f^{cde} f^{abe'} f^{cde'} =
2 f^{abe} f^{cde} f^{ace'} f^{bde'} = N_c^2 (N_c^2 -1) = 72$,
\beqa
&&
\sum_{\rm colors} G_s^2 = \frac{72 g_s^4}{s^2} ,
\qquad\qquad\qquad\qquad\,\,\,
\sum_{\rm colors} G_t^2 = \frac{72 g_s^4}{(t - m^2_{\tilde g})^2} ,
\\
&&
\sum_{\rm colors} G_u^2 = \frac{72 g_s^4}{(u - m^2_{\tilde g})^2} ,
\qquad\qquad\qquad
\sum_{\rm colors} G_s G_t = \frac{36 g_s^4}{s(t - m^2_{\tilde g})} ,
\\
&&
\sum_{\rm colors} G_s G_u = -\frac{36 g_s^4}{s(u - m^2_{\tilde g})} ,
\qquad\qquad
\sum_{\rm colors} G_t G_u =
\frac{36 g_s^4}{(t - m^2_{\tilde g})(u - m^2_{\tilde g})} .
\phantom{xxx}
\eeqa
Putting all the factors together, and averaging over the initial
state colors and spins, we have:
\beqa
\frac{d\sigma}{dt} &=&
\frac{1}{16\pi s^2}
\Biggl (
\frac{1}{64}\sum_{\rm colors}
\frac{1}{4}\sum_{\rm spins} |{\cal M}|^2
\Biggr ) \nonumber
\\[6pt]
&=& \frac{9 \pi \alpha_s^2}{4 s^4}
\Biggl [
2 (t - m_{\tilde g}^2)(u - m_{\tilde g}^2)
-3 s^2 - 4 m_{\tilde g}^2 s
+ \frac{s^2(s+2 m_{\tilde g}^2)^2}{
(t - m_{\tilde g}^2)(u - m_{\tilde g}^2)}
- \frac{4 m_{\tilde g}^4 s^4}{(t -
 m_{\tilde g}^2)^2(u - m_{\tilde g}^2)^2}
\Biggr ] ,\nonumber \\
\phantom{xxxx}
\eeqa
which agrees with the result of \cite{ggggpapers,Dawson:1983fw} (after some
rearrangement).
Note that in the center-of-momentum frame,
the Mandelstam variable $t$ is
related to the scattering
angle $\theta$ between an initial state gluon and a final state gluino by:
\beqa
t = m_{\tilde g}^2 + \frac{s}{2}\Bigl (
\cos\theta \, \sqrt{1- 4 m_{\tilde g}^2/s} - 1
\Bigr )\,.
\label{t-gggg}
\eeqa
Since the final state has identical particles, the total
cross-section can now be obtained by:
\beqa
\sigma = \frac{1}{2} \int_{t_-}^{t_+} \frac{d\sigma}{dt} dt\,,
\eeqa
where $t_\pm$ are obtained by inserting $\cos\theta = \pm 1$ into
\eq{t-gggg}.

\subsection{R-parity violating stau decay:
\texorpdfstring{${\stilde\tau}^+_R \ra e^+\nubar_\mu$}{\texttau\texttilde\textscr\textplussuperior\textrightarrow e\textplussuperior {\textnu\textoverline}\textuinferior}}
\label{rpv-decay1}
\setcounter{equation}{0}
\setcounter{figure}{0}
\setcounter{table}{0}

In an R-parity-violating extension of the MSSM (denoted henceforth
by RPV-MSSM),
new Yukawa couplings can arise [see \eqst{rpvyuk1}{rpvyuk3}]
that violate either a global U(1)
lepton number $L$ or baryon number $B$.  The corresponding
Feynman rules are derived in \app{L}.
Consider the decay of a right-handed scalar tau via an
$L$-violating $LL\Ebar$ coupling governed by \eq{rpvyuk1}.
This is particularly relevant when the scalar tau is the
lightest supersymmetric particle (LSP)
\cite{Allanach:2003eb,Allanach:2006st} and in the case of
resonant slepton production \cite{res-slep,Dreiner:2006sv}.
Note that in R-parity violation the LSP need not be the lightest
neutralino and in a minimal supergravity embedding often it is not
\cite{Bernhardt:2008jz,Dreiner:2008ca}. The Feynman diagram is shown
in \fig{RPVstau}, where we have also defined the momenta and the
helicities of the fermions.
\begin{figure}[hb!]
\begin{center}
\begin{picture}(200,80)(0,8)
\DashArrowLine(10,40)(60,40)5
\ArrowLine(100,70)(60,40)
\ArrowLine(100,10)(60,40)
\Text(30,30)[]{$\widetilde \tau_{R}^+$}
\Text(123,12)[]{${e}^\dagger(k_e,\lam_e)$}
\Text(131,75)[]{$\nu_\mu^\dagger(k_{\nubar_\mu},
\lam_{\nubar_\mu})$}
\end{picture}
\end{center}
\label{RPVstau}
\caption{Feynman diagram for the R-parity-violating
decay ${\stilde\tau}^+_R \ra e^+ \nubar_\mu$}
\end{figure}

The amplitude for the R-parity-violating $\tilde\tau_R^+$ decay is given by:
\beq
i{\cal M}=-i\lam y_e y_{\nubar_\mu}\,.
\eeq
Here we have defined $\lam\equiv\lam_{123}$, and
the external wave functions are denoted by $y_e\equiv
y(\boldsymbol{\vec k}_e,\lam_e),$ and $y_{\nubar_\mu}
\equiv y(\boldsymbol {\vec k}_{\nubar_\mu},\lam_{\nubar_\mu})$,
respectively.
Using eq.~(\ref{eq:conbil}),
the amplitude squared is
\beqa
|{\cal M}|^2 &=&|\lam|^2 y_e y_{\nubar_\mu}  y^\dagger_{\nubar_\mu}  y^\dagger_e .
\eeqa
Summing over the fermion spins using \eq{yydagsummed} gives:
\beqa
\sum_{\lambda_e,\lambda_{\nubar_\mu}}
|{\cal M}|^2 &=&|\lam|^2
{\rm Tr}[k_e\newcdot\sigma \,
k_{\nubar_\mu}\newcdot\sigmabar ]
=
|\lam|^2 m_{\tilde\tau_R}^2\,,
\eeqa
where in the last step we have used the trace formula
eq.~(\ref{trssbar}), and neglected the mass of the
electron and the neutrino. The total decay rate is then
given by
\beq
\Gamma = \frac{1}{16\pi m_{\tilde \tau_R}}
\biggl ( \sum_{\lambda_e,\lambda_{\nubar_\mu}} |{\cal M}|^2 \biggr )
=
\frac{|\lam|^2}{16\pi} m_{\tilde\tau_R}\,,
\eeq
which agrees with the computation in refs.~\cite{Dimopoulos:1988fr,
Dreiner:1999qz, Richardson:2000nt}. Completely analogously we can
obtain the total rate for the decays ${\stilde\nu}_\mu\ra \tau^- e^+$
and $\stilde e^-_L\ra\tau^-\nubar_\mu$, which proceed via the same
operator, by replacing $m_{\tilde \tau_R}\ra(m_{\tilde
e_L},m_{\tilde\nu_\mu})$, respectively.

In general the two-body decay rate of a sfermion $\stilde f$
via the $L$-violating $LQ\Dbar$ coupling governed by \eq{rpvyuk2}
or the $B$-violating $\Ubar\Dbar\Dbar$ coupling governed
by \eq{rpvyuk3} is given by:
\beq
\Gamma (\tilde f\ra f_1 f_2) =\frac{C|\lam|^2 }{16\pi}
m_{\tilde f}\,,
\eeq
where we have neglected the masses $m_{1,2}$ of the final state
fermions. The factor $C$ denotes the color factor. For the slepton
decays via the $LQ\Dbar$ coupling which are summed over the final state quark
colors, $C=\delta^{ij}\delta_{ij}=3$, where $i,j=1,2,3$ and
$\delta_{ij}$ is the symmetric
invariant tensor of color SU(3).  For the squark decays via
the $LQ\Dbar$ where the
initial state color is averaged over and the final state color is
summed, $C=1$. For the squark decays via the
${\Ubar}{\Dbar}{\Dbar}$ coupling,
$C=\frac{1}{3}\epsilon^{ijk}\epsilon_{ijk}=2$, where the Levi-Civita
tensor, $\epsilon^{ijk}=\epsilon_{ijk}$, is the
antisymmetric invariant tensor of color SU(3).
In realistic cases, one must also include the effects of
mixing for the third-family sfermions,
which we have omitted here for simplicity.

\subsection{R-parity-violating neutralino decay:
\texorpdfstring{$\stilde N_i\ra \mu^- u \dbar$}{N\texttilde\textiinferior\textrightarrow\textmu\textminussuperior u{d\textoverline}}}
\label{rpv-decay2}
\setcounter{equation}{0}
\setcounter{figure}{0}
\setcounter{table}{0}

\begin{figure}[t!]
\begin{center}
\begin{picture}(150,100)(-100,100)
\ArrowLine(-50,160)(-100,160)
\ArrowLine(-50,160)(-10,190)
\DashArrowLine(-10,130)(-50,160)5
\ArrowLine(-10,130)(30,160)
\ArrowLine(-10,130)(30,100)
\Text(-80,174)[]{${\chi^{0\,\dagger}_i}\,(p_i, \lam_i)$}
\Text(23,170)[]{$u \,(k_u,\lam_u)$}
\Text(-5,100)[]{$\bar d\,(k_d,\lam_d)$}
\Text(-3,197)[]{$\mu \, (k_\mu,\lam_\mu)$}
\put(-50,130){$\stilde\mu_L$}
\end{picture}
\hspace{2.5cm}
\begin{picture}(150,100)(0,-25)
\ArrowLine(50,35)(0,35)
\ArrowLine(50,35)(90,65)
\DashArrowLine(50,35)(90,5)5
\ArrowLine(90,5)(130,35)
\ArrowLine(90,5)(130,-25)
\Text(20,49)[]{${\chi^{0\,\dagger}_i}\,(p_i,\lam_i)$}
\Text(120,44)[]{$\mu\,(k_\mu,\lam_\mu)$}
\Text(95,-21)[]{$u \,(k_u,\lam_u)$}
\Text(97,72)[]{$\bar d\,(k_d,\lam_d)$}
\put(50,5){$\stilde d_R$}
\end{picture}
\end{center}
\vspace{0.5cm}
\begin{center}
\begin{picture}(150,97)(100,100)
\ArrowLine(150,160)(100,160)
\ArrowLine(150,160)(190,190)
\DashArrowLine(190,130)(150,160)5
\ArrowLine(190,130)(230,160)
\ArrowLine(190,130)(230,100)
\Text(120,174)[]{${\chi^{0\,\dagger}_i} \,(p_i,\lam_i)$}
\Text(220,170)[]{$\mu\,(k_\mu,\lam_\mu)$}
\Text(195,100)[]{$\bar d\,(k_d,\lam_d)$}
\Text(196,197)[]{$u \,(k_u,\lam_u)$}
\put(155,132){$\stilde u_L$}
\end{picture} \phantom{xxx}
\end{center}
\caption{Feynman diagrams for the R-parity
violating decay $\Ni \ra \mu^- u \dbar$.}
\label{rpv-neutdeca}
\end{figure}

Next we consider the R-parity-violating three-body decay of a
neutralino $\Ni \ra \mu^- u \dbar$, which arises
due to the $L$-violating $LQ\Dbar$
coupling governed by \eq{rpvyuk2}. This is of
particular interest when the neutralino is the LSP, since it
determines the final state signatures
\cite{Dreiner:1991pe,Dawson:1985vr,Allanach:1997sa}.
The three Feynman diagrams are shown in \fig{rpv-neutdeca},
including the definitions of the momenta and helicities.
We have neglected sfermion mixing, i.e. we assume $\stilde
\mu_L$, $\stilde u_L$, and $\stilde d_R$ are mass eigenstates. Using
the Feynman rules given in \figs{LQD-rules}{nqsq} (or \ref{nqsqmixed}),
we obtain the corresponding contributions to the decay amplitude,
\beqa
i{\cal M}_1 &=&
\left (i \lambda^{\prime *}\right )
\left [\frac{i}{\sqrt{2}}(g N_{i2} + g' N_{i1}) \right ]
\left [
\frac{\BDpos i}{(p_i - k_\mu)^2 \BDminus m_{\tilde \mu_L}^2}
\right ]
 y^\dagger_i  x^\dagger_\mu  x^\dagger_u  x^\dagger_d\,,
\\
i{\cal M}_2 &=&
\left (i \lambda^{\prime *}\right )
\left [-\frac{i \sqrt{2}}{3} g' N_{i1} \right ]
\left [
\frac{\BDpos i}{(p_i - k_d)^2 \BDminus m_{\tilde d_R}^2}
\right ]
 y^\dagger_i  x^\dagger_d  x^\dagger_\mu  x^\dagger_u\,,
\\
i{\cal M}_3 &=&
\left (i \lambda^{\prime *}\right )
\left [-\frac{i}{\sqrt{2}}(g N_{i2} + g' N_{i1}/3) \right ]
\left [
\frac{\BDpos i}{(p_i - k_u)^2 \BDminus m_{\tilde u_L}^2}
\right ]
 y^\dagger_i  x^\dagger_u  x^\dagger_d  x^\dagger_\mu\,.
\eeqa
Here we have defined $\lambda' \equiv\lambda'_{211}$, and
the external wave functions are denoted by $ y^\dagger_i \equiv
 y^\dagger (\boldsymbol{\vec p}_i ,\lam_i)$, $ x^\dagger_\mu \equiv  x^\dagger
(\boldsymbol{\vec k}_\mu ,\lam_\mu)$, $ x^\dagger_u \equiv  x^\dagger
(\boldsymbol{\vec k}_u ,\lam_u)$, and $ x^\dagger_d \equiv  x^\dagger
(\boldsymbol{\vec k}_d ,\lam_d)$, respectively.
In the following, we will neglect all of the
final state fermion masses. The results will be expressed in terms of
the kinematic variables
\beqa
z_\mu &\equiv& \BDpos 2 p_i \newcdot k_\mu/m_{\tilde N_i}^2
= 2 E_\mu/m_{\tilde N_i},
\label{eq:RPVNdecayzmu}
\\
z_d &\equiv& \BDpos 2 p_i \newcdot k_d/m_{\tilde N_i}^2
= 2 E_d/m_{\tilde N_i} ,
\\
z_u &\equiv& \BDpos 2 p_i \newcdot k_u/m_{\tilde N_i}^2
= 2 E_u/m_{\tilde N_i},
\label{eq:RPVNdecayzu}
\eeqa
which satisfy $z_\mu + z_d + z_u = 2$. Then we can rewrite the total
matrix element as:
\beqa
{\cal M} =
c_1  y^\dagger_i  x^\dagger_\mu  x^\dagger_u  x^\dagger_d
+
c_2  y^\dagger_i  x^\dagger_d  x^\dagger_\mu  x^\dagger_u
+
c_3  y^\dagger_i  x^\dagger_u  x^\dagger_d  x^\dagger_\mu\,,
\eeqa
where
\beqa
c_1 &\equiv& \frac{1}{\sqrt{2}} \lambda^{\prime *}
(g N_{i2} + g' N_{i1})/[
m_{\tilde \mu_L}^2 - m_{\tilde N_i}^2 (1 - z_\mu)] ,
\label{eq:RPVNdecaycone}
\\
c_2 &\equiv& -\frac{\sqrt{2}}{3} \lambda^{\prime *} g' N_{i1}/[
m_{\tilde d_R}^2 - m_{\tilde N_i}^2 (1 - z_d)] ,
\\
c_3 &\equiv& -\frac{1}{\sqrt{2}} \lambda^{\prime *}
(g N_{i2} + g' N_{i1}/3)/[
m_{\tilde u_L}^2 - m_{\tilde N_i}^2 (1 - z_u)] .
\label{eq:RPVNdecaycthree}
\eeqa

Before squaring the amplitude, it is convenient to use the Fierz
identity [\eq{eq:twocompfierztwo}] to reduce the number of terms:
\beqa
{\cal M} =
(c_1 - c_3)  y^\dagger_i  x^\dagger_\mu  x^\dagger_u  x^\dagger_d
+
(c_2 - c_3)  y^\dagger_i  x^\dagger_d  x^\dagger_\mu  x^\dagger_u .
\eeqa
Using eq.~(\ref{eq:conbil}), we obtain
\beqa
|{\cal M}|^2 &=& |c_1 - c_3|^2  y^\dagger_i  x^\dagger_\mu x_\mu y_i
                x^\dagger_u  x^\dagger_d x_d x_u
+ |c_2 - c_3|^2  y^\dagger_i  x^\dagger_d x_d y_i
                 x^\dagger_\mu  x^\dagger_u x_u x_\mu
\nonumber \\ &&
- 2 {\rm Re}[(c_1 - c_3) (c_2^* - c_3^*)
 y^\dagger_i  x^\dagger_\mu x_\mu x_u  x^\dagger_u  x^\dagger_d x_d y_i]\,,
\eeqa
where eq.~(\ref{zonetwo}) was used on the last term.
Summing over the fermion spins using
\eqst{xxdagsummed}{ydagxdagsummed}, we obtain:
\beqa
\sum_{\rm spins} |{\cal M}|^2 &=&
|c_1 - c_3|^2 {\rm Tr}[k_\mu \newcdot \sigmabar p_i \newcdot \sigma]
              {\rm Tr}[k_d \newcdot \sigmabar k_u \newcdot \sigma]
+
|c_2 - c_3|^2 {\rm Tr}[k_d \newcdot \sigmabar p_i \newcdot \sigma]
              {\rm Tr}[k_u \newcdot \sigmabar k_\mu \newcdot \sigma]
\nonumber \\ &&
- 2 {\rm Re} \bigl [(c_1 - c_3) (c_2^* - c_3^*)
{\rm Tr}[ k_\mu \newcdot \sigmabar k_u \newcdot \sigma k_d \newcdot \sigmabar
          p_i \newcdot \sigma] \bigr ]\,.
\eeqa
Applying the trace formulae, \eqs{trssbar}{trsbarssbars},
we obtain
\beqa
\sum_{\rm spins} |{\cal M}|^2
&=& 4 |c_1 - c_3|^2 p_i \newcdot k_\mu \, k_d \newcdot k_u
  + 4 |c_2 - c_3|^2 p_i \newcdot k_d \, k_\mu \newcdot k_u
\nonumber \\ &&
  -4  {\rm Re} \bigl [(c_1 - c_3) (c_2^* - c_3^*)]
  (k_\mu \newcdot k_u\, p_i \newcdot k_d
  + p_i \newcdot k_\mu \,k_d \newcdot k_u
  - k_\mu \newcdot k_d \, p_i \newcdot k_u)
\nonumber \\[3pt]
&=& m_{\tilde N_i}^4 \Bigl [
    |c_1|^2 z_\mu (1 - z_\mu)
    + |c_2|^2 z_d (1 - z_d)
    + |c_3|^2 z_u (1 - z_u)
\nonumber \\ &&
    - 2{\rm Re}[c_1 c_2^*] (1 - z_\mu)(1 - z_d)
    - 2{\rm Re}[c_1 c_3^*] (1 - z_\mu)(1 - z_u)
\nonumber \\ &&
    - 2{\rm Re}[c_2 c_3^*] (1 - z_d)(1 - z_u)
    \Bigr ]\,,
\eeqa
where in the last equality we have used
\eqst{eq:RPVNdecayzmu}{eq:RPVNdecayzu} and
\beqa
\BDpos 2 k_\mu \newcdot k_d
=
(1 - z_u) m_{\tilde N_i}^2 ,
\quad\quad
\BDpos 2 k_\mu \newcdot k_u
=
(1 - z_d) m_{\tilde N_i}^2 ,
\quad\quad
\BDpos 2 k_d \newcdot k_u  &=& (1 - z_\mu) m_{\tilde N_i}^2 .
\phantom{xxxx}
\eeqa
The differential decay rate follows:
\beqa
\frac{d^2 \Gamma}{dz_\mu d z_d}
=
\frac{N_c m_{\tilde N_i}}{2^8 \pi^3}
\biggl (\frac{1}{2} \sum_{\rm spins} |{\cal M}|^2 \biggr ) ,
\eeqa
where a factor of $N_c = 3$ has been included for the sum over colors,
a factor of $1/2$ to average over the neutralino spin, and the
kinematic limits are
\beqa
0 < &z_\mu &< 1,\\
1 - z_\mu < & z_d & <1.
\eeqa
In the limit of heavy sfermions, the integrations over $z_d$ and then
$z_\mu$
are simple, with
the result for the total decay width:
\beqa
\Gamma = \frac{N_c m_{\tilde N_i}^5}{2^{11}\cdot3 \pi^3}
\left ( |c_1'|^2 + |c_2'|^2 + |c_3'|^2 - {\rm Re}[
c_1' c_2^{\prime *} + c_1' c_3^{\prime *}
+ c_2' c_3^{\prime *}] \right ),
\eeqa
where the $c_i'$ are obtained from $c_i$ of
\eqst{eq:RPVNdecaycone}{eq:RPVNdecaycthree}
by neglecting $m_{\tilde N_i}^2$ in the denominators.
Our results agree with the complete computation (which includes
mixing) given in
refs.~\cite{Baltz:1997gd,Dreiner:1999qz,Richardson:2000nt}.
Earlier calculations with some
simplifications are given in refs.~\cite{Dawson:1985vr,Butterworth:1992tc}.

\subsection{Top-quark condensation from a
Nambu-Jona-Lasinio model gap equation}
\setcounter{equation}{0}
\setcounter{figure}{0}
\setcounter{table}{0}

The previous examples have involved renormalizable field theories.
However, there are cases in which it is preferable to use effective
four-fermion interactions. The obvious historical example is
the four-fermion Fermi theory of weak decays. This has been superseded
by a more complete and accurate theory of the weak interactions but
is still useful for leading order calculations of low-energy processes.
Another case of some interest is the use of strong coupling
four-fermion interactions to drive symmetry breaking
via a Nambu-Jona-Lasinio model \cite{NJL}, as in the
top quark condensate
approach \cite{Nambu,topquarkcondensates,BHL,CR,topcolor}
to electroweak symmetry breaking.

Consider an effective four-fermion
Lagrangian involving the top quark \cite{BHL},
written in two-component fermion form as:
\beqa
\mathscr{L} =
i t^\dagger \sigmabar^\mu \partial_\mu t
+ i {\bar t}^\dagger \sigmabar^\mu \partial_\mu \bar t
+ \frac{G}{\Lambda^2} (t\bar t) ( t^\dagger {\bar t}^\dagger) .
\eeqa
Here the Standard Model gauge interactions have been suppressed; the
quantities within parentheses are color singlets. Note also that there
is no top quark Yukawa coupling to a Higgs scalar boson, nor a top
quark mass term, which would normally appear in the form $- m_t (t\bar t
+ t^\dagger {\bar t}^\dagger)$. Instead, the effective top quark mass is
supposed to be driven by a non-perturbatively large and positive
dimensionless coupling $G$, with $\Lambda$ the cutoff scale at which
$G$ arises from some more fundamental physics such as topcolor
\cite{topcolor}.

The Feynman rule for the four-fermion interaction can be derived from
the mode expansion results of \sec{sec:externalfermions},
and is given in \fig{fig:fourfermion}.
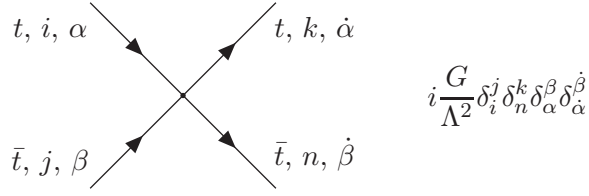
\begin{figure}[t!]
\begin{center}
\begin{picture}(200,62)(0,9)
\ArrowLine(50,35)(85,70)
\ArrowLine(50,35)(85,0)
\ArrowLine(15,70)(50,35)
\ArrowLine(15,0)(50,35)
\Vertex(50,35)1
\Text(0,60)[]{$t$, $i$, $\alpha$}
\Text(0,10)[]{$\bar t$, $j$, $\beta$}
\Text(100,60)[]{$t$, $k$, $\dot{\alpha}$}
\Text(100,13)[]{$\bar t$, $n$, $\dot{\beta}$}
\Text(173,35)[]{$\displaystyle{i \frac{G}{\Lambda^2} \delta_i^j \delta_n^k
\delta_\alpha^\beta \delta_{\dot{\alpha}}^{\dot{\beta}}}$}
\end{picture}
\end{center}
\caption{\label{fig:fourfermion} Feynman rule for the four-fermion
interaction in the top quark condensate model. The indices
$i,j,k,n = 1,2,3$ are for color in the fundamental representation
of $SU(3)$, and the indices
$\alpha,\beta,\dot{\alpha},\dot{\beta}$
are two-component spinor indices.}
\end{figure}
The resulting
gap equation for the dynamically generated top quark mass is shown in
\fig{fig:topgap}.
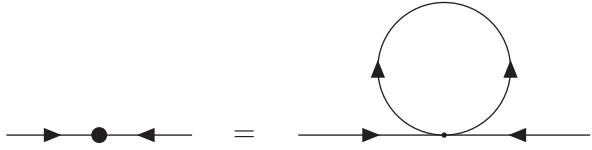
\begin{figure}[t]
\begin{center}
\begin{picture}(200,58)(0,15)
\ArrowLine(0,10)(35,10)
\ArrowLine(70,10)(35,10)
\Vertex(35,10)3
\Text(90,10)[]{$=$}
\ArrowLine(110,10)(165,10)
\ArrowLine(220,10)(165,10)
\Vertex(165,10)1
\CArc(165,35)(25,0,360)
\ArrowLine(140,34.99)(140,35.01)
\ArrowLine(190,34.99)(190,35.01)
\end{picture}
\end{center}
\caption{\label{fig:topgap} The Nambu-Jona-Lasinio gap equation
for a possible dynamically generated top quark mass $m_t$.}
\end{figure}
Evaluating this using the Feynman rules of
\figs{fig:massinsertion}{fig:Diracpropagators}, one finds:
\beq
-im_t \delta_i^j \delta_\alpha^\beta =
(-1) \int^\Lambda \frac{d^4 k}{(2\pi)^4}
\>\left (i \frac{G}{\Lambda^2} \delta_i^j
\delta_n^k \delta_\alpha^\beta \delta_{\dot{\alpha}}^{\dot{\beta}} \right )
\>
\left (
\delta_k^n \delta_{\dot{\beta}}^{\dot{\alpha}}
\frac{\BDpos i m_t}{k^2 \BDminus m_t^2 \BDplus i \epsilon}
\right ) .
\eeq
Here $i,j,k,n$ are color indices of the fundamental representation of
$SU(3)$, and $\alpha,\beta,\dot{\alpha},\dot{\beta}$ are two-component
spinor indices. The factor of $(-1)$ on the right-hand side is due to
the presence of a fermion loop.

Euclideanizing the loop integration over $k^\mu$ by
$k^2 \rightarrow \BDneg k_E^2$ and
$\int d^4 k \rightarrow
i\int d^4 k_E$, and then rewriting the integration in terms of
$x = k_E^2$, this amounts to \cite{BHL}:
\beq
m_t = \frac{2 N_c G m_t}{16 \pi^2 \Lambda^2}
\int_0^{\Lambda^2} dx/(1 + m_t^2/x)
= \frac{3 G m_t}{8 \pi^2}
[1 - (m_t^2/\Lambda^2) \ln (\Lambda^2/m_t^2) + \ldots]\,,
\eeq
where $N_c =3$ is the number of colors, and a factor of two arises from
the sum over dotted spinor indices of $\delta_{\dot{\alpha}}^{\dot{\beta}}
\delta_{\dot{\beta}}^{\dot{\alpha}}$.

For small or negative $G$, only the trivial solution $m_t=0$ is possible.
However, for $G \geq G_{\mbox{critical}} = 8 \pi^2/3 \approx 26$,
there is
a positive solution for $m_t^2/\Lambda^2$ \cite{BHL}.
It is now known that this minimal
version
of the model cannot explain the top quark mass
and the observed features of
electroweak symmetry breaking, but extensions of it may be
viable \cite{Hill:2002ap}.

\subsection{Electroweak vector boson self-energies from fermion loops}
\setcounter{equation}{0}
\setcounter{figure}{0}
\setcounter{table}{0}

In this subsection, we consider the contributions to the self-energy
functions of the Standard Model electroweak vector bosons coming
from quark and lepton loops. (For a derivation of equivalent results in
the four-component fermion formalism, see for example Section 21.3 of
\cite{Peskin:1995ev}.)
The independent self-energies are given by
$\Pi^{WW}_{\mu\nu}$, $\Pi^{ZZ}_{\mu\nu}$,
$\Pi^{\gamma Z}_{\mu\nu} = \Pi^{Z\gamma}_{\mu\nu}$,
and $\Pi^{\gamma\gamma}_{\mu\nu}$, as shown in
\figs{fig:PIWW}{fig:PiVV}.  In each case, $i\Pi_{\mu\nu}$ is equal to
the sum of Feynman diagrams
for two-point functions with amputated external legs, and is implicitly
a function of the external momentum $p^\mu$.
\begin{figure}[t!]
\begin{center}
\begin{picture}(48,50)(0,5)
\Text(0,25)[c]{$i \Pi^{WW}_{\mu\nu}(p) \,\,=\,\phantom{x}$}
\end{picture}
\begin{picture}(101,50)(-8,5)
\Photon(-3,25)(23,25){3.5}{4}
\Photon(85,25)(59,25){3.5}{4}
\CArc(41,25)(18,0,180)
\CArc(41,25)(18,180,360)
\ArrowLine(41,43)(40.99,43)
\ArrowLine(41,7)(41.01,7)
\Text(-17,26)[c]{$W^+$}
\Text(102,26)[c]{$W^+$}
\Text(41,52)[c]{$f$}
\Text(41,-1)[c]{$f'$}
\Text(7,44)[c]{$p$}
\Text(73,44)[c]{$p$}
\LongArrow(65,35)(78,35)
\LongArrow(-2,35)(13,35)
\Text(5,13)[c]{$\mu$}
\Text(75,13)[c]{$\nu$}
\end{picture}
\end{center}
\caption{\label{fig:PIWW} Contributions
to the self-energy function for the $W$ boson
in the Standard Model, from loops
involving the left-handed quark and lepton pairs
$(f,f') = (e, \nu_e)$, $(\mu, \nu_\mu)$,
$(\tau, \nu_\tau)$, $(d,u)$, $(s,c)$, and $(b,t)$.
The momentum of the positively charged $W^+$ flows from
left to right.
}
\end{figure}
\begin{figure}[t!p]
\begin{center}
\begin{picture}(52,55)(0,7)
\Text(24,25)[c]{$i \Pi^{VV'}_{\mu\nu} \,=$}
\end{picture}
\begin{picture}(82,50)(0,7)
\Photon(0,25)(23,25){3.5}{3.5}
\Photon(82,25)(59,25){3.5}{3.5}
\CArc(41,25)(18,0,180)
\CArc(41,25)(18,180,360)
\ArrowLine(41,43)(40.99,43)
\ArrowLine(41,7)(41.01,7)
\Text(41,52)[c]{$f$}
\Text(41,-0.5)[c]{$f$}
\Text(2,13)[c]{$\mu$}
\Text(80,13)[c]{$\nu$}
\Text(2,37)[c]{$V$}
\Text(80,37.5)[c]{$V'$}
\end{picture}
\begin{picture}(13,50)(0,7)
\Text(6.5,25)[c]{$+$}
\end{picture}
\begin{picture}(82,50)(0,7)
\Photon(0,25)(23,25){3.5}{3.5}
\Photon(82,25)(59,25){3.5}{3.5}
\CArc(41,25)(18,0,180)
\CArc(41,25)(18,180,360)
\ArrowLine(41,43)(40.99,43)
\ArrowLine(41,7)(41.01,7)
\Text(41,52)[c]{$\bar f$}
\Text(41,-1)[c]{$\bar f$}
\Text(2,13)[c]{$\mu$}
\Text(80,13)[c]{$\nu$}
\Text(2,37)[c]{$V$}
\Text(80,37.5)[c]{$V'$}
\end{picture}
\begin{picture}(13,50)(0,7)
\Text(6.5,25)[c]{$+$}
\end{picture}
\begin{picture}(82,50)(0,7)
\Photon(0,25)(23,25){3.5}{3.5}
\Photon(82,25)(59,25){3.5}{3.5}
\CArc(41,25)(18,0,180)
\CArc(41,25)(18,180,360)
\ArrowLine(53.7279,37.7279)(53.7379,37.7179)
\ArrowLine(53.7279,12.2721)(53.7179,12.2621)
\ArrowLine(28.2721,37.7279)(28.2621,37.7179)
\ArrowLine(28.2721,12.2721)(28.2821,12.2621)
\Text(26,47)[c]{$f$}
\Text(25,4)[c]{$f$}
\Text(61,47)[c]{$\bar f$}
\Text(59,4)[c]{$\bar f$}
\Text(2,13)[c]{$\mu$}
\Text(80,13)[c]{$\nu$}
\Text(2,37)[c]{$V$}
\Text(80,37.5)[c]{$V'$}
\end{picture}
\begin{picture}(13,50)(0,7)
\Text(6.5,25)[c]{$+$}
\end{picture}
\begin{picture}(82,50)(0,7)
\Photon(0,25)(23,25){3.5}{3.5}
\Photon(82,25)(59,25){3.5}{3.5}
\CArc(41,25)(18,0,180)
\CArc(41,25)(18,180,360)
\ArrowLine(53.7279,37.7279)(53.7379,37.7179)
\ArrowLine(53.7279,12.2721)(53.7179,12.2621)
\ArrowLine(28.2721,37.7279)(28.2621,37.7179)
\ArrowLine(28.2721,12.2721)(28.2821,12.2621)
\Text(27,47)[c]{$\bar f$}
\Text(25,4)[c]{$\bar f$}
\Text(60,47)[c]{$f$}
\Text(58,4)[c]{$f$}
\Text(2,13)[c]{$\mu$}
\Text(80,13)[c]{$\nu$}
\Text(2,37)[c]{$V$}
\Text(80,37.5)[c]{$V'$}
\end{picture}
\end{center}
\caption{\label{fig:PiVV} Contributions
to the diagonal and off-diagonal self-energy functions for the neutral
vector bosons $V, V' = \gamma, Z$ in the Standard Model, from loops
involving the three generations of leptons and quarks:
$f = e,\nu_e, \mu, \nu_\mu, \tau, \nu_\tau, d,u,s,c,b,t$.
}
\end{figure}

First consider the self-energy function for the $W$ boson, shown in
\fig{fig:PIWW}. The $W$ boson only couples to left-handed
fermions, so there is only one Feynman diagram for each Standard model
weak isodoublet. Taking the external momentum flowing from left to
right to be $p$, and the loop momentum flowing counterclockwise in the
upper fermion line ($f$)
to be $k$, we have from the Feynman rules of \fig{SMintvertices}:
\beqa
i\Pi^{WW}_{\mu\nu} &=&  (-1) \,
\mu^{2\epsilon} \int \frac{d^dk}{(2\pi)^d}
\>\sum_{(f,f')} N_c^f
{\rm Tr}
\biggl [
\Bigl ( \BDneg i \frac{g}{\sqrt{2}} \sigmabar_\mu \Bigr )
\biggl ( \frac{i k\newcdot \sigma}{k^2 \BDminus m_f^2} \biggr )
\Bigl ( \BDneg i \frac{g}{\sqrt{2}} \sigmabar_\nu \Bigr )
\biggl ( \frac{i (k+p)\newcdot \sigma}{(k+p)^2 \BDminus m_{f'}^2}
\biggr )
\biggr ].
\phantom{.}
\nonumber \\ &&
\label{eq:PiWW}
\eeqa
Here $\mu$ is a regularization scale for dimensional regularization
in $d \equiv 4 - 2 \epsilon$
dimensions. The sum in \eq{eq:PiWW} is over the six isodoublet pairs
$(f,f') = (e, \nu_e)$, $(\mu, \nu_\mu)$,
$(\tau, \nu_\tau)$, $(d,u)$, $(s,c)$, and $(b,t)$ with CKM mixing neglected,
and
\beq
N_c^f=\begin{cases} 3\,,\quad &  f = \textrm{quarks}\,,
\\ 1\,,\quad   & f= \textrm{leptons}\,.\end{cases}
\eeq
The first factor of $(-1)$ in \eq{eq:PiWW}
is due to the presence of a closed fermion loop.
The trace is taken over the two-component dotted spinor indices.
Using eq.~(\ref{APPtrsbarssbars}), it follows that
\beqa
\Pi_{\mu\nu}^{WW} = \frac{g^2}{32 \pi^2} \sum_f N_c^f
I_{\mu\nu} (m_f^2, m_{f'}^2)\,,
\label{eq:PiWWresult}
\eeqa
where we have defined
\beqa
I_{\mu\nu} (x, y) =
i (16 \pi^2) \,
\mu^{2\epsilon} \int \frac{d^dk}{(2\pi)^d}\>
\frac{
4 k_\mu k_\nu + 2 k_\mu p_\nu + 2 k_\nu p_\mu - 2 k\newcdot (k+p) \,\metric_{\mu\nu}
}{
(k^2 \BDminus x)[(k+p)^2 \BDminus y]
}\,.
\eeqa
We do not explicitly exhibit above the
term proportional to $\epsilon_{\mu\nu\alpha\beta}$, as it
integrates to zero.
The integral $I_{\mu\nu}$ can be evaluated by the standard dimensional
regularization methods \cite{Peskin:1995ev,Ramond},
\beq
I_{\mu\nu} (x, y) =
(p^2 \metric_{\mu\nu} - p_\mu p_\nu ) I_1 (\BDpos p^2; x,y)
\BDplus \metric_{\mu\nu} I_2 (\BDpos p^2; x,y)\,,
\label{eq:defImunu}
\eeq
where
\beqa
\!\!\!\!\!\!I_1(s;x,y) &=&
-\frac{2}{3\epsilon} +
\frac{2}{3 s^2} \biggl\{
(2x-2y - s) A(x)
+ (2y-2x - s) A(y)
\nonumber \\ &&
+ \Bigl[ 2 (x-y)^2 -s (x+y) - s^2 \Bigr] B(s;x,y)
-s (x+y) + s^2/3
\biggr\},
\\[6pt]
\!\!\!\!\!\!I_2(s; x,y) &=& \frac{x+y}{\epsilon}
- \frac{1}{s} \biggl\{
(x-y) \bigl[A(x) - A(y)\bigr] + \Bigl[(x-y)^2 -s (x+y)\Bigr] B(s;x,y)
\biggr\}\,,\phantom{xxx}
\label{eq:defItwo}
\eeqa
after neglecting terms that vanish as $\epsilon \rightarrow 0$.
The functions
\beqa
A(x) &\equiv& x \ln (x/Q^2) - x,
\label{eq:defAPV}
\\[6pt]
B(s;x,y) &\equiv& -\int_0^1 dt\, \ln \left(\frac{t x +
(1-t) y - t (1-t) s - i \varepsilon}{Q^2}\right) \,,
\label{eq:defBPV}
\eeqa
are the finite parts of one-loop Passarino-Veltman
functions~\cite{PassVelt},
with the renormalization scale $Q$ related to the regularization scale
$\mu$ by the modified minimal subtraction relation
\begin{eqnarray}
\mu^2 = Q^2 e^{\gamma}/4\pi ,
\label{eq:relatemuQ}
\end{eqnarray}
where $\gamma = 0.577216\ldots$ is Euler's constant.

The photon and $Z$ boson have mixed self-energy functions, defined
in \fig{fig:PiVV}. Applying the pertinent Feynman rules from
\fig{SMintvertices}, we obtain:
\beqa
&& \hspace{-0.3in} i\Pi^{VV'}_{\mu\nu}
\,=\,
(-1)
\mu^{2\epsilon} \int \frac{d^dk}{(2\pi)^d}
\sum_f N_c^f
 {\rm Tr}
\Biggl\{
\bigl ( \BDneg i G_V^f \sigmabar_\mu \Bigr )
\biggl ( \frac{ i k\newcdot \sigma}{k^2 \BDminus m_f^2} \biggr )
\bigl (\BDneg i G_{V'}^f \sigmabar_\nu \Bigr )
\biggl ( \frac{i (k+p)\newcdot \sigma}{(k+p)^2 \BDminus m_{f}^2}
\biggr )
\nonumber \\ &&\hspace{1.5in}
+
\bigl ( \BDneg i G_V^{\bar f} \sigmabar_\mu \Bigr )
\biggl ( \frac{ i k\newcdot \sigma}{k^2 \BDminus m_f^2} \biggr )
\bigl ( \BDneg i G_{V'}^{\bar f} \sigmabar_\nu \Bigr )
\biggl ( \frac{ i (k+p)\newcdot \sigma}{(k+p)^2 \BDminus m_{f}^2}
\biggr )
\nonumber \\ &&\hspace{1.5in}
+
\bigl ( \BDneg i G_V^{f} \sigmabar_\mu \Bigr )
\biggl ( \frac{\BDpos i m_f}{k^2 \BDminus m_f^2} \biggr )
\bigl ( \BDpos i G_{V'}^{\bar f} \sigma_\nu \Bigr )
\biggl ( \frac{\BDpos i m_f}{(k+p)^2 \BDminus m_{f}^2}
\biggr )
\nonumber \\ && \hspace{1.5in}+
\bigl ( \BDneg i G_V^{\bar f} \sigmabar_\mu \Bigr )
\biggl ( \frac{\BDpos i m_f}{k^2 \BDminus m_f^2} \biggr )
\bigl ( \BDpos i G_{V'}^{f} \sigma_\nu \Bigr )
\biggl ( \frac{\BDpos i m_f}{(k+p)^2 \BDminus m_{f}^2}
\biggr )
\Biggr\},
\label{eq:PiVV}
\eeqa
where $V$ and $V'$ can each be either $\gamma$ or $Z$, and
$\sum_f$ is taken over the 12 Standard Model fermions.
The corresponding $Vff$ and $V\bar f \bar f$ couplings
are:\footnote{Note that there is no contribution from
the left-handed two-component antineutrino fields,
$\bar\nu_{e}$, $\bar\nu_{\mu}$, $\bar\nu_{\tau}$,
which do not exist in the Standard Model.}
\beqa
&&G_\gamma^f = - G_\gamma^{\bar f} = e Q_f,\\
&&G_Z^f = \frac{g}{c_W} (T_3^f - s_W^2 Q_f), \qquad\quad
G_Z^{\bar f} = \frac{g}{c_W} s_W^2 Q_f .
\eeqa
The four terms in eq.~(\ref{eq:PiVV}) correspond to the four
diagrams in \fig{fig:PiVV}, in the same order.

The first two terms in eq.~(\ref{eq:PiVV})
are computed exactly as for $\Pi^{WW}_{\mu\nu}$,
while in the last two terms we use eq.~(\ref{APPtrssbar}) to compute
the trace. It follows that the neutral electroweak vector boson self-energy
function matrix, after dropping terms that vanish as $\epsilon\to 0$,
is given by
\beq
\Pi_{\mu\nu}^{VV'} =
\frac{1}{16\pi^2}
\sum_f N_c^f \Bigl [
(G_V^f G_{V'}^f + G_V^{\bar f} G_{V'}^{\bar f}) I_{\mu\nu}(m_f^2, m_f^2)
\BDplus\metric_{\mu\nu}
(G_V^f G_{V'}^{\bar f} + G_V^{\bar f} G_{V'}^{f})
m_f^2  I_3 (m_f^2, m_f^2)
\Bigr ],
\label{eq:PiVVresult}
\eeq
where $I_{\mu\nu}(x,y)$ was defined in
\eqst{eq:defImunu}{eq:defItwo}, and
we have defined the function
\beqa
I_3 (x,y) =
-i (16 \pi^2)
\>
\mu^{2\epsilon} \int \frac{d^dk}{(2\pi)^d}
\> \frac{2}{
(k^2 \BDminus x)[(k+p)^2 \BDminus y]
}
=
\frac{2}{\epsilon} + 2 B(\BDpos p^2;x,y)  .\phantom{xxx}
\eeqa

The photon self-energy function is a simple special case of
eq.~(\ref{eq:PiVVresult}):
\beq
\Pi^{\gamma\gamma}_{\mu\nu}=
\frac{1}{16\pi^2} \sum_f 2 N_c^f (e Q_f)^2
\bigl[
I_{\mu\nu}(m_f^2,m_f^2)
\BDminus \metric_{\mu\nu} m_f^2 I_3 (m_f^2, m_f^2)
\bigr]\,.
\eeq
Evaluating the integrals $I_{\mu\nu}$ and $I_3$ yields
\beq
\Pi^{\gamma\gamma}_{\mu\nu}=
\frac{\alpha}{3\pi} \sum_f N_c^f Q_f^2
\left (p^2 \metric_{\mu\nu} - p_{\mu}p_{\nu}
\right )
\biggr \{ -\frac{1}{\epsilon}
+ \frac{1}{3}
\BDminus \frac{2}{p^2} \left[A(m_f^2) + m_f^2\right]
- \biggl ( 1 \BDplus \frac{2m_f^2}{p^2} \biggr )
B(\BDpos p^2;m_f^2,m_f^2) \biggr \},
\label{eq:Pigammagamma}
\eeq
in agreement with the result given in, for example,
eq.~(7.90) of \cite{Peskin:1995ev}.
This formula satisfies $
p^\mu \Pi_{\mu\nu}^{\gamma\gamma} = p^\nu \Pi_{\mu\nu}^{\gamma\gamma} = 0
$ as required by the Ward identity of QED, and is regular in
the limit $p^2 \rightarrow 0$.

In each of eqs.~(\ref{eq:PiWWresult}),
(\ref{eq:PiVVresult}), and (\ref{eq:Pigammagamma}), there are $1/\epsilon$
poles, contained in the loop integral functions. In the $\MSbar$
renormalization scheme,
these poles are simply removed by counterterms, which have no other effect.

In eqs.~(\ref{eq:PiWW}) and (\ref{eq:PiVV}),
we chose to write a
$\sigmabar_\mu$ for the left vertex in the Feynman diagram in each case.
This is an arbitrary choice;
we could also have chosen to use instead
$-\sigma_\mu$ for the left vertex in
any given diagram, as mentioned in the
caption for \fig{SMintvertices}. This would have dictated the
replacements
$\sigmabar \leftrightarrow -\sigma$ throughout the expression for
the diagram, including for the fermion propagators, as was indicated
in \fig{fig:Diracpropagators}. It is not hard to check that
the result after computing the spinor index
traces is unaffected. Note that the contribution proportional to
$\epsilon_{\mu\nu\rho\kappa}$ from eq.~(\ref{APPtrssbarssbar})
or eq.~(\ref{APPtrsbarssbars}) vanishes;
this is clear
because the self-energy function is symmetric under interchange of
vector indices, and there is only one independent momentum in the problem.

\subsection{Self-energy and pole mass of the top quark}
\label{subsec:toppole}
\setcounter{equation}{0}
\setcounter{figure}{0}
\setcounter{table}{0}

We next consider the one-loop calculation of the self-energy and the
pole mass of the top quark in the Standard Model, including the
effects of the gauge interactions and the top and bottom quark Yukawa
couplings.  As in \sec{tdecay}, we treat this as a one-generation
problem, neglecting CKM mixing.  Consequently, the corresponding Yukawa
couplings $Y_t$ and $Y_b$ are real and positive (by a suitable phase
redefinition of the Higgs field\footnote{As shown in
\sec{subsec:generalmass}, after the fermion mass matrix diagonalization
procedure, the tree-level fermion masses are real and non-negative.
If CKM mixing is neglected, it follows from \eq{mvy} that the
corresponding diagonal Yukawa couplings are real and positive if the
phase of the Higgs field is chosen such that the neutral Higgs vacuum
expectation value $v>0$.}). Using the formalism of
\sec{subsec:selfenergies} for Dirac fermions, the independent 1PI
self-energy functions are given by\footnote{Since the Yukawa couplings can
be chosen real (in the one-generation model), $\Sigmabar_{Lt}=
\Sigma_{Lt}$. Note that after suppressing the color degrees of
freedom, $\Sigma_{Lt}$, $\Sigma_{Rt}$ and $\Sigma_{Dt}$ are
one-dimensional matrices, so we do not employ boldface letters in this
case.} $\Sigma_{Lt}$, $\Sigma_{Rt}$ and $\Sigma_{Dt}$ (defined in
\fig{fig:diracselfenergies}) as shown in \fig{fig:topself}.
\begin{figure}[ht!]
\begin{flushleft}
\begin{picture}(70,47)(0,0)
\Text(35,47)[c]{$p$}
\LongArrow(50,42)(20,42)
\ArrowLine(25,25)(0,25)
\ArrowLine(70,25)(45,25)
\GCirc(35,25){10}{0.85}
\Text(35,5)[c]{$\BDneg i p\newcdot \sigmabar\, \Sigma_{Lt}$}
\end{picture}
\begin{picture}(26,47)(0,0)
\Text(13,25)[c]{$=$}
\end{picture}
\begin{picture}(70,47)(0,0)
\ArrowLine(20,15)(0,15)
\ArrowLine(50,15)(20,15)
\ArrowLine(70,15)(50,15)
\Text(5,7)[c]{$t$}
\Text(35,7)[c]{$t$}
\Text(65,7)[c]{$t$}
\Text(35,40)[c]{$g,\gamma,Z$}
\PhotonArc(35,15)(15,0,180){2.5}{6.5}
\end{picture}
\begin{picture}(12,47)(0,0)
\Text(6,25)[c]{$+$}
\end{picture}
\begin{picture}(70,47)(0,0)
\ArrowLine(20,15)(0,15)
\ArrowLine(50,15)(20,15)
\ArrowLine(70,15)(50,15)
\Text(5,7)[c]{$t$}
\Text(35,7)[c]{$b$}
\Text(65,7)[c]{$t$}
\Text(35,40)[c]{$W^+$}
\PhotonArc(35,15)(15,0,180){2.5}{6.5}
\end{picture}
\begin{picture}(12,47)(0,0)
\Text(6,25)[c]{$+$}
\end{picture}
\begin{picture}(70,47)(0,0)
\ArrowLine(20,15)(0,15)
\ArrowLine(20,15)(50,15)
\ArrowLine(70,15)(50,15)
\Text(5,7)[c]{$t$}
\Text(35,7)[c]{$\bar t$}
\Text(65,7)[c]{$t$}
\Text(35,40)[c]{$h_{\rm SM}, G^0$}
\DashCArc(35,15)(15,0,180){3.5}
\end{picture}
\begin{picture}(12,47)(0,0)
\Text(6,25)[c]{$+$}
\end{picture}
\begin{picture}(70,47)(0,0)
\ArrowLine(20,15)(0,15)
\ArrowLine(20,15)(50,15)
\ArrowLine(70,15)(50,15)
\Text(5,7)[c]{$t$}
\Text(35,7)[c]{$\bar b$}
\Text(65,7)[c]{$t$}
\Text(35,40)[c]{$G^+$}
\DashArrowArcn(35,15)(15,180,0){3.5}
\end{picture}
\end{flushleft}
%
%
\begin{flushleft}
\begin{picture}(70,50)(0,0)
\Text(35,47)[c]{$p$}
\LongArrow(50,42)(20,42)
\ArrowLine(0,25)(25,25)
\ArrowLine(45,25)(70,25)
\GCirc(35,25){10}{0.85}
\Text(35,5)[c]{$\BDneg i p\newcdot \sigma \,\Sigma_{Rt}$}
\end{picture}
\begin{picture}(26,50)(0,0)
\Text(13,25)[c]{$=$}
\end{picture}
\begin{picture}(70,50)(0,0)
\ArrowLine(0,15)(20,15)
\ArrowLine(20,15)(50,15)
\ArrowLine(50,15)(70,15)
\Text(5,7)[c]{$\bar t$}
\Text(35,7)[c]{$\bar t$}
\Text(65,7)[c]{$\bar t$}
\Text(35,40)[c]{$g,\gamma,Z$}
\PhotonArc(35,15)(15,0,180){2.5}{6.5}
\end{picture}
\begin{picture}(12,50)(0,0)
\Text(6,25)[c]{$+$}
\end{picture}
\begin{picture}(70,50)(0,0)
\ArrowLine(0,15)(20,15)
\ArrowLine(50,15)(20,15)
\ArrowLine(50,15)(70,15)
\Text(5,7)[c]{$\bar t$}
\Text(35,7)[c]{$t$}
\Text(65,7)[c]{$\bar t$}
\Text(35,40)[c]{$h_{\rm SM}, G^0$}
\DashCArc(35,15)(15,0,180){3.5}
\end{picture}
\begin{picture}(12,50)(0,0)
\Text(6,25)[c]{$+$}
\end{picture}
\begin{picture}(70,50)(0,0)
\ArrowLine(0,15)(20,15)
\ArrowLine(50,15)(20,15)
\ArrowLine(50,15)(70,15)
\Text(5,7)[c]{$\bar t$}
\Text(35,7)[c]{$b$}
\Text(65,7)[c]{$\bar t$}
\Text(35,40)[c]{$G^+$}
\DashArrowArcn(35,15)(15,180,0){3.5}
\end{picture}
\end{flushleft}
%
%
\begin{flushleft}
\begin{picture}(70,50)(0,0)
\Text(35,47)[c]{$p$}
\LongArrow(50,42)(20,42)
\ArrowLine(0,25)(25,25)
\ArrowLine(70,25)(45,25)
\GCirc(35,25){10}{0.85}
\Text(32,4)[c]{$-i \Sigma_{Dt}$}
\end{picture}
\begin{picture}(26,50)(0,0)
\Text(13,25)[c]{$=$}
\end{picture}
\begin{picture}(70,50)(0,0)
\ArrowLine(0,15)(20,15)
\ArrowLine(20,15)(35,15)
\ArrowLine(50,15)(35,15)
\ArrowLine(70,15)(50,15)
\Text(5,7)[c]{$\bar t$}
\Text(27,7)[c]{$\bar t$}
\Text(43,7)[c]{$t$}
\Text(65,7)[c]{$t$}
\Text(35,40)[c]{$g,\gamma,Z$}
\PhotonArc(35,15)(15,0,180){2.5}{6.5}
\end{picture}
\begin{picture}(12,50)(0,0)
\Text(6,25)[c]{$+$}
\end{picture}
\begin{picture}(70,50)(0,0)
\ArrowLine(0,15)(20,15)
\ArrowLine(35,15)(20,15)
\ArrowLine(35,15)(50,15)
\ArrowLine(70,15)(50,15)
\Text(5,7)[c]{$\bar t$}
\Text(27,7)[c]{$t$}
\Text(43,7)[c]{$\bar t$}
\Text(65,7)[c]{$t$}
\Text(35,40)[c]{$h_{\rm SM}, G^0$}
\DashCArc(35,15)(15,0,180){3.5}
\end{picture}
\begin{picture}(12,50)(0,0)
\Text(6,25)[c]{$+$}
\end{picture}
\begin{picture}(70,50)(0,0)
\ArrowLine(0,15)(20,15)
\ArrowLine(35,15)(20,15)
\ArrowLine(35,15)(50,15)
\ArrowLine(70,15)(50,15)
\Text(5,7)[c]{$\bar t$}
\Text(27,7)[c]{$b$}
\Text(43,7)[c]{$\bar b$}
\Text(65,7)[c]{$t$}
\Text(35,40)[c]{$G^+$}
\DashArrowArcn(35,15)(15,180,0){3.5}
\end{picture}
\begin{picture}(14,50)(0,0)
\Text(7,25)[c]{$+$}
\end{picture}
\begin{picture}(50,50)(0,0)
\ArrowLine(0,15)(26,15)
\ArrowLine(52,15)(26,15)
\DashLine(26,15)(26,36){3.2}
\Text(5,7)[c]{$\bar t$}
\Text(48.5,7)[c]{$t$}
\Text(37.5,26)[c]{$h_{\rm SM}$}
\GCirc(26,44){10}{0.85}
\end{picture}
\end{flushleft}
\caption{\label{fig:topself} One-loop contributions to the
1PI self-energy functions for the top quark in
the Standard Model.  The external momentum of the physical top quark,
$p^\mu$, flows from the right to the left. The loop momentum $k^\mu$ in
the text is taken to flow clockwise. Spinor and color indices are
suppressed. The external legs are amputated. The last diagram
contains one-loop tadpole contributions.}
\end{figure}
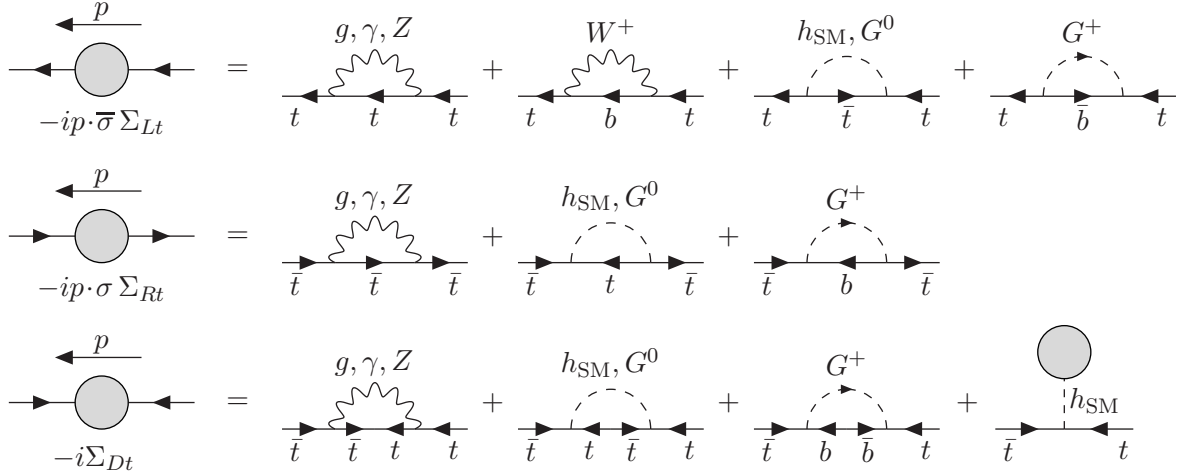

Note that in these diagrams, the physical top quark moves from right
to left, carrying momentum $p^\mu$.  Then according to the general
formula obtained in \eq{eq:diracpolemass}, 
the complex pole squared mass of the top quark is given by:
\beq
M_t^2 - i \Gamma_t M_t = \frac{(m_t + \Sigma_{Dt})^2}
{(1-\Sigma_{Lt})(1 - \Sigma_{Rt})}\,,
\eeq
where $m_t$ is the tree-level mass.
Working consistently to one-loop order, this yields\footnote{It would 
be just as valid to substitute in $s = M_t^2 + i \varepsilon$ here,
as two-loop order effects are being neglected.}
\beq
M_t^2  - i \Gamma_t M_t =
\left[m_t^2 (1+ \Sigma_{Lt} + \Sigma_{Rt}) + 2 m_t \Sigma_{Dt}\right]\Bigl
|_{s = m_t^2 + i \varepsilon}
\Bigr. .
\label{eq:toppoleoneloop}
\eeq

It remains to calculate the self-energy functions $\Sigma_{Lt}$,
$\Sigma_{Rt}$ and $\Sigma_{Dt}$.
Two regularization procedures will be used simultaneously---the $\MSbar$
scheme \cite{MSbar} based on dimensional regularization \cite{dimreg}
and the $\DRbar$
scheme based on dimensional reduction \cite{DRbar}. This is accomplished
by integrating over the loop momentum in
\beq
d\equiv 4 - 2 \epsilon
\eeq
dimensions, but with the vector bosons possessing
\beq
D \equiv 4 - 2 \epsilon \delta_{\MSbar}
\eeq
components, where
\beqa
\delta_{\MSbar} \equiv
\Biggl \{
  \begin{array}{ll}
     1 \phantom{xxx} & \mbox{for}\quad\MSbar\,, \\
     0 \phantom{xxx} & \mbox{for}\quad\DRbar\,.
  \end{array}
\label{eq:defdeltaMSbar}
\eeqa
In other words, the metric $\metric^{\mu\nu}$ appearing explicitly in the
vector propagator is treated as four dimensional in $\DRbar$, but as
$d$-dimensional in $\MSbar$. The renormalization scale $Q$ is related to
the regularization scale $\mu$ in both cases by the modified minimal
subtraction relation of \eq{eq:relatemuQ}.

The calculation of the non-tadpole contributions to the self-energy functions
will be performed below in a general $R_\xi$ gauge, with a vector boson propagator
as in \fig{fig:bosonprops}. There are different ways to treat the tadpole
contributions, corresponding to different choices for the Higgs vacuum expectation value around which
the tree-level Lagrangian is expanded. If one chooses to expand around the
minimum of the tree-level Higgs potential, then there are no tree-level tadpoles,
but there will be non-zero contributions from the last diagram shown in
\fig{fig:topself}. (This corresponds to the treatment given, for example,
in ref.~\cite{Hempfling:1994ar}.)
Alternatively, one can choose to expand around the Higgs vacuum expectation value $v$
that minimizes the one-loop Landau
gauge\footnote{This procedure
is considerably more involved outside of Landau gauge,
because the propagators mix the longitudinal
components of the vector boson with the
Nambu-Goldstone bosons for $\xi\neq 0$ if one expands around a Higgs
vacuum expectation value that does not
minimize the tree-level potential. This is the same reason the effective
potential is traditionally calculated specifically in Landau gauge.}
effective potential. In that case, the one-loop tadpole contribution is precisely
canceled by the tree-level Higgs tadpole, as shown in \fig{fig:tadpolescancel}.
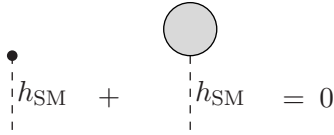
\begin{figure}
\begin{center}
\begin{picture}(30,40)(0,5)
\DashLine(15,0)(15,28){3.2}
\Text(26.5,14)[c]{$h_{\rm SM}$}
\Vertex(15,28){2}
\end{picture}
\begin{picture}(30,40)(0,5)
\Text(18,12)[c]{$+$}
\end{picture}
\begin{picture}(30,40)(0,5)
\DashLine(15,0)(15,28){3.2}
\Text(26.5,14)[c]{$h_{\rm SM}$}
\GCirc(15,38){10}{0.85}
\end{picture}
\begin{picture}(30,40)(0,5)
\Text(26,12)[c]{$=\> 0$}
\end{picture}
\end{center}
\caption{\label{fig:tadpolescancel} The tree-level Higgs tadpole cancels
against the one-loop Higgs tadpole, provided that one expands
around a Higgs vacuum expectation value that minimizes the one-loop effective potential
(rather than the tree-level Higgs potential, which would yield no
tree-level tadpole).}
\end{figure}
Here, we have in mind the latter prescription; the calculation
for the pole mass is therefore complete without tadpole contributions
provided that the tree-level top quark mass is taken to be
\beq
m_t = Y_t v,
\eeq
where $Y_t$ is the $\MSbar$ or $\DRbar$ Yukawa coupling, and
$v$ is the Higgs vacuum expectation value at the minimum of the one-loop effective potential
in Landau gauge. To be consistent with this choice,
$\xi=0$ should be taken in all formulae below
that involve electroweak gauge bosons or Goldstone bosons. (The gluon
contribution is naturally independent of $\xi$ because
the gauge symmetry is unbroken, providing
a check of gauge-fixing invariance.) Nevertheless, for the sake of
generality we will keep the dependence on
$\xi$ in the computation of the
individual non-tadpole self-energy diagrams below.

Consider the one-loop calculation of the self-energy $\Sigma_{Lt}$,
which is the sum of individual diagram contributions
$
\Sigma_{Lt} =
[\Sigma_{Lt}]_{g}
+[\Sigma_{Lt}]_{\gamma}
+[\Sigma_{Lt}]_{Z}
+[\Sigma_{Lt}]_{W}
+[\Sigma_{Lt}]_{h_{\rm SM}}
+[\Sigma_{Lt}]_{G^0}
+[\Sigma_{Lt}]_{G^+} .
$
First, consider the diagrams involving exchanges of the scalars
$\phi=h_{\rm SM},G^0,G^\pm$. These contributions all have the same form
\beqa
\BDneg i p\newcdot \sigmabar \, [\Sigma_{Lt}]_\phi  \,=\,
\mu^{2\epsilon} \int \frac{d^dk}{(2\pi)^d}\>
( -i Y^* )
\biggl ( \frac{i (k+p)\newcdot \sigmabar}
{(k+p)^2 \BDminus m_f^2} \biggr )
( -i Y )
\biggl ( \frac{\BDpos i}{k^2 \BDminus m_\phi^2} \biggr ) ,
\phantom{xxx}
\eeqa
where the loop momentum $k^\mu$ flows clockwise, and the couplings and
propagator masses are, using the Feynman rules of
\figs{qqhiggs}{fig:SMNGbosons},
\beqa
&\mbox{ for } \phi = h_{\rm SM}:\qquad
&Y = Y_t/\sqrt{2}; \qquad \;m_f = m_t; \qquad m_\phi^2 = m^2_{h_{\rm SM}},
\label{eq:Atthrules}
\\
&\mbox{ for } \phi = G^0:\phantom{i}\qquad
&Y = iY_t/\sqrt{2}; \qquad m_f = m_t;  \qquad m^2_\phi = \xi m^2_{Z},
\label{eq:AttGzerorules}
\\
&\mbox{ for } \phi = G^\pm:\phantom{.}\qquad
&Y = Y_b; \qquad \;\;\;\;\;\;\;m_f = m_b;  \qquad m^2_\phi = \xi m^2_{W}.
\label{eq:AttGpmrules}
\eeqa
Multiplying both sides by $p\newcdot \sigma$ and taking the
trace over spinor indices using eq.~(\ref{APPtrssbar}),
one finds
\beqa
[\Sigma_{Lt}]_\phi  \,=\,
i |Y|^2 \frac{\mu^{2\epsilon}}{p^2} \int \frac{d^d k}{(2 \pi)^d}
\frac{p \newcdot (k+p)}{[(k+p)^2 \BDminus m_f^2][k^2 \BDminus m_\phi^2]}
\,.
\eeqa
Performing the loop momentum integration in the standard way
\cite{Peskin:1995ev,Ramond}, and expanding in
$\epsilon$ up to constant terms,
one finds that in each case
\beq
[\Sigma_{Lt}]_{\phi} = -\frac{1}{16 \pi^2} |Y|^2 \, I_{FS}(s; m_f^2, m_\phi^2).
\label{eq:Aphitt}
\eeq
Here we have introduced some notation for the loop integral:
\beq
I_{FS}(s;x,y) \equiv
\frac{1}{2\epsilon} + \frac{(s+x-y) B(s;x,y) + A(x) - A(y)}{2s}\,,
\label{eq:defIFS}
\eeq
where the Passarino-Veltman functions $A(x)$ and $B(s;x,y)$ were defined
in eqs.~(\ref{eq:defAPV}) and (\ref{eq:defBPV}).  These functions depend
on the renormalization scale $Q$, which is related to $\mu$ via
\eq{eq:relatemuQ}. It can be checked that $I_{FS}(s;x,y)$ has a smooth
limit as $s \rightarrow 0$.

Next, let us consider the contributions to $\Sigma_{Lt}$ involving the vector
bosons $V = g,\gamma,Z,W$. These have the common form:
\beqa
\BDneg i p\newcdot \sigmabar [\Sigma_{Lt}]_V  \,=\,
\mu^{2\epsilon} \int \frac{d^dk}{(2\pi)^d}\>
\left ( \BDneg i G\, \sigmabar_\mu \right )
\biggl ( \frac{i (k+p)\newcdot \sigma}{(k+p)^2 \BDminus m_f^2} \biggr )
\left ( \BDneg i G\, \sigmabar_\nu \right )
\phantom{,}
&&
\nonumber \\
\left ( \frac{-i}{k^2 \BDminus m_V^2} \right )
\left (\metric^{\mu\nu} + \frac{(\xi-1) k^\mu k^\nu}{k^2 \BDminus \xi m_V^2}
\right ),
&&
\eeqa
where again the loop momentum $k$ flows clockwise, and, using the
rules of \figs{SMintvertices}{fig:SUSYQCDgluonrules}:
\beqa
&\mbox{ for } V = g:\qquad
&G= g_s T^a\,, \qquad \qquad \qquad \quad \;\;\, m_f = m_t,
\label{eq:Attgrules}
\\
&\mbox{ for } V = \gamma:\qquad
&G= e Q_t\,, \qquad \qquad \qquad \qquad \; m_f = m_t,
\label{eq:Attgammarules}
\\
&\mbox{ for } V = Z:\qquad
&G= g (T_3^t - s_W^2 Q_t)/c_W\,, \qquad m_f = m_t,
\label{eq:AttZrules}
\\
&\mbox{ for } V = W:\qquad
&G= g/\sqrt{2}\,, \qquad \qquad \qquad \quad\;\;  m_f = m_b.
\label{eq:AttWrules}
\eeqa
In the case of gluon exchange ($V=g$), the $T^a$ are the $SU(3)_C$
generators (with color indices suppressed). The adjoint representation
index $a$ is summed over, producing a factor of the Casimir invariant
$(T^a T^a)_{ij} = C_F\delta_{ij} = \nicefrac{4}{3}\delta_{ij}$. We now use
$\sigmabar_\mu \,\sigma_\rho \,\sigmabar_\nu \,\metric^{\mu\nu} =
\BDneg (D-2 )\, \sigmabar_\rho$ [see \eq{eq:genfierzfour}]; note that this
introduces a difference between the $\MSbar$ and $\DRbar$ schemes.
Also, we use $k\newcdot \sigmabar (k+p)\newcdot \sigma k\newcdot \sigmabar =
\BDpos (k^2 + 2 k\newcdot p) k\newcdot \sigmabar
\BDminus k^2 p \newcdot \sigmabar$,
which follows from eq.~(\ref{eq:simplifysbarssbar}).
One
therefore obtains, after multiplying by $p\newcdot \sigma$ and taking the
trace over spinor indices:
\beqa
[\Sigma_{Lt}]_V  &=&
-i\, G^2 \frac{\mu^{2\epsilon}}{p^2} \int \frac{d^dk}{(2\pi)^d}\>
\frac{1}{[(k+p)^2 \BDminus m_f^2][k^2 \BDminus m_V^2]}
\biggl [
(2-D) p\newcdot (k+p)
\nonumber \\ &&
+ \left (k^2 k\newcdot p + 2 (k \newcdot p)^2  - k^2 p^2 \right )
\frac{(\xi-1)}{k^2 \BDminus \xi m_V^2}
\biggr ]
\,.
\eeqa
Performing the loop momentum integration, one finds that
\beqa
[\Sigma_{Lt}]_V = -\frac{1}{16\pi^2} G^2 I_{FV}(s; m_f^2, m_V^2) ,
\label{eq:Avtt}
\eeqa
where we have introduced the notation
\beqa
I_{FV}(s; x,y) &=&
\frac{\xi}{\epsilon} +
[(s+x-y) B(s;x,y) + A(x) - A(y)]/s - \delta_{\MSbar}
+ \bigl \{ (s-x)[A(y) - A(\xi y)]\phantom{xx}
\nonumber \\ &&
\!\!\!\!\!\!\!\!\!\!\!\!\!\!\!\!
+ [(s-x)^2 - y (s+x)] B(s;x,y)
- [(s-x)^2 - \xi y (s+x)] B(s;x,\xi y)
\bigr \}/2ys,
\label{eq:defIFV}
\eeqa
after dropping terms that vanish as $\epsilon \rightarrow 0$. Combining
the results of eqs.~(\ref{eq:Aphitt}) and (\ref{eq:Avtt}):
\beqa
&& \hspace{-0.6in} \Sigma_{Lt} = -\frac{1}{16\pi^2} \Bigl [
\Bigl (g_s^2 C_F + e^2 Q_t^2 \Bigr ) I_{FV}(m_t^2; m_t^2,0)
+ [g (T_3^t - s_W^2 Q_t)/c_W]^2 I_{FV}(m_t^2; m_t^2,m_Z^2)
\nonumber \\[6pt] &&
+ \frac{1}{2} g^2 I_{FV}(m_t^2; m_b^2, m_W^2)
+ \frac{1}{2} Y_t^2 I_{FS} (m_t^2; m_t^2, m_{h_{\rm SM}}^2)
\nonumber \\[6pt] &&
+ \frac{1}{2} Y_t^2 I_{FS} (m_t^2; m_t^2, \xi m_{Z}^2)
+ Y_b^2 I_{FS} (m_t^2; m_b^2, \xi m_{W}^2)
\Bigr ] ,
\label{eq:Attresult}
\eeqa
where we have now substituted $s = m_t^2$.
It is useful to note that for massless gauge bosons,
\beq
I_{FV}(x; x,0) = \xi
\biggl [\frac{1}{\epsilon} - \ln(x/Q^2) + 2\biggr ] + 1 - \delta_{\MSbar}.
\label{eq:IFVxx0}
\eeq

The contributions to $\Sigma_{Rt} =
[\Sigma_{Rt}]_g
+ [\Sigma_{Rt}]_\gamma
+ [\Sigma_{Rt}]_Z
+ [\Sigma_{Rt}]_{h_{\rm SM}}
+ [\Sigma_{Rt}]_{G^0}
+ [\Sigma_{Rt}]_{G^\pm}
$ are obtained similarly.
[Note that there is no $W$ boson contribution, since the right-handed
top quark is an $SU(2)_L$ singlet.] For the scalar exchange diagrams
with $\phi = h_{\rm SM}, G^0, G^\pm$, the general form is:
\beqa
\BDneg i p\newcdot \sigma [\Sigma_{Rt}]_\phi  \,=\,
\mu^{2\epsilon} \int \frac{d^dk}{(2\pi)^d}\>
(-iY)
\biggl ( \frac{i (k+p)\newcdot \sigma}{(k+p)^2 \BDminus m_f^2} \biggr )
(-iY^*)
\left ( \frac{\BDpos i}{k^2 \BDminus m_\phi^2} \right )\,,
\phantom{xxx}
\eeqa
which yields
\beq
[\Sigma_{Rt}]_{\phi} = -\frac{1}{16 \pi^2} |Y|^2 \,
I_{FS}(s; m_f^2, m_\phi^2)  .
\eeq
Here the couplings and propagator masses for $h_{\rm SM}$ and $G^0$
are the same as in eqs.~(\ref{eq:Atthrules}),
(\ref{eq:AttGzerorules}), but now instead of eq.~(\ref{eq:AttGpmrules}),
\beqa
\mbox{ for } \phi = G^\pm:\phantom{.}\qquad
Y = -Y_t\,, \qquad m_f = m_b\,,  \qquad m_\phi^2 = \xi m_{W}^2\,,
\eeqa
from \fig{fig:SMNGbosons}. For the contributions due to exchanges of
vectors $v=g,\gamma,Z$, the general form is given by
\beqa
\BDneg i p\newcdot \sigma [\Sigma_{Rt}]_V  \,=\,
\mu^{2\epsilon} \int \frac{d^dk}{(2\pi)^d}\>
\left (\BDpos i G \,\sigma_\mu \right )
\biggl ( \frac{i (k+p)\newcdot \sigmabar}
{(k+p)^2 \BDminus m_f^2} \biggr )
\left (\BDpos i G \,\sigma_\nu \right )
\phantom{,}
&&
\nonumber \\
\left ( \frac{-i}{k^2 \BDminus m_V^2} \right )
\left (\metric^{\mu\nu} + \frac{(\xi-1) k^\mu k^\nu}{k^2 \BDminus \xi m_V^2}
\right ),
&&
\eeqa
where
\beqa
&\mbox{ for } V = g:\qquad
&G= -g_s \boldsymbol{T^a}\,, 
\\
&\mbox{ for } V = \gamma:\qquad
&G= -e Q_t\,, 
\\
&\mbox{ for } V = Z:\qquad
&G= g s_W^2 Q_t/c_W\,, 
\eeqa
after using the rules of \figs{SMintvertices}{fig:SUSYQCDgluonrules}
with $m_f = m_t$ in each case.  We then make use of
$\sigma_\mu \,\sigmabar_\rho \,\sigma_\nu\, g^{\mu\nu} =
\BDneg (D-2)\, \sigma_\rho$ [cf.~\eq{eq:genfierzthree}] and
$k\newcdot \sigma (k+p)\newcdot \sigmabar k\newcdot \sigma =
\BDpos (k^2 + 2 k\newcdot p) k\newcdot \sigma
\BDminus k^2 p \newcdot \sigma$
[cf.~eq.~(\ref{eq:simplifyssbars})].  After
multiplying by $p\newcdot \sigmabar$ and taking the trace over
spinor indices [using eq.~(\ref{APPtrssbar})], we obtain
\beq
[\Sigma_{Rt}]_V = -\frac{1}{16\pi^2} G^2 I_{FV}(s; m_t^2, m_V^2)\,,
\eeq
in terms of the same function appearing in
eqs.~(\ref{eq:defIFV}) and (\ref{eq:IFVxx0}).
Adding up these contributions and taking $s = m_t^2$ yields
\beqa
\hspace{-0.2in}
\Sigma_{Rt} &=& -\frac{1}{16\pi^2} \Bigl [
\Bigl (g_s^2 C_F + e^2 Q_t^2 \Bigr ) I_{FV}(m_t^2; m_t^2,0)
+ (g^2 Q_t^2 s_W^4/c_W^2) I_{FV}(m_t^2; m_t^2,m_Z^2)
\nonumber \\[4pt] &&
+ \frac{1}{2} Y_t^2 I_{FS} (m_t^2; m_t^2, m_{h_{\rm SM}}^2)
+ \frac{1}{2} Y_t^2 I_{FS} (m_t^2; m_t^2, \xi m_{Z}^2)
+ Y_t^2 I_{FS} (m_t^2; m_b^2, \xi m_{W}^2)
\Bigr ] .\phantom{xxx}
\label{eq:Atctcresult}
\eeqa

Next, consider the contributions to
$\Sigma_{Dt} =
[\Sigma_{Dt}]_g
+ [\Sigma_{Dt}]_\gamma
+ [\Sigma_{Dt}]_Z
+ [\Sigma_{Dt}]_{h_{\rm SM}}
+ [\Sigma_{Dt}]_{G^0}
+ [\Sigma_{Dt}]_{G^\pm},
$
ignoring the tadpole contribution for now.
The diagrams involving the exchange of scalars $\phi = h_{\rm SM}, G^0, G^\pm$
have the form:
\beqa
-i [\Sigma_{Dt}]_\phi  \,=\,
\mu^{2\epsilon} \int \frac{d^dk}{(2\pi)^d}\>
\left ( -i Y_1 \right )
\biggl ( \frac{\BDpos i m_f}{(k+p)^2 \BDminus m_f^2} \biggr )
\left ( -i Y_2 \right )
\biggl ( \frac{\BDpos i}{k^2 \BDminus m_\phi^2} \biggr ) ,
\phantom{xxx}
\eeqa
so that
\beqa
[\Sigma_{Dt}]_\phi  &=&
i m_f Y_1 Y_2 \mu^{2\epsilon} \int \frac{d^dk}{(2\pi)^d}\>
\frac{1}{[(k+p)^2 \BDminus m_f^2][k^2 \BDminus m_\phi^2]}
\nonumber \\
&=& \frac{1}{16\pi^2} m_f Y_1 Y_2
I_{\Fbar  S} (s; m_f^2, m^2_\phi)\,,
\label{eq:Bphittc}
\eeqa
where we have introduced the notation:
\beqa
I_{\Fbar  S} (s; x,y) \equiv
-\frac{1}{\epsilon} - B(s;x,y) ,
\label{eq:defIfS}
\eeqa
after dropping terms that vanish as $\epsilon\to 0$.
The relevant couplings and masses are, from
\figs{qqhiggs}{fig:SMNGbosons}:
\beqa
&\mbox{ for } \phi = h_{\rm SM}:\qquad
&Y_1 = Y_2 = Y_t/\sqrt{2}\,,\,\,\,\,\,\,
\qquad m_f = m_t\,, \qquad m^2_\phi = m^2_{h_{\rm SM}},
\\
&\mbox{ for } \phi = G^0:\qquad
&Y_1 = Y_2 = iY_t/\sqrt{2},\,\, \qquad\,\,\, m_f = m_t\,, \qquad m^2_\phi = \xi m^2_{Z},
\\
&\mbox{ for } \phi = G^\pm:\qquad
&Y_1 = Y_b,\>\>\>\> Y_2 = -Y_t,\qquad m_f = m_b\,, \qquad m^2_\phi = \xi m^2_{W}.
\eeqa

The contributions from vector boson exchanges are of the form
\beqa
-i [\Sigma_{Dt}]_V  \,=\,
\mu^{2\epsilon} \int \frac{d^dk}{(2\pi)^d}\>
\left (\BDpos i G_1 \sigma_\mu \right )
\biggl ( \frac{\BDpos i m_f}{(k+p)^2 \BDminus m_f^2} \biggr )
\left (\BDneg i G_2 \sigmabar_\nu \right )
\phantom{,}
&&
\nonumber \\
\left ( \frac{-i}{k^2 \BDminus m_V^2} \right )
\left (\metric^{\mu\nu} +  \frac{(\xi-1)k^\mu k^\nu}{k^2 \BDminus \xi m_V^2}
\right ),
&&
\eeqa
Using $\sigma_\mu\sigmabar_\nu g^{\mu\nu} = \BDpos D$
[see \eq{eq:genfierzone}] and
$k\newcdot \sigma k \newcdot \sigmabar = \BDpos k^2$
[from eq.~(\ref{eq:ssbarsym})] yields
\beqa
[\Sigma_{Dt}]_V  &=& im_f G_1 G_2
\mu^{2\epsilon} \int \frac{d^dk}{(2\pi)^d}\>
\frac{1}{[(k+p)^2 \BDminus m_f^2][k^2 \BDminus m_V^2]}
\left [D + \frac{(\xi - 1) k^2}{k^2 \BDminus \xi m_V^2} \right ]
\nonumber\\[6pt]
&=& \frac{1}{16\pi^2} m_f G_1 G_2  I_{\Fbar  V} (s; m_f^2, m^2_V)\,,
\label{eq:Bvttc}
\eeqa
where
\beqa
I_{\Fbar  V} (s; x,y) \equiv -\frac{3 + \xi}{\epsilon} -3 B(s;x,y)
-\xi B(s;x,\xi y)
+ 2 \delta_{\MSbar},
\label{eq:defIfV}
\eeqa
after dropping terms that vanish as $\epsilon\to 0$.
It is useful to note that for massless gauge bosons
\beqa
I_{\Fbar  V} (x; x,0) \equiv -\frac{3 + \xi}{\epsilon}
+ (3 + \xi) [\ln(x/Q^2) -2]
+ 2 \delta_{\MSbar}.
\eeqa
The relevant couplings are obtained from the rules of
\figs{SMintvertices}{fig:SUSYQCDgluonrules}:
\beqa
&\mbox{ for } V = g:\qquad
&G_1 = -G_2 = g_s T^a,\\
&\mbox{ for } V = \gamma:\qquad
&G_1 = -G_2 = e Q_t,\\
&\mbox{ for } V = Z:\qquad
&G_1 = g (T_3^t - s_W^2 Q_t)/c_W, \qquad
G_2 = g s_W^2 Q_t/c_W,
\eeqa
and $m_f = m_t$ in each case.
Adding up these contributions and taking $s = m_t^2$, we have:
\beqa
\Sigma_{Dt} &= &
\frac{m_t}{16\pi^2} \biggl\{
g^2 \left[(T_3^t - s_W^2 Q_t)s_W^2 Q_t/c_W^2\right]
I_{\Fbar V}(m_t^2; m_t^2,m_Z^2)
-(g_s^2 C_F + e^2 Q_t^2 ) I_{\Fbar V}(m_t^2; m_t^2,0)
\phantom{xx}
\nonumber \\[6pt] &&
+ \half Y_t^2 I_{\Fbar S} (m_t^2; m_t^2, m_{h_{\rm SM}}^2)
- \half Y_t^2 I_{\Fbar S} (m_t^2; m_t^2, \xi m_{Z}^2)
- Y_b^2 I_{\Fbar S} (m_t^2; m_b^2, \xi m_{W}^2)
\biggr\}\, ,
\label{eq:Bttcresult}
\eeqa
where $Y_t = m_t Y_b/m_b$ was used on the last term.

In each of the self-energy functions above, there are poles in
$1/\epsilon$, contained within the functions $I_{FV}$, $I_{FS}$,
$I_{\Fbar V}$ and $I_{\Fbar S}$. In the $\MSbar$ or $\DRbar$
schemes, these poles are simply canceled by counterterms, which have no
other effect at one-loop order. The one-loop top quark pole mass can now
be obtained by plugging eqs.~(\ref{eq:Attresult}), (\ref{eq:Atctcresult}),
and (\ref{eq:Bttcresult}) into eq.~(\ref{eq:toppoleoneloop}) with $\xi=0$,
as discussed earlier. It is not hard to check that the terms
from massless Nambu-Goldstone boson exchange just cancel against
the terms
from the vector exchange diagrams that came from $\xi m_W^2$
and $\xi m_Z^2$.

As a simple example, consider the one-loop pole mass with only QCD effects
included. Then the result of eq.~(\ref{eq:toppoleoneloop}) has no
imaginary part. Taking the square root (and dropping a two-loop order
part) yields the well-known result \cite{Tarrach:1980up}:
\beqa
M_{t,{\rm pole}} &=& m_t (1 +\half \Sigma_{Lt}+\half \Sigma_{Rt})
 + \Sigma_{Dt} \nonumber \\[8pt]
&=&
m_t \Bigl (
1 -\frac{C_F g_s^2}{16\pi^2} \Bigl[
I_{FV}(m_t^2; m_t^2,0) + I_{\Fbar V}(m_t^2; m_t^2,0) \Bigr] \Bigr )
\nonumber \\[6pt]
&=& m_t \Bigr (
1 + \frac{\alpha_s}{4\pi} C_F \Bigl [5 - \delta_{\MSbar} - 3
\ln (m_t^2/Q^2 )\Bigr]\Bigr ).
\eeqa
As another check, consider the imaginary part of the 
pole squared mass of the top quark.
At leading order, \eq{eq:toppoleoneloop} implies:
\beqa
\Gamma_t &=& -{\rm Im}[m_t (\Sigma_{Lt} + \Sigma_{Rt}) + 2 \Sigma_{Dt}]
\nonumber \\[8pt]
&=&
\frac{m_t}{16 \pi^2} {\rm Im} \Bigl [ \frac{g^2}{2} I_{FV}(m_t^2; m_b^2, m_W^2) +
(Y_t^2 + Y_b^2)  I_{FS}(m_t^2; m_b^2, \xi m_W^2)
+ 2 Y_b^2  I_{\Fbar S}(m_t^2; m_b^2, \xi m_W^2)\Bigr ] \nonumber
\\[8pt]
&=&
\frac{1}{32 \pi^2 m_t} \left
\{
(g^2 + Y_t^2 + Y_b^2)(m_t^2 + m_b^2 - m_W^2) - 4 Y_b^2 m_t^2
\right \}
{\rm Im} [ B(m_t^2; m_b^2, m_W^2)] .
\label{eq:gammatfromoneloop}
\eeqa
The fact that the $\xi$ dependence canceled here
is a successful check of gauge-fixing invariance, since
the tadpole diagram in \fig{fig:topself} does not contribute
to the absorptive part of the self-energy.
One can express ${\rm Im}[B(s;x,y)]$ in terms of the triangle
function [cf.~\eq{eq:deftrianglefunction}],
\beq
{\rm Im}[B(s;x,y)] =
\Biggl \{
  \begin{array}{ll}
     0 \phantom{xxx} & \mbox{for}\quad s \leq (\sqrt{x} + \sqrt{y})^2, \\
     \pi \lambda^{1/2}(s,x,y)/s\phantom{xxx} &
      \mbox{for}\quad s > (\sqrt{x}+\sqrt{y})^2 .
  \end{array}
\eeq
\Eq{eq:gammatfromoneloop} then reproduces the result of
eq.~(\ref{eq:gammat}) for the top quark width at leading order.

\subsection{Self-energy and pole mass of the gluino}
\label{subsec:gluinopole}
\setcounter{equation}{0}
\setcounter{figure}{0}
\setcounter{table}{0}

The Feynman diagrams for the gluino self-energy are shown in
\fig{fig:gluinoself}.  Since the gluino is a Majorana fermion, we can
use the general formalism of \sec{subsec:selfenergies}. We will
compute the self-energy functions $\Xi_{\tilde g}\equiv \Xi_{\tilde
g}{}^{\tilde g}$ and $\Omega_{\tilde g}\equiv\Omega^{\tilde g\tilde g}$
defined in \fig{fig:selfenergies}, and infer
$\Omegabar _{\tilde g}\equiv \Omegabar _{\tilde g\tilde g}$
from the latter by replacing all Lagrangian parameters by their complex
conjugates.\footnote{Suppressing the color degrees of freedom, $\Xi$,
$\Omega$ and $\Omegabar $ are one-dimensional matrices,
so we do not employ boldface letters
in this case.}
At one-loop order, it follows from the general result of \eq{eq:polemass}
that the complex pole squared mass of the gluino is related to the tree-level
mass $m_{\tilde g}$ by
\beq
M^2_{\tilde g} - i M_{\tilde g} \Gamma_{\tilde g} =
\left[m^2_{\tilde g} (1 + 2 \Xi_{\tilde g}) +
m_{\tilde g} (\Omega_{\tilde g} + {\Omegabar }_{\tilde g})\right]
\Bigl |_{s = m_{\tilde g}^2 + i \varepsilon}
\Bigr.\,.
\label{eq:polegluinooneloop}
\eeq

It is convenient to split the self-energy functions into
gluon/gluino loop and squark/quark loop
contributions, as
\beq
\Xi_{\tilde g} = [\Xi_{\tilde g}]_{g} +
\sum_{q}\sum_{x=1,2}\,
[\Xi_{\tilde g}]_{\tilde q_x}\,,\qquad
{\rm and}\qquad
\Omega_{\tilde g} = [\Omega_{\tilde g}]_{g} + \sum_{q}\sum_{x=1,2}\,
[\Omega_{\tilde g}]_{\tilde q_x}\,,
\eeq
where the sum over $q$ runs over the six squark flavors $u,d,s,c,b,t$,
and $x=1,2$
corresponds to the two squark mass eigenstates [i.e., the two
appropriate linear combinations (for fixed squark flavor) of $\widetilde
q_L$ and $\widetilde q_R$]. The gluon exchange contributions, following
from the Feynman rules of \fig{fig:SUSYQCDgluonrules}, are:
\beqa
\BDneg i p\newcdot \sigmabar \,[\Xi_{\tilde g}]_{g}\,
\delta^{ab}
&=&
\mu^{2\epsilon} \int \frac{d^dk}{(2\pi)^d}\>
\left (\BDneg g_s f^{aec} \sigmabar_\mu \right )
\biggl ( \frac{i (k+p)\newcdot \sigma}{
(k+p)^2 \BDminus m_{\tilde g}^2} \biggr )
\left (\BDneg g_s f^{ebc} \sigmabar_\nu \right )
\nonumber \\ &&
\qquad \left ( \frac{-i}{k^2} \right )
\left ( \metric^{\mu\nu} + (\xi-1) \frac{k^\mu k^\nu}{k^2} \right )
,
\\[6pt]
-i \, [\Omega_{\tilde g}]_{g}\, \delta^{ab}
&=&
\mu^{2\epsilon} \int \frac{d^dk}{(2\pi)^d}\>
\left (\BDpos g_s f^{eac} \sigma_\mu \right )
\biggl ( \frac{\BDpos i m_{\tilde g}}{
(k+p)^2 \BDminus m_{\tilde g}^2} \biggr )
\left (\BDneg g_s f^{ebc} \sigmabar_\nu \right )
\nonumber \\ &&
\qquad \left ( \frac{-i}{k^2} \right )
\left ( \metric^{\mu\nu} + (\xi-1) \frac{k^\mu k^\nu}{k^2} \right )
 .
\eeqa
The internal gluon and gluino lines carry $SU(3)_c$ adjoint representation
indices $c$ and $e$ respectively, while the external
gluinos on the left and right carry indices
$a$ and $b$ respectively.  The gluino external momentum $p^\mu$ flows from
right to left,
and the loop momentum $k^\mu$ flows clockwise.
Comparing with the derivations of eqs.~(\ref{eq:Avtt}) and
(\ref{eq:Bvttc}) in the previous subsection,
and using $-f^{aec} f^{ebc} = f^{eac} f^{ebc} = \delta^{ab} C_A$
[with $C_A = 3$ for $SU(3)_c$],
we can immediately conclude that
\beq
[\Xi_{\tilde g}]_{g} =
-\frac{\alpha_s}{4\pi} C_A I_{FV} (s; m_{\tilde g}^2,0) ,
\eeq
\beq
[\Omega_{\tilde g}]_{g} = -\frac{\alpha_s}{4\pi} C_A
m_{\tilde g} I_{\Fbar V} (s; m_{\tilde g}^2,0) ,
\eeq
where the loop integral functions $I_{FV}$ and $I_{\Fbar V}$ were
defined in eqs.~(\ref{eq:defIFV}) and (\ref{eq:defIfV}).

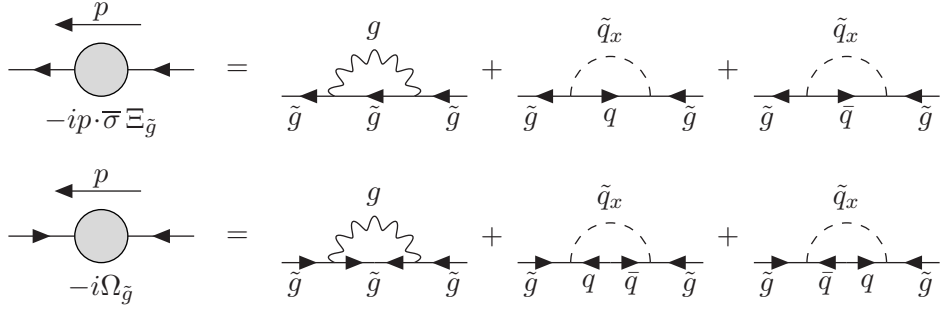
\begin{figure}[t]
\begin{center}
\begin{picture}(70,47)(0,0)
\Text(35,47)[c]{$p$}
\LongArrow(50,42)(20,42)
\ArrowLine(25,25)(0,25)
\ArrowLine(70,25)(45,25)
\GCirc(35,25){10}{0.85}
\Text(35,5)[c]{$\BDneg i p\newcdot \sigmabar \,\Xi_{\tilde g}$}
\end{picture}
\begin{picture}(26,47)(0,0)
\Text(13,25)[c]{$=$}
\end{picture}
\begin{picture}(70,47)(0,0)
\ArrowLine(20,15)(0,15)
\ArrowLine(50,15)(20,15)
\ArrowLine(70,15)(50,15)
\Text(5,7)[c]{$\tilde g$}
\Text(35,7)[c]{$\tilde g$}
\Text(65,7)[c]{$\tilde g$}
\Text(35,40)[c]{$g$}
\PhotonArc(35,15)(15,0,180){2.5}{6.5}
\end{picture}
\begin{picture}(12,47)(0,0)
\Text(6,25)[c]{$+$}
\end{picture}
\begin{picture}(70,47)(0,0)
\ArrowLine(20,15)(0,15)
\ArrowLine(20,15)(50,15)
\ArrowLine(70,15)(50,15)
\Text(5,7)[c]{$\tilde g$}
\Text(35,7)[c]{$q$}
\Text(65,7)[c]{$\tilde g$}
\Text(35,40)[c]{$\tilde q_x$}
\DashCArc(35,15)(15,0,180){3.5}
\end{picture}
\begin{picture}(12,47)(0,0)
\Text(6,25)[c]{$+$}
\end{picture}
\begin{picture}(70,47)(0,0)
\ArrowLine(20,15)(0,15)
\ArrowLine(20,15)(50,15)
\ArrowLine(70,15)(50,15)
\Text(5,7)[c]{$\tilde g$}
\Text(35,7)[c]{$\bar q$}
\Text(65,7)[c]{$\tilde g$}
\Text(35,40)[c]{$\tilde q_x$}
\DashCArc(35,15)(15,0,180){3.5}
\end{picture}
\end{center}
%
%
\begin{center}
\begin{picture}(70,50)(0,0)
\Text(35,47)[c]{$p$}
\LongArrow(50,42)(20,42)
\ArrowLine(0,25)(25,25)
\ArrowLine(70,25)(45,25)
\GCirc(35,25){10}{0.85}
\Text(35,5)[c]{$-i \Omega_{\tilde g}$}
\end{picture}
\begin{picture}(26,50)(0,0)
\Text(13,25)[c]{$=$}
\end{picture}
\begin{picture}(70,50)(0,0)
\ArrowLine(0,15)(20,15)
\ArrowLine(20,15)(35,15)
\ArrowLine(50,15)(35,15)
\ArrowLine(70,15)(50,15)
\Text(5,7)[c]{$\tilde g$}
\Text(35,7)[c]{$\tilde g$}
\Text(65,7)[c]{$\tilde g$}
\Text(35,40)[c]{$g$}
\PhotonArc(35,15)(15,0,180){2.5}{6.5}
\end{picture}
\begin{picture}(12,50)(0,0)
\Text(6,25)[c]{$+$}
\end{picture}
\begin{picture}(70,50)(0,0)
\ArrowLine(0,15)(20,15)
\ArrowLine(35,15)(20,15)
\ArrowLine(35,15)(50,15)
\ArrowLine(70,15)(50,15)
\Text(5,7)[c]{$\tilde g$}
\Text(27,7)[c]{$q$}
\Text(43,7)[c]{$\bar q$}
\Text(65,7)[c]{$\tilde g$}
\Text(35,40)[c]{$\tilde q_x$}
\DashCArc(35,15)(15,0,180){3.5}
\end{picture}
\begin{picture}(12,50)(0,0)
\Text(6,25)[c]{$+$}
\end{picture}
\begin{picture}(70,50)(0,0)
\ArrowLine(0,15)(20,15)
\ArrowLine(35,15)(20,15)
\ArrowLine(35,15)(50,15)
\ArrowLine(70,15)(50,15)
\Text(5,7)[c]{$\tilde g$}
\Text(27,7)[c]{$\bar q$}
\Text(43,7)[c]{$q$}
\Text(65,7)[c]{$\tilde g$}
\Text(35,40)[c]{$\tilde q_x$}
\DashCArc(35,15)(15,0,180){3.5}
\end{picture}
\end{center}
\caption{\label{fig:gluinoself} Self-energy functions for the gluino
in supersymmetry.
The external momentum $p^\mu$ flows from the right to the left.
The loop momentum $k^\mu$ in the text
is taken to flow clockwise.  Spinor and color indices are
suppressed. The index $x=1,2$ labels the two squark mass eigenstates of
a given flavor $q=u,d,s,c,b,t$. Both $x$ and $q$ must be summed over.
The external legs are amputated.}
\end{figure}

Next consider the virtual squark exchange diagrams contributing to
$\Xi_{\tilde g}$.
Labeling the quark and squark with color
indices $j,k$ respectively, we have for each squark mass eigenstate:
\beqa
&&
\!\!\!
\BDneg i p\newcdot \sigmabar \,
[\Xi_{\tilde g}]_{\tilde q_x}\, \delta^{ab}
=
\mu^{2\epsilon} \!\int\! \frac{d^dk}{(2\pi)^d}
\bigl (-i \sqrt{2} g_s T_j^{ak} L_{\tilde q_x} \bigr )
\left ( \frac{i (k+p)\newcdot \sigmabar}{
(k+p)^2 \BDminus m_{q}^2} \right )
\bigl (-i \sqrt{2} g_s T_k^{bj} L_{\tilde q_x}^* \bigr )
\biggl ( \frac{\BDpos i }{k^2 \BDminus m_{\tilde q_x}^2 } \biggr )
\nonumber
\\[4pt]
&&
\qquad\quad +
\mu^{2\epsilon} \int \frac{d^dk}{(2\pi)^d}\>
\left (i \sqrt{2} g_s T_k^{aj} R^*_{\tilde q_x} \right )
\left ( \frac{i (k+p)\newcdot \sigmabar}{
(k+p)^2 \BDminus m_{q}^2} \right )
\left (i \sqrt{2} g_s T_j^{bk} R_{\tilde q_x} \right )
\biggl ( \frac{\BDpos i }{k^2 \BDminus m_{\tilde q_x}^2 } \biggr ).
\phantom{xxxxxx}
\eeqa
This uses the Feynman rules shown in \fig{fig:SUSYQCDsquarkmasseigrules},
given in terms of the squark mixing parameters $L_{\tilde q_x}$ and
$R_{\tilde q_x}$ defined in \eq{eq:sfermionmix}.
Using ${\rm Tr}[T^a T^b] = \half\delta^{ab}$ and
$|L_{\tilde q_x}|^2 + |R_{\tilde q_x}|^2 = 1$, and comparing to the
derivation of eq.~(\ref{eq:Aphitt}) of the previous subsection,
we obtain:
\beq
[\Xi_{\tilde g}]_{\tilde q_x} =
-\frac{\alpha_s}{4\pi} I_{FS} (s; m_q^2, m^2_{\tilde q_x}).
\eeq
Similarly,
for the last two diagrams of \fig{fig:gluinoself},
we obtain:
\beqa
&&\!\!\!\!\!
-i [\Omega_{\tilde g}]_{\tilde q_x}\, \delta^{ab}
=
\mu^{2\epsilon} \int \frac{d^dk}{(2\pi)^d}\>
\left (-i \sqrt{2} g_s T_k^{aj} L^*_{\tilde q_x} \right )
\left ( \frac{\BDpos i m_q}{
(k+p)^2 \BDminus m_{q}^2} \right )
\left (i \sqrt{2} g_s T_j^{bk} R_{\tilde q_x} \right )
\biggl ( \frac{\BDpos i }{k^2 \BDminus m_{\tilde q_x}^2 } \biggr )
\nonumber \\[4pt] &&
\quad\quad\, +
\mu^{2\epsilon} \int \frac{d^dk}{(2\pi)^d}\>
\left (i \sqrt{2} g_s T_j^{ak} R_{\tilde q_x} \right )
\left ( \frac{\BDpos i m_q}{
(k+p)^2 \BDminus m_{q}^2} \right )
\left (-i \sqrt{2} g_s T_k^{bj} L^*_{\tilde q_x} \right )
\biggl ( \frac{\BDpos i }{k^2 \BDminus m_{\tilde q_x}^2 } \biggr ) ,
\phantom{xxxxx}
\eeqa
again using the Feynman rules shown in
\fig{fig:SUSYQCDsquarkmasseigrules}.
As before, $j$ and $k$ are the color indices for the quark and the
squark, respectively.
Comparing to the
derivation
of eq.~(\ref{eq:Bphittc}) of the previous subsection,
we obtain:
\beq
[\Omega_{\tilde g}]_{\tilde q_x}
= -\frac{\alpha_s}{2\pi}  L^*_{\tilde q_x} R_{\tilde q_x}
m_q I_{\Fbar  S} (s; m_q^2, m^2_{\tilde q_x})\,.
\eeq

Summing up the results obtained above, and taking $s = m_{\tilde g}^2$, we have:
\beqa
\Xi_{\tilde g} &=&
-\frac{\alpha_s}{4\pi}
\biggl [ C_A I_{FV}(m_{\tilde g}^2; m^2_{\tilde g},0)
+ \sum_q \sum_{x=1,2} I_{FS}(m_{\tilde g}^2; m_q^2,m^2_{\tilde q_x})
\biggr ] \label{eq:gluinoAresult}
,
\\[6pt]
\Omega_{\tilde g} &=&
-\frac{\alpha_s}{4\pi}
\biggl [ C_A m_{\tilde g} I_{\Fbar V}(m_{\tilde g}^2; m^2_{\tilde
g},0)
+ 2 \sum_q \sum_{x=1,2}
L^*_{\tilde q_x} R_{\tilde q_x}
m_q I_{\Fbar S}(m_{\tilde g}^2; m_q^2,m^2_{\tilde q_x})
\biggr ]
\,.
\label{eq:gluinoBresult}
\eeqa
As previously noted,
we can now write down $\Omegabar _{\tilde g}$ by replacing
the Lagrangian parameters of \eq{eq:gluinoBresult} by their complex
conjugates:
\beq
\Omegabar _{\tilde g} =
-\frac{\alpha_s}{4\pi}
\biggl [ C_A m_{\tilde g} I_{\Fbar V}(m_{\tilde g}^2;
m^2_{\tilde g},0)
+ 2 \sum_q \sum_{x=1,2}
L_{\tilde q_x} R^*_{\tilde q_x}
m_q I_{\Fbar S}(m_{\tilde g}^2; m_q^2,m^2_{\tilde q_x})
\biggr ]\,.
\label{eq:gluinoBbarresult}
\eeq
Inserting the results of \eqst{eq:gluinoAresult}{eq:gluinoBbarresult} into
eq.~(\ref{eq:polegluinooneloop}), one obtains the result \cite{MV,Pierce}:
\beqa
M^2_{\tilde g} - i M_{\tilde g} \Gamma_{\tilde g} &=&
m_{\tilde g}^2 \biggl [
1 + \frac{\alpha_s}{2\pi} \Bigl\{
C_A \left[5 - \delta_{\MSbar}
- 3 \ln\bigl(m_{\tilde g}^2/Q^2\bigr)\right]
\nonumber \\[4pt] &&
\!\!\!\!\!\!\!\!\!\!\!\!\!\!\!\!\!\!\!\!\!
- \sum_q \sum_{x=1,2} \Bigl [
I_{FS} (m_{\tilde g}^2; m_q^2, m_{\tilde q_x}^2)
+2 {\rm Re}[L^*_{\tilde q_x}R_{\tilde q_x}]
\frac{m_q}{m_{\tilde g}} I_{\Fbar S}
(m_{\tilde g}^2; m_q^2, m_{\tilde q_x}^2)
\Bigr ]\Bigr\}\biggr ],
\eeqa
with $\delta_{\MSbar}$ defined in
eq.~(\ref{eq:defdeltaMSbar}).

\subsection{Triangle anomaly from chiral fermion loops}
\label{sec:anomaly}
\setcounter{equation}{0}
\setcounter{figure}{0}
\setcounter{table}{0}

As our final example, we consider the anomaly in chiral symmetries for
fermions,
arising from the triangle diagram involving three currents carrying
vector indices.\footnote{The discussion here parallels that given in
ref.~\cite{Weinberg2}, Section 22.3.}
Since the anomaly is independent of the fermion masses, we simplify
the computation by setting all fermion masses to zero.
In four-component notation,\footnote{For an excellent review of the
computation of the chiral anomaly via four-component
massless and massive spinor triangle loops, see \Ref{chill}.}
the treatment of the anomaly requires care because of
the difficulty in defining a consistent and unambiguous
$\gamma_5$ and the epsilon tensor
in dimensional regularization~\cite{novotony,jegerlehner}.
The same subtleties arise in two-component
language, of course, but in a slightly different form since
$\gamma_5$ does not appear explicitly.

We shall assemble all the $(\half,0)$ [left-handed]
two-component fermion fields of the theory into a
multiplet $\psi_j$.  For example, the fermions of the Standard Model
are:
$\psi_j=(\ell_k\,,\,\bar\ell_k\,,\,\nu_k\,,\,q_{i\ell}\,,\,\bar q_{i\ell})$,
where $k=1,2,3$ and $i=1,2,\ldots,6$ are flavor labels and
$\ell=1,2,3$ are color labels [see Table~\ref{tab:nomenclature}].
The two-component spinor indices are suppressed here.
Let the symmetry generators be given by hermitian matrices
$\boldsymbol{T^a}$, so that the $\psi_j$ transform as:
\beqa
\delta \psi_j = -i \theta^a (\boldsymbol{T^a})_j{}^k \psi_k ,
\eeqa
for infinitesimal parameters $\theta^a$. The matrices $\boldsymbol{T^a}$
form a representation $R$ of the generators of the Lie algebra of the symmetry group.
In general $R$ will be reducible, in which case the $\boldsymbol{T^a}$
have a block diagonal structure, where each
block separately transforms (irreducibly) the corresponding field of $\psi_j$
according to its symmetry transformation properties.
Some or all of these symmetries may be gauged. The Feynman rule
for the corresponding currents is the same as for external gauge
bosons, as in \fig{fig:Gaugevertexrules} (but without the
gauge couplings), and is shown in \fig{fig:currentrule}.
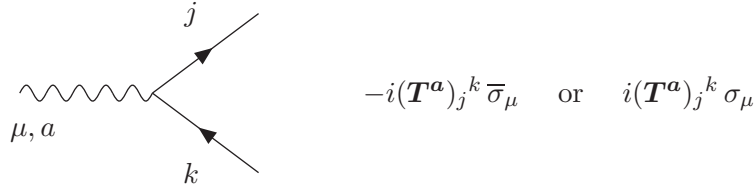
\begin{figure}[t!]
\begin{center}
\begin{picture}(300,57)(0,15)
\Photon(60,40)(10,40){3}{5}
\ArrowLine(60,40)(100,70)
\ArrowLine(100,10)(60,40)
\Text(15,25)[]{$\mu, a$}
\Text(75,10)[]{$k$}
\Text(75,70)[]{$j$}
\Text(140,40)[l]{$\BDneg i (\boldsymbol{T^a})_j{}^k \,
\sigmabar_\mu$
\quad \mbox{or} \quad $\BDpos i (\boldsymbol{T^a})_j{}^k\,
\sigma_{\mu}$}
\end{picture}
\end{center}
\caption{\label{fig:currentrule}
{Feynman rule for the coupling of a current carrying vector index
$\mu$ and corresponding to
the symmetry generator $\boldsymbol{T^a}$
acting on $(\half,0)$ [left-handed] fermions. Spinor indices are suppressed.}}
\end{figure}

\Fig{fig:anomaly} exhibits
the two Feynman diagrams that contribute at one-loop
to the three-point function of the
symmetry currents.
Applying the $\sigmabar$-version of the Feynman rule for the currents
given in \fig{fig:currentrule}, and employing the Feynman rules
of \fig{fig:neutprop} (with $m=0$)
for the propagators [traversing the loop in the direction dictated
by \eq{rule2}],
the sum of the two triangle diagrams shown in \fig{fig:anomaly}
can be evaluated.
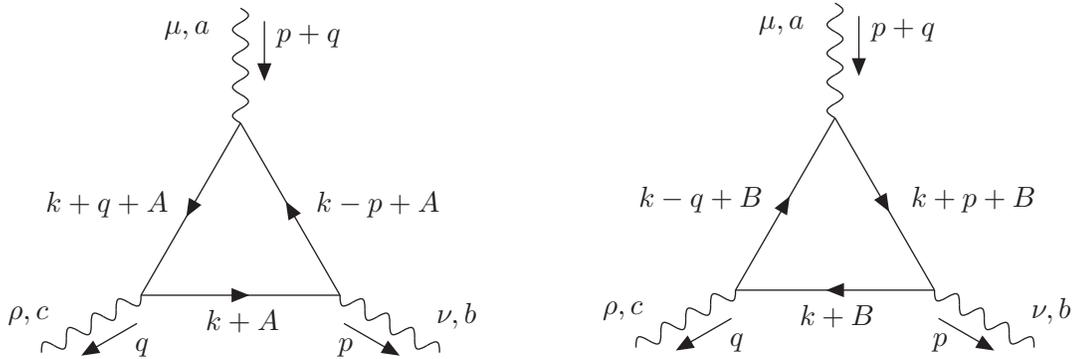
\begin{figure}[hb!]
\begin{center}
\begin{picture}(150,135)(5,8)
\ArrowLine(75,86.6025)(37.5,21.651)
\ArrowLine(112.5,21.651)(75,86.6025)
\ArrowLine(37.5,21.651)(112.5,21.651)
\Photon(0,0)(37.5,21.651){3}{4}
\Photon(150,0)(112.5,21.651){-3}{4}
\Photon(75,129.904)(75,86.6025){3}{4}
\Text(55,122)[c]{$\mu,a$}
\Text(157,14)[c]{$\nu,b$}
\Text(-5,14)[c]{$\rho,c$}
\Text(76,12)[c]{$k + A$}
\Text(104,56)[l]{$k - p + A$}
\Text(48,56)[r]{$k + q + A$}
\Text(89,120)[l]{$p+q$}
\Text(115,2)[c]{$p$}
\Text(38,2)[c]{$q$}
\LongArrow(84,125)(84,105)
\LongArrow(114,11)(132.75,0.1745)
\LongArrow(36,11)(17.25,0.1745)
\end{picture}
\hspace{2.2cm}
\begin{picture}(150,129.904)(0,6)
\ArrowLine(37.5,21.651)(75,86.6025)
\ArrowLine(75,86.6025)(112.5,21.651)
\ArrowLine(112.5,21.651)(37.5,21.651)
\Photon(0,0)(37.5,21.651){3}{4}
\Photon(150,0)(112.5,21.651){-3}{4}
\Photon(75,129.904)(75,86.6025){3}{4}
\Text(55,122)[c]{$\mu,a$}
\Text(157,14)[c]{$\nu,b$}
\Text(-5,14)[c]{$\rho,c$}
\Text(76,12)[c]{$k + B$}
\Text(104,56)[l]{$k + p  + B$}
\Text(48,56)[r]{$k - q  + B$}
\Text(89,120)[l]{$p+q$}
\Text(115,2)[c]{$p$}
\Text(38,2)[c]{$q$}
\LongArrow(84,125)(84,105)
\LongArrow(114,11)(132.75,0.1745)
\LongArrow(36,11)(17.25,0.1745)
\end{picture}
\end{center}
\caption{\label{fig:anomaly} Triangle Feynman diagrams leading to
the chiral fermion anomaly. Fermion spinor and flavor indices are suppressed.
The fermion momenta, as labeled, flow in the arrow directions.}
\end{figure}

The resulting sum of loop integrals is
\beqa
&&\!\!\! i\Gamma^{abc}_{\mu\nu\rho}=
(-1) \int \frac{d^4 k}{(2\pi)^4}
{\rm Tr} \Biggl\{
(\BDneg i \sigmabar_\mu \boldsymbol{T}^a)
\frac{i (k-p +A) \newcdot \sigma}{(k-p+A)^2}
(\BDneg i \sigmabar_\nu \boldsymbol{T}^b)
\frac{i (k+A) \newcdot \sigma}{(k + A)^2}
(\BDneg i \sigmabar_\rho \boldsymbol{T}^c)
\frac{i(k+q+A) \newcdot \sigma}{(k+q+A)^2}
\nonumber \\[8pt]
&&
\hspace{0.5in}
+(\BDneg i \sigmabar_\mu \boldsymbol{T}^a)
\frac{i (k-q +B) \newcdot \sigma}{(k-q+B)^2}
(\BDneg i \sigmabar_\rho \boldsymbol{T}^c)
\frac{i (k+B) \newcdot \sigma}{(k + B)^2}
(\BDneg i \sigmabar_\nu \boldsymbol{T}^b)
\frac{i (k+p+B) \newcdot \sigma}{(k+p+B)^2}
\Biggr\}\,,
\label{eq:anomalyGamma}
\eeqa
where the overall  factor of $(-1)$ is due to
the presence of a closed fermion loop.
The trace is taken over fermion flavor/group and spinor indices,
both of which are suppressed.
Because the individual integrals are linearly divergent,
we must allow for arbitrary constant four-vectors
$A^\mu$ and $B^\mu$ as offsets for the loop momentum
when defining the loop integrations for the
two diagrams~\cite{anomaly1,elias}.

The persistence of the symmetry in the quantum theory for the
currents labeled by $\mu,a$ and $\nu,b$ and $\rho,c$ implies the
naive Ward identities:\footnote{The derivation of the Ward identities
is most easily achieved by writing the three-point function in
position space as a vacuum expectation value of the time-ordered
product of three currents.  After taking
the divergence (with respect to the position of any one of the three
currents) of the time-ordered product and using the fact
that the currents are conserved ($\partial_\mu j^{a\mu} =0$), the surviving
terms can be evaluated using the equal-time commutation relations,
$\delta(x^0-y^0)[j^{a0}(x),j^{b\nu}(y)]= i f^{abc}j^{c\nu} (x)\delta^4(x-y)$.
Fourier-transforming the result yields the terms on the right-hand
side of \eqst{ward1}{ward3}.  See refs.~\cite{bertlmann,bilal}
for further details.}
\beqa
(p+q)^\mu\,i \Gamma^{abc}_{\mu\nu\rho}(-p-q,p,q) &=& f^{abd}\Pi^{dc}_{\nu\rho}(q)
+f^{acd}\Pi^{bd}_{\nu\rho}(p)\,,
\label{ward1}
\\
-p^\nu\,i \Gamma^{abc}_{\mu\nu\rho}(-p-q,p,q) &=&
f^{bcd}\Pi^{da}_{\rho\mu}(p+q)+f^{bad}\Pi^{cd}_{\rho\mu}(q)\,,
\label{ward2}
\\
-q^\rho\,i \Gamma^{abc}_{\mu\nu\rho}(-p-q,p,q) &=&
f^{cad}\Pi^{db}_{\mu\nu}(p) + f^{cbd}\Pi^{ad}_{\mu\nu}(p+q)\,,\label{ward3}
\eeqa
where $i\Pi^{ab}_{\mu\nu}(p)$ is the one-loop
current-current two-point function shown in \fig{fig:PIab}.
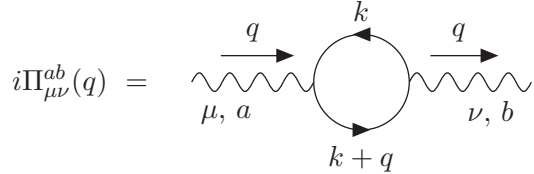
\begin{figure}[ht!]
\begin{center}
\begin{picture}(48,50)(0,5)
\Text(0,25)[c]{$i \Pi^{ab}_{\mu\nu}(q) \,\,=\,\phantom{x}$}
\end{picture}
\begin{picture}(101,50)(-8,5)
\Photon(-23,25)(23,25){3.5}{4}
\Photon(105,25)(59,25){3.5}{4}
\CArc(41,25)(18,0,180)
\CArc(41,25)(18,180,360)
\ArrowLine(41,43)(40.99,43)
\ArrowLine(41,7)(41.01,7)
\Text(41,52)[c]{$k$}
\Text(41,-4)[c]{$k+q$}
\Text(0,44)[c]{$q$}
\Text(78,44)[c]{$q$}
\LongArrow(67,35)(90,35)
\LongArrow(-12,35)(13,35)
\Text(-10,13)[c]{$\mu$, $a$}
\Text(90,13)[c]{$\nu$, $b$}
\end{picture}
\end{center}
\caption{\label{fig:PIab}
The one-loop contribution to the current-current two-point function.  The
fermion momenta, as labeled, flow along the corresponding arrow directions.}
\end{figure}%

By Lorentz covariance, $\Pi^{ab}_{\mu\nu}(p)$ is a rank-two symmetric
tensor that is an even function of the four-momentum $p$
[cf.~\eq{ipiab}].
In \eqst{ward1}{ward3}, we have employed a convention in which the arguments of
$i\Gamma$ correspond to the \textit{outgoing} momentum
of the external legs of the corresponding
one-loop Feynman diagrams, and the order of the momentum arguments
matches the order of the indices.

It is convenient to define the \textit{symmetrized} three-point
function by symmetrizing over the indices $a$, $b$ and $c$:
\beqa
{\cal A}_{\mu\nu\rho}^{abc} = \nicefrac{1}{6}
i\Gamma^{abc}_{\mu\nu\rho} + [\mbox{five permutations of $a,b,c$}].
\eeqa
In terms of the symmetrized three-point function, the naive Ward identities
imply
 \beq \label{naivew}
(p+q)^\mu \mathcal{A}^{abc}_{\mu\nu\rho} = 0\,,\qquad
-p^\nu \mathcal{A}^{abc}_{\mu\nu\rho} = 0\,,\quad {\rm and} \quad
-q^\rho \mathcal{A}^{abc}_{\mu\nu\rho} = 0\,.
\eeq

We now perform the explicit diagrammatic computation to show that
the naive Ward identities exhibited in \eq{naivew} are violated due to
a quantum anomaly.
Although the
symmetrized three-point function is ultraviolet finite,
the individual
loop momentum integrals are divergent, and must be defined with
care. We do not regularize them by the usual procedure of
continuing to $d=4-2\epsilon$
dimensions, because the trace over sigma matrices crucially
involves the antisymmetric tensor with four indices,
brought in by eqs.~(\ref{APPtrssbarssbar}) and (\ref{APPtrsbarssbars}),
for which there is no consistent and unambiguous
generalization outside of
four dimensions. (This is related to the difficulty of
defining $\gamma_5$ in the four-component spinor formalism.)
The existence of the vectors $A$ and $B$ corresponds to an ambiguity in the
regulation procedure, which can be fixed to preserve some of the symmetries,
as we will see below.

Starting from \eq{eq:anomalyGamma}, it follows from \eq{dabcstr}
that the symmetrized three-point function is proportional to the
group theory factor (often called the anomaly coefficient),
\beqa
D^{abc} = \half {\rm Tr}[\{\boldsymbol{T^a}, \boldsymbol{T^b}\} \boldsymbol{T^c}]\,,
\eeqa
where the numerical values of the $D^{abc}$ depend on the representation $R$.
As discussed in
\app{E}, $D^{abc}$ vanishes for all simple Lie groups, with the exception
of SU($N$) for $N\geq 3$.  The $D^{abc}$ are also non-vanishing in general
for any non-semisimple
compact Lie group, which contains at least one U(1) factor.

First, consider the result for $(p+q)^\mu {\cal A}^{abc}_{\mu\nu\rho}$.
This can be simplified by rewriting
\beqa
(p+q)^\mu &=& (k+q+A)^\mu - (k-p+A)^\mu\,,
\label{eq:anomalytrickone}
\\
(p+q)^\mu &=& (k+p+B)^\mu - (k-q+B)^\mu\,,
\eeqa
in the first and second diagram terms, respectively, and then applying
the formulae
\beq
v\newcdot \sigma \,v \newcdot \sigmabar \,=\, \BDpos v^2\,,\qquad\quad
v\newcdot \sigmabar\, v \newcdot \sigma \,=\, \BDpos v^2\,,
\eeq
which follow from \eqs{identityseven}{identityeight}.
After rearranging the terms using the cyclic property of the
trace, we obtain:
\beqa
(p+q)^\mu {\cal A}_{\mu\nu\rho}^{abc} \,&=&\,
-D^{abc}\,
{\rm Tr} [\sigma_\kappa \sigmabar_\nu \sigma_\lambda \sigmabar_\rho ]
\,
X^{\kappa\lambda},\nonumber\\
&=&-2D^{abc}\left[X_{\nu\rho}+X_{\rho\nu}-g_{\nu\rho}X_\lambda{}^\lambda+
i\epsilon_{\kappa\nu\lambda\rho}X^{\kappa\lambda}\right]\,,
\label{eq:anomalysep}
\eeqa
after applying eq.~(\ref{APPtrssbarssbar}).  (In our conventions, $\epsilon_{0123}=-1$.)
The integral $X^{\kappa\lambda}$ is given by:
\beqa
X^{\kappa\lambda} = \int \frac{d^4 k}{(2\pi)^4}
\,
\biggl [
\frac{(k-p+A)^\kappa}{(k-p+A)^2}
\frac{(k+A)^\lambda}{(k+A)^2}
-
\frac{(k+q+A)^\kappa}{(k+q+A)^2}
\frac{(k+A)^\lambda}{(k+A)^2}
\phantom{x.}
&&
\nonumber \\[8pt]
+
\frac{(k+B)^\kappa}{(k+B)^2}
\frac{(k-q+B)^\lambda}{(k-q+B)^2}
-
\frac{(k+B)^\kappa}{(k+B)^2}
\frac{(k+p+B)^\lambda}{(k+p+B)^2}
\biggr ] .
&&
\eeqa

Naively, this integral appears to vanish,
because the first term is equal to
the negative of the fourth term after a momentum shift
$k \rightarrow k - p + A - B$, and the second term is equal to
the negative of the third term after $k \rightarrow k + q + A - B$.
However, these momentum shifts are not valid for the individually
divergent integrals. Instead, $X^{\kappa\lambda}$ can be evaluated
by a Wick rotation to Euclidean space, followed by isolating
the terms that contribute for large $k^2$ and are responsible for the
integral not vanishing, and then employing the divergence (Gauss')
theorem in four dimensions to rewrite $X^{\kappa\lambda}$ as
an angular integral over a three-sphere with radius tending to infinity.
This integral is initially evaluated at large but finite
Euclidean $k$, with the limit $k\to\infty$ taken at the end of the
computation.
For example, consider a smooth function  $f(k)$ of
the four-momentum $k$ with the property that the integral
\beq \label{qdiv}
\int d^4 k f(k)
\eeq
is at worst quadratically divergent.  We define the even and odd
parts of $f(k)$, respectively, by:
\beq
f_e(k)\equiv \half\left[f(k)+f(-k)\right]\,,\qquad\quad
f_o(k)\equiv \half\left[f(k)-f(-k)\right]\,.
\eeq
It then follows
that~\cite{bertlmann,pugh,jauch}
\beq \label{lindiv}
\int \frac{d^4 k}{(2\pi)^4}
\,\left[f(k+a)-f(k)\right]=\frac{i}{(2\pi)^4}\biggl[2\pi^2a_\mu
\lim_{k\to\infty} k^\mu k^2 f_o(k)
+\pi^2 a_\mu a_\nu\lim_{k\to\infty} k^\mu k^2\frac{\partial}
{\partial k_\nu} f_e(k)\biggr]
\eeq
has a finite limit.\footnote{If \eq{qdiv} is linearly
divergent, then the second term on the right-hand side of \eq{lindiv}
is zero.  If \eq{qdiv} is logarithmically divergent or finite, then
the right-hand side of \eq{lindiv} vanishes.}
In deriving this result, we have
expanded $f(k+a)$ in a Taylor expansion and follow the procedure
outlined above \eq{qdiv}.  Note that the angular integration
removes the even parts of $f(k)$ and $\partial f/\partial
k^\nu\equiv 2k_\nu \,\partial f/\partial k^2$ from the right-hand side of
\eq{lindiv}.  The ``limits'' in \eq{lindiv} actually correspond
to an average over the three-sphere  at large Euclidean $k$,
and thus should be interpreted by the use of:
\beqa
\lim_{k\to\infty} \frac{k^\mu k^\nu}{k^2}&=&\quarter g^{\mu\nu}
\,,\label{kklim1} \\[7pt]
\lim_{k\to\infty} \frac{k^\mu k^\nu k^\rho k^\lambda}{(k^2)^2}&=&
\frac{1}{24}\left(g^{\mu\nu} g^{\rho\lambda}+g^{\mu\rho}g^{\nu\lambda}
+g^{\mu\lambda} g^{\nu\rho}\right)\,.\label{kklim2}
\eeqa
For example, if
\beq
f(k)=\frac{(k-p+A)^\kappa (k+A)^\lambda}{(k-p+A)^2
(k+A)^2}\,,
\eeq
then in evaluating \eq{lindiv}, it is sufficient to write:
\beqa
f_o(k)&\simeq &\half (k-p+A)^\kappa (k+A)^\lambda\left[\frac{1}{(k^2)^2}+
\frac{2k\newcdot (p-2A)}{(k^2)^3}\right]-(k\rightarrow -k)
\nonumber \\[8pt]
&\simeq& \frac{k^\kappa A^\lambda-k^\lambda(p-A)^\kappa}{(k^2)^2}
+\frac{2k^\kappa k^\lambda\,k\newcdot(p-2A)}{(k^2)^3}\,,
\eeqa
where we have dropped terms that do not contribute to \eq{lindiv} in
the limit of $k\to\infty$.  Similarly,
\beq
\frac{\partial f_e}{\partial k_\nu}\simeq\frac{g^{\kappa\nu}k^\lambda
+g^{\lambda\nu} k^\kappa}{(k^2)^2}-\frac{4k^\kappa k^\lambda k^\nu}
{(k^2)^3}\,.
\eeq

The evaluation of $X^{\kappa\lambda}$ is now straightforward
[after using \eqs{kklim1}{kklim2}]:
\beqa
X^{\kappa\lambda}
&=&
\frac{i}{96\pi^2}
\Bigl [
\metric^{\kappa\lambda} (p+q) \newcdot (A+B)
+ (A - 2B)^\kappa (p+q)^\lambda
+ (p+q)^\kappa (B - 2 A)^\lambda
\Bigr ] .
\eeqa
Hence, \eq{eq:anomalysep} yields the result for the anomaly in the current
labeled by $\mu,a$:
\beqa
(p+q)^\mu {\cal A}_{\mu\nu\rho}^{abc}
&=&
\frac{i}{48\pi^2} D^{abc}
\Bigl [
(p+q)_\nu (A+B)_\rho +  (A+B)_\nu (p+q)_\rho
+ g_{\nu\rho} (p+q)\newcdot (A+B)
\phantom{xx.}
\nonumber \\[6pt] &&
\qquad\qquad\qquad
- 3 i \epsilon_{\nu\rho\kappa\lambda} (p+q)^\kappa (A-B)^\lambda
\Bigr ] .
\label{eq:anomalya}
\eeqa
Repeating all of the steps starting with eq.~(\ref{eq:anomalytrickone}),
we similarly obtain:\footnote{Alternatively, one can
simply note that eq.~(\ref{eq:anomalyb}) follows from
eq.~(\ref{eq:anomalya}) by making the replacements $\mu\rightarrow \nu$,
$\nu\rightarrow \rho$, $\rho\rightarrow \mu$,
$A \rightarrow A+q$, $B\rightarrow B-q$, $p \rightarrow q$, and
$q \rightarrow -p-q$, while
eq.~(\ref{eq:anomalyc}) follows from
eq.~(\ref{eq:anomalya}) by making the replacements
$\mu\rightarrow \rho$,
$\nu\rightarrow \mu$, $\rho\rightarrow \nu$,
$A \rightarrow A-p$, $B\rightarrow B+p$, $p \rightarrow -p-q$, and
$q \rightarrow p$.}
\beqa
-p^\nu {\cal A}_{\mu\nu\rho}^{abc}
&=&
-\frac{i}{48\pi^2} D^{abc}
\Bigl [
p_\rho (A+B)_\mu +  p_\mu (A+B)_\rho
+ g_{\mu\rho} p \newcdot (A+B)
- 3 i \epsilon_{\rho\mu\kappa\lambda} p^\kappa (A-B + 2 q)^\lambda
\Bigr ] ,\nonumber \\
&&\phantom{line}
\label{eq:anomalyb}
\\
-q^\rho {\cal A}_{\mu\nu\rho}^{abc}
&=&
-\frac{i}{48\pi^2} D^{abc}
\Bigl [
q_\mu (A+B)_\nu +  q_\nu (A+B)_\mu
+ g_{\mu\nu} q \newcdot (A+B)
- 3 i \epsilon_{\mu\nu\kappa\lambda} q^\kappa (A-B -2p)^\lambda
\Bigr ] .\nonumber \\
&&\phantom{line}
\label{eq:anomalyc}
\eeqa

Non-chiral anomalies will arise for all three of the
currents (assuming $D^{abc}$ is non-vanishing),
unless we choose the arbitrary constant vectors $A$ and $B$
such that
\beq
A+B=0\,,
\eeq
with the result:
\beqa
(p+q)^\mu {\cal A}_{\mu\nu\rho}^{abc}
\,&=&\,
\frac{1}{8\pi^2} D^{abc}
\epsilon_{\nu\rho\kappa\lambda} (p+q)^\kappa A^\lambda
,\label{wardanom1}
\\[5pt]
-p^\nu {\cal A}_{\mu\nu\rho}^{abc}
\,&=&\,
-\frac{1}{8\pi^2} D^{abc}
\epsilon_{\rho\mu\kappa\lambda} p^\kappa (A + q)^\lambda
,\label{wardanom2}
\\[5pt]
-q^\rho {\cal A}_{\mu\nu\rho}^{abc}
\,&=&\,
-\frac{1}{8\pi^2} D^{abc}
\epsilon_{\mu\nu\kappa\lambda} q^\kappa (A -p)^\lambda
. \label{wardanom3}
\eeqa
If $D^{abc}$ is non-vanishing, it is not possible to
avoid an anomaly simultaneously in all
three symmetries, but one can still arrange for two of the
symmetries to be non-anomalous.  If one wants an anomaly to arise only in
the current labeled by $\mu,a$ (for example, if the symmetries labeled by
$b$, $c$ are gauged), one must now choose
$A = p-q$. The standard result follows:
\beqa
(p+q)^\mu {\cal A}_{\mu\nu\rho}^{abc}
\,&=&\,
-\frac{1}{4\pi^2} D^{abc}
\epsilon_{\nu\rho\kappa\lambda} p^\kappa q^\lambda
,\label{anomvva1}
\\[5pt]
-p^\nu {\cal A}_{\mu\nu\rho}^{abc}
\,&=&\,
0
,\label{anomvva2}
\\[5pt]
-q^\rho {\cal A}_{\mu\nu\rho}^{abc}
\,&=&\,
0 .\label{anomvva3}
\eeqa
In particular, one cannot gauge all three symmetries labeled by $a,b,c$ unless
$D^{abc} = 0$.

If all three currents are identical, then by Bose symmetry the anomalies
of the three currents
must coincide.  This can be achieved by choosing $A=\frac{1}{3}(p-q)$,
in which case,
\beqa
(p+q)^\mu {\cal A}_{\mu\nu\rho}^{abc}
\,&=&\,
-\frac{1}{12\pi^2} D^{abc}
\epsilon_{\nu\rho\kappa\lambda} p^\kappa q^\lambda
,\label{wardanomsym1}
\\[5pt]
-p^\nu {\cal A}_{\mu\nu\rho}^{abc}
\,&=&\,
-\frac{1}{12\pi^2} D^{abc}
\epsilon_{\rho\mu\kappa\lambda} p^\kappa q^\lambda
,\label{wardanomsym2}
\\[5pt]
-q^\rho {\cal A}_{\mu\nu\rho}^{abc}
\,&=&\,
-\frac{1}{12\pi^2} D^{abc}
\epsilon_{\mu\nu\kappa\lambda} p^\kappa q^\lambda
. \label{wardanomsym3}
\eeqa

Returning briefly to the original naive Ward identities given in
\eqst{ward1}{ward3}, the analysis above shows that these identities
must be modified by an additional additive contribution given
by the right-hand side of \eqst{wardanom1}{wardanom3}.  In particular,
there is no anomalous contribution proportional to $f^{abc}$.
This can be checked explicitly by a diagrammatic computation of
the two-point and three-point functions that appear in \eqs{ward1}{ward3}.
We use \eqs{cubicindex}{trttt} to write
\beq
\Tr(\boldsymbol{T^a}\boldsymbol{T^b}\boldsymbol{T^c})=D^{abc}(R)+\frac{i}{2}
I_2(R)f^{abc}\,,
\eeq
where $I_2(R)$ is the index defined in \eq{index2} and $R$ is the representation
of the generators $\boldsymbol{T^a}$.  For example, inserting this result
in \eq{eq:anomalyGamma}, it follows that:
\beq
(p+q)^\mu \,i\Gamma_{\mu\nu\rho}^{abc} \,=\,
-\left[D^{abc}\,X^{\kappa\lambda}+\frac{i}{2}I_2(R) f^{abc}\,Y^{\kappa\lambda}\right]
{\rm Tr} [\sigma_\kappa \sigmabar_\nu \sigma_\lambda \sigmabar_\rho ]
\,,
\label{eq:pqiGamma}
\eeq
where the integral $Y^{\kappa\lambda}$ is given
by:\footnote{Here $Y^{\kappa\lambda}$
is obtained from $X^{\kappa\lambda}$
by setting $A=B=0$, since we can use dimensional regularization
for this part of the computation as explained below \eq{yint}.}
\beq \label{yint}
Y^{\kappa\lambda} = \int \frac{d^4 k}{(2\pi)^4}
\,
\biggl [
\frac{(k-p)^\kappa}{(k-p)^2}
\frac{k^\lambda}{k^2}
-
\frac{(k+q)^\kappa}{(k+q)^2}
\frac{k^\lambda}{k^2}
-
\frac{k^\kappa}{k^2}
\frac{(k-q)^\lambda}{(k-q)^2}
+
\frac{k^\kappa}{k^2}
\frac{(k+p)^\lambda}{(k+p)^2}
\biggr ] .
\eeq
By letting $k\to -k$ in the third and fourth term in the integrand
of \eq{yint}, we see that $Y^{\kappa\lambda}=Y^{\lambda\kappa}$, and hence
by \eq{APPtrssbarssbar},
\beq
-\frac{i}{2}I_2(R)f^{abc}\,Y^{\kappa\lambda}
{\rm Tr} [\sigma_\kappa \sigmabar_\nu \sigma_\lambda \sigmabar_\rho ]=
-iI_2(R)f^{abc}\left[2Y_{\nu\rho}-g_{\nu\rho}Y_\lambda{}^\lambda\right]\,.
\eeq
Since no $\epsilon$-tensor appears, we can evaluate this integral
in $d\neq 4$ dimensions using the standard techniques of dimensional
regularization.

One can check that this result matches the diagrammatic
calculation of the right-hand side of \eq{wardanom1}.   In particular,
\fig{fig:PIab} yields
\beqa
i\Pi^{ab}_{\mu\nu}(q)&=&(-1)
\int\frac{d^4 k}{(2\pi)^4} \Tr \left [
(\BDneg i\sigmabar_\mu\boldsymbol{T^a})
\frac{ik\newcdot\sigma}{k^2}
(\BDneg i\sigmabar_\nu\boldsymbol{T^b})
\frac{i(k+q)\newcdot\sigma}{(k+q)^2} \right ]
\nonumber \\[8pt]
&=&
-I_2(R)\delta^{ab}\Tr(\sigmabar_\mu\sigma_\rho\sigmabar_\nu\sigma_\lambda)
\int\frac{d^4 k}{(2\pi)^4}\frac{k^\rho (k+q)^\lambda}{k^2(k+q)^2}\,,
\eeqa
where we have used \eq{index2}.  Lorentz covariance implies that
\beq \label{ipiab}
i\Pi^{ab}_{\mu\nu}(q)
=\delta^{ab}\left[C_1(q^2)g_{\mu\nu}+C_2(q^2)q_\mu q_\nu\right]\,,
\eeq
for some scalar functions $C_1$ and $C_2$.  It follows that
$\Pi^{ab}_{\mu\nu}(q)=\Pi^{ab}_{\mu\nu}(-q)$ and
$\Pi^{ab}_{\mu\nu}(q)=\Pi^{ab}_{\nu\mu}(q)$.  Consequently, we can write:
\beq
\Pi^{ab}_{\mu\nu}(q)=\frac{i}{2}
I_2(R)\delta^{ab}\Tr(\sigmabar_\mu\sigma_\rho\sigmabar_\nu\sigma_\lambda
+\sigmabar_\nu\sigma_\rho\sigmabar_\mu\sigma_\lambda)
\int\frac{d^4 k}{(2\pi)^4}\frac{k^\rho (k+q)^\lambda}{k^2(k+q)^2}\,,
\label{eq:nouseforalabel}
\eeq
and so no $\epsilon$-tensor appears in the evaluation of the trace.
As above, we are now free to evaluate the integral in $d\neq 4$ dimensions.
Comparing eqs.~(\ref{eq:pqiGamma}) and (\ref{yint}) to eq.~(\ref{eq:nouseforalabel}),
and using eq.~(\ref{wardanom1}),
the end result is
\beq
(p+q)^\mu\,i\Gamma^{abc}_{\mu\nu\rho}(-p-q,p,q)=I_2(R)f^{abc}\left[\Pi_{\nu\rho}(q)
-\Pi_{\nu\rho}(p)\right]+\frac{1}{8\pi^2}D^{abc}(R)\epsilon_{\nu\rho\kappa\lambda}
(p+q)^\kappa A^\lambda\,,
\eeq
where we have written $\Pi^{ab}_{\nu\rho}\equiv I_2(R)\delta^{ab}\Pi_{\nu\rho}$.
Indeed the terms on the right-hand side proportional to $f^{abc}$ match those
of the naive Ward identity given in \eq{ward1}.  As previously asserted,
the anomaly only resides in the contributions to the Ward identity
proportional to $D^{abc}$.

In writing down eq.~(\ref{eq:anomalyGamma}), we chose to use
the rules with $\sigmabar$ matrices for the current vertices and
$\sigma$ matrices for the massless fermion propagators. If we had chosen
the opposite prescription (i.e., $\sigma$ matrices for the current vertices and
$\sigmabar$ matrices for the massless fermion propagators), then
the order of the factors inside the trace of
eq.~(\ref{eq:anomalyGamma}) would have been reversed.\footnote{The
arrowed fermion lines in the loop must be traversed in the direction
parallel [antiparallel] to the arrow directions when  the
$\sigmabar$ [$\sigma$] versions of the propagator rule are employed,
as indicated in \eq{rule2} [and in the discussion that follows].
This rule determines the order of the factors inside the spinor trace.}
Instead of
\eq{eq:anomalysep}, we would have obtained
\beq
(p+q)^\mu {\cal A}_{\mu\nu\rho}^{abc} \,=\,
-D^{abc}\,
{\rm Tr} [\sigmabar_\kappa \sigma_\nu \sigmabar_\lambda \sigma_\rho ]
\,
\bar X^{\kappa\lambda}
=-2D^{abc}\left[\bar{X}_{\nu\rho}+\bar{X}_{\rho\nu}-g_{\nu\rho}\bar{X}_\lambda{}^\lambda-
i\epsilon_{\kappa\nu\lambda\rho}\bar{X}^{\kappa\lambda}\right]\,,
\label{eq:anomalysep2}
\eeq
after applying \eq{APPtrsbarssbars}.
The integral $\bar{X}^{\kappa\lambda}$
is simply related to $X^{\kappa\lambda}$ by:
\beq \label{xxbar}
\bar{X}^{\kappa\lambda} = X^{\lambda\kappa}\,.
\eeq
Inserting \eq{xxbar} into \eq{eq:anomalysep2}, we immediately
reproduce the result of \eq{eq:anomalysep}, as expected.

It is instructive to examine the case of massless QED.  The
terms of the Lagrangian involving the electron fields
is given by
\beq \label{qed0}
\mathscr{L}= i\chi^\dagger\sigmabar^\mu D_\mu\chi
+ i\eta^\dagger\sigmabar^\mu D_\mu\eta\,,
\eeq
where $D_\mu\equiv\partial_\mu \BDplus iQA_\mu$ is the covariant
derivative, and $Q$ is the charge operator.
Here, we identify $\chi$ as the two-component (left-handed) electron
field and $\eta$ as the two-component (left-handed) positron
field.  The corresponding eigenvalues of the charge operator are:
$Q\chi=-e\chi$ and $Q\eta=+e\eta$ (where $e>0$ is the electromagnetic
gauge coupling constant, or equivalently the electric
charge of the positron).

At the classical level, the massless QED Lagrangian [\eq{qed0}] is invariant
under a U(1)$_V\times$U(1)$_A$ global symmetry.  Under a
U(1)$_V\times$U(1)$_A$ transformation specified by the
infinitesimal parameters $\theta_V$ and $\theta_A$,
\beqa
&&{\rm U}(1)_V:\qquad \delta\chi=ie\theta_V\chi\,,\qquad\quad\,\,\,
\delta\eta=-ie\theta_V\eta\,,\\[8pt]
&&{\rm U}(1)_A:\qquad \delta\chi=i\theta_A\chi\,,\qquad\qquad
\delta\eta=i\theta_A\eta\,.
\eeqa
We can combine these equations into a two-dimensional matrix equation,
\beq
\delta\psi_j=-i\theta_a(\boldsymbol{T_a})_j{}^k\psi_k\,,\qquad
{\rm where}\qquad
\psi=\begin{pmatrix} \chi \\ \eta \end{pmatrix}\,,
\eeq
and the index $a$ takes on two values, $a=V$, $A$.  It follows
that the U(1)$_V\times$U(1)$_A$ generators are given by
\beqa
T_V&=&e\begin{pmatrix} -1 & \phm 0 \phm\\ \phm 0 & \phm 1\phm
\end{pmatrix}\,,\qquad {\rm for}
\quad {\rm U}(1)_V\,,\\
T_A&=&\begin{pmatrix} -1 & \phm 0\phm \\ \phm 0 &  -1\phm\end{pmatrix}\,,
\qquad\quad {\rm for}
\quad {\rm U}(1)_A\,.
\eeqa

The classically conserved Noether currents corresponding to the
U(1)$_V\times$U(1)$_A$ global symmetry are the
vector and axial currents:\footnote{Note that the interaction
Lagrangian for massless QED is $\mathscr{L}_{\rm int}=\BDneg J^\mu_V A_\mu$,
as expected. This accounts for the factor of $e$ in the definition of
the vector current.  The axial vector current does not couple to the
photon field; hence no coupling constant is included in its definition.}
\beqa
J^\mu_V&=&
- e( \chi^\dagger\sigmabar^\mu\chi-
\eta^\dagger\sigmabar^\mu\eta)\,,\\
J^\mu_A&=&
- \chi^\dagger\sigmabar^\mu\chi
- \eta^\dagger\sigmabar^\mu\eta\,.
\eeqa
Since the U(1)$_V$ symmetry is gauged, we demand that this symmetry
should be anomaly free.  Thus, we make use of
\eqst{anomvva1}{anomvva3}, where we identify the index pair $\mu$, $a$
with the axial vector current and the index pairs $\nu$, $b$ and
$\rho$, $c$ with the vector current.  Thus, we compute:
\beq
D^{AVV}=\Tr\left(T_A T_V T_V\right)=-2e^2\,.
\eeq
Moreover, for an abelian symmetry group, $f^{abc}=0$.  Hence, using
\eq{anomvva1} [which also applies in this case
to the unsymmetrized three-point function],
the U(1) axial vector anomaly equation reads:
\beq \label{qedanom}
(p+q)^\mu \,i\Gamma^{AVV}_{\mu\nu\rho}
\,=\,
\frac{e^2}{2\pi^2}
\epsilon_{\nu\rho\kappa\lambda} p^\kappa q^\lambda\,,
\eeq
in agreement with the well-known result.\footnote{This result
was first obtained by Adler~\cite{adler}.  In comparing \eq{qedanom}
with Adler's result, note that the
normalization of the triangle amplitude in \Ref{adler} differs
by a factor of $(2\pi)^4$ and the opposite sign convention for
$\epsilon_{0123}$ is employed.}

We now convert \eq{qedanom} into an operator equation.
Consider the process of two photon production
by an axial vector current source~\cite{Huang}.
First, we note that $\partial_\mu J^\mu_A(x)
=i[P^\mu,J_{A\mu}(x)]$, where $P^\mu$ is the momentum operator.
It follows that:
\beq \label{jamu}
\vev{p\,,\,q\,|\,\partial_\mu J^\mu_A(0)\,|\,0}=i
\vev{p\,,\,q\,|\,[P^\mu\,,\,J_{A\mu}(0)]\,|\,0}=i(p+q)^\mu
\vev{p\,,\,q\,|\,J_{A\mu}(0)|\,0}\,.
\eeq
We identify the S-matrix amplitude for the two photon production as:
\beq \label{igamavv}
i\Gamma_{\mu\nu\rho}^{AVV}\,\varepsilon^{\nu\,*}(p)
\varepsilon^{\rho\,*}(q)=\vev{p\,,\,q\,|\,-iJ_{A\mu}(0)|\,0}\,,
\eeq
where $\varepsilon(p)$ and $\varepsilon(q)$
are the polarization vectors for the final state photons.  Note that
the factor of $-i$ on the right-hand side of \eq{igamavv}
has been inserted to be consistent with the
Feynman rule for the axial vector current insertion given
in \fig{fig:currentrule}.  Thus, using \eqst{qedanom}{igamavv},
we end up with~\cite{Peskin:1995ev}:
\beqa
\vev{p\,,\,q\,|\,\partial_\mu J^\mu_A(0)\,|\,0}&=& -\,\frac{e^2}{2\pi^2}
\epsilon_{\nu\rho\kappa\lambda}\varepsilon^{\nu\,*}(p)
\varepsilon^{\rho\,*}(q)
p^\kappa q^\lambda \nonumber \\
&=&-\,\frac{e^2}{16\pi^2}\langle p\,,\,q\,|\,\epsilon_{\kappa\nu\lambda\rho}
F^{\kappa\nu}F^{\lambda\rho}(0)\,|0\rangle\,,\label{epsff}
\eeqa
where $\epsilon_{\kappa\nu\lambda\rho}F^{\kappa\nu}F^{\lambda\rho}=
4\epsilon_{\kappa\nu\lambda\rho}(\partial^\kappa A^\nu)
(\partial^\lambda A^\rho)$ has been used to
eliminate the photon fields in favor of
a product of electromagnetic field strength tensors.  In deriving
\eq{epsff}, an additional factor
of two arises due to two possible contractions of the photon fields
with the
external states.  We thus obtain the operator form for the axial vector
anomaly:\footnote{\label{epsign}%
In the literature, \eq{opanomaly} often occurs with
the opposite sign due to a sign convention for the Levi-Civita
$\epsilon$-tensor that is opposite to the one employed
in this review.  Here, we have reproduced the form given in
\Ref{Peskin:1995ev}.}
\beq \label{opanomaly}
\partial_\mu J^\mu_A=-\,\frac{e^2}{8\pi^2} F^{\lambda\rho}\widetilde
F_{\lambda\rho}\,,
\eeq
where the dual electromagnetic field strength tensor is defined by
$\widetilde F_{\lambda\rho}\equiv\half
\epsilon_{\kappa\nu\lambda\rho} F^{\kappa\nu}$.

As a final example, we examine the anomalous baryon number
and lepton number currents in
the theory of electroweak
interactions~\cite{baryonanomaly,bnumber2,nair}.
For simplicity of notation, we consider
a one-generation model.  The baryon number current is a vector
current given by:
\beq
J^\mu_B=\nicefrac{1}{3}\left[{u}^\dagger\sigmabar^\mu u
+ {d}^\dagger\sigmabar^\mu d - {\bar u}^\dagger\sigmabar^\mu \bar u
-{\bar d}^\dagger\sigmabar^\mu \bar d\right]\,,
\eeq
following the particle naming conventions of Table~\ref{tab:nomenclature}.
Consider the process of gauge boson pair production by a baryon number
current source.  It is convenient to work in the interaction
basis of gauge fields, $\{W^{\mu\,a}\,,\,B^\mu\}$, where $W^{\mu\,a}$ is
an SU(2)-triplet of gauge fields and $B^\mu$ is a U(1)$_{\rm Y}$ hypercharge
gauge field.  We consider triangle diagrams where one generation of
quarks runs in the loop.  The external vertices consist of the baryon
number current source and the two gauge bosons.

The generators corresponding to the SU(2) gauge boson vertices are
given in block diagonal form by:
\beq
\boldsymbol{T^b}= g\,{\rm diag} \left(\frac{\tau^b}{2}\otimes
\mathds{1}_{3\times 3}\,,\,\,0\,,\,0\right)\,,
\eeq
where the $\tau^b$ are the Pauli matrices, $\mathds{1}_{3\times 3}$
is the identity matrix in color space, and $\otimes$ is the Kronecker
product.\footnote{The Kronecker product of an $n\times n$ matrix and an
$m\times m$ matrix is an $nm\times nm$ matrix.  In addition, the
following two properties of the Kronecker product are
noteworthy~\cite{matrixref,handbook}:
(i) $(A\otimes B)(C\otimes D)=AC\otimes BD$, and
(ii)~$\Tr(A\otimes B)=\Tr A\,\Tr B$.}
We have included a factor of the weak SU(2) coupling $g$ in the
definition of $\boldsymbol{T^b}$, since the Feynman rule given by
\fig{fig:currentrule} does not explicitly include the gauge coupling.
Likewise, the generators  corresponding to the U(1)$_{\rm Y}$
gauge boson vertices are given in block diagonal form by
(cf.~Table~\ref{SMfermions}):
\beq
\boldsymbol{Y}=g'\,{\rm diag}\left(
\nicefrac{1}{6}\mathds{1}_{2\times 2}\otimes\mathds{1}_{3\times 3}\,,\,
-\nicefrac{2}{3}\mathds{1}_{3\times 3}\,,\,
\third\mathds{1}_{3\times 3}\right)\,,
\eeq
where $\mathds{1}_{2\times 2}$ is the identity matrix in weak isospin
space, and $g'$ is the U(1)$_{\rm Y}$ hypercharge gauge coupling.
Finally, the generator corresponding to the baryon number current
source is given in block diagonal form
by:
\beq
\boldsymbol{B}=\third{\rm diag}\left(
\mathds{1}_{2\times 2}\otimes\mathds{1}_{3\times 3}\,,\,
-\mathds{1}_{3\times 3}\,,\,-\mathds{1}_{3\times 3}\right)\,.
\eeq

Consider first the production of two SU(2)-triplet gauge
fields. We put $\boldsymbol{T^a}=
\boldsymbol{B}$ and associate
the indices $b$ and $c$ with the SU(2)-triplet gauge bosons.
A simple calculation yields
\beq
D^{Bbc}= g^2\Tr(\boldsymbol{B T^b T^c})=\half g^2\delta^{bc}\,,
\eeq
where the superscript index $B$ refers to the baryon number current.
Since the gauged weak SU(2) and hypercharge U(1)$_{\rm Y}$ currents
must be anomaly free for the mathematical consistency of the
electroweak theory, it follows that \eqst{anomvva1}{anomvva3} apply.
That is, the symmetrized amplitude for
the production of SU(2) gauge boson pairs by a baryon number source
is anomalous:
\beq \label{anombtt}
(p+q)^\mu {\cal A}_{\mu\nu\rho}^{Bbc}=
-\frac{g^2}{8\pi^2} \delta^{bc}
\epsilon_{\nu\rho\kappa\lambda} p^\kappa q^\lambda\,.
\eeq

Next, consider the production of two U(1)$_{\rm Y}$ hypercharge
gauge fields.  A simple calculation yields
\beq
D^{BYY}=g^{\prime\,2}\Tr(\boldsymbol{BY^2})=-\half g^{\prime\,2}\,.
\eeq
That is, the symmetrized amplitude for
the production of U(1)$_{\rm Y}$ gauge boson pairs by a baryon number source
is anomalous:
\beq \label{anombyy}
(p+q)^\mu {\cal A}_{\mu\nu\rho}^{BYY}=
\frac{g^{\prime\,2}}{8\pi^2}
\epsilon_{\nu\rho\kappa\lambda} p^\kappa q^\lambda\,.
\eeq
Finally,
the symmetrized amplitude for the associated production of an
SU(2)-triplet and U(1)$_{\rm Y}$ hypercharge
gauge field exhibits no anomaly as the corresponding
$D^{BYc}=gg'\Tr(\boldsymbol{BYT^c})=0$.

The symmetrized amplitudes of the triangle diagrams involving
a baryon number current source and a pair of SU(2) or U(1)$_{\rm Y}$
gauge bosons are anomalous.  Since the baryon number current is
a vector current, we conclude that the source of the anomaly is
a VVA triangle diagram in which one of the gauge boson currents
is vector~(V) and the other gauge boson current
is axial vector~(A).  Nevertheless, the gauge boson axial vector
current must be conserved, as noted above.  Hence, the
baryon number vector current must be
anomalous~\cite{baryonanomaly}.  In \eqst{qedanom}{epsff},
we showed how to derive the operator form of the anomaly equation from
the anomalous non-conservation of the symmetrized triangle amplitude.
Following the same set of steps starting with
\eqs{anombtt}{anombyy}, one obtains the anomalous
non-conservation of the baryon number vector current,
in a model with $N_g$ quark generations~\cite{bnumber2,Pokorski,nair}:
\beq \label{smanom}
\partial_\mu J^\mu_B=\frac{g^2 N_g}{32\pi^2}W^{\lambda\rho b}
\widetilde{W}^b_{\lambda\rho}-\frac{g^{\prime\,2}N_g}{32\pi^2}
B^{\lambda\rho}\widetilde{B}_{\lambda\rho}\,,
\eeq
where $B_{\lambda\rho}$ and
\beq
W_{\lambda\rho}^b=\partial_\lambda W^b_\rho-\partial_\rho W_\lambda^b
\BDminus g\epsilon^{bca} W_\lambda^c W_\rho^a\,,
\eeq
are the field strength
tensors for the hypercharge U(1)$_{\rm Y}$ gauge
boson and SU(2) gauge boson
fields, respectively.\footnote{We again caution the reader that a
different overall sign in \eq{smanom} often appears in the literature
due to a sign convention for the Levi-Civita
$\epsilon$-tensor that is opposite to the one employed
in this review.  Here, we have chosen $\epsilon^{0123}=+1$.}
Note that for
the non-abelian SU(2) gauge fields $W^a_\mu$,
\beqa
W^{\lambda\rho b}\,\widetilde{W}^b_{\lambda\rho}
&=& 2\epsilon^{\kappa\nu\lambda\rho}\left[
(\partial_\kappa W_\nu^b)(\partial_\lambda W_\rho^b)
\BDminus g\epsilon^{abc}(\partial_\kappa W_\nu^a)W_\lambda^b W_\rho^c
\right]\nonumber \\[7pt]
&=&
2\epsilon^{\kappa\nu\lambda\rho}\partial_\kappa\left[
W_\nu^b(\partial_\lambda W_\rho^b)
\BDminus\third g\epsilon^{abc} W_\nu^a W_\lambda^b W_\rho^c\right]\,.
\eeqa
Strictly speaking, the triangle graphs yield only the terms on
the right-hand side of \eq{smanom} that are quadratic in the gauge fields.
To obtain the
corresponding terms that are cubic in the gauge terms,
one must compute the anomalies that arise
from VVVA and VAAA box diagrams~\cite{bardeen,bertlmann}.

For completeness, we re-express the anomalous non-conservation
of the baryon number current in
terms of the mass eigenstate SU(2)$\times$U(1)$_{\rm Y}$ gauge fields:
\beq
\partial_\mu J^\mu_B=\frac{g^2 N_g}{16\pi^2}W^{\lambda\rho+}
\widetilde{W}^-_{\lambda\rho}-\frac{g^2N_g}{32\pi^2 c^2_W}
Z^{\lambda\rho}\widetilde{Z}_{\lambda\rho}-\frac{egN_g}{32\pi^2 c_W}
\left[Z^{\lambda\rho}
\widetilde{F}_{\lambda\rho}+F^{\lambda\rho}
\widetilde{Z}_{\lambda\rho}\right]\,,
\eeq
where $c_W\equiv\cos\theta_W$, and $W_{\lambda\rho}^{\pm}$,
$Z_{\lambda\rho}$ and $F_{\lambda\rho}$ are the $W^\pm$, $Z$ and
the electromagnetic field strength tensors, respectively.

By a similar analysis, one can also compute the anomalous
non-conservation of the lepton number current,
\beq \label{jlepton}
J_L^\mu= \ell^\dagger\sigmabar^\mu \ell+ \nu^\dagger
\sigmabar^\mu\nu-{\bar\ell}^\dagger\sigmabar^\mu\bar\ell\,,
\eeq
due to triangle diagrams with $N_g$ generations of leptons running in
the loop.  In the one-generation calculation, the relevant generators are:
\beq
\boldsymbol{T^b}= g\,{\rm diag}\left(\frac{\tau^b}{2}\,,\,\,0\right)
\,,\qquad
\boldsymbol{Y}=g'\,{\rm diag}\left(
-\half\mathds{1}_{2\times 2}\,,\,1 \right)\,,\qquad
\boldsymbol{L}={\rm diag}\left(
\mathds{1}_{2\times 2}\,,\, -1\right)\,.
\eeq
Thus, we end up with:
\beq
D^{Lbc}=\half g^2\delta^{bc}\,,\qquad\quad
D^{LYY}=-\half g^{\prime\,2}\,,\qquad\quad D^{LYc}=0\,.
\eeq
Thus, in the Standard Model with $N_g$ generations of quarks and leptons,
\beq \label{BminusL}
\partial_\mu J^\mu_L=\partial_\mu J^\mu_B\,.
\eeq
Hence, the $B-L$ current is conserved.  However, $B-L$ is \textit{not} anomaly free, due to the fact
that the lepton number current exhibited in \eq{jlepton} has both vector and axial vector pieces.  In particular, the symmetrized amplitude
of the triangle diagrams with three lepton number current sources is anomalous.  To avoid this anomaly, one
can add a right-handed neutrino to the Standard Model, in which case the leptonic current, 
$J_L^\mu= \ell^\dagger\sigmabar^\mu \ell+ \nu^\dagger
\sigmabar^\mu\nu-{\bar\ell}^\dagger\sigmabar^\mu\bar\ell-\bar{\nu}^\dagger\sigmabar^\mu\bar{\nu}$, is a
vector current.   In this case, \eq{BminusL} still holds, and the $B-L$ current is conserved and anomaly free.

\bigskip
\addcontentsline{toc}{section}{Acknowledgments}
\subsection*{Acknowledgments}

H.K.D. would like to thank the Santa Cruz Institute for Particle
Physics and the Stanford Linear Accelerator Center for their
hospitality during numerous visits to collaborate on this review.
H.E.H. would also like to thank the Physikalisches Institut der
Universit\"at Bonn and Northern Illinois University for their
hospitality during his visits to collaborate on this review.  H.K.D.
and H.E.H gratefully acknowledge the support and hospitality of the
Aspen Center for Physics, the CERN Theory group, the Institute for
Particle Physics Phenomenology at the University of Durham, and the Kavli
Institute for Theoretical Physics at the University of California,
Santa Barbara, where parts of this review were written.  H.E.H. and S.P.M.
appreciate the support and hospitality of Fermi National Accelerator
Laboratory at various stages of this work.

We greatly appreciate many valuable conversations with our
colleagues, Jonathan Bagger, Thomas Banks,  Andrew Cohen,
Athanasios (Sakis) Dedes, Michael Dine, Lance Dixon,
Manuel Drees, Rainald Flume, Paul Langacker, Christopher Hill, Daniel Maitre,
Hans Peter Nilles, Michael Peskin,
Stefano Profumo, Janusz Rosiek, Mark Srednicki, Lorenzo Ubaldi,
Pascal Vaudrevange, Carlos Wagner, and Dieter Zeppenfeld.
We are also grateful for e-mail comments and suggestions received from
Garrett Lisi, Chris Oakley,
Nicholas Setzer, Jos\'e Valle, and the anonymous referee.
We thank Michaela M\"uhl (Physikalisches Institut, Bonn) 
and Antje Daum (DESY, Hamburg) for their efforts in helping us locate
some of the relevant earliest references.

Finally, we would like to acknowledge the funding agencies that
supported this work.  H.K.D. is supported in part
by SFB Transregio 33 (``The Dark Universe''), by BMBF
grant 05 HT6PDA, and by the Helmholtz Alliance HA-101 (``Physics at
the Terascale'').
H.E.H. is supported in part by U.S. Department of
Energy grant number DE-FG02-04ER41268 and in part by a
Humboldt Research Award sponsored by the Alexander von Humboldt Foundation. 
The work of S.P.M. is
supported in part by the National Science Foundation
grants PHY-0140129, PHY-0456635, and PHY-0757325.

\bigskip\bigskip\bigskip

\begin{appendices}

\section{\texorpdfstring{Metric and sigma matrix conventions}{Metric and sigma matrix conventions}}
\label{appendix:A}
\renewcommand{\theequation}{A.\arabic{equation}}
\renewcommand{\thefigure}{A.\arabic{figure}}
\renewcommand{\thetable}{A.\arabic{table}}
\setcounter{equation}{0}
\setcounter{figure}{0}
\setcounter{table}{0}

In this review, the metric tensor of four-dimensional Minkowski space
is taken to be:\footnote{An otherwise identical version of this paper
with the opposite metric signature
is also available; see footnote~\ref{metricsign}.}
\beq
\label{app:signofmetric}
g_{\mu\nu}=
g^{\mu\nu}={\rm diag}(\BDplus 1 , \BDminus 1, \BDminus 1, \BDminus 1)\, ,
\eeq
where $\mu, \nu= 0,1,2,3$ are spacetime vector indices.
Contravariant four-vectors (e.g. positions, momenta, gauge fields and
currents) are defined with raised
indices, and covariant four-vectors (e.g. derivatives) with
lowered indices:
\beqa
x^\mu &=& (t\,;\,\mathbold{\vec x})\,,\label{conven4}\\
p^\mu &=& (E\,;\,\mathbold{\vec p})\,,\\
A^\mu(x) &=& (\Phi(\mathbold{\vec x},t)\,;\,
\mathbold{\vec A}(\mathbold{\vec x},t))\,,\\
J^\mu(x) &=& (\rho(\mathbold{\vec x},t)\,;\,
\mathbold{\vec J}(\mathbold{\vec x},t))\,,\\
\partial_\mu & \equiv &\frac{\partial}{\partial x^\mu}
= (\partial/\partial t\,;\,\mathbold{\vec \nabla})\,,\label{partialmu}
\eeqa
in units with $c=1$.  The totally antisymmetric pseudo-tensor
$\eps^{\mu\nu\rho\sigma}$ is defined such that
\beq \label{app:eps0123}
\eps^{0123}=-\eps\ls{0123}=+1\,.
\eeq
\Eqst{conven4}{app:eps0123} are taken to be
independent of the metric signature convention.

The sigma matrices are defined with a raised (contravariant)
index to be independent of the metric signature convention,
\beq \label{paulidown}
\sigma^\mu=(\mathds{1}_{2\times 2}\,;
\,\mathbold{\vec\sigma})
\,,\qquad\qquad\qquad
\sigmabar^\mu=(\mathds{1}_{2\times 2}\,;\,-\mathbold{\vec\sigma})\,,
\eeq
where the three-vector of Pauli matrices is given by
$\boldsymbol{\vec\sigma}\equiv (\sigma^1\,,\,\sigma^2\,,\,\sigma^3)$
[cf.~\eq{pauli}]
and $\mathds{1}_{2\times 2}$ is the $2\times 2$ identity matrix.
The corresponding quantities with lower
(covariant) index are:
\beq \label{pauliup}
\sigma_\mu=g_{\mu\nu}\sigma^\nu=
(\BDpos \mathds{1}_{2\times 2}\,;
\,\BDneg \mathbold{\vec\sigma})
\,,\qquad\qquad\quad
\sigmabar_\mu=g_{\mu\nu}\sigmabar^\nu=
(\BDpos \mathds{1}_{2\times 2}\,;\,
\BDpos \mathbold{\vec\sigma})\,.
\eeq
Various identities involving products of sigma matrices are given in \app{B}.
The generators of the $(\half,0)$ and $(0,\half)$ representations of
the Lorentz group are, respectively, 
given by:
\beq \label{Lgenerators}
\sigma^{\mu\nu} \equiv \frac{i}{4}(\sigma^\mu\sigmabar^\nu-\sigma^\nu
\sigmabar^\mu)\,,\qquad\quad
\sigmabar^{\mu\nu} \equiv \frac{i}{4}(\sigmabar^\mu\sigma^\nu-\sigmabar^\nu
\sigma^\mu)\,.
\eeq

In adopting the above definition of the sigma matrices, we differ from
the corresponding conventions of Wess and Bagger~\cite{WessBagger} and
Bilal~\cite{bilalsusy}.  The
Wess/Bagger and Bilal (WBB) definition
of the sigma matrices can be written
(with lowered index $\mu$) as:\footnote{Although Wess/Bagger and Bilal employ
opposite metric signatures of $g_{00}=-1$ and $g_{00}=+1$, respectively,
their definitions of $\sigma_\mu$ and $\sigmabar_\mu$ (with
covariant index $\mu$) coincide.
Note that the spinor structure of the $\sigma$ and
$\sigmabar$ matrices and the definitions of the various (two-index
and four-index) epsilon tensors [cf.~\eqs{epssign}{app:eps0123}]
are identical in both the WBB conventions and in our conventions.}
\beqa
(\sigma^{\rm WBB})_{\mu\,\alpha\dot{\beta}}
&=&
\BDpos\sigma_{0\alpha\dot{\gamma}}\sigmabar_{\mu}^{\dot{\gamma}\delta}
\sigma_{0\delta\dot{\beta}}=
(\mathds{1}_{2\times 2}\,;\,\mathbold{\vec\sigma})\,,
\label{sigswap1}
\\
(\sigmabar^{\rm{WBB}})_\mu^{\dot{\alpha}\beta}
&=&
\BDpos\sigmabar_0^{\dot{\alpha}\gamma}\sigma_{\mu\gamma\dot{\delta}}
\sigmabar_0^{\dot{\delta}\beta}=
(\mathds{1}_{2\times 2}\,;\,-\mathbold{\vec\sigma})\, .
\label{sigswap2}
\eeqa
One consequence of the WBB definition of $\sigma$ and $\sigmabar$
is that $\gamma\ls{5}=\rm{diag}(\mathds{1}_{2\times 2}\,,\,
-\mathds{1}_{2\times 2})$ in the chiral representation [cf.~\eq{gamma4}].
This associates a lowered undotted [raised dotted] two-component spinor with
a right-handed [left-handed] four-component spinor
[cf.~\eqs{general4comp}{plprdefs}].  Indeed, this was the common
convention in the older literature (e.g., see
refs.~\cite{LifshitzI,Cbook,Novozhilov,LifshitzII,
Scheck,Akhiezer}).\footnote{This convention persists in the
literature of the spinor helicity
method (cf.~footnote~\ref{fnwarned} in \app{I.2}).}
However, in the modern formulation of electroweak theory in terms
of left-handed fermions, it is now more common to associate a lower
undotted [raised dotted] two-component spinor with
a left-handed [right-handed] four-component spinor.  This is the
motivation for our conventions for the sigma matrices given
in \eqs{paulidown}{pauliup}.

In order to facilitate the comparison with the metric signature with
$g_{00}=\BDminus 1$, we provide the key ingredients
needed for translating
between Minkowski metrics of opposite signature.  In our conventions
[cf.~\eqst{conven4}{pauliup}], each of the following objects
(with the Lorentz index heights as shown) is
defined
\textit{independently} of the metric signature:
\beqa \label{poslist}
&& x^\mu
\,,\,
p^\mu
\,,\,
J^{\mu\nu}\,,\,
J_{\mu\nu}\,,\,
\partial_\mu
\,,\,
\sigma^\mu
\,,\,
\sigmabar^\mu
\,,\,
S^\mu
\,,\,
J^\mu
\,,\,
A^\mu
\,,\,
D_\mu
\,,\,
G^\mu{}_\nu
\,,\,
\gamma^{\mu}
\,,\,
\gamma_5
\,,\,
\delta^\mu_{\nu}
\,,\,
\epsilon^{\mu\nu\rho\sigma}
\,,\,
\epsilon_{\mu\nu\rho\sigma}
\,,\,
\theta^\mu{}_{\nu}
\nonumber \\
&&\hspace{4.3in}  [\mbox{no sign change}],
\eeqa
whereas the following objects change sign when the Minkowski metric
signature is reversed:
\beq \label{neglist}
g_{\mu\nu}
\,,\,
g^{\mu\nu}
\,,\,
x_\mu
\,,\,
p_\mu
\,,\,
\partial^\mu
\,,\,
\sigma_\mu
\,,\,
\sigmabar_\mu
\,,\,
S_\mu
\,,\,
J_\mu
\,, \,
A_\mu
\,,\,
D^\mu
\,,\,
G^{\mu\nu}
\,,\,
G_{\mu\nu}
\,,\,
\gamma\ls{\mu}
\,,\,
\theta^{\mu\nu}
\,,\,
\theta_{\mu\nu}
\qquad [\mbox{sign change}].
\eeq
Here, $J^{\mu\nu}$ is the angular momentum tensor, the spin four-vector $S^\mu$ is defined in \eq{fixedsvect},
$J^\mu$ is any conserved current,
$A^\mu$ is any gauge vector potential, and $D_\mu$ and
$G_{\mu\nu}$ are the corresponding covariant derivative and antisymmetric
tensor field strength, respectively.
The Dirac gamma matrices are defined in \eq{gamma4}, and the 
tensor $\theta^{\mu\nu}$ parameterizes Lorentz transformations [cf.~eqs.~(\ref{lambda44}), (\ref{generallorentzmatrix}), and (\ref{Linf})--(\ref{MDinf})].
The list of eq.~(\ref{neglist}) can be deduced from eq.~(\ref{poslist})
by using the metric tensor and its inverse to lower and raise
Lorentz indices, simply because each metric or inverse metric changes
sign when the metric signature is reversed.
Given any other object not included in \eqs{poslist}{neglist},
it is straightforward to make the appropriate assignment by considering
how the object is defined.
For example, we must assign $\sigma_{\mu\nu}$, $\sigmabar_{\mu\nu}$,
$\sigma^{\mu\nu}$ and $\sigmabar^{\mu\nu}$ to the list of
\eq{poslist}, based on the definitions given
in \eqs{sigmamunu}{sigmabarmunu}.
In general, objects that do not carry Lorentz vector indices
(including all fermion spinor fields and spinor wave functions) are
defined
to be the same in the two metric signatures, with the obvious exception
of scalar quantities formed from an odd number of objects from the
list of
eq.~(\ref{neglist}).  For example, the dot product of two
four-vectors may or may not change sign when the Minkowski metric
signature is reversed.  By writing out the dot product explicitly using
the metric tensor to contract the indices, one can use
\eqs{poslist}{neglist} to determine the behavior of a dot product under
the reversal of the metric signature.
In particular, $p\newcdot A$ changes sign
whereas $\sigma\newcdot\partial$ and $\sigmabar\newcdot\partial$ do not change sign, when the Minkowski metric
signature is reversed.

The translation between Minkowski metrics of opposite signatures
is now straightforward.
Given any relativistic covariant quantity or equation in
the convention where $g_{00}=\BDplus 1$, one need only employ
\eqs{poslist}{neglist} to
obtain the same quantity or equation in the convention where
$g_{00}=\BDminus 1$, and vice versa.\footnote{Note that for any
relativistic covariant term appearing additively in a valid equation,
the \textit{relative} sign that results from changing between
Minkowski metrics of opposite signature is simply given by
$\mathcal{S}=(-1)^{\mathcal{N}}$, where $\mathcal{N}\equiv
N_m+N_d+N_G+\ldots$.  Here $N_m$ is the number of metric tensors
appearing either explicitly or implicitly through
contracted upper and lower indices, $N_d$ is the number of spacetime
and/or covariant derivatives, $N_G$ is the number of gauge field
strength tensors, and the ellipsis ($\ldots$) accounts for any
additional quantities whose contravariant forms (with all Lorentz
indices raised) appear in the list of \eq{neglist}.}

As an example, let us verify that under the reversal
of the Minkowski metric signature
the gauge covariant derivative $D_\mu$ does not change
sign and the gauge field
strength tensor $G^{\mu\nu}$ changes sign.
In the metric signature with $g_{00}=\BDplus 1$, we define
\beq
\label{covderplus}
D_\mu\equiv \boldsymbol{I}_{d_R}\partial_\mu \BDplus igA_\mu
\,,\qquad \qquad (g_{00}=\BDplus 1)\,,
\eeq
where
$A_\mu\equiv A_\mu^a{\boldsymbol{T^a}}$ is the matrix gauge field
for a representation $R$ of dimension $d_R$,
and
$\boldsymbol{I}_{d_R}$ is the $d_R\times d_R$ identity matrix.
Since under the reversal of the metric signature,
$\partial_\mu$ does not
change sign [according to \eq{poslist}] whereas $A_\mu$ changes
sign [according to \eq{neglist}], it follows that
the quantity defined in the metric signature where $g_{00}=\BDminus
1$,
\beq
\label{covderminus}
D_\mu\equiv \boldsymbol{I}_{d_R}\partial_\mu \BDminus igA_\mu
\,,\qquad \qquad (g_{00}=\BDminus 1)\,
\eeq
has the same overall sign as eq.~(\ref{covderplus}).
It follows that when the metric signature is reversed,
$D_\mu$ does not change sign whereas $D^\mu\equiv g^{\mu\nu}D_\nu$
does change sign,
as indicated in \eqs{poslist}{neglist}.
Next, consider the matrix gauge field
strength tensor $G_{\mu\nu}\equiv G_{\mu\nu}^a{\boldsymbol{T^a}}$,
defined by
\beq
G^{\mu\nu}\equiv \frac{\BDneg i}{g}\left[D^\mu\,,\,D^\nu\right]
=\partial^\mu A^\nu - \partial^\nu A^\mu \BDplus
ig[A^\mu\,,\,A^\nu]\,,\qquad
\quad (g_{00}=\BDplus 1)\,,
\eeq
where the commutator
$[D^\mu\,,\,D^\nu]$ is an operator that acts on
fields that transform with respect to an arbitrary representation
$R$.
In the metric signature with $g_{00}=\BDminus 1$, we define the gauge
field strength tensor as a commutator of covariant derivatives
with the \textit{opposite} overall sign:
\beq
G^{\mu\nu}\equiv \frac{\BDpos i}{g}\left[D^\mu\,,\,D^\nu\right]
=\partial^\mu A^\nu-\partial^\nu A^\mu \BDminus
ig[A^\mu\,,\,A^\nu]\,,\qquad
\quad (g_{00}=\BDminus 1)\,,
\eeq
where $D^\mu$ is now defined as in eq.~(\ref{covderminus}).
Since under a reversal of the metric signature, $A^\mu$ does not 
change sign [according to \eq{poslist}] whereas $\partial^\mu$ changes
sign [according to \eq{neglist}], it follows that $G^{\mu\nu}$
and $G_{\mu\nu}\equiv g_{\mu\rho}g_{\nu\sigma}G^{\rho\sigma}$ do indeed change
sign when the metric signature is reversed, as stated in \eq{neglist}.

As another simple illustration,
consider the $\sigma$-matrix identity,
\beq \label{sigidpmmm}
\sigmabar^\mu \sigma^\nu \sigmabar^\rho =
\BDpos \metric^{\mu\nu} \sigmabar^\rho
\BDminus  \metric^{\mu\rho} \sigmabar^\nu
\BDplus \metric^{\nu\rho} \sigmabar^\mu
\BDminus i \epsilon^{\mu\nu\rho\kappa}\sigmabar_\kappa\,,\qquad
(g_{00}=\BDplus 1)\,,
\eeq
In the opposite metric signature
with $g_{00}= \BDminus 1$, we apply the results of \eqs{poslist}{neglist}
and then multiply both sides of the equation by $-1$ to obtain:
\beq \label{sigidmppp}
\sigmabar^\mu \sigma^\nu \sigmabar^\rho =
\BDneg \metric^{\mu\nu} \sigmabar^\rho
\BDplus  \metric^{\mu\rho} \sigmabar^\nu
\BDminus \metric^{\nu\rho} \sigmabar^\mu
\BDplus i \epsilon^{\mu\nu\rho\kappa}\sigmabar_\kappa\,,\qquad
(g_{00}=\BDminus 1)\,.
\eeq
Finally, in
the sigma matrix conventions of Wess/Bagger~\cite{WessBagger}
and Bilal~\cite{bilalsusy}, both
\eqs{sigidpmmm}{sigidmppp} are modified by
changing the overall sign of $i \epsilon^{\mu\nu\rho\kappa}\sigmabar_\kappa$.
In general, to convert the identities of \app{B} to
the conventions of WBB, one must first convert (if necessary) to the
appropriate metric signature, and then
interchange $\sigma\leftrightarrow\BDpos\sigmabar$
[cf.~\eqs{sigswap1}{sigswap2}].

We end this Appendix with a
brief summary of our conventions for
four-dimensional Euclidean space.   The Euclidean components of
the coordinates [represented in Minkowski space by the
contravariant four-vector,
$x^\mu=(x^0\,;\,\boldsymbol{\vec x})$, for $\mu=0,1,2,3$],
are defined as
\beq
x^\mu_E=x\ls{E\,\mu}= (\boldsymbol{\vec x}\,,\,x\ls{E}^4)\,,
\qquad\quad x\ls{E}^4=x_{E4}\equiv ix^0\,,
\qquad\quad (\mu=1,2,3,4)\,.
\eeq
The four-momentum
operator in Minkowski space is $p^\mu=\BDpos i\partial^\mu= i
(\partial/\partial t\,,\,-\boldsymbol{\vec\nabla})$.
Following the conventions of \Ref{makeenko}, the Euclidean counterpart
of the momentum operator is
\beq
p^\mu_E=p_{E\mu}=(\boldsymbol{\vec p}\,,\,p^4_E)=
- i\partial^\mu_E= - i(\boldsymbol{\vec\nabla}\,,\,
\partial/\partial x_E^4)\,,\qquad p^4_E=p_{E4}=ip^0\,,
\eeq
The Minkowski space
Green functions are obtained from Euclidean space Green functions
by means of a Wick rotation~\cite{wick,montvay,makeenko} of $x^4_E \equiv ix^0$
in a counterclockwise sense.\footnote{Expressing the corresponding
Green functions as Fourier transforms of momentum-space Green functions,
one must simultaneously Wick-rotate $p^4_E \equiv ip^0$ in
a clockwise sense to avoid singularities in the complex
$p^0$-plane.}
Scalar products of
Euclidean four-vectors are carried out by employing the Euclidean
metric tensor
$\delta_{\mu\nu}=\delta^{\mu\nu}={\rm diag}(1\,,\,1\,,\,1\,,\,1)$.
For example, the Euclidean counterpart of $-p\newcdot x=\BDneg p^0 x^0
\BDplus\boldsymbol{\vec p\newcdot\vec x}$ is
$p^\mu_E x^\mu_E = \boldsymbol{\vec p\newcdot\vec x}+p^4_E x^4_E$, etc.
Given any tensorial equation in Euclidean space, the heights of
the indices is irrelevant.  Consequently, one can simply place all
indices at the same height (either all raised or all lowered),
with an implied sum over a pair of repeated indices.

\clearpage

One can also introduce Euclidean sigma matrices~\cite{rubakov}:
\beq \label{esigma}
\sigma^\mu_E \equiv (-i\boldsymbol{\vec\sigma}\,,\,\sigma_E^{\,4})\,,
\qquad\quad
\sigmabar^\mu_E \equiv
(i\boldsymbol{\vec\sigma}\,,\,\sigmabar_E^{\,4})\,,
\qquad {\rm where}\,\,\,
\sigma^{\,4}_E=\sigmabar^{\,4}_E\equiv\mathds{1}_{2\times 2}\,,
\eeq
which satisfy:\footnote{It is seemingly more natural to define
$\sigma^\mu_E\equiv (\boldsymbol{\vec\sigma}\,,\,\sigma_E^4)$
and $\sigmabar^\mu_E\equiv (-\boldsymbol{\vec\sigma}\,,\,\sigmabar_E^4)$
where $\sigma^4_E=\sigmabar^4_E\equiv i\mathds{1}_{2\times 2}$,
in which case one must replace
$\delta^{\mu\nu}$ with $-\delta^{\mu\nu}$ in \eq{sigdelta}.
Nevertheless we prefer \eq{esigma}, which
avoids an overall minus
sign in the respective anticommutation relations of the
Euclidean sigma and gamma matrices
[cf.~footnote~\ref{fncliff}].
\label{fnsigmaE}}
\beq \label{sigdelta}
\sigma^\mu_E\sigmabar^\nu_E+\sigma^\nu_E\sigmabar^\mu_E
=2\delta^{\mu\nu}\,,
\qquad\qquad
\sigmabar^\mu_E\sigma^\nu_E+\sigmabar^\nu_E\sigma^\mu_E
=2\delta^{\mu\nu}\,.
\eeq
The four-dimensional rotation group in Euclidean space is
SO(4), which is locally equivalent to SU(2)$\times$SU(2).  It
possesses two independent pseudo-real
two-dimensional spinor representations $(\half,0)$ and
$(0,\half)$ [not related by hermitian conjugation in contrast to
the Lorentz group],
with corresponding hermitian generators $\sigma_E^{\mu\nu}$ and
$\sigmabar_E^{\mu\nu}$, respectively:
\beq
\sigma_E^{\mu\nu}=\frac{i}{4}\left(\sigma_E^\mu\sigmabar_E^\nu
-\sigma_E^\nu\sigmabar_E^\mu\right)\,,\qquad\qquad
\sigmabar_E^{\mu\nu}=\frac{i}{4}\left(\sigmabar_E^\mu\sigma_E^\nu
-\sigmabar_E^\nu\sigma_E^\mu\right)\,.
\eeq
These tensors are anti-self-dual and self-dual,
respectively~\cite{instanton3},
\beq
\sigma_E^{\mu\nu}= -\half \epsilon^{\mu\nu\rho\tau}\sigma_E^{\rho\tau}\,,
\qquad\quad
\sigmabar_E^{\mu\nu}=\half \epsilon^{\mu\nu\rho\tau}\sigmabar_E^{\rho\tau}\,,
\eeq
where the totally antisymmetric Levi-Civita tensor is
defined in Euclidean space such that $\epsilon^{1234}=\epsilon_{1234}=+1$.
One can express $\sigma_E^{\mu\nu}$ and $\sigmabar_E^{\mu\nu}$
in terms of the 't~Hooft eta symbols~\cite{etasymbols},
\beq
\sigma^{\mu\nu}_E=-\half\overline\eta^{k\mu\nu}\sigma^k\,,
\qquad\quad
\sigmabar^{\mu\nu}_E=-\half\eta^{k\mu\nu}\sigma^k\,,
\eeq
where $\mu,\nu=1,2,3,4$ and there is an implicit sum over $k=1,2,3$.
Equivalently,
\beq
\sigma^\mu_E\sigmabar^\nu_E=\delta^{\mu\nu}
+i\overline\eta^{k\mu\nu}\sigma^k\,,\qquad\qquad
\sigmabar^\mu_E\sigma^\nu_E=\delta^{\mu\nu}
+i\eta^{k\mu\nu}\sigma^k\,.
\eeq
The 't~Hooft symbols $\eta$ and $\overline\eta$ satisfy self-duality
and anti-self-duality properties, respectively:
\beq
\eta^{k\mu\nu}=\half\epsilon^{\mu\nu\rho\lambda}\eta^{k\rho\lambda}\,,
\qquad\qquad
\overline\eta^{k\mu\nu}=-\half\epsilon^{\mu\nu\rho\lambda}
\overline\eta^{k\rho\lambda}\,,
\eeq
and are explicitly given by:
\beq
\eta^{kij}=\overline\eta^{kij}=\epsilon^{ijk}\,,\qquad
\eta^{kj4}=-\eta^{k4j}=\overline\eta^{k4j}
=-\overline\eta^{kj4}=\delta^{kj}\,,\qquad
\eta^{k44}=\overline\eta^{k44}=0\,.
\eeq
For a more comprehensive treatment of two-component spinors in
Euclidean space, see \Ref{sherry}.

\section{\texorpdfstring{Sigma matrix identities and Fierz identities}{Sigma matrix identities and Fierz identities}}
\label{appendix:B}
\renewcommand{\theequation}{B.\arabic{equation}}
\renewcommand{\thefigure}{B.\arabic{figure}}
\renewcommand{\thetable}{B.\arabic{table}}
\setcounter{equation}{0}
\setcounter{figure}{0}
\setcounter{table}{0}

In \sec{sec:notations}, we derived a number of identities
involving $\sigma^\mu$, $\sigmabar^\mu$, $\sigma^{\mu\nu}$ and
$\sigmabar^{\mu\nu}$.  When considering a theory regularized by
dimensional continuation~\cite{dimreg}, one must give meaning
to the sigma matrices and their respective identities in $d\neq 4$
dimensions.  In many cases, it is possible to reinterpret the sigma
matrix identities for $d\neq 4$.  However, the  Fierz identities, 
which depend on the completeness of $\{\mathds{1}_{2\times 2}$\,, 
$\sigma^i\}$ in the vector space of $2\times 2$ matrices, do not 
 have a consistent, unambiguous meaning outside of four
dimensions (e.g., see refs.~\cite{Siegel:1980qs,Avdeev:1982xy,Blatter,
gammafive} and references therein).  In \app{B.1}, we exhibit a
comprehensive list of identities from which many generalized Fierz
identities can be derived.  In \app{B.2}, we examine the class of
sigma matrix identities that can unambiguously be 
extended to $d\neq 4$ dimension and thus can be employed 
in the context of dimensional regularization.
 
\subsection{Two-component spinor Fierz identities}
\renewcommand{\theequation}{B.1.\arabic{equation}}
\renewcommand{\thefigure}{B.1.\arabic{figure}}
\renewcommand{\thetable}{B.1.\arabic{table}}
\setcounter{equation}{0}
\setcounter{figure}{0}
\setcounter{table}{0}

We begin with the basic identity for $2\times 2$ matrices~\cite{Bailin}, 
\beq \label{basicid}
\delta_{ab}\delta_{cd}=\half\left[\delta_{ad}\delta_{cb}+
\sigma^i_{ad}\sigma^i_{cb}\right]\,,
\eeq
where there is an implicit sum over the
repeated superscript $i=1,2,3$.  \Eq{basicid} is a consequence of the
completeness of $\{\mathds{1}_{2\times 2}$\,, 
$\sigma^i\}$ in the four-dimensional vector space of $2\times 2$ matrices.
Applying these considerations to matrices that possess two indices,
either undotted and/or dotted, one can establish four isomorphic 
four-dimensional vector spaces, each of which is spanned by four
linearly independent hermitian matrices,
\beq \label{fourspace}
\mathcal{V}=\{\delta_\alpha{}^\beta\,,\,i(\sigma^{0i})_\alpha{}^\beta\}\,,
\qquad \overline{\mathcal{V}}=\{\delta^{\dot\alpha}{}_{\dot\beta}\,,\,
i(\sigmabar^{0i})^{\dot\,\alpha}{}_{\dot\beta}\}\,,\qquad
\mathcal{V}^{\,\prime}=\{\sigma^\mu_{\alpha\dot\beta}\}\,,\qquad
\overline{\mathcal{V}}^{\,\prime}=\{\sigmabar^{\mu\,\dot\alpha\beta}\}\,.
\eeq
Note that the  $\sigma^{jk}$ and $\sigmabar^{jk}$ 
are completely determined by the six matrices
$\sigma^{0i}$ and $\sigmabar^{0i}$ (where $i,j,k=1,2,3$)
due to the self-duality relations given by
\eq{eq:selfduality}.  

It is therefore convenient to consider the set of matrices,
\beq \label{Gammacomplete}
\Gamma\equiv
\left\{\delta_{\alpha}{}^\beta\,,\,\sigma^\mu_{\alpha\dot{\beta}}\,,
\,\sigma^{\mu\nu}{}_\alpha{}^\beta\,,\,
\delta^{\dot\alpha}{}_{\dot\beta}\,,\,\sigma^{\mu\,\dot{\alpha}\beta}\,,
\,\sigmabar^{\mu\nu\,\dot{\alpha}}{}_{\dot{\beta}}\right\}\,.
\eeq
Elements of $\Gamma$ will be denoted by $\Gamma^{(n)}$
($n=1,2,\ldots, 6$).
Starting from \eq{basicid}, one can establish a set of 21
identities of the following form:
\beq \label{basicfierz}
(\Gamma^{(k)})^I_{AB}(\Gamma^{(n)})^J_{CD}=\sum_{p,q,K,L} 
(C^{kn}_{pq})^{IJ}_{KL}\,(\Gamma^{(p)})^K_{AD}(\Gamma^{(q)})^L_{CB}\,,
\eeq
where each label $I$, $J$, $K$ and $L$ can represent zero, one or two
Lorentz spacetime indices, and
$A$,~$B$, $C$ and $D$ represent two-component spinor indices, each of 
which may be undotted or dotted and in the lowered or raised position
as appropriate.
The sum in \eq{basicfierz} is taken over the  
matrices specified in \eq{Gammacomplete}, and the 
$C^{kn}_{pq}$ are numerical coefficients [cf.~\eqst{Fierz1}{Fierz21}].

Let us multiply \eq{basicfierz} by four (commuting or anticommuting) 
two-component spinors
$Z_{1A} Z_{2B} Z_{3C} Z_{4D}$, where $Z_i$ stands for either 
the undaggered or daggered spinor $z_i$ 
or $z_i^\dagger$,
depending on whether the corresponding spinor index is undotted or dotted.
This procedure yields generalized Fierz 
identities of the form~\cite{Bailin,Terning,MullerKirsten,Drees}:
\beq \label{fierz21}
(Z_1\Gamma^{(k)I} Z_2)(Z_3\Gamma^{(n)J} Z_4)=(-1)^A\sum_{p,q,K,L}
 (C^{kn}_{pq})^{IJ}_{KL}(Z_1\Gamma^{(p)K} Z_4)(Z_3\Gamma^{(q)L} Z_2)\,,
\eeq
where $(-1)^A=+1$ [$-1$] for commuting [anticommuting] spinors.\footnote{It 
is often convenient to reverse the order of the spinors $Z_2$ and $Z_3$ 
on the right-hand side of \eq{fierz21} by using
\eqst{zonetwo}{europeanvacation} and (\ref{zsigmunuz1})--(\ref{zsigmunuz2})
to eliminate the factor of $(-1)^A$ [cf.~\eq{twocompfierza}].
\label{fnfierzorder}}

The explicit expressions for the 21
identities represented by \eq{basicfierz} are as follows.
\beqa
\hspace{-0.5in}
\delta_{\alpha}{}^{\beta} \delta^{\dot{\beta}}{}_{\dot{\alpha}}&=&
\BDpos\half\sigma^\mu_{\alpha\dot{\alpha}} \sigmabar_\mu^{\dot{\beta}\beta}
\,,
\label{Fierz1}
\\ 
\hspace{-0.6in}
\delta_\alpha{}\rsup{\beta}\delta\ls{\gamma}{}\rsup{\tau}&=&
\half\left[\delta_\alpha{}\rsup{\tau}\delta\ls{\gamma}{}\rsup{\beta}+
(\sigma^{\mu\nu})_\alpha{}\rsup{\tau} 
(\sigma_{\mu\nu})\ls{\gamma}{}\rsup{\beta}\right]\,,
\label{Fierz2}\\
\hspace{-0.6in}
\delta^{\dot{\alpha}}{}_{\dot{\beta}}\delta^{\dot{\gamma}}{}_{\dot{\tau}}&=&
\half\left[\delta^{\dot{\alpha}}{}_{\dot\tau}\delta^{\dot\gamma}{}_{\dot\beta}
+(\sigmabar^{\mu\nu})^{\dot{\alpha}}{}_{\dot{\tau}} 
(\sigmabar_{\mu\nu})^{\dot{\gamma}}{}_{\dot{\beta}}\right]\,,
\label{Fierz3}\\
\hspace{-0.6in}
\delta_{\alpha}{}^{\beta}\sigma^\mu_{\gamma\dot\alpha}&=&\half
\sigma^\mu_{\alpha\dot\alpha}\delta_\gamma{}^\beta
\BDminus i\sigma_{\nu\,\alpha\dot\alpha}(\sigma^{\mu\nu})_\gamma{}^\beta\,,
\label{Fierz4}\\
\hspace{-0.6in}
\delta_{\alpha}{}^{\beta}\sigmabar^{\mu\,\dot{\beta}\gamma}&=&\half
\delta_\alpha{}^\gamma\sigmabar^{\mu\,\dot{\beta}\beta}
\BDplus i(\sigma^{\mu\nu})_\alpha{}^\gamma\sigmabar_\nu^{\dot{\beta}\beta}\,,
\label{Fierz5}\\
\hspace{-0.6in}
\delta^{\dot{\alpha}}{}_{\dot{\beta}}\sigma^\mu_{\beta\dot\gamma}&=&
\half\delta^{\dot\alpha}{}_{\dot\gamma}\sigma^\mu_{\beta\dot\beta}
\BDplus i(\sigmabar^{\mu\nu})^{\dot\alpha}{}_{\dot\gamma}
\sigma_{\nu\,\beta\dot{\beta}}\,,
\label{Fierz6}\\
\hspace{-0.6in}
\delta^{\dot\alpha}{}_{\dot\beta}\sigmabar^{\mu\,\dot{\gamma}\alpha}&=&\half
\sigmabar^{\mu\,\dot{\alpha}\alpha}\delta^{\dot\gamma}{}_{\dot\beta}
\BDminus i\sigmabar_{\nu}^{\dot{\alpha}\alpha}
(\sigmabar^{\mu\nu})^{\dot\gamma}{}_{\dot\beta}\,,
\label{Fierz7}\\
\hspace{-0.6in}
\delta_{\alpha}{}^{\beta}(\sigma^{\mu\nu})_\gamma{}^\tau&=&
\half\biggl\{(\sigma^{\mu\nu})_\alpha{}^\tau\delta_\gamma{}^\beta
+\delta_\alpha{}^\tau(\sigma^{\mu\nu})_\gamma{}^\beta
\BDminus ig_{\rho\kappa}\left[(\sigma^{\mu\kappa})_\alpha{}^\tau
(\sigma^{\nu\rho})_\gamma{}^\beta-(\sigma^{\nu\kappa})_\alpha{}^\tau
(\sigma^{\mu\rho})_\gamma{}^\beta\right]\biggr\},\nonumber \\
\phantom{line}\label{Fierz8}\\
\hspace{-0.6in}
\delta_{\alpha}{}^{\beta}(\sigmabar^{\mu\nu})^{\dot\beta}{}_{\dot\alpha}&=&
-\quarter i\left[\sigma^\mu_{\alpha\dot\alpha}\sigmabar^{\nu\,\dot{\beta}\beta}
-\sigma^\nu_{\alpha\dot\alpha}\sigmabar^{\mu\,\dot{\beta}\beta}
+i\epsilon^{\mu\nu\rho\kappa}\sigma_{\rho\,\alpha\dot\alpha}
\sigmabar_\kappa^{\dot{\beta}\beta}\right]\,,
\label{Fierz9}\\
\hspace{-0.6in}
\delta^{\dot\alpha}{}_{\dot\beta}(\sigma^{\mu\nu})_\beta{}^\alpha&=&
-\quarter i\left[\sigmabar^{\mu\,\dot{\alpha}\alpha}\sigma^\nu_{\beta\dot\beta}
-\sigmabar^{\nu\,\dot{\alpha}\alpha}\sigma^\mu_{\beta\dot\beta}
-i\epsilon^{\mu\nu\rho\kappa}\sigmabar_\rho^{\dot{\alpha}\alpha}
\sigma_{\kappa\,\beta\dot\beta}\right]\,,
\label{Fierz10}\\
\hspace{-0.6in}
\delta^{\dot\alpha}{}_{\dot\beta}(\sigmabar^{\mu\nu})^{\dot\gamma}{}_{\dot\tau}
&=&\half\biggl\{(\sigmabar^{\mu\nu})^{\dot\alpha}{}_{\dot\tau}
\delta^{\dot\gamma}{}_{\dot\beta}+\delta^{\dot\alpha}{}_{\dot\tau}
(\sigmabar^{\mu\nu})^{\dot\gamma}{}_{\dot\beta}
\BDminus ig_{\rho\kappa}\left[(\sigmabar^{\mu\kappa})^{\dot\alpha}{}_{\dot\tau}
(\sigmabar^{\nu\rho})^{\dot\gamma}{}_{\dot\beta}-
(\sigmabar^{\nu\kappa})^{\dot\alpha}{}_{\dot\tau}
(\sigmabar^{\mu\rho})^{\dot\gamma}{}_{\dot\beta}\right]\biggr\},\nonumber \\
\phantom{line}\label{Fierz11}\\
\hspace{-0.5in}
\sigma^\mu_{\alpha\dot{\alpha}}\sigma^\nu_{\beta\dot{\beta}} &=&
\half\left[\sigma^\mu_{\alpha\dot{\beta}}\sigma^\nu_{\beta\dot{\alpha}}
+\sigma^\nu_{\alpha\dot{\beta}}\sigma^\mu_{\beta\dot{\alpha}}
-g^{\mu\nu}\sigma^\lambda_{\alpha\dot{\beta}}
\sigma_{\lambda\,\beta\dot{\alpha}}+i\epsilon^{\mu\nu\rho\kappa}
\sigma_{\rho\,\alpha\dot{\beta}}\,\sigma_{\kappa\,\beta\dot{\alpha}}\right]\,,
\label{Fierz12}\\[6pt]
\sigmabar^{\mu\dot{\alpha}\alpha} \sigmabar^{\nu\dot{\beta}\beta} &=&
\half\left[\sigmabar^{\mu\dot{\alpha}\beta}
\sigmabar^{\nu\dot{\beta}\alpha}
+\sigmabar^{\nu\dot{\alpha}\beta}\sigmabar^{\mu\dot{\beta}\alpha}
-g^{\mu\nu}\sigmabar^{\lambda\,\dot{\alpha}\beta}
\sigmabar_\lambda^{\dot{\beta}\alpha}-i\epsilon^{\mu\nu\rho\kappa}
\sigmabar_\rho^{\dot{\alpha}\beta}\,\sigmabar_\kappa^{\dot{\beta}\alpha}
\right]\,,
\label{Fierz13}\\[6pt]
\sigma^\mu_{\alpha\dot{\alpha}}\sigmabar^{\nu\dot{\beta}\beta} &=&
\BDpos \half\left[
g^{\mu\nu}\delta_\alpha{}^\beta\delta^{\dot{\beta}}{}_{\dot{\alpha}}
\BDminus 2i(\sigma^{\mu\nu})_\alpha{}^\beta
\delta^{\dot{\beta}}{}_{\dot{\alpha}}
\BDplus 2i\delta_\alpha{}^\beta
(\sigmabar^{\mu\nu})^{\dot{\beta}}{}_{\dot{\alpha}}
-4g_{\rho\kappa}(\sigma^{\mu\kappa})_\alpha{}^\beta
(\sigmabar^{\nu\rho})^{\dot{\beta}}{}_{\dot{\alpha}}\right]\,,\nonumber\\
\phantom{line}
\label{Fierz14}\\[-8pt]
(\sigma^{\mu\nu})_\alpha{}^\beta\sigma^\rho_{\gamma\dot\alpha}
&=&\half\left[\sigma^\nu_{\alpha\dot\alpha}(\sigma^{\mu\rho})_\gamma{}^\beta
-\sigma^\mu_{\alpha\dot\alpha}(\sigma^{\nu\rho})_\gamma{}^\beta
+i\epsilon^{\mu\nu\kappa}{}_\lambda
\sigma_{\kappa\,\alpha\dot\alpha}(\sigma^{\lambda\rho})_\gamma{}^\beta
\right.\nonumber \\
&&\qquad\qquad \left.\BDminus\half i\,
\left(g^{\mu\rho}\sigma^\nu_{\alpha\dot\alpha}
-g^{\nu\rho}\sigma^\mu_{\alpha\dot{\alpha}}-i\epsilon^{\mu\nu\rho\kappa}
\sigma_{\kappa\,\alpha\dot\alpha}\right)\delta_\gamma{}^\beta\right]
\,, \label{Fierz15} \\[6pt]
\sigmabar^{\rho\,\dot{\alpha}\beta}(\sigma^{\mu\nu})_\gamma{}^\alpha
&=&\half\left[\sigmabar^{\nu\,\dot{\alpha}\alpha}
(\sigma^{\mu\rho})_\gamma{}^\beta-\sigmabar^{\mu\,\dot{\alpha}\alpha}
(\sigma^{\nu\rho})_\gamma{}^\beta+i\epsilon^{\mu\nu\kappa}{}_\lambda
\sigmabar_\kappa^{\dot{\alpha}\alpha}(\sigma^{\lambda\rho})_\gamma{}^\beta
\right.\nonumber \\ 
&&\qquad\qquad \left.
\BDplus\half i\,\left(g^{\mu\rho}\sigmabar^{\nu\,\dot{\alpha}\alpha}
-g^{\nu\rho}\sigmabar^{\mu\,\dot{\alpha}\alpha}-i\epsilon^{\mu\nu\rho\kappa}
\sigmabar_\kappa^{\dot{\alpha}\alpha}\right)\delta_\gamma{}^\beta\right]\,,
 \label{Fierz16}
\\[6pt]
\sigma^\rho_{\alpha\dot\beta}
(\sigmabar^{\mu\nu})^{\dot{\gamma}}{}_{\dot\alpha}&=&
\half\left[\sigma^\nu_{\alpha\dot\alpha}
(\sigmabar^{\mu\rho})^{\dot\gamma}{}_{\dot\beta}
-\sigma^\mu_{\alpha\dot\alpha}
(\sigmabar^{\nu\rho})^{\dot\gamma}{}_{\dot\beta}
-i\epsilon^{\mu\nu\kappa}{}_\lambda \sigma_{\kappa\,\alpha\dot\alpha}
(\sigmabar^{\lambda\rho})^{\dot\gamma}{}_{\dot\beta}\right. \nonumber \\
&&\qquad\qquad \left. \BDplus\half i\left(g^{\mu\rho}
\sigma^\nu_{\alpha\dot\alpha}-g^{\nu\rho}\sigma^\mu_{\alpha\dot\alpha}
+i\epsilon^{\mu\nu\rho\kappa}\sigma_{\kappa\,\alpha\dot\alpha}\right)
\delta^{\dot\gamma}{}_{\dot\beta}\right]\,,
 \label{Fierz17}
 \\[6pt]
(\sigmabar^{\mu\nu})^{\dot\alpha}{}_{\dot\beta}
\sigmabar^{\rho\,\dot{\gamma}\alpha}&=&
\half\left[\sigmabar^{\nu\,\dot{\alpha}\alpha}
(\sigmabar^{\mu\rho})^{\dot\gamma}{}_{\dot\beta}-
\sigmabar^{\mu\,\dot{\alpha}\alpha}
(\sigmabar^{\nu\rho})^{\dot\gamma}{}_{\dot\beta}
-i\epsilon^{\mu\nu\kappa}{}_\lambda\sigmabar_\kappa^{\dot{\alpha}\alpha}
(\sigmabar^{\lambda\rho})^{\dot\gamma}{}_{\dot\beta}\right.\nonumber \\
&&\qquad\qquad \left.
\BDminus\half i\left(g^{\mu\rho}\sigmabar^{\nu\,\dot{\alpha}\alpha}-
g^{\nu\rho}\sigmabar^{\mu\,\dot{\alpha}\alpha}
+i\epsilon^{\mu\nu\rho\kappa}\sigmabar_\kappa^{\dot{\alpha}\alpha}
\right)\delta^{\dot\gamma}{}_{\dot\beta}\right]\,,
\label{Fierz18}
\eeqa
\beqa
(\sigma^{\mu\nu})_\alpha{}^\beta(\sigma^{\rho\kappa})_\gamma{}^\tau&=&
\half(\sigma^{\mu\nu})_\alpha{}^\tau(\sigma^{\rho\kappa})_\gamma{}^\beta
+\nicefrac{1}{8}\delta_\alpha{}^\tau\delta_\gamma{}^\beta
\left(g^{\mu\rho}g^{\nu\kappa}-g^{\mu\kappa}g^{\nu\rho}
-i\epsilon^{\mu\nu\rho\kappa}\right)\nonumber \\
&& 
\BDplus\quarter i\delta_\alpha{}^\tau\left(g^{\mu\rho}\sigma^{\nu\kappa}
+g^{\nu\kappa}\sigma^{\mu\rho}-g^{\nu\rho}\sigma^{\mu\kappa}
-g^{\mu\kappa}\sigma^{\nu\rho} \right)_\gamma{}^\beta \nonumber \\
&& 
\BDminus\quarter i\delta_\gamma{}^\beta\left(g^{\mu\rho}\sigma^{\nu\kappa}
+g^{\nu\kappa}\sigma^{\mu\rho}-g^{\nu\rho}\sigma^{\mu\kappa}
-g^{\mu\kappa}\sigma^{\nu\rho} \right)_\alpha{}^\tau \nonumber \\
&&
+\quarter
\left[(\sigma^{\mu\rho})_\alpha{}^\tau(\sigma^{\nu\kappa})_\gamma{}^\beta
+(\sigma^{\nu\kappa})_\alpha{}^\tau(\sigma^{\mu\rho})_\gamma{}^\beta
-(\sigma^{\nu\rho})_\alpha{}^\tau(\sigma^{\mu\kappa})_\gamma{}^\beta
-(\sigma^{\mu\kappa})_\alpha{}^\tau(\sigma^{\nu\rho})_\gamma{}^\beta\right]
\nonumber \\
&&+\quarter
g_{\lambda\sigma}\left[g^{\mu\kappa}(\sigma^{\rho\sigma})_\alpha{}^\tau
(\sigma^{\nu\lambda})_\gamma{}^\beta
+g^{\nu\rho}(\sigma^{\kappa\sigma})_\alpha{}^\tau
(\sigma^{\mu\lambda})_\gamma{}^\beta\right. \nonumber \\
&&\qquad\quad\left.
-g^{\nu\kappa}(\sigma^{\rho\sigma})_\alpha{}^\tau
(\sigma^{\mu\lambda})_\gamma{}^\beta
-g^{\mu\rho}(\sigma^{\kappa\sigma})_\alpha{}^\tau
(\sigma^{\nu\lambda})_\gamma{}^\beta\right]\,,
\label{Fierz19}\\[6pt]
(\sigmabar^{\mu\nu})^{\dot\alpha}{}_{\dot\beta}
(\sigmabar^{\rho\kappa})^{\dot\gamma}{}_{\dot\tau}&=&
\half(\sigmabar^{\mu\nu})^{\dot\alpha}{}_{\dot\tau}
(\sigmabar^{\rho\kappa})^{\dot\gamma}{}_{\dot\beta}
+\nicefrac{1}{8}\delta^{\dot\alpha}{}_{\dot\tau}
\delta^{\dot\gamma}{}_{\dot\beta}
\left(g^{\mu\rho}g^{\nu\kappa}-g^{\mu\kappa}g^{\nu\rho}
+i\epsilon^{\mu\nu\rho\kappa}\right)\nonumber \\
&& 
\BDplus\quarter i\delta^{\dot\alpha}{}_{\dot\tau}
\left(g^{\mu\rho}\sigmabar^{\nu\kappa}
+g^{\nu\kappa}\sigmabar^{\mu\rho}-g^{\nu\rho}\sigmabar^{\mu\kappa}
-g^{\mu\kappa}\sigmabar^{\nu\rho} \right)^{\dot\gamma}{}_{\dot\beta} 
\nonumber \\
&& 
\BDminus\quarter i\delta^{\dot\gamma}{}_{\dot\beta}
\left(g^{\mu\rho}\sigmabar^{\nu\kappa}
+g^{\nu\kappa}\sigmabar^{\mu\rho}-g^{\nu\rho}\sigmabar^{\mu\kappa}
-g^{\mu\kappa}\sigmabar^{\nu\rho} \right)^{\dot\alpha}{}_{\dot\tau} 
\nonumber \\
&&
+\quarter
\left[(\sigmabar^{\mu\rho})^{\dot\alpha}{}_{\dot\tau}
(\sigmabar^{\nu\kappa})^{\dot\gamma}{}_{\dot\beta}
+(\sigmabar^{\nu\kappa})^{\dot\alpha}{}_{\dot\tau}
(\sigmabar^{\mu\rho})^{\dot\gamma}{}_{\dot\beta}
-(\sigmabar^{\nu\rho})^{\dot\alpha}{}_{\dot\tau}
(\sigmabar^{\mu\kappa})^{\dot\gamma}{}_{\dot\beta}
-(\sigmabar^{\mu\kappa})^{\dot\alpha}{}_{\dot\tau}
(\sigmabar^{\nu\rho})^{\dot\gamma}{}_{\dot\beta}\right]
\nonumber \\
&&+\quarter
g_{\lambda\sigma}\left[
g^{\mu\kappa}(\sigmabar^{\rho\sigma})^{\dot\alpha}{}_{\dot\tau}
(\sigmabar^{\nu\lambda})^{\dot\gamma}{}_{\dot\beta}
+g^{\nu\rho}(\sigmabar^{\kappa\sigma})^{\dot\alpha}{}_{\dot\tau}
(\sigmabar^{\mu\lambda})^{\dot\gamma}{}_{\dot\beta}\right. \nonumber \\
&&\qquad\quad\left.
-g^{\nu\kappa}(\sigmabar^{\rho\sigma})^{\dot\alpha}{}_{\dot\tau}
(\sigmabar^{\mu\lambda})^{\dot\gamma}{}_{\dot\beta}
-g^{\mu\rho}(\sigmabar^{\kappa\sigma})^{\dot\alpha}{}_{\dot\tau}
(\sigmabar^{\nu\lambda})^{\dot\gamma}{}_{\dot\beta}\right]\,,
\label{Fierz20}\\[6pt]
(\sigma^{\mu\nu})_\alpha{}^\beta
(\sigmabar^{\rho\kappa})^{\dot\beta}{}_{\dot\alpha}&=&\BDpos
\nicefrac{1}{8}\left[(g^{\mu\rho}g^{\nu\kappa}-g^{\mu\kappa}g^{\nu\rho})
\sigma^\lambda_{\alpha\dot\alpha}\sigmabar_\lambda^{\dot{\beta}\beta}
\right.\nonumber \\
&&
+i\epsilon^{\mu\nu\rho\lambda}\sigma_{\lambda\,\alpha\dot\alpha}
\sigmabar^{\kappa\,\dot{\beta}\beta}
-i\epsilon^{\mu\nu\kappa\lambda}\sigma_{\lambda\,\alpha\dot\alpha}
\sigmabar^{\rho\,\dot{\beta}\beta}
-i\epsilon^{\mu\rho\kappa\lambda}\sigma^\nu_{\alpha\dot\alpha}
\sigmabar_\lambda^{\dot{\beta}\beta}
+i\epsilon^{\nu\rho\kappa\lambda}\sigma^\mu_{\alpha\dot\alpha}
\sigmabar_\lambda^{\dot{\beta}\beta}
\nonumber \\
&&
-g^{\mu\rho}(\sigma^\kappa_{\alpha\dot\alpha}\sigmabar^{\nu\,\dot{\beta}\beta}
+\sigma^\nu_{\alpha\dot\alpha}\sigmabar^{\kappa\,\dot{\beta}\beta})
+g^{\nu\rho}(\sigma^\kappa_{\alpha\dot\alpha}\sigmabar^{\mu\,\dot{\beta}\beta}
+\sigma^\mu_{\alpha\dot\alpha}\sigmabar^{\kappa\,\dot{\beta}\beta})
\nonumber \\
&&\left.
+g^{\mu\kappa}(\sigma^\rho_{\alpha\dot\alpha}\sigmabar^{\nu\,\dot{\beta}\beta}
+\sigma^\nu_{\alpha\dot\alpha}\sigmabar^{\rho\,\dot{\beta}\beta})
-g^{\nu\kappa}(\sigma^\rho_{\alpha\dot\alpha}\sigmabar^{\mu\,\dot{\beta}\beta}
+\sigma^\mu_{\alpha\dot\alpha}\sigmabar^{\rho\,\dot{\beta}\beta})\right]\,.
\label{Fierz21}
\eeqa
{}From \eqst{Fierz1}{Fierz21}, 
one immediately obtains the corresponding
21 Fierz identities represented by \eq{fierz21}.  Eleven of these
identities also appear in Appendix A of \Ref{Bailin}.\footnote{Note
that in \Ref{Bailin}, $\epsilon^{\mu\nu\rho\kappa}$ has the opposite
sign with respect to our conventions,
and $\sigma^{\mu\nu}$ is defined without an overall factor of $i$.
Taking these differences into account, we have confirmed that the
results of Appendix A of \Ref{Bailin} match the corresponding results
obtained here.}

The derivation of the 21 identities listed above is straightforward.
\Eqst{Fierz1}{Fierz3} are equivalent to the completeness relation of
\eq{basicid}.  The next eight identities
[\eqst{Fierz4}{Fierz11}] are easily derived starting from 
\eqst{Fierz1}{Fierz3}.  As a simple example,
using the results of \eqs{Fierz2}{sigsigid1}, it follows that
\beqa
\delta_\alpha{}^\beta\sigma^\mu_{\gamma\dot\alpha}&=&
\delta_\alpha{}^\beta\delta_\gamma{}^\tau\sigma^\mu_{\tau\dot\alpha}
=\half\left[\delta_\alpha{}\rsup{\tau}\delta\ls{\gamma}{}\rsup{\beta}+
(\sigma^{\rho\kappa})_\alpha{}\rsup{\tau} 
(\sigma_{\rho\kappa})\ls{\gamma}{}\rsup{\beta}\right]
\sigma^\mu_{\tau\dot\alpha}\nonumber \\
&=&\half\left[\sigma^\mu_{\alpha\dot\alpha}\delta\ls{\gamma}{}\rsup{\beta}
+(\sigma^{\rho\kappa}\sigma^\mu)_{\alpha\dot\alpha}
(\sigma_{\rho\kappa})_\gamma{}^\beta\right]\nonumber\\
&=&\half\left[\sigma^\mu_{\alpha\dot\alpha}\delta\ls{\gamma}{}\rsup{\beta}
\BDplus\half i(\metric^{\kappa\mu}\sigma^\rho
-\metric^{\rho\mu}\sigma^\kappa
+i\epsilon^{\rho\kappa\mu\nu}\sigma_\nu)_{\alpha\dot\alpha}
(\sigma_{\rho\kappa})_\gamma{}^\beta\right] \nonumber \\
&=&\half
\delta_\alpha{}^\gamma\sigmabar^{\mu\,\dot{\beta}\beta}
\BDplus i(\sigma^{\mu\nu})_\alpha{}^\gamma\sigmabar_\nu^{\dot{\beta}\beta}\,,
\eeqa
where \eq{eq:selfduality} was employed in the final step.
We can now use
\eqst{Fierz4}{Fierz7} to derive \eqst{Fierz12}{Fierz18} by
a similar technique.  Finally, starting from
\eqst{Fierz8}{Fierz11} we may employ the same technique once more
to derive \eqst{Fierz19}{Fierz21}.\footnote{In particular, 
the identities given in
\eqs{app:sigmunueps1}{app:sigmunueps2} are especially useful in the
evaluation of \eqst{Fierz15}{Fierz20}.}
A useful check of the last three identities can be carried out
by multiplying these results by $g_{\mu\rho}g_{\nu\kappa}$
and summing over the two repeated Lorentz index pairs.  We then find:
\beqa
(\sigma^{\mu\nu})_\alpha{}^\beta(\sigma_{\mu\nu})_\gamma{}^\tau
&=& -\half (\sigma^{\mu\nu})_\alpha{}^\tau(\sigma_{\mu\nu})_\gamma{}^\beta
+\nicefrac{3}{2}\delta_\alpha{}^\tau\delta_\gamma{}^\beta\,,
\label{sigsigsum1}\\
(\sigmabar^{\mu\nu})^{\dot\alpha}{}_{\dot\beta}
(\sigmabar_{\mu\nu})^{\dot\gamma}{}_{\dot\tau}
&=& -\half (\sigmabar^{\mu\nu})^{\dot\alpha}{}_{\dot\tau}
(\sigmabar_{\mu\nu})^{\dot\gamma}{}_{\dot\beta}+\nicefrac{3}{2}
\delta^{\dot\alpha}{}_{\dot\tau}\delta^{\dot\gamma}{}_{\dot\beta}\,,
\label{sigsigsum2} \\
(\sigma^{\mu\nu})_\alpha{}^\beta(\sigmabar_{\mu\nu})^{\dot\gamma}{}_{\dot\tau}
&=&0\,.\label{sigsigsum3}
\eeqa
\Eq{sigsigsum3} has already been recorded in \eq{ssfierzc}.
To verify \eqs{sigsigsum1}{sigsigsum2}, we first
rewrite these equations with the interchange of
$\beta\leftrightarrow\tau$ and $\dot\beta\leftrightarrow\dot\tau$.
Inserting the resulting equations back into \eqs{sigsigsum1}{sigsigsum2}
then yields the previously obtained \eqs{ssfierza}{ssfierzb}
[or equivalently, \eqs{Fierz2}{Fierz3}].

A similar check can be performed on \eqst{Fierz12}{Fierz14} by
multiplying these results by $g_{\mu\nu}$ and summing over the
repeated Lorentz index pair [with assistance from \eq{sigsigsum3}]:
\beqa
\sigma^\mu_{\alpha\dot\alpha}\sigma_{\mu\,\beta\dot\beta}&=&
-\sigma^\mu_{\alpha\dot\beta}\sigma\ls{\mu\,\beta\dot\alpha}\,,
\label{sigmasum1}\\
\sigmabar^{\mu\,\dot{\alpha}\alpha}\sigmabar_\mu^{\dot\beta\beta}&=&
-\sigmabar^{\mu\,\dot{\alpha}\beta}\sigmabar_\mu^{\dot{\beta}\alpha}\,,
\label{sigmasum2}\\
\sigma^\mu_{\alpha\dot\alpha}\sigmabar_\mu^{\dot\beta\beta}&=&\BDpos
2\,\delta_\alpha{}^\beta\delta^{\dot\beta}{}_{\dot\alpha}\,.\label{sigmasum3}
\eeqa
It follows that:
\beqa
\phantom{line}\nonumber \\[-28pt]
\sigma^\mu_{\alpha\dot\alpha}\sigma_{\mu\,\beta\dot\beta}&=&\BDpos 2\,
\epsilon_{\alpha\beta}\epsilon_{\dot{\alpha}\dot{\beta}}\,,
\label{sigmasum4}\\[3pt]
\sigmabar^{\mu\,\dot{\alpha}\alpha}\sigmabar_\mu^{\dot\beta\beta}&=&\BDpos 2\,
\epsilon^{\alpha\beta}\epsilon^{\dot{\alpha}\dot{\beta}}\,,\label{sigmasum5}
\eeqa
since \eqs{sigmasum1}{sigmasum2} 
are antisymmetric
under the separate interchanges of $\alpha\leftrightarrow\beta$ and
$\dot{\alpha}\leftrightarrow\dot{\beta}$.  The coefficients
in \eqs{sigmasum4}{sigmasum5} are
determined by substituting $\alpha=\dot\alpha=1$
and $\beta=\dot\beta=2$.  Thus, we have confirmed the results
previously obtained in \eqst{mainfierz}{mainfierz3}.

\Eqst{Fierz1}{Fierz3} can also be used to derive four additional
identities, which yield Fierz identities of a different form.
Simply multiply each of these equations by two $\epsilon$ symbols
(with appropriately chosen undotted and/or dotted spinor indices), and use
\eqs{sigsig1}{sigmunurel1}.   Two of the resulting identities
coincide with \eqs{sigmasum4}{sigmasum5}, while the other two are:
\beqa
\epsilon_{\alpha\beta}\epsilon^{\gamma\tau}
&=&-\half\bigl[\delta_\alpha{}^\gamma\delta_\beta{}^\tau-
(\sigma^{\mu\nu})_\alpha{}^\gamma
(\sigma_{\mu\nu})_\beta{}^\tau\bigr]\,,\label{epseps1} \\[6pt]
\epsilon^{\dot{\alpha}\dot{\beta}}\epsilon_{\dot{\gamma}\dot{\tau}}
&=&-\half
\left[\delta^{\dot\alpha}{}_{\dot\gamma}\delta^{\dot\beta}{}_{\dot\tau}-
(\sigmabar^{\mu\nu})^{\dot\alpha}{}_{\dot\gamma} 
(\sigmabar_{\mu\nu})^{\dot\beta}{}_{\dot\tau}\right]\,.\label{epseps2}
\eeqa
One can check that \eqs{epseps1}{epseps2} are equivalent to 
the previously obtained \eqs{ssfierza}{ssfierzb}.
Multiplying \eqst{sigmasum4}{epseps2} by four (commuting or anticommuting) 
two-component spinors $Z_{1A} Z_{2B} Z_{3C} Z_{4D}$ yields the
corresponding Fierz identities of the form:
\beq \label{fierzalt}
(Z_1\Gamma^{(k)I} Z_2)(Z_3\Gamma^{(n)J} Z_4)=(-1)^A\sum_{p,q,K,L}
(C^{kn}_{pq})^{IJ}_{KL} (Z_1\Gamma^{(p)K}Z_3)(Z_2\Gamma^{(q)L} Z_4)\,,
\eeq
which differs from \eq{fierz21} in the ordering of the spinors on the
right-hand side.

Finally, we note that the Schouten identities,
\beq
\eps_{\alpha\beta}\eps_{\gamma\delta}+\eps_{\alpha\gamma}\eps_{\delta\beta}
+\eps_{\alpha\delta}\eps_{\beta\gamma}=0\,, \qquad\qquad
\eps_{\dot{\alpha}\dot{\beta}}\eps_{\dot{\gamma}\dot{\delta}}
+\eps_{\dot{\alpha}\dot{\gamma}}\eps_{\dot{\delta}\dot{\beta}}
+\eps_{\dot{\alpha}\dot{\delta}}\eps_{\dot{\beta}\dot{\gamma}}=0\,,
\label{app:schouten}
\eeq
are the basis for Fierz identities given by
\eqs{eq:twocompfierzone}{eq:twocompfierztwo}, which 
do not assume the simple forms of either \eqor{fierz21}{fierzalt}.

\subsection{Sigma matrix identities in \texorpdfstring{$d\neq 4$}{d\textneq 4} dimensions}
\renewcommand{\theequation}{B.2.\arabic{equation}}
\renewcommand{\thefigure}{B.2.\arabic{figure}}
\renewcommand{\thetable}{B.2.\arabic{table}}
\setcounter{equation}{0}
\setcounter{figure}{0}
\setcounter{table}{0}

When considering a theory regularized by
dimensional continuation~\cite{dimreg},
one must be careful in treating cases with
contracted spacetime vector indices $\mu,\nu,\kappa,\rho, \ldots$. Instead of
taking on four possible values, these vector indices formally run over $d$
values, where $d$ is infinitesimally different from 4.
This means that some identities that would hold in unregularized
four-dimensional theories are inconsistent and must not be used; other
identities remain valid if $d$ replaces 4 in the appropriate spots;
and still other identities hold without modification.

Two important identities that do hold in $d\neq 4$ dimensions are:
\beqa
&&{[\sigma^\mu\sigmabar^\nu + \sigma^\nu \sigmabar^\mu ]_\alpha}^\beta
= \BDpos 2\metric^{\mu\nu}
\delta_\alpha{}^\beta \,,
\label{identityseven}
\\
&&[\sigmabar^\mu\sigma^\nu + \sigmabar^\nu \sigma^\mu ]
^{\dot{\alpha}}{}_{\dot{\beta}} = \BDpos 2\metric^{\mu\nu}
\delta^{\dot{\alpha}}{}_{\dot{\beta}}\,.
\label{identityeight}
\eeqa
Equivalently,
\beqa
(\sigma^\mu\sigmabar^\nu)_\alpha{}^\beta &=&  \BDpos
g^{\mu\nu}\delta_\alpha{}^\beta
-2i(\sigma^{\mu\nu})_\alpha{}^\beta\,,\label{smunu}\\
(\sigmabar^\mu\sigma^\nu)^{\dot{\alpha}}{}_{\dot{\beta}}
&=&  \BDpos g^{\mu\nu}\delta^{\dot{\alpha}}{}_{\dot{\beta}}
-2i(\sigmabar^{\mu\nu})^{\dot{\alpha}}{}_{\dot{\beta}}\,,\label{sbarmunu}
\eeqa
where $\sigma^{\mu\nu}$ and $\sigmabar^{\mu\nu}$ are defined in \eq{Lgenerators}.
The trace identities,
\beqa
{\rm Tr}[\sigma^\mu \sigmabar^\nu ] =
{\rm Tr}[\sigmabar^\mu \sigma^\nu ] &=&
\BDpos 2 \metric^{\mu\nu}\,, \label{APPtrssbar} \\
\Tr\sigma^{\mu\nu}=\Tr\sigmabar^{\mu\nu}&=&0\,,\label{trsigmunu}
\eeqa
then follow.  We also note that the spinor index trace identity,
\beq
{\rm Tr}[\mathds{1}] =
\delta_\alpha^\alpha = \delta_{\dot\alpha}^{\dot\alpha}= 2\,,
\eeq
continues to hold in dimensional continuation regularization methods.
In contrast, the Fierz identities of \app{B.1} 
do not have a consistent, unambiguous meaning outside of four
dimensions~\cite{Siegel:1980qs,Avdeev:1982xy,Blatter,
gammafive}.  However, the following identities
that are implied by \eq{Fierz1} do consistently 
generalize to $d\neq 4$ spacetime dimensions:
\beqa
&& [\sigma^\mu\sigmabar_\mu ]_\alpha{}^\beta =
\BDpos d \delta_\alpha^\beta\,,
\label{eq:genfierzone}
\\
&&
[\sigmabar^\mu\sigma_\mu ]^{\dot{\alpha}}{}_{\dot{\beta}}
= \BDpos d \delta^{\dot{\alpha}}_{\dot{\beta}}\,.
\label{eq:genfierztwo}
\eeqa
Using \eqs{eq:genfierzone}{eq:genfierztwo}
along with the repeated use
of \eqs{identityseven}{identityeight} then yields:
\beqa
&&
[\sigma^\mu\sigmabar^\nu \sigma_\mu ]_{\alpha\dot\beta} =
\BDneg (d-2) \sigma^\nu_{\alpha\dot\beta}\,,
\label{eq:genfierzthree}
\\
&&
[\sigmabar^\mu \sigma_\nu \sigmabar_\mu ]^{\dot\alpha\beta} =
\BDneg (d-2) \sigmabar_\nu^{\dot\alpha\beta}\,,
\label{eq:genfierzfour}
\\
&&
[\sigma^\mu \sigmabar^\nu \sigma^\rho\sigmabar_\mu
]_\alpha{}^\beta
= 4 \metric^{\nu\rho} \delta_\alpha^\beta
\BDminus (4-d) [\sigma^\nu \sigmabar^\rho]_\alpha{}^\beta\,,
\\
&&
[\sigmabar^\mu \sigma^\nu \sigmabar^\rho\sigma_\mu
]^{\dot{\alpha}}{}_{\dot{\beta}}
= 4 \metric^{\nu\rho} \delta^{\dot{\alpha}}_{\dot{\beta}}
\BDminus (4-d) [\sigmabar^\nu \sigma^\rho]^{\dot\alpha}{}_{\dot\beta} \,,
\\
&&
[\sigma^\mu \sigmabar^\nu
\sigma^\rho\sigmabar^\kappa \sigma_\mu]_{\alpha\dot\beta}
=
\BDneg 2 [\sigma^\kappa \sigmabar^\rho \sigma^\nu]_{\alpha\dot\beta}
\BDplus (4-d)[\sigma^\nu \sigmabar^\rho \sigma^\kappa]_{\alpha\dot\beta}
\,,\\
&&
[\sigmabar^\mu \sigma^\nu
\sigmabar^\rho\sigma^\kappa \sigmabar_\mu]^{\dot\alpha\beta}
=
\BDneg 2 [\sigmabar^\kappa \sigma^\rho \sigmabar^\nu]^{\dot\alpha\beta}
\BDplus (4-d)[\sigmabar^\nu \sigma^\rho
\sigmabar^\kappa]^{\dot\alpha\beta}\,.
\eeqa

Identities that involve the (explicitly and
inextricably four-dimensional)
$\epsilon^{\mu\nu\rho\kappa}$ symbol
\beqa
&&\hspace{-0.7in} \sigmabar^\mu \sigma^\nu \sigmabar^\rho =
\BDpos \metric^{\mu\nu} \sigmabar^\rho
\BDminus  \metric^{\mu\rho} \sigmabar^\nu
\BDplus \metric^{\nu\rho} \sigmabar^\mu
\BDminus i \epsilon^{\mu\nu\rho\kappa}\sigmabar_\kappa\,,
\label{sig3a}
\\
&&\hspace{-0.7in} \sigma^\mu \sigmabar^\nu \sigma^\rho =
\BDpos \metric^{\mu\nu} \sigma^\rho
\BDminus \metric^{\mu\rho} \sigma^\nu
\BDplus \metric^{\nu\rho} \sigma^\mu
\BDplus i \epsilon^{\mu\nu\rho\kappa}\sigma_\kappa\,,
\label{sig3b}
\\
&&\hspace{-0.7in}\epsilon^{\mu\nu\kappa}{}_\lambda\sigma^{\lambda\rho}=
-i\left(\metric^{\kappa\rho}\sigma^{\mu\nu}
-\metric^{\nu\rho}\sigma^{\mu\kappa}
+\metric^{\mu\rho}\sigma^{\nu\kappa}\right)\,,
\label{app:sigmunueps1}
\\
&&\hspace{-0.7in}\epsilon^{\mu\nu\kappa}{}_\lambda\sigmabar^{\lambda\rho}=
i\left(\metric^{\kappa\rho}\sigmabar^{\mu\nu}
-\metric^{\nu\rho}\sigmabar^{\mu\kappa}
+\metric^{\mu\rho}\sigmabar^{\nu\kappa}\right)\,,
\label{app:sigmunueps2}
\\
&&\hspace{-0.7in} \sigma^{\mu\nu}\sigma^\rho = \BDpos\nicefrac{1}{2}i\left(
\metric^{\nu\rho}\sigma^\mu
-\metric^{\mu\rho}\sigma^\nu
+i\epsilon^{\mu\nu\rho\kappa}\sigma_\kappa\right)\,,
\label{sigsigid1}
\\
&&\hspace{-0.7in} \sigmabar^{\mu\nu}\sigmabar^\rho = \BDpos\nicefrac{1}{2}i
\left(\metric^{\nu\rho}\sigmabar^\mu
-\metric^{\mu\rho}\sigmabar^\nu
-i\epsilon^{\mu\nu\rho\kappa}\sigmabar_\kappa\right)\,,
\label{sigsigid2}
\\
&&\hspace{-0.7in} \sigmabar^\mu\sigma^{\nu\rho}=\BDpos\nicefrac{1}{2}i\left(
\metric^{\mu\nu}\sigmabar^\rho
-\metric^{\mu\rho}\sigmabar^\nu
-i\epsilon^{\mu\nu\rho\kappa}\sigmabar_\kappa\right)\,,
\label{sigsigid3}
\\
&&\hspace{-0.7in} \sigma^\mu\sigmabar^{\nu\rho}=\BDpos\nicefrac{1}{2}i\left(
\metric^{\mu\nu}\sigma^\rho
-\metric^{\mu\rho}\sigma^\nu
+i\epsilon^{\mu\nu\rho\kappa}\sigma_\kappa\right)\,,
\label{sigsigid4} \\
&&\hspace{-0.7in} 
\sigma^{\mu\nu}\sigma^{\rho\kappa}=-\quarter(\metric^{\nu\rho}
\metric^{\mu\kappa}-\metric^{\mu\rho}\metric^{\nu\kappa}
+i\epsilon^{\mu\nu\rho\kappa})
\BDplus \nicefrac{1}{2}i(\metric^{\nu\rho}\sigma^{\mu\kappa}
+\metric^{\mu\kappa}\sigma^{\nu\rho}-\metric^{\mu\rho}\sigma^{\nu\kappa}
-\metric^{\nu\kappa}\sigma^{\mu\rho})\,,
\\
&&\hspace{-0.7in} 
\sigmabar^{\mu\nu}\sigmabar^{\rho\kappa}=-\quarter(\metric^{\nu\rho}
\metric^{\mu\kappa}-\metric^{\mu\rho}\metric^{\nu\kappa}
-i\epsilon^{\mu\nu\rho\kappa})
\BDplus \nicefrac{1}{2}i(\metric^{\nu\rho}\sigmabar^{\mu\kappa}
+\metric^{\mu\kappa}\sigmabar^{\nu\rho}-\metric^{\mu\rho}\sigmabar^{\nu\kappa}
-\metric^{\nu\kappa}\sigmabar^{\mu\rho})\,,\label{sigsigid8}
\eeqa
are also only meaningful in exactly four dimensions.
This applies as well to the trace identities which follow from
them.\footnote{This is
analogous to the statement that ${\rm Tr}~(\gamma_5\gamma^\mu
\gamma^\nu\gamma^\rho\gamma^\kappa)=-4i\epsilon^{\mu\nu\rho\kappa}$
[in our convention where $\epsilon^{0123}=+1$, and $\gamma_5$
is defined by
eq.~(\ref{gamma4})] is only meaningful in
$d=4$ dimensions.  In two-component notation, the equivalent result
is ${\rm Tr}[\sigma^\mu \sigmabar^\nu \sigma^\rho \sigmabar^\kappa
-\sigmabar^\mu \sigma^\nu \sigmabar^\rho \sigma^\kappa]
=4i\epsilon^{\mu\nu\rho\kappa}$.
In the literature various schemes have been
proposed for defining the properties of $\gamma_5$ in $d\neq 4$
dimensions~\cite{jegerlehner,gammafive}.  In two-component notation, this
would translate into a procedure for dealing with general traces
involving four or more $\sigma$/$\sigmabar$ matrices.}
For example,
\beqa
&&
{\rm Tr}[\sigma^\mu \sigmabar^\nu \sigma^\rho \sigmabar^\kappa ] =
2 \left ( \metric^{\mu\nu} \metric^{\rho\kappa} - \metric^{\mu\rho}
\metric^{\nu\kappa} + \metric^{\mu\kappa} \metric^{\nu\rho} + i
\epsilon^{\mu\nu\rho\kappa} \right )
,
\label{APPtrssbarssbar}
\\
&&
{\rm Tr}[\sigmabar^\mu \sigma^\nu \sigmabar^\rho \sigma^\kappa ] =
2 \left ( \metric^{\mu\nu} \metric^{\rho\kappa} - \metric^{\mu\rho}
\metric^{\nu\kappa} + \metric^{\mu\kappa} \metric^{\nu\rho} - i
\epsilon^{\mu\nu\rho\kappa} \right )\,.
\label{APPtrsbarssbars}
\eeqa
This could lead to ambiguities in loop computations where it is
necessary to perform the computation in $d\neq 4$ dimensions (until
the end of the calculation where the limit $d\to 4$ is taken).
However, in practice one typically finds that the above expressions
appear multiplied by the metric and/or other external tensors (such as
four-momenta appropriate to the problem at hand).  In almost all such cases,
two of the indices appearing in \eqs{APPtrssbarssbar}{APPtrsbarssbars}
are symmetrized which eliminates the $\epsilon^{\mu\nu\rho\kappa}$
term, rendering the resulting expressions unambiguous.
Similarly, the sum of the above
trace identities can be assigned an unambiguous
meaning in $d\neq 4$ dimensions:
\beq
{\rm Tr}[\sigma^\mu \sigmabar^\nu \sigma^\rho \sigmabar^\kappa
]+{\rm Tr}[\sigmabar^\mu \sigma^\nu \sigmabar^\rho \sigma^\kappa ] =
4 \left ( \metric^{\mu\nu} \metric^{\rho\kappa} - \metric^{\mu\rho}
\metric^{\nu\kappa} + \metric^{\mu\kappa} \metric^{\nu\rho} \right ).
\label{symmtracessss}
\eeq
One can recursively
derive trace formulae for products of six or more $\sigma$/\,$\sigmabar$
matrices by using the results of \eqs{sig3a}{sig3b} to reduce the
number of $\sigma$/$\,\sigmabar$ matrices by two.  For example,
\beqa
{\rm Tr}[\sigma^\mu \sigmabar^\nu \sigma^\rho \sigmabar^\kappa
\sigma^\lambda \sigmabar^\delta]&=&
\BDpos\metric^{\mu\nu}{\rm Tr}[\sigma^\rho \sigmabar^\kappa \sigma^\lambda
\sigmabar^\delta]
\BDminus\metric^{\mu\rho}{\rm Tr}[\sigma^\nu \sigmabar^\kappa \sigma^\lambda
\sigmabar^\delta]
\BDplus\metric^{\nu\rho}{\rm Tr}[\sigma^\mu \sigmabar^\kappa \sigma^\lambda
\sigmabar^\delta]
\nonumber \\
&& \BDplus i\epsilon^{\mu\nu\rho\epsilon}
{\rm Tr}[\sigma_\epsilon \sigmabar^\kappa
\sigma^\lambda\sigmabar^\delta]\,, \\[8pt]
{\rm Tr}[\sigmabar^\mu \sigma^\nu \sigmabar^\rho \sigma^\kappa
\sigmabar^\lambda \sigma^\delta]&=&
\BDpos\metric^{\mu\nu}{\rm Tr}[\sigmabar^\rho \sigma^\kappa \sigmabar^\lambda
\sigma^\delta]
\BDminus\metric^{\mu\rho}{\rm Tr}[\sigmabar^\nu \sigma^\kappa \sigmabar^\lambda
\sigma^\delta]
\BDplus\metric^{\nu\rho}{\rm Tr}[\sigmabar^\mu \sigma^\kappa \sigmabar^\lambda
\sigma^\delta]
\nonumber \\
&& \BDminus i\epsilon^{\mu\nu\rho\epsilon}{\rm Tr}[\sigmabar_\epsilon
\sigma^\kappa \sigmabar^\lambda\sigma^\delta]\,.
\eeqa
We then use \eqs{APPtrssbarssbar}{APPtrsbarssbars} to evaluate
the remaining traces over
four $\sigma$/\,$\sigmabar$ matrices.

\section{\texorpdfstring{Explicit forms for the two-component spinor wave functions}{Explicit forms for the two-component spinor wave functions}}
\label{appendix:C}
\renewcommand{\theequation}{C.\arabic{equation}}
\renewcommand{\thefigure}{C.\arabic{figure}}
\renewcommand{\thetable}{C.\arabic{table}}
\setcounter{equation}{0}
\setcounter{figure}{0}
\setcounter{table}{0}

In this Appendix, we construct the explicit forms for the eigenstates
of the spin operator $\half{\boldsymbol{\vec\sigma\newcdot\hat s}}$,
and examine their properties.  For massive fermions, it is possible
to transform to the rest frame, and quantize the spin along a fixed
axis in space.  The corresponding spinor wave functions will
be called fixed-axis spinors.  For either massive or massive fermions,
one can quantize the spin along the direction of momentum.  The
corresponding spinor wave functions are helicity spinors.
Helicity spinor wave functions are most conveniently applied
to massless fermions or fermions in the relativistic limit of high energy
$E\gg m$.  Fixed-axis spinors are most conveniently applied to massive
fermions in the non-relativistic limit.

\subsection{Fixed-axis spinor wave functions}
\renewcommand{\theequation}{C.1.\arabic{equation}}
\renewcommand{\thefigure}{C.1.\arabic{figure}}
\renewcommand{\thetable}{C.1.\arabic{table}}
\setcounter{equation}{0}
\setcounter{figure}{0}
\setcounter{table}{0}

Consider a spin-1/2 fermion in
its rest frame and quantize the spin along a fixed axis specified by
the unit vector
\beq
{\boldsymbol{\hat s}}\equiv(\sin\theta\cos\phi\,,\,
\sin\theta\sin\phi\,,\,\cos\theta)\,,
\eeq
with polar angle $\theta$ and
azimuthal angle $\phi$ with respect to a fixed $z$-axis.
The corresponding spin states will be called fixed-axis spin states.
The relevant basis of
two-component fixed-axis spinors $\chi\ls{s}$ are eigenstates
of $\half{\boldsymbol{\vec\sigma\newcdot\hat s}}$, i.e.,
\beq
\half {\boldsymbol{\vec\sigma\newcdot\hat s}}\,\chi\ls{s} =
s\chi\ls{s}\,,
  \qquad s = \pm\half\,.\label{app:chiess}
\eeq

In order to construct the eigenstates of
$\half{\boldsymbol{\vec\sigma\newcdot\hat s}}$, we first consider the \pagebreak
case where ${\boldsymbol{\hat s}}={\boldsymbol{\hat z}}$.  In this
case, we define the eigenstates of $\half\sigma^{3}$ to be:
\beq
  \chi\ls{1/2}({\boldsymbol{\hat z}}) =
\begin{pmatrix} 1 \\ 0 \end{pmatrix},
  \qquad \chi\ls{-1/2} ({\boldsymbol{\hat z}}) =
\begin{pmatrix} 0 \\ 1 \end{pmatrix}\,.\label{twocompz}
\eeq
By convention, we have set an arbitrary overall multiplicative phase factor
for each spinor of \eq{twocompz} to unity.
We then determine $\chi\ls{s}(\boldsymbol{\hat s})$ from
$\chi\ls{s}(\boldsymbol{\hat z})$ by employing the spin-1/2 rotation
operator that corresponds to a rotation from $\boldsymbol{\hat z}$ to
$\boldsymbol{\hat s}$.  This rotation is represented by a $3\times 3$
matrix $\mathcal{R}$ such that $\boldsymbol{\hat s}=
\mathcal{R}\boldsymbol{\hat z}$.
However, this rotation operator is not unique.  In its most general form, the
rotation operator can be parameterized in terms of three Euler
angles (e.g., see \refs{tung}{Rao}):
\beq \label{calrdef}
\mathcal{R}(\phi\,,\,\theta\,,\,\gamma)\equiv R({\boldsymbol{\hat z}}\,,\,\phi)\,
R({\boldsymbol{\hat y}}\,,\,\theta)\,R({\boldsymbol{\hat z}}\,,\,\gamma)\,,
\eeq
The Euler angles can be chosen to lie in the range
$0\leq\theta\leq\pi$ and $0\leq \phi\,,\,\gamma<2\pi$.
In general, $R({\boldsymbol{\hat n}}\,,\,\theta)$ is a $3\times 3$
orthogonal matrix of unit determinant that represents a rotation by an angle~$\theta$ about
a fixed axis $\boldsymbol{\hat n}$,
\beq \label{rij}
R^{ij}({\boldsymbol{\hat n}}\,,\,\theta)=\exp(-i\theta\boldsymbol{\hat{n}\newcdot
\vec{\mathcal{S}}})=
n^i n^j+(\delta^{ij}-n^i n^j)\cos\theta-\epsilon^{ijk}n^k\sin\theta\,,
\eeq
where the $\boldsymbol{\vec{\mathcal{S}}}=(\mathcal{S}^1\,,\,\mathcal{S}^2\,,\,\mathcal{S}^3)$
are three $3\times 3$ matrices whose matrix elements are given by
$(\mathcal{S}^i)^{jk}=-i\epsilon^{ijk}$ [cf.~\eq{explicitsmunu}].

However, the angle $\gamma$ is
arbitrary, since $R({\boldsymbol{\hat z}}\,,\,\gamma)
{\boldsymbol{\hat z}}={\boldsymbol{\hat z}}$.  Thus,
\beq \label{rotcon1}
\boldsymbol{\hat s}=\mathcal{R}\boldsymbol{\hat z}=
(\sin\theta\cos\phi\,,\,\sin\theta\sin\phi\,,\,\cos\theta)\,,
\eeq
independently of the choice of $\gamma$.
For $\theta=0$ or $\theta=\pi$, where ${\boldsymbol{\hat s}}$
is parallel to the $z$-axis, the azimuthal angle $\phi$ is undefined.
Since ${\boldsymbol{\hat s}}\to -{\boldsymbol{\hat s}}$ corresponds in
general to $\theta\to \pi-\theta$ and $\phi\to \phi+\pi$ (mod $2\pi$),
we shall adopt a convention whereby:
\beq \label{negativez}
\phi=\begin{cases} 0\,, & \quad \textrm{for}~{\boldsymbol{\hat s}}
={\boldsymbol{\hat z}}\,,\quad (\theta=0)\,,
\\ \pi\,, & \quad \textrm{for}~{\boldsymbol{\hat s}}
=-{\boldsymbol{\hat z}}\,,\quad (\theta=\pi)\,.
\end{cases}
\eeq

Using the spin-1/2 rotation operator corresponding to
$\mathcal{R}(\phi\,,\,\theta\,,\,\gamma)$,
one can compute $\chi_s({\boldsymbol{\hat s}})$,
\beq \label{chidefinition}
\chi\ls{s}(\boldsymbol{\hat s}) = \mathcal{D}(\phi\,,\,\theta\,,\,\gamma)
\,\chi\ls{s}({\boldsymbol{\hat z}})\,,
\eeq
where $\mathcal{D}$ is the spin-1/2 unitary
representation matrix~\cite{rose}
\beq
\mathcal{D}(\phi\,,\,\theta\,,\,\gamma)
\equiv D(\boldsymbol{\hat z},\phi)\,D(\boldsymbol{\hat y},\theta)\,
D(\boldsymbol{\hat z},\gamma)\,,
\eeq
and $D$ is the $2\times 2$ unitary matrix
\beq \label{dhalf}
D(\boldsymbol{\hat n},\theta)\equiv\exp\left({-i\theta
\boldsymbol{\hat n\newcdot\vec\sigma}/2}\right)=
\cos\frac{\theta}{2}-i\boldsymbol{\hat n\newcdot\vec\sigma}
\sin\frac{\theta}{2}\,.
\eeq
\Eq{chidefinition}
yields explicit forms for the eigenstates of
$\half{\boldsymbol{\vec\sigma\newcdot\hat s}}$:
\beq \label{twocomp}
  \chi\ls{1/2}({\boldsymbol{\hat s}}) =
\begin{pmatrix} e^{-i(\phi+\gamma)/2}\cos
\displaystyle{\frac{\theta}{2}} \\[7pt]
     e^{i(\phi-\gamma)/2} \sin\displaystyle{\frac{\theta}{2}} \end{pmatrix},
  \qquad \chi\ls{-1/2} ({\boldsymbol{\hat s}}) = \begin{pmatrix}%
       -e^{-i(\phi-\gamma)/2} \sin\displaystyle{\frac{\theta}{2}} \\[7pt]
e^{i(\phi+\gamma)/2}\cos \displaystyle{\frac{\theta}{2}}\end{pmatrix}\,.
\eeq

The well-known two-to-one mapping between SU(2) and SO(3) implies
that for a given rotation matrix $\mathcal{R}$ there are two
corresponding spin-1/2 rotation matrices $\mathcal{D}$.  In particular,
\beq \label{twoone}
\mathcal{D}(\phi+2\pi\,,\,\theta\,,\,\gamma)=
-\mathcal{D}(\phi\,,\,\theta\,,\,\gamma)\,,
\eeq
which implies that
a rotation of a spinor by $2\pi$ yields an overall change of
sign in the spinor wave function (an effect that can be
observed in quantum interference experiments!).  Strictly speaking, we should
take the range of the Euler angles to be $0\leq \phi< 4\pi$,
$0\leq\theta\leq \pi$ and $0\leq\gamma<2\pi$.  However, when
constructing the spinor wave function of a spin-1/2 particle whose
spin quantization axis is given by \eq {rotcon1}, we will
fix the overall sign of the spinor wave function by convention.

More generally, the overall phase of the spinor wave function
is unphysical.  Noting that $D(\boldsymbol{\hat z},\gamma)
\,\chi\ls{s}({\boldsymbol{\hat z}})
=e^{-is\gamma}\chi\ls{s}({\boldsymbol{\hat z}})$,
the choice of $\gamma$ is also a matter of convention.
First, we will require that
when $\boldsymbol{\hat s}=\boldsymbol{\hat z}$, \eq{chidefinition}
should reproduce the spinor wave functions given in \eq{twocompz}.
This implies that:
\beq \label{gammatheta0}
\gamma=0\,, \qquad \textrm{for}~\boldsymbol{\hat s}=\boldsymbol{\hat z}\,,
\quad (\theta=\phi=0)\,.
\eeq
For $\boldsymbol{\hat s}=-\boldsymbol{\hat z}$, we
use \eq{negativez} to obtain:
\beq \label{zminus}
\chi\ls{s}({\boldsymbol{-\hat z}}) =ie^{-is\gamma(\boldsymbol{-\hat z})}\,
\chi\ls{-s}({\boldsymbol{\hat z}})\,,\qquad\qquad s=\pm\half\,,
\eeq
where the notation $\gamma(\boldsymbol{-\hat z})$ has been employed
to allow the possibility that the
convention for $\gamma$ depends on the direction indicated by
its argument.

Two different conventions are commonly employed
in the literature.  In the first convention, one chooses
$\gamma=-\phi$.  This choice has the good feature that
$\mathcal{R}(\phi\,,\,0\,,\,-\phi)=\mathds{1}_{3\times 3}$, independently
of the angle $\phi$, which is undefined when $\theta=0$.\footnote{However,
$\mathcal{R}(\phi\,,\,\pi\,,\,-\phi)\neq\mathds{1}_{3\times 3}$
even though $\phi$ is also undefined when $\theta=\pi$.}
Moreover, the rotation matrix $\mathcal{R}(\phi\,,\,\theta\,,\,-\phi)$
and the corresponding spin-1/2 rotation matrix
$\mathcal{D}(\phi\,,\,\theta\,,\,-\phi)$
can be expressed simply as a
single rotation by an angle $\theta$ about a fixed axis that points along
a unit vector in the azimuthal direction:
\beq
\boldsymbol{\hat \varphi}\equiv(-\sin\phi\,,\,\cos\phi\,,\,0)\,,
\eeq
In particular,
\beqa
R(\boldsymbol{\hat \varphi}\,,\,\theta)&=&
R(\boldsymbol{\hat z}\,,\,\phi)\,R(\boldsymbol{\hat y}\,,\,\theta)\,
R(\boldsymbol{\hat z}\,,\,-\phi)\,,\\
D(\boldsymbol{\hat \varphi}\,,\,\theta)&=&
\mathcal{D}(\phi\,,\,\theta\,,\,-\phi)\,.
\eeqa
Hence, in this convention $\chi\ls{s}(\boldsymbol{\hat s})=
D(\boldsymbol{\hat \varphi}\,,\,\theta)\chi\ls{s}(\boldsymbol{\hat z})$, which is the most
common choice for the spinor wave function~\cite{Cbook,auvil,Carruthers}.

In the second convention, one chooses $\gamma=0$.  One motivation for this
choice is that the corresponding rotation matrix is somewhat simpler:
\beq \label{rotationmat}
\mathcal{R}(\phi\,,\,\theta\,,\,0)
=R(\boldsymbol{\hat z}\,,\,\phi)
\,R(\boldsymbol{\hat y}\,,\,\theta)=
\begin{pmatrix} \cos\theta\cos\phi & \quad -\sin\phi & \quad
  \sin\theta\cos\phi \\[5pt] \cos\theta\sin\phi & \quad \phm\cos\phi & \quad
  \sin\theta\sin\phi \\[5pt] -\sin\theta & \quad 0 &\quad \cos\theta
\end{pmatrix}\,.
\eeq
Employing the corresponding spin-1/2 rotation operator
$\mathcal{D}(\phi\,,\,\theta\,,\,0)$ in \eq{chidefinition}
yields a slightly more symmetrical
form for the spinor wave function~\cite{leader}.

Explicit forms for the spinor wave functions in the two conventions
are obtained from \eq{twocomp} by taking
$\gamma(\boldsymbol{\hat s})=-\phi$ and $\gamma(\boldsymbol{\hat s})=0$,
respectively.  For example, \eq{zminus} reduces to:
\beq
\chi\ls{s}(-\boldsymbol{\hat z})=\begin{cases}
-2s\chi\ls{-s}(\boldsymbol{\hat z}) & \textrm{for}~
\gamma(-\boldsymbol{\hat z})=-\phi=-\pi\,,\\[6pt]
\phm i\chi\ls{-s}(\boldsymbol{\hat z}) &
\textrm{for}~\gamma(-\boldsymbol{\hat z})=0\,,
\end{cases}\qquad\qquad s=\pm\half\,,
\eeq
in the convention specified by \eq{negativez}.

Many of the properties of the spinor wave functions are
independent of the choice of the Euler angle
$\gamma$.  The spinor wave functions $\chi\ls{s}$ defined
by \eq{chidefinition} are normalized such that
\beq \label{spinornorm}
\chi^\dagger_s({\boldsymbol{\hat
      s}})\chi_{s'}({\boldsymbol{\hat s}})=\delta_{ss'}\,,
\eeq
and satisfy the following completeness relation:
\beq \label{completeness}
\sum_s\chi\ls{s}({\boldsymbol{\hat s}})
\chi\ls{s}^\dagger({\boldsymbol{\hat s}})= \left(\begin{array}{cc} 1&
    0\\ 0&  1\end{array}\right)\,.
\eeq
The spinor wave functions $\chi\ls{s}({\boldsymbol{\hat s}})$ and
$\chi\ls{-s}({\boldsymbol{\hat s}})$ are connected by the following
relation:
\beq
\chi\ls{-s}({\boldsymbol{\hat s}})= -2si\sigma^{2}\,
\chi^\ast\ls{s}({\boldsymbol{\hat s}})\,.\label{twocomppropa}
\eeq

Consider a spin-1/2 fermion with four-momentum
$p^\mu=(E\,,\,{\boldsymbol{\vec p}})$, with
$E=(|{\boldsymbol{\vec p}}|^2+m^2)^{1/2}$,
and the direction of ${\boldsymbol{\vec p}}$ given by
\beq \label{pdirection}
{\boldsymbol{\hat p}}=(\sin\theta_p\cos\phi_p\,,\,\sin\theta_p\sin\phi_p
\,,\,\cos\theta_p)\,.
\eeq
Using \eqs{sqpsigma}{sqpsigmabar}, one can
employ \eqst{explicitxa}{explicityb} to obtain explicit
expressions for
the two-component spinor wave functions $x({\boldsymbol{\vec p}},s)$,
$y({\boldsymbol{\vec p}},s)$, $ x^\dagger({\boldsymbol{\vec p}},s)$ and
$ y^\dagger({\boldsymbol{\vec p}},s)$.

Additional properties of the $\chi_s$ can be derived by
introducing an orthonormal set of unit three-vectors
${\boldsymbol{\hat s}^a}$ that
provide a basis for a right-handed coordinate system.  Explicitly,
\beqa
{\boldsymbol{{\hat s}^a}}{\boldsymbol{\newcdot}}{\boldsymbol{{\hat s}^b}}
&=&\delta^{ab}\,, \label{hatsdef1}\\
{\boldsymbol{{\hat s}^a \times {\hat s}^b}}&=&\epsilon^{abc}
{\boldsymbol{{\hat s}^c}}\,. \label{hatsdef2}
\eeqa
We shall identify
\beq
{\boldsymbol{{\hat s}^3}}\equiv{\boldsymbol{{\hat s}}}
\eeq
as the quantization axis
used in defining the third component of the
spin of the fermion in its rest frame.  The unit vectors
${\boldsymbol{{\hat s}^1}}$ and ${\boldsymbol{{\hat s}^2}}$ are then chosen
such that \eqs{hatsdef1}{hatsdef2} are satisfied.
To explicitly construct the ${\boldsymbol{{\hat s}^a}}$, we begin with
the orthonormal set $\{{\boldsymbol{\hat x}}\,,\,{\boldsymbol{\hat y}}
\,,\,{\boldsymbol{\hat z}}\}$, and employ the \textit{same} rotation
operator $\mathcal{R}$
used to define $\chi\ls{s}({\boldsymbol{\hat s}})$. That is,
\beq \label{saexplicit}
({\boldsymbol{{\hat s}^1}}\,,\,{\boldsymbol{{\hat s}^2}}\,,\,
{\boldsymbol{{\hat s}^3}})=
(\mathcal{R}{\boldsymbol{{\hat x}}}\,,\,\mathcal{R}
{\boldsymbol{{\hat y}}}\,,\,\mathcal{R}{\boldsymbol{{\hat z}}})\,,
\qquad {\rm where}
\quad \mathcal{R}\equiv \mathcal{R}(\phi\,,\,\theta\,,\,\gamma)\,,
\eeq
and $\phi$, $\theta$ and $\gamma$ are the Euler angles used to
define the spinor wave function in \eq{chidefinition}.
From \eq{saexplicit}, one can immediately derive the completeness
relation (as a consequence of $\mathcal{R}\mathcal{R}^{\T}=\mathds{1}$),
\beq \label{scomplete}
\boldsymbol{{\hat s}^a}^i\boldsymbol{{\hat s}^a}^j=\delta^{ij}\,,
\eeq
where $i$ and $j$ label the space components of the three-vector
$\boldsymbol{{\hat s}^a}$.

We can use the ${\boldsymbol{\hat s^{a}}}$ to extend
the defining equation of $\chi_s$ [\eq{app:chiess}]:
\beq \label{tauchi}
\half\, {\boldsymbol{\vec\sigma\newcdot{\hat s}^a}}
\chi_{s'}(\boldsymbol{\hat s})=
\half \tau^a_{ss'}\chi_s(\boldsymbol{\hat s})\,,
\eeq
where the $\tau^a_{ss'}$ are the matrix elements of the
Pauli matrices.\footnote{\label{fnpauli}%
We use the symbol $\tau$ rather than
$\sigma$ to emphasize that the indices of the Pauli matrices $\tau^a$ are
spin labels $s$, $s'$ and \textit{not}
spinor indices $\alpha$, $\dot{\alpha}$.
The first (second) row
and column of the $\tau$-matrices correspond to $s=1/2\ (-1/2)$.
For example,
$\tau^3_{ss^\prime}=2s\delta_{ss^\prime}$ (no sum over $s$).\label{fntau}}
That is, $\half{\boldsymbol{\vec\sigma}\newcdot}
({\boldsymbol{s^1}}\pm i{\boldsymbol{s^2}})$ serve as ladder operators
that connect the spinor wave functions $\chi\ls{1/2}$ and $\chi\ls{-1/2}$.
Using \eq{spinornorm}, it follows that \eq{tauchi} is equivalent to:
\beq \label{tautheorem}
\chi^\dagger_s(\boldsymbol{\hat s})\,
{\boldsymbol{\vec\sigma\newcdot{\hat s}^a}}\chi_{s'}(\boldsymbol{\hat s})=
\tau^a_{ss'}\,.
\eeq

It is instructive to prove \eq{tautheorem} directly.
Employing \eq{chidefinition} and using the fact that $\mathcal{D}$ is a
unitary matrix,
\beq
\chi^\dagger_s({\boldsymbol{\hat s}})\,
{\boldsymbol{\vec\sigma\newcdot{\hat s}^a}}\chi_{s'}({\boldsymbol{\hat s}})=
\chi^\dagger_s({\boldsymbol{\hat z}})\,
[\mathcal{D}(\phi\,,\theta\,,\,\gamma)]^{-1}
{\boldsymbol{\vec\sigma\newcdot{\hat s}^a}}\,
\mathcal{D}(\phi\,,\theta\,,\,\gamma)
\chi_{s'}({\boldsymbol{\hat z}}) \,.
\eeq
The above result can be simplified by a repeated use of the following identity,
\beq
e^{i\theta
\boldsymbol{\hat n\newcdot\vec\sigma}/2}\,\sigma^{j}\,
e^{-i\theta
\boldsymbol{\hat n\newcdot\vec\sigma}/2}
=R^{jk}({\boldsymbol{\hat n}}\,,\,\theta)\sigma^{k}\,,
\eeq
which is valid for any fixed axis $\boldsymbol{\hat n}$,
where $R({\boldsymbol{\hat n}}\,,\,\theta)$ is the rotation matrix
defined in \eq{rij}.  It follows that
\beq \label{dsigd}
[\mathcal{D}(\phi\,,\theta\,,\,\gamma)]^{-1}\,\sigma^{j}\,
\mathcal{D}(\phi\,,\theta\,,\,\gamma)
=\mathcal{R}^{jk}(\phi\,,\theta\,,\,\gamma)\,\sigma^{k}\,,
\eeq
where $\mathcal{R}(\phi\,,\theta\,,\,\gamma)$ is defined in \eq{calrdef}.
Since $\mathcal{R}^{\T}=\mathcal{R}^{-1}$,
\beq
\chi^\dagger_s({\boldsymbol{\hat s}})\,
{\boldsymbol{\vec\sigma\newcdot{\hat s}^a}}\chi_{s'}({\boldsymbol{\hat s}})=
\chi^\dagger_s({\boldsymbol{\hat z}})\,
{\boldsymbol{\vec\sigma\newcdot}} \left[
\mathcal{R}^{-1}{\boldsymbol{{\hat s}^a}}\right]\,
\chi_{s'}({\boldsymbol{\hat z}}) \,.
\eeq
\Eq{saexplicit} implies that $(\mathcal{R}^{-1}{\boldsymbol{{\hat s}^1}}\,,\,
\mathcal{R}^{-1}{\boldsymbol{{\hat s}^2}}\,,\,
\mathcal{R}^{-1}{\boldsymbol{{\hat s}^3}})
=({\boldsymbol{\hat x}}\,,
\,{\boldsymbol{\hat y}}\,,\,{\boldsymbol{\hat z}})$, and it follows that
\beq
{\boldsymbol{\vec\sigma\newcdot}} \left[
\mathcal{R}^{-1}{\boldsymbol{{\hat s}^a}}\right]=
\sigma^{a}\,.
\eeq
Consequently, we end up with
\beq \label{tautheorem2}
\chi^\dagger_s({\boldsymbol{\hat s}})\,
{\boldsymbol{\vec\sigma\newcdot{\hat s}^a}}\chi_{s'}({\boldsymbol{\hat s}})=
\chi^\dagger_s({\boldsymbol{\hat z}})\sigma^{a}
\chi_{s'}({\boldsymbol{\hat z}})\equiv\tau^a_{ss'}\,,
\eeq
which defines the matrix elements of the Pauli matrices, and our proof
of \eq{tautheorem} is complete.

Using the completeness relation given by \eq{scomplete},
we can rewrite \eq{tautheorem} as
\beq \label{tautheorem3}
\chi^\dagger_s(\boldsymbol{\hat s})\, \sigma^i\chi_{s'}(\boldsymbol{\hat s})=
\tau^a_{ss'}{\boldsymbol{{\hat s}^{a}}^i}\,.
\eeq
Taking the hermitian conjugate of \eq{tautheorem3} is
equivalent to interchanging $s\leftrightarrow s'$, since the $\sigma^i$ are
hermitian matrices and $(\tau^a_{ss'})^*=\tau^a_{s's}$.
To evaluate expressions similar to \eq{tautheorem3} that contain products of
$\sigma$-matrices, it is sufficient to use the relation
$\sigma^i\sigma^j=\delta^{ij}\mathds{1}+i\epsilon^{ijk}\sigma^k$
as many times as
needed to reduce the final expression to terms containing at most one
$\sigma$-matrix.  For example, using \eqs{spinornorm}{tautheorem3}, it
follows that
\beq \label{tautheorem4}
\chi^\dagger_s(\boldsymbol{\hat s})\,
\sigma^i\sigma^j\chi_{s'}(\boldsymbol{\hat s})=\delta_{ss'}\delta^{ij}
+i\epsilon^{ijk}\tau^a_{ss'}\boldsymbol{{\hat s}^{a}}^k\,.
\eeq

It is sometimes useful to have a more explicit representation of
the ${\boldsymbol{{\hat s}^a}}$.  In the
convention where $\gamma=-\phi$,
\eq{saexplicit} yields:
\beqa
{\boldsymbol{\hat s^{1}}} &=& (1-2\cos^2\phi\,\sin^2 \halftheta\,,\,
-\sin 2\phi\,\sin^2\halftheta\,,\,-\sin\theta\cos\phi)\,,
\nonumber \\[5pt]
{\boldsymbol{\hat s^{2}}}  &=& (-\sin 2\phi\,\sin^2\halftheta\,,\,
1-2\sin^2\phi\,\sin^2\halftheta\,,\,-\sin\theta\sin\phi)\,,
\nonumber \\[5pt]
{\boldsymbol{\hat s^{3}}} &=& (\sin\theta\cos\phi\,,\,
\sin\theta\sin\phi\,,\,\cos\theta)\,.\label{s123}
\eeqa
The explicit forms for the ${\boldsymbol{\hat s^a}}$ are somewhat
simpler in the convention where $\gamma=0$.
In this case, \eqs{rotationmat}{saexplicit} yield:
\beqa
{\boldsymbol{\hat s^{1}}} &=& (\cos\theta\cos\phi,\,
\cos\theta\sin\phi,\,-\sin\theta)\,,
\nonumber \\[5pt]
{\boldsymbol{\hat s^{2}}}  &=& (-\sin\phi,\,\cos\phi,\,0)\,,
\nonumber \\[5pt]
{\boldsymbol{\hat s^{3}}} &=& (\sin\theta\cos\phi,\,
\sin\theta\sin\phi,\,\cos\theta)\,.\label{s123p}
\eeqa

\subsection{Fixed-axis spinors in the non-relativistic limit}
\renewcommand{\theequation}{C.2.\arabic{equation}}
\renewcommand{\thefigure}{C.2.\arabic{figure}}
\renewcommand{\thetable}{C.2.\arabic{table}}
\setcounter{equation}{0}
\setcounter{figure}{0}
\setcounter{table}{0}

Consider an on-shell
massive fermion of three-momentum~$\boldsymbol{\vec p}$, mass~$m$ and
spin quantum number~$s$, where $s=\pm\half$ are the
possible projections of
the spin vector (in units of $\hbar$)
along the fixed $\boldsymbol{\hat s}$
direction [cf.~\eq{app:chiess}].   The spinor
wave functions, $x$, $y$, and their hermitian conjugates are given by
\eqst{explicitxa}{explicityb}.  In the non-relativistic limit,
\beqa
\sqrt{p\cdot\sigma} & \simeq &
\sqrt{m}\left(\mathds{1}-
\frac{\boldsymbol{\vec{\sigma}\newcdot\vec{p}}}{2m}\right)\,,\\[6pt]
\sqrt{p\cdot\sigmabar}&\simeq&
\sqrt{m}\left(\mathds{1}
+\frac{\boldsymbol{\vec{\sigma}\newcdot\vec{p}}}{2m}\right)\,,
\eeqa
where we keep terms only up to $\mathcal{O}(|\boldsymbol{\vec p}|/m)$.
Inserting the above results into \eqst{explicitxa}{explicityb} yields:
\beqa
x_\alpha(\vec{p},s)&\simeq&
\sqrt{m}\left(\mathds{1}-\frac{\boldsymbol{\vec{\sigma}\newcdot\vec{p}}}
{2m}\right)\chi_s(\boldsymbol{\hat s})\,,\label{fa1}\\
x^\alpha(\vec{p},s)&\simeq&
-2s\sqrt{m}\,\chi_{-s}^\dagger(\boldsymbol{\hat s})\left(\mathds{1}+
\frac{\boldsymbol{\vec{\sigma}\newcdot\vec{p}}}{2m}\right)\,,\\
y_\alpha(\vec{p},s)&\simeq&
2s\sqrt{m}\left(\mathds{1}-
\frac{\boldsymbol{\vec{\sigma}\newcdot\vec{p}}}{2m}\right)
\chi_{-s}(\boldsymbol{\hat s})\,,\\
y^\alpha(\vec{p},s)&\simeq&
\sqrt{m}\,\chi_s^\dagger(\boldsymbol{\hat s})\left(\mathds{1}+
\frac{\boldsymbol{\vec{\sigma}\newcdot\vec{p}}}{2m}\right)\,,
\eeqa
for the undotted spinor wave functions and
\beqa
x^{\dagger\dot{\alpha}}(\vec{p},s)&\simeq& -2s\sqrt{m}\,\left(\mathds{1}
+\frac{\boldsymbol{\vec{\sigma}\newcdot\vec{p}}}{2m}\right)
\chi_{-s}(\boldsymbol{\hat s})\,,\\
x^\dagger_{\dot{\alpha}}(\vec{p},s)&\simeq& \sqrt{m}\,
\chi_s^\dagger(\boldsymbol{\hat s})\left(\mathds{1}
-\frac{\boldsymbol{\vec{\sigma}\newcdot\vec{p}}}{2m}\right)\,,\\
y^{\dagger\dot{\alpha}}(\vec{p},s)&\simeq&\sqrt{m}\left(\mathds{1}+
\frac{\boldsymbol{\vec{\sigma}\newcdot\vec{p}}}{2m}\right)
\chi_s(\boldsymbol{\hat s})\,,\\
y^\dagger_{\dot{\alpha}}(\vec{p},s)&\simeq&
2s\sqrt{m}\,\chi_{-s}^\dagger(\boldsymbol{\hat s})\left(\mathds{1}
-\frac{\boldsymbol{\vec{\sigma}\newcdot\vec{p}}}{2m}\right)\,,\label{fa8}
\eeqa
for the dotted spinor wave functions.

In the computation of the $S$-matrix amplitudes for
scattering and decay processes, one typically
must evaluate a bilinear product of spinors, i.e.~quantities of the form
\beq \label{zzGamma}
z_1(\boldsymbol{\vec{p}}_1,s_1)\,\Gamma\,z_2(\boldsymbol{\vec{p}}_2,s_2)\,,
\eeq
where $z_1$ and $z_2$ represent one of the two-component spinor
wave functions $x$, $y$, $x^\dagger$ or
$y^\dagger$, and $\Gamma$ is a $2\times 2$ matrix (in spinor
space) that is either the identity matrix, or is made up of
alternating products of $\sigma$ and $\sigmabar$.
In the non-relativistic limit, these bilinears take on rather simple
forms.  In what follows, we
work to first order in $|\boldsymbol{\vec{p}}_i|/m_i$.
For example,
\beqa
y^\alpha(\boldsymbol{\vec{p}}_1,s_1)
x_\alpha(\boldsymbol{\vec{p}}_2,s_2)&\simeq&
\sqrt{m_1m_2}\;\chi^\dagger_{s_1}(\boldsymbol{\hat s})\left(\mathds{1}+
\frac{\boldsymbol{\vec{\sigma}\newcdot\vec{p}}}{2m_1}
-\frac{\boldsymbol{\vec{\sigma}\newcdot\vec{p}}}{2m_2}
\right)\chi_{s_2}(\boldsymbol{\hat s}) \nonumber \\
&\simeq& \sqrt{m_1m_2}
\left[\delta_{s_1,s_2}+\left(\frac{\boldsymbol{\vec{p}}_1
}{2m_1} -\frac{\boldsymbol{\vec{p}}_2}{2m_2}
\right)\boldsymbol{\newcdot\hat{s}^a}\tau^a_{s_1,s_2} \right]\,,
\eeqa
where we have used the results of \eqs{spinornorm}{tautheorem3}.  Similarly,
\beqa
\hspace{-0.5in}
y^\alpha(\boldsymbol{p_1},s_1) \sigma^\mu_{\alpha\dot{\beta}}
y^{\dagger\dot{\beta}}(\boldsymbol{p_2},s_2)
&\simeq& \sqrt{m_1m_2}\,\chi^\dagger_{s_1}(\boldsymbol{\hat s})\left[
\sigma^\mu + \frac{\boldsymbol{\vec{\sigma}\newcdot\vec{p}_1}}{2m_1}
\sigma^\mu +\sigma^\mu \frac{\boldsymbol{\vec{\sigma}\newcdot\vec{p}_2}}{2m_2}
\right]\chi_{s_2}(\boldsymbol{\hat s}) \nonumber \\[8pt]
&\simeq& \sqrt{m_1m_2}\,
Z^\mu_{s_1,s_2}(\boldsymbol{\vec p_1},\boldsymbol{\vec p_2})\,,
\eeqa
where\footnote{We also define $Z^\mu_{s's}
(\boldsymbol{\vec p_2},\boldsymbol{\vec p_1})$ as the expression
given by \eq{Zssdef} with the interchange of
$\{s\,,\,\boldsymbol{\vec p_1}\,,\,m_1\}$ and
$\{s'\,,\,\boldsymbol{\vec p_2}\,,\,m_2\}$\,.\label{fnchange}}
\beq \label{Zssdef}
Z^\mu_{ss'}(\boldsymbol{\vec p_1},\boldsymbol{\vec p_2})
\equiv \begin{cases} \displaystyle \delta_{ss'}
+\left(\frac{\boldsymbol{\vec{p}}_1}{2m_1}
+\frac{\boldsymbol{\vec{p_2}}}{2m_2}\right)\boldsymbol{\newcdot}
\boldsymbol{\hat{s}^a}\tau^a_{ss'}
\,, & \text{for}~\mu=0\,,\\[8pt]
\displaystyle \boldsymbol{\hat{s}^a}^i\tau^a_{ss'}
+\left(\frac{p_1^i}{2m_1}
+\frac{p_2^i}{2m_2}\right)\delta_{ss'}
+\left(\frac{p_2^j}{2m_2}
-\frac{p_1^j}{2m_1}\right)i\epsilon^{ijk}\boldsymbol{\hat{s}^a}^k
\tau_{ss'}^a\,, & \text{for}~\mu=i=1,2,3\,,\end{cases}
\eeq
is obtained after using the results of \eqs{tautheorem3}{tautheorem4}.

In summary, we list the non-relativistic forms of the spinor
bilinears. Referring to \eq{zzGamma}, 
if $\Gamma=\mathds{1}$, then
\beqa
x^\alpha(\boldsymbol{\vec{p}}_1,s_1)x_\alpha(\boldsymbol{\vec{p}}_2,s_2)
&\simeq& 2s_2\sqrt{m_1m_2}
\left[\delta_{-s_2,s_1}+\left(\frac{\boldsymbol{\vec{p}}_1
}{2m_1} -\frac{\boldsymbol{\vec{p}}_2}{2m_2}
\right)\boldsymbol{\newcdot\hat{s}^a}\tau^a_{-s_2,s_1} \right],
\label{nrxx}\\
y^\alpha(\boldsymbol{\vec{p}}_1,s_1)
y_\alpha(\boldsymbol{\vec{p}}_2,s_2)&\simeq&
2s_2\sqrt{m_1m_2}
\left[\delta_{s_1,-s_2}+\left(\frac{\boldsymbol{\vec{p}}_1
}{2m_1} -\frac{\boldsymbol{\vec{p}}_2}{2m_2}
\right)\boldsymbol{\newcdot\hat{s}^a}\tau^a_{s_1,-s_2} \right],
\label{nryy}\\
x^\alpha(\boldsymbol{\vec{p}}_1,s_1)
y_\alpha(\boldsymbol{\vec{p}}_2,s_2)&\simeq&
\sqrt{m_1m_2}
\left[-\delta_{s_2,s_1}+\left(\frac{\boldsymbol{\vec{p}}_1
}{2m_1} -\frac{\boldsymbol{\vec{p}}_2}{2m_2}
\right)\boldsymbol{\newcdot\hat{s}^a}\tau^a_{s_2,s_1} \right],
\label{nrxy}\\
y^\alpha(\boldsymbol{\vec{p}}_1,s_1)
x_\alpha(\boldsymbol{\vec{p}}_2,s_2)&\simeq&
\sqrt{m_1m_2}\left[\delta_{s_1,s_2}+\left(\frac{\boldsymbol{\vec{p}}_1
}{2m_1} -\frac{\boldsymbol{\vec{p}}_2}{2m_2}
\right)\boldsymbol{\newcdot\hat{s}^a}\tau^a_{s_1,s_2} \right],
\label{nryx}
\eeqa
where we have used
\beq \label{tauid}
\tau^a_{s's}=-4ss'\tau^a_{-s,-s'}\,,\qquad\qquad s,s'=\pm\half\,,
\eeq
to arrive at the final forms given in \eqs{nrxx}{nrxy}.
However, in using the above results, one must now pay close attention
to the ordering of the subscript indices of the $\tau^a$.
The corresponding formulae for dotted spinor wave function bilinears
are obtained by taking the hermitian conjugates of
\eqst{nrxx}{nryx}, which complex-conjugates the $\tau^a$ that appear
on the right-hand side of these equations.  Since $(\tau^a_{ss'})^*=
\tau^a_{s's}$, we obtain
\beqa
x^\dagger_{\dot{\alpha}}(\boldsymbol{\vec{p}}_1,s_1)
x^{\dagger\dot{\alpha}}(\boldsymbol{\vec{p}}_2,s_2)
&\simeq& 2s_1\sqrt{m_1m_2}
\left[\delta_{s_2,-s_1}-\left(\frac{\boldsymbol{\vec{p}}_1
}{2m_1} -\frac{\boldsymbol{\vec{p}}_2}{2m_2}
\right)\boldsymbol{\newcdot\hat{s}^a}\tau^a_{s_2,-s_1} \right],
\label{nrxxb}\\
y^\dagger_{\dot\alpha}(\boldsymbol{\vec{p}}_1,s_1)
y^{\dagger\dot{\alpha}}(\boldsymbol{\vec{p}}_2,s_2)&\simeq&
2s_1\sqrt{m_1m_2}
\left[\delta_{-s_1,s_2}-\left(\frac{\boldsymbol{\vec{p}}_1
}{2m_1} -\frac{\boldsymbol{\vec{p}}_2}{2m_2}
\right)\boldsymbol{\newcdot\hat{s}^a}\tau^a_{-s_1,s_2} \right],
\label{nryyb}\\
y^\dagger_{\dot{\alpha}}(\boldsymbol{\vec{p}}_1,s_1)
x^{\dagger\dot{\alpha}}(\boldsymbol{\vec{p}}_2,s_2)&\simeq&
-\sqrt{m_1m_2}
\left[\delta_{s_2,s_1}+\left(\frac{\boldsymbol{\vec{p}}_1
}{2m_1} -\frac{\boldsymbol{\vec{p}}_2}{2m_2}
\right)\boldsymbol{\newcdot\hat{s}^a}\tau^a_{s_2,s_1} \right],
\label{nryxb}\\
x^\dagger_{\dot{\alpha}}(\boldsymbol{\vec{p}}_1,s_1)
y^{\dagger\dot{\alpha}}(\boldsymbol{\vec{p}}_2,s_2)&\simeq&
\sqrt{m_1m_2}\left[\delta_{s_1,s_2}-\left(\frac{\boldsymbol{\vec{p}}_1
}{2m_1} -\frac{\boldsymbol{\vec{p}}_2}{2m_2}
\right)\boldsymbol{\newcdot\hat{s}^a}\tau^a_{s_1,s_2} \right].
\label{nrxyb}
\eeqa

Likewise, if $\Gamma=\sigma^\mu$, then
\beqa
x^\alpha(\boldsymbol{p_1},s_1) \sigma^\mu_{\alpha\dot{\beta}}
x^{\dagger\dot{\beta}}(\boldsymbol{p_2},s_2)
&\simeq& 4s_1s_2\sqrt{m_1m_2}
\,Z^\mu_{-s_1,-s_2}(\boldsymbol{\vec p_1},\boldsymbol{\vec p_2})
\,,\label{xxdnr}\\[6pt]
y^\alpha(\boldsymbol{p_1},s_1) \sigma^\mu_{\alpha\dot{\beta}}
y^{\dagger\dot{\beta}}(\boldsymbol{p_2},s_2)
&\simeq& \sqrt{m_1m_2}
\,Z^\mu_{s_1,s_2}(\boldsymbol{\vec p_1},\boldsymbol{\vec p_2})
\,,\label{yydnr}\\[6pt]
x^\alpha(\boldsymbol{p_1},s_1) \sigma^\mu_{\alpha\dot{\beta}}
y^{\dagger\dot{\beta}}(\boldsymbol{p_2},s_2)
&\simeq& -2s_1\sqrt{m_1m_2}
\,Z^\mu_{-s_1,s_2}(\boldsymbol{\vec p_1},\boldsymbol{\vec p_2})
\,,\label{xydnr}\\[6pt]
y^\alpha(\boldsymbol{p_1},s_1) \sigma^\mu_{\alpha\dot{\beta}}
x^{\dagger\dot{\beta}}(\boldsymbol{p_2},s_2)
&\simeq& -2s_2\sqrt{m_1m_2}
\,Z^\mu_{s_1,-s_2}(\boldsymbol{\vec p_1},\boldsymbol{\vec p_2})\,,\label{yxdnr}
\eeqa
where $Z^\mu_{ss'}(\boldsymbol{\vec p_1},\boldsymbol{\vec p_2})$
is defined in \eq{Zssdef}.  If
$\Gamma=\sigmabar^\mu$, one can use $z_1\sigma^\mu
z_2^\dagger=z_2^\dagger\sigmabar^\mu z_1$
[i.e.~\eq{europeanvacation} for commuting spinors]
to obtain the corresponding formulae for the spinor wave function
bilinears (cf.~footnote~\ref{fnchange}):
\beqa
x^\dagger_{\dot{\alpha}}(\boldsymbol{p_1},s_1) \sigmabar^{\mu\dot{\alpha}\beta}
x_\beta(\boldsymbol{p_2},s_2)
&\simeq& 4s_1s_2\sqrt{m_1m_2}
\,Z^\mu_{-s_2,-s_1}(\boldsymbol{\vec p_2},\boldsymbol{\vec p_1})
\,,\label{xxbdnr}\\[6pt]
y^\dagger_{\dot{\alpha}}(\boldsymbol{p_1},s_1) \sigmabar^{\mu\dot{\alpha}\beta}
y_\beta(\boldsymbol{p_2},s_2)
&\simeq &\sqrt{m_1m_2}
\,Z^\mu_{s_2,s_1}(\boldsymbol{\vec p_2},\boldsymbol{\vec p_1})\,,
\label{yybdnr} \\[6pt]
y^\dagger_{\dot{\alpha}}(\boldsymbol{p_1},s_1) \sigmabar^{\mu\dot{\alpha}\beta}
x_\beta(\boldsymbol{p_2},s_2)
&\simeq& -2s_2\sqrt{m_1m_2}
\,Z^\mu_{-s_2,s_1}(\boldsymbol{\vec p_2},\boldsymbol{\vec p_1})
\,,\label{xybdnr}\\[6pt]
x^\dagger_{\dot{\alpha}}(\boldsymbol{p_1},s_1) \sigmabar^{\mu\dot{\alpha}\beta}
y_\beta(\boldsymbol{p_2},s_2)
&\simeq& -2s_1\sqrt{m_1m_2}
\,Z^\mu_{s_2,-s_1}(\boldsymbol{\vec p_2},\boldsymbol{\vec p_1})\,.
\label{yxbdnr}
\eeqa
These results can also be derived directly from
\eqst{fa1}{fa8}, after employing \eq{tauid}.

It is straightforward to evaluate the spinor wave function bilinears when
$\Gamma$ is a product of two or more $\sigma/\sigmabar$ matrices.
As the corresponding expressions are considerably more complicated,
we shall not write them out explicitly here.

\subsection{Helicity spinor wave functions}
\renewcommand{\theequation}{C.3.\arabic{equation}}
\renewcommand{\thefigure}{C.3.\arabic{figure}}
\renewcommand{\thetable}{C.3.\arabic{table}}
\setcounter{equation}{0}
\setcounter{figure}{0}
\setcounter{table}{0}

All the results of \app{C.1} apply to the helicity spinors
$\chi\ls\lambda$, which are defined to be eigenstates
of $\half{\boldsymbol{\vec\sigma\newcdot\hat p}}$, i.e.,
\beq
\half {\boldsymbol{\vec\sigma\newcdot\hat p}}\,
\chi\ls{\lambda}({\boldsymbol{\hat p}}) =
\lambda\chi\ls{\lambda}({\boldsymbol{\hat p}})\,,
  \qquad \lambda = \pm\half\,,\label{chihelicity}
\eeq
where ${\boldsymbol{\hat p}}=(\sin\theta_p\cos\phi_p\,,\,\sin\theta_p\sin\phi_p
\,,\,\cos\theta_p)$.
It follows that:
\beq
\sqrt{\BDpos p\newcdot\sigma}\,\chi\ls{\lambda}({\boldsymbol{\hat p}})
=\omega\ls{-\lambda}(\boldsymbol{\vec p})
\,\chi\ls{\lambda}({\boldsymbol{\hat p}})\,,
\qquad\qquad
\sqrt{\BDpos p\newcdot\sigmabar}\,\chi\ls{\lambda}({\boldsymbol{\hat p}})
=\omega\ls{\lambda}(\boldsymbol{\vec p})
\,\chi\ls{\lambda}({\boldsymbol{\hat p}})\,,
\eeq
where $\omega\ls{\lambda}(\boldsymbol{\vec p})
\equiv (E+2\lambda|{\boldsymbol{\vec p}}|\,)^{1/2}$ and $E=\sqrt{|\boldsymbol{\vec p}|^2+m^2}$.
As a result, the explicit forms
for the two-component helicity spinor wave functions
[cf.~\eqst{explicitxa}{explicityb}] simplify:
\beqa
x_\alpha(\boldsymbol{\vec p},\lambda)
&=&\omega\ls{-\lambda}
\,\chi\ls{\lambda}({\boldsymbol{\hat p}})\,,
\qquad\qquad\quad\qquad\quad\,
x^\alpha(\boldsymbol{\vec p},\lambda)
=-2\lambda\,\omega\ls{-\lambda}\,
\chi^\dagger\ls{-\lambda}({\boldsymbol{\hat p}})\,,
\label{explicithelxa} \\
y_\alpha(\boldsymbol{\vec p},\lambda)&=&2\lambda\,
\omega\ls{\lambda}
\,\chi\ls{-\lambda}({\boldsymbol{\hat p}})
\,,\qquad\quad\qquad\qquad
y^\alpha(\boldsymbol{\vec p},\lambda)=
\omega\ls{\lambda}
\,\chi^\dagger\ls{\lambda}({\boldsymbol{\hat p}})\,,
\label{explicithelya} \\
 x^{\dagger\dot{\alpha}}(\boldsymbol{\vec p},\lambda)&=&
-2\lambda\,\omega\ls{-\lambda}
\,\chi\ls{-\lambda}({\boldsymbol{\hat p}})
\,,\qquad\qquad\qquad\!\!
 x^\dagger_{\dot{\alpha}}(\boldsymbol{\vec p},\lambda)
=\omega\ls{-\lambda}
\,\chi^\dagger\ls{\lambda}({\boldsymbol{\hat p}})\,,\\
\label{explicithelxb}
 y^{\dagger\dot{\alpha}}(\boldsymbol{\vec p},\lambda)&=&
\omega\ls{\lambda}
\,\chi\ls{\lambda}({\boldsymbol{\hat p}})
\,,\qquad\qquad\qquad\qquad\quad\!
 y^\dagger_{\dot{\alpha}}(\boldsymbol{\vec p},\lambda)
=2\lambda\,\omega_{\lambda}\,
\chi^\dagger\ls{-\lambda}({\boldsymbol{\hat p}})\,,
\label{explicithelyb}
\eeqa
where $\omega\ls{\pm\lambda}=\omega\ls{\pm\lambda}(\boldsymbol{\vec p})$.
\clearpage

In analogy with the ${\boldsymbol{{\hat s}^a}}$,
it is convenient to introduce an orthonormal set of unit three-vectors
${\boldsymbol{{\hat p}^a}}$ such that
${\boldsymbol{{\hat p}^3}}={\boldsymbol{\hat p}}$.
Then,
\eqst{hatsdef1}{tautheorem} apply as well to the two-component
helicity spinors after taking
${\boldsymbol{{\hat s}^a}}={\boldsymbol{{\hat p}^a}}$.

In scattering processes, it is often convenient to work in the rest
frame of the incoming particles, in which the
corresponding incoming
fermion three-momenta are denoted by ${\boldsymbol{\vec p}}$ and
$-{\boldsymbol{\vec p}}$, respectively.  The helicity
spinor wave function of the second fermion depends on the definition of
$\chi\ls{\lambda}(-{\boldsymbol{\hat p}})$.  In this
review, we follow a convention\footnote{An alternative convention
(called the \textit{second-particle convention})
advocated by Jacob and Wick~\cite{jacobwick}
is to define $\chi\ls{\lambda}(-\boldsymbol{\hat p})$ by starting with
$\chi\ls{-\lambda}(\boldsymbol{\hat z})$ and then rotating the spinor
by polar and azimuthal angles $\theta_p$ and $\phi_p$.  In this case,
$\chi\ls{\lambda}(-\boldsymbol{\hat p})
=\chi\ls{-\lambda}(\boldsymbol{\hat p})$,
and the extra phase factors of \eq{cphase} is absent, i.e.
$\xi\ls{\lambda}({\boldsymbol{\hat p}})=1$ in \eq{cphase}.
However, this convention is less suited to
scattering processes involving final states with more than two
fermions.  Hence, we do not adopt the second-particle
convention in this review.}
in which $\chi\ls{\lambda}
(-\boldsymbol{\hat p})$ is defined to be the spinor wave function
obtained from $\chi\ls{\lambda}(\boldsymbol {\hat
z})$ via a rotation by a polar angle $\pi-\theta_p$ and an azimuthal angle
$\phi_p+\pi$ with respect to the ${\boldsymbol{\hat z}}$-direction.  Then,
\beq \label{chidchi}
\chi\ls{\lambda}(-{\boldsymbol{\hat p}})=
\mathcal{D}(\phi_p+\pi\,,\,\pi-\theta_p\,,\,\gamma(-\boldsymbol{\hat p}))
\,\chi\ls{\lambda}({\boldsymbol{\hat z}})\,,
\eeq
where we have exhibited the possible dependence of $\gamma$ on the
direction $-\boldsymbol{\hat p}$.  Using the properties of the
spin-1/2 rotation matrices, one can derive
\beq \label{dgamgam}
\mathcal{D}(\phi_p+\pi\,,\,\pi-\theta_p\,,\,\gamma(-\boldsymbol{\hat p}))=
-\mathcal{D}(\phi_p\,,\,\theta_p\,,\,\gamma(\boldsymbol{\hat p}))\,
D(\boldsymbol{\hat z},-\gamma(\boldsymbol{\hat p})
-\gamma(-\boldsymbol{\hat p}))\,D(\boldsymbol{\hat x},\pi)\,.
\eeq
Inserting this result in \eq{chidchi} and using the relation
\beq
D(\boldsymbol{\hat x},\pi)\chi\ls{\lambda}({\boldsymbol{\hat z}})
=-i\sigma^{1} \chi\ls{\lambda}({\boldsymbol{\hat z}})
=-i\chi\ls{-\lambda}({\boldsymbol{\hat z}})\,,
\eeq
we obtain
\beq
\chi\ls{\lambda}(-{\boldsymbol{\hat p}})= \xi_{-\lambda}({\boldsymbol{\hat p}})
\chi\ls{-\lambda}({\boldsymbol{\hat p}})\,,
\label{cphase}
\eeq
where the phase factor $\xi_{\lambda}({\boldsymbol{\hat p}})$
is given by
\beq \label{xip}
\xi_\lambda(\boldsymbol{\hat p}) =ie^{i\lambda[\gamma(\boldsymbol{\hat p})
+\gamma(-\boldsymbol{\hat p})]}\,,\qquad\quad \lambda=\pm\half\,.
\eeq
Since $\gamma$ is a real angle, it follows that:
\beq \label{xistarprops}
\xi\ls{\lambda}^*({\boldsymbol{\hat p}})=
\frac{1}{\xi\ls{\lambda}({\boldsymbol{\hat p}})}=-
\xi\ls{-\lambda}({\boldsymbol{\hat p}})\,.
\eeq
Using \eq{xip}, we note that
$\chi\ls{\lambda}(\boldsymbol{\hat p})$ possesses the
peculiar property that:
\beq \label{doublep}
\chi\ls{\lambda}(-(-\boldsymbol{\hat p}))=-\chi\ls{\lambda}
(\boldsymbol{\hat p})\,.
\eeq
This is a consequence of the fact that the result of
two successive inversions is
equivalent to $\phi_p\to\phi_p+2\pi$, which yields an overall change
of sign of a spinor wave function
[cf.~\eq{twoone}].\footnote{A slightly modified procedure
(not adopted in this review) is
to take the azimuthal angle of $-\boldsymbol{\hat p}$ to be
$\phi_p\pm\pi$, where the $\pm$ sign is chosen according to which of
the two conditions $0\leq\phi_p\pm\pi<2\pi$ is true.  This procedure
would yield an extra minus sign in the definition of
$\xi\ls{\lambda}(\boldsymbol{\hat p})$ when $\pi\leq\phi_p<2\pi$.
In this convention, two successive inversions are equivalent to the identity
rotation so that $\chi\ls{\lambda}(-(-\boldsymbol{\hat p}))=\chi\ls{\lambda}
(\boldsymbol{\hat p})$.}

For example, corresponding to the two conventional choices
for $\gamma$,
\beq \label{xphase}
\xi_\lambda(\boldsymbol{\hat p}) =\begin{cases}
(-1)\rrsup{{\frac{1}{2}}-\lambda}\,
  e^{-2i\lambda\phi_p} &
  \textrm{for}~~\gamma(\boldsymbol{\hat p})=-\phi_p\,, \quad
\gamma(-\boldsymbol{\hat p})=-\phi_p-\pi\,,
\\[6pt]
  \qquad i & \textrm{for}~~\gamma(\boldsymbol{\hat p})=
\gamma(-\boldsymbol{\hat p})=0\,,
\end{cases}
\eeq
with the proviso that for
${\boldsymbol{\hat p}}=\pm {\boldsymbol{\hat z}}$, we define $\phi_p$
according to \eq{negativez}.

Suppose that the two fermions considered above have equal mass.
In the center-of-mass frame, if the four-momentum of one of the
fermions is $p^\mu=(E\,;\,{\boldsymbol{\vec p}})$, then the
four-momentum of the other fermion is
${\widetilde p}^{\,\mu}\equiv (E\,;\,-\boldsymbol{\vec p})$.
The following \textit{numerical} identities are then satisfied:
$\sigma\newcdot\widetilde p=\sigmabar\newcdot p$ and
$\sigmabar\newcdot\widetilde p=\sigma\newcdot p$.  However, in order to
maintain covariance with respect to the undotted and dotted spinor
indices, we shall write these identities as:
\beq \label{sigpbar12}
\widetilde p\newcdot\sigma_{\alpha\dot{\beta}}=\sigma^0_{\alpha\dot{\alpha}}
(p\newcdot\sigmabar^{\dot{\alpha}\beta})\,
\sigma^0_{\beta\dot{\beta}}\,,\qquad\quad
\widetilde p\newcdot\sigmabar^{\dot{\alpha}\beta}=\sigmabar^{0\dot{\alpha}\alpha}
(p\newcdot\sigma_{\alpha\dot{\beta}})\,\sigmabar^{0\dot{\beta}\beta}\,.
\eeq
Taking the matrix square root of both sides of the equations above
removes one of the factors of $\sigma^0$ and $\sigmabar^0$,
respectively [cf.~\eqst{sqpsigma1}{squaring2}].
Thus, using \eqs{explicitxa}{cphase},
\beq
x_\alpha(-{\boldsymbol{\vec p}},-\lambda)=
\sqrt{\BDpos \widetilde p\newcdot\sigma}
\chi\ls{-\lambda}(-{\boldsymbol{\hat p}})
=
\sigma^0\sqrt{\BDpos p\newcdot\sigmabar}\,
\xi\ls{\lambda}(\boldsymbol{\hat p})\,\chi\ls{\lambda}({\boldsymbol{\vec p}})=
\sigma^0_{\alpha\dot{\beta}}\,\xi\ls{\lambda}(\boldsymbol{\hat p})\,
{y}^{\dagger\dot{\beta}}({\boldsymbol{\vec p}},\lambda)\,.
\eeq

In this way, we can
derive all relations of this kind for the helicity spinor wave functions:
\beqa
x_\alpha(-{\boldsymbol{\vec p}},-\lambda)&=&
\xi\ls{\lambda}\sigma^0_{\alpha\dot{\beta}}\,
{ y}^{\dagger\dot{\beta}}({\boldsymbol{\vec p}},\lambda)
=\omega\ls{\lambda}\xi\ls{\lambda}
\,\chi\ls{\lambda}({\boldsymbol{\hat p}})\,,
\label{pminus1}\\
\hspace{-0.3in}
y_\alpha(-{\boldsymbol{\vec p}},-\lambda)&=&
\xi\ls{-\lambda}\sigma^0_{\alpha\dot{\beta}}\,
{ x}^{\dagger\dot{\beta}}({\boldsymbol{\vec p}},\lambda)
=-2\lambda\,\omega\ls{-\lambda}\xi\ls{-\lambda}\,
\chi\ls{-\lambda}({\boldsymbol{\hat p}})
\,,
\label{pminus2}\\
\hspace{-0.3in}
{x}^{\dagger\dot{\alpha}}(-{\boldsymbol{\vec p}},-\lambda)&=&
\xi\ls{-\lambda}\sigmabar^{0\dot{\alpha}\beta}\,
y_\beta({\boldsymbol{\vec p}},\lambda)
=2\lambda\,\omega\ls{\lambda}\xi\ls{-\lambda}\,
\chi\ls{-\lambda}({\boldsymbol{\hat p}})\,,
\label{pmisu3}\\
\hspace{-0.3in}
{y}^{\dagger\dot{\alpha}}(-{\boldsymbol{\vec p}},-\lambda)&=&
\xi\ls{\lambda}\sigmabar^{0\dot{\alpha}\beta}\,
x_\beta({\boldsymbol{\vec p}},\lambda)
=\omega\ls{-\lambda}\xi\ls{\lambda}\,
\chi\ls{\lambda}({\boldsymbol{\hat p}})\,,
\label{pminus4}
\eeqa
where $\omega\ls{\pm\lambda}\equiv\omega\ls{\pm\lambda}(\boldsymbol{\vec p})$
and $\xi\ls{\lambda}\equiv\xi\ls{\lambda}(\boldsymbol{\hat p})$.  Raising [lowering]
the undotted [dotted] index 
yields:
\beqa
x^\alpha(-{\boldsymbol{\vec p}},-\lambda)&=&
-{ y}^\dagger_{\dot{\beta}}({\boldsymbol{\vec p}},\lambda)\,
\xi\ls{\lambda}\sigmabar^{0\dot{\beta}\alpha}
=-2\lambda\,\omega\ls{\lambda}\xi\ls{\lambda}
\chi^\dagger\ls{-\lambda}({\boldsymbol{\hat p}})\,,\label{pmunus5}\\
y^\alpha(-{\boldsymbol{\vec p}},-\lambda)&=&
-{ x}^\dagger_{\dot{\beta}}({\boldsymbol{\vec p}},\lambda)\,
\xi\ls{-\lambda}\sigmabar^{0\dot{\beta}\alpha}
=-\omega\ls{-\lambda}\xi\ls{-\lambda}\,
\chi^\dagger\ls{\lambda}({\boldsymbol{\hat p}})\,,\label{pminus6} \\
{x}^\dagger_{\dot{\alpha}}(-{\boldsymbol{\vec p}},-\lambda)&=&
-y^\beta({\boldsymbol{\vec p}},\lambda)\,
\xi\ls{-\lambda}\sigma^0_{\beta\dot{\alpha}}
=-\omega\ls{\lambda}\xi\ls{-\lambda}\,
\chi^\dagger\ls{\lambda}({\boldsymbol{\hat p}})\,,\label{pminus7} \\
\quad\qquad
{y}^\dagger_{\dot{\alpha}}(-{\boldsymbol{\vec p}},-\lambda)&=&
-x^\beta({\boldsymbol{\vec p}},\lambda)\,
\xi\ls{\lambda}\sigma^0_{\beta\dot{\alpha}}
=2\lambda\,\omega\ls{-\lambda}\xi\ls{\lambda}\,
\chi^\dagger\ls{-\lambda}({\boldsymbol{\hat p}})\,.\label{pminus8}
\eeqa
\Eqst{pminus1}{pminus8} can also be obtained directly from
\eqst{explicithelxa}{explicithelyb}.

\section{\texorpdfstring{Matrix decompositions for mass matrix diagonalization}{Matrix decompositions for mass matrix diagonalization}}
\label{appendix:D}
\renewcommand{\theequation}{D.\arabic{equation}}
\renewcommand{\thefigure}{D.\arabic{figure}}
\renewcommand{\thetable}{D.\arabic{table}}
\setcounter{equation}{0}
\setcounter{figure}{0}
\setcounter{table}{0}

In scalar field theory, the diagonalization of
the scalar squared-mass matrix $M^2$ is straightforward.
For a theory of $n$ complex scalar fields, $M^2$ is an hermitian
$n\times n$ matrix, which can be
diagonalized by a unitary matrix $W$:
\beq \label{app:wtrans}
W^\dagger M^2 W=m^2={\rm diag}(m_1^2,m_2^2,\ldots,m_n^2)\,.
\eeq
For a theory of $n$ real scalar fields, $M^2$
is a real symmetric $n\times n$
matrix, which can be diagonalized by an orthogonal
matrix $Q$:
\beq \label{app:qtrans}
Q^{\T}M^2Q=m^2={\rm diag}(m_1^2,m_2^2,\ldots,m_n^2)\,.
\eeq
In both cases, the eigenvalues $m_k^2$ of $M^2$ are real.  These are
the standard matrix diagonalization problems that are treated in all
elementary linear algebra textbooks.

In spin-1/2 fermion field theory,
the most general fermion mass matrix, obtained from the
Lagrangian, written in terms of two-component spinors, is complex
and symmetric [cf.~\sec{subsec:generalmass}].
If the Lagrangian exhibits a U(1) symmetry, then
a basis can be found such that fields that are charged under the U(1)
pair up into Dirac fermions.  The fermion mass matrix then decomposes
into the direct sum of a complex Dirac fermion mass matrix and a
complex symmetric neutral fermion mass matrix.
In this Appendix, we review the
linear algebra theory relevant for the matrix decompositions associated
with the general charged and neutral spin-1/2 fermion mass
matrix diagonalizations.  The diagonalization of
the Dirac fermion mass matrix is governed by the singular value
decomposition of a complex matrix, as shown in \app{D.1}.  In contrast, the
diagonalization of a neutral fermion mass matrix
is governed by the Takagi diagonalization of a complex symmetric
matrix, which is treated in \app{D.2}.\footnote{One may choose not to work in a
basis where the fermion fields are eigenstates of the U(1) charge
operator.  In this case, all fermions are governed by a complex
symmetric mass matrix, which can be Takagi-diagonalized according to the
procedure described in \app{D.2}.}  These two techniques are compared
and contrasted in \app{D.3}.
Dirac fermions can also arise in the
case of a pseudo-real representation of fermion fields.  As shown in
\sec{subsec:generalmass}, this latter case requires the reduction of a
complex antisymmetric fermion mass matrix to real normal form.  The
relevant theorem and its proof are given in \app{D.4}.

\subsection{Singular value decomposition}
\renewcommand{\theequation}{D.1.\arabic{equation}}
\renewcommand{\thefigure}{D.1.\arabic{figure}}
\renewcommand{\thetable}{D.1.\arabic{table}}
\setcounter{equation}{0}
\setcounter{figure}{0}
\setcounter{table}{0}

The diagonalization of the charged (Dirac) fermion mass matrix
requires the singular value decomposition of an arbitrary
complex matrix $M$.

{\bf Theorem:}
For any complex [or real] $n\times n$ matrix $M$,
unitary [or real orthogonal] matrices $L$ and $R$ exist such that
\beq \label{app:svd}
L^{\T} M R= M_D={\rm diag}(m_1,m_2,\ldots,m_n),
\eeq
where the $m_k$ are real and non-negative.  This is called the singular
value decomposition of the matrix~$M$~(e.g., see refs.~\cite{horn,horn2}).

In general, the $m_k$ are
\textit{not} the eigenvalues of $M$.  Rather, the $m_k$ are the
\textit{singular values}
of the general complex matrix $M$, which are
defined to be the non-negative square roots of the eigenvalues
of $M^\dagger M$ (or equivalently of $MM^\dagger$).
An equivalent definition of the singular values can be
established as follows.  Since $M^\dagger M$ is an hermitian
non-negative matrix, its eigenvalues are real and non-negative and its
eigenvectors, $v_k$, defined by $M^\dagger M v_k = m_k^2 v_k$,
can be chosen to be orthonormal.\footnote{We define
the inner product of two vectors
to be $\ip{v}{w}\equiv v^\dagger w$.  Then, $v$ and $w$ are
orthonormal if $\ip{v}{w}=0$.
The norm of a vector is
defined by $\|v\,\|=\ip{v}{v}^{1/2}$.}
Consider first the eigenvectors corresponding to the non-zero
eigenvalues of  $M^\dagger M$.  Then, we
define the vectors $w_k$ such that
\beq
M v_k= m_k w^*_k\,.
\eeq
It follows that $m_k^2 v_k=M^\dagger M v_k= m_k M^\dagger w^*_k$,
which yields: $M^\dagger w^*_k=m_k v_k$.  Note that these equations
also imply that $MM^\dagger w^*_k=m^2_k w^*_k$.
The orthonormality of the $v_k$ implies the
orthonormality of the~$w_k$, and vice versa.  For example,
\beq \label{dww}
\delta_{jk}=\ip{v_j}{v_k}=\frac{1}{m_j m_k}
\ip{M^\dagger w^*_j}{M^\dagger w^*_k}=\frac{1}{m_j m_k}
\ip{w_j}{MM^\dagger w^*_k}=\frac{m_k}{m_j}\ip{w^*_j}{w^*_k}\,,
\eeq
which yields $\ip{w_k}{w_j}=\delta_{jk}$.
If $M$ is a real matrix, then the eigenvectors $v_k$ can be chosen to
be real, in which case the corresponding $w_k$ are also real.

If $v_i$ is an eigenvector of $M^\dagger M$ with zero eigenvalue, then
$0=v_i^\dagger M^\dagger M v_i=\ip{ M v_i}{M v_i}$, which implies that
$M v_i=0$.
Likewise, if $w^*_i$ is an eigenvector of $MM^\dagger$ with zero
eigenvalue, then $0=w_i^{\T} M M^\dagger w^*_i=
\ip{ M^{\T} w_i}{M^{\T}
w_i}^*$, which implies that $M^{\T} w_i=0$. Because the eigenvectors of
$M^\dagger M$
[$MM^\dagger$] can be chosen orthonormal, the eigenvectors
corresponding to the zero eigenvalues of $M$ [$M^{\dagger}$] can be taken to
be orthonormal.\footnote{This analysis shows that the number of
linearly independent eigenvectors of $M^\dagger M$ [$MM^\dagger$]
with zero eigenvalue coincides with the number
of linearly independent eigenvectors of $M$ [$M^{\dagger}$] with zero
eigenvalue.}
Finally, these eigenvectors are also orthogonal
to the eigenvectors corresponding to the non-zero eigenvalues of
$M^\dagger M$ [$MM^\dagger$].  That is, if the indices $i$~and $j$
run over the
eigenvectors corresponding to the zero and non-zero eigenvalues of
$M^\dagger M$ [$MM^\dagger$],
respectively, then
\beq \label{ortho2}
\ip{v_j}{v_i}=\frac{1}{m_j}\ip{M^\dagger w^*_j}{v_i}
=\frac{1}{m_j}\ip{w^*_j}{Mv_i} =0\,,
\eeq
and similarly $\ip{w_j}{w_i}=0$.

Thus, we can define the singular values of a general complex
$n\times n$ matrix
$M$ to be the simultaneous solutions (with real non-negative $m_k$)
of:\footnote{One can always find a solution to \eq{singvals} such
that the $m_k$ are real and non-negative. Given a solution where $m_k$
is complex, we simply write $m_k=|m_k|e^{i\theta}$ and redefine $w_k
\to w_k e^{i\theta}$ to remove the phase $\theta$.}
\beq \label{singvals}
Mv_k=m_k w^*_k\,,\qquad\quad w_k^{\T} M =m_k v_k^\dagger\,.
\eeq
The corresponding $v_k$ ($w_k$), normalized to have unit norm, are
called the right (left) singular vectors of $M$.  In particular,
the number of linearly independent $v_k$
coincides with the number of linearly independent $w_k$ and is equal
to $n$.

{\textbf{Proof of the singular value decomposition theorem:}}
\Eqs{dww}{ortho2} imply
that the right [left] singular vectors can be chosen to be orthonormal.
Consequently, the unitary matrix $R$ [$L$] can be constructed such
that its $k$th column is given by the right [left] singular vector
$v_k$ [$w_k$].  It then follows from \eq{singvals} that:
\beq \label{svdpr}
w_k^{\T} M v_\ell=m_k\delta_{k\ell}\,,\qquad\qquad ({\rm no~sum~over}~k).
\eeq
In matrix form, \eq{svdpr} coincides with \eq{app:svd}, and the
singular value decomposition is established.  If $M$ is real, then
the right and left singular vectors, $v_k$ and $w_k$, can be chosen to
be real, in which case \eq{app:svd} holds for real orthogonal matrices
$L$ and $R$.

The singular values of a complex matrix $M$ are unique (up to
ordering), as they correspond to the eigenvalues of $M^\dagger M$ (or
equivalently the eigenvalues of $MM^\dagger$).  The unitary matrices
$L$ and $R$ are not unique.
The matrix $R$ must satisfy
\beq \label{rmatrix}
R^\dagger M^\dagger M R=M_D^2\,,
\eeq
which follows directly from \eq{app:svd} by computing
$M_D^\dagger M_D=M_D^2$.
That is, $R$ is a
unitary matrix that diagonalizes the non-negative
definite matrix $M^\dagger M$.  Since the eigenvectors of
$M^\dagger M$ are
orthonormal, each $v_k$ corresponding to a non-degenerate
eigenvalue of $M^\dagger M$ can be multiplied by an arbitrary phase
$e^{i\theta_k}$.  For the case of degenerate eigenvalues, any
orthonormal linear combination of the corresponding eigenvectors is
also an eigenvector of $M^\dagger M$. It follows that within the subspace
spanned by the eigenvectors corresponding to
non-degenerate eigenvalues, $R$ is uniquely determined up to
multiplication on the right by an arbitrary diagonal unitary matrix.
Within the subspace spanned by the eigenvectors of $M^\dagger M$
corresponding to a degenerate eigenvalue, $R$ is determined up
to multiplication on the right by an arbitrary unitary
matrix.

Once $R$ is fixed,
$L$ is obtained from
\eq{app:svd}:
\beq \label{lmtrmd}
L=(M^{\T})^{-1}R^* M_D\,.
\eeq
However, if some of the diagonal elements of $M_D$ are zero, then
$L$ is not uniquely defined.  Writing $M_D$ in $2\times 2$ block
form such that the upper left block is a diagonal matrix with positive
diagonal elements and the other three blocks are equal to the zero
matrix of the appropriate dimensions,
it follows that, $M_D=M_D W$, where
\beq \label{veeform}
W=\left(\begin{array}{c @{\vdashline}c} \mathds{1} \,&
\mathds{O} \\[3pt] \hdashline
    \\[-18pt] \mathds{O} \,&  W_0\end{array}\right)\,,
\eeq
$W_0$ is an arbitrary unitary matrix whose dimension is
equal to the number of zeros that appear in the diagonal elements
of $M_D$, and $\mathds{1}$ and $\mathds{O}$ are
respectively the identity matrix
and zero matrix of the appropriate size.
Hence, we can multiply both sides of \eq{lmtrmd} on the right by $W$,
which means that
$L$ is only determined up to multiplication on the right by
an arbitrary unitary matrix whose form is given by
\eq{veeform}.\footnote{Of course, one can reverse the above procedure by
first determining the unitary matrix $L$. \Eq{app:svd} implies that
$L^{\T} MM^\dagger L^*=M_D^2$,
in which case $L$ is determined up to multiplication on the
right by an arbitrary [diagonal] unitary matrix within the
subspace spanned by the eigenvectors corresponding to the
degenerate [non-degenerate] eigenvalues of $MM^\dagger$.
Having fixed $L$, one can obtain $R=M^{-1}L^*M_D$ from \eq{app:svd}.
As above, $R$ is only determined up to multiplication on the right by
a unitary matrix whose form is given by \eq{veeform}.}

If $M$ is a real matrix, then the singular value
decomposition of $M$ is given by Eq. (D.1.1), where $L$ and $R$ are
real orthogonal matrices.  This result is easily established 
by replacing ``phase'' with ``sign'' and replacing ``unitary''
by ``real orthogonal'' in the above proof.

\subsection{Takagi diagonalization}
\renewcommand{\theequation}{D.2.\arabic{equation}}
\renewcommand{\thefigure}{D.2.\arabic{figure}}
\renewcommand{\thetable}{D.2.\arabic{table}}
\setcounter{equation}{0}
\setcounter{figure}{0}
\setcounter{table}{0}

The mass matrix of neutral fermions (or a system of
two-component fermions in a generic basis) is complex and symmetric.
This mass matrix must be diagonalized in order to identify the physical
fermion mass eigenstates and to compute their masses.
However, the fermion mass matrix is \textit{not} diagonalized by the
standard unitary similarity transformation.  Instead a
different diagonalization equation is employed
that was discovered by Takagi
\cite{takagi}, and rediscovered many times since
\cite{horn}.\footnote{Subsequently, it was recognized in
Ref.~\cite{horn2} that the Takagi diagonalization was first
established for nonsingular complex symmetric matrices by Autonne
\cite{autonne}. In the physics literature, the first proof
of \eq{app:takagi} was given in \Ref{zum}.
Applications of Takagi diagonalization to the study of neutrino
mass matrices can be found in \refs{Valle-Schechter1}{Pal}.}

{\bf Theorem:}
For any complex symmetric $n\times n$ matrix $M$,
there exists a unitary matrix $\Omega$ 
such that:
\begin{eqnarray}
\label{app:takagi}
\Omega^{\T} M\, \Omega = M_D = {\rm diag}(m_1,m_2,\ldots,m_n)\,,
\end{eqnarray}
where the $m_k$ are real and non--negative.  This is the Takagi
diagonalization\footnote{In Ref.~\cite{horn}, \eq{app:takagi} is
called the Takagi factorization of a complex symmetric matrix.  We
choose to refer to this as Takagi \textit{diagonalization} to
emphasize and contrast this with the more standard diagonalization of
normal matrices by a unitary similarity transformation.  In
particular, not all \textit{complex} symmetric matrices are
diagonalizable by a similarity transformation, whereas complex
symmetric matrices are \textit{always} Takagi-diagonalizable.} of the
complex symmetric matrix $M$.

In general, the $m_k$ are \textit{not} the eigenvalues of $M$.
Rather, the $m_k$ are the singular values of the symmetric matrix
$M$. From
\eq{app:takagi} it follows that:
\begin{eqnarray} \label{app:diagmm}
\Omega^\dagger  M^\dagger M \Omega= M_D^2={\rm diag}(m^2_1,m^2_2,
     \ldots,m^2_n)\,.
\end{eqnarray}
If all of the singular values $m_k$ are non-degenerate, then one can
find a solution to eq.~(\ref{app:takagi})
for $\Omega$
from \eq{app:diagmm}.  This is no longer true if
some of the singular values are degenerate.  For example, if $M=
\bigl(\begin{smallmatrix}0\,\, & m \\ m\,\, & 0\end{smallmatrix}\bigr)$,
then the singular value $|m|$ is doubly--degenerate, but \eq{app:diagmm}
yields $\Omega^\dagger \Omega= \mathds{1}_{2\times 2}$, which does not
specify
$\Omega$. That is, in the degenerate case, the physical fermion states
\textit{cannot} be determined by the diagonalization of $M^\dagger M$.
Instead, one must make direct use of \eq{app:takagi}.
Below, we shall present a constructive method for determining $\Omega$
that is applicable in both the non-degenerate and the degenerate cases.

\Eq{app:takagi} can be rewritten as $M\Omega=\Omega^*M_D$, where
the columns of $\Omega$ are orthonormal.  If we denote the $k$th
column of $\Omega$ by $v_k$, then,
\begin{eqnarray} \label{mvks}
Mv_k=m_k v_k^*\,,
\end{eqnarray}
where the $m_k$ are the singular values and the vectors $v_k$
are normalized to have unit norm.  Following Ref.~\cite{takcompute},
the $v_k$ are called the {\it Takagi vectors} of the complex symmetric
$n\times n$ matrix~$M$.  The Takagi vectors corresponding to
non--degenerate non--zero [zero] singular
values are unique up to an overall sign [phase].
Any orthogonal [unitary] linear
combination of Takagi vectors corresponding to a set of degenerate
non--zero [zero] singular values is also a Takagi vector
corresponding to the same singular value.  Using these results,
one can determine the degree of non--uniqueness of the matrix $\Omega$.
For definiteness,
we fix an ordering of the diagonal elements of $M_D$.\footnote{Permuting
the order of the singular values is equivalent to permuting the order
of the columns of $\Omega$.}
If the singular values of $M$ are distinct, then the matrix
$\Omega$ is uniquely determined up to multiplication by a diagonal
matrix whose entries are either $\pm 1$ (i.e., a diagonal
orthogonal matrix).  If there are
degeneracies corresponding to non--zero singular values,
then within the degenerate subspace, $\Omega$
is unique up to multiplication on the right by an arbitrary orthogonal
matrix.  Finally, in the subspace corresponding to zero singular
values, $\Omega$
is unique up to multiplication on the right by an arbitrary unitary
matrix.

For a real symmetric matrix $M$, the Takagi
diagonalization [\eq{app:takagi}] still holds for a unitary matrix $\Omega$,
which is easily determined as follows.  Any real symmetric
matrix $M$ can be diagonalized by a real orthogonal matrix $Z$,
\beq \label{zmz}
Z^{\T}MZ={\rm diag}(\varepsilon_1 m_1\,,\,\varepsilon_2 m_2\,,\,\ldots\,,\,
\varepsilon_n m_n)\,,
\eeq
where the $m_k$ are real and non-negative and the $\varepsilon_k m_k$
are the real eigenvalues of $M$ with
corresponding signs
$\varepsilon_k=\pm 1$.\footnote{In the case of $m_k=0$, we conventionally
choose the corresponding $\varepsilon_k=+1$.}
Then, the Takagi diagonalization of $M$ is achieved by taking:
\beq \label{oz}
\Omega_{ij}=\varepsilon_j^{1/2} Z_{ij}\,,\qquad\quad \text{no sum over $j$}\,.
\eeq

{\bf Proof of the Takagi diagonalization}.  To prove the existence of
the Takagi diagonalization of a complex symmetric matrix, it is
sufficient to provide an algorithm for constructing
the orthonormal Takagi vectors $v_k$ that make up the columns of
$\Omega$. This is achieved by rewriting the $n\times n$ complex
matrix equation $Mv=mv^*$ [with $m$ real and non--negative]
as a $2n\times 2n$ real matrix equation \cite{dreeschoi,chkz}:
\beqa \label{eigprob}
M\ls{R}\, \left(\begin{array}{c} \Re v \\ \Im v\end{array}\right)\equiv
\left(\begin{array}{cc} \phm\Re M & \quad -\Im M \\ -\Im M & \quad -\Re M
\end{array}\right)\,\left(\begin{array}{c} \Re v \\ \Im v\end{array}\right) =
m \left(\begin{array}{c} \Re v \\ \Im v\end{array}\right)
\,,\ \ {\rm where}~m\geq 0\,.
\eeqa
Since $M=M^{\T}$, the $2n\times 2n$ matrix $M\ls{R}
\equiv\bigl(\begin{smallmatrix}\phm\Re M & \quad -\Im M \\
-\Im M & \quad -\Re M\end{smallmatrix}\bigr)$
is a real symmetric matrix.\footnote{The $2n\times 2n$ matrix
$M\ls{R}$ is a real representation of the $n\times n$ complex matrix $M$.}
In particular, $M\ls{R}$ is diagonalizable by a real
orthogonal similarity transformation, and its eigenvalues are real.
Moreover, if $m$ is an eigenvalue of $M\ls{R}$
with eigenvector $(\Re v\,,\,\Im v)$, then $-m$ is an eigenvalue of
$M\ls{R}$ with (orthogonal) eigenvector $(-\Im v\,,\,\Re v)$.
This observation implies that $M\ls{R}$ has an equal number of positive and
negative eigenvalues and an even number of zero
eigenvalues.\footnote{ \label{fn1}%
Note that
$(-\Im v\,,\,\Re v)$ corresponds to replacing $v_k$ in \eq{mvks}
by $i v_k$.  However, for $m<0$ these solutions are not relevant for
Takagi diagonalization (where the $m_k$ are by definition non--negative).
The case of $m=0$ is considered in footnote~\ref{fn0}.}
Thus, \eq{mvks} has been converted into an ordinary
eigenvalue problem for a real symmetric matrix.  Since $m\geq 0$, we
solve the eigenvalue problem $M\ls{R} u = mu$ for the
real eigenvectors $u\equiv(\Re v\,,\,\Im v)$
corresponding to the non--negative
eigenvalues of $M_R$,\footnote{\label{fn0}%
For $m=0$, the corresponding vectors $(\Re v\,,\,\Im v)$ and
$(-\Im v\,,\,\Re v)$ are two
linearly independent eigenvectors of $M\ls{R}$; but these yield only one
independent Takagi vector $v$ (since $v$ and $iv$ are
linearly dependent).}
which then immediately yields the complex Takagi vectors, $v$.
It is straightforward to prove that the total number of linearly independent
Takagi vectors is equal to $n$.  Simply note that the orthogonality of
$(\Re v_1\,,\,\Im v_1)$ and $(-\Im v_1\,,\,\Re v_1)$ with
$(\Re v_2\,,\,\Im v_2)$ implies that $v_1^\dagger v_2=0$.

Thus, we have derived a constructive method for obtaining the Takagi
vectors $v_k$.  If there are degeneracies, one can always choose the
$v_k$ in the degenerate subspace to be orthonormal.  The Takagi
vectors then make up the columns of the matrix $\Omega$ in
\eq{app:takagi}.  A numerical package for performing the Takagi
diagonalization of a complex symmetric matrix has recently been
presented in \Ref{hahn} (see also
refs.~\cite{takcompute,takcompute2} for
previous numerical approaches to Takagi diagonalization).

\subsection{Relation between Takagi diagonalization and singular
value decomposition}
\renewcommand{\theequation}{D.3.\arabic{equation}}
\renewcommand{\thefigure}{D.3.\arabic{figure}}
\renewcommand{\thetable}{D.3.\arabic{table}}
\setcounter{equation}{0}
\setcounter{figure}{0}
\setcounter{table}{0}

The Takagi diagonalization is a special case of the singular value
decomposition.  If the complex matrix $M$ in \eq{app:svd} is
symmetric, $M=M^{\T}$, then the Takagi diagonalization corresponds to
$\Omega=L=R$. In this case, the right and left
singular vectors coincide ($v_k=w_k$) and are identified
with the Takagi vectors defined in \eq{mvks}. However as 
previously noted, the
matrix $\Omega$ cannot be determined from \eq{app:diagmm} in cases
where there is a degeneracy among the singular values.\footnote{This
is in contrast to the singular value decomposition, where $R$ can be
determined from \eq{rmatrix} modulo right multiplication by a
[diagonal] unitary matrix in the [non-]degenerate subspace and
$L$ is then determined by \eq{lmtrmd} modulo multiplication on the
right by \eq{veeform}.}  
For example, one possible singular value decomposition of
the matrix
$M=\bigl(\begin{smallmatrix} 0 & m\\
m&0\end{smallmatrix}\bigr)$ [with $m$ assumed real and positive] can
be obtained by choosing
$R= \bigl(\begin{smallmatrix} 1 & 0\\ 0&1\end{smallmatrix}\bigr)
$ and
$L=\bigl(\begin{smallmatrix} 0 & 1\\ 1& 0\end{smallmatrix}\bigr)$,
in which case $L^{\T} M R =
\bigl(\begin{smallmatrix} m & 0\\ 0&m\end{smallmatrix}\bigr)
= M_D$.
Of course, this is not a Takagi diagonalization because $L\neq R$.
Since
$R$ is only defined modulo the multiplication on the right by an
arbitrary $2\times 2$ unitary matrix $\mathcal{O}$, then at least
one singular value decomposition exists that is also a Takagi
diagonalization. For the example under consideration, it is not
difficult to deduce the Takagi diagonalization: $\Omega^{\T}
M\Omega=M_D
$, where
\beq \label{exomega}
\Omega=\frac{1}{\sqrt
2}\begin{pmatrix} 1&\phm i \\ 1 & -i\end{pmatrix} \mathcal{O}\,,
\eeq
and $\mathcal{O}$ is any $2\times 2$ orthogonal matrix.

Since the Takagi diagonalization is a special case of the singular
value decomposition, it seems plausible that one can prove the
former from the latter.  This turns out to be correct; for
completeness, we provide the proof below.  Our second proof
depends on the following lemma:

{\bf Lemma:} For any symmetric unitary matrix $V$, there exists
a unitary matrix $U$ such that $V=U^{\T} U$.

{\bf Proof of the Lemma:} For any $n\times n$
unitary matrix $V$, there exists an hermitian
matrix $H$ such that $V=\exp{(iH)}$ (this is the polar decomposition of
$V$).  If $V=V^{\T}$ then $H=H^{\T}=H^*$ (since $H$ is hermitian);
therefore $H$ is real symmetric.  But, any
real symmetric matrix can be diagonalized by an orthogonal
transformation.  It follows that $V$ can also be diagonalized by an orthogonal
transformation.  Since the eigenvalues of any unitary matrix are pure
phases, there exists a real orthogonal matrix $Q$
such that
$Q^{\T}VQ=\rm{diag}~(e^{i\theta_1}\,,\,e^{i\theta_2}\,,\,\ldots
\,,\,e^{i\theta_n})$.  Thus, the unitary matrix,
\beq
U={{\rm diag}}~(e^{i\theta_1/2}\,,\,e^{i\theta_2/2}\,,\,\ldots
\,,\,e^{i\theta_n/2})\, Q^{\T}\,,
\eeq
satisfies $V=U^{\T} U$ and the lemma is proved.  Note that $U$ is unique
modulo multiplication on the left by an arbitrary real orthogonal matrix.

{\bf Second Proof of the Takagi diagonalization.} Starting from the
singular value decomposition of $M$, there exist unitary matrices $L$
and $R$ such that $M=L^*M_DR^\dagger$, where $M_D$ is the diagonal
matrix of singular values. Since $M=M^{\T}=R^*M_D L^{\dagger}$, we have
two different singular value decompositions for $M$.  However, as
noted below \eq{rmatrix}, $R$ is unique modulo multiplication on the
right by an arbitrary [diagonal] unitary matrix, $V$, within the
[non-]degenerate subspace.  Thus, it follows that a
[diagonal] unitary matrix $V$
exists such that $L=RV$. Moreover, $V=V^{\T}$. This
is manifestly true within the non-degenerate subspace where $V$ is
diagonal.  Within the degenerate subspace, $M_D$ is proportional to
the identity matrix so that $L^*R^\dagger=R^*L^{\dagger}$. Inserting
$L=RV$ then yields $V^{\T}=V$.  Using the Lemma proved above, there
exists a unitary matrix $U$ such that $V=U^{\T} U$.  That is,
\beq \label{rlutu}
L=RU^{\T} U\,,
\eeq
for some unitary matrix
$U$.  Moreover, it is now straightforward to show that
\beq \label{mduumd}
M_DU^*=U^* M_D\,.
\eeq
To see this, note that
within the degenerate subspace,
\eq{mduumd} is trivially true since $M_D$ is proportional to the
identity matrix.  Within the non-degenerate subspace $V$ is
diagonal; hence we may choose $U=U^{\T}=V^{1/2}$, so that
\eq{mduumd} is true since diagonal matrices commute. Using
\eqs{rlutu}{mduumd}, we can write the singular value decomposition
of $M$ as follows
\beq \label{mlmr}
M=L^*M_DR^\dagger
=R^* U^\dagger U^*M_D R^\dagger=(RU^{\T})^* M_D U^*R^\dagger
=\Omega^* M_D\Omega^\dagger\,,
\eeq
where $\Omega\equiv RU^{\T}$
is a unitary matrix. Thus the existence of the Takagi
diagonalization of an arbitrary complex symmetric matrix
[\eq{app:takagi}] is once again proved.

In the diagonalization of the
two-component fermion mass matrix, $M$,
the eigenvalues of $M^\dagger M$ typically
fall into two classes---non-degenerate
eigenvalues corresponding to neutral fermion mass eigenstates and
degenerate pairs corresponding to charged (Dirac) mass eigenstates.
In this case, the sector of the neutral fermions corresponds to a
non-degenerate subspace of the space of fermion fields.
Hence, in order to identify the neutral fermion mass eigenstates,
it is sufficient to diagonalize $M^\dagger M$ with a unitary matrix $R$
[as in \eq{rmatrix}], and then adjust the overall phase of each
column of $R$ so that the resulting matrix $\Omega$ satisfies
$\Omega^{\T} M\Omega=M_D$, where $M_D$ is a diagonal matrix of the
non-negative fermion masses.
This last result is a consequence of \eqst{rlutu}{mlmr}, where
$\Omega=RV^{1/2}$ and $V$ is a diagonal matrix of phases.

\subsection{Reduction of a complex antisymmetric matrix to
real normal form}
\renewcommand{\theequation}{D.4.\arabic{equation}}
\renewcommand{\thefigure}{D.4.\arabic{figure}}
\renewcommand{\thetable}{D.4.\arabic{table}}
\setcounter{equation}{0}
\setcounter{figure}{0}
\setcounter{table}{0}

In the case of two-component fermions that transform under a
pseudo-real representation of a compact Lie group
[cf.~\eq{LagPseudoreal}], the corresponding
mass matrix is in general complex and antisymmetric.  In this case,
one needs the antisymmetric analogue of the Takagi diagonalization of a
complex symmetric matrix~\cite{horn}.

\textbf{Theorem:} For any complex [or real]
antisymmetric $n\times n$ matrix $M$,
there exists a unitary [or real orthogonal] matrix $U$ such that:
\beq \label{antinormal}
U^{\T} M U = N\equiv {\rm diag} \left\{\begin{pmatrix} \phm 0 & m_1 \\ -m_1 & 0
\end{pmatrix}\,,\,\begin{pmatrix} \phm 0 & m_2 \\ -m_2 & 0
\end{pmatrix}\,,\,\cdots \,,\begin{pmatrix} \phm 0 & m_p \\ -m_p & 0
\end{pmatrix}\,,\,\mathds{O}_{n-2p}\right\}\,,
\eeq
where $N$ is written in block diagonal form with
$2\times 2$ matrices appearing along the diagonal, followed by an
$(n-2p)\times (n-2p) $ block of zeros (denoted by $\mathds{O}_{n-2p}$),
and the $m_j$ are real and positive. $N$ is called the
\textit{real normal form} of an antisymmetric matrix~\cite{zum,Becker,Greub}.

\textbf{Proof:} A number of proofs can be found in the
literature~\cite{zum,Becker,Napoly,Greub,Li}.  Here we provide a proof
inspired by \Ref{Becker}.  Following \app{D.1}, we first consider the
eigenvalue equation for $M^\dagger M$:
\beq \label{MMwu}
M^\dagger M v_k = m_k^2 v_k\,,\qquad m_k>0\,,\qquad {\rm and} \qquad
M^\dagger M u_k = 0\,,
\eeq
where we have distinguished the eigenvectors corresponding to positive
eigenvalues and zero eigenvalues, respectively.  The quantities
$m_k$ are the positive singular values of $M$.
Noting that  $u^\dagger_k M^\dagger Mu_k=\vev{Mu_k\,|
\,Mu_k}=0$, it follows that
\beq \label{Mu}
Mu_k=0\,,
\eeq
so that the $u_k$ are the eigenvectors corresponding to the
zero eigenvalues of $M$.  For each eigenvector of $M^\dagger M$
with $m_k\neq 0$, we
define a new vector
\beq \label{vkdef}
w_k\equiv \frac{1}{m_k} M^* v_k^*\,.
\eeq
It follows that $m_k^2v_k=M^\dagger Mv_k=m_kM^\dagger w_k^*$, which
yields $M^\dagger w_k^*=m_k v_k$.
Comparing with \eq{singvals}, we identify
$v_k$ and $w_k$ as the right and left singular vectors, respectively,
corresponding to the non-zero singular values of $M$.
For any antisymmetric matrix, $M^{\dagger}=-M^*$.  Hence,
\beq \label{Mvw}
Mv_k=m_k w_k^*\,,\qquad\qquad Mw_k=-m_k v_k^*\,,
\eeq
and
\beq
M^\dagger M w_k = -m_k M^\dagger v_k^*=m_k M^*v_k^*=m_k^2 w_k\,,
\qquad m_k>0\,.
\eeq
That is, the $w_k$ are also eigenvectors of $M^\dagger M$.

The key observation is that for fixed $k$ the vectors $v_k$ and $w_k$ are
orthogonal, since  \eq{Mvw}
implies that:
\beq \label{vwortho}
\vev{w_k|v_k}=\vev{v_k|w_k}^*=-\frac{1}{m_k^2}\langle Mw_k|Mv_k\rangle=
-\frac{1}{m_k^2}\langle w_k|M^\dagger Mv_k\rangle=-\vev{w_k|v_k}\,,
\eeq
which yields $\vev{w_k|v_k}=0$.  Thus, if all the $m_k$ are distinct,
it follows that $m_k^2$ is a doubly degenerate eigenvalue of $M^\dagger M$,
with corresponding linearly independent eigenvectors $v_k$ and $w_k$, where
$k=1,2,\ldots,p$ (and $p\leq \half n$).
The remaining zero eigenvalues are $(n\!-\!2p)$-fold
degenerate, with corresponding eigenvectors
$u_k$ (for $k=1,2,\ldots,n-2p$).
If some of the $m_k$ are degenerate, these conclusions still
apply.  For example, suppose that $m_j=m_k$ for $j\neq k$,
which means that $m_k^2$ is
at least a three-fold degenerate eigenvalue of $M^\dagger M$.
Then, there must exist an eigenvector $v_j$ that is orthogonal to
$v_k$ and $w_k$ such that $M^\dagger Mv_j=m_k^2 v_j$.  We now construct
$w_j\equiv M^*v^*_j/m_k$ according to \eq{vkdef}.  According to
\eq{vwortho}, $w_j$ is orthogonal to $v_j$.  However, we still must show that
$w_j$ is also orthogonal to $v_k$ and $w_k$.  But this is
straightforward:
\beqa
\vev{w_j|w_k}&=&\vev{w_k|w_j}^*=
\frac{1}{m_k^2}\langle M v_k|Mv_j\rangle=
\frac{1}{m_k^2}\langle v_k|M^\dagger M v_j\rangle
=\vev{v_k|v_j}=0\,,\\
\vev{w_j|v_k}&=&\vev{v_k|w_j}^*=-\frac{1}{m_k^2}\langle Mw_k|Mv_j\rangle=
-\frac{1}{m_k^2}\langle w_k|M^\dagger Mv_j\rangle=-\vev{w_k|v_j}=0\,,
\eeqa
where we have used the assumed orthogonality of $v_j$ with $v_k$ and
$w_k$, respectively.  It follows that $v_j$, $w_j$, $v_k$ and $w_k$
are linearly independent eigenvectors
corresponding to a four-fold degenerate eigenvalue $m_k^2$ of
$M^\dagger M$.  Additional degeneracies are treated in the same way.

Thus, the number of non-zero eigenvalues of $M^\dagger M$ must be an
even number, denoted by $2p$ above.  Moreover, one can always choose the
complete set of eigenvectors $\{u_k\,,\,v_k\,,\,w_k\}$
of $M^\dagger M$ to be orthonormal.
These orthonormal vectors can be used to
construct a unitary matrix $U$ with matrix elements:
\beqa \label{umats}
U_{\ell\,,\,2k-1}&=& (w_k)_\ell\,,\qquad U_{\ell\,,\,2k}
=(v_k)_\ell\,,\qquad \quad
\quad k=1\,,2\,,\ldots\,,p\,,\nonumber \\
U_{\ell\,,\,k+2p}&=& (u_k)_\ell\,,
\qquad \qquad\qquad\qquad\qquad\qquad \,\,\,\,
k=1\,,2\,,\ldots\,,n-2p\,,
\eeqa
for $\ell=1\,,2\,,\ldots\,, n$, where
e.g., $(v_k)_\ell$ is the $\ell$th component of the vector $v_k$ with respect
to the standard orthonormal basis.  The orthonormality of
$\{u_k\,,\,v_k\,,\,w_k\}$ implies that $(U^\dagger
U)_{\ell k}=\delta_{\ell k}$ as required.
\Eqs{Mu}{Mvw} are thus equivalent to
the matrix equation $MU=U^*N$, which immediately yields \eq{antinormal},
and the theorem is proven.  If $M$ is a real antisymmetric matrix,
then all the eigenvectors of $M^\dagger M$ can be chosen to be real,
in which case $U$ is a real orthogonal matrix.

Finally, we address the non-uniqueness of the matrix $U$.  For
definiteness, we fix an ordering of the $2\times 2$ blocks containing
the $m_k$ in the matrix $N$.  In the subspace corresponding to
a non-zero singular
value of degeneracy $d$, $U$ is unique up to multiplication on the
right by a $2d\times 2d$ unitary matrix $S$ that satisfies:
\beqa \label{symplectic}
S^{\T} J S=J\,,
\eeqa
where the $2d\times 2d$ matrix $J$, defined by
\beq \label{Jdef}
J={\rm diag} \left\{\begin{pmatrix} \phm 0 & \quad 1 \\ -1 & \quad 0
\end{pmatrix}\,,\,\begin{pmatrix} \phm 0 & \quad 1 \\ -1 & \quad 0
\end{pmatrix}\,,\,\cdots \,,\begin{pmatrix} \phm 0 & \quad 1 \\ -1 &
\quad 0\end{pmatrix}\right\}\,,
\eeq
is a block diagonal matrix with $d$ blocks of $2\times 2$ matrices.
A unitary matrix $S$ that satisfies \eq{symplectic}
is an element of the unitary symplectic group,
Sp($d$).  If there are no degeneracies among the $m_k$, then $d=1$.
Identifying Sp(1)$\,\iso\,$SU(2), it follows that within the subspace
corresponding to a non-degenerate singular value, $U$ is unique up to
multiplication on the right by an arbitrary SU(2) matrix.  Finally, in
the subspace corresponding to the zero eigenvalues
of~$M$, $U$ is unique up to multiplication on the right by an
arbitrary unitary matrix.

\section{\texorpdfstring{Lie group theoretical techniques for gauge theories}{Lie group theoretical techniques for gauge theories}}
\renewcommand{\theequation}{E.\arabic{equation}}
\renewcommand{\thefigure}{E.\arabic{figure}}
\renewcommand{\thetable}{E.\arabic{table}}
\setcounter{equation}{0}
\setcounter{figure}{0}
\setcounter{table}{0}

\subsection{Basic facts about Lie groups, Lie algebras and
their representations}
\renewcommand{\theequation}{E.1.\arabic{equation}}
\renewcommand{\thefigure}{E.1..\arabic{figure}}
\renewcommand{\thetable}{E.1.\arabic{table}}
\setcounter{equation}{0}
\setcounter{figure}{0}
\setcounter{table}{0}

Consider a compact connected Lie Group $G$~\cite{gilmore}.
The most general form for $G$ is
a direct product of
compact simple groups and U(1) groups.  If no U(1) factors are
present, then $G$ is semisimple.  For any $U\in G$,
\beq \label{Uexp}
U=\exp(-i\theta^a \boldsymbol{T^a})\,,
\eeq
where the
$\boldsymbol{T^a}$ are called the generators of $G$, and the $\theta^a$
are real numbers that parameterize the elements of $G$.
The corresponding real Lie algebra $\mathfrak{g}$ consists of arbitrary
real linear combinations of the generators, $\theta^a\boldsymbol{T^a}$.
The Lie group generators $\boldsymbol{T^a}$ satisfy the
commutation relations:
\beq \label{fabcdef}
[\boldsymbol{T^a},\boldsymbol{T^b}]= if_c^{ab}\boldsymbol{T^c}\,,
\eeq
where the real structure constants $f_c^{ab}$ define the compact Lie algebra.
The generator indices run over $a$, $b$, $c=1,2,\ldots, d_G$, where
$d_G$ is the dimension of the Lie algebra.
For compact Lie algebras, the Killing form $g^{ab}=\Tr(
\boldsymbol{T^a}\boldsymbol{T^b})$ is positive definite, so one can
always choose a basis for the Lie algebra in which $g^{ab}\propto\delta^{ab}$
(where the proportionality constant is a positive real number).
With respect to this new basis, the structure constants
$f^{abc}\equiv g^{ad}f_d^{bc}$ are totally antisymmetric with respect to
the interchange of the indices $a$, $b$ and $c$.
Henceforth, we shall always assume that such a \textit{preferred}
basis of generators has been chosen.

The elements of the compact Lie group $G$
act on a multiplet of fields that transform under some
$d_R$-dimensional representation $R$ of $G$.
The group
elements $U\in G$ are represented by $d_R\times d_R$ unitary matrices,
$D_R(U)=\exp(-i\theta^a \boldsymbol{T^a_R})$,
where the $\boldsymbol{T_R^a}$ are $d_R\times d_R$ hermitian
matrices that satisfy \eq{fabcdef} and thus provide a
representation of the Lie group generators.
For any representation $R$ of a semisimple group,
$\Tr \boldsymbol{T^a_R}=0$ for all $a$.
A representation $R'$ is unitarily equivalent to~$R$ if there exists a fixed
unitary matrix $S$ such that $D_{R'}(U)=S^{-1}D_R(U)S$ for
all $U\in G$.  Similarly, the corresponding generators satisfy
$\boldsymbol{T_{R'}^a}=S^{-1}\boldsymbol{T_R^a} S$ for all $a=1,2,\dots, d_G$.

For compact semisimple Lie groups, two representations
are noteworthy.  If $G$ is one of the classical groups, SU($N$) [for
$N\geq 2$],
SO($N$) [for $N\geq 3$] or
Sp($N/2$) [the latter is defined by \eqs{symplectic}{Jdef} for
even $N\geq 2$],
then the $N\times N$ matrices that define these groups comprise
the \textit{fundamental (or defining) representation} $F$,
with $d_F=N$.
For example,
the fundamental representation of SU($N$) consists of $N\times N$
unitary matrices with determinant equal to one, and the corresponding
generators comprise
a suitably chosen basis for the $N\times N$ traceless hermitian matrices.
Every Lie group $G$ also possesses an
\textit{adjoint
representation} $A$, with $d_A=d_G$.  The matrix elements of the generators
in the adjoint representation are given by\footnote{\label{fnadjoint}%
Since the
$f^{abc}$ are real, the $i\boldsymbol{T^a_A}$ are real antisymmetric
matrices. The heights of the adjoint labels $a$, $b$ and $c$ are not
significant, as they can be lowered by the inverse Killing form given by
$g_{ab}\propto\delta_{ab}$ in the preferred basis.}
\beq
(\boldsymbol{T_A^a})^{bc}=-if^{abc}\,.
\eeq

Given the unitary representation matrices $D_R(U)$ of the
representation $R$ of $G$,
the matrices $[D_R(U)]^*$ constitute the
\textit{conjugate} representation $R^*$.
Equivalently, if the $\boldsymbol{T_R^a}$ comprise a representation of the Lie
algebra $\mathfrak{g}$, then the
$-(\boldsymbol{T_R^a})^*=-(\boldsymbol{T_R^a})^{\T}$ comprise
a representation $R^*$ of $\mathfrak{g}$ of the same dimension $d_R$.
If $R$ and $R^*$ are unitarily equivalent representations, then we say that
the representation $R$ is \textit{self-conjugate}.  Otherwise, we say that
the representation $R$ is \textit{complex}, or ``strictly complex'' in the
language of \Ref{oraif}.
However, the representation matrices $D_R(U)$ of a self-conjugate
representation can also be complex.
We can then define two classes
of self-conjugate representations.  If $R$ and $R^*$ are
unitarily equivalent to a
representation $R'$ that satisfies the reality property
$[D_{R'}(U)]^*=[D_{R'}(U)]$ for all
$U\in G$ (equivalently, the matrices
$i\boldsymbol{T_{R'}^a}$ are real for all $a$),
then $R$ is said to be \textit{real}, or ``strictly real'' in the
language of \Ref{oraif}.  If $R$ and $R^*$ are unitarily equivalent
representations, but neither is unitarily equivalent to a representation that
satisfies the reality property above, then $R$ is said to be
\textit{pseudo-real}.

Henceforth, we drop the adjective ``strictly''
and simply refer to real, pseudo-real and complex representations.
Self-conjugate representations are either real or pseudo-real.
An important theorem states that
for self-conjugate representations,
there exists a constant unitary matrix $W$ such that~\cite{oraif}
\beq \label{selfconjugate}
[D_R(U)]^*=W D_R(U)W^{-1}\,,\quad \text{or equivalently,}\quad
(i\boldsymbol{T^a_R})^*=W(i\boldsymbol{T^a_R})W^{-1}\,,
\eeq
where
\beqa
&& WW^*=\mathds{1}\,,\quad\qquad W^{\T}=W\,,\qquad\quad
\text{for real representations}\,,\label{realrep}\\
&& WW^*=-\mathds{1}\,,\qquad \,W^{\T}=-W\,,\qquad
\text{for pseudo-real representations}\,,\label{prealrep}
\eeqa
and $\mathds{1}$ is the $d_R\times d_R$ identity matrix.
Taking the determinant of \eq{prealrep}, and using the fact
that $W$ is unitary (and hence invertible),
it follows that $1=(-1)^{d_R}$.  Therefore,
a pseudo-real representation must be even-dimensional.

If we redefine the basis for the Lie group generators by
$\boldsymbol{T^a_R}\to V^{-1}\boldsymbol{T^a_R}\,V$,
where $V$ is unitary, then
$W\to V^{\T}WV$.  We can make use of this change of basis
to transform $W$ to a canonical form.  Since $W$ is unitary, its
singular values (i.e.~the positive square roots of the eigenvalues
of $W^\dagger W$) are all equal to 1.  Hence, in the two cases
corresponding to $W^{\T}=\pm W$, respectively, \eqs{app:takagi}{antinormal}
yield the following canonical forms (for an appropriately chosen $V$),
\beqa
W&=&\mathds{1}\,,\qquad
\text{for a real representation}~ R\,,
\label{realcan}\\[6pt]
W&=& J\,,\qquad
\text{for a pseudo-real representation}~R\,,
\label{prealcan}
\eeqa
where $J\equiv{\rm diag}\left\{\left(\begin{smallmatrix} \phm 0 & \,\,1 \\
-1 &\,\, 0 \end{smallmatrix}\right)\,,\,
\left(\begin{smallmatrix} \phm 0 & \,\,1 \\
-1 &\,\, 0 \end{smallmatrix}\right)\,,\,\cdots\,,\,\left(\
\begin{smallmatrix} \phm 0 & \,\,1 \\
-1 &\,\, 0 \end{smallmatrix}\right)\right\}$
is a $d_R\times d_R$ matrix (and $d_R$ is even).

There are many examples of complex, real and pseudo-real
representations in mathematical physics.
For example,
the fundamental representation of SU($N$) is complex for $N\geq 3$.
The adjoint representation of any compact
Lie group is real [cf.~footnote~\ref{fnadjoint}].
The simplest example of a pseudo-real
representation is the two-dimensional representation of
SU(2),\footnote{No unitary matrix $W$ exists such that
the $W i\tau^a W^{-1}$ are real for all $a=1,2,3$.  Thus, the two-dimensional
representation of SU(2) is not real.  However, $(i\tau^a)^*=(i\tau^2)
(i\tau^a)(i\tau^2)^{-1}$ for $a=1,2,3$, which proves that the two-dimensional
representation of SU(2) is pseudo-real.}
where $\boldsymbol{T^a}=\half\tau^a$ (and the $\tau^a$ are the usual
Pauli matrices).
More generally, the generators of a pseudo-real representation
must satisfy
\beq \label{unitaryC}
(i\boldsymbol{T^a_R})^*=C^{-1}(i\boldsymbol{T^a_R})C\,,
\eeq
for some fixed unitary antisymmetric matrix $C$ [previously
denoted by $W^{-1}$ in \eqs{selfconjugate}{prealrep}].
For the doublet representation of SU(2) just given,
$C^{ab}=(i\tau^2)^{ab}\equiv\epsilon^{ab}$ is the familiar
SU(2)-invariant tensor.

Finally, we note that for U(1), all irreducible representations are
one-dimensional.  The structure constants vanish and any
$d$-dimensional representation of the U(1)-generator is given by the
$d\times d$ identity matrix multiplied by the corresponding U(1)-charge.
For a Lie group that is a direct product of a semisimple group and
U(1) groups, $\Tr \boldsymbol{T_R^a}$ is non-zero when $a$ corresponds to
one of the U(1)-generators, unless the sum of the corresponding
U(1)-charges of the states of the representation $R$ vanishes.

\subsection{The quadratic and cubic index and Casimir operator}
\renewcommand{\theequation}{E.2.\arabic{equation}}
\renewcommand{\thefigure}{E.2.\arabic{figure}}
\renewcommand{\thetable}{E.2.\arabic{table}}
\setcounter{equation}{0}
\setcounter{figure}{0}
\setcounter{table}{0}

In this section, we define the index and Casimir operator of a
representation of a compact semisimple Lie algebra $\mathfrak{g}$.
The \textit{index} $I_2(R)$ of the representation $R$ is defined
by~\cite{gilmore,indexdef,indexbook,groupfactors}
\beq \label{index2}
\Tr(\boldsymbol{T_R^a}\boldsymbol{T_R^b})=I_2(R)\delta^{ab}\,,
\eeq
where $I_2(R)$ is a positive real number that depends on $R$.
Once $I_2(R)$ is defined for one representation, its value is
uniquely fixed for any other representation.  In the case of
a simple compact Lie algebra $\mathfrak{g}$, it is traditional
to normalize the generators of the fundamental (or defining)
representation $F$ according to\footnote{In the literature,
the index is often defined as the ratio $I_2(R)/I_2(F)$,
where $I_2(F)$ is fixed by some convention.
This has the advantage that the index of $R$ is independent
of the normalization convention of the generators.  In this Appendix,
we will simply refer to $I_2(R)$ as the index.
}
\beq \label{normgenerator}
\Tr(\boldsymbol{T_F^a}\boldsymbol{T_F^b})=\half\delta^{ab}\,.
\eeq

If the representation
$R$ is reducible, it can be decomposed into the direct sum of irreducible
representations,
$R=\sum_k R_k$.  In this case, the index of $R$ is given by
\beq
I_2(R)=\sum_k I_2(R_k)\,.
\eeq
The index of a tensor product of two representations
$R_1$ and $R_2$ is given by~\cite{indexdef}
\beq
I_2(R_1\otimes R_2)=d_{R_1}I_2(R_2)+d_{R_2}I_2(R_1)\,.
\eeq
Finally, we note that if $R^*$ is the complex conjugate of the
representation $R$, then
$I_2(R^*)=I_2(R)$.

A Casimir operator of a Lie algebra $\mathfrak{g}$ is an operator
that commutes with all the generators $\boldsymbol{T^a}$.  If the
representation of the $\boldsymbol{T^a}$ is \textit{irreducible},
then Schur's lemma implies that
the Casimir operator is a multiple of the identity.  The proportionality
constant depends on the representation $R$.
The quadratic
Casimir operator of an \textit{irreducible} representation $R$ is given by
\beq
(\boldsymbol{T_R^2})_i{}^j\equiv (\boldsymbol{T_R^a})_i{}^k
(\boldsymbol{T_R^a})_k{}^j = C_R \delta_i{}^j
\,,
\eeq
where the sum over the repeated indices are implicit and $i$, $j$, $k=1,2\ldots d_R$.
A simple computation then yields the eigenvalue of the quadratic Casimir
operator, $C_R$,
\beq \label{casimir}
C_R=\frac{I_2(R) d_G}{d_R}\,.
\eeq
For a simple Lie algebra (where the adjoint representation is irreducible),
it immediately follows that $C_A=I_2(A)$.
For a reducible representation, $\boldsymbol{T^2}$ is a block
diagonal matrix consisting of $d_{R_k}\times d_{R_k}$
blocks given by $C_{R_k}\mathds{1}$
for each irreducible component $R_k$ of $R$.

The example of the simple Lie algebra $\mathfrak{su}(N)$ is well known.
The dimension of this Lie algebra (equal to the number of generators) is
given by $N^2-1$.  As previously noted, $d_F=N$ and $I_2(F)=\half$.
It then follows that $C_F=(N^2-1)/(2N)$.  One can also check that
$C_A=I_2(A)=N$.

The Lie algebras $\mathfrak{su}(N)$ [$N\geq 3$]
are the only simple Lie algebra that possesses a cubic
Casimir operator.  First, we define the symmetrized trace of three
generators~\cite{groupfactors,groupfactors2}:
\beq \label{dabcstr}
D^{abc}\equiv{\rm Str}~(\boldsymbol{T^a}\boldsymbol{T^b}\boldsymbol{T^c})
=\nicefrac{1}{6}\Tr(\boldsymbol{T^a}\boldsymbol{T^b}
\boldsymbol{T^c}+{\rm perm.})\,,
\eeq
where ``perm.'' indicates five other terms obtained by permuting the
indices $a$, $b$ and $c$ in all possible ways.  Due to the properties of the
trace, it follows that for a given representation $R$,
\beq  \label{Dabcr}
D^{abc}(R)=\half\Tr\left[
\{\boldsymbol{T_R^a},\boldsymbol{T_R^b}\}\boldsymbol{T_R^c}\right]\,.
\eeq
For the $N$-dimensional defining representation of $\mathfrak{su}(N)$,
it is conventional to define
\beq \label{dabcdef}
d^{abc}\equiv 2\Tr\left[\{\boldsymbol{T_F^a},\boldsymbol{T_F^b}\}
\boldsymbol{T_F^c}\right]\,.
\eeq
One important property of the $d^{abc}$ is~\cite{okubo1,sudbery}:
\beq \label{dabcdabc}
d^{abc}d^{abc}=\frac{(N^2-1)(N^2-4)}{N}\,.
\eeq
In general, $D^{abc}(R)$ is proportional to $d^{abc}$.
In particular, the \textit{cubic index} $I_3(R)$ of
a representation $R$ is
defined such that~\cite{banks,okubo1,groupfactors},
\beq \label{cubicindex}
D^{abc}(R)=I_3(R)d^{abc}\,.
\eeq

Having fixed $I_3(F)=\quarter$, the cubic index is
uniquely determined for all representations of
$\mathfrak{su}(N)$~\cite{banks,okubo1,patera,baha}.
As in the case of the quadratic index
$I_2(R)$, we have:
\beq
I_3(R)=\sum_k I_3(R_k)\,,
\eeq
for a reducible representation $R=\sum_k R_k$.
The cubic index of a tensor product of two representations
$R_1$ and $R_2$ is given by~\cite{banks}
\beq
I_3(R_1\otimes R_2)=d_{R_1}I_3(R_2)+d_{R_2}I_3(R_1)\,.
\eeq
If the generators of the representation $R$ are $\boldsymbol{T_R^a}$, then the generators
of the complex conjugate representation
$R^*$ are $-\boldsymbol{T_R^a}^{\T}$.  It then follows that
$I_3(R^*)=-I_3(R)$.
In particular, the cubic index of a self-conjugate representation vanishes.
Note that the converse is not true.  That is, it is possible for the cubic
index of a complex representation of $\mathfrak{su}(N)$
to vanish in special circumstances~\cite{baha}.

One can show that among the simple Lie groups, $D^{abc}=0$ except for the
case of SU($N$), when $N\geq 3$~\cite{okubo1}.
For any non-semisimple Lie group (i.e., a Lie group that is a direct product of
simple Lie groups and at least one U(1) factor), $D^{abc}$ is generally non-vanishing.
For example, suppose that the $\boldsymbol{T^a_R}$ constitute an irreducible representation
of the generators of $G\times$U(1), where $G$ is a semisimple Lie group.
Then the U(1) generator (which we denote by setting $\boldsymbol{a}=\boldsymbol{Q}$)
is $\boldsymbol{T^Q_R}\equiv
q\mathds{1}$, where $q$ is the
corresponding U(1)-charge.  It then follows that $D^{Qab}=qI_2(R)\delta^{ab}$.
More generally, for a compact non-semisimple Lie group,
$D^{abc}$ can be non-zero when either one or
three of its indices corresponds to a U(1) generator.

In the computation of the anomaly [cf.~\sec{sec:anomaly}],
the quantity $\Tr(\boldsymbol{T_R^a}
\boldsymbol{T_R^b}\boldsymbol{T_R^c})$ appears.   We can evaluate this
trace using \eqs{fabcdef}{cubicindex}:
\beq \label{trttt}
\Tr(\boldsymbol{T_R^a}
\boldsymbol{T_R^b}\boldsymbol{T_R^c})=I_{3}(R)d^{abc}+\frac{i}{2}I_2(R) f^{abc}\,.
\eeq

The cubic Casimir operator of an \textit{irreducible} representation $R$ is given by
\beq
(\boldsymbol{T_R^3})_i{}^j\equiv d^{abc}(\boldsymbol{T_R^a}\boldsymbol{T_R^b}
\boldsymbol{T_R^c})_i{}^j = C_{3R} \delta_i{}^j\,.
\eeq
Using \eqs{dabcdabc}{cubicindex}, we obtain a relation between the
eigenvalue of the cubic
Casimir operator, $C_{3R}$ and the cubic index~\cite{okubo1}:
\beq \label{casimir3}
C_{3R}=\frac{(N^2-1)(N^2-4)I_3(R)}{Nd_R}\,.
\eeq
Again, we provide two examples.  For the fundamental representation of $\mathfrak{su}(N)$,
$I_3(F)=\quarter$ and $C_{3F}=(N^2-1)(N^2-4)/(4N^2)$.  For the adjoint
representation, $I_3(A)=C_{3A}=0$, since the adjoint representation is
self-conjugate.  A general formula for the eigenvalue of the cubic Casimir operator
in an arbitrary $\mathfrak{su}(N)$ representation
[or equivalently the cubic index $I_3(R)$, which is related
to $C_{3R}$ by \eq{casimir3}] can be found
in refs.~\cite{banks,okubo1,patera, baha}.

\section{\texorpdfstring{Path integral treatment of two-component fermion propagators}{Path integral treatment of two-component fermion propagators}}
\renewcommand{\theequation}{F.\arabic{equation}}
\renewcommand{\thefigure}{F.\arabic{figure}}
\renewcommand{\thetable}{F.\arabic{table}}
\setcounter{equation}{0}
\setcounter{figure}{0}
\setcounter{table}{0}

In \sec{subsec:fermionprops} we derived the two-component
fermion propagators in momentum space, which are the Fourier
transforms of the free-field expectation values of time-ordered
products of two two-component fermion fields, for example,
\beq
\bra{0}T\xi_\alpha(x) \xi^\dagger_{\dot{\beta}}(y)\ket{0}_{\rm FT}
\equiv\int
d^4 w\bra{0}T\xi_\alpha(x)\xi^\dagger_{\dot{\beta}}(y)\ket{0}
e^{\BDpos ip\cdot w}\,,\qquad w\equiv x-y
\,,
\eeq
where the (translationally invariant) expectation values such as
$\bra{0}T\xi_\alpha(x)\xi^\dagger_{\dot{\beta}}(y)\ket{0}$ are functions
of the coordinate difference $w\equiv x-y$.  In \sec{subsec:fermionprops},
the Fourier transforms of these quantities
were computed by using the free-field expansion obtained from the
canonical quantization procedure, and then evaluating the
resulting spin sums.
In this Appendix, we provide a derivation of the
same result by employing path integral techniques.
We follow the analysis
given in Appendix~C of ref.~\cite{dedes2} (with a few minor changes
in notation).  For a similar
textbook treatment of two-component fermion propagators see for
example \Ref{Ramond}. For the analogous treatment of the four-component
fermion propagator, see for example \Ref{Peskin:1995ev}.

We first consider the action for a
single massive neutral two-component fermion
$\xi_\alpha(x)$, coupled to an anticommuting
two-component fermionic source term $J_\alpha(x)$
[cf.~\eq{lagsingleMajorana}]:
\beq
S=  \int d^4x \,(\mathscr{L}+ J\xi +\xi^\dagger J^\dagger)=
\int d^4x \,\Bigr\{\half \left[
i\xi^\dagger\sigmabar^\mu\partial_\mu\xi
+ i\xi\sigma^\mu\partial_\mu\xi^\dagger
-  m (\xi \xi + \xi^\dagger \xi^\dagger )\right]
+ J\xi +\xi^\dagger J^\dagger\Bigl\}\,,
\label{app-action}
\eeq
where we have split the kinetic energy term symmetrically into two terms. The
generating functional is given by
\beq
W[J,\, J^\dagger\,]= N\int\mathcal{D}\xi\,\mathcal{D}\xi^\dagger\;
e^{iS[\xi,\,\xi^\dagger,\,J,\, J^\dagger]}\,,
\label{app-functal}
\eeq
where $N$ is a normalization factor chosen such that $W[0,0]=1$
and $\mathcal{D}\xi\,\mathcal{D}
\xi^\dagger$ is the integration measure. It is convenient to
Fourier transform the fields $\xi(x),\,\xi^\dagger(x)$ and sources
$J(x),\, J^\dagger(x)$ in \eq{app-functal}, and rewrite the action in terms of the
corresponding Fourier coefficients
$\widehat\xi(p),\,\widehat{\xi}^\dagger(p)\,,
\widehat J(p)$ and $\widehat{J}^{\,\dagger}(p)$:
\beqa
\xi_\alpha(x) &=& \int \frac{d^4p}{(2\pi)^4} e^{\BDneg i p\newcdot x} \widehat\xi_\alpha(p)\,,\qquad
\xi^\dagger_{\dot{\alpha}}(x) = \int \frac{d^4p}{(2\pi)^4} e^{\BDpos i p\newcdot x}
\widehat{\xi}^{\,\dagger}_{\dot\alpha}(p)\,, \label{ftxi}\\
J_\alpha(x) &=& \int \frac{d^4p}{(2\pi)^4} e^{\BDneg i p\newcdot x} \widehat J_
\alpha(p)\,,\qquad
J^\dagger_{\dot{\alpha}}(x) = \int \frac{d^4p}{(2\pi)^4} e^{\BDpos
i p\newcdot x}
\widehat{J}^{\,\dagger}_{\dot\alpha}(p)\,.\label{ftj}
\eeqa
Furthermore, we introduce the integral representation of the delta
function:
\beq
\delta^{(4)}(x-x') = \int \frac{d^4p}{(2\pi)^4} e^{\BDneg ip\newcdot(x-x')}\,.
\eeq

In order to rewrite \eq{app-functal} in a more convenient matrix form,
we introduce the following definitions:
\beq \label{OXM}
\Omega(p)\equiv \left(\begin{array}{c}
\widehat{\xi}^{\,\dagger\dot{\alpha}}(-p)\\[2mm]
\widehat{\xi}_\alpha(p) \end{array}\right)\,,
\qquad
X(p)\equiv \left(\begin{array}{c}\widehat{J}_{{\alpha}}(p) \\[2mm]
\widehat{J}^{\,\dagger\dot\alpha} (-p)\end{array}\right)\,, \qquad
\mathcal{M}(p)\equiv \left(\begin{array}{cc}
\BDpos p\newcdot\sigma_{\alpha\dot{\beta}}
 & \quad -m\,\delta_\alpha{}^\beta \\[2mm] \!\!
-m \,\delta^{\dot{\alpha}}{}_{\dot{\beta}}
 & \quad \!
\BDpos p\newcdot\sigmabar^{\dot{\alpha}\beta}
\end{array}
\right) \,.
\eeq
Note that $\mathcal{M}$ is an hermitian matrix. We can then rewrite the action
[\eq{app-action}] in the following matrix form [after using
\eqs{zonetwo}{barzonetwo} to write the product of the spinor field and
the source in a symmetrical fashion]:
\beq
S=\frac{1}{2}\int \frac{d^4 p}{(2\pi)^4}\,
\left(\Omega^\dagger \mathcal{M}\Omega + \Omega^\dagger X
  +X^\dagger\Omega\right) \,.
\label{separate}
\eeq
The linear term in the field $\Omega$ can be removed by a field
redefinition
\beq
\Omega'=\Omega+\mathcal{M}^{-1}X\,.
\label{trasfn}
\eeq
In terms of $\Omega^\prime$, the action now takes the convenient form:
\beq
S=\frac{1}{2}\int \frac{d^4p}{(2\pi)^4}\,
\left(\Omega^{\prime\dagger}\mathcal{M}\Omega'
-X^\dagger\mathcal{M}^{-1}X\right)\,,
\label{action-matrix}
\eeq
where the inverse of the matrix $\mathcal{M}$ is given by
\beq
\mathcal{M}^{-1}=\frac{1}{p^2 \BDminus m^2}
\left(\begin{array}{cc}
p\newcdot\sigmabar^{\dot{\alpha}{\beta}}
 &\quad
\BDpos m\,\delta^{\dot{\alpha}}{}_{\dot{\beta}} \\[2mm]
\;
\BDpos m \,\delta_{{\alpha}}{}^{{\beta}} & \quad \;
p\newcdot\sigma_{{\alpha}\dot{\beta}}
\end{array}\right)\,.
\eeq

The Jacobian of the field transformation given in \eq{trasfn} is
unity.  Hence, one can insert
the new action, \eq{action-matrix}, in the
generating functional, \eq{app-functal} to obtain (after dropping the
primes on the two-component fermion fields):
\beqa
W[\widehat{J},\,\widehat{J}^{\,\dagger}\,]&=&
N \int \mathcal{D}\xi \,\mathcal{D} \xi^\dagger
\exp\left\{\frac{i}{2} \int \frac{d^4p}{(2\pi)^4}\,
\left(\Omega^{\dagger} \mathcal{M}\Omega
    -X^\dagger\mathcal{M}^{-1}X\right)\right\} \nonumber \\[6pt]
&=& N \left[ \int \mathcal{D}\xi\,\mathcal{D}\xi^\dagger \, \exp\left\{
    \frac{i}{2}\Omega^\dagger \mathcal{M}\Omega \right\} \right]
\,\exp\left\{-\frac{i}{2}
  \int \frac{d^4p}{(2\pi)^4}\, X^\dagger \mathcal{M}^{-1}X \right\} \nonumber \\[6pt]
&=&\exp\left\{-\frac{i}{2} \int \frac{d^4p}{(2\pi)^4}\, X^\dagger
  \mathcal{M}^{-1}X \right\}\,, \label{last-term}
\eeqa
where we have defined the normalization constant $N$ such that $W[0,0]=1$.
Inserting the explicit forms for $X$ and $\mathcal{M}$ into
\eq{last-term}, we obtain
\beqa
\hspace{-0.4in}
W[\widehat{J},\,\widehat{J}^{\,\dagger}\,]
&=&\exp\Biggl\{-\frac{1}{2}\int \frac{d^4p}{(2\pi)^4}\,
\Biggl(\widehat J^\alpha(-p)
\frac{ip\newcdot\sigma_{\alpha\dot{\beta}}}{p^2 \BDminus m^2}
\widehat{J}^{\,\dagger \dot\beta}(-p)
+\widehat{J}^{\,\dagger}_{\dot\alpha}(p)
\frac{ip\newcdot\sigmabar^{\dot{\alpha}\beta}}{p^2 \BDminus m^2} \widehat{J}_{\beta}(p)
 \nonumber\\[2mm]
&& \qquad\qquad\qquad\quad
+\widehat J^\alpha(-p) \frac{\BDpos im\delta_\alpha{}^\beta}
{p^2 \BDminus m^2}\widehat{J}_\beta(p)
+\widehat{J}^{\,\dagger}_{\dot\alpha}(p)
\frac{\BDpos im\delta^{\dot\alpha}{}_{\dot \beta}}{p^2 \BDminus m^2}
\widehat{J}^{\,\dagger\dot\beta}(-p)\Biggr)\Biggr\}\,.
\label{xyxy2}
\eeqa
Using \eq{europeanvacation}, it is convenient to rewrite
the first two terms of the integrand on the right-hand side
of \eq{xyxy2} in two different ways:
\beqa \label{xyxy3}
&&\frac{1}{2}\int \frac{d^4p}{(2\pi)^4}\,
\left[\widehat J^\alpha(-p)
\frac{ip\newcdot\sigma_{\alpha\dot
{\beta}}}{p^2 \BDminus m^2} \widehat{J}^{\,\dagger\dot\beta}(-p)
+\widehat{J}^{\,\dagger}_{\dot\alpha}(p)
\frac{ip\newcdot\sigmabar^{\dot{\alpha}\beta}}{p^2 \BDminus m^2}
\widehat{J}_{\beta}(p)\right]\nonumber \\[6pt]
&&\qquad\qquad =\int \frac{d^4p}{(2\pi)^4}\,
\widehat J^\alpha(-p)
\frac{ip\newcdot\sigma_{\alpha\dot{\beta}}}{p^2 \BDminus m^2}
\widehat{J}^{\,\dagger\dot\beta}(-p)
= \int \frac{d^4p}{(2\pi)^4}\,\widehat{J}^{\,\dagger}_{\dot\alpha}(p)
\frac{ip\newcdot\sigmabar^{\dot{\alpha}
\beta}}{p^2 \BDminus m^2} \widehat{J}_{\beta}(p)\,,
\eeqa
where we have changed integration variables from $p\to -p$ in relating
the two terms above.
The vacuum expectation value of the time-ordered product of two spinor
fields in configuration space is obtained by taking two
functional derivatives of the generating functional
with respect to the sources $J$ and $ J^\dagger$ and then setting
$J= J^\dagger=0$ at the end of the computation
(e.g., see \Ref{Peskin:1995ev}).  For example,
\beqa
\left.\left(-i\frac{\overrightarrow\delta}{\delta J^\alpha (x_1)}\right)
W[J, J^\dagger]\left(-i\frac{\overleftarrow\delta}
{\delta  J\l2sup{\dagger\dot\beta}(x_2)}\right)
\right|_{J={J}^\dagger= 0} &=&  N\int\mathcal{D}\xi\,\mathcal{D}\xi^\dagger\;
\xi_\alpha(x_1) {\xi}^\dagger_{\dot\beta}(x_2) \exp
\left ( i\int d^4 x\,
\mathscr{L}
\right )
\nonumber \\[7pt]
&=&\langle0| T \xi_\alpha(x_1) {\xi}^\dagger_{\dot\beta}(x_2) |0\rangle\,,
\eeqa
where the functional derivatives act in the indicated direction
(which ensures that no extra minus signs are generated due to the
anticommutativity properties of the sources and their functional
derivatives).  To obtain the two-point functions involving the product
of two spinor fields with different combinations of
dotted and undotted spinors, it may be more convenient to write
$J\xi=\xi J$ and/or $\xi^\dagger J^\dagger= J^\dagger \xi^\dagger$
in \eq{app-functal}.  One can then easily verify the following
expressions for the four possible two-point functions:
\beqa
\langle0| T \xi_\alpha(x_1) {\xi}^\dagger_{\dot\beta}(x_2) |0\rangle
&=&\left.\left(-i\frac{\overrightarrow\delta}{\delta J^\alpha (x_1)}\right)
W[J, J^\dagger]\left(-i\frac{\overleftarrow\delta}
{\delta J\l2sup{\dagger\dot\beta}(x_2)}\right)
\right|_{J={J}^\dagger= 0}\,, \label{jj1} \\[6pt]
\langle0| T {\xi}^{\dagger\dot\alpha}(x_1) \xi^\beta(x_2) |0\rangle
&=&\left.\left(-i\frac{\overrightarrow\delta}
{\delta  J^\dagger_{\dot\alpha}(x_1)}\right)W[J, J^\dagger]
\left(-i\frac{\overleftarrow\delta}{\delta J_\beta(x_2)}\right)
\right|_{J={J}^\dagger= 0}\,, \label{jj2} \\[6pt]
\langle0| T {\xi}^{\dagger\dot\alpha}(x_1)
{\xi}^\dagger_{\dot\beta}(x_2) |0\rangle
&=&\left.\left(-i\frac{\overrightarrow\delta}
{\delta  J^\dagger_{\dot\alpha}(x_1)}\right)W[J, J^\dagger]
\left(-i\frac{\overleftarrow\delta}{\delta J^{\dagger\dot\beta}(x_2)}\right)
\right|_{J={J}^\dagger= 0}\,, \label{jj3} \\[6pt]
\langle0| T \xi_\alpha(x_1) \xi^\beta(x_2) |0\rangle
&=&\left.\left(-i\frac{\overrightarrow\delta}{\delta J^\alpha (x_1)}\right)
W[J, J^\dagger]\left(-i\frac{\overleftarrow\delta}
{\delta J_\beta(x_2)}\right) \right|_{J={J}^\dagger= 0}\,. \label{jj4}
\eeqa

As an example, we provide details for the evaluation of \eq{jj1}.
Using \eqs{xyxy2}{xyxy3},
we obtain:
\beq \label{toprod}
\langle0| T \xi_\alpha(x_1) {\xi}^\dagger_{\dot\beta}(x_2) |0\rangle
=\frac{\overrightarrow\delta}{\delta J^\alpha (x_1)}
\left(\int \frac{d^4p}{(2\pi)^4}\,
\widehat J^\alpha(-p)\frac{ip\newcdot
\sigma_{\alpha\dot{\beta}}}{p^2 \BDminus m^2}
\widehat{J}^{\,\dagger\dot\beta}(-p)\right)
\frac{\overleftarrow\delta}{\delta  J^{\dagger\dot\beta}(x_2)}\,.
\eeq
The chain rule for functional differentiation and the inverse
Fourier transforms of \eq{ftj} yield:
\beqa
\frac{\delta}{\delta J^\alpha (x_1)}&=&
\int d^4p_1\,\frac{\delta \widehat{J}^\beta(-p_1)}
{\delta J^\alpha(x_1)}
\frac{\delta}{\delta \widehat J^\beta (-p_1)}
= \int d^4p_1\, e^{\BDneg ip_1\newcdot x_1}
\frac{\delta}{\delta \widehat J^\alpha(-p_1)}\,,\label{chain1}\\[6pt]
\frac{\delta}{\delta J^{\dagger\dot\beta}(x_2)}&=&
\int d^4p_2\,\frac{\delta \widehat{ J}^{\,\dagger\dot\alpha}(-p_2)}
{\delta J^{\dagger\dot\beta}(x_2)}
\frac{\delta}{\delta\widehat{J}^{\,\dagger\dot\alpha}(-p_2)}
= \int d^4p_2\, e^{\BDpos ip_2\newcdot x_2}
\frac{\delta}{\delta \widehat{ J}^{\,\dagger\dot\beta}(-p_2)}\,.
\label{chain2}
\eeqa
Applying \eqs{chain1}{chain2} to \eq{toprod}, we obtain:
\beq
\langle0| T \xi_\alpha(x_1) {\xi}^\dagger_{\dot\beta}(x_2) |0\rangle
= \int \frac{d^4p}{(2\pi)^4}\,e^{\BDneg ip\newcdot(x_1-x_2)}
\frac{ip\newcdot\sigma_{\alpha\dot{\beta}}}{p^2 \BDminus m^2}\,,
\eeq
which is equivalent to \eq{ft1} of \sec{subsec:fermionprops}.
With the same  methods applied to \eqst{jj2}{jj4},
one can easily reproduce the results of \eqst{ft2}{ft4}.

We next consider the action for a single massive Dirac two-component
fermion. We shall work in a basis of fields where the action,
including external anticommuting sources, is given by
\beqa
S[\chi,\chi^\dagger,\eta,\eta^\dagger,J_\chi, J^\dagger_{\chi},J_\eta,
 J^\dagger_{\eta}] &=& \int\! d^4x \Bigl [
i\chi^\dagger\sigmabar^\mu\partial_\mu\chi +
  i\eta^\dagger\sigmabar^\mu\partial_\mu\eta
  -m(\chi\eta+\chi^\dagger\eta^\dagger)
\nonumber \\ &&
+J_\chi\chi+\chi^\dagger J^\dagger_{\chi}+J_\eta\eta+
\eta^\dagger J^\dagger_{\eta}\Bigr ].
\label{diracaction}
\eeqa
Following the techniques employed above, we introduce
Fourier coefficients for all the fields and sources and define
\beqa \label{OXM4}
\phantom{line} \nonumber \\[-30pt]
\Omega_c(p)\equiv \left(\begin{array}{c}
\widehat{\eta}^{\,\dagger\dot{\alpha}}(-p)\\[2mm]
 \widehat\chi_\alpha (p)\end{array}\right)\,,
\qquad\qquad
X_c(p)\equiv \left(\begin{array}{c}
\widehat{J}_{\eta\alpha}(p)
\\[2mm] \widehat{ J}^{\,\dagger\dot\alpha}_{\chi}(-p)
 \end{array}\right)\,.
\eeqa
The action functional, \eq{diracaction}, can then rewritten in matrix
form as before (but with no overall factor of $1/2$):
\beq
S=\int \frac{d^4 p}{(2\pi)^4}\,
\left(\Omega_c^\dagger \mathcal{M}\Omega_c + \Omega_c^\dagger X_c
  +X_c^\dagger\Omega_c\right) \,,
\label{separatedirac}
\eeq
where $\mathcal{M}$ is again given by \eq{OXM}.
The remaining calculation proceeds
as before with few modifications, and yields the Dirac
two-component fermion free-field propagators given in
\eqst{chprop1}{chprop4}.

\section{\texorpdfstring{Correspondence to four-component spinor notation}{Correspondence to four-component spinor notation}}

\subsection{Dirac gamma matrices and four-component spinors}
\label{Diracgamma}
\renewcommand{\theequation}{G.1.\arabic{equation}}
\renewcommand{\thefigure}{G.1.\arabic{figure}}
\renewcommand{\thetable}{G.1.\arabic{table}}
\setcounter{equation}{0}
\setcounter{figure}{0}
\setcounter{table}{0}

\indent
In four-dimensional Minkowski space,
four-component spinor notation employs four-component Dirac spinor
fields and the $4\times 4$ Dirac gamma matrices, whose
defining property is:
\beq \label{diracgamma}
\{\gamma^\mu,\gamma^\nu\}= \BDpos 2g^{\mu\nu}\mathds{1}\,,
\eeq
where $\mathds{1}$ is the $4\times 4$ identity matrix.

The correspondence between the two-component spinor notation
and the
four-component Dirac spinor notation is most easily exhibited
in the basis in which $\gamma_5$ is diagonal (this is called the
\textit{chiral} representation\footnote{For a review of other
representations of the Dirac gamma matrices and their properties,
see e.g.~refs.~\cite{Diracmatrices,Uschersohn}.}).
In 2$\times$2 blocks,
the gamma matrices are given by:\footnote{Employing the conventions
for the sigma matrices described in \app{A}, it follows that the
definition of $\gamma^\mu$ is independent of the choice of metric
signature, whereas $\gamma_\mu \equiv g_{\mu\nu} \gamma^\nu$
changes sign under a reversal of the metric signature.  In the
metric signature convention with $g_{00}=+1$, our gamma matrix
conventions follow those of \Ref{Peskin:1995ev}, whereas in the
convention with $g_{00}=-1$, our gamma matrix
conventions follow those of \Ref{Srednicki}.
}
\beq \label{gamma4}
\gamma^\mu
\equiv \begin{pmatrix} 0 & \sigma^\mu_{\alpha{\dot{\beta}}}\\
\sigmabar^{\mu\,{\dot{\alpha}}\beta} & 0\end{pmatrix}
\,,
\qquad\qquad
\gamma_5 \equiv i\gamma^0\gamma^1\gamma^2\gamma^3=\begin{pmatrix}
-\delta_\alpha{}^\beta & 0\\ 0 & \delta^{\dot{\alpha}}{}_{\dot{\beta}}
\end{pmatrix}
\,,
\eeq
and the $4\times 4$ identity matrix that appears in \eq{diracgamma}
can be written as:
\beq
\mathds{1}=\begin{pmatrix} \delta_\alpha{}^\beta & \,\, 0 \\
0 & \,\, \delta^{\dot{\alpha}}{}_{\dot{\beta}}\end{pmatrix}\,.
\eeq
In addition, we identify the generators of the Lorentz group
in the $(\half,0)\oplus (0,\half)$ representation:\footnote{In most textbooks,
$\Sigma^{\mu\nu}$ is called $\sigma^{\mu\nu}$.  Here, we use the
former symbol so that there is no confusion with 
$\sigma^{\mu\nu}{}_\alpha{}^{\beta}$ given in \eq{sigmamunu}.}
\beq \label{Sigmamunu}
\half\Sigma^{\mu\nu}\equiv\frac{i}{4}[\gamma^\mu,\gamma^\nu]=
\begin{pmatrix} \sigma^{\mu\nu}{}_\alpha{}^\beta & \,\, 0\\
0 & \,\,\sigmabar^{\mu\nu}{}^{\dot{\alpha}}{}_{\dot{\beta}}\end{pmatrix}\,,
\eeq
where $\Sigma^{\mu\nu}$ satisfies the duality relation,
\beq
\gamma\ls{5}\Sigma^{\mu\nu}=\half i \epsilon^{\mu\nu\rho\tau}\Sigma_{\rho\tau}\,.
\eeq
\Eq{Sigmamunu} yields the spin-$\half$ angular momentum matrix 
representation $\{\half\Sigma^i\}$, $i=1,2,3$, where
\beq
\Sigma^i\equiv \half\epsilon^{ijk}\Sigma_{jk}=\gamma^0 \gamma^i \gamma\ls{5}=
\begin{pmatrix} -(\sigma^0\sigmabar^i)_\alpha{}^{\beta} 
&\,\,\, 0 \\ 0 & \,\,\,
(\sigmabar^0\sigma^i)^{\dot\alpha}{}_{\dot\beta}
\end{pmatrix}\,.
\eeq

A four-component \textit{Dirac spinor} field, $\Psi(x)$, is made up of two
mass-degenerate two-component spinor fields, $\chi_\alpha(x)$ and
$\eta_\alpha(x)$, of opposite U(1)-charge as follows:
\beq
\Psi(x)\equiv\begin{pmatrix} \chi_\alpha(x)
\\[4pt] \eta^{\dagger\dot{\alpha}}(x)\end{pmatrix}\,.
\label{general4comp}
\eeq
We next introduce the chiral projections operators,\footnote{In the earlier
literature, a different set of conventions for the sigma matrices
in which the roles of $\sigma$ and $\sigmabar$ were reversed
[e.g, as in \eqs{sigswap1}{sigswap2}] resulted in
$\gamma\ls{5}=\rm{diag}(\mathds{1}_{2\times 2}\,,\,
-\mathds{1}_{2\times 2})$ in the chiral representation, which differs
from our convention by an overall sign [cf.~\eq{gamma4}].  As a
result, in this latter convention, $P_L$ [$P_R$] projects out the
raised dotted [lowered undotted] two-component spinor field.  This
latter convention is still prevalent in the literature of the spinor
helicity method (see footnote~\ref{fnwarned} in \app{I.2}).}
\beq \label{PLPR}
P_L\equiv \half(\mathds{1}-\gamma_5)=\begin{pmatrix} \delta_\alpha{}^\beta
& \,\, 0 \\ 0 & \,\, 0\end{pmatrix}
\,, \qquad {\rm and}
\qquad P_R\equiv \half(\mathds{1}+\gamma_5)=\begin{pmatrix} 0
& \,\, 0 \\ 0 & \,\,  \delta^{\dot{\alpha}}{}_{\dot{\beta}}\end{pmatrix}\,,
\eeq
and the (left and right-handed) \textit{Weyl spinor} fields,
$\Psi_L(x)$ and $\Psi_R(x)$, which are defined by:
\beq \label{plprdefs}
\Psi_L(x)\equiv P_L\Psi(x) = \begin{pmatrix} \chi_\alpha(x)
\\[4pt] 0\end{pmatrix}\,,\qquad\qquad \Psi_R(x)\equiv P_R\Psi(x)=
\begin{pmatrix}0 \\[4pt]
\eta^{\dagger\dot{\alpha}}(x)\end{pmatrix}\,.
\eeq
Equivalently, one can define
the Weyl spinors $\Psi_L$ and $\Psi_R$ as
the four-component spinor eigenstates
of $\gamma\ls{5}$ with corresponding eigenvalues $-1$ and $+1$, respectively
(i.e., $\gamma\ls{5}\Psi_{L,R}=\mp\Psi_{L,R}$).

The Dirac conjugate field $\Psibar(x)$ 
and the charge conjugate field $\Psi^C(x)$ are defined by:
\beqa
\Psibar(x)&\equiv&\Psi^\dagger A =
\left(\eta^\alpha(x),\> \chi^\dagger_{\dot{\alpha}}(x)\right)\,,
\label{psibar}\\[6pt]
\Psi^C(x)&\equiv&C\Psibar\llsup{\,\T}(x)=
\begin{pmatrix} \eta_\alpha(x) \\[4pt] \chi^{\dagger\dot{\alpha}}(x)
\end{pmatrix}\,, \label{CConj}
\eeqa
where the Dirac conjugation matrix $A$ and the charge conjugation
matrix $C$ satisfy~\cite{abcmatrices,plonge,bailinweak}:
\beq \label{acdef4}
A\gamma_\mu A^{-1}=\gamma_\mu^\dagger\,,\qquad\qquad\qquad
C^{-1} \gamma_\mu C=-\gamma_\mu^{\T}\,.
\eeq
It is convenient to introduce a notation for left and right-handed 
charge-conjugated fields (which are also Weyl spinor fields)
following the conventions
of \refs{Lang}{langacker},\footnote{The reader is
warned that the opposite convention is often employed
in the literature (e.g., see \Ref{MajWeyl})
in which $\Psi\llsup{C}\ls{L}$ is a right-handed field and
$\Psi\llsup{C}\ls{R}$ is a left-handed field. \label{fnlang}}
\beqa
\Psi^C\ls{L}(x)\equiv P_L\Psi^C(x)&=&C\Psibar_R^{\T}(x)
=[\Psi\ls{R}(x)]^C\,, \label{psiLC} \\
\Psi^C\ls{R}(x)\equiv P_R\Psi^C(x)&=&C\Psibar_L^{\T}(x)
=[\Psi\ls{L}(x)]^C\,. \label{psiRC}
\eeqa

To fix the properties of $A$ and $C$,
it is conventional to impose two additional conditions:
\beq \label{extraconditions}
\Psi=A^{-1}\Psibar^\dagger\,, \qquad\qquad (\Psi^C)^C=\Psi\,.
\eeq
The first of these conditions together with \eq{psibar} is equivalent
to the statement that $\Psibar\Psi$ is hermitian.
The second condition corresponds to the statement
that the (discrete)
charge conjugation transformation applied twice is equal to the
identity operator.
Using \eqs{acdef4}{extraconditions}
and the defining property of the gamma matrices [\eq{diracgamma}],
one can show (independently of the gamma matrix representation)
that the matrices $A$ and $C$ must satisfy:
\beq
 A^\dagger = A\,,\quad\qquad  C^{\T}=-C\,,\qquad\quad  (AC)^{-1}=(AC)^*\,.
\label{acproperties}
\eeq

Following \Ref{sohniusapp}, it is convenient to introduce a
matrix $D$ such that
\beq \label{Dmatrix}
D\equiv CA^{\T}\,,\qquad\qquad D^{-1}\gamma_\mu D=-\gamma_\mu^\ast\,,
\eeq
and $D^\ast D=DD^\ast=\mathds{1}$.
The charge-conjugated four-component spinor is then given by:

\beq \label{ccd}
\Psi^C(x)\equiv D\Psi^\ast(x)\,.
\eeq
A four-component \textit{Majorana spinor} field,
$\Psi_M(x)$, is defined by imposing
the constraint $\Psi^C(x)=\Psi(x)$ on a four-component Dirac spinor,
which sets $\eta=\chi$.
That is, the \textit{Majorana condition} is
\beq \label{psiCD}
\Psi_M(x)=D\Psi_M^*(x)=
\begin{pmatrix}\chi_\alpha(x)\\ \chi^{\dagger\dot{\alpha}}(x)
\end{pmatrix}\,.
\eeq
For a review of the Majorana field and its
properties, see e.g.~\refs{Case}{MajReview}.
\clearpage

For completeness, we also introduce a matrix $B$ that
satisfies~\cite{WPauli,abcmatrices,plonge,bailinweak}:
\beq \label{bdef4}
B\equiv -C^{-1}\gamma\ls{5}\,,\qquad\quad
B\gamma_\mu B^{-1}=\gamma_\mu^{\T}\,.
\eeq
The matrix $B$ arises in the study of time reversal invariance of the
Dirac equation.
In the chiral representation, $A$, $B$, $C$ and $D$ are explicitly given by
\beqa
A&=&\begin{pmatrix} 0 & \quad \delta^{\dot{\alpha}}{}_{\dot{\beta}} \\
\delta_\alpha{}^\beta & \quad0\end{pmatrix}\,,\qquad\qquad\qquad
C =\begin{pmatrix} \epsilon_{\alpha\beta}& \quad 0\\
                 0 & \quad\epsilon^{\dot{\alpha}\dot{\beta}}\end{pmatrix}\,,
\label{acmatrix} \\[8pt]
B &=& \begin{pmatrix} \epsilon^{\alpha\beta}& \quad 0\\
                 0 & \quad -\epsilon_{\dot{\alpha}\dot{\beta}}
\end{pmatrix}\,, \qquad\qquad\qquad\!\!\!
D = \begin{pmatrix}0 &  \quad\epsilon_{\alpha\beta}\\
                \epsilon^{\dot{\alpha}\dot{\beta}} & \quad 0\end{pmatrix}
\,.\label{bdmatrix}
\eeqa
Note the \textit{numerical} equalities,
$A=\gamma^0$, $B=\gamma^1\gamma^3$,
$C=i\gamma^0\gamma^2$ and $D=-i\gamma^2$.
However these identifications do not respect either
the structure of the undotted and dotted spinor indices specified in
\eqs{acmatrix}{bdmatrix}, or the four-component
spinor index structure introduced below 
[cf.~\eqs{deltaGamma}{ABCDinv}].\footnote{When treated
as ordinary $4\times 4$ matrices $A$, $B$, $C$
and $D$ are unitary.
But when written in $2\times 2$ block form [noting that
$\delta^{\dot{\alpha}}{}_{\dot{\beta}}=
(\delta^\alpha{}_\beta)^\ast$ and $\epsilon^{\dot{\alpha}\dot{\beta}}\equiv
(\epsilon^{\alpha\beta})^\ast$, as indicated  below 
\eqs{epssign}{deltaK}], 
the products $AA^\dagger$,
$BB^\dagger$, $CC^\dagger$ and $DD^\dagger$ are not
covariant with respect to the dotted and undotted
two-component spinor indices.  
Similarly, these matrix products are not covariant with
respect to the four-component spinor indices.
In practice,
only covariant combinations of $A$, $B$, $C$, $D$
and the four-component spinor fields arise in typical calculations.}
In translating between
two-component and four-component spinor notation, 
\eqs{acmatrix}{bdmatrix} should always be used.  In 
practical four-component spinor calculations,
there is often no harm in employing the numerical
values for $A$, $B$, $C$ and~$D$.

Using \eqst{acdef4}{bdef4}, the following results are easily derived:
\beqa
A\Gamma A^{-1} &=& \eta\ls{\Gamma}^A\Gamma^\dagger\,,\qquad\quad
\eta\ls{\Gamma}^A=\begin{cases} +1\,, &
\text{\quad for $\Gamma=\mathds{1}\,,\,\gamma^\mu\,,\,
\gamma^\mu\gamma\ls{5}\,,\,\Sigma^{\mu\nu}$\,,}\\  -1\,, &
\text{\quad for $\Gamma=\gamma\ls{5}
\,,\,\Sigma^{\mu\nu}\gamma\ls{5}$\,,}\end{cases} \label{aagamma}
\\[6pt]
B\Gamma B^{-1} &=& \eta\ls{\Gamma}^B\Gamma^{\T}\,,\qquad\quad
\eta\ls{\Gamma}^B=\begin{cases} +1\,, &
\text{\quad for $\Gamma=\mathds{1}\,,\,\gamma\ls{5}\,,\,\gamma^\mu$\,,}
\\  -1\,, &
\text{\quad for $\Gamma=\gamma^\mu\gamma\ls{5}\,,\,\Sigma^{\mu\nu}
\,,\,\Sigma^{\mu\nu}\gamma\ls{5}$\,,}\end{cases} \label{bbgamma} \\[6pt]
C^{-1}\Gamma C &=& \eta\ls{\Gamma}^C \Gamma^{\T}\,,\qquad\quad
\eta\ls{\Gamma}^C=\begin{cases} +1\,, &
\text{\quad for $\Gamma=\mathds{1}\,,\,\gamma\ls{5}\,,\,
\gamma^\mu\gamma\ls{5}$\,,} \\  -1\,, &
\text{\quad for $\Gamma=\gamma^\mu\,,\,\Sigma^{\mu\nu}
\,,\,\Sigma^{\mu\nu}\gamma\ls{5}$\,,}\end{cases} \label{ccgamma} 
\eeqa
\beqa
D^{-1}\Gamma D &=& \eta\ls{\Gamma}^D \Gamma^{\ast}\,,\qquad\quad
\eta\ls{\Gamma}^D=\begin{cases} +1\,, &
\text{\quad for $\Gamma=\mathds{1}\,,\,\gamma^\mu\gamma\ls{5}\,,\,
\Sigma^{\mu\nu}\gamma\ls{5}$\,,} \\  -1\,, &
\text{\quad for $\Gamma=\gamma^\mu\,,\,\gamma\ls{5}
\,,\,\Sigma^{\mu\nu}$\,.}\end{cases} \label{ddgamma}
\eeqa

The Lorentz transformation properties of the four-component spinor
field can be determined from those of the two-component
spinor fields given in \sec{sec:notations}.  The $4\times 4$ representation
matrices of the Lorentz group in the 
$(\half,0)\oplus(0,\half)$ representation
are given by
\beq \label{dsM}
\mathds{M}=\begin{pmatrix} M & \quad 0 \\
0 & \quad (M^{-1})^\dagger \end{pmatrix}
=\exp\left(-\frac{i}{4}\theta_{\mu\nu}\Sigma^{\mu\nu}\right)
\simeq \mathds{1}_{4\times 4}-\quarter i\theta_{\mu\nu}\Sigma^{\mu\nu}\,,
\eeq
where the infinitesimal forms of $M$ and $(M^{-1})^\dagger$ are given
in \eqs{Minf}{MDinf}.  Two useful identities that follow from 
\eqss{aagamma}{ccgamma}{dsM} are:\footnote{Note that \eq{C1MC}
is a direct consequence of the
identities in two-component spinor notation given in \eqs{covar1}{covar2}.}
\beqa 
A\,\mathds{M}A^{-1}&=&(\mathds{M}^{-1})^{\dagger}\,,\label{A1MA} \\
C^{-1}\mathds{M}C&=&(\mathds{M}^{-1})^{\T}\,.\label{C1MC}
\eeqa

The four-component Dirac or Majorana spinor, $\Psi_a\,,$ is assigned
a lowered spinor index $a$, and is defined in terms of two-component
spinors by eqs.~(\ref{general4comp}) or (\ref{psiCD}), respectively.  
Four-component spinor indices,
which will be chosen in general from the beginning of the lower case Roman 
alphabet, $a,b,c,\ldots$,
can assume integer values $1,2,3,4$.
Under a Lorentz
transformation, $\Psi_a$ transforms as
\beq \label{LT1}
\Psi_a\to \mathds{M}_a{}^b\,\Psi_b\,.
\eeq
In analogy with the conventions for two-component spinor indices, we 
sum implicitly over a pair of repeated indices
consisting of a raised and a lowered spinor index.
The transformation law for the Dirac conjugate spinor (often called
the Dirac adjoint spinor),
$\overline\Psi=\Psi^\dagger A$, is
obtained from \eq{LT1} after employing $A^\dagger=A$ and \eq{A1MA},
\beq \label{LT2}
\overline\Psi\llsup{\,a}\to\overline\Psi\llsup{\,b}\,(\mathds{M}^{-1})_b{}\rsup{a}\,.
\eeq
In particular, $\overline\Psi\Psi\equiv\overline\Psi\llsup{\,a}\Psi_a$ 
is a Lorentz scalar, which justifies the assignment of a raised
spinor index for the Dirac conjugate spinor $\overline\Psi\llsup{\,a}$.

It is convenient to introduce \textit{barred} four-component spinor
indices~\cite{Brandt} in the transformation laws of the
hermitian-conjugated four-component spinors,\footnote{Of course, 
\eqst{LT1}{LT4} can also be derived directly from the corresponding
two-component spinor transformation laws of \sec{sec:notations}.}
\beqa\,
\Psi\rsup{\dagger}\ls{\bar{a}}&\to& \Psi\rsup\dagger\ls{\bar{b}}\,
(\mathds{M}^\dagger)\rsup{\bar{b}}{}_{\bar{a}}\,,\label{LT3}\\
\overline\Psi\llsup{\,\dagger\,\bar{a}}&\to& 
[(\mathds{M}^{-1})^\dagger]\rsup{\bar{a}}{}_{\bar{b}}\,
\overline\Psi\llsup{\,\dagger\,\bar{b}}\,,\label{LT4} \\[-24pt]
\phantom{line}\nonumber
\eeqa
where there is an implicit sum over the repeated lowered and raised
barred spinor indices, and
\beq \label{Psidagger}
\Psi^\dagger_{\bar a}\equiv (\Psi_a)^\dagger\,,\qquad \qquad\quad
\overline\Psi\llsup{\dagger\,\bar{a}}
\equiv(\overline\Psi\llsup{\,a})^\dagger\,.
\eeq
The spinor index structure of the Dirac conjugation matrix $A$ 
is then fixed by noting that
the Dirac conjugate spinor, 
$\overline\Psi\llsup{\,b}\equiv \Psi^\dagger_{\bar a} A^{\bar a b}$,
has a raised unbarred spinor
index, whereas the hermitian-conjugated spinor has a lowered barred
spinor index.

The charge conjugation matrix can be used to raise and lower
four-component spinor indices~\cite{Brandt},
which we shall employ in defining the spinors $\Psi^a$, $\Psi^{\dagger\,\bar a}$,
$\overline{\Psi}_a$ and $\overline{\Psi}\llsup{\,\dagger}\ls{\bar a}$,%
\footnote{In contrast 
to the epsilon symbols of the two-component spinor formalism,
here we prefer to explicitly exhibit the inverse symbols in
$(C^{-1})^{ab}$ and $(C^{-1})^{\bar a\bar b}$ 
[cf.~footnote~\ref{fntransparent}].}
\beqa
\Psi_a&=& C_{ab}\Psi^b\,,\qquad\qquad\quad
\Psi^a=(C^{-1})^{ab}\Psi_b\,,\label{Craise}\\
\Psi^\dagger_{\bar a}&=& C_{{\bar a}{\bar b}}\Psi^{\dagger\,\bar b}\,,
\qquad\qquad
\Psi^{\dagger\,\bar a}= (C^{-1})^{{\bar a}{\bar b}}
\Psi\rsup{\dagger}\ls{\bar b}\,,\label{Clower}
\eeqa
where 
\beq
C_{\bar a \bar b}\equiv (C_{ab})^*\,,\qquad\qquad 
(C^{-1})^{\bar a\bar b}\equiv [(C^{-1})^{ab}]^*\,.
\eeq
\Eqs{Craise}{Clower} also apply to $\overline\Psi\llsup{\,a}$,
$\overline\Psi_a$ and their hermitian conjugates. In particular, 
one can identify
the Dirac conjugate spinor with a lowered spinor index 
($\overline{\Psi}_a$) 
as the charge-conjugated spinor, $\Psi^C\equiv
C\overline\Psi\llsup{\,\T}$, and 
the Dirac spinor with a raised spinor index ($\Psi^a$) 
as the Dirac conjugate of the charge-conjugated spinor, 
$\overline{\Psi^C}=-\Psi^{\T}\,C^{-1}$.  That is,\footnote{For a Dirac
spinor field defined in \eq{general4comp}, $\overline{\Psi}_a(x)=\Psi^C_a(x)$ 
is given in terms of two-component spinors by \eq{CConj}, and
$\Psi^a(x)=\overline{\Psi^C}\llsup{\,a}(x)=
\left(\chi^\alpha(x)\,,\,\eta^\dagger_{\dot\alpha}(x)\right)$.} 
\beq \label{Psica}
\Psi^C_a \equiv\overline\Psi_a=C_{ab}\overline\Psi\llsup{\,b}\,,\qquad
\qquad
\overline{\Psi^C}\llsup{\,a}=\Psi^a=(C^{-1})^{ab}\Psi_b\,.
\eeq
The rules for raising and lowering spinor indices are consistent with 
the Lorentz transformation properties of \eqst{LT1}{LT4}, as
a consequence of \eq{C1MC}.  In particular, the condition for 
a self-conjugate four-component (Majorana) spinor,
$\overline{\Psi}_a\equiv\Psi^C_a=\Psi_a$, is Lorentz covariant.

Using eqs.~(\ref{acproperties}), (\ref{Craise}), (\ref{Clower}),
and the definition of the Dirac conjugate spinor, 
it then follows that:
\beqa
\phantom{line}\nonumber \\[-32pt]
\overline\Psi_a&=&(A^{-1})_{a\bar b}\Psi^{\dagger\,\bar b}\,,
\qquad\qquad\quad
\overline\Psi\llsup{\,a}=\Psi^\dagger_{\bar b}\,A^{\bar ba}\,,\label{Araise} \\
\Psi^{\,\dagger}_{\bar a} &=&
\overline\Psi\llsup{\,b}(A^{-1})_{b\bar a}\,,\qquad\qquad \quad
\Psi^{\dagger\,\bar a}=A^{\bar a b}\,\overline\Psi_b\,.\label{Alower}
\eeqa
One can check that \eqs{Araise}{Alower} are consistent with 
the Lorentz transformation properties of \eqst{LT1}{LT4}, as
a consequence of \eq{A1MA}.

In addition to the Lorentz scalar
$\overline\Psi\Psi\equiv\overline\Psi\llsup{\,a}\Psi_a$, one can construct two
additional independent Lorentz scalar quantities,\footnote{A fourth
possible Lorentz scalar, $\Psi^a\overline{\Psi}_a=(C^{-1})^{ab}C_{ac}
\Psi_b\overline{\Psi}\llsup{\,c}=-\Psi_c\overline{\Psi}\llsup{\,c}=\
\overline{\Psi}\llsup{\,c}\Psi_c$, is not independent.  Here,
we have used $C^{\T}=-C$ and the anticommutativity of the spinors.
Equivalently, $\overline{\Psi^C}\Psi\lllsup{C}=\overline{\Psi}\Psi$.
\label{fnCC}} 
\beq \label{LS1} 
-\Psi^{\T}C^{-1}\,\Psi\equiv -\Psi_a(C^{-1})^{ab}\Psi_b=\Psi^a\Psi_a\,,
\eeq
and its hermitian conjugate,
\beq
\overline\Psi\, C\,\overline\Psi\llsup{\,\T}\equiv
\overline\Psi\llsup{\,a} C_{ab}\overline\Psi\llsup{\,b}=
\overline{\Psi}\llsup{\,a}\overline{\Psi}_a=\Psi^\dagger_{\bar
  a}\Psi^{\dagger\,\bar a}=(\Psi^a\Psi_a)^\dagger\,,
\label{LS2}
\eeq
after using $C^{-1}$ and $C$ to raise and lower the appropriate
spinor indices, respectively.  The penultimate 
equality in \eq{LS2} is a consequence of \eq{Alower}.
The Lorentz invariance of $\overline{\Psi}\llsup{\,a}\Psi_a$, 
$\Psi^a\Psi_a$ and $\Psi^\dagger_{\bar a}\Psi^{\dagger\,\bar a}
\!=\!\overline{\Psi}\llsup{\,a}\overline{\Psi}_a$ is manifest and demonstrates
the power of the four-component spinor index notation developed above.
After invoking \eq{Psica}, 
we note that [analogous to \eq{suppressionrule}]
\textit{descending} contracted unbarred spinor indices and \textit{ascending}
contracted barred spinor indices can be suppressed 
in spinor-index-contracted
products.  For example,
\beq
\overline{\Psi}\llsup{\,a}\Psi_a\equiv\overline{\Psi}\Psi\,\qquad\quad
\Psi^a\Psi_a=\overline{\Psi^C}\llsup{\,a}\Psi_a\equiv\overline{\Psi^C}\Psi\,
\qquad\quad
\overline{\Psi}\llsup{\,a}\overline{\Psi}_a=\overline{\Psi}\llsup{\,a}
\Psi^C_a\equiv\overline{\Psi}\Psi\lllsup{C}\,,
\eeq
where the suppression of barred spinor indices is implicit in the definition of
$\overline\Psi\llsup{\,b}\equiv \Psi^\dagger\ls{\bar a}A^{\bar a b}$.
\clearpage

The charge-conjugated spinor can also be written as
$\Psi^C_a\equiv D_a{}^{\bar c}\Psi^\dagger_{\bar c}$ [cf.~\eq{ccd}].
The spinor index structure of $D$ (and its inverse) derives from:
\beq
D_a{}^{\bar c}\equiv C_{ab}(A^{\T})^{b\bar c}=C_{ab}A^{\bar c b}\,,
\qquad\qquad
(D^{-1})_{\bar a}{}^c\equiv (C^*A)_{\bar a}{}^c=
C_{\bar a \bar b}A^{\bar b c}\,,
\eeq
where we have used $D^{-1}=D^*$.  
Combining the results of eqs.~(\ref{Craise}), (\ref{Clower}), 
(\ref{Araise}) and (\ref{Alower}) then yields:
\beqa
\overline{\Psi}_a&=&D_a{}^{\bar c}\,\Psi^\dagger\ls{\bar c}\,,\qquad\qquad
\qquad\quad
\Psi^\dagger\ls{\bar a}=(D^{-1})_{\bar a}{}^c\,\overline\Psi_c\,,\\
\overline{\Psi}\llsup{\,a}&=&-\Psi^{\dagger\,\bar c}(D^{-1})_{\bar c}{}^a
\,,\qquad\qquad \Psi^{\dagger\,\bar a}=-\overline{\Psi}\llsup{\,c}\,
D_c{}^{\bar a}\,.
\eeqa

In summary, 
a four-component spinor $\Psi_a$ and its charge-conjugated spinor $\Psi^C_a$
possess a lowered unbarred spinor index, whereas the
corresponding Dirac conjugates, 
$\overline\Psi\llsup{\,a}$ and $\overline{\Psi^C}\llsup{\,a}$,
possess a raised unbarred spinor index.
The corresponding hermitian-conjugated spinors exhibit 
barred spinor indices (with the height of each spinor index unchanged).
Following \eqs{Craise}{Clower},
one can also lower or raise a four-component unbarred or barred spinor
index by multiplying by the appropriate matrix $C$, $C^{-1}$,
$C^*$ or $(C^{-1})^*$, respectively.

The identity matrix, the gamma matrices and their products are denoted
collectively by $\Gamma$.  The spinor index structure of 
these matrices and their inverses is given by:
\beq \label{deltaGamma}
\delta_a^b\,,\,\Gamma_a{}^b\,,\,(\Gamma^{-1})_a{}^b\,,
\eeq
where the $\delta_a^b$ are the matrix elements of
the identity matrix $\mathds{1}$.  In this case, the rows are labeled by
the lowered index and the columns are labeled by the raised index.
Note that the
quantities $\overline\Psi\llsup{\,a}\,\Gamma_a{}^b\,\Psi_b$, $\Psi^a\,
\Gamma_a{}^b\,\Psi_b$, and $\overline\Psi\llsup{\,a}\,\Gamma_a{}^b\,
\overline\Psi\llsup{\,b}$ transform as Lorentz tensors, whose rank is
equal to the number of (suppressed) spacetime indices of $\Gamma$.

For the matrices $A$, $B$, $C$, $D$ and their inverses, the spinor
index structure is given by:
\beq \label{ABCDinv}
A^{\bar a b}\,,\,(A^{-1})_{a\bar b}\,,\,
B^{ab}\,,\,(B^{-1})_{ab}\,,C_{ab}\,,\,(C^{-1})^{ab}\,,D_a{}^{\bar b}\,,\,
(D^{-1})_{\bar a}{}^b\,.
\eeq
The corresponding complex-conjugated matrices exhibit the analogous
spinor index structure with unbarred spinor indices changed to barred
spinor indices and vice versa.  
Matrix transposition interchanges rows and columns.  For example,
\beq
(\Gamma^{\T})^a{}_b\equiv\Gamma_b{}^a\,,\qquad
(A^{\T})^{a\bar b}\equiv A^{\bar b a}\,,\qquad (C^{\T})_{ab}=C_{ba}\,,
\qquad (D^{\T})^{\bar a}{}_b\equiv D_b{}^{\bar a}\,.
\eeq
Hermitian conjugation is complex conjugation followed by
matrix transposition.  For example,
\beq
(\Gamma^\dagger)^{\bar a}{}_{\bar b}\equiv (\Gamma_b{}^a)^*\,,\qquad
(A^\dagger)^{\bar a b}\equiv (A^{\bar b a})^*\,,\qquad
(C^\dagger)_{\bar a \bar b}=(C_{ba})^*\,,\qquad
(D^\dagger)^a{}_{\bar b}\equiv  (D_b{}^{\bar a})^*\,.
\eeq
Using the above results, it is straightforward to identify the
four-component spinor index structure of \eqst{diracgamma}{C1MC}.
For example, 
specifying the four-component spinor indices of 
\eq{C1MC} yields:
\beq \label{CMC}
(C^{-1})^{ab}\mathds{M}_b{}^cC_{cd}=[(\mathds{M}^{-1})^{\T}]^a{}_d
\equiv (\mathds{M}^{-1})_d{}\rsup{a}\,.
\eeq

To complete the spinor index formalism, we introduce hybrid quantities
$L$, $\overline{L}$, $R$ and $\overline{R}$ that contain an unbarred
four-component spinor index
and a two-component undotted or dotted spinor index~\cite{Setzer}:
\beqa
L_\beta{}^b&=&\left(\mathds{1}_{2\times 2}\quad
\mathds{O}_{2\times 2}\right)\,,
\qquad\quad
R^{\dot{\beta}\,b}=\left(
\mathds{O}_{2\times 2}\quad \mathds{1}_{2\times 2}\right)\,,\label{LRdef1}
\\[8pt]
\overline{L}_b{}^\beta &=& \begin{pmatrix}\mathds{1}_{2\times 2} \\
\mathds{O}_{2\times 2}\end{pmatrix}\,,\qquad\qquad\quad
\overline{R}_{b\dot{\beta}} = \begin{pmatrix}\mathds{O}_{2\times 2} \\
\mathds{1}_{2\times 2}\end{pmatrix}\,.\label{LRdef2}
\eeqa
These quantities satisfy:
\beqa
\overline{L}_a{}^\alpha L_\alpha{}^b&=&(P_L)_a{}^b\,,\qquad\qquad
L_{\alpha}{}^a\overline{L}_a{}^\beta=
\delta_\alpha{}^\beta\,,\label{setz1}\\
\overline{R}_{a\dot{\alpha}} R^{\dot{\alpha}\,b}&=&(P_R)_a{}^b\,,\qquad\qquad
R^{\dot{\alpha}\,a}\overline{R}_{a\dot{\beta}}
=\delta^{\dot{\alpha}}{}_{\dot{\beta}}\,,\label{setz2}
\eeqa
where $P_L$ and $P_R$ are the chiral projection operators defined in
\eq{PLPR}.  It then follows that:
\beqa
L_\alpha{}^a(P_L)_a{}^b&=&L_\alpha{}^b\,,\qquad\qquad
(P_L)_a{}^b\overline{L}_b{}^\beta=\overline{L}_a{}^\beta\,,\\
R^{\dot{\alpha}\,a}(P_R)_a{}^b&=&R^{\dot{\alpha}\,b}\,,\qquad\qquad
(P_R)_a{}^b\overline{R}_{b\dot{\beta}}=\overline{R}_{a\dot{\beta}}\,.
\eeqa

The hybrid quantities $L$, $\overline{L}$, $R$ and $\overline{R}$
connect objects with four-component and two-component spinor indices.
For the Dirac spinor defined in \eq{general4comp}, it follows
that:
\beqa
\chi\ls{\alpha}&=&L_\alpha{}^b\Psi_b\,,\qquad\qquad
\eta^\alpha=\Psibar^b\,\overline{L}_b{}^\alpha\,,\label{LL}\\
\eta^{\dagger\dot{\alpha}}&=& R^{\dot{\alpha}\,b}\Psi_b\,,\qquad\qquad
\chi^\dagger\ls{\dot{\alpha}}=\Psibar^b\,\overline{R}_{b\dot{\alpha}}\,.
\label{RR}
\eeqa
The corresponding inverse relations are:
\beqa
(P_L)_a{}^b\,\Psi_b&=&\overline{L}_a{}^\beta\chi\ls{\beta}\,,\qquad\quad\,\,
\Psibar^a(P_L)_a{}^b=\eta^\beta L_\beta{}^b\,, \label{ILL}\\
(P_R)_a{}^b\,\Psi_b&=&\overline{R}_{a\dot{\beta}}\eta^{\dagger\dot{\beta}}\,,
\qquad\quad \Psibar^a(P_R)_a{}^b
=\chi^\dagger_{\dot{\beta}}R^{\dot{\beta}\,b}\,.
\label{IRR}
\eeqa
One can use \eqss{gamma4}{Sigmamunu}{acmatrix} to identify:
\beqa
\sigma^\mu_{\alpha\dot{\beta}}&=& L_\alpha{}^a(\gamma^\mu)_a{}^b
\overline{R}_{b\dot{\beta}}\,,\qquad\qquad\qquad\quad\,
\sigmabar^{\mu\dot{\alpha}\beta}= R^{\dot{\alpha}\,a}(\gamma^\mu)_a{}^b
\overline{L}_b{}^\beta\,,\label{LRsigma}\\
\sigma^{\mu\nu}{}_\alpha{}^\beta&=& L_\alpha{}^a(\half \Sigma^{\mu\nu})_a{}^b
\overline{L}_b{}^\beta\,,\qquad\qquad\quad\,\,\,
\sigmabar^{\mu\nu\dot{\alpha}}{}_{\dot{\beta}}=
R^{\dot{\alpha}\,a}(\half\Sigma^{\mu\nu})_a{}^b
\overline{R}_{b\dot{\beta}}\,,\label{LLsigma}\\
\delta_\alpha{}^\beta&=& -L_\alpha{}^a(\gamma\ls{5})_a{}^b
\overline{L}_b{}^\beta\,,\qquad\qquad\qquad\quad
\delta^{\dot{\alpha}}{}_{\dot{\beta}}=
R^{\dot{\alpha}\,a}(\gamma\ls{5})_a{}^b
\overline{R}_{b\dot{\beta}}\,,\\
\epsilon_{\alpha\beta} &=& L_\alpha{}^a C_{ab}L_\beta{}^b\,,
\qquad\qquad\qquad\qquad\quad
\epsilon^{\dot{\alpha}\dot{\beta}} = R^{\dot{\alpha}\,a}C_{ab}
R^{\dot{\beta}\,b}\,,\\
\epsilon^{\alpha\beta} &=& \overline{L}_a{}^\alpha (C^{-1})^{ab}
\overline{L}_b{}^\beta\,,\qquad\qquad\qquad\quad
\epsilon_{\dot{\alpha}\dot{\beta}} = \overline{R}_{a\dot{\alpha}}
(C^{-1})^{ab}\overline{R}_{b\dot{\beta}}\,.
\eeqa
Inverting these results yields:
\beqa
(\gamma^\mu P_L)_c{}^d &=& \overline{R}_{c\dot{\alpha}}
\sigmabar^{\mu\dot{\alpha}\beta}L_\beta{}^d\,,\qquad\qquad\qquad\,\,\,\,\,\,
(\gamma^\mu P_R)_c{}^d = \overline{L}_c{}^\alpha
\sigma^\mu_{\alpha\dot{\beta}}R^{\dot{\beta}d}\,,\label{invert1}\\
\half(\Sigma^{\mu\nu} P_L)_c{}^d &=& \overline{L}_c{}^\alpha
\sigma^{\mu\nu}{}_\alpha{}^\beta L_\beta{}^d\qquad\qquad\qquad
\half(\Sigma^{\mu\nu} P_R)_c{}^d =  \overline{R}_{c\dot{\alpha}}
\sigmabar^{\mu\nu\dot{\alpha}}{}_{\dot{\beta}}R^{\dot{\beta}d}\,,
\label{invert2} \\
(AP_L)_c{}^d &=& \overline{R}_c{}^\beta L_\beta{}^d\,,
\qquad\qquad\qquad\,\,\,\,\,\qquad\,\,\,
(AP_R)_c{}^d=\overline{L}_{c\dot{\beta}}R^{\dot{\beta}d}\,,\label{invert3}
\eeqa
\beqa
(P_L C)_{cd}&=&\epsilon_{\alpha\beta}
\overline{L}_c{}^\alpha\overline{L}_d{}^\beta\,,\qquad\qquad\qquad\qquad
(P_R C)_{cd}=\epsilon^{\dot{\alpha}\dot{\beta}}
\overline{R}_{c\dot{\alpha}}\overline{R}_{d\dot{\beta}}\,,\label{invert4}\\
(C^{-1}P_L)^{cd}&=&\epsilon^{\alpha\beta}L_\alpha{}^c L_\beta{}^d\,,
\qquad\qquad\qquad\quad
(C^{-1}P_R)^{cd}=
\epsilon_{\dot{\alpha}\dot{\beta}}R^{\dot{\alpha}c}
R^{\dot{\beta}d}\,.
\label{invert5}
\eeqa

Likewise, one can introduce $L^\dagger$, $\overline{L}\llsup\dagger$,
$R^\dagger$ and $\overline{R}\llsup\dagger$, which are hybrid quantities
that contain a barred four-component spinor index and a two-component
undotted or dotted spinor index:
\beqa
&& (L^\dagger)^{\bar a}{}_{\dot{\beta}}\equiv (L_\beta{}^{\,a})^*\,,
\qquad\qquad\qquad
(R^\dagger)^{{\bar a}\beta}\equiv (R^{\dot{\beta}a})^*\,,\label{altdef1}\\
&& (\overline{L}\llsup\dagger)^{\dot{\beta}}{}_{\bar a}
\equiv (\overline{L}_a{}^\beta)^*\,,\qquad\qquad\qquad\,\,
(\overline{R}\llsup\dagger)_{\beta\bar a}\equiv (\overline{R}_{a\dot{\beta}})^*\,.
\label{altdef2}
\eeqa
In particular, using \eqs{ILL}{IRR}, one can relate the 
quantities $L$, $\overline{L}$, $R$ and $\overline{R}$ and their
hermitian conjugates:
\beqa
(L^\dagger)^{\bar a}{}_{\dot\beta}&=&A^{\bar a b}\overline{R}_{b\dot{\beta}}\,,
\qquad\qquad\qquad\quad
(R^\dagger)^{{\bar a}\beta}=A^{\bar a b}\overline{L}_b{}^\beta
\,,\label{LLRR1}\\
(\overline{L}\llsup\dagger)^{\dot{\beta}}{}_{\bar a}&=& R^{\dot{\beta}b}
(A^{-1})_{b\bar a}\,,\qquad\qquad\quad\,
(\overline{R}\llsup\dagger)_{\beta\bar a}=L_\beta{}^b(A^{-1})_{b\bar a}\,,
\label{LLRR2}
\eeqa
after employing $AP_L=P_R^\dagger A$ [cf.~\eq{aagamma}] and $A^\dagger=A$.
The set of equations analogous to \eqst{setz1}{invert5}
involving the corresponding hermitian-conjugated quantities
can also be obtained.   However, such formulae will rarely be needed  
in practice.

\Eqst{setz1}{invert5} [and their hermitian conjugates]
can be employed to translate any expression
involving two-component spinors into the corresponding expression
involving four-component spinors, and vice versa.  With a little
practice, both two-component and four-component spinor indices can be
suppressed, which greatly simplifies the manipulation of the spinor
quantities.  In particular, by treating the four-component spinors
$\Psi_a$ and $\Psi^C_a$
as column vectors and their hermitian (Dirac) conjugates
$\Psi^\dagger_{\bar a}$ and $\Psi^{C\,\dagger}\ls{\bar a}$ 
($\overline{\Psi}\llsup{\,a}$ and $\overline{\Psi^C}\llsup{\,a}$) as row
vectors, all equations in the four-component spinor formalism
have a natural
interpretation as products of matrices and vectors.
Henceforth, we shall suppress all four-component spinor
indices.

Multiple species of fermions are indicated with a flavor index
such as $i$ and $j$.  Dirac fermions are constructed from
two-component fields of opposite charge, $\chi\ls{\,i}$ and
$\eta^{\,i}$ (hence the opposite flavor index heights).  Thus,
we establish the following conventions for the flavor indices
of four-component Dirac fermions:
\beq \label{diracflavorind}
\Psi_i(x)\equiv\begin{pmatrix} \chi_{\alpha i}(x)\\
\eta^{\dagger\dot{\alpha}}_{\,i}(x)\end{pmatrix}\,,\qquad
\Psibar\llsup{i}(x)=\left(\eta^{\alpha i}(x)\,,\,\,
\chi^{\dagger\,i}_{\dot{\alpha}}(x)\right)\,,
\qquad
\Psi^{C\,i}(x)\equiv\begin{pmatrix} \eta^{\,i}_\alpha(x)\\
\chi^{\dagger\dot{\alpha}i}(x)\end{pmatrix}\,.
\eeq
Note that $\chi^{\dagger\,i}=(\chi_i)^\dagger$
and $\eta^\dagger_i\equiv (\eta^i)^\dagger$ following
the conventions established in \sec{subsec:generalmass}.
Raised flavor indices can only be contracted with lowered flavor indices and
vice versa.
In contrast, Majorana fermions are neutral so that there is no
a priori distinction between raised and lowered flavor indices.
That is,
\beq \label{majflavorind}
\Psi_{Mi}(x)=\Psi_M^i(x)=\Psi^{C}_{Mi}(x)=\Psi_M^{C\,i}(x)
\equiv\begin{pmatrix} \xi_{\alpha i}(x)\\
\xi^{\dagger\dot{\alpha}i}(x)\end{pmatrix},\qquad
\Psibar_{Mi}(x)=\Psibar\llsup{i}_M(x)\equiv\left(\xi^\alpha_{\,i}(x)\,,\,\,
\xi^{\dagger\,i}_{\dot{\alpha}}(x)\right).
\eeq
In this case, the contraction of two repeated flavor indices is
allowed in all cases,
irrespective of the heights of the two indices.
In the convention adopted in \sec{subsec:generalmass}, in which all
neutral left-handed $(\half,0)$ [right-handed $(0,\half)$]
fermions have lowered [raised] flavor indices, the height of the
flavor index of a four-component Majorana fermion field is
meaningful when multiplied by a left-handed or right-handed
projection operator.  Thus,
the height of the flavor index for Majorana fermions
can be consistently chosen according to one of the following four cases:
\beq \label{majheights}
P_L\Psi_{Mi}\,,\qquad \Psibar_{Mi}P_L\,,\qquad
P_R\Psi_{M}\llsup{\,i}\,,\qquad \Psibar_{M}\llsup{\,i}P_R\,.
\eeq

Bilinear covariants are quantities that are quadratic in the
spinor fields and transform irreducibly as Lorentz tensors.
We first construct a
translation table between the two-component form and the
four-component form for the bilinear covariants made up of a pair
of Dirac fields.  Using \eqs{ILL}{IRR}
to convert the four-component spinor fields into the
corresponding two-component spinor fields, and employing the appropriate
identities involving products of the hybrid quantities
$L$, $\overline{L}$, $R$ and $\overline{R}$, the following results are
then obtained:
\beqa
\Psibar\llsup{i} P_L \Psi_j &=& \eta^{\,i}\chi_j\,, \label{twotofoura}\\[4pt]
\Psibar\llsup{i} P_R \Psi_j &=& \chi^{\dagger\,i}\eta^\dagger_j
      \,, \label{twotofourb}\\[4pt]
\Psibar\llsup{i} \gamma^\mu P_L\Psi_j &=&\chi^{\dagger\,i}\sigmabar^\mu\chi_j
      \,, \label{twotofourc}\\[4pt]
\Psibar\llsup{i}\gamma^\mu P_R\Psi_j &=& \eta^{\,i}\sigma^\mu\eta^\dagger_j
      \,,\label{twotofourd}
\eeqa
\beqa
\Psibar\llsup{i} \Sigma^{\mu\nu}P_L \Psi_j
      &=& 2\,\eta^{\,i}\sigma^{\mu\nu}\chi_j\,, \label{twotofoure}\\[4pt]
\Psibar\llsup{i} \Sigma^{\mu\nu}P_R \Psi_j
      &=& 2\,\chi^{\dagger\,i}\sigmabar^{\mu\nu}\eta^\dagger_j\,.
\label{twotofourf}
\eeqa
The first two results above follow immediately after using
\eqs{setz1}{setz2}, respectively, and the last four results
are a consequence of \eqs{LRsigma}{LLsigma}.

\Eqst{twotofoura}{twotofourf} apply to both commuting and
anticommuting fermion fields.\footnote{\label{fnetasig}%
In the case of anticommuting spinors, it
is often useful to apply \eq{europeanvacation} to
\eqss{twotofourd}{vecbilinear}{axbilinear} and
rewrite $\eta^{\,i}\sigma^\mu\eta^\dagger_j=
-\eta^{\dagger}_j\sigmabar^\mu\eta^{\,i}$.}
These results can then be used to express the standard four-component
spinor bilinear covariants in terms of two-component spinor bilinears:
\beqa
\Psibar\llsup{i} \Psi_j &=& \eta^{\,i}\chi_j +
\chi^{\dagger\,i}\eta^\dagger_j  \label{scalarbilinear}\\[4pt]
\Psibar\llsup{i}\gamma_5\Psi_j &=&
-\eta^{\,i}\chi_j + \chi^{\dagger\,i}\eta^\dagger_j
\label{pseudoscalarbilinear} \\[4pt]
\Psibar\llsup{i}\gamma^\mu\Psi_j &=& \chi^{\dagger\,i}\sigmabar^\mu\chi_j
       +\eta^{\,i}\sigma^\mu \eta^\dagger_j \label{vecbilinear}
\label{bV}\\[4pt]
\Psibar\llsup{i}\gamma^\mu\gamma_5\Psi_j
       &=& -\chi^{\dagger\,i}\sigmabar^\mu\chi_j
       +\eta^{\,i}\sigma^\mu \eta^\dagger_j \label{axbilinear}\\[4pt]
\Psibar\llsup{i}\Sigma^{\mu\nu}\Psi_j &=& 2(\eta^{\,i} \sigma^{\mu\nu}
       \chi_j + \chi^{\dagger\,i} \sigmabar^{\mu\nu} \eta^\dagger_j)
       \label{tensorbilinear} \\[4pt]
\Psibar\llsup{i}\Sigma^{\mu\nu}\gamma_5 \Psi_j &=&
       2(-\eta^{\,i} \sigma^{\mu\nu}
       \chi_j + \chi^{\dagger\,i} \sigmabar^{\mu\nu} \eta^\dagger_j)
       \label{ptensorbilinear}\,.
\eeqa

Additional identities can be derived that involve the charge-conjugated
four-component Dirac fermion fields.  As an example,
we may use $C^{\T}=-C$ and $\overline{\Psi^C}= -\Psi^{\T} C^{-1}$
to prove that
\beq \label{cinteractions}
\overline{\Psi^C_{\,i}}\,\Gamma\,\Psi^{C\,j}
=-(-1)^A\Psibar\llsup{\,j} C\,\Gamma^{\T}\,
C^{-1}\Psi_i=-(-1)^A\eta^C\ls{\,\Gamma\,}\Psibar\llsup{\,j}\,\Gamma\, \Psi_i\,,
\eeq
where the sign $\eta^C\ls{\Gamma}$ is given in \eq{ccgamma}.
The factor of $(-1)^A=\pm 1$ [for commuting/anticommut\-ing
fermion fields, respectively] arises at the second
step above after reversing the order of the terms by matrix transposition.
Identities involving just one charge-conjugated four-component field
can also be easily obtained.  For example, using \eqs{LL}{invert5},
\beq
\overline{\Psi^C_{\,i}} P_L\Psi_j=-\Psi^{\T}_{\,i}C^{-1}P_L\Psi_j
=-\Psi_{ai}\epsilon^{\alpha\beta}L_\alpha{}^aL_\beta{}^b\Psi_{bj}
=-\epsilon^{\alpha\beta}\chi\ls{\alpha i}\chi\ls{\beta j}=
\chi\ls{i}\chi\ls{j}\,.
\eeq
In general, if one replaces $\Psi_k$ with $\Psi^{C\,k}$
in \eqst{twotofoura}{ptensorbilinear}, then in the
corresponding two-component expression one simply interchanges
$\chi_k\leftrightarrow\eta^k$ and
$\chi^{\dagger\,k}\leftrightarrow\eta^\dagger_k$.

\Eqst{twotofoura}{ptensorbilinear}
also apply to four-component Majorana
spinors, $\Psi_{Mi}$, by setting $\chi_i=\eta^i\equiv\xi_i$,
and $\chi^{\dagger\,i}=\eta^\dagger_i\equiv\xi^{\dagger\,i}$.
This implements the Majorana
condition, $\Psi_{Mi}=D\Psi_{Mi}^*$,
and imposes additional
restrictions on the Majorana bilinear covariants.
For example, \eqs{ccgamma}{cinteractions} imply that
\textit{anticommuting} Majorana four-component fermions
satisfy:\footnote{Here, one is free to choose all flavor indices
to be in the lowered position [cf.~\eq{majflavorind}].}
\beqa
\Psibar_{Mi}\Psi_{Mj}&=&
\phm\Psibar_{Mj}\Psi_{Mi}\,, \label{majidentity1}\\
\Psibar_{Mi}\gamma\ls{5} \Psi_{Mj}&=&
\phm\Psibar_{Mj}\gamma\ls{5} \Psi_{Mi}\,, \label{majidentity2}\\
\Psibar_{Mi}\gamma^\mu \Psi_{Mj}&=&
-\Psibar_{Mj}\gamma^\mu \Psi_{Mi}\,, \label{majidentity3}\\
\Psibar_{Mi}\gamma^\mu\gamma\ls{5} \Psi_{Mj}&=&
\phm\Psibar_{Mj}\gamma^\mu\gamma\ls{5} \Psi_{Mi}\,, \label{majidentity4}\\
\Psibar_{Mi}\Sigma^{\mu\nu}\Psi_{Mj}&=&
-\Psibar_{Mj}\Sigma^{\mu\nu}\Psi_{Mi}\,, \label{majidentity5} \\
\Psibar_{Mi}\Sigma^{\mu\nu}\gamma\ls{5}\Psi_{Mj}&=&
-\Psibar_{Mj}\Sigma^{\mu\nu}\gamma\ls{5}\Psi_{Mi}
\,.\label{majidentity6}
\eeqa
By setting $i=j$, it follows that $\Psibar_{M}\gamma^\mu\Psi_{M}
=\Psibar_{M}\Sigma^{\mu\nu}\Psi_{M}=
\Psibar_{M}\Sigma^{\mu\nu}\gamma\ls{5}\Psi_{M}=0$.
One additional useful result is:
\beq
\label{majidentitya}
\Psibar_{M}\llsup{i}\gamma^\mu P_L\Psi_{Mj}=
-\Psibar_{Mj}\gamma^\mu P_R\Psi_{M}^i\,,
\eeq
which follows immediately from \eqs{majidentity3}{majidentity4}.
Note that in \eq{majidentitya}, the heights of the flavor indices
follow the convention established in \eq{majheights}.

In the four-component spinor formalism, Fierz identities (first introduced
in \Ref{Fierz}) consist of relations among products of two 
bilinear covariants, in which the fermion fields appear in two 
different orders.  The corresponding 
two-component spinor Fierz identities are treated in detail
in \app{B.1}.
In principle, the latter can be converted into 
four-component spinor Fierz identities using the techniques developed  
in this Appendix.  However, it is easier 
to derive the four-component spinor Fierz identities 
directly using the properties of
the gamma matrix algebra~\cite{WPauli,Diracmatrices}.  

Instead of \eqst{Fierz1}{Fierz3}, 
the equivalent identity relevant for four-component spinors~is:
\beqa \label{fierzfour}
\hspace{-0.5in}
\delta_a^b\delta_c^d=\nicefrac{1}{4}\left[\delta_a^d\delta_c^b
+(\gamma\ls{5})_a{}^d(\gamma\ls{5})_c{}^b
\BDplus(\gamma^\mu)_a{}^d(\gamma_\mu)_c{}^b
\BDminus(\gamma^\mu\gamma\ls{5})_a{}^d(\gamma_\mu\gamma\ls{5})_c{}^b
+\half(\Sigma^{\mu\nu})_a{}^d(\Sigma_{\mu\nu})_c{}^b\right]\,.
\eeqa

This is the fundamental identity from which many other such
identities can be derived (cf.~the Appendix of \Ref{bailinweak}).
One of many possible Fierz identities can be obtained by 
multiplying \eq{fierzfour} by 
$\Psibar_1^a\Psi_{2b}\Psibar_3^c\Psi_{4d}=(-1)^A\,\Psibar_1^a\Psi_{4d}
\Psibar_3^c\Psi_{2b}$, where $(-1)^A=+1$ [$-1$] for commuting
[anticommuting] Dirac, Majorana or Weyl spinors. 
More generally~\cite{Diracmatrices,DeVries:1970cg,Nieves}, 
\beqa \label{reorder}
&&\phantom{line} \nonumber \\[-22pt]
&&(\Psibar_1\Gamma^{(k)I}\Psi_2)(\Psibar_3\Gamma^{(k)}_I\Psi_4)
=(-1)^A\sum_{n=1}^5
F^k{}_n\,(\Psibar_1\Gamma^{(n)J}\Psi_4)(\Psibar_3\Gamma^{(n)}_J\Psi_2)\,,
\eeqa
where the sum is taken over the $4\times 4$ matrices,
$\Gamma^{(n)}\in\Gamma$, which
have been ordered as follows,\footnote{The 16 matrices of 
$\Gamma$ constitute
a complete set that spans the sixteen-dimensional vector space of $4\times 4$ 
matrices.}
\beqa \label{spanning}
&&\phantom{line}\nonumber \\[-26pt]
&&\Gamma=\{\mathds{1}\,,\,\gamma^\mu\,,\,\Sigma^{\mu\nu}\,\,
(\mu<\nu)\,,\,\gamma^\mu\gamma\ls{5}\,,
\,\gamma\ls{5}\}\,,\\[-26pt]
&&\phantom{line}\nonumber 
\eeqa
$I$, $J$ represent zero, one or two 
spacetime indices (sums over repeated $I$ and $J$ are implied),
and
\beq
F=\frac{1}{4}\begin{pmatrix} \phm 1 & \BDposs 1 & \phm\half & \BDnegg
  1 &\phm 1 \\
\BDposs 4 & -2 & \phm 0 & -2 & \BDnegg 4 \\
\phm 12 & \phm 0 & -2 & \phm 0 & \phm 12 \\
\BDnegg 4 & -2 & \phm 0 & -2 & \BDposs 4 \\ \phm 1 & \BDnegg 1 & \phm \half & \BDposs 1 & \phm 1 \end{pmatrix}\,.
\eeq 
For example, taking $k=1$ in \eq{reorder} yields a result equivalent
to \eq{fierzfour}:
\beqa
(\Psibar_1\Psi_2)(\Psibar_3\Psi_4)&=&
\quarter(-1)^A\left[(\Psibar_1\Psi_4)(\Psibar_3\Psi_2)
+(\Psibar_1\gamma\ls{5}\Psi_4)(\Psibar_3\gamma\ls{5}\Psi_2)
\BDplus(\Psibar_1\gamma^\mu\Psi_4)(\Psibar_3\gamma_\mu\Psi_2)\right. 
\nonumber \\
&&\qquad\left.
\BDminus(\Psibar_1\gamma^\mu\gamma\ls{5}\Psi_4)
(\Psibar_3\gamma_\mu\gamma\ls{5}\Psi_2)
+\half(\Psibar_1\Sigma^{\mu\nu}\Psi_4)
(\Psibar_3\Sigma_{\mu\nu}\Psi_2)\right]\,.
\eeqa

For a  comprehensive treatment of all possible 
four-component spinor Fierz identities, see \Ref{Takahashi}.
Simple derivations of generalized Fierz identities have also 
been
given in refs.~\cite{Nieves,Nishi}.  
A Mathematica package for performing Fierz
transformations is available in~\Ref{Gran:2001yh}.

\subsection{Free-field four-component fermion Lagrangians}
\label{fourLag}
\renewcommand{\theequation}{G.2.\arabic{equation}}
\renewcommand{\thefigure}{G.2.\arabic{figure}}
\renewcommand{\thetable}{G.2.\arabic{table}}
\setcounter{equation}{0}
\setcounter{figure}{0}
\setcounter{table}{0}

The free-field Lagrangian density in four-component spinor notation can be obtained
from the corresponding two-component fermion Lagrangian by employing
the relevant identities for the bilinear covariants
given in \eqst{twotofoura}{ptensorbilinear}.
First, consider a collection of free anticommuting four-component Majorana
fields, $\Psi_{Mi}=\Psi^C_{Mi}$.
The free-field Lagrangian (in terms of mass eigenstate fields)
may be obtained from \eq{lagMajorana} by
converting to four-component spinor notation using
\eqs{scalarbilinear}{vecbilinear} with $\chi=\eta\equiv\xi$, which 
yields~\cite{Majorana}:
\beq \label{LagMaj4}
\mathscr{L}=\half i\Psibar_{Mi}\gamma^\mu\partial_\mu\Psi_{Mi}
-\half m_i\Psibar_{Mi}\Psi_{Mi}\,,
\eeq
where the sum over $i$ is implicit.  The corresponding free-field
equation for $\Psi_{Mi}$ is
the Dirac equation:
\beq
(i\gamma^\mu\partial_\mu-m)\Psi_{Mi}=0\,.
\eeq

For simplicity, we focus on a theory of a single four-component
Majorana fermion field, $\Psi_M(x)=\Psi_M^C(x)$.
One can rewrite the free-field Majorana
fermion Lagrangian in terms of a single Weyl fermion,
$\Psi\ls{L}(x)\equiv P_L\Psi(x)$, where $\Psi(x)$ is a four-component fermion
field whose lower two components (in the chiral representation)
are irrelevant for the present
discussion.
The Majorana and Weyl fields are related by:
\beq \label{MLR}
\Psi\ls{M}(x)=\Psi\ls{L}(x)+\Psi^C\ls{R}(x)\,,
\eeq
where $\Psi^C_R(x)$ is defined in \eq{psiRC}.
The corresponding Dirac conjugate field is given by
$\Psibar\ls{M}(x)=\Psibar\ls{L}(x)+\overline{\Psi\llsup{C}\ls{R}}(x)$,
where
\beqa
\Psibar\ls{L}(x)&\equiv& [P_L\Psi(x)]^\dagger A=\Psibar(x)P\ls{R}\,,\\
\overline{\Psi\llsup{C}\ls{R}}(x) &\equiv &\overline{\Psi^C}(x)P_L=
-\Psi^{\T}(x)C^{-1}P\ls{L}
=-\Psi_L^{\T}(x)C^{-1}\,.\label{psiRCbar}
\eeqa
Using the identity:\footnote{In deriving \eq{psiLRC}, we have used
\eq{ccgamma} and the anticommutativity of the spinor fields.
The total divergence can be dropped from the Lagrangian, as it
does not contribute to the field equations.}
\beq \label{psiLRC}
\overline{\Psi\llsup{C}\ls{R}}\gamma^\mu\partial_\mu \Psi^C\ls{R}=
-\Psi^{\T}C^{-1}P_L\gamma^\mu\partial_\mu P_R C\Psibar\llsup{\,\T}=
\Psibar\ls{L}\gamma^\mu\partial_\mu \Psi\ls{L}+\hbox{\rm total divergence}\,,
\eeq
the Lagrangian for a single Majorana field can be written in terms
of a single Weyl field:\footnote{Using
\eq{acproperties}, it follows that 
$(\Psi^{\T}C^{-1}\Psi)^\dagger=-\overline\Psi A^{-1}
C^{-1\,*}A^{-1\,*}\overline\Psi\llsup{\,\T}=
-\overline\Psi\, C\,\overline\Psi\llsup{\,\T}$.}
\beq \label{singleWeyl}
\mathscr{L}=i\Psibar_{L}\gamma^\mu\partial_\mu\Psi_{L}
+\half m\left(\Psi\ls{L}^{\T}C^{-1}\Psi\ls{L}
-\Psibar_L C \Psibar_L^{\T}\right)\,.
\eeq
The corresponding free-field equation is
\beq
i\gamma^\mu\partial_\mu\Psi_L=mC\Psibar_L^{\T}\,,
\eeq
where we have used $(\Psibar_L C)^{\T}=-C\Psibar_L^{\T}$ and the
anticommutativity of $\Psi_L$, $\Psibar_L$.
The generalization of \eqst{MLR}{singleWeyl} to the case of a multiplet of
four-component Majorana fields is straightforward and is left as an
exercise for the reader.

Of course, one could have chosen instead to rewrite the four-component
Majorana fermion Lagrangian in terms of a single Weyl fermion,
$\Psi\ls{R}(x)\equiv  P_R\Psi(x)$, in which case the upper two components
(in the chiral representation) of $\Psi(x)$
are not relevant.  In this case, the Majorana and Weyl fields
are related by:\footnote{If $\Psi$ is an unconstrained four-component
spinor, then $\Psi_L$ and $\Psi_R$ are
independent Weyl fields, in which case $\Psi_L+\Psi_R^C$ and $\Psi_R+\Psi_L^C$
are independent self-conjugate fields.}
\beq \label{MRL}
\Psi\ls{M}(x)=\Psi\ls{R}(x)+\Psi^C\ls{L}(x)\,,
\eeq
where $\Psi^C_L(x)$ is defined in \eq{psiLC}.
The corresponding Dirac conjugate field is given by
$\Psibar\ls{M}(x)=\Psibar\ls{R}(x)+\overline{\Psi\llsup{C}\ls{L}}(x)$,
where
\beqa
\Psibar\ls{R}(x)&\equiv& [P_R\Psi(x)]^\dagger A=\Psibar(x)P\ls{L}\,,\\
\overline{\Psi\llsup{C}\ls{L}}(x) &\equiv &\overline{\Psi^C}(x)P_R=
-\Psi^{\T}(x)C^{-1}P\ls{R}
=-\Psi_R^{\T}(x)C^{-1}\,.\label{psiLCbar}
\eeqa%
The corresponding Weyl fermion Lagrangian is given by \eq{singleWeyl} with $L$
replaced by $R$.

Thus, a Majorana fermion can be represented either
by a four-component self-conjugate field $\Psi\ls{M}(x)$ or
by a single Weyl field [either $\Psi\ls{L}(x)$ or $\Psi\ls{R}(x)$].
Both descriptions are unitarily equivalent~\cite{MajWeyl,Serpe}; i.e.,
one can construct a unitary similarity transformation that
connects a Majorana field operator and a Weyl field operator (and vice versa).
Of course, this is hardly a surprise in the two-component spinor
formalism, where both the Majorana and Weyl forms of the Lagrangian
correspond to the same field theory of
a single two-component spinor field $\xi_\alpha(x)$.

For $m\!\neq\!0$, the Weyl Lagrangian given by \eq{singleWeyl}
possesses no global symmetry, and hence no conserved charge.
In contrast, for $m=0$ the Weyl Lagrangian exhibits a U(1)
chiral symmetry.  In a theory of massless neutrinos, the
U(1) chiral charge of the neutrino is correlated with its lepton number $L$,
and one is free to use either a Majorana or Weyl
description.  In the former, the neutrino is a neutral self-conjugate
fermion, which is not an eigenstate of $L$.  In the latter,
$\Psi\ls{L}(x)$ corresponds to the left-handed neutrino and
$\Psi^C\ls{R}(x)$ corresponds to the right-handed antineutrino,
which are eigenstates of $L$ with opposite sign lepton numbers.
No experimental observable can distinguish between these two descriptions.

We now consider a collection of free anticommuting four-component Dirac
fields, $\Psi_{i}$.
The free-field Lagrangian (in terms of mass eigenstate fields)
may be obtained from \eq{lagDiracdiag} by
converting to four-component spinor notation. We then obtain the standard
textbook result:
\beq
\mathscr{L}=i\Psibar^{i}\gamma^\mu\partial_\mu\Psi_{i}
- m_i\Psibar^{i}\Psi_{i}\,.
\eeq
By writing $\Psi=\Psi\ls{L}+\Psi\ls{R}$, we see that
the Lagrangian for a single Dirac field can be written in terms
of two Weyl fields:
\beq
\mathscr{L}=i\Psibar_{L}\gamma^\mu\partial_\mu\Psi_{L}+
i\Psibar_{R}\gamma^\mu\partial_\mu\Psi_{R}
-m\left(\Psibar\ls{L}\Psi\ls{R}+\Psibar\ls{R}\Psi\ls{L}\right)\,.
\eeq
The corresponding free-field equations are:
\beq
i\gamma^\mu\partial_\mu\Psi_L=m\Psi_R\,,\qquad\qquad
i\gamma^\mu\partial_\mu\Psi_R=m\Psi_L\,.
\eeq
Summing these two equations yields the Dirac equation,
$(i\gamma^\mu\partial_\mu-m)\Psi=0$.

As a pedagogical example in which both Dirac and Majorana mass terms
are present, we perform the diagonalization of the neutrino mass
matrix in a one-generation seesaw model\footnote{
In \app{J.2}, the seesaw model of neutrino masses is introduced using
the two-component spinor formalism.}
using the four-component spinor formalism.
Following Appendix~A of \Ref{dhr},  we first introduce a four-component
anticommuting neutrino field $\nu_D$, and the corresponding Weyl fields,
\beq
\nu_L\equiv P_L\nu_D\,, \qquad \nu_L^C\equiv
P_L\nu_D^C\,,\qquad \nu_R\equiv P_R\nu_D\,,
\quad {\rm and} \quad \nu_R^C\equiv P_R\nu_D^C\,.
\eeq
Note that \eqs{psiLC}{psiRCbar} imply
that the anticommuting Weyl fermion fields satisfy:
\beq \label{nulr}
\overline{\nu_R^C}\nu_L^C=\overline{\nu_R}\nu_L\,,
\qquad\qquad
\overline{\nu_L^C}\nu_R^C=
\overline{\nu_L}\nu_R\,.
\eeq

A Dirac mass term for the neutrinos in the one-generation seesaw model
couples $\nu_L$ and~$\nu_L^C$ (and by hermiticity of the Lagrangian,
$\nu_R^C$ and $\nu_R$), and can be written equivalently as:
\beq
m_D(\nu_L^{\T}C^{-1}\nu_L^C+\nu_R^{\T}\,C^{-1}\nu_R^C)=
-m_D(\overline{\nu_R^C}\nu_L^C+\overline{\nu_L^C}\nu_R^C)=
-m_D(\overline{\nu_R}\nu_L+\overline{\nu_L}\nu_R)=-m_D\overline\nu_D \nu_D\,,
\eeq
after making use of \eq{nulr}.  The Majorana mass term for the neutrinos
in the one-generation seesaw model couples $\nu_L^C$ to itself (and
by hermiticity of the Lagrangian, $\nu_R$ to itself),
and can be written equivalently as:
\beq
\half M(\nu_L^{C\,\T}C^{-1}\nu_L^C+\nu_R^{\T}\,C^{-1}\nu_R)
=-\half M(\overline{\nu_R}\nu_L^C+\overline{\nu_L^C}\nu_R)\,.
\eeq
We shall define the phases of the neutrino fields such that the
parameters $m_D$ and $M$ are real and non-negative.

Thus, the
mass terms of the one-generation neutrino seesaw Lagrangian, given in
\eq{eq:mterms} in terms
of two-component fermion fields, translates
in four-component spinor notation to
\beqa
\hspace{-1in} \mathscr{L}_{\rm mass}&=&-\half
m_D(\overline{\nu_L}\nu_R+\overline{\nu_R}\nu_L
+\overline{\nu^C_L}\nu_R^C+\overline{\nu_R^C}\nu_L^C) -\half
M(\overline{\nu_R}\nu_L^C+\overline{\nu_L^C}\nu_R) \nonumber \\[8pt]
&=&-\half\left(\begin{array}{cc}\overline{\nu_R^C} &
\quad\overline{\nu_R}\end{array}\right)\,
\left(\begin{array}{cc} 0 & \quad m_D\\ m_D & \quad M
\end{array}\right)\,\left(\begin{array}{c} \nu_L \\ \nu_L^C\end{array}\right)
- \half\left(\begin{array}{cc}\overline{\nu_L} &
\quad\overline{\nu_L^C}\end{array}\right)\,
\left(\begin{array}{cc} 0 & \quad m_D\\ m_D & \quad M
\end{array}\right)\,\left(\begin{array}{c} \nu_R^C \\ \nu_R\end{array}\right)
\nonumber \\[8pt]
&=&\half\left(\begin{array}{cc}{\nu_L^\T} &
\quad {\nu_L^{C\,\T}}\end{array}\right)\,C^{-1}
\left(\begin{array}{cc} 0 & \quad m_D\\ m_D & \quad M
\end{array}\right)\,\left(\begin{array}{c} \nu_L \\ \nu_L^C\end{array}\right)
+{\rm h.c.}\,,
\label{fn4}
\eeqa
where we have used \eq{nulr} to write the first line of \eq{fn4} in a
symmetrical fashion and \eqs{psiLC}{psiRCbar} to obtain the final form above.
Note that if $M=0$, then one can write $\mathscr{L}_{\rm mass}=-m_D
\overline\nu_D\nu_D$ and identify $\nu_D$ as a four-component
massive Dirac neutrino.

The Takagi diagonalization of the neutrino mass matrix yields two
mass eigenstates, which we designate
by $\nu\ls{\ell}$ and $\nu\ls{h}$, where $\ell$ and $h$ stand for
\textit{light} and \textit{heavy}, respectively.
The mass eigenstate Weyl neutrino fields are related to the
interaction eigenstate Weyl neutrino fields via
\beq
\left(\begin{array}{c} \nu_L \\ \nu_L^C\end{array}\right)=\mathcal{U}
\left(\begin{array}{c} P_L \nu\ls{\ell} \\ P_L\nu\ls{h}^C\end{array}\right)\,,
\eeq
where $\mathcal{U}$ is a $2\times 2$ unitary matrix that is chosen
such that
\beq
\mathcal{U}^{\T} \left(\begin{array}{cc} 0 & m_D\\ m_D & M
\end{array}\right)\,\mathcal{U}=\left(\begin{array}{cc} m_{\nu\ls{\ell}} &
  0 \\ 0 & m_{\nu\ls{h}}\end{array}\right)\,.
\eeq
For $M\neq 0$, the neutrino mass eigenstates are \textit{not} Dirac fermions.
In the seesaw limit of $M\gg m_D$, the corresponding neutrino masses are
$m_{\nu\ls{\ell}}\simeq m_D^2/M$ and $m_{\nu\ls{h}}\simeq M+m_D^2/M$,
with $m_{\nu\ls{\ell}}\ll m_{\nu\ls{h}}$.
In terms of the mass eigenstates, the
neutrino mass Lagrangian is:
\beq \label{majmass}
\mathscr{L}_{\rm mass}=
\half \left[m_{\nu\ls{\ell}}\nu\ls{\ell}^{\T} C^{-1}P_L\nu\ls{\ell}
 +m_{\nu\ls{h}}\nu\ls{h}^{C\,\T} C^{-1}P_L \nu\ls{h}^C\right]+{\rm h.c.}\,,
\eeq
after using \eq{ccgamma}.
We now define four-component self-conjugate Majorana neutrino fields,
denoted by $\Psi_\ell$ and $\Psi_h$ respectively,
according to \eqs{MLR}{MRL},
\beqa
\Psi_\ell &\equiv &
P_L\nu\ls{\ell}+P_R \,C\overline\nu\ls{\ell}^{\T}\,,
\qquad\qquad
\overline\Psi_\ell\equiv \overline\nu\ls{\ell} P_R
-\nu\ls{\ell}^{\T} C^{-1}P_L\,, \label{lightmaj} \\
\Psi_h &\equiv &
P_R \nu\ls{h}+P_L C\overline\nu\ls{h}^{\T}\,,
\qquad\qquad
\overline\Psi_h\equiv \overline\nu\ls{h} P_L
-\nu\ls{h}^{\T} C^{-1}P_R\,.\label{heavymaj}
\eeqa
Then, \eq{majmass} reduces to the expected form:
\beq
\mathscr{L}_{\rm mass} =
-\half\left[m_{\nu\ls{\ell}}\overline\Psi_\ell\Psi\ls{\ell}
+m_{\nu\ls{h}}\overline{\Psi}_h\Psi_h\right]\,.
\eeq
A comparison with the analysis of the neutrino mass matrix given
in \app{J.2} exhibits the power and the simplicity of the
two-component spinor formalism, as compared to the rather
awkward four-component spinor analysis presented above.

\subsection{Gamma matrices and spinors in spacetimes of diverse dimensions and signatures}
\label{Diracgammadiverse}
\renewcommand{\theequation}{G.3.\arabic{equation}}
\renewcommand{\thefigure}{G.3.\arabic{figure}}
\renewcommand{\thetable}{G.3.\arabic{table}}
\setcounter{equation}{0}
\setcounter{figure}{0}
\setcounter{table}{0}

\indent
The translation from two-component to four-component spinor notation
given in \app{G.1} is specific to
$3+1$ spacetime dimensions.  In $d=4$ Euclidean space dimensions
(independently of the choice of convention for the Minkowski metric), the
Dirac gamma matrix algebra is defined by
$\{\gamma_E^\mu\,,\,\gamma_E^\nu\}=2\delta^{\mu\nu}\mathds{1}$,
where $\delta^{\mu\nu}\equiv{\rm diag}(1\,,\,1\,,\,1\,,\,1)$.
Using \eqs{esigma}{gamma4},
the Euclidean gamma matrices (defined for
$\mu$, $\nu=1\,,\,\ldots\,,\,4$) are hermitian and given by
$\gamma^k_E\equiv -i\gamma^k$ ($k=1,2,3)$,
$\gamma^4_E\equiv \gamma^0$
and $\gamma\ls{5E}\equiv
-\gamma^1_E\gamma^2_E\gamma^3_E\gamma^4_E=\gamma\ls{5}$
(e.g, see Appendix A.1.2 of \Ref{bdj}).\footnote{One can also choose
to define the Euclidean Dirac algebra by
$\{\gamma_E^\mu\,,\,\gamma_E^\nu\}=-2\delta^{\mu\nu}\mathds{1}$ (simply by
multiplying all gamma matrices by a factor of $i$),
in which case the Euclidean gamma matrices,
$\gamma^k_E\equiv \gamma^k$ and
$\gamma^4_E\equiv i\gamma^0$ are anti-hermitian, and $\gamma\ls{5E}\equiv
-\gamma^1_E\gamma^2_E\gamma^3_E\gamma^4_E=\gamma\ls{5}$ is hermitian
(e.g., see \Ref{euclid}).  These conventions arise more naturally
in the general treatment of gamma matrices in $d$ spacetime
dimensions as defined in \eq{cliff}.  The corresponding Euclidean
sigma matrices would then be defined as in
footnote~\ref{fnsigmaE}.\label{fncliff}}
The four-dimensional
reducible (Dirac) spinor representation corresponds to the
$(\half,0)\otimes (0,\half)$
representation of SO(4), although the  $(\half,0)$ and $(0,\half)$
representations are independent pseudo-real representations of SO(4)
not related by hermitian conjugation, as noted at the end of
\sec{sec:notations}.  A complete treatment of Euclidean two-component
spinors can be found in \Ref{sherry}.

The Euclidean space formalism for fermions is necessary for a rigorous
definition of the path integral in quantum field
theory~\cite{zinn,fujikawa}.  Using the
Euclidean Dirac gamma matrices introduced above, one can
express the four-component Dirac Lagrangian directly in
Euclidean space~\cite{Ramond}.
Carrying out the same procedure for the four-component Majorana
Lagrangian is problematical.
Because the $(\half,0)$ and $(0,\half)$ representations of SO(4)
are not hermitian conjugates of each other,
a self-conjugate Euclidean Majorana fermion does not exist.
Nevertheless,  it is possible to devise a continuous Wick rotation
from Minkowski spacetime to
Euclidean space for Dirac, Majorana and Weyl spinor fields
and the gamma matrices.   In particular, one can construct a non-hermitian
Euclidean action for a single Majorana or Weyl field
whose Green functions are related to the usual Minkowski space
Green functions by analytic continuation and a Wick rotation of the
spinor fields. Further details can be found
in refs.~\cite{mehta,waldron,wet}.\footnote{Previous attempts in the literature to
define Euclidean Majorana field theories can be
found in \Ref{Osterwalder:1973kn}.}

The two-component spinor technology of this review is specifically
designed to treat spinors in three space and one time dimension.
In theories of $d$ spacetime dimensions (where $d$ is any positive
integer), more general techniques are required.  By considering
spinors in this more general setting, one gains insight into the
concepts of Majorana, Weyl and Dirac spinors and their distinguishing features.

The mathematics of spinors~\cite{brauer} in spacetimes of dimension $d=t+s$
(where $t$ is the number of time dimensions and $s$ is the number
of space dimensions) is most easily treated
by introducing higher-dimensional analogues of the gamma matrices,
$\Gamma^\mu$, which satisfy the Clifford
algebra~\cite{Trautman,Benn,wet,scherk,vanN,coq,kugo,sohniusapp,vanN2,andrade,tanii,vanpro,Brandt,Braden:1982em},\footnote{This
  includes the Euclidean case~\cite{thacker}
corresponding to $t=0$ and $s=d$ [cf.~footnote~\ref{fncliff}],
and the Minkowski case corresponding to $t=1$ and $s=d-1$.}
\beq \label{cliff}
\{\Gamma^\mu\,,\,\Gamma^\nu\}=\BDpos 2\eta^{\mu\nu}\mathds{1}\,,\qquad
\quad
\eta^{\mu\nu}={\rm diag}(\underbrace{\BDplus\BDplus\cdots\BDplus}_{t}\,,\,
\underbrace{\BDminus\BDminus\cdots\BDminus}_s)\,,
\eeq
where the identity matrix $\mathds{1}$ and the $\Gamma^\mu$
are $2^{[d/2]}\times 2^{[d/2]}$ matrices, and
$[d/2]$ is the integer part of $d/2$,
\beq
[d/2]\equiv\begin{cases} d/2\,, & \text{for $d$ even,} \\
(d-1)/2\,, & \text{for $d$ odd.}
\end{cases}
\eeq
The choice of $(s,t)$ denotes the signature of the
spacetime.
One can choose $\Gamma^{\mu\,\dagger}=\Gamma^\mu$
for $\mu=1,2,\ldots, t$ and $\Gamma^{\mu\,\dagger}=-\Gamma^\mu$
for $\mu=t+1\,,\,t+2\,,\,\ldots\,,\, d$.  We identify
$\half \Sigma^{\mu\nu}\equiv\frac{1}{4}\,i\,[\Gamma^\mu\,,\,\Gamma^\nu]$
as the generators of SO$(s,t)$ in the spinor representation.
Next, we introduce the $[d/2]$-component (complex) Dirac spinor $\Psi$
and its Dirac conjugate $\overline\Psi\equiv \Psi^\dagger A$,
where $A=\Gamma^1\Gamma^2\cdots\Gamma^t$
is a unitary matrix that satisfies:\footnote{In $d$-dimensional Euclidean
space (where $t=0$), $\Gamma^{\mu\,\dagger}=-\Gamma^{\mu}$ for
all $\mu=1,2,\ldots,d$.  As a result, we may choose $A=\mathds{1}$,
in which case $\overline\Psi=\Psi^\dagger$.}
\beq
A\Gamma^\mu A^{-1}=(-1)^{t+1}\Gamma^{\mu\,\dagger}\,,\qquad\quad
A^\dagger=(-1)^{t(t-1)/2}A\,.
\eeq
One can now build SO($s,t$)-covariant bilinears,
$\overline\Psi \Gamma\Psi$,
where $\Gamma$ is a product
of gamma matrices.  Biquadratic spinor Fierz identities involving
quantities such as $(\Psibar_1\Gamma^I\Psi_2)(\Psibar_3\Gamma\ls{I}\Psi_4)$
can also be derived~\cite{fierzD}, where the 
$\Gamma^I=\bigl\{{\mathds{1}\,,\,\Gamma^\mu,\Gamma^{\mu\nu}\,\,(\mu<\nu})\,,\,
\Gamma^\mu\Gamma^\nu\Gamma^\lambda\,\,(\mu<\nu<\lambda)\,,\,\ldots,
\Gamma^1\Gamma^2\cdots\Gamma^{2[d/2]}\bigr\}$ 
are a complete set of $2^{2[d/2]}$ linearly independent
matrices [which generalizes \eq{spanning}].

If $d$ is even,
one can also introduce the $d$-dimensional analogue of
$\gamma\ls{5}$ by defining\footnote{For $t=1$ and $d$ even, one traditionally
takes $\mu=0,1,2,\ldots,d-1$ (where 0 is the time index), in
which case, $\Gamma_{d+1}\equiv i^{{(d-2)/2}}\,\Gamma^0\Gamma^1
\cdots\Gamma^{d-1}$.}
\beq
\Gamma_{d+1}\equiv i^{{(s-t)/2}}\,\Gamma^1\Gamma^2\cdots\Gamma^d\,,
\eeq
which is hermitian and satisfies $(\Gamma_{d+1})^2=\mathds{1}$ and
$\{\Gamma^\mu\,,\,\Gamma_{d+1}\}=0$.
In the case of even-dimensional spacetimes, there are two possible
choices for
the charge-conjugated spinor $\Psi^C$,\footnote{In four-dimensional
Minkowski spacetime, we identify $D=B_{-}^{-1}$ [cf.~\eq{ccd}]
and $\gamma\ls{5}D=B_+^{-1}$.}
\beq \label{dConj}
\Psi^C=B_\eta^{-1}\Psi^*\,\,,\qquad {\rm where}\quad \eta=\pm 1\,,
\eeq
and the $B_\eta$ are unitary matrices that satisfy:
\beq \label{Betadef}
B_\eta\Gamma^\mu B_\eta^{-1}=\eta\Gamma^{\mu\,*}\,,\quad\qquad  \eta=\pm 1\,.
\eeq
For even $d$, a convenient choice is
$B_{+}=B_{-}\Gamma_{d+1}$~\cite{sohniusapp}.

If $d$ is odd with signature $(s,t)$, then the
$2^{(d-1)/2}\times 2^{(d-1)/2}$ gamma matrices
$\Gamma^\mu$ ($\mu=1,2,\ldots,d$) consist
of $\left\{\Gamma^1, \Gamma^2,\ldots,\Gamma^{d-1},
\pm i\Gamma_{d+1}\right\}$ of the $(d-1)$-dimensional theory of
signature $(s-1,t)$.  By assumption, $\mu=d$ is a space index, so that
$\Gamma^d\equiv \pm i\Gamma_{d+1}$ is anti-hermitian.
In the case of odd $d$, only one sign choice for $\eta$, namely
$\eta=(-1)^{(s-t+1)/2}$, is consistent with \eq{Betadef}
as applied to $\Gamma^d$.\footnote{The two
sign choices for $\Gamma^d$ correspond to two inequivalent
representations of the Clifford algebra [\eq{cliff}] for $d$ odd.
Nevertheless, the corresponding $\Sigma^{\mu\nu}$
yield equivalent spinor representations of SO($s,t$).}
Consequently, only
one definition of the charge-conjugated spinor is viable,
namely $\Psi^C=B_-^{-1}\Psi^*$
for $s-t=1$, 5 (mod 8) and  $\Psi^C=B_+^{-1}\Psi^*$ for
$s-t=3$, 7 (mod~8).

One important property of the $B_\eta$ is~\cite{wet,scherk,kugo,andrade,tanii}:
\beq \label{Bproperty}
B^*_\eta B_\eta=\varepsilon_{\eta}\,,\qquad \varepsilon_\eta=\pm 1\,,
\eeq
for $\eta=\pm 1$ in even-dimensional spacetimes and
$\eta=(-1)^{(s-t+1)/2}$ in odd-dimensional spacetimes.
In particular~\cite{kugo},\footnote{For $d$ even, one can use
$B_{+}=B_{-}\Gamma_{d+1}$ and
$B_\eta\Gamma_{d+1}B_{\eta}^{-1}=(-1)^{(s-t)/2}\Gamma_{d+1}^*$
to derive $\varepsilon_{+}=(-1)^{(s-t)/2}\,\varepsilon_{-}$.}
\beq \label{signeps}
\varepsilon_{-}=\begin{cases}
+1\,, & \text{for $s-t=0,1,2$ (mod 8)}\,,\\
-1\,, & \text{for $s-t=4,5,6$ (mod 8)}\,,\end{cases}
\qquad\quad
\varepsilon_{+}=\begin{cases}
+1\,, & \text{for $s-t=0,6,7$ (mod 8)}\,,\\
-1\,, & \text{for $s-t=2,3,4$ (mod 8)}\,.\end{cases}
\eeq

Using the charge-conjugated spinor defined in \eq{dConj},
one can define a self-conjugate spinor, $\Psi^C=\Psi$.
Two cases arise depending on the sign of
$\eta$~\cite{kugo,andrade,tanii,vanpro},
\beqa
\text{Majorana spinor:}&\qquad &\Psi=B^{-1}_{-}\Psi^*\,,
\label{mspinor} \\
\text{pseudo-Majorana spinor:}&\qquad &\Psi=B^{-1}_{+}\Psi^*\,.
\label{pmspinor}
\eeqa
Due to the reality conditions [\eqs{mspinor}{pmspinor}],
the (pseudo-)Majorana spinor possesses
$2^{[d/2]}$ real degrees of freedom.
Using \eq{Bproperty}, one immediately sees that \eqs{mspinor}{pmspinor}
are respectively consistent
if and only if $\varepsilon_\eta=+1$.\footnote{If $\varepsilon_\eta=-1$
then one can introduce a generalized reality condition
[cf.~\eq{genreality}], which constrains
the structure of a multiplet of Dirac fermions that
transforms under a pseudo-real representation of the flavor group.
In this case, the corresponding
(generalized) self-conjugate spinors are called
symplectic (pseudo-)Majorana spinors, as discussed below
\eq{wwt}.}
The possible existence of Majorana [pseudo-Majorana] spinors
in $d$-dimensional spacetime
depends on the choice of $s-t$ such that $\varepsilon_{-}=+1$
[$\varepsilon_{+}=+1]$.  Using \eq{signeps}, it
follows that Majorana spinors can only exist
in spacetimes where $s-t=0$, 1, 2 (mod 8), and
pseudo-Majorana can only exist in spacetimes where $s-t=0$, 6, 7
(mod 8).\footnote{As shown in \Ref{kugo,andrade}, no SO$(s,t)$-invariant
mass term is allowed for a pseudo-Majorana spinor.}
In particular, a Majorana spinor cannot exist in
four-dimensional Euclidean space.

Given a choice of sign for $\eta=\pm 1$, one can define a
corresponding charge conjugation matrix
$C_\eta$, which is unitary and
is defined by\footnote{In four-dimensional Minkowski spacetime,
we identify $C=(C_{-}^{\T})^{-1}=C_{-}^*$ [cf.~\eq{CConj}]
and $B=C_+$ [cf.~\eq{bdef4}].  In this case, one cannot use $C_{+}$ to
consistently define a self-conjugate spinor, as the corresponding
$\varepsilon_{+}=-1$.}
\beq
C_\eta\equiv B_\eta^{\T}A\,,\qquad {\rm where}\quad
C_\eta\Gamma^\mu C^{-1}_\eta=\eta (-1)^{t+1}\Gamma^{\mu\,\T}\,.
\eeq
\Eq{dConj} then yields $\Psi^{C}=C_\eta^*\,\overline{\Psi}^{\T}$.
The unitary matrices $A$, $B_\eta$ and $C_\eta$ satisfy the following useful
identities\cite{kugo,andrade}:
\beq
B_\eta^{\T}=\varepsilon_\eta B_\eta\,,\qquad
C_\eta^{\T}=\varepsilon_\eta \eta^t(-1)^{t(t-1)/2} C_\eta\,,\qquad
A^* B_\eta =\eta^t B_\eta A\,,\qquad
A^{\T}C_\eta=\eta^t C_{\eta}A^{-1}\,.
\eeq

In the case of even $d$,
one can define left and right-handed chiral projection operators:
\beq
P_L\equiv\half(1-\Gamma_{d+1})\,,\qquad\quad
P_R\equiv\half(1+\Gamma_{d+1})\,,
\eeq
and introduce Weyl fermions,  $\Psi_L$ and $\Psi_R$,
which satisfy $\Gamma_{d+1}\Psi_{R,L}=\pm\Psi_{R,L}$.
Equivalently,
\beq
\Psi_L\equiv P_L\Psi\,,\qquad\qquad \Psi_R\equiv P_R\Psi\,,
\eeq
so that $\Psi_L$ (and likewise $\Psi_R$) possesses $2^{(d-2)/2}$
complex degrees of freedom.
It is possible for a spinor to be simultaneously a (pseudo) Majorana
and a Weyl spinor if the spinor and its charge conjugate have the same
chirality, in which case $B_\eta\Gamma_{d+1}B_\eta^{-1}=\Gamma_{d+1}^\ast$
(for even $d$).  The latter condition holds when $i^{s-t}=1$ or
equivalently $s-t=0$ (mod 4).  Combining this
requirement with the condition for
the existence of a (pseudo) Majorana spinor, it follows that a (pseudo)
Majorana-Weyl spinor, which possesses $2^{(d-2)/2}$
real degrees of freedom, can only exist in spacetimes where
$s-t=0$ (mod 8).
For further details, see
refs.~\cite{Benn,scherk,vanN,coq,kugo,
andrade,tanii,vanpro,Brandt}.

As in \sec{subsec:generalmass}, one can also consider a multiplet of
fermions $\Psi_i$ that transforms under a complex, real or pseudo-real
representation $R$ of the flavor group $G$ as
\beq \label{DR}
\Psi_i\to (D_R)_i{}^j\Psi_j\,,\qquad \quad
D_R=\exp(-i\theta^a\boldsymbol{T_R^a})\,,\qquad\quad
i,j=1,2,\ldots,d_R\,,
\eeq
where $D_R$ is unitary and the corresponding generators
$\boldsymbol{T_R^a}$ are hermitian.
The dimension of $R$ is denoted by $d_R$,
which must be even in the pseudo-real case.
In both the real and pseudo-real cases, one can also
impose a reality condition
that generalizes the Majorana conditions
of \eqs{mspinor}{pmspinor},
\beq \label{genreality}
(\Psi_i)^*\equiv\Psi^{*\,i}=W^{ij}B_\eta\Psi_j\,,
\eeq
where $W$ is a unitary matrix and
$B_\eta$ acts on the (suppressed) spinor indices of $\Psi_j$.
Additional constraints on the form of $W$ are obtained as follows.
First, taking the complex conjugate of \eq{genreality} and inserting the
result back into the same equation, it follows that
\beq \label{ww}
W^*W=\varepsilon_\eta\mathds{1}\,,
\eeq
after making use of \eq{Bproperty}.  Second, \eq{genreality} must
hold true if $\Psi$ is replaced by
$D_R\Psi$
on both sides of the
equation, in order to be compatible with the flavor symmetry group
transformation law [\eq{DR}].  This
latter requirement combined with \eq{ww} yields:
\beq \label{Dreal}
D_R=\varepsilon_\eta W^*D_R^* W = W^{-1} D_R^* W\,.
\eeq
\Eq{Dreal} can be expressed in terms of the flavor group generators,
\beq
i\boldsymbol{T^a_R}=W^{-1} (i\boldsymbol{T^a_R})^*\,W\,.
\eeq
Comparing with \eqst{selfconjugate}{prealrep}, we conclude that
the unitary matrix $W$ satisfies:
\beq \label{wwt}
W=\varepsilon_\eta W^{\T}\,,\qquad\qquad
\varepsilon_\eta=\begin{cases}
+1\,, & \text{$R$ is a real representation}\,, \\
-1\,, & \text{$R$ is a pseudo-real representation}\,.\end{cases}
\eeq

When $R$ is a real representation, $W=W^{\T}$, and a basis for the
flavor group
generators can be chosen such that $W=\mathds{1}$
[cf.~\eq{realcan}], in which case $D_R$ is a real
orthogonal matrix.   Since
$\varepsilon_\eta=+1$, \eq{genreality} yields (pseudo-)Majorana spinors
(depending on the sign of $\eta$) as
defined previously in \eqs{mspinor}{pmspinor}.

When $R$ is a pseudo-real representation, $W=-W^{\T}$, and
a basis for the flavor group
generators can be chosen such that
$W=J\equiv{\rm diag}\left\{\left(\begin{smallmatrix} \phm 0 & \,\,1 \\
-1 &\,\, 0 \end{smallmatrix}\right)\,,\,
\left(\begin{smallmatrix} \phm 0 & \,\,1 \\
-1 &\,\, 0 \end{smallmatrix}\right)\,,\,\cdots\,,\,\left(\
\begin{smallmatrix} \phm 0 & \,\,1 \\
-1 &\,\, 0 \end{smallmatrix}\right)\right\}$
is a $d_R\times d_R$ matrix, where $d_R$ is even [cf.~\eq{prealcan}].
In this case, $D_R^{\T}JD_R=J$, which implies that $D_R$ is
a unitary symplectic matrix~\cite{vanN2}.  Moreover,
$\varepsilon_\eta=-1$, which was incompatible with the reality
conditions of \eqs{mspinor}{pmspinor}, but is compatible
with the generalized reality condition of \eq{genreality}.

Therefore, we define \textit{symplectic} (pseudo-)Majorana
spinors~\cite{wetterich,kugo,vanN2,vanpro,Brandt} to be spinors that
transform as a pseudo-real representation under some flavor group
and satisfy the generalized reality condition of
\eq{genreality}, where $W$ is a unitary antisymmetric matrix,
depending on the choice of
$\eta=\pm 1$ (with $\eta=-1$ yielding the ``pseudo'' designation).
As suggested by \eqst{LagPseudoreal}{lagDiracpseudo}, $2d_R$ symplectic
(pseudo-)Majorana spinors are equivalent to $d_R$ Dirac fermions.
The possible existence of symplectic
(pseudo-)Majorana spinors in a $d$-dimensional spacetime is governed
by \eq{signeps}.  Requiring that $\varepsilon_\eta=-1$ implies that
symplectic Majorana spinors exist in spacetimes where $s-t=4$, 5, 6 (mod 8),
and symplectic pseudo-Majorana spinors exist  in spacetimes
where $s-t=2$, 3, 4 (mod 8).  Using this nomenclature, the
fermions described by the four-dimensional Minkowski space Lagrangian given in
\eq{LagPseudoreal} are symplectic pseudo-Majorana spinors.

\subsection{Four-component spinor wave functions}
\renewcommand{\theequation}{G.4.\arabic{equation}}
\renewcommand{\thefigure}{G.4.\arabic{figure}}
\renewcommand{\thetable}{G.4.\arabic{table}}
\setcounter{equation}{0}
\setcounter{figure}{0}
\setcounter{table}{0}

In four-dimensional Minkowski space,
the free four-component Majorana field
can be expanded in a Fourier series;
each positive [negative] frequency mode is
multiplied by a {\it commuting} spinor wave function
$u(\boldsymbol{\vec p},s)$ [$v(\boldsymbol{\vec p},s)$]
as in \eq{twocompmodes},\footnote{Some subtleties arise
in the choice of relative phases of the creation and annihilation
operators, which are related to the C, CP and CPT transformation
properties of the Majorana field.  For further details, see \Ref{kayser}.}
\beq \label{Majmodes}
\Psi_{Mi}(x)=
\sum_s\int\,\frac{d^3 \boldsymbol{\vec p}}
{(2\pi)^{3/2}(2E_{\boldsymbol{p}i})^{1/2}}
\left[u(\boldsymbol{\vec p},s)a_i(\boldsymbol{\vec
p},s)e^{\BDneg ip\newcdot x}+v(\boldsymbol{\vec p},s)a_i^\dagger(
\boldsymbol{\vec p},s)e^{\BDpos ip\newcdot x}\right]\,,
\eeq
where $E_{\boldsymbol{p}i}\equiv (|{\boldsymbol{\vec p}}|^2+m_i^2)^{1/2}$,
and the creation operators $a_i^\dagger$
and the annihilation operators $a_i$
satisfy anticommutation relations:
\beq \label{etacrabM}
\{a_i(\boldsymbol{\vec p},s),a_j^\dagger(\boldsymbol{\vec
p}^{\,\,\prime},s^\prime)\}=\delta^3(\boldsymbol{\vec p}
-\boldsymbol{\vec p}^{\,\prime})
\delta_{ss'}\delta_{ij}\,,
\eeq
with all other anticommutation relations vanishing.
We employ covariant normalization of the one-particle states given
by \eq{xistate}.  It then follows that
\beqa
\bra{0}\Psi_M(x)\ket{\boldsymbol{\vec p},s} &=&
u({\boldsymbol{\vec p}},s)e^{\BDneg ip\newcdot x}\,,
\qquad\quad\,\,
\bra{0}
\Psibar_M(x)\ket{\boldsymbol{\vec p},s}=
\bar{v}({\boldsymbol{\vec p}},s)e^{\BDneg ip\newcdot x}\,,
\label{instateM}\\
\bra{{\boldsymbol{\vec p}},s}\Psibar_M(x)\ket{0} &=&
\bar{u}({\boldsymbol{\vec p}},s)e^{\BDpos ip\newcdot x}\,,
\qquad\qquad
\bra{{\boldsymbol{\vec p}},s}\Psi_M(x)\ket{0}=
v({\boldsymbol{\vec p}},s)e^{\BDpos ip\newcdot x}
\,.
\label{outstateM}
\eeqa
These results are the four-component spinor versions of
\eqs{instate}{outstate}.

Likewise, the
free Dirac field can be expanded in a Fourier series,
\beq \label{fourcompmodes}
\Psi_{i}(x)=
\sum_s\int\,\frac{d^3 \boldsymbol{\vec p}}
{(2\pi)^{3/2}(2E_{\boldsymbol{p}i})^{1/2}}
\left[u(\boldsymbol{\vec p},s)a_i(\boldsymbol{\vec
p},s)e^{\BDneg ip\newcdot x}+v(\boldsymbol{\vec p},s)b_i^\dagger(
\boldsymbol{\vec p},s)e^{\BDpos ip\newcdot x}\right]\,,
\eeq
where the creation operators $a_i^\dagger$ and $b_i^\dagger$
and the annihilation operators $a_i$ and $b_i$
satisfy anticommutation relations:
\beqa \label{etacrab}
\{a_i(\boldsymbol{\vec p},s),a_j^\dagger(\boldsymbol{\vec
p}^{\,\,\prime},s^\prime)\}&=&\delta^3(\boldsymbol{\vec p}
-\boldsymbol{\vec p}^{\,\prime})
\delta_{ss'}\delta_{ij}\,,\\
\{b_i(\boldsymbol{\vec p},s),b_j^\dagger(\boldsymbol{\vec
p}^{\,\,\prime},s^\prime)\}&=&\delta^3(\boldsymbol{\vec p}
-\boldsymbol{\vec p}^{\,\prime})
\delta_{ss'}\delta_{ij}\,,
\eeqa
with all other anticommutation relations vanishing.
We employ covariant normalization of the
fermion ($F)$ and antifermion ($\overline{F}$) one-particle states given
by \eq{chietastate}.  It then follows that
\beqa
\bra{0}\Psi(x)\ket{\boldsymbol{\vec p},s;F} &=&
u({\boldsymbol{\vec p}},s)e^{\BDneg ip\newcdot x}\,,
\qquad\quad\,\,
\bra{0}
\Psibar(x)\ket{\boldsymbol{\vec p},s;\overline{F}}=
\bar{v}({\boldsymbol{\vec p}},s)e^{\BDneg ip\newcdot x}\,,
\label{instateD}\\
\bra{{\boldsymbol{\vec p}},s;F}\Psibar(x)\ket{0} &=&
\bar{u}({\boldsymbol{\vec p}},s)e^{\BDpos ip\newcdot x}\,,
\qquad\qquad
\bra{{\boldsymbol{\vec p}},s;\overline{F}}\Psi(x)\ket{0}=
v({\boldsymbol{\vec p}},s)e^{\BDpos ip\newcdot x}
\,,
\label{outstateD}
\eeqa
and the four other single-particle matrix elements vanish.
These results are the four-component spinor versions of
\eqst{chainstate}{chboutstate}.
The Fourier expansion of the
charge-conjugated free Dirac field $\Psi_i^C(x)=C\Psibar^{\T}_i(x)$
is given by:
\beq \label{Cfourcompmodes}
\Psi^C_{i}(x)=
\sum_s\int\,\frac{d^3 \boldsymbol{\vec p}}
{(2\pi)^{3/2}(2E_{\boldsymbol{p}i})^{1/2}}
\left[u(\boldsymbol{\vec p},s)b_i(\boldsymbol{\vec
p},s)e^{\BDneg ip\newcdot x}+v(\boldsymbol{\vec p},s)a_i^\dagger(
\boldsymbol{\vec p},s)e^{\BDpos ip\newcdot x}\right]\,,
\eeq
where we have used \eq{uvspinrelation1}.  That is,
the charge conjugation transformation
interchanges the annihilation and creation operators,
$a_i\leftrightarrow b_i$ and $a^\dagger_i\leftrightarrow b^\dagger_i$.
Thus, if $\Psi^C(x)=\Psi(x)$, then we must identify $a=b$ and
$a^\dagger=b^\dagger$, corresponding to the free Majorana field given
in \eq{Majmodes}.

The two-component spinor momentum space wave functions are
related to the traditional
four-component spinor wave functions according to:
\beqa
&&u(\boldsymbol{\vec p},s) = \begin{pmatrix}x_\alpha(\boldsymbol{\vec p},s)
\\[4pt]  y^{\dagger\dot{\alpha}}(\boldsymbol{\vec p},s)\end{pmatrix}
\,,\qquad\qquad
\ubar(\boldsymbol{\vec p},s) = (y^\alpha(\boldsymbol{\vec p},s),\>
 x^\dagger_{\dot{\alpha}}(\boldsymbol{\vec p},s))
\,,\label{uspin4}\\[6pt]
&&v(\boldsymbol{\vec p},s) = \begin{pmatrix} y_\alpha(\boldsymbol{\vec p},s)
\\[4pt]   x^{\dagger\dot{\alpha}}(\boldsymbol{\vec p},s)\end{pmatrix}
\,, \qquad\qquad
\vbar(\boldsymbol{\vec p},s) = (x^\alpha(\boldsymbol{\vec p},s),\>
 y^\dagger_{\dot{\alpha}}(\boldsymbol{\vec p},s))
\,,\label{vspin4}
\eeqa
where the $u$ and $v$-spinors are related by
\beqa
v(\boldsymbol{\vec p},s) &=& C\ubar(\boldsymbol{\vec p},s)^{\T}
\,,\qquad\qquad\quad
u(\boldsymbol{\vec p},s) = C\vbar(\boldsymbol{\vec p},s)^{\T}\,,
\label{uvspinrelation1} \\
\vbar(\boldsymbol{\vec p},s) &=& -u(\boldsymbol{\vec p},s)^{\T}C^{-1}
\,,\qquad\quad\,
\ubar(\boldsymbol{\vec p},s) = -v(\boldsymbol{\vec p},s)^{\T}C^{-1}\,.
\label{uvspinrelation2}
\eeqa

The spin quantum number takes on values $s=\pm\half$, and refers
either to the component of the spin as measured in the rest frame with
respect to a fixed axis or to the helicity (as discussed in
\sec{subsec:singleWeyl} and \app{C}).  Note that the
$u$ and $v$-spinors also satisfy:
\beq \label{uvrelat}
v({\boldsymbol{\vec p}},s)=-2s\gamma\ls{5}u({\boldsymbol{\vec p}},-s)\,,
\qquad\qquad
u({\boldsymbol{\vec p}},s)=2s\gamma\ls{5}v({\boldsymbol{\vec p}},-s)\,,
\eeq
which follows from \eq{xyrelation}.
Explicit forms for the four-component
spinor wave functions in the chiral representation can be obtained using
\eqst{explicitxa}{explicityb}, where $\chi\ls{s}({\boldsymbol{\hat s}})$
is given in \eq{twocomp}.  For helicity spinors, further
simplifications result by employing \eqst{explicithelxa}{explicithelyb}.

One can check that $u$ and $v$
satisfy the Dirac equations
\beqa
&& (\slashchar{p} \BDminus m)\, u(\boldsymbol{\vec p},s) =
(\slashchar{p} \BDplus m)\, v(\boldsymbol{\vec p},s)= 0\,, \label{fourdiraceq1}
\\[6pt]
&& \ubar(\boldsymbol{\vec p},s)\, (\slashchar{p} \BDminus m)  =
\vbar(\boldsymbol{\vec p},s)\, (\slashchar{p} \BDplus m) = 0 \, ,
\label{fourdiraceq2}
\eeqa
corresponding to \eqst{onshellone}{onshellfour}, and
\beqa
&& (2s\gamma\ls{5}\slashchar{S}\BDminus 1)\,u(\boldsymbol{\vec p},s)=
(2s\gamma\ls{5}\slashchar{S}\BDminus 1)\,v(\boldsymbol{\vec p},s)=0\,,
 \label{fourspineq1}
\\[5pt]
&& \ubar(\boldsymbol{\vec p},s)\,(2s\gamma\ls{5}\slashchar{S}\BDminus 1)=
\vbar(\boldsymbol{\vec p},s)\,(2s\gamma\ls{5}\slashchar{S}\BDminus 1)=0\,,
 \label{fourspineq2}
\eeqa
corresponding to \eqst{spinone}{spinfour}, where the
spin vector $S^\mu$ is defined in \eq{fixedsvect}.\footnote{We
use the standard Feynman slash
notation: $\slashchar{p}\equiv \gamma_\mu p^\mu$ and
$\slashchar{S} \equiv\gamma_\mu S^\mu$.}
For massive fermions, \eqst{xxdagmassive}{ydagxdagmassive}
correspond to
\beqa
&&\phantom{line} \nonumber \\[-23pt]
&& u({\boldsymbol{\vec p}},s) \ubar({\boldsymbol{\vec p}},s)
= \half(1\BDplus 2s\gamma\ls{5}\slashchar{S})\,(\BDpos \slashchar{p}+m)\,,
\label{projectionops1}
\\[5pt]
&& v({\boldsymbol{\vec p}},s) \vbar({\boldsymbol{\vec p}},s)
= \half(1\BDplus 2s\gamma\ls{5}\slashchar{S})\,(\BDpos \slashchar{p}-m)\,.
\label{projectionops2}
\phantom{line} \nonumber \\[-23pt]
\eeqa

To apply the above formulae to the massless case
we must employ helicity states, where $s$ is replaced by
the helicity quantum number $\lambda$, and $S^\mu$ is defined by
\eq{spinvec}.  In particular,
in the $m\to 0$ limit, $S^\mu= p^\mu/m+{\mathcal O}(m/E)$.  Inserting
this result in \eqs{fourspineq1}{fourspineq2} and using the Dirac equations,
it follows that the massless helicity spinors are eigenstates of
$\gamma\ls{5}$,
\beq \label{mzerospinors}
  \gamma\ls{5} u({\boldsymbol{\vec p}},\lambda) = 2\lambda
u({\boldsymbol{\vec p}},\lambda)\,,
\qquad\qquad\quad
  \gamma\ls{5} v({\boldsymbol{\vec p}},\lambda) = -2\lambda
v({\boldsymbol{\vec p}},\lambda)\,.
\eeq
Combining these results with \eq{uvrelat}
[with $s$ replaced by $\lambda$] yields:
\beq \label{vequalu}
  v(p,\lambda) = -2\lambda \gamma\ls{5} u(p,-\lambda)=
  u(p,-\lambda)\,,
\qquad\qquad \lambda=\pm\half\,,
\eeq
and we see that the massless $u$ and $v$ spinors of opposite
helicity are the same.

Applying the above $m\to 0$ limiting procedure to
\eqs{projectionops1}{projectionops2} and using the mass-shell condition
($\slashchar{p}\slashchar{p}= \BDpos p^2= m^2$),
one obtains the massless helicity projection operators
corresponding to  \eqst{xxdagmassless}{ydagxdagmassless}:
\beqa
&&\phantom{line} \nonumber \\[-23pt]
u({\boldsymbol{\vec p}},\lambda) \ubar({\boldsymbol{\vec p}},\lambda) &=&
\BDpos
\half(1+2\lambda\gamma\ls{5})\,\slashchar{p}\,,
\label{masslessprojection1}\\[5pt]
v({\boldsymbol{\vec p}},\lambda) \vbar({\boldsymbol{\vec p}},\lambda) &=&
\BDpos
\half(1-2\lambda\gamma\ls{5})\,\slashchar{p}\,.
\label{masslessprojection2}
\phantom{line}\nonumber \\[-23pt]
\eeqa

Summing over the spin degree of freedom, we obtain the
spin-sum identities corresponding to \eqst{xxdagsummed}{ydagxdagsummed},
\beqa
&&\phantom{line} \nonumber \\[-30pt]
\sum_s u({\boldsymbol{\vec p}},s) \ubar({\boldsymbol{\vec p}},s)
 &=& \BDpos \slashchar{p} + m\,, \\[5pt]
\sum_s v({\boldsymbol{\vec p}},s) \vbar({\boldsymbol{\vec p}},s)
 &=& \BDpos \slashchar{p} - m\,, \\ [5pt]
 \sum_s u({\boldsymbol{\vec p}},s) v^{\T}({\boldsymbol{\vec p}},s)
 &=& (\BDpos \slashchar{p} + m)C^{\T}\,, \\[5pt]
\sum_s \ubar^{\T}({\boldsymbol{\vec p}},s) \vbar({\boldsymbol{\vec p}},s)
 &=&  C^{-1}(\BDpos \slashchar{p} - m)\,,\\ [5pt]
\sum_s \vbar^{\T}({\boldsymbol{\vec p}},s) \ubar({\boldsymbol{\vec p}},s)
 &=& C^{-1}(\BDpos \slashchar{p} + m)\,, \\[5pt]
\sum_s v({\boldsymbol{\vec p}},s) u^{\T}({\boldsymbol{\vec p}},s)
 &=&  (\BDpos \slashchar{p} - m)C^{\T}\,,
\eeqa
which are valid for both the massive case and the massless $m\to 0$ limit.

As previously noted, the results for the bilinear covariants obtained
in \eqst{twotofoura}{ptensorbilinear} can also be applied to expressions
involving the commuting spinor wave functions.  Various relations
among the possible bilinear covariants can be established by using
\eqs{uvspinrelation1}{uvspinrelation2}.  As an example, for
$\Gamma=\mathds{1}\,,\,\gamma\ls{5}\,,\,\gamma^\mu\,,\,\gamma^\mu\gamma\ls{5}
\,,\,\Sigma^{\mu\nu}\,,\,\Sigma^{\mu\nu}\gamma\ls{5}$,
\beqa
\ubar(\boldsymbol{\vec p}_1,s_1)\Gamma v(\boldsymbol{\vec p}_2,s_2)
&=& -v(\boldsymbol{\vec p}_1,s_1)^{\T}C^{-1}\Gamma C
\ubar(\boldsymbol{\vec p}_2,s_2)^{\T}=
-\eta^C\ls{\Gamma}\ubar(\boldsymbol{\vec p}_2,s_2)\Gamma
v(\boldsymbol{\vec p}_1,s_1)\,,\label{uGammav} \\
\ubar(\boldsymbol{\vec p}_1,s_1)\Gamma u(\boldsymbol{\vec p}_2,s_2)
&=& -v(\boldsymbol{\vec p}_1,s_1)^{\T}C^{-1}\Gamma C
\vbar(\boldsymbol{\vec p}_2,s_2)^{\T}=
-\eta^C\ls{\Gamma}\vbar(\boldsymbol{\vec p}_2,s_2)\Gamma
v(\boldsymbol{\vec p}_1,s_1)\,, \label{uGammau}
\eeqa
where the sign $\eta^C\ls{\Gamma}$ [defined in \eq{ccgamma}] arises after
taking the transpose and applying \eq{ccgamma}.
In particular, the (commuting) $u$ and $v$ spinors satisfy the
following relations:
\beqa
\ubar(\boldsymbol{\vec p}_1,s_1)P_L v(\boldsymbol{\vec p}_2,s_2) &=&
-\ubar(\boldsymbol{\vec p}_2,s_2)P_L
v(\boldsymbol{\vec p}_1,s_1)\,, \label{uva} \\[4pt]
\ubar(\boldsymbol{\vec p}_1,s_1)P_R v(\boldsymbol{\vec p}_2,s_2) &=&
-\ubar(\boldsymbol{\vec p}_2,s_2)P_R
v(\boldsymbol{\vec p}_1,s_1)\,, \label{uvb}\\[4pt]
\ubar(\boldsymbol{\vec p}_1,s_1)\gamma^\mu
P_L v(\boldsymbol{\vec p}_2,s_2) &=&
\ubar(\boldsymbol{\vec p}_2,s_2)\gamma^\mu
P_R v(\boldsymbol{\vec p}_1,s_1)\,, \label{uvc} \\[4pt]
\ubar(\boldsymbol{\vec p}_1,s_1)\gamma^\mu
P_R v(\boldsymbol{\vec p}_2,s_2) &=&
\ubar(\boldsymbol{\vec p}_2,s_2)\gamma^\mu
P_L v(\boldsymbol{\vec p}_1,s_1)\,, \label{uvd}
\eeqa
and four similar relations obtained by interchanging
$v(\boldsymbol{\vec p}_2,s_2)\leftrightarrow u(\boldsymbol{\vec p}_2,s_2)$.

\subsection{Feynman rules for four-component fermions}
\renewcommand{\theequation}{G.5.\arabic{equation}}
\renewcommand{\thefigure}{G.5.\arabic{figure}}
\renewcommand{\thetable}{G.5.\arabic{table}}
\setcounter{equation}{0}
\setcounter{figure}{0}
\setcounter{table}{0}

We now illustrate some basic applications of the above formalism.
In particular, we shall establish a set of Feynman rules for
four-component fermions that treat both Dirac and Majorana fermions on
the same footing.  These rules generalize the standard Feynman rules
for four-component Dirac fermions found in most quantum field theory
textbooks.  Two advantages of the rules presented here are: (i) no
factors of the charge conjugation matrix $C$ are required for
fermion interaction vertices and propagators, and (ii) the
\textit{relative} sign between different
diagrams corresponding to the same physical process
is simply determined.  Our rules
have been obtained by translating our two-component
fermion Feynman rules
into the four-component spinor language.  The resulting Feynman
rules for four-component Majorana fermions are equivalent to the
set of rules independently obtained in \Ref{Denner:1992me}
(see also \refs{Gates:1987ay}{Kleiss:2009hu}).

Consider first the Feynman rule for the four-component fermion
propagator.
Virtual Dirac fermion lines can either correspond to $\Psi$ or
$\Psi^C$.  Here, there is no ambiguity in the propagator Feynman rule,
since for free Dirac fermion fields,\footnote{%
In deriving \eq{CCprop}, we have used 
$\mathcal{C}\,\Psi_a\,\mathcal{C}^{-1}=\eta_c\Psi_a^C$
and $\mathcal{C}\,\overline\Psi\llsup{\,a}\,\mathcal{C}^{-1}
=\eta^*_c\,\overline{\Psi^C}\llsup{\,a}$, where
$\mathcal{C}$ is the 
charge conjugation \textit{operator} that acts on
the quantum Hilbert space 
and $\eta_c$ is a convention-dependent phase factor~\cite{Cbook,Novozhilov}.
Note that $\mathcal{C}$ is a unitary operator and
$\mathcal{C}\ket{0}=\ket{0}$ in the free-field vacuum.}
\beq \label{CCprop}
\left\langle 0\right|T[\Psi_a(x)\Psibar\llsup{\,b}(y)]
\left|0\right\rangle=
\left\langle 0\right |T[\Psi^C_a(x)\overline{\Psi^C}\llsup{\,b}(y)]
\left|0\right\rangle\,,
\eeq
so that the Feynman rules for the propagator of a $\Psi$ and $\Psi^C$
line, exhibited in \fig{fig:diracprop}, are identical.
The same rule also applies to a four-component Majorana fermion.
\begin{figure}[ht!]
\centerline{
\begin{picture}(200,50)(-135,-6)
\thicklines
\LongArrow(-110,25)(-70,25)
\ArrowLine(-130,15)(-50,15)
\put(-90,30){$p$}
\put(-56,2){$a$}
\put(-130,2){$b$}
\put(20,10){$\displaystyle
 {\frac{i(\slashchar{p}\BDplus m)_a{}^b}
 {p^2 \BDminus m^2 \BDplus i\epsilon}}$}
\end{picture}
}
\caption[0]{\label{fig:diracprop} Feynman rule for the propagator of
a four-component fermion with mass $m$.   The same rule applies to
a Majorana, Dirac and charge-conjugated Dirac fermion.
The four-component spinor labels $a$ and $b$ are specified.
}
\end{figure}

Using \eq{gamma4}, the four-component
fermion propagator Feynman rule can be expressed as
a partitioned matrix of $2\times 2$ blocks,
\beqa
\qquad
\begin{picture}(47.5,21)(-10,10.5)
\Text(-20,32)[c]{$p$}
\LongArrow(2,24)(-40,24)
\ArrowLine(13,12)(-55,12)
\Text(30,12)[c]{$=$}
\Text(-58,4)[c]{$a$}
\Text(15,3)[c]{$b$}
\end{picture}
&& \begin{pmatrix}
\mbox{\begin{picture}(47.5,21)(-10,10.5)
\SetScale{0.7}
\ArrowLine(15,25)(-15,25)
\ArrowLine(15,25)(45,25)
\end{picture}}
\mbox{\begin{picture}(47.5,21)(-10,10.5)
\SetScale{0.7}
\ArrowLine(53,25)(-3,25)
\end{picture}}
\,\,\\
\mbox{\begin{picture}(47.5,21)(-10,10.5)
\SetScale{0.7}
\ArrowLine(-15,25)(40,25)
\end{picture}}
\mbox{\begin{picture}(47.5,21)(-10,10.5)
\SetScale{0.7}
\ArrowLine(-5,25)(30,25)
\ArrowLine(55,25)(30,25)
\end{picture}}
\end{pmatrix}
\,\,=\,\,
\frac{i}{p^2 \BDminus m^2 \BDplus i\epsilon}\begin{pmatrix}
\BDpos m\,\delta_{\alpha}{}^{\beta} & \quad
p\newcdot\sigma_{\alpha\dot{\beta}}
 \\[8pt]
p\newcdot\sigmabar^{\dot{\alpha}\beta} & \quad
\BDpos m\,\delta^{\dot{\alpha}}{}_{\dot\beta}
\end{pmatrix}
\,,\label{twofourprop}
\eeqa
where $a$ and $b$ are four-component spinor indices.
That is, \eq{twofourprop} is a partitioned matrix whose blocks consist of
two-component fermion propagators defined in \fig{fig:neutprop},
with the undotted and dotted $\alpha$ [$\beta$]
indices on the left [right] and with the momentum flowing from right to left.

The derivation of the four-component Dirac fermion propagator is
treated in most modern textbooks of quantum field theory
(see, e.g., \Ref{Peskin:1995ev}).  Here, we briefly sketch the
path integral derivation of the four-component fermion propagator by
exploiting the path integral treatment of the two-component
fermion propagators outlined in \app{F}.
Consider a single massive Dirac fermion
$\Psi(x)$ coupled to an anticommuting four-component
Dirac fermionic source term
\beq
J_\psi(x)\equiv\begin{pmatrix} J_{\eta\alpha}(x) \\
 J_{\chi}^{\dagger \dot{\alpha}}(x)
\end{pmatrix}\,.
\eeq
The corresponding action [\eq{app-action}] in four-component
notation is given by
\beq
S=  \int d^4x \,(\mathscr{L}+ \Jbar_\psi\Psi +\Psibar\,J_\psi)=
\int d^4x \,\left[
\Psibar(i\slashchar{\partial}-m)\Psi
+ \Jbar_\psi\Psi +\Psibar\,J_\psi\right]\,.
\label{app-actionfour}
\eeq
Introducing the momentum space Fourier coefficients:
\beq
\Psi(x) = \int \frac{d^4p}{(2\pi)^4}
e^{\BDneg i p\newcdot x} \widehat\Psi(p)
\,,\qquad
J_\psi(x) = \int \frac{d^4p}{(2\pi)^4}
e^{\BDneg i p\newcdot x} \widehat J_\psi(p)\,,
\eeq
we can identify the following four-component quantities with
matrices of two-component quantities given in \eqs{OXM}{OXM4}:
\beq
\widehat\Psi(p)=A^{-1}\Omega_c(p)\,,\qquad
\widehat J_\psi(p)=X_c(p)\,,\qquad
\slashchar{p} \BDminus m = \BDpos \mathcal{M}(p)A\,,
\eeq
where $A$ is the Dirac conjugation matrix defined in \eqs{psibar}{acdef4}.
Using the results of \app{F}, one easily derives:
\beq
\langle0| T (\Psi(x_1) \Psibar(x_2)) |0\rangle
=\left.\left(-i\frac{\overrightarrow\delta}
{\delta \Jbar_\psi (x_1)}\right) W[J,\Jbar]\left(-i\frac{\overleftarrow\delta}
{\delta J_\psi(x_2)}\right) \right|_{J_\psi=\Jbar_\psi= 0}\,,
\eeq
where
\beq \label{functd4}
W[J_\psi\,,\,\Jbar_\psi]=\exp\left\{-i
\int\frac{d^4 p}{(2\pi)^4} \widehat{\Jbar}_\psi(p)
\frac{\slashchar{p} \BDplus m}{p^2 \BDminus m^2}\widehat J_\psi(p)\right\}\,.
\eeq
Using the analogues of \eqs{chain1}{chain2}, we end up with the expected
result
\beq \label{funprop4}
\langle0| T(\Psi(x_1) \Psibar(x_2)) |0\rangle
=
\int \frac{d^4p}{(2\pi)^4}\,e^{\BDneg i p\newcdot(x_1-x_2)}
\,
\frac{i (\slashchar{p} \BDplus m)}{p^2 \BDminus m^2}\,.
\eeq

In principle, the analogous computation can be carried out for a
single four-component Majorana fermion field $\Psi_M(x)$
coupled to a Majorana fermionic source, $J_\xi(x)$.  The corresponding
action is similar to that of \eq{app-actionfour}, with an extra
overall factor of $1/2$.  However, in evaluating the functional
derivative in \eq{functd4}, one must take into account that the
Majorana fermionic source $J_\xi(x)$ satisfies
$J_\xi^C\equiv C\Jbar\lsup{\,\T}_\xi=J_\xi$.  Consequently, the
functional derivative with respect to $\Jbar_\xi$ is related to
the corresponding functional derivative with respect to $J_\xi$.
Hence, the calculation of \eq{functd4} will yield two equal terms
that will cancel the overall factor of $1/2$, resulting
again in \eq{funprop4}.  Nevertheless, this computation is somewhat
awkward using four-component spinor notation, in contrast to the
straightforward calculation of \app{F}.

We next examine the various interactions involving  four-component fermions.
First, we consider the interactions of a neutral scalar $\phi$ or a
gauge boson $A^a_{\mu}$ with a pair of Majorana fermions
To obtain the interactions of the four-component fermion fields, we first
identify the neutral two-component fermion
mass eigenstate neutral fields $\xi_i$.
Using \eqs{concrete}{eq:lintG2Maj},
the interaction Lagrangian in two-component form is given by:
\beq
\label{lintexpanded}
\mathscr{L}_{\rm int} = -\half(\lambda^{ij}\xi_i\xi_j+\lambda_{ij}
\xi^{\dagger i}\xi^{\dagger j})\phi
\BDminus (G^a)_i{}^j\,\xi^{\dagger i}\sigmabar^\mu\xi_j A^a_\mu\,,
\eeq
where $\lambda$ is a complex symmetric matrix with
$\lambda^{ij}\equiv\lambda_{ij}^*$
[cf.~\eq{eq:complexindexconvention}],
the $A_\mu^a$ are the mass eigenstate gauge fields, and the
corresponding hermitian matrices $G^a$ are defined in \eq{gadefMaj}.
It is now simple to convert this result into
four-component notation:
\beq \label{lint4}
\mathscr{L}_{\rm int} = -\half(\lambda^{ij}\Psibar_{Mi} P_L\Psi_{Mj}
+\lambda_{ij}\Psibar_{M}\llsup{i}P_R\Psi_{M}\llsup{j})\phi
\BDminus (G^a)_i{}^j\Psibar_{M}\llsup{\,i}\gamma^\mu P_L\Psi_{Mj} A^a_\mu\,,
\eeq
where the $\Psi_{Mj}$ are a set of (neutral) Majorana
four-component fermions.
It is convenient to use \eq{majidentitya} to
rewrite the term proportional to
$(G^a)_i{}^j$ in \eq{lint4} as follows
\beqa
(G^a)_i{}^j\Psibar_{M}\llsup{\,i}\gamma^\mu P_L\Psi_{Mj}&=&\half
(G^a)_i{}^j\left[\Psibar_{M}\llsup{\,i}\gamma^\mu P_L\Psi_{Mj}
-\Psibar_{Mj}\gamma^\mu P_R\Psi_{M}\llsup{\,i}\right] \nonumber \\
&=& \half\Psibar_{Mi}\gamma^\mu\left[(G^a)_i{}^j P_L-(G^a)_j{}^i P_R\right]
\Psi_{Mj}\,.
\eeqa
In the last step above, we have lowered the flavor indices of the
four-component Majorana fermion fields, as the heights of these
indices can be arbitrarily chosen [cf.~\eq{majflavorind}].
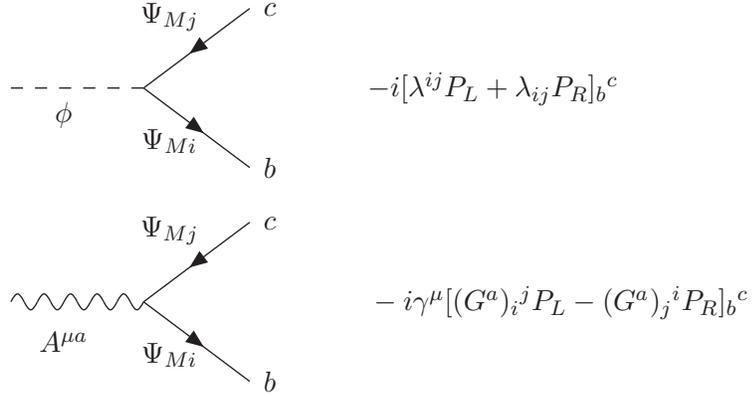
\begin{figure}[t!]
\begin{center}
\begin{picture}(200,58)(0,20)
\DashLine(10,40)(60,40)5
\ArrowLine(100,70)(60,40)
\ArrowLine(60,40)(100,10)
\Text(108,70)[]{$c$}
\Text(108,10)[]{$b$}
\Text(30,30)[]{$\phi$}
\Text(70,20)[]{$\Psi_{Mi}$}
\Text(70,67)[]{$\Psi_{Mj}$}
\Text(145,40)[l]{$-i[\lambda^{ij}P_L+\lambda_{ij} P_R]_b{}^c$}
\end{picture}
\end{center}
\begin{center}
\begin{picture}(200,58)(0,30)
\Photon(60,40)(10,40){3}{5}
\ArrowLine(100,70)(60,40)
\ArrowLine(60,40)(100,10)
\Text(108,70)[]{$c$}
\Text(108,10)[]{$b$}
\Text(30,25)[]{$A^{\mu a}$}
\Text(70,20)[]{$\Psi_{Mi}$}
\Text(70,67)[]{$\Psi_{Mj}$}
\Text(145,40)[l]{$\phantom{\BDpos}
\BDneg i\gamma^\mu[(G^a)_i{}^j P_L-(G^a)_j{}^i P_R]_b{}^c$}
\end{picture}
\end{center}
\caption{Feynman rules for
neutral scalar and gauge boson interactions with a pair of
four-component Majorana fermions (labeled by four-component spinor
indices $b$ and $c$).
The $G^a$ are defined in \eq{gadefMaj}.
The index $a$ runs over the neutral
(mass eigenstate) gauge bosons.
}
\label{intvertices4}
\end{figure}

Using standard four-component spinor methods,
the corresponding four-component spinor Feynman rules
are displayed in \fig{intvertices4}.
A Majorana fermion is neutral under all conserved
charges (and thus equal to its own antiparticle).  Hence, an arrow on a
Majorana fermion line simply reflects the structure of the interaction
Lagrangian; i.e., $\Psibar_M$ [$\Psi_M$] is represented by
an arrow pointing out of [into] the vertex.  These arrows are then used
for determining the placement of the $u$ and $v$ spinors in an
invariant amplitude, according to the rules of \app{G.6}.
In particular, the four-component spinor labels of \fig{intvertices4}
indicate that one should traverse any continuous fermion
line by moving antiparallel to the direction of the fermion arrows.

Next, we consider the interactions of a (possibly complex) scalar $\Phi$
or a gauge boson $A_{\mu}^a$ with a pair of Dirac fermions.
The Dirac fermions are charged with respect to some global or local
U(1) symmetry, which is assumed to be a symmetry of the Lagrangian.
To obtain the interactions of the four-component fermion fields, we first
identify the
mass-degenerate oppositely charged pairs $\chi_j$ and $\eta_j$
(with U(1)-charges $q_j$ and $-q_j$, respectively) that combine to
form the mass eigenstate Dirac fermions.  The scalar field $\Phi$
carries a U(1)-charge $q\ls{\Phi}$.
We also identify the
gauge boson mass eigenstates of definite U(1)-charge by
$A_\mu^a$ as described in \sec{subsec:fermioninteractions}
(cf.~footnote~\ref{vectormass}).
Using \eqs{concrete}{massgaugedirac}, the interaction Lagrangian
in two-component form is given by:
\beq
\label{lintchargedDirac}
\mathscr{L}_{\rm int} =
-\kappa^i{}_j\chi_i\eta^j\Phi
-\kappa_i{}^j\chi^{\dagger i}\eta_j^\dagger\Phi^\dagger
\BDminus\left[(G_L^a)_i{}^j\chi^{\dagger i}\sigmabar^\mu\chi_j
-(G_R^a)_j{}^i\eta^{\dagger}_i\sigmabar^\mu\eta^j\right]
A_\mu^a\,,
\eeq
where $\kappa_i{}^j\equiv (\kappa^i{}_j)^*$ [cf.~\eq{Mdagger}] and
$\kappa$ is an arbitrary complex matrix coupling, subject to the conditions
that $\kappa^i{}_j=0$ unless
$q\ls{\Phi}=q_j-q_i$.  For the gauge boson couplings, we follow
the notation of \eqs{gachidef}{gaetadef}.  In particular,
$A_\mu^a G^a_L$ and $A_\mu^a G^a_R$ are hermitian matrix-valued
gauge fields, which when summed over $a$ can
contain both neutral and charged [with respect to U(1)] mass eigenstate
gauge boson fields.
Converting to four-component notation yields:
\beq \label{lintchargedDirac4}
\mathscr{L}_{\rm int} =
-\kappa^i{}_j\Psibar\llsup{\,j}P_L\Psi_{i}\Phi
-\kappa_i{}^j\Psibar\llsup{i}P_R\Psi_{j}\Phi^\dagger
\BDminus \left[(G_L^a)_i{}^j\Psibar\llsup{\,i}\gamma^\mu P_L\Psi_{j}
+(G_R^a)_i{}^j\Psibar\llsup{\,i}\gamma^\mu P_R\Psi_{j}\right]A^a_\mu\,,
\eeq
where the $\Psi_{j}$ are a set of Dirac four-component fermions.
If $\Phi$ is a real (neutral) scalar field, then
we shall write $\phi\equiv\Phi=\Phi^\dagger$.
The corresponding four-component spinor Feynman rules are exhibited
in \fig{cDDintvertices4}.
The rules involving the charge-conjugated Dirac fields have been 
obtained by using \eq{cinteractions}.
Note that the arrows on the charged scalar and Dirac fermion lines
depict the flow of the conserved U(1)-charge.
\begin{figure}[t!]
\begin{center}
\begin{picture}(200,78)(0,0)
\DashLine(-100,40)(-50,40)5
\ArrowLine(-50,40)(-10,70)
\ArrowLine(-10,10)(-50,40)
\Text(-2,70)[]{$b$}
\Text(-2,10)[]{$c$}
\Text(-80,30)[]{$\phi$}
\Text(-40,20)[]{$\Psi_{i}$}
\Text(-40,67)[]{$\Psi_{j}$}
\Text(20,40)[l]{or}
\DashLine(60,40)(110,40)5
\ArrowLine(150,70)(110,40)
\ArrowLine(110,40)(150,10)
\Text(158,70)[]{$c$}
\Text(158,10)[]{$b$}
\Text(80,30)[]{$\phi$}
\Text(120,16)[]{$\Psi^{C\,i}$}
\Text(120,67)[]{$\Psi^{C\,j}$}
\Text(190,40)[l]{$-i(\kappa^i{}_j P_L+\kappa_j{}^iP_R)_b{}^c$}
\end{picture}
\end{center}
\begin{center}
\begin{picture}(200,78)(0,0)
\DashArrowLine(-100,40)(-50,40)5
\ArrowLine(-50,40)(-10,70)
\ArrowLine(-10,10)(-50,40)
\Text(-2,70)[]{$b$}
\Text(-2,10)[]{$c$}
\Text(-80,30)[]{$\Phi$}
\Text(-40,20)[]{$\Psi_{i}$}
\Text(-40,67)[]{$\Psi_{j}$}
\Text(20,40)[l]{or}
\DashArrowLine(60,40)(110,40)5
\ArrowLine(150,70)(110,40)
\ArrowLine(110,40)(150,10)
\Text(158,70)[]{$c$}
\Text(158,10)[]{$b$}
\Text(80,30)[]{$\Phi$}
\Text(120,16)[]{$\Psi^{C\,i}$}
\Text(120,67)[]{$\Psi^{C\,j}$}
\Text(190,40)[l]{$-i\kappa^i{}_j (P_L)_b{}^c$}
\end{picture}
\end{center}
\begin{center}
\begin{picture}(200,68)(0,0)
\DashArrowLine(-50,40)(-100,40)5
\ArrowLine(-10,70)(-50,40)
\ArrowLine(-50,40)(-10,10)
\Text(-2,70)[]{$c$}
\Text(-2,10)[]{$b$}
\Text(-80,30)[]{$\Phi$}
\Text(-40,20)[]{$\Psi_{i}$}
\Text(-40,67)[]{$\Psi_{j}$}
\Text(20,40)[l]{or}
\DashArrowLine(110,40)(60,40)5
\ArrowLine(110,40)(150,70)
\ArrowLine(150,10)(110,40)
\Text(158,70)[]{$b$}
\Text(158,10)[]{$c$}
\Text(80,30)[]{$\Phi$}
\Text(120,16)[]{$\Psi^{C\,i}$}
\Text(120,67)[]{$\Psi^{C\,j}$}
\Text(190,40)[l]{$-i\kappa_i{}^j (P_R)_b{}^c$}
\end{picture}
\end{center}
\begin{center}
\begin{picture}(200,68)(0,0)
\Photon(30,40)(-20,40){3}{5}
\ArrowLine(70,70)(30,40)
\ArrowLine(30,40)(70,10)
\Text(78,70)[]{$c$}
\Text(78,10)[]{$b$}
\Text(0,25)[]{$A^{\mu a}$}
\Text(40,20)[]{$\Psi_i$}
\Text(40,67)[]{$\Psi_j$}
\Text(145,40)[l]{$\phantom{\BDpos}
\BDneg i\gamma^\mu[(G_L^a)_i{}^j P_L+(G_R^a)_i{}^j P_R]_b{}^c$}
\Text(30,-10)[]{or}
\Text(210,-10)[]{or}
\end{picture}
\end{center}
\begin{center}
\begin{picture}(200,92)(0,0)
\Photon(30,40)(-20,40){3}{5}
\ArrowLine(30,40)(70,70)
\ArrowLine(70,10)(30,40)
\Text(78,70)[]{$b$}
\Text(78,10)[]{$c$}
\Text(0,25)[]{$A^{\mu a}$}
\Text(40,16)[]{$\Psi^{Ci}$}
\Text(40,65)[]{$\Psi^{Cj}$}
\Text(145,40)[l]{$\phm\phantom{\BDneg}\BDpos
i\gamma^\mu[(G_R^a)_i{}^j P_L+ (G_L^a)_i{}^j P_R]_b{}^c$}
\end{picture}
\end{center}
\caption{Feynman rules for neutral scalar ($\phi$),
charged scalar ($\Phi$) and gauge boson ($A^{\mu a}$)
interactions with a pair of four-component Dirac fermions
(labeled by four-component spinor indices $b$ and $c$).
In each case, one has two choices for the corresponding
Feynman rule: one involving $\Psi$ and one involving
the oppositely charged $\Psi^{C}$ (with the arrows of the corresponding
$\Psi$ and $\Psi^C$ lines pointing in opposite directions).
The arrows
indicate the direction of flow of the U(1)-charges of the Dirac fermion
and charged scalar fields.
The index $a$ runs over both neutral
and charged (mass eigenstate) gauge bosons, consistent with
charge conservation at the vertex.}
\label{cDDintvertices4}
\end{figure}
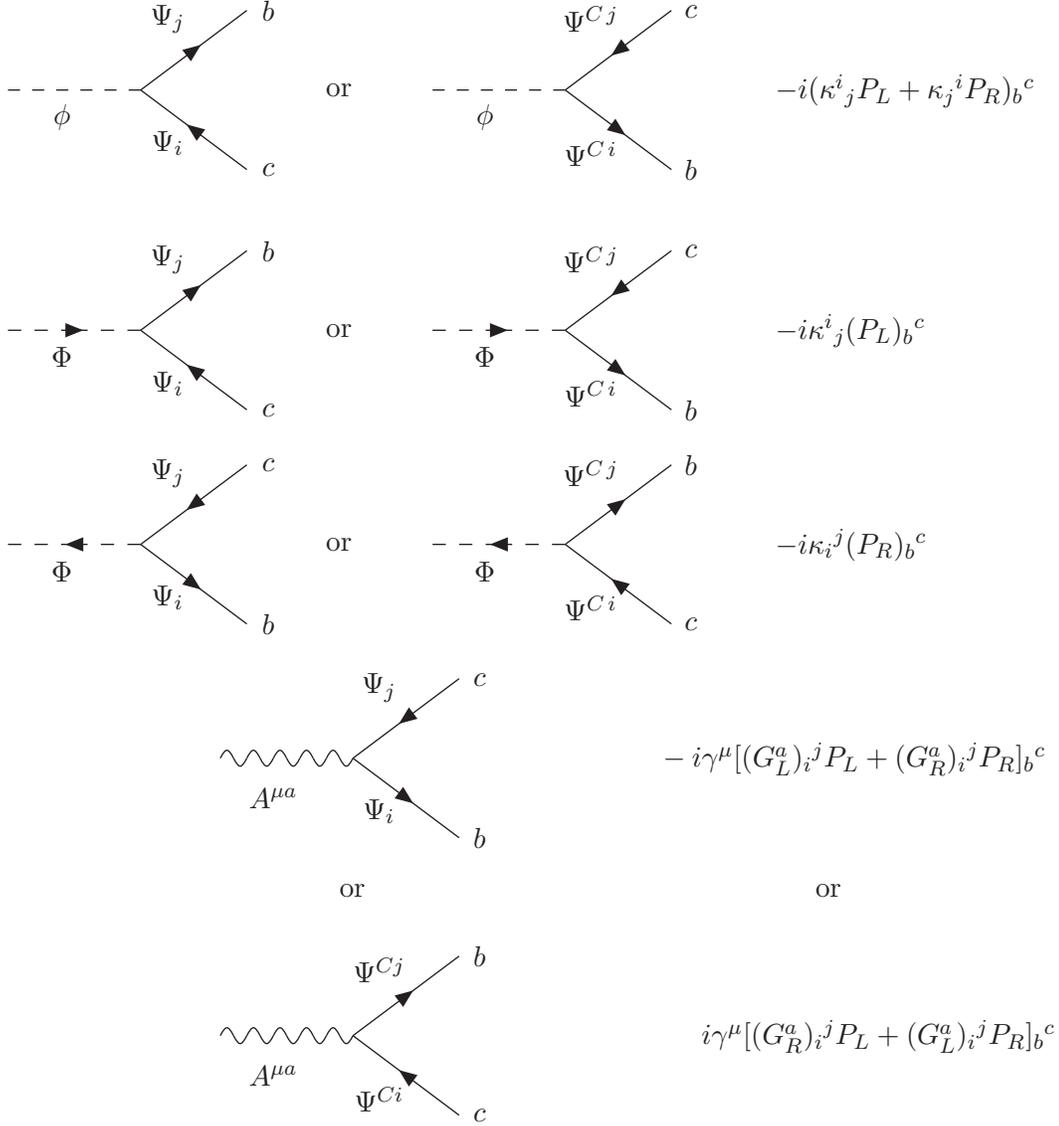

Finally, we treat the interaction of a charged scalar boson
$\Phi$ (with U(1)-charge $q\ls{\Phi}$) or a charged vector
boson $W$ (with U(1)-charge $q\ls{W}$) with a fermion pair
consisting of one Majorana and one Dirac fermion.
We denote the neutral fermion mass eigenstate
fields by $\xi_i$ and pairs of
oppositely charged fermion mass eigenstate
fields by $\chi_j$ and $\eta^j$ (with U(1)-charges $q_j$ and $-q_j$,
respectively).
Using \eqs{concrete}{lintwplus}, the interaction Lagrangian
is given by:
\beqa
\hspace{-0.3in}
\mathscr{L}_{\rm int} &=&
-(\kappa_1)^i{}_j\xi_i\eta^j
+(\kappa_2)_{ij}\xi^{\dagger i} \chi^{\dagger j}]\Phi
-[(\kappa_2)^{ij}\xi_i\chi_j
+(\kappa_1)_i{}^j\xi^{\dagger i}_i \eta^{\dagger}_j]\Phi^\dagger
\nonumber \\
&&
\BDminus [(G_1)_j{}^i\chi^{\dagger j}\sigmabar^\mu\xi_i
-(G_2)_{ij}\xi^{\dagger i}\sigmabar^\mu \eta^j]W_\mu
\BDminus [(G_1)^j{}_i\xi^{\dagger i}\sigmabar^\mu\chi_j
-(G_2)^{ij}\eta^{\dagger}_j\sigmabar^\mu\xi_i]W_\mu^\dagger\,,
\eeqa
where
$\kappa_1$, $\kappa_2$, $G_1$, and $G_2$ are arbitrary
complex coupling matrices, subject to the conditions that
$(\kappa_1)^i{}_j=(\kappa_2)_{ij}=0$ unless $q\ls{\Phi}=q_j$, and
$(G_1)_j{}^i=(G_2)_{ij}=0$ unless $q\ls{W}=q_j$.
Converting to four-component spinor notation yields:
\beqa \label{lintc4}
\mathscr{L}_{\rm int} &=&
-\left[(\kappa_1)^i{}_j\Psibar\llsup{\,j}P_L\Psi_{Mi}
+(\kappa_2)_{ij}\Psibar\llsup{j}P_R\Psi_{M}^i\right]\Phi
\nonumber \\
&&
\BDminus \left[(G_1)_j{}^i\Psibar\llsup{\,j}\gamma^\mu P_L\Psi_{Mi}
+(G_2)_{ij}\Psibar\llsup{\,j}\gamma^\mu P_R\Psi_{M}^i\right]W_\mu
+{\rm h.c.}
\eeqa
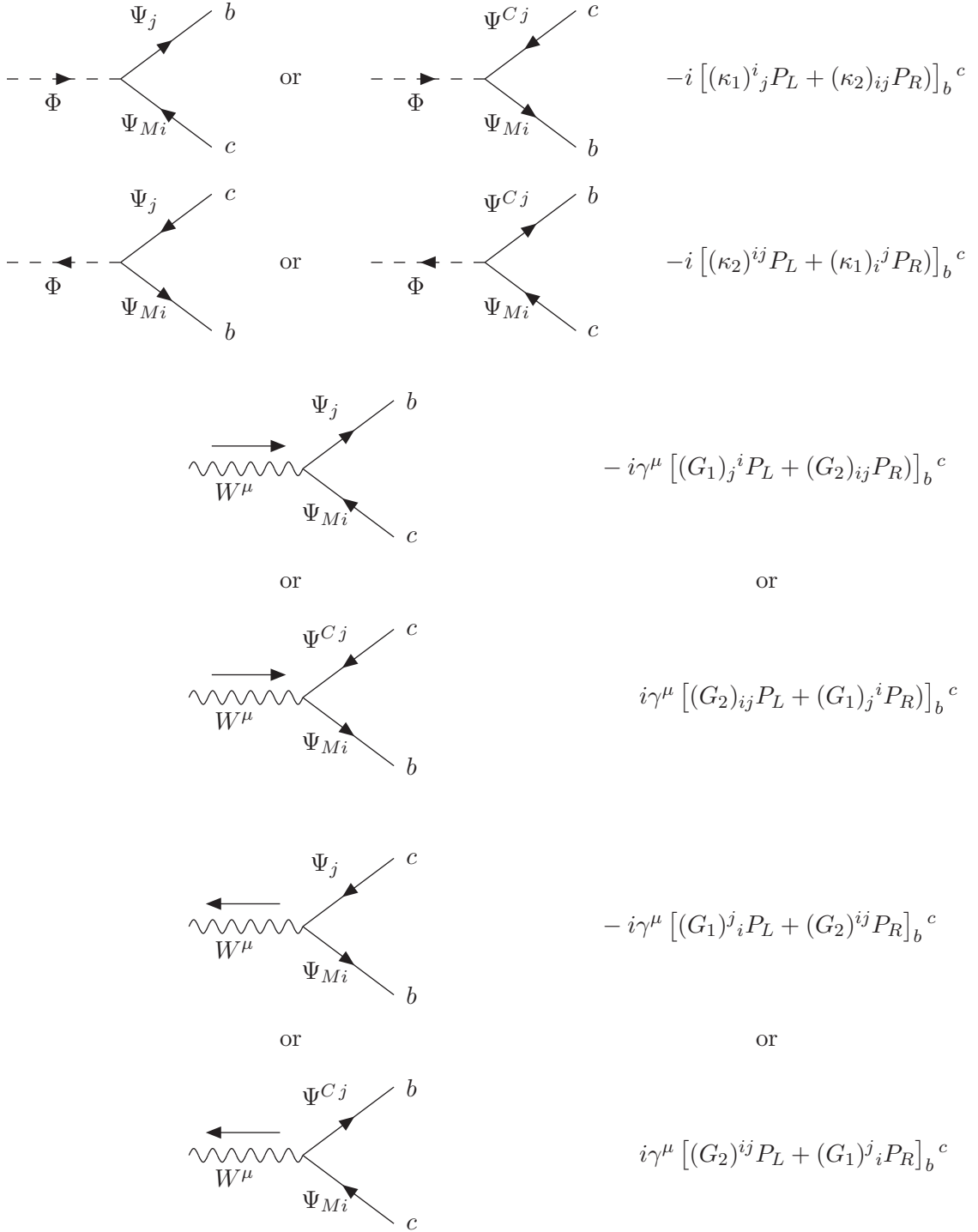
\begin{figure}[pt!]
\begin{center}
\begin{picture}(200,78)(0,0)
\DashArrowLine(-100,40)(-50,40)5
\ArrowLine(-50,40)(-10,70)
\ArrowLine(-10,10)(-50,40)
\Text(-2,70)[]{$b$}
\Text(-2,10)[]{$c$}
\Text(-80,30)[]{$\Phi$}
\Text(-40,20)[]{$\Psi_{Mi}$}
\Text(-40,67)[]{$\Psi_{j}$}
\Text(20,40)[l]{or}
\DashArrowLine(60,40)(110,40)5
\ArrowLine(150,70)(110,40)
\ArrowLine(110,40)(150,10)
\Text(158,70)[]{$c$}
\Text(158,10)[]{$b$}
\Text(80,30)[]{$\Phi$}
\Text(120,20)[]{$\Psi_{Mi}$}
\Text(120,67)[]{$\Psi^{C\,j}$}
\Text(190,40)[l]{$-i\left
[(\kappa_1)^i{}_j P_L+(\kappa_2)_{ij} P_R)\right]_b{}^c$}
\end{picture}
\end{center}
\begin{center}
\begin{picture}(200,68)(0,0)
\DashArrowLine(-50,40)(-100,40)5
\ArrowLine(-10,70)(-50,40)
\ArrowLine(-50,40)(-10,10)
\Text(-2,70)[]{$c$}
\Text(-2,10)[]{$b$}
\Text(-80,30)[]{$\Phi$}
\Text(-40,20)[]{$\Psi_{Mi}$}
\Text(-40,67)[]{$\Psi_{j}$}
\Text(20,40)[l]{or}
\DashArrowLine(110,40)(60,40)5
\ArrowLine(110,40)(150,70)
\ArrowLine(150,10)(110,40)
\Text(158,70)[]{$b$}
\Text(158,10)[]{$c$}
\Text(80,30)[]{$\Phi$}
\Text(120,20)[]{$\Psi_{Mi}$}
\Text(120,67)[]{$\Psi^{C\,j}$}
\Text(190,40)[l]{$-i
\left[(\kappa_2)^{ij}P_L+(\kappa_1)_i{}^j P_R)\right]_b{}^c$}
\end{picture}
\end{center}
\begin{center}
\begin{picture}(200,78)(0,0)
\Photon(-20,40)(30,40){3}{6}
\ArrowLine(30,40)(70,70)
\ArrowLine(70,10)(30,40)
\LongArrow(-10,50)(20,50)
\Text(78,70)[]{$b$}
\Text(78,10)[]{$c$}
\Text(0,30)[]{$W^\mu$}
\Text(40,20)[]{$\Psi_{Mi}$}
\Text(40,67)[]{$\Psi_{j}$}
\Text(160,40)[l]{$\phantom{\BDpos}\BDneg i\gamma^\mu
\left[(G_1)_j{}^iP_L+(G_2)_{ij} P_R)\right]_b{}^c$}
\Text(20,-10)[l]{\rm or}
\Text(230,-10)[l]{\rm or}
\end{picture}
\end{center}
\begin{center}
\begin{picture}(200,68)(0,20)
\Photon(-20,40)(30,40){3}{6}
\ArrowLine(70,70)(30,40)
\ArrowLine(30,40)(70,10)
\LongArrow(-10,50)(20,50)
\Text(78,70)[]{$c$}
\Text(78,10)[]{$b$}
\Text(0,30)[]{$W^\mu$}
\Text(40,20)[]{$\Psi_{Mi}$}
\Text(40,67)[]{$\Psi^{C\,j}$}
\Text(170,40)[l]{$\phantom{\BDneg}\BDpos i\gamma^\mu
\left[(G_2)_{ij} P_L+ (G_1)_j{}^i P_R)\right]_b{}^c$}
\end{picture}
\end{center}
\begin{center}
\begin{picture}(200,88)(0,20)
\Photon(-20,40)(30,40){3}{6}
\ArrowLine(70,70)(30,40)
\ArrowLine(30,40)(70,10)
\LongArrow(20,50)(-10,50)
\Text(78,70)[]{$c$}
\Text(78,10)[]{$b$}
\Text(0,30)[]{$W^\mu$}
\Text(40,20)[]{$\Psi_{Mi}$}
\Text(40,67)[]{$\Psi_{j}$}
\Text(20,-10)[l]{\rm or}
\Text(230,-10)[l]{\rm or}
\Text(160,40)[l]{$\phantom{\BDpos}\BDneg i\gamma^\mu\left[(G_1)^j{}_i P_L+
(G_2)^{ij} P_R\right]_b{}^c$}
\end{picture}
\end{center}
\begin{center}
\begin{picture}(200,88)(0,20)
\Photon(-20,40)(30,40){3}{6}
\ArrowLine(30,40)(70,70)
\ArrowLine(70,10)(30,40)
\LongArrow(20,50)(-10,50)
\Text(78,70)[]{$b$}
\Text(78,10)[]{$c$}
\Text(0,30)[]{$W^\mu$}
\Text(40,20)[]{$\Psi_{Mi}$}
\Text(40,67)[]{$\Psi^{C\,j}$}
\Text(170,40)[l]{$\phantom{\BDneg}\BDpos i\gamma^\mu\left[(G_2)^{ij} P_L+
(G_1)^j{}_i P_R\right]_b{}^c$}
\end{picture}
\end{center}
\caption{Feynman rules for charged scalar and vector boson
interactions with a fermion pair consisting of
one Majorana and one Dirac four-component fermion
(labeled by four-component spinor indices $b$ and $c$).
In each case, one has two choices for the corresponding
Feynman rule: one involving $\Psi$ and one involving
the oppositely charged $\Psi^{C}$ (with the arrows
of the $\Psi$ and $\Psi^C$ lines pointing in opposite directions).
The arrows of the Dirac fermion
and charged bosons indicate the direction of flow of the
corresponding U(1)-charges.
That is, the charge of the boson
(either $\Phi$ or $W$ above) must
coincide with the charge of $\Psi_j$.  The arrows of the Majorana fermions
satisfy the requirement that the fermion line arrow directions flow
continuously through the vertex.
}
\label{cintvertices4}
\end{figure}
The corresponding four-component spinor Feynman rules are exhibited
in \fig{cintvertices4}.

There is an equivalent form for the interactions given
by \eqs{lintchargedDirac}{lintc4} where
$\mathscr{L}_{\rm int}$ is written in terms of charge-conjugated
Dirac fields [after using \eq{cinteractions}].
The Feynman rules involving
Dirac fermions can take two possible forms, as
shown in \figs{cDDintvertices4}{cintvertices4}.
As previously noted, the direction of an
arrow on a Dirac fermion line indicates the
direction of the fermion charge flow (whereas the arrow on
the Majorana fermion line is unconnected to charge flow).
However, we are free to choose either a
$\Psi$ or $\Psi^C$ line to represent a Dirac fermion at any place in a
given Feynman graph.\footnote{Since the charge of $\Psi^C$ is opposite
in sign to the charge
of $\Psi$, the corresponding arrow directions of the $\Psi$ and $\Psi^C$
lines must point in opposite directions.}
For any decay or scattering process,
a suitable choice of either the $\Psi$-rule or the $\Psi^C$-rule
at each vertex (the choice can be different at different vertices),
will guarantee that
the arrow directions on fermion lines flow continuously through
the Feynman diagram.  Then, to evaluate an invariant amplitude,
one should traverse \textit{any} continuous fermion
line (either $\Psi$ or $\Psi^C$)
by moving antiparallel to the direction of the fermion arrows,
as indicated by the order of the four-component spinor labels in
the Feynman rules of \figs{cDDintvertices4}{cintvertices4}.
Examples will be provided in \app{G.6}.

\subsection{Applications of four-component spinor Feynman rules}
\renewcommand{\theequation}{G.6.\arabic{equation}}
\renewcommand{\thefigure}{G.6.\arabic{figure}}
\renewcommand{\thetable}{G.6.\arabic{table}}
\setcounter{equation}{0}
\setcounter{figure}{0}
\setcounter{table}{0}

For a given process, there may be a number of distinct
choices for the arrow directions on the Majorana fermion lines,
which may depend on whether one represents a given Dirac fermion by
$\Psi$ or $\Psi^C$.
However, different choices do {\it not} lead to independent Feynman
diagrams.\footnote{In contrast, the two-component Feynman rules
developed in \sec{sec:externalfermionrules}
require that two vertices differing by the
direction of the arrows on the two-component fermion lines must both be
included in the calculation of the matrix element.}
When computing an invariant amplitude, one
first writes down the relevant
Feynman diagrams with no arrows on any Majorana
fermion line.  The number of distinct graphs contributing to the
process is then determined.  Finally, one makes some choice for
how to distribute the arrows on the Majorana fermion lines
and how to label Dirac fermion lines (either as the field $\Psi$ or its
charge conjugate $\Psi^C$) in a manner consistent
with the rules of \figs{intvertices4}{cintvertices4}.
The end result for the invariant
amplitude (apart from an overall unobservable phase)
does not depend on the choices
made for the direction of the fermion arrows.

Using the above procedure, the Feynman rules for the
external fermion wave functions are the same for Dirac and Majorana fermions:
\begin{itemize}
\item
$u(\boldsymbol{\vec p},s)$: incoming $\Psi$ [or $\Psi^C$]
with momentum $\boldsymbol{\vec p}$ parallel to the arrow direction,
\item
$\ubar(\boldsymbol{\vec p},s)$: outgoing $\Psi$ [or $\Psi^C$] with
momentum $\boldsymbol{\vec p}$ parallel to the arrow direction,
\item
$v(\boldsymbol{\vec p},s)$: outgoing $\Psi$ [or $\Psi^C$] with
momentum $\boldsymbol{\vec p}$ antiparallel to the arrow direction,
\item
$\vbar(\boldsymbol{\vec p},s)$: incoming $\Psi$ [or $\Psi^C$] with
momentum $\boldsymbol{\vec p}$ antiparallel to the arrow direction.
\end{itemize}
The proof that the above rules for external wave functions apply
unambiguously to
Majorana fermions is straightforward.  Simply insert the
plane wave expansion of the Majorana field given by \eq{Majmodes}
into \eq{lint4}, and evaluate matrix elements for, e.g., the decay of
a scalar or vector particle into a pair of Majorana fermions.

We now reconsider the matrix elements for scalar and vector
particle decays into fermion pairs and $2\to 2$ elastic scattering
of a fermion off a scalar and vector boson, respectively.
We shall compute the matrix elements using the Feynman rules of
\fig{intvertices4}, and check that the results agree with the ones
obtained by two-component methods in \sec{subsec:simpleapps}.

Consider first
the decay of a neutral scalar boson $\phi$ into a pair of Majorana
fermions, $\phi\to\Psi_{Mi}(\boldsymbol{\vec
p}_1,s_1)\Psi_{Mj}(\boldsymbol{\vec p}_2,s_2)$, of flavor $i$ and $j$,
respectively.  Here,
$\Psi_{Mi}(\boldsymbol{\vec p},s)$
denotes the one-particle state given by \eq{xistate}.
The matrix element for the decay is given by
\beq \label{phitopsipsi}
i\mathcal{M}=-i\ubar(\boldsymbol{\vec p}_1,s_1)(\lambda^{ij} P_L+\lambda_{ij}
P_R) v(\boldsymbol{\vec p}_2,s_2)\,.
\eeq
One can easily check that this result matches with \eq{scalardecay},
which was derived using two-component spinor techniques.  Note that if one
had chosen to switch the two final states (equivalent to switching the
directions of the Majorana fermion arrows), then the resulting matrix
element would simply exhibit an overall sign change [due to the
results of \eqs{uva}{uvb}]. This overall sign change is a
consequence of the Fermi-Dirac statistics, and corresponds to changing
which order one uses to construct the two-particle final state.

Consider next the decay of a (neutral or charged) scalar boson
$\Phi$ into a pair of
Dirac fermions,
$\Phi\to F_i(\boldsymbol{\vec p}_1,s_1)
\overline{F}\llsup{\,j}(\boldsymbol{\vec p}_2,s_2)$,
where by $F(\boldsymbol{\vec p},s)$ and $\overline{F}(\boldsymbol{\vec p},s)$
we mean the one-particle states given by \eq{chietastate}.
The matrix element for the decay is given by
\beq \label{phitopsipsibar}
i\mathcal{M}=-i\ubar(\boldsymbol{\vec p}_1,s_1)(\kappa^j{}_i P_L+\kappa_i{}^j
P_R) v(\boldsymbol{\vec p}_2,s_2)\,,
\eeq
which is equivalent to \eq{scalardiracdecay}, which was
derived using two-component spinor techniques.

For the decay of a neutral vector boson (denoted by $A_\mu$)
into a pair of Majorana fermions,
$A_\mu\to\Psi_{Mi}(\boldsymbol{\vec p}_1,s_1)
\Psi_{Mj}(\boldsymbol{\vec p}_2,s_2)$,
we use the Feynman rules of \fig{intvertices4} to obtain:
\beq \label{Amutopsi1psi2}
i\mathcal{M}=
\BDneg i\ubar(\boldsymbol{\vec p}_1,s_1)\gamma^\mu
\left[G_i{}^j P_L-G_j{}^i P_R\right]v(\boldsymbol{\vec
p}_2,s_2)\varepsilon_\mu\,,
\eeq
The above result is equivalent to \eq{vectordecay}, which was
derived using two-component spinor techniques.
Again, we note that if
one had chosen to switch the two final states (equivalent to switching
the directions of the Majorana fermion arrows), then the resulting
matrix element would simply exhibit an overall sign change [due to the
results of \eqs{uvc}{uvd}].

For $i=j$, \eq{Amutopsi1psi2} simplifies to
\beq \label{Amutopsipsi}
i\mathcal{M}=
\BDpos i{G}\ubar(\boldsymbol{\vec p}_1,s_1)\gamma^\mu\gamma_5
v(\boldsymbol{\vec p}_2,s_2)\varepsilon_\mu\,,
\eeq
where $G\equiv G_i{}^i$.  The absence of a vector coupling of
the vector boson to a pair of identical Majorana fermions is a
consequence of the identity $\Psibar_M\gamma^\mu\Psi_M=0$ noted
below \eq{majidentity6}.

For the decay of a (neutral or charged) vector
particle $A_\mu$
into a fermion pair consisting of a Dirac fermion and antifermion,
$A_\mu\to F_i(\boldsymbol{\vec p}_1,s_1)
\overline{F}\llsup{\,j}(\boldsymbol{\vec p}_2,s_2)$,
the matrix element is given by:
\beq \label{Amutopsibarpsi}
i\mathcal{M}=
\BDneg i\ubar(\boldsymbol{\vec p}_1,s_1)\gamma^\mu
\left[(G_L)_i{}^j P_L+(G_R)_i{}^j P_R\right]
v(\boldsymbol{\vec p}_2,s_2)\varepsilon_\mu\,,
\eeq
which matches the result of \eq{vectordecay2}.

Finally, we consider the decay of a charged
boson to a fermion pair consisting of one Dirac fermion and
one Majorana fermion.  Using the Feynman rules of
\fig{cintvertices4}, we see that we have a choice of two rules for
each decay process.  As an example, consider the decay
$W\to \Psi_{Mi}(\boldsymbol{\vec p}_1,s_1) F_j(\boldsymbol{\vec p}_2,s_2)$.
If we apply the $W\Psi_M\Psi$ Feynman rule of \fig{cintvertices4},
we obtain:
\beq \label{WFM}
i\mathcal{M}=-i\ubar(\boldsymbol{\vec p}_2,s_2)\left[(G_1)_j{}^i P_L
+(G_2)_{ij} P_R\right]v(\boldsymbol{\vec p}_1,s_1)\,.
\eeq
If we apply the corresponding  $W\Psi_M\Psi^C$ Feynman rule, we obtain the
negative of
\eq{WFM} with  $P_L\leftrightarrow P_R$ and
$(\boldsymbol{\vec p}_1,s_1)\leftrightarrow
(\boldsymbol{\vec p}_2,s_2)$.  Using
\eqs{uvc}{uvd}, the resulting amplitude is the negative of \eq{WFM},
as expected since the order of the spinor wave functions in the two
computations is reversed.  A similar conclusion is obtained for
the decay $\Phi\to \Psi_{Mi} F_j$.

Turning to the elastic scattering of a Majorana fermion and
a neutral
scalar, we shall examine two equivalent ways for computing the
amplitude.  Following the rules previously stated, there are two
possible choices for the direction of arrows on the Majorana fermion
lines.
Thus, we may evaluate either one of the following two diagrams:

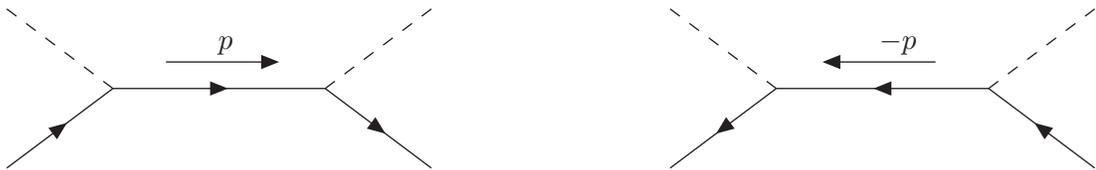
\begin{picture}(400,92)(0,-28)
\thicklines
\ArrowLine(40,15)(120,15)
\DashLine(40,15)(0,45)5
\ArrowLine(0,-15)(40,15)
\DashLine(120,15)(160,45)5
\ArrowLine(120,15)(160,-15)
\LongArrow(60,25)(100,25)
\put(80,30){$p$}
\ArrowLine(370,15)(290,15)
\DashLine(290,15)(250,45)5
\ArrowLine(290,15)(250,-15)
\DashLine(370,15)(410,45)5
\ArrowLine(410,-15)(370,15)
\LongArrow(350,25)(310,25)
\put(330,30){$-p$}
\end{picture}

\noindent
plus a second set of diagrams (not shown)
where the initial and final state scalars are crossed.

Evaluating the first diagram above, the matrix element for
$\phi\Psi_M\to\phi\Psi_M$ is given by:
\beqa \label{phiPsiscatter}
i\mathcal{M}&=&\frac{-i}{s-m^2}\,
\ubar(\boldsymbol{\vec p}_2,s_2)(\lambda P_L+\lambda^* P_R)
(\BDpos \slashchar{p}+m)
(\lambda P_L+\lambda^* P_R) u(\boldsymbol{\vec p}_1,s_1)+{(\rm crossed)}
\nonumber \\[5pt]
&=& \frac{-i}{s-m^2}\,
\ubar(\boldsymbol{\vec p}_2,s_2)\left[
\BDpos |\lambda|^2 \slashchar{p}+
\left(\lambda^2 P_L+(\lambda^*)^2 P_R\right)m\right]
u(\boldsymbol{\vec p}_1,s_1)+{(\rm crossed)}\,,
\eeqa
where $m$ is the Majorana fermion mass and $\sqrt{s}$ is the
center-of-mass energy.  Using \eqs{gamma4}{uspin4}, one
recovers the results of \eq{pxscatter}.  Had we chosen to evaluate the
second diagram instead, the resulting amplitude would have been given
by:
\beq \label{phiPsiscatter2}
i\mathcal{M}= \frac{-i}{s-m^2}\,
\vbar(\boldsymbol{\vec p}_1,s_1)\left[
\BDneg |\lambda|^2 \slashchar{p}+
\left(\lambda^2 P_L+(\lambda^*)^2 P_R\right)m\right]
v(\boldsymbol{\vec p}_2,s_2)+{(\rm crossed)}\,.
\eeq
Using \eq{uGammau}, it follows that:
\beqa
\vbar(\boldsymbol{\vec p}_1,s_1) v(\boldsymbol{\vec p}_2,s_2)&=&
-\ubar(\boldsymbol{\vec p}_2,s_2) u(\boldsymbol{\vec p}_1,s_1)\,,
 \label{uvrelations1} \\
\vbar(\boldsymbol{\vec p}_1,s_1)\gamma^\mu v(\boldsymbol{\vec p}_2,s_2)&=&
\ubar(\boldsymbol{\vec p}_2,s_2)\gamma^\mu u(\boldsymbol{\vec p}_1,s_1)\,.
 \label{uvrelations2}
\eeqa
Consequently, the amplitude computed in \eq{phiPsiscatter2} is just
the negative of \eq{phiPsiscatter}.  This is expected, since the order
of spinor wave functions in  \eq{phiPsiscatter2} is an odd
permutation of the order of spinor wave functions
in \eq{phiPsiscatter} [$(12)$ and $(21)$, respectively].
As in the two-component Feynman rules, the
overall sign of the amplitude is arbitrary, but the relative signs of
any pair of diagrams is unambiguous.
This relative sign is positive [negative] if the permutation of
the order of spinor wave functions of
one diagram relative to the other diagram is even [odd].

Next, we consider the elastic scattering of a charged fermion and a
neutral scalar.  Again, we examine two equivalent ways for computing
the amplitude.  Following our rules, there are two
possible choices for the directions of the fermion line arrows,
depending on whether we represent the fermion by $\Psi$ or $\Psi^C$.
Thus, we may evaluate either one of the following two
diagrams:

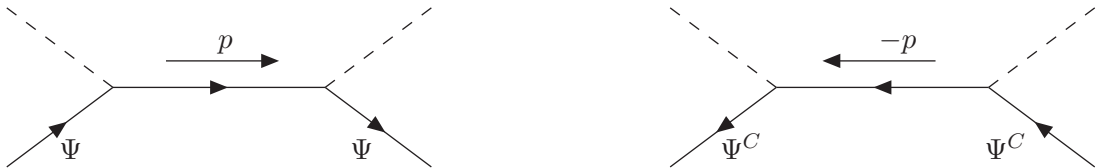
\begin{picture}(400,92)(0,-28)
\thicklines
\ArrowLine(40,15)(120,15)
\DashLine(40,15)(0,45)5
\ArrowLine(0,-15)(40,15)
\DashLine(120,15)(160,45)5
\ArrowLine(120,15)(160,-15)
\LongArrow(60,25)(100,25)
\put(80,30){$p$}
\put(20,-12){$\Psi$}
\put(130,-12){$\Psi$}
\ArrowLine(370,15)(290,15)
\DashLine(290,15)(250,45)5
\ArrowLine(290,15)(250,-15)
\DashLine(370,15)(410,45)5
\ArrowLine(410,-15)(370,15)
\LongArrow(350,25)(310,25)
\put(330,30){$-p$}
\put(270,-12){$\Psi^C$}
\put(370,-12){$\Psi^C$}
\end{picture}

\noindent
plus a second set of diagrams (not shown)
where the initial and final state scalars are crossed.
Evaluating the first diagram above, the matrix element for
$\phi F\to\phi F$ is given by
\eq{phiPsiscatter}, with $\lambda$ replaced by $\kappa$.
Had we chosen to evaluate the second diagram instead,
the resulting amplitude would have been
given by \eq{phiPsiscatter2}, with $\lambda$ replaced by $\kappa$.
Thus, the discussion above in the case of neutral fermion scattering
processes also applies to charged fermion scattering processes.

In processes that only involve vertices with two Dirac fields, it is
never necessary to use charge-conjugated Dirac fermion lines.  In
contrast, consider the following process that involves a vertex with
one Dirac and one Majorana fermion.  Specifically, we examine the
scattering of a Dirac fermion and a charged scalar
into its charge-conjugated final state, via the
exchange of a Majorana fermion: $\Phi^\dagger F\to \Phi
\overline{F}$.  If one
attempts to draw the relevant Feynman diagram employing Dirac fermion
lines but with no charge-conjugated Dirac fermion lines, one finds
that there is no possible choice of arrow direction for the Majorana
fermion that is consistent with the vertex rules of
\fig{cintvertices4}.
The resolution is simple: one can
choose the incoming line to be $\Psi$ and the outgoing line to be
$\Psi^C$ or vice versa.  Thus, the two possible choices are given by:

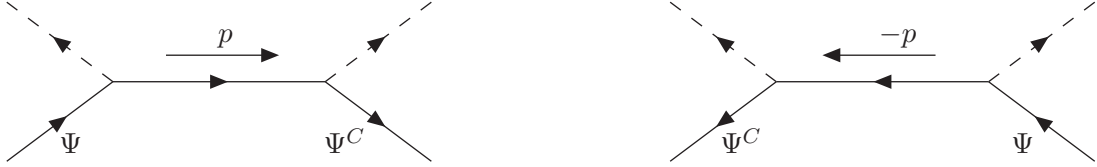
\begin{picture}(400,92)(0,-28)
\thicklines
\ArrowLine(40,15)(120,15)
\DashArrowLine(40,15)(0,45)5
\ArrowLine(0,-15)(40,15)
\DashArrowLine(120,15)(160,45)5
\ArrowLine(120,15)(160,-15)
\LongArrow(60,25)(100,25)
\put(80,30){$p$}
\put(20,-12){$\Psi$}
\put(120,-12){$\Psi^C$}
\ArrowLine(370,15)(290,15)
\DashArrowLine(290,15)(250,45)5
\ArrowLine(290,15)(250,-15)
\DashArrowLine(370,15)(410,45)5
\ArrowLine(410,-15)(370,15)
\LongArrow(350,25)(310,25)
\put(330,30){$-p$}
\put(270,-12){$\Psi^C$}
\put(380,-12){$\Psi$}
\end{picture}

\noindent
plus a second diagram in each case (not shown)
in which the initial and final scalars
are crossed.  If we evaluate the first diagram,
the resulting amplitude is given by:
\beqa \label{phicPsiscatter}
i\mathcal{M}&=&\frac{-i}{s-m^2}\,
\ubar(\boldsymbol{\vec p}_2,s_2)(\kappa_2 P_L+\kappa_1^* P_R)
(\BDpos \slashchar{p}+m)
(\kappa_2 P_L+\kappa_1^* P_R) u(\boldsymbol{\vec p}_1,s_1)+{(\rm crossed)}
\nonumber \\[5pt]
&=& \frac{-i}{s-m^2}\,
\ubar(\boldsymbol{\vec p}_2,s_2)\left[
\BDpos \kappa_1^*\kappa_2 \slashchar{p}+
\left(\kappa_2^2 P_L+(\kappa_1^*)^2 P_R\right)m\right]
u(\boldsymbol{\vec p}_1,s_1)+{(\rm crossed)}\,,
\eeqa
where $m$ is the Majorana fermion mass.  This result is
equivalent to \eq{chsfscatter} obtained via the two-component
spinor methods.  Had we chosen to evaluate the second diagram instead, 
the resulting amplitude [after 
using \eqs{uvrelations1}{uvrelations2}] would have been found to be
the negative of \eq{phicPsiscatter}, as
expected.  As before, the relative sign between diagrams for the same
process is unambiguous.

In the literature, there are a number of alternative methods for
dealing with scattering processes involving Majorana particles.  For
example, one can define a fermion number violating propagator for
four-component fermions (see, \textit{e.g.}, \Ref{HaberKane}).
Using the methods of \Ref{HaberKane}, factors of the charge
conjugation matrix $C$
appear both in fermion-number-violating propagators and vertices.
However, all such factors of $C$ eventually cancel out by the end of the
computation of any $S$-matrix amplitude.  Moreover, such
methods often involve subtle choices of signs that require
first-principles computations to verify.  As previously noted,
our four-component fermion diagrammatic techniques do not suffer
from either of these complications.

In the case of elastic scattering of a Majorana fermion and
a neutral vector boson,
the two contributing diagrams include
the following diagram:
\vskip -0.2in
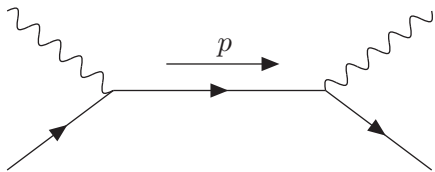
\begin{picture}(200,92)(0,-28)
\thicklines
\ArrowLine(140,15)(220,15)
\Photon(140,15)(100,45){3}{5}
\ArrowLine(100,-15)(140,15)
\Photon(220,15)(260,45){3}{5}
\ArrowLine(220,15)(260,-15)
\LongArrow(160,25)(200,25)
\put(180,30){$p$}
\end{picture}
\vskip -0.25in

\noindent
plus a second diagram (not shown) where the initial and final state
vector bosons are crossed.  
\clearpage

\noindent
Consider first the scattering of a neutral
Majorana fermion of mass $m$.  Using the Feynman rules of
\fig{intvertices4}, the Feynman rule for the
$A_\mu \Psibar_M\Psi_M$
vertex is given by $\BDpos iG\gamma^\mu\gamma\ls{5}$.
Hence, the corresponding matrix element is given by
\beq
i\mathcal{M}=\frac{\BDneg iG^2}{s-m^2}
\ubar(\boldsymbol{\vec p}_2,s_2)\,\gamma\newcdot\varepsilon\ls{2}^*
\,(\slashchar{p}\BDminus m)\,\gamma\newcdot\varepsilon\ls{1}
u(\boldsymbol{\vec p}_1,s_1)+{(\rm crossed)}\,,
\eeq
where we have used $\gamma^\nu\gamma\ls{5}
(\slashchar{p}\BDplus m)\gamma^\mu\gamma\ls{5}=\gamma^\nu
(\slashchar{p}\BDminus m)\gamma^\mu$.  Using
\eqs{gamma4}{uspin4}, one easily recovers the results of
\eq{gxscatter}.

The scattering of a Dirac fermion of mass $m$
and a neutral vector boson can be similarly treated.  The relevant
Feynman graphs are the same as in the previous case, and the matrix
element is given by
\beqa
i\mathcal{M}&=&\frac{\BDneg i}{s-m^2}
\ubar(\boldsymbol{\vec p}_2,s_2)\,\gamma\newcdot\varepsilon\ls{2}^*
\,(G_L P_L+G_R
P_R)(\slashchar{p}\BDplus m)\,\gamma\newcdot\varepsilon\ls{1}
\,(G_L P_L+G_R
P_R)u(\boldsymbol{\vec p}_1,s_1)+{(\rm crossed)}\nonumber \\[5pt]
&=&
\frac{\BDneg i}{s-m^2}
\ubar(\boldsymbol{\vec p}_2,s_2)\,\gamma\newcdot\varepsilon\ls{2}^*
\left[(G_L^2 P_L+G_R^2 P_R)
\slashchar{p} \BDplus G_L G_R m\right]
\,\gamma\newcdot\varepsilon\ls{1}u(\boldsymbol{\vec p}_1,s_1)
+{(\rm crossed)}\,.
\eeqa
One can easily check that this result coincides with that of
\eq{gfscatter}.

Finally, we examine the elastic scattering of two identical Majorana
fermions via scalar exchange.  The three contributing diagrams are:

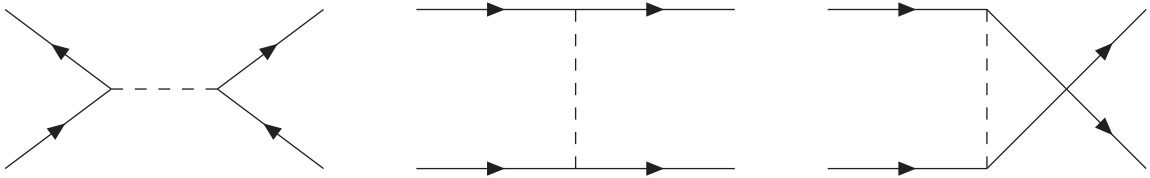
\centerline{
\begin{picture}(450,85)(15,-25)
\thicklines
\ArrowLine(60,15)(20,45)
\ArrowLine(20,-15)(60,15)
\DashLine(60,15)(100,15){5}
\ArrowLine(100,15)(140,45)
\ArrowLine(140,-15)(100,15)
\ArrowLine(175,-15)(235,-15)
\ArrowLine(235,-15)(295,-15)
\ArrowLine(175,45)(235,45)
\ArrowLine(235,45)(295,45)
\DashLine(235,45)(235,-15){5}
\ArrowLine(330,-15)(390,-15)
\Line(390,-15)(420,15)
\ArrowLine(420,15)(450,45)
\ArrowLine(330,45)(390,45)
\Line(390,45)(420,15)
\ArrowLine(420,15)(450,-15)
\DashLine(390,-15)(390,45){5}
\end{picture}
}

\noindent
and the corresponding matrix element is given by
\beqa
i\mathcal{M}&=&
\frac{-i}{s-m_\phi^2}\left[\vbar_1(\lambda P_L+
\lambda^\ast P_R)u_2\,\ubar_3(\lambda P_L+\lambda^\ast P_R)v_4\right]
\nonumber\\[5pt]
&+&(-1)\frac{-i}{t-m_\phi^2}\left[\ubar_3(\lambda P_L+
\lambda^\ast P_R)u_1\,\ubar_4(\lambda P_L+\lambda^\ast P_R)u_2\right]
\nonumber \\[5pt]
&+&\frac{-i}{u-m_\phi^2}\left[\ubar_4(\lambda P_L+
\lambda^\ast P_R)u_1\,\ubar_3(\lambda P_L+\lambda^\ast P_R)u_2\right]\,,
\label{eq:ffscattfour}
\eeqa
where $u_i\equiv u(\boldsymbol{\vec p}_i,s_i)$,
$v_j\equiv v(\boldsymbol{\vec p}_j,s_j)$ and
$m_\phi$ is the exchanged scalar mass.  The relative minus sign
of the $t$-channel graph relative to the
$s$ and $u$-channel graphs is obtained by
noting that $3142$ [$4132$] is an odd [even] permutation of $1234$.
Using \eqss{PLPR}{uspin4}{vspin4}, one easily recovers the results of
\eq{eq:ffscatttwo}.

\subsection{Self-energy functions and pole masses for
four-component fermions}
\renewcommand{\theequation}{G.7.\arabic{equation}}
\renewcommand{\thefigure}{G.7.\arabic{figure}}
\renewcommand{\thetable}{G.7.\arabic{table}}
\setcounter{equation}{0}
\setcounter{figure}{0}
\setcounter{table}{0}

In this section, we examine the self-energy functions
and the pole masses for a set of four-component fermions.  We first consider
four-component Dirac fermion fields $\Psi_{a i}$, where
$a$ is the four-component spinor index and $i$ is the flavor index.
The full, loop-corrected Feynman propagators with four-momentum $p^\mu$
are defined by the Fourier transforms [cf.~footnote \ref{footnotefnft}]
of vacuum expectation values of
time-ordered products of bilinears of the fully interacting four-component
fermion fields:
\beq
\label{ftfullfour}
\bra{0}T\Psi_{ai}(x)\Psibar\llsup{\,bj}(y)\ket{0}_{\rm FT} =
i(\boldsymbol{S}_a{}^b)_i{}^j(p)\,,
\eeq
with~\cite{Donoghue:1979jq,Talon:1982bp,Denner:1990yz,Kniehl:1996bd,
Pierce:1997wu,Kiyoura:1998yt,Gambino:1999ai,mixingrenormalization}
\beq \label{essex}
\boldsymbol{S}(p)\equiv
\BDpos \slashchar{p}\left[P_L{\boldsymbol{S_L^{\T}}}(p^2)
+P_R{\boldsymbol{S_R}}(p^2)\right]
+P_L{\boldsymbol{\Sbar_D^{\T}}}(p^2)
+P_R{\boldsymbol{S_D}}(p^2)\,,
\eeq
where the
four-component spinor indices $a$ and $b$ and the
flavor indices $i$ and $j$ have been
suppressed.  As in
\sec{subsec:selfenergies}, we shall organize the computation of the
full propagator in terms of the 1PI self-energy
function\cite{Kniehl:1996bd}:\footnote{Our notation in \eq{sigmaex}
differs from that of \Ref{Kniehl:1996bd}, as we employ
$\boldsymbol{\Sigma^{\T}_R}$ instead of $\boldsymbol{\Sigma_R}$.  Our
motivation for this choice is that in the case of Majorana fermions
[cf.~\eq{majselfenergy}],
we simply have $\boldsymbol{\Sigma_L}=\boldsymbol{\Sigma_R}$,
without an extra transpose (or conjugation).  We have also
chosen to employ $\boldsymbol{S_L^{\T}}$ in \eq{essex} for similar reasons.}
\beq \label{sigmaex}
\boldsymbol{\Sigma}(p)\equiv
\BDpos \slashchar{p}\left[P_L{\boldsymbol{\Sigma_L}}(p^2)
+P_R{\boldsymbol{\Sigma_R^{\T}}}(p^2)\right]
+P_L{\boldsymbol{\Sigma_D}}(p^2)
+P_R{\boldsymbol{\Sigmabar_D^{\T}}}(p^2)\,.
\eeq
Diagrammatically, $i\boldsymbol{S}$ and $-i\boldsymbol{\Sigma}$
are shown in \fig{fig:fullprops4}.
\begin{figure}[ht!]
\begin{center}
\begin{picture}(80,70)(0,8)
\ArrowLine(28,40)(0,40)
\ArrowLine(80,40)(52,40)
\GBox(28,28)(52,52){0.85}
\Text(76,49)[]{$b$}
\Text(6,48)[]{$a$}
\Text(74,32)[]{$j$}
\Text(6,32)[]{$i$}
\Text(40,72)[c]{$p$}
\LongArrow(56,64)(24,64)
\Text(40,8)[c]{$i({\boldsymbol S}_a{}^b)_i{}^j(p)$}
\end{picture}
\hspace{3.1cm}
\begin{picture}(80,70)(0,8)
\Text(40,72)[c]{$p$}
\LongArrow(56,64)(24,64)
\ArrowLine(28,40)(0,40)
\ArrowLine(80,40)(52,40)
\GCirc(40,40){12}{0.85}
\Text(74,48)[]{$b$}
\Text(6,48)[]{$a$}
\Text(6,32)[]{$i$}
\Text(74,32)[]{$j$}
\Text(40,8)[c]{$-i({\boldsymbol \Sigma}_a{}^b)_i{}^j(p)$}
\end{picture}
\end{center}
\caption{The full, loop-corrected propagator for four-component
Dirac fermions, $i({\boldsymbol S}_a{}^b)_i{}^j(p)$,
 is denoted by the
shaded box, which represents the sum of all connected
Feynman diagrams, with external legs included.
The self-energy function for four-component Dirac fermions,
$-i({\boldsymbol \Sigma}_a{}^b)_i{}^j(p)$, is denoted
by the shaded circle, which represents
the sum of all one-particle irreducible, connected Feynman diagrams
with the external legs amputated.  In both cases,
The four-momentum $p$ flows from right to left.
}
\label{fig:fullprops4}
\end{figure}
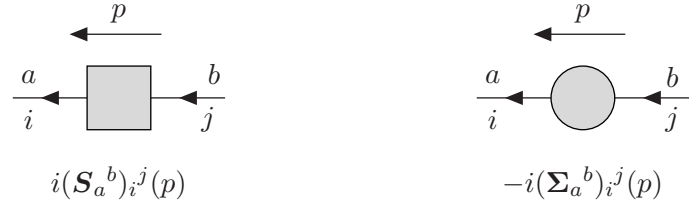

The hermiticity of the effective action implies that $\boldsymbol{S}$
and $\boldsymbol{\Sigma}$ satisfy
hermiticity conditions~\cite{Aoki:1982ed,bdj}
\beq \label{hcond}
[{\boldsymbol S}^{\T}]^\star= A{\boldsymbol S}A^{-1}\,,\qquad\qquad
[{\boldsymbol \Sigma}^{\T}]^\star= A{\boldsymbol \Sigma}A^{-1}\,,
\eeq
where $A$ is the Dirac conjugation matrix [$A=\gamma^0$ in the
standard representations; see \eq{acmatrix}
and the text that follows]
and the star symbol was defined
in the paragraph below \eq{ctid}.
Applying \eq{hcond} to \eqs{essex}{sigmaex} then yields the following
conditions for the complex matrix functions:
\beq
[\boldsymbol{S_L^{\T}}]^\star=\boldsymbol{S_L}\,,\qquad
[\boldsymbol{S_R^{\T}}]^\star=\boldsymbol{S_R}\,, \qquad
\boldsymbol{\Sbar_D}=\boldsymbol{S_D^{\,\star}}\,,
\eeq
\beq
[\boldsymbol{\Sigma_L^{\T}}]^\star=\boldsymbol{\Sigma_L}\,,\qquad
[\boldsymbol{\Sigma_R^{\T}}]^\star=\boldsymbol{\Sigma_R}\,, \qquad
\!\!{\boldsymbol{\Sigmabar_D}}=\boldsymbol{\Sigma_D^{\,\star}}\,.
\eeq

Starting at tree level and comparing with
\fig{fig:diracprop}, the full propagator function is given by:
\beq \label{treeprop4}
{\boldsymbol S}_i{}^j(p)=
(\slashchar{p} \BDplus m)\delta_i^j
/(p^2 \BDminus m_i^2)+ \ldots\,,
\eeq
with no sum over $i$ implied.  The full loop-corrected propagator can
be expressed diagrammatically in terms of the 1PI self-energy
function:
\beqa \label{diagsum}
&&
\begin{picture}(215,120)(135,-60)
\ArrowLine(48,40)(20,40)
\ArrowLine(100,40)(72,40)
\GBox(48,28)(72,52){0.85}
\Text(94,49)[]{$b$}
\Text(26,48)[]{$a$}
\Text(94,32)[]{$j$}
\Text(26,32)[]{$i$}
\Text(120,40)[]{$=$}
\Text(426,49)[]{$b$}
\Text(265,48)[]{$a$}
\Text(426,32)[]{$j$}
\Text(150,32)[]{$i$}
\Text(300,49)[]{$c$}
\Text(150,48)[]{$a$}
\Text(300,32)[]{$k$}
\Text(265,32)[]{$i$}
\Text(350,49)[]{$d$}
\Text(350,31)[]{$\ell$}
\Text(210,49)[]{$b$}
\Text(210,32)[]{$j$}
\ArrowLine(218,40)(140,40)
\Text(238,40)[]{$+$}
\ArrowLine(306,40)(258,40)
\GCirc(318,40){12}{0.85}
\ArrowLine(366,40)(330,40)
\GBox(366,28)(390,52){0.85}
\ArrowLine(430,40)(390,40)
\end{picture}
\nonumber \\[-100pt]
&&\phantom{line}
\eeqa
As in \sec{subsec:selfenergies}, the
algebraic representation of \eq{diagsum} can be written as
[cf.~footnote~\ref{fniter}]:
\beq \label{tsigs}
{\boldsymbol{S}}={\boldsymbol{T}}+{\boldsymbol{T\Sigma S}}
=(\boldsymbol{T}^{-1}-\boldsymbol{\Sigma})^{-1}\,,
\eeq
where
$\boldsymbol{T}_i{}^j\equiv
(\slashchar{p} \BDplus m)\delta_i^j/(p^2 \BDminus m_i^2)$
is the tree-level contribution to
$\boldsymbol{S}$ given in \eq{treeprop4}.  By writing the
expressions for $\boldsymbol{S}$ and $\boldsymbol{\Sigma}$
given in \eqs{essex}{sigmaex} and $\boldsymbol{T}$
in block matrix form using \eq{gamma4}, one can verify that
\eq{tsigs} is equivalent to \eq{matfullDiracprop}.
Consequently, the complex pole masses of the corresponding Dirac
fermions are again determined from \eq{eq:diracpolemass}.

In the special case of a parity-conserving vectorlike
theory of Dirac fermions (such as QED or QCD), the pseudoscalar
and pseudovector parts of $\boldsymbol{S}(p)$ and
$\boldsymbol{\Sigma}(p)$ must be absent.  Thus, the
following relations must hold among the loop-corrected propagator functions
and self-energy functions, respectively:
\beqa
&& \boldsymbol{S_R}
=\boldsymbol{S_L^{\T}}\,,
\qquad\qquad \boldsymbol{S_D}=
[\boldsymbol{S\llsup{\,\T}_D}]^\star\,,\label{app:vectorlike1}\\
&& \boldsymbol{\Sigma_L}=\boldsymbol{\Sigma_R^{\T}}\,,
\qquad\qquad \boldsymbol{\Sigma_D}
=[\boldsymbol{\Sigma_D\llsup{\,\T}}]^\star
\,,\label{app:vectorlike2}
\eeqa
in agreement with \eqs{vectorlike1}{vectorlike2}.

In the case of a set of four-component Majorana fermion fields, we can
still use the results of \eqst{essex}{tsigs}.  However, one obtains additional
constraints on the full propagator and self-energy matrix functions
due to the Majorana condition $\Psi_{Mi}=C\Psibar_{Mi}^{\T}$.
Inserting this result into \eq{ftfullfour}, and making use of the
anticommutativity of the fermion fields, one easily derives:
\beq
\label{majftfullfour}
\bra{0}T\Psi_{Mai}(x)\Psibar_{Mj}\llsup{\,b}(y)\ket{0}_{\rm FT} =
C_{ae}\bra{0}T\Psi_{Mdi}(x)\Psibar_{Mj}\llsup{\,e}(y)\ket{0}_{\rm FT}
(C^{-1})^{db}\,.
\eeq
Consequently,
\beq
C\boldsymbol{S^{\T}}C^{-1}=\boldsymbol{S}\,,\qquad\qquad
C\boldsymbol{\Sigma^{\T}}C^{-1}=\boldsymbol{\Sigma}\,.
\eeq
Inserting the expressions for $\boldsymbol{S}$ and $\boldsymbol{\Sigma}$
[\eqs{essex}{sigmaex}] and using the result of \eq{cinteractions},
it follows that:
\beqa
\boldsymbol{S_L}&=&\boldsymbol{S_R}\,,\qquad
\boldsymbol{S_D}=\boldsymbol{S_D^{\T}}\,,\qquad
\boldsymbol{\Sbar_D}=\boldsymbol{\Sbar_D^{\T}}\,,\\
\boldsymbol{\Sigma_L}&=&\boldsymbol{\Sigma_R}\,,\qquad
\boldsymbol{\Sigma_D}=\boldsymbol{\Sigma_D^{\T}}\,,\qquad
\boldsymbol{\Sigmabar_D}=\boldsymbol{\Sigmabar_D^{\T}}\,.
\label{majselfenergy}
\eeqa
As expected, with these constraints the form of \eq{matfullDiracprop}
matches precisely with the form of \eq{matfullprop}, corresponding to
the equation for the full propagator functions for a theory of
generic two-component fermion fields.  In the notation of
\sec{subsec:selfenergies}, we can therefore identify:
$\boldsymbol{C}\equiv \boldsymbol{S_L}=\boldsymbol{S_R}$\,,\,
$\boldsymbol{D}\equiv\boldsymbol{S_D}$\,,\, $\boldsymbol{\Xi}
\equiv\boldsymbol{\Sigma_L}
=\boldsymbol{\Sigma_R}$, and $\boldsymbol{\Omega}\equiv\boldsymbol{\Sigma_D}$.

\section{\texorpdfstring{Covariant spin operators and the Bouchiat-Michel formulae}{Covariant spin operators and the Bouchiat-Michel formulae}}
\renewcommand{\theequation}{H.\arabic{equation}}
\renewcommand{\thefigure}{H.\arabic{figure}}
\renewcommand{\thetable}{H.\arabic{table}}
\setcounter{equation}{0}
\setcounter{figure}{0}
\setcounter{table}{0}

\noindent
Bouchiat and Michel derived a useful set of formulae~\cite{bouchmich}
that generalize
the spin projection operators used in four-component spinor computations.  In
this Appendix, we establish the two-component analogues of the
Bouchiat-Michel formulae, and demonstrate their equivalence
to the corresponding four-component spinor formulae.

\subsection{The covariant spin operators for a spin-1/2 fermion}
\renewcommand{\theequation}{H.1.\arabic{equation}}
\renewcommand{\thefigure}{H.1.\arabic{figure}}
\renewcommand{\thetable}{H.1.\arabic{table}}
\setcounter{equation}{0}
\setcounter{figure}{0}
\setcounter{table}{0}

Consider a massive spin-1/2 fermion of mass $m$ and four-momentum
$p$. We define
a set of three four-vectors $S^{a\mu}$ ($a=1,2,3$) such that the $S^{a\mu}$
and $p^\mu/m$ form an orthonormal set of four-vectors.
In the rest frame of the fermion, where
$p^\mu=(m\,;\,{\boldsymbol{\vec 0}})$, we can define
\beq \label{sarestdef}
S^{a\mu}\equiv(0\,;\,{\boldsymbol{{\hat s}^a}})\,,\qquad a=1,2,3\,,
\eeq
where the unit vectors ${\boldsymbol{{\hat s}^a}}$ are a mutually
orthonormal set of unit three-vectors that
form a basis for a right-handed coordinate system. Explicit forms for
the ${\boldsymbol{{\hat s}^a}}$ depend on the Euler angle $\gamma$
used to define the spinor wave function $\chi\ls{s}(\boldsymbol{\hat s})$.
Two common choices corresponding to $\gamma=-\phi$ and $\gamma=0$ are given
in \eqs{s123}{s123p}, respectively.
Using \eq{lorentzboost},
the three four-vectors $S^{a\mu}$, in a reference frame in
which the four-momentum of the fermion is
$p^\mu=(E\,;\,\boldsymbol{\vec p})$, are given by:
\beq \label{smuboosted}
S^{a\mu}= \left(\frac{\boldsymbol{\vec{p}\newcdot\hat{s}^a}}{m}\,;\,
{\boldsymbol{\hat{s}^a}}+\frac{\boldsymbol{(\vec{p}\newcdot\hat{s}^a})
\,\boldsymbol{\vec{p}}}{m(E+m)}\right)\,,\qquad a=1,2,3\,.
\eeq
As discussed in \app{C}, we
identify ${\boldsymbol{{\hat s}}}={\boldsymbol{{\hat s}^3}}$
as the quantization axis used in defining the third component of the
spin of the fermion in its rest frame.
It then follows that the
spin four-vector, previously introduced in \eq{fixedsvect} is given by
$S^\mu=S^{3\mu}$.

The orthonormal set of four four-vectors $p^\mu/m$ and the $S^{a\mu}$ satisfy
the following Lorentz-covariant relations:
\beqa \label{orthonormalset}
      p\newcdot S^a &=& 0\,,\label{oset1}
\\[5pt]
      S^a\newcdot S^b &=& \BDneg\delta^{ab}\,,\label{oset2}
\\[5pt]
\epsilon^{\mu\nu\lambda\sigma}p_\mu S^1_\nu S^2_\lambda S^3_\sigma&=&-m\,,
\label{oset3}
\\[5pt]
S_\mu^a S_\nu^b-S_\nu^a S_\mu^b&=& \epsilon^{abc}
\epsilon_{\mu\nu\rho\sigma} S^{c\rho} \,\frac{p^\sigma}{m}
\,,\label{oset4}\\[5pt]
S^a_\mu\,S^a_\nu &=&
\BDneg g_{\mu\nu} + \frac{p_\mu p_\nu}{m^2}\,,\label{oset5}
\eeqa
where the sum over the repeated indices is implicit.
It is convenient to define a matrix-valued spin four-vector
$\mathscr{S}^\mu$, whose matrix elements are given by:
\beq \label{sssdef}
\mathscr{S}^\mu_{ss'}\equiv S^{a\mu}\tau^a_{ss'}\,,
\qquad s,s'=\pm \half\,,
\eeq
where $\tau^a_{ss'}$ are the matrix elements of the Pauli matrices
(see footnote \ref{fnpauli}).  Then, we can rewrite \eqs{oset2}{oset4} as:
\beqa
\third\,g_{\mu\nu} \mathscr{S}^\mu \mathscr{S}^\nu &=&
\BDneg\mathds{1}_{2\times 2}\,,\label{scasimir} \\
\mathscr{S}^\mu \mathscr{S}^\nu-\mathscr{S}^\nu \mathscr{S}^\mu
&=&\frac{2i}{m}\epsilon^{\mu\nu\rho\sigma}\mathscr{S}_\rho
p_\sigma\,, \label{scomm}
\eeqa
where the product $\mathscr{S}^\mu \mathscr{S}^\nu$ corresponds to ordinary
$2\times 2$ matrix multiplication.
The $\mathscr{S}^\mu$ serve as covariant spin operators for a spin-1/2 fermion.
In particular, in the
rest frame, the $\half\mathscr{S}^i$ satisfy the usual SU(2)
commutation relations, with
$(\half{\bold{\vec{\mathscr{S}}}}\,)^2=\frac{3}{4}$
as expected for a spin-1/2 particle.

It is often desirable to work with helicity states.
In this case, we choose:
\beq \label{sshat3}
{\boldsymbol{{\hat s}^a}}={\boldsymbol{{\hat p}^a}}\,,
\eeq
where the ${\boldsymbol{{\hat p}^a}}$ are an orthonormal triad of
unit three-vectors
with ${\boldsymbol{{\hat p}^3}}\equiv{\boldsymbol{\hat p}}$.
Moreover, since
${\boldsymbol{{\hat p}^a}\newcdot\boldsymbol{\hat p}}=0$ for $a\neq 3$,
it follows
that $S^{a\mu}=(0\,;\,{\boldsymbol{\hat p^a}})$ for $a=1,2$ in \text{all}
reference frames obtained from the rest frame
by a boost in the ${\boldsymbol{\hat p}}$ direction.
Hence, in a reference frame where $p^\mu=(E\,;\,{\boldsymbol{\vec p}})$,
\eq{smuboosted} yields,
\beqa
S^{1\mu} &=& (0\,;\,{\boldsymbol{\hat p^1}})\,,
\label{sa1}\\[5pt]
S^{2\mu} &=& (0\,;\, {\boldsymbol{\hat p^2}})\,, \label{sa2}\\[7pt]
S^{3\mu} &=& \left(\frac{|\boldsymbol{\vec p}\,|}{m}\,;\,\frac{E}{m}
\,{\boldsymbol{\hat p}} \right)\,,\phantom{,\cos\phi\sin\phi,-\sin\theta,}
\label{sa3}
\eeqa
in a coordinate system where
${\boldsymbol{\hat p}}=
(\sin\theta\cos\phi,\,\sin\theta\sin\phi,\,\cos\theta)$.
One can check that \eqst{sarestdef}{oset5} are also satisfied by
the $S^{a\mu}$ defined in \eqst{sa1}{sa3}.

As expected, $S^{3\mu}$ is the spin four-vector for helicity states
obtained in \eq{spinvec}.  In the high energy limit ($E\gg m$),
\beq
mS^{a\mu}=p^\mu\,\delta^{a3}+\mathcal{O}(m)\,.
\eeq
Explicit forms for ${\boldsymbol{\hat p^1}}$ and ${\boldsymbol{\hat p^2}}$
are convention dependent and depend on the conventional choice of
the Euler angle $\gamma$.
For example, consider the quantities:
\beq \label{s12s21}
S^\mu_{-}\equiv \half S^{a\mu}\tau^a_{\hhalf \,,\,-\hhalf}
=\half(S^{1\mu}-iS^{2\mu})\,,\qquad\qquad
S^\mu_{+}\equiv \half S^{a\mu}\tau^a_{-\hhalf\,,\,\hhalf}
=\half(S^{1\mu}+iS^{2\mu})\,.
\eeq
Using \eqst{sshat3}{sa3} and employing
\eq{saexplicit} with $\mathcal{R}$ given by \eq{calrdef},
\beq \label{sigs12}
\sigma\newcdot S_{-} = e^{i\gamma}\begin{pmatrix}
\BDpos \half \sin\theta & \quad
\BDpos e^{-i\phi}\sin^2\displaystyle\frac{\theta}{2} \\[10pt]
\BDneg e^{i\phi}\cos^2\displaystyle\frac{\theta}{2} &
\quad \BDneg \half \sin\theta
\end{pmatrix} \,,\quad\
\sigmabar\newcdot S_{+}=e^{-i\gamma}\begin{pmatrix}
\BDpos \half \sin\theta & \quad
\BDneg e^{-i\phi}\cos^2\displaystyle\frac{\theta}{2} \\[10pt]
\BDpos e^{i\phi}\sin^2\displaystyle\frac{\theta}{2} &
\quad \BDneg \half \sin\theta
\end{pmatrix} \,.
\eeq
In the convention of
\eq{s123} [\eq{s123p}], we take $\gamma=-\phi$ [$\gamma=0$], respectively.

\subsection{Two-component spinor wave function relations}
\renewcommand{\theequation}{H.2.\arabic{equation}}
\renewcommand{\thefigure}{H.2.\arabic{figure}}
\renewcommand{\thetable}{H.2.\arabic{table}}
\setcounter{equation}{0}
\setcounter{figure}{0}
\setcounter{table}{0}

\indent

In \sec{subsec:singleWeyl}, we wrote down explicit forms for the
undotted spinor wave functions
\beqa
x_\alpha(\boldsymbol{\vec p},s)
&=&\sqrt{\BDpos p\newcdot\sigma}\,\chi\ls{s}\,,
\qquad\qquad\quad\,\,
x^\alpha(\boldsymbol{\vec p},s)
=-2s\chi^\dagger\ls{-s}\sqrt{\BDpos p\newcdot\sigmabar}\,,
\label{app:explicitxa} \\
y_\alpha(\boldsymbol{\vec p},s)&=&2s
\sqrt{\BDpos p\newcdot\sigma}\,\chi\ls{-s}\,,\qquad\quad\,\,\,\,
y^\alpha(\boldsymbol{\vec p},s)=
\chi^\dagger\ls{s}\sqrt{\BDpos p\newcdot\sigmabar}\,,
\label{app:explicitya}
\eeqa
and the dotted spinor wave functions
\beqa
 x^{\dagger\dot{\alpha}}(\boldsymbol{\vec p},s)&=&
-2s\sqrt{\BDpos p\newcdot\sigmabar}\,\chi\ls{-s}\,,\qquad\qquad
 x^\dagger_{\dot{\alpha}}(\boldsymbol{\vec p},s)
=\chi^\dagger\ls{s}\sqrt{\BDpos p\newcdot\sigma}\,,\\
\label{app:explicitxb}
 y^{\dagger\dot{\alpha}}(\boldsymbol{\vec p},s)&=&
\sqrt{\BDpos p\newcdot\sigmabar}\,\chi\ls{s}\,,\qquad\qquad\qquad\,\,
 y^\dagger_{\dot{\alpha}}(\boldsymbol{\vec p},s)
=2s\chi^\dagger\ls{-s}\sqrt{\BDpos p\newcdot\sigma}\,,
\label{app:explicityb}
\eeqa
where $\sqrt{\BDpos p\newcdot\sigma}$ and
$\sqrt{\BDpos p\newcdot\sigmabar}$ are defined  by
\eqs{sqpsigma}{sqpsigmabar}, respectively.

As shown in \app{C}, the two-component spinors $\chi\ls{s}$ satisfy:
\beq
\half\, {\boldsymbol{\vec\sigma\newcdot{\hat s}^a}}\chi_{s'}=
\half \tau^a_{ss'}\chi_s\,,\qquad
\chi^\dagger_s({\boldsymbol{\hat
      s}})\chi_{s'}({\boldsymbol{\hat s}})=\delta_{ss'}\,,
  \qquad s\,,s' = \pm\half\,.\label{echiess}
\eeq

Next, we use \eq{ssr12} to obtain:
\beqa
\sqrt{\BDpos p\newcdot\sigma}\,
S^a\newcdot\sigmabar\,
\sqrt{\BDpos p\newcdot\sigma} &=& \BDpos m\,
{\boldsymbol{\vec\sigma\newcdot{\hat s}^a}}\,,\label{psap1} \\
\sqrt{\BDpos p\newcdot\sigmabar}\,
S^a\newcdot\sigma\,\sqrt{\BDpos p\newcdot\sigmabar} &=& \BDneg m\,
{\boldsymbol{\vec\sigma\newcdot{\hat s}^a}}\,,
\label{psap2}
\eeqa
which extends the results of \eq{psp12}.  As a result,
we obtain a generalization of \eqst{spinone}{spinfour}:

\beqa
&&\!\!
(S^a\newcdot\sigmabar)^{\dot{\alpha}\beta}
x_{\beta}({\boldsymbol{\vec p}},s') = \BDpos \tau^a\ls{ss'}
 y^{\dagger\dot{\alpha}}({\boldsymbol{\vec p}},s)
\, \,, \qquad
(S^a\newcdot\sigma)_{\alpha\dot{\beta}}
 y^{\dagger\dot{\beta}}({\boldsymbol{\vec p}},s') =
\BDneg \tau^a\ls{ss'} x_\alpha({\boldsymbol{\vec p}},s)
\,,
\label{spinoneprime} \\[5pt] &&\!\!
(S^a\newcdot\sigma)_{\alpha\dot{\beta}}
 x^{\dagger\dot{\beta}}({\boldsymbol{\vec p}},s') = \BDneg
 \tau^a\ls{s's} y_\alpha({\boldsymbol{\vec p}},s)
\, \,, \qquad\!\!
(S^a\newcdot\sigmabar)^{\dot{\alpha}\beta}
 y_{\beta}({\boldsymbol{\vec p}},s')  =
\BDpos \tau^a\ls{s's}
 x^{\dagger\dot{\alpha}}({\boldsymbol{\vec p}},s)
\> \,,
\label{spintwoprime} \\[5pt] &&
x^{\alpha}({\boldsymbol{\vec p}},s')
(S^a\newcdot\sigma)_{\alpha\dot{\beta}}
 = \BDneg \tau^a\ls{s's}
 y^\dagger_{\dot{\beta}}({\boldsymbol{\vec p}},s)
\,, \qquad\!\!
 y^\dagger_{\dot{\alpha}}({\boldsymbol{\vec p}},s')
(S^a\newcdot\sigmabar)^{\dot{\alpha}\beta} =
\BDpos \tau^a\ls{s's} x^{\beta}({\boldsymbol{\vec p}},s)
\,,
\label{spinthreeprime}
\\[5pt]  &&
 x^\dagger_{\dot{\alpha}}({\boldsymbol{\vec p}},s')
(S^a\newcdot\sigmabar)^{\dot{\alpha}\beta} =
\BDpos \tau^a\ls{ss'} y^{\beta}({\boldsymbol{\vec p}},s)
\,, \qquad\,\,
y^{\alpha}({\boldsymbol{\vec p}},s')
(S^a\newcdot\sigma)_{\alpha\dot{\beta}} =
\BDneg \tau^a\ls{ss'}
 x^\dagger_{\dot{\beta}}({\boldsymbol{\vec p}},s)
\,, \label{spinfourprime}
\eeqa
where there are implicit sums over the repeated labels $s=\pm\half$.
As expected, the case of $a=3$ simply reproduces
the results of \eqst{spinone}{spinfour} obtained previously.
The above equations also apply to helicity wave functions
$x({\boldsymbol{\vec p}},\lambda)$ and $y({\boldsymbol{\vec p}},\lambda)$
by replacing $s$, $s'$ with $\lambda$, $\lambda^\prime$
and defining the $S^{a\mu}$ by \eqst{sa1}{sa3}.

The derivation of \eqst{spinoneprime}{spinfourprime} for arbitrary $a$
closely follows the corresponding derivation for $a=3$ previously given.
For example, using \eqs{psap1}{psap2} and the definitions for
$x_\alpha({\boldsymbol{\vec{p}}},s)$ and
$ y^{\dagger\dot{\alpha}}({\boldsymbol{\vec{p}}},s)$,
we find (suppressing spinor indices),
\beq \label{firstderiv}
\sqrt{\BDpos p\newcdot\sigma}\,
S^a\newcdot\sigmabar\,x({\boldsymbol{\vec{p}}},s')=
\sqrt{\BDpos p\newcdot\sigma}\,S^a\newcdot\sigmabar\,
\sqrt{\BDpos p\newcdot\sigma}\,\chi\ls{s'}
= \BDpos m{\boldsymbol{\vec\sigma\newcdot{\hat{s}^a}}}\,\chi\ls{s'}
= \BDpos m\tau^a_{ss'}\,\chi\ls{s}\,,
\eeq
after using \eq{echiess}.
Multiplying both sides of \eq{firstderiv} by
$\sqrt{\BDpos p\newcdot\sigmabar}$,
we end up with
\beq
S^a\newcdot\sigmabar\,x({\boldsymbol{\vec{p}}},s')=
\BDpos \tau^a_{ss'}\sqrt{\BDpos p\newcdot\sigmabar}\,\chi\ls{s}
= \BDpos \tau^a_{ss'}\, y^\dagger({\boldsymbol{\vec{p}}},s)\,.
\eeq
Similarly,
\beq
S^a\newcdot\sigma {x}^\dagger({\boldsymbol{\vec{p}}},s')=
\BDpos 2s'\tau^a_{-s,-s'}\sqrt{\BDpos p\newcdot\sigma}\,\chi\ls{-s}
= \BDneg \tau^a_{s's}\,y({\boldsymbol{\vec{p}}},s)\,,
\eeq
where we have used:
\beq \label{tauss}
4ss'\tau^a_{-s,-s'}=-\tau^a_{s's}\,,\qquad\quad {\rm for}\quad s,s'=\pm 1/2\,.
\eeq
All the results of \eqst{spinoneprime}{spinfourprime}
can be derived in this manner.

\subsection{Two-component Bouchiat-Michel formulae}
\label{bouchiat2}
\renewcommand{\theequation}{H.3.\arabic{equation}}
\renewcommand{\thefigure}{H.3.\arabic{figure}}
\renewcommand{\thetable}{H.3.\arabic{table}}
\setcounter{equation}{0}
\setcounter{figure}{0}
\setcounter{table}{0}

To establish the Bouchiat-Michel formulae, we begin with the following
identity:
\beq \label{bmstart}
\half(\delta_{ss'}+{\boldsymbol{\vec{\sigma}
\newcdot\hat
s}}^a\,\tau^a_{ss'})\sum_{t=\pm 1/2}\chi\ls{t}\chi^\dagger\ls{t}
=\chi\ls{s'}\chi\ls{s}^\dagger\,.
\eeq
To verify \eq{bmstart}, we first make use of \eq{echiess}
to write ${\boldsymbol{\vec{\sigma}\newcdot\hat s}}^a\chi\ls{t}
=\tau^a_{t't}\chi\ls{t'}$ and evaluate the product of two Pauli matrices:
\beq
\tau^a_{ss'}\tau^a_{t't}=2\,\delta_{st}\delta_{s't'}
-\delta_{ss'}\delta_{tt'}\,.
\eeq
We then use \eq{psap1} and
the completeness relation given in \eq{completeness} to rewrite
\eq{bmstart}
in terms of $\mathscr{S}^\mu_{ss'}$ defined in \eq{sssdef},
\beq
\chi_{s'}\chi_s^\dagger=\half\left(\delta_{ss'}
\BDplus \frac{1}{m}\,
\sqrt{\BDpos p\newcdot\sigma}\,\mathscr{S}_{ss'}
\newcdot\sigmabar\,\sqrt{\BDpos p\newcdot\sigma}\right)\,.
\eeq
Hence, with both spinor indices in the lowered position,
\beqa
x(\boldsymbol{\vec p},s') x^\dagger(\boldsymbol{\vec p},s)&=&
\sqrt{\BDpos p\newcdot\sigma}\,\chi\ls{s'}\chi^\dagger\ls{s}\,
\sqrt{\BDpos p\newcdot\sigma} \nonumber \\
&=& \half\sqrt{\BDpos p\newcdot\sigma}\left[\delta_{ss'}
\BDplus \frac{1}{m}\,\sqrt{\BDpos p\newcdot\sigma}
\,\mathscr{S}_{ss'}\newcdot\sigmabar\,
\sqrt{\BDpos p\newcdot\sigma}\right]
\sqrt{\BDpos p\newcdot\sigma}
\nonumber \\
&=&\half\left[\BDpos p\newcdot\sigma \delta_{ss'}
\BDplus
\frac{1}{m}\, p\newcdot\sigma\, \mathscr{S}_{ss'}\newcdot\sigmabar\,
p\newcdot\sigma\right] \nonumber \\[4pt]
&=& \half \left(\BDpos p\newcdot\sigma \delta_{ss'}
\BDminus m\mathscr{S}_{ss'}
\newcdot\sigma\right)\,.\label{bouchmich}
\eeqa
In the final step of \eq{bouchmich}, we simplified the product of three
dot products by noting that $p\newcdot S^a=0$ implies that
$\mathscr{S}_{ss'}\newcdot\sigmabar\; p\newcdot\sigma
=-p\newcdot\sigmabar\;\mathscr{S}_{ss'}\newcdot\sigma$.
\Eq{bouchmich} is the two-component version of one of the Bouchiat-Michel
formulae.  We list below a complete set of Bouchiat-Michel formulae,
which can be derived by similar techniques:
\beqa
&&x_\alpha({\boldsymbol{\vec p}},s')  x^\dagger_{\dot{\beta}}
({\boldsymbol{\vec p}},s) = \half
(\BDpos p\,\delta\ls{ss'} \BDminus m\mathscr{S}\ls{ss'})
\newcdot\sigma_{\alpha\dot{\beta}}
\,,
\label{xxdagmassivehel}
\\[5pt]
&& y^{\dagger\dot{\alpha}}({\boldsymbol{\vec p}},s') y^\beta
({\boldsymbol{\vec p}},s) = \half
(\BDpos p\,\delta\ls{ss'} \BDplus m \mathscr{S}\ls{ss'})
\newcdot\sigmabar^{\dot{\alpha}{\beta}}
\,,
\\[5pt]
&&x_\alpha({\boldsymbol{\vec p}},s')
y^\beta({\boldsymbol{\vec p}},s)
 = \half\left[  m \delta\ls{ss'}\delta_\alpha{}^\beta -
[(\sigma\newcdot \mathscr{S}\ls{ss'})\, (\sigmabar  \newcdot
p)]_\alpha{}^\beta\right]
\,,
\\[5pt]
&& y^{\dagger\dot{\alpha}}
({\boldsymbol{\vec p}},s')
 x^\dagger_{\dot{\beta}}({\boldsymbol{\vec p}},s)
 = \half \left[m \delta\ls{ss'}
\delta^{\dot{\alpha}}{}_{\dot{\beta}} +
[(\sigmabar\newcdot \mathscr{S}\ls{ss'})\,
(\sigma \newcdot p)]^{\dot{\alpha}}{}_{\dot{\beta}}\right]
\,.
\label{ydagxdagmassivehel}
\eeqa
If we set $s=s'$, we recover
\eqst{xxdagmassive}{ydagxdagmassive} as expected.
The Bouchiat-Michel formulae can also be verified directly by using
the explicit forms for the two-component spinor wave functions
[\eq{twocomp}] and the $\mathscr{S}^\mu_{ss'}$ [defined in \eq{sssdef}].
The latter depends
on the explicit form of the
$\boldsymbol{\hat s^a}$ via \eq{smuboosted}.

An equivalent set of Bouchiat-Michel formulae can be obtained
by raising and/or lowering the appropriate free spinor indices using
\eqs{sigsig1}{sigmunurel1}:
\beqa
&& x^{\dagger\dot{\alpha}}({\boldsymbol{\vec p}},s')  x^\beta
({\boldsymbol{\vec p}},s) = \half
(\BDpos p\,\delta\ls{s's} \BDminus m\mathscr{S}\ls{s's})
\newcdot\sigmabar^{\dot{\alpha}{\beta}}
\,,
\label{xxdagmassivehel2}
\\[5pt]
&&y_\alpha({\boldsymbol{\vec p}},s')  y^\dagger_{\dot{\beta}}
({\boldsymbol{\vec p}},s) = \half
(\BDpos p\,\delta\ls{s's} \BDplus m \mathscr{S}\ls{s's})
\newcdot\sigma_{\alpha\dot{\beta}}
\,,
\\[5pt]
&&y_\alpha({\boldsymbol{\vec p}},s')
x^\beta({\boldsymbol{\vec p}},s)
 = -\half\left[  m \delta\ls{s's}\delta_\alpha{}^\beta +
[(\sigma\newcdot \mathscr{S}\ls{s's})\, (\sigmabar  \newcdot
p)]_\alpha{}^\beta \right]
\,,
\\[5pt]
&& x^{\dagger\dot{\alpha}}
({\boldsymbol{\vec p}},s')
 y^\dagger_{\dot{\beta}}({\boldsymbol{\vec p}},s)
 = -\half\left[  m \delta\ls{s's}
\delta^{\dot{\alpha}}{}_{\dot{\beta}} -
[(\sigmabar\newcdot \mathscr{S}\ls{s's})\,
(\sigma \newcdot p)]^{\dot{\alpha}}{}_{\dot{\beta}}\right]
\,.
\label{ydagxdagmassivehel2}
\eeqa
 In this derivation,
the spin labels in
\eqst{xxdagmassivehel2}{ydagxdagmassivehel2} are reversed relative
to those in \eqst{xxdagmassivehel}{ydagxdagmassivehel}
due to \eq{tauss}.
Eight additional relations of the Bouchiat-Michel type
can be obtained by replacing one $x$-spinor
with a $y$-spinor (or vice versa).
Recalling that the $x$ and $y$
spinors are related by [cf.~\eq{xyrelation}],
\beq \label{app:xyrelation}
y({\boldsymbol{\vec p}},s)=2s x({\boldsymbol{\vec p}},-s)\,,\qquad\qquad
 y^\dagger({\boldsymbol{\vec p}},s)=2s  x^\dagger({\boldsymbol{\vec p}},-s)\,,
\eeq
all possible spinor bilinears can be obtained
from \eqst{xxdagmassivehel}{ydagxdagmassivehel2}.

Note that \eqst{xxdagmassivehel}{ydagxdagmassivehel2}
also apply to helicity
spinor wave functions
$x({\boldsymbol{\vec p}},\lambda)$ and $y({\boldsymbol{\vec p}},\lambda)$
after replacing $s,s'$ with $\lambda$,~$\lambda^\prime$
and using the $S^{a\mu}$ as defined in \eqst{sa1}{sa3}.  Strictly speaking, all
results involving the spinor wave functions
obtained up to this point apply in the case of a massive spin-1/2 fermion.
If we take the massless limit, then the
four-vector $S^{3\mu}$ does not exist, as its definition depends on the
existence of a rest frame.  (In contrast, the four-vectors $S^{1\mu}$ and
$S^{2\mu}$ do exist in the massless limit.)
Nevertheless, massless helicity spinor wave
functions are well defined; explicit forms can be found in
\eqst{helexplicitxa}{helexplicityb}.  Using these forms, one can
derive the Bouchiat-Michel formulae for a massless spin-1/2 fermion:
\beqa
&& x_\alpha({\boldsymbol{\vec p}},\lambda^\prime)  x^\dagger_{\dot{\beta}}
({\boldsymbol{\vec p}},\lambda) =
\BDpos (\half-\lambda)\,\delta_{\lambda\lambda^\prime}\,
p \newcdot \sigma_{\alpha \dot{\beta}}
\,,
\label{app:xxdagmassless}
\\
&&
 y^{\dagger\dot{\alpha}}({\boldsymbol{\vec p}},\lambda^\prime) y^\beta
({\boldsymbol{\vec p}},\lambda) =
\BDpos (\half +\lambda)\,\delta_{\lambda\lambda^\prime}\,
p \newcdot \sigmabar^{\dot{\alpha}\beta}
\,,
\\
&&
x_\alpha({\boldsymbol{\vec p}},\lambda^\prime)
y^\beta({\boldsymbol{\vec p}},\lambda)
= -(\half-\lambda^\prime)(\half+\lambda)
\left[(\sigma\newcdot S_{-})
(\sigmabar\newcdot p)\right]_\alpha{}^\beta
\,,
\\
&&
 y^{\dagger\dot{\alpha}}({\boldsymbol{\vec p}},\lambda^\prime)
 x^\dagger_{\dot{\beta}}({\boldsymbol{\vec p}},\lambda)=
(\half+\lambda^\prime)(\half-\lambda)
\left[(\sigmabar\newcdot S_{+})
(\sigma\newcdot p)\right]^{\dot{\alpha}}{}_{\dot{\beta}}
\,,
\label{app:ydagxdagmassless}
\eeqa
where $S^\mu_{-}$ and $S^\mu_{+}$ are defined in \eq{s12s21}.
The equivalent set of Bouchiat-Michel formulae, obtained by raising and/or
lowering the appropriate free spinor indices, is given by:

\beqa
&&
 x^{\dagger\dot{\alpha}}({\boldsymbol{\vec p}},\lambda^\prime) x^\beta
({\boldsymbol{\vec p}},\lambda) =
\BDpos (\half-\lambda)\,\delta_{\lambda\lambda^\prime}\,
p \newcdot \sigmabar^{\dot{\alpha}\beta}
\,,
\label{app:xxdagmassless2}
\\
&&
y_\alpha({\boldsymbol{\vec p}},\lambda^\prime)  y^\dagger_{\dot{\beta}}
({\boldsymbol{\vec p}},\lambda) =
\BDpos (\half +\lambda)\,\delta_{\lambda\lambda^\prime}\,
p \newcdot \sigma_{\alpha \dot{\beta}}
\,,
\\
&&
y_\alpha({\boldsymbol{\vec p}},\lambda^\prime)
x^\beta({\boldsymbol{\vec p}},\lambda)
=  -(\half+\lambda^\prime)(\half-\lambda)
\left[(\sigma\newcdot S_{-})
(\sigmabar\newcdot p)\right]_\alpha{}^\beta
\,,
\\
&&
 x^{\dagger\dot{\alpha}}({\boldsymbol{\vec p}},\lambda^\prime)
 y^\dagger_{\dot{\beta}}({\boldsymbol{\vec p}},\lambda)=
(\half-\lambda^\prime)(\half+\lambda)
\left[(\sigmabar\newcdot S_{+})
(\sigma\newcdot p)\right]^{\dot{\alpha}}{}_{\dot{\beta}}
\,.
\label{app:ydagxdagmassless2}
\eeqa
Eight additional relations of the Bouchiat-Michel type
can be obtained by replacing one $x$-spinor
with a $y$-spinor (or vice versa), using the results of \eq{app:xyrelation}.
As a check, one can verify that the above results follow from
\eqst{xxdagmassivehel}{ydagxdagmassivehel2}
by replacing $s$ with $\lambda$, setting
$mS^{a\mu}=p^\mu\,\delta^{a3}$,
applying the mass-shell condition ($p^2= \BDpos m^2$),
and taking the $m\to 0$ limit at the
end of the computation.

We now demonstrate how to use the Bouchiat-Michel formulae to evaluate
helicity amplitudes involving two equal mass spin-1/2 fermions.  A
typical amplitude involving a fermion-antifermion pair, evaluated in
the center-of-mass frame of the pair has the generic structure:
\beq \label{zform}
z({\boldsymbol{\vec{p}}},\lambda)\,\Gamma \,z^\prime
({-\boldsymbol{\vec{p}}},\lambda^\prime)\,,
\eeq
where $z$ is one of the two-component spinor wave functions
$x$, $x^\dagger$,
$y$, or $y^\dagger$, and $\Gamma$ is a $2\times 2$ matrix (in spinor
space) that is either the identity matrix, or is made up of
alternating products of $\sigma$ and $\sigmabar$.
As an illustration, we evaluate:
\beq
{x}^\dagger_{\dot{\alpha}}({\boldsymbol{\vec p}},\lambda)
\,\Gamma^{\dot{\alpha}\beta}\,
y_\beta(-{\boldsymbol{\vec p}},\lambda^\prime)=
2\lambda^\prime \,\Gamma^{\dot{\alpha}\beta}\,
x_\beta(-{\boldsymbol{\vec p}},-\lambda^\prime)
{x}^\dagger_{\dot{\alpha}}({\boldsymbol{\vec p}},\lambda)
= 2\lambda^\prime\,\xi\ls{\lambda^\prime}(\boldsymbol{\hat p})\,
\,\Gamma^{\dot{\alpha}\beta}\sigma^0_{\beta\dot{\beta}}\,
{y}^{\dagger\dot{\beta}}({\boldsymbol{\vec p}},\lambda^\prime)
{x}^\dagger_{\dot{\alpha}}({\boldsymbol{\vec p}},\lambda)\,,
\eeq
where $\xi\ls{\lambda^\prime}(\boldsymbol{\hat p})$ is defined in \eq{xphase},
and we have used \eqs{pminus1}{xyrelation}.  We can now
employ the Bouchiat-Michel formula
to convert the above result into a trace.
By a similar computation, all expressions of the form of \eq{zform}
can be expressed as a trace:
\beqa
{x}^\dagger_{\dot{\alpha}}({\boldsymbol{\vec p}},\lambda)
\,\Gamma^{\dot{\alpha}\beta}\,
y_\beta(-{\boldsymbol{\vec p}},\lambda^\prime)
&=&\lambda^\prime\,\xi\ls{\lambda^\prime}(\boldsymbol{\hat p})\,
\,{\rm Tr}\left[\Gamma\,\sigma^0(m\delta_{\lambda\lambda^\prime}
+\sigmabar\newcdot \mathscr{S}_{\lambda\lambda^\prime}
\,\sigma\newcdot p)\right]\,,
\label{helamps1}
\\
y^\alpha({\boldsymbol{\vec p}},\lambda)
\,\Gamma_{\alpha\dot{\beta}}\,
{x}^{\dagger\dot{\beta}}(-{\boldsymbol{\vec p}},\lambda^\prime)
&=&-\lambda^\prime\,\xi\ls{\lambda^\prime}(\boldsymbol{\hat p})\,
\,{\rm Tr}\left[\Gamma\,\sigmabar^0(m\delta_{\lambda\lambda^\prime}
-\sigma\newcdot \mathscr{S}_{\lambda\lambda^\prime}\,
\sigmabar\newcdot p)\right]\,,
\label{helamps2}
\\
y^\alpha({\boldsymbol{\vec p}},\lambda)
\,\Gamma_{\alpha}{}^\beta\,
y_\beta(-{\boldsymbol{\vec p}},\lambda^\prime)
&=& \BDpos \lambda^\prime\,\xi\ls{\lambda^\prime}(\boldsymbol{\hat p})\,
\,{\rm Tr}\left[\Gamma\,\sigma^0(\sigmabar\newcdot p\,
\delta_{\lambda\lambda^\prime}
+m\sigmabar\newcdot \mathscr{S}_{\lambda\lambda^\prime})\right]\,,
\label{helamps3}
\\
{x}^\dagger_{\dot{\alpha}}({\boldsymbol{\vec p}},\lambda)
\,\Gamma^{\dot{\alpha}}{}_{\dot{\beta}}\,
{x}^{\dagger\dot{\beta}}(-{\boldsymbol{\vec p}},\lambda^\prime)
&=& \BDneg \lambda^\prime\,\xi\ls{\lambda^\prime}(\boldsymbol{\hat p})\,
\,{\rm Tr}\left[\Gamma\,\sigmabar^0(\sigma\newcdot p\,
\delta_{\lambda\lambda^\prime}
-m\sigma\newcdot \mathscr{S}_{\lambda\lambda^\prime})\right]\,,
\label{helamps4}
\eeqa
after making use of \eqs{xxdagmassivehel}{ydagxdagmassivehel}.
Similarly, there are four additional results that make use of
\eqs{xxdagmassivehel2}{ydagxdagmassivehel2}:
\beqa
{y}^\dagger_{\dot{\alpha}}({\boldsymbol{\vec p}},\lambda)
\,\Gamma^{\dot{\alpha}\beta}\,
x_\beta(-{\boldsymbol{\vec p}},\lambda^\prime)
&=&\lambda^\prime\,\xi\ls{-\lambda^\prime}(\boldsymbol{\hat p})\,
\,{\rm Tr}\left[\Gamma\,\sigma^0(m\delta_{\lambda^\prime\lambda}
-\sigmabar\newcdot \mathscr{S}_{\lambda^\prime\lambda}
\,\sigma\newcdot p)\right]\,,
\label{helampp1}
\\
x^\alpha({\boldsymbol{\vec p}},\lambda)
\,\Gamma_{\alpha\dot{\beta}}\,
{y}^{\dagger\dot{\beta}}(-{\boldsymbol{\vec p}},\lambda^\prime)
&=&-\lambda^\prime\,\xi\ls{-\lambda^\prime}(\boldsymbol{\hat p})\,
\,{\rm Tr}\left[\Gamma\,\sigmabar^0(m\delta_{\lambda^\prime\lambda}
+\sigma\newcdot \mathscr{S}_{\lambda^\prime\lambda}
\,\sigmabar\newcdot p)\right]\,,
\label{helampp2}
\\
x^\alpha({\boldsymbol{\vec p}},\lambda)
\,\Gamma_{\alpha}{}^\beta\,
x_\beta(-{\boldsymbol{\vec p}},\lambda^\prime)
&=&
\BDneg \lambda^\prime\,\xi\ls{-\lambda^\prime}(\boldsymbol{\hat p})\,
\,{\rm Tr}\left[\Gamma\,\sigma^0(\sigmabar\newcdot p\,
\delta_{\lambda^\prime\lambda}
-m\sigmabar\newcdot \mathscr{S}_{\lambda^\prime\lambda})\right]\,,
\label{helampp3}
\\
{y}^\dagger_{\dot{\alpha}}({\boldsymbol{\vec p}},\lambda)
\,\Gamma^{\dot{\alpha}}{}_{\dot{\beta}}\,
{y}^{\dagger\dot{\beta}}(-{\boldsymbol{\vec p}},\lambda^\prime)
&=&
\BDpos \lambda^\prime\,\xi\ls{-\lambda^\prime}(\boldsymbol{\hat p})\,
\,{\rm Tr}\left[\Gamma\,\sigmabar^0(\sigma\newcdot p\,
\delta_{\lambda^\prime\lambda}
+m\sigma\newcdot \mathscr{S}_{\lambda^\prime\lambda})\right]
\,.\label{helampp4}
\eeqa
For amplitudes involving equal mass fermions (or equal mass
antifermions), other combinations of
spinor bilinears appear in which one $x$-spinor above is replaced by a
$y$-spinor or vice versa.  These amplitudes can be reduced to one of
the eight listed above by using \eq{xyrelation}.

In the massless limit, one can again put $mS^{a\mu}=p^\mu\delta^{a3}$,
set $p^2= \BDpos m^2$ and take $m\to 0$ at the end of the computation.
Alternatively, one can repeat the derivation of \eqst{helamps1}{helampp4}
using the results of \eqs{app:xxdagmassless}{app:ydagxdagmassless2}.
For completeness, we record the end result here.
\beqa
{x}^\dagger_{\dot{\alpha}}({\boldsymbol{\vec p}},\lambda)
\,\Gamma^{\dot{\alpha}\beta}\,
y_\beta(-{\boldsymbol{\vec p}},\lambda^\prime)
&=& (\half+\lambda^\prime)(\half-\lambda)\,
\xi\ls{\lambda^\prime}(\boldsymbol{\hat p})\,\Tr(\Gamma\, \sigma^0
\sigmabar\newcdot S_{-}\,\sigma\newcdot p)\,,
\label{mzerohelamps1}
\\
y^\alpha({\boldsymbol{\vec p}},\lambda)
\,\Gamma_{\alpha\dot{\beta}}\,
{x}^{\dagger\dot{\beta}}(-{\boldsymbol{\vec p}},\lambda^\prime)
&=&-(\half-\lambda^\prime)(\half+\lambda)
\,\xi\ls{\lambda^\prime}(\boldsymbol{\hat p})\,\Tr(\Gamma\,\sigmabar^0
\sigma\newcdot S_{-}\,\sigmabar\newcdot p)\,,
\label{mzerohelamps2}
\\
y^\alpha({\boldsymbol{\vec p}},\lambda)
\,\Gamma_{\alpha}{}^\beta\,
y_\beta(-{\boldsymbol{\vec p}},\lambda^\prime)
&=&
\BDpos (\half+\lambda)\,\delta_{\lambda\lambda^\prime}
\,\xi\ls{\lambda^\prime}(\boldsymbol{\hat p})\,\Tr(\Gamma\,\sigma^0
\,\sigmabar\newcdot p)
\,,\label{mzerohelamps3}
\\
{x}^\dagger_{\dot{\alpha}}({\boldsymbol{\vec p}},\lambda)
\,\Gamma^{\dot{\alpha}}{}_{\dot{\beta}}\,
{x}^{\dagger\dot{\beta}}(-{\boldsymbol{\vec p}},\lambda^\prime)
&=&
\BDpos (\half-\lambda)\,\delta_{\lambda\lambda^\prime}
\,\xi\ls{\lambda^\prime}(\boldsymbol{\hat p})\,\Tr(\Gamma\,\sigmabar^0
\,\sigma\newcdot p)
\,.\label{mzerohelamps4}
\eeqa
The equivalent set of formulae, obtained by raising and/or
lowering the appropriate free spinor indices as before, is given by:
\beqa
{y}^\dagger_{\dot{\alpha}}({\boldsymbol{\vec p}},\lambda)
\,\Gamma^{\dot{\alpha}\beta}\,
x_\beta(-{\boldsymbol{\vec p}},\lambda^\prime)
&=&(\half-\lambda^\prime)(\half+\lambda)
\,\xi\ls{-\lambda^\prime}(\boldsymbol{\hat p})\,\Tr(\Gamma\, \sigma^0
\sigmabar\newcdot S_{+}\,\sigma\newcdot p)\,,
\label{mzerohelampp1}
\\
x^\alpha({\boldsymbol{\vec p}},\lambda)
\,\Gamma_{\alpha\dot{\beta}}\,
{y}^{\dagger\dot{\beta}}(-{\boldsymbol{\vec p}},\lambda^\prime)
&=&-(\half+\lambda^\prime)(\half-\lambda)
\,\xi\ls{-\lambda^\prime}(\boldsymbol{\hat p})\,\Tr(\Gamma\,\sigmabar^0
\sigma\newcdot S_{+}\,\sigmabar\newcdot p)\,,
\label{mzerohelampp2}
\\
x^\alpha({\boldsymbol{\vec p}},\lambda)
\,\Gamma_{\alpha}{}^\beta\,
x_\beta(-{\boldsymbol{\vec p}},\lambda^\prime)
&=&
\BDpos (\half-\lambda)\,\delta_{\lambda\lambda^\prime}
\,\xi\ls{-\lambda^\prime}(\boldsymbol{\hat p})\,\Tr(\Gamma\,\sigma^0
\,\sigmabar\newcdot p)
\,,\label{mzerohelampp3}
\\
{y}^\dagger_{\dot{\alpha}}({\boldsymbol{\vec p}},\lambda)
\,\Gamma^{\dot{\alpha}}{}_{\dot{\beta}}\,
{y}^{\dagger\dot{\beta}}(-{\boldsymbol{\vec p}},\lambda^\prime)
&=&
\BDpos (\half+\lambda)\,\delta_{\lambda\lambda^\prime}
\,\xi\ls{-\lambda^\prime}(\boldsymbol{\hat p})\,\Tr(\Gamma\,\sigmabar^0
\,\sigma\newcdot p)
\,.\label{mzerohelampp4}
\eeqa

The traces are easily evaluated using the results of \app{B}.
Here, we apply the above results to the amplitude for the decay
$Z^0\to f\fbar$ [see \sec{zff}].  The corresponding center-of-mass
frame helicity amplitude is a linear combination of \eqs{helamps1}{helamps2}
with $\Gamma=\sigmabar$ and $\Gamma=\sigma$, respectively.  Evaluating
the corresponding terms, we find for $\Gamma=\sigmabar$,
\beq \label{zffhel1}
{x}^\dagger({\boldsymbol{\vec p}},\lambda)
\sigmabar^\mu y(-{\boldsymbol{\vec p}},\lambda^\prime)
=
2\lambda^\prime\,\xi\ls{\lambda^\prime}(\boldsymbol{\hat p})\,
\left[\BDpos mg^{\mu 0}\delta_{\lambda\lambda^\prime}
+ p^\mu \mathscr{S}^0_{\lambda\lambda^\prime}
- p^0 \mathscr{S}^\mu_{\lambda\lambda^\prime}
- 2m(\mathscr{S}^\mu \mathscr{S}^0-\mathscr{S}^0
\mathscr{S}^\mu)_{\lambda\lambda^\prime}
\right]\,,
\eeq
where we have used \eq{scomm} to replace the term with the Levi-Civita
tensor.  Similarly, we calculate for $\Gamma=\sigma$,
\beq \label{zffhel2}
y({\boldsymbol{\vec p}},\lambda)
\sigma^\mu {x}^\dagger(-{\boldsymbol{\vec p}},\lambda^\prime)
=
2\lambda^\prime\,\xi\ls{\lambda^\prime}(\boldsymbol{\hat p})\,
\left[\BDneg mg^{\mu 0}\delta_{\lambda\lambda^\prime}
+ p^\mu \mathscr{S}^0_{\lambda\lambda^\prime}
- p^0 \mathscr{S}^\mu_{\lambda\lambda^\prime}
+ 2m(\mathscr{S}^\mu \mathscr{S}^0-\mathscr{S}^0
\mathscr{S}^\mu)_{\lambda\lambda^\prime}
\right]\,.
\eeq
\Eqs{zffhel1}{zffhel2} provide explicit forms for the $Z^0\to f\fbar$
decay helicity amplitudes defined in \eqs{zffamp1}{zffamp2}.

The above method is not applicable if the two fermions have unequal
mass.  In order to compute the helicity amplitudes of the form given
by \eq{zform} for unequal masses, a generalization of the
above techniques is required.  Some methods for four-component spinor
wave functions have been proposed in \Ref{vega}. We leave it as an
exercise for the reader to translate these techniques so that they are
applicable to helicity amplitudes expressed in terms of two-component
spinor wave functions.  An alternative approach, which is applicable
to the computation of
helicity amplitudes for processes involving multi-fermion final
states of arbitrary mass, is reviewed in \app{I.1}.

\subsection{Four-component Bouchiat-Michel formulae}
\renewcommand{\theequation}{H.4.\arabic{equation}}
\renewcommand{\thefigure}{H.4.\arabic{figure}}
\renewcommand{\thetable}{H.4.\arabic{table}}
\setcounter{equation}{0}
\setcounter{figure}{0}
\setcounter{table}{0}

\indent

Using the results of \app{G},
the translation of the results of \app{H.3} into
four-component spinor notation is straightforward.
First, we consider a massive spin-1/2 fermion.
\Eqst{spinoneprime}{spinfourprime} yield~\cite{bailinweak}:
\beqa
\gamma\ls{5}\slashchar{S}^a\,u({\boldsymbol{\vec p}},s')
&=&
\BDpos \tau^a_{ss'}\,u({\boldsymbol{\vec p}},s)\,,\qquad\quad
\gamma\ls{5}\slashchar{S}^a\,v({\boldsymbol{\vec p}},s')
=
\BDpos \tau^a_{s's}\,v({\boldsymbol{\vec p}},s)\,,\label{dollar1}
\\
\,\ubar({\boldsymbol{\vec p}},s')\,\gamma\ls{5}\slashchar{S}^a
&=&
\BDpos \tau^a_{ss'}\,\ubar({\boldsymbol{\vec p}},s)\,,\qquad\quad
\vbar({\boldsymbol{\vec p}},s')\,\gamma\ls{5}\slashchar{S}^a
=
\BDpos \tau^a_{s's}\,\vbar({\boldsymbol{\vec p}},s)\,.\label{dollar2}
\eeqa
In the case of $a=3$, \eqs{dollar1}{dollar2} reduce to those of
\eqs{fourspineq1}{fourspineq2}.

The  four-component Bouchiat-Michel
formulae~\cite{bouchmich,ssiheh,vega} can be obtained
from \eqst{xxdagmassivehel}{ydagxdagmassivehel2}:
\beqa
u({\boldsymbol{\vec p}},s')\ubar({\boldsymbol{\vec p}},s)
&=&\half\left[\delta_{ss'}
\BDplus \gamma\ls{5}\gamma\ls{\mu}\mathscr{S}^\mu_{ss'}\right]
(\BDpos \slashchar{p}+m)\,,\label{bouchiat41} \\
v({\boldsymbol{\vec p}},s')\vbar({\boldsymbol{\vec p}},s)
&=&\half\left[\delta_{s's}
\BDplus \gamma\ls{5}\gamma\ls{\mu}\mathscr{S}^\mu_{s's}\right]
(\BDpos \slashchar{p}-m)\,,\label{bouchiat42}
\eeqa
where $\mathscr{S}^\mu_{ss'}\equiv S^{a\mu}\tau^a_{ss'}$.  As expected,
the above results for $s=s'$
correspond to the spin projection operators given in
\eqs{projectionops1}{projectionops2}.  Related formulae involving
products of $u$ and $v$-spinors can be obtained by using
[cf.~\eq{uvrelat}]:
\beq \label{app:uvrelat}
v({\boldsymbol{\vec p}},s)=-2s\gamma\ls{5}u({\boldsymbol{\vec p}},-s)\,,
\qquad\qquad
u({\boldsymbol{\vec p}},s)=2s\gamma\ls{5}v({\boldsymbol{\vec p}},-s)\,.
\eeq

\Eqst{dollar1}{bouchiat42} also apply to helicity $u$ and
$v$-spinors, after replacing $s,s'$ with $\lambda$,~$\lambda^\prime$
and using the $S^a$ as defined in \eq{sa3}.
The four-component versions of
\eqst{pminus1}{pminus4} yield:
\beqa
u(-{\boldsymbol{p}}\,,\,-\lambda)&=&\xi\ls{\lambda}
(\boldsymbol{\hat p})\,\gamma^0\,
u({\boldsymbol{p}}\,,\,\lambda)\,,
\qquad\qquad\quad\,\,
v(-{\boldsymbol{p}}\,,\,-\lambda)=\xi\ls{-\lambda}
(\boldsymbol{\hat p})\,\gamma^0\,
v({\boldsymbol{p}}\,,\,\lambda)\,,\label{uminus}
\\
\ubar(-{\boldsymbol{p}}\,,\,-\lambda)&=&
-\ubar({\boldsymbol{p}}\,,\,\lambda)\,
\gamma^0\,\xi\ls{-\lambda}(\boldsymbol{\hat p})\,,\qquad\qquad
\vbar(-{\boldsymbol{p}}\,,\,-\lambda)= -\vbar({\boldsymbol{p}}\,,\,\lambda)
\,\gamma^0\,\xi\ls{\lambda}(\boldsymbol{\hat p})\,,\label{ubarminus}
\eeqa
where the phase $\xi\ls{\lambda}(\boldsymbol{\hat p})$ was defined in
\eq{xip}.
In order to consider the massless limit, one must employ helicity
spinors, as discussed in \app{H.3}.  For $a=1,2$,
\eqs{dollar1}{dollar2} apply in the $m\to 0$ limit as written.
The corresponding massless limit for the case of $a=3$ is smooth and
results in \eq{mzerospinors}.  Similarly, the massless
limit of the Bouchiat-Michel formulae for helicity
spinors can be obtained by  setting
$mS^{a\mu}=p^\mu\,\delta^{a3}$, applying the mass-shell condition
($p^2= \BDpos m^2$), and taking the $m\to 0$ limit at the
end of the computation.  The end result is given by
\beqa
u(p,\lambda^\prime)\ubar(p,\lambda) &=&
\BDpos \half
(1+2\lambda\gamma\ls{5})\,\slashchar{p}\,
\delta_{\lambda\lambda^\prime}
+\half\gamma\ls{5}[\slashchar{S}^1\tau^1_{\lambda\lambda^\prime}
+\slashchar{S}^2\tau^2_{\lambda\lambda^\prime}]\,\slashchar{p}
\,, \label{mzerobm1}\\
v(p,\lambda^\prime)\vbar(p,\lambda) &=&
\BDpos \half
(1-2\lambda\gamma\ls{5})\,\slashchar{p}\,\delta_{\lambda^\prime\lambda}
+\half\gamma\ls{5}[\slashchar{S}^1\tau^1_{\lambda^\prime\lambda}
+\slashchar{S}^2\tau^2_{\lambda^\prime\lambda}]\,\slashchar{p}\,.
\label{mzerobm2}
\eeqa
As expected, when $\lambda=\lambda^\prime$, we recover the helicity
projection operators for massless spin-1/2 particles given in
\eqs{masslessprojection1}{masslessprojection2}.

As before, we can use the Bouchiat-Michel formulae to evaluate
helicity amplitudes involving two equal mass spin-1/2 fermions.  A
typical amplitude involving a fermion-antifermion pair, evaluated in
the center-of-mass frame of the pair, has the generic structure:
\beq \label{wform}
\overline{w}({\boldsymbol{\vec{p}}},\lambda)\,\Gamma\,w^\prime
(-{\boldsymbol{\vec{p}}},\lambda^\prime)\,,
\eeq
where $w$ is either a $u$ or $v$ spinor,  $w^\prime$ is respectively
either a $v$ or $u$ spinor, and $\Gamma$ is a product of
Dirac gamma matrices. For example,
\beq
\ubar({\boldsymbol{\vec{p}}},\lambda)\,\Gamma\,v
(-{\boldsymbol{\vec{p}}},\lambda^\prime)=-2\lambda^\prime
\ubar({\boldsymbol{\vec{p}}},\lambda)\,\Gamma\,\gamma\ls{5}\,
u(-{\boldsymbol{\vec{p}}},-\lambda^\prime)
=-2\lambda^\prime\,\xi\ls{\lambda^\prime}(\boldsymbol{\hat p})\,
\ubar({\boldsymbol{\vec{p}}},\lambda)\,\Gamma\,\gamma\ls{5}\,\gamma^0\,
u({\boldsymbol{\vec{p}}},\lambda^\prime)\,,
\eeq
where we have used the results of \eqs{app:uvrelat}{uminus}.
We can now
employ the Bouchiat-Michel formula
to convert the above result into a trace.
By a similar computation, all expressions of the form of \eq{wform}
can be expressed as a trace:
\beqa
\ubar({\boldsymbol{\vec{p}}},\lambda)\,\Gamma\,v
(-{\boldsymbol{\vec{p}}},\lambda^\prime)&=&
-\lambda^\prime\,\xi\ls{\lambda^\prime}(\boldsymbol{\hat p})\,
{\rm Tr}\left[\Gamma\gamma\ls{5}\gamma^0
(\delta_{\lambda\lambda^\prime}
\BDplus \gamma\ls{5}\gamma\ls{\mu}\mathscr{S}^\mu_{\lambda\lambda^\prime})
(\BDpos \slashchar{p}+m)\right],\label{helampu1}\\
\vbar({\boldsymbol{\vec{p}}},\lambda)\,\Gamma\,u
(-{\boldsymbol{\vec{p}}},\lambda^\prime)&=&
\lambda^\prime\,\xi\ls{-\lambda^\prime}(\boldsymbol{\hat p})\,
{\rm Tr}\left[\Gamma\gamma\ls{5}\gamma^0
(\delta_{\lambda^\prime\lambda}
\BDplus \gamma\ls{5}\gamma\ls{\mu}\mathscr{S}^\mu_{\lambda^\prime\lambda})
(\BDpos \slashchar{p}-m)\right].\label{helampu2}
\eeqa
These results are the four-component analogues of \eqst{helamps1}{helamps4}
and \eqst{helampp1}{helampp4}, respectively.
For amplitudes that involve a pair of equal mass fermions
[or equal mass antifermions], $w$ and $w^\prime$
in \eq{wform} are both $u$-spinors [or $v$-spinors].  Using
\eq{uvrelat},
these amplitudes can then be evaluated using the results of
\eqs{helampu1}{helampu2} above.

In the massless limit, one can again put $mS^{a\mu}=p^\mu\delta^{a3}$,
set $p^2= \BDpos m^2$ and take $m\to 0$ at the end of the computation.
Alternatively, one can repeat the derivation of \eqst{helampu1}{helampu2}
using the results of \eqs{mzerobm1}{mzerobm2}.
For completeness, we record the end result here.
\beqa
\ubar({\boldsymbol{\vec{p}}},\lambda)\,\Gamma\,v
(-{\boldsymbol{\vec{p}}},\lambda^\prime)&=&
\xi\ls{\lambda^\prime}(\boldsymbol{\hat p})
\biggl\{
\BDpos \half\delta_{\lambda\lambda^\prime}\,\Tr[\Gamma\gamma^0
(1+2\lambda\gamma_5)\slashchar{p}]+\lambda^\prime\,\Tr[\Gamma\gamma^0
(\slashchar{S}^1\tau^1_{\lambda\lambda^\prime}
+\slashchar{S}^2\tau^2_{\lambda\lambda^\prime})\,\slashchar{p}]\biggr\},
\nonumber \\
&&\phantom{line}\label{mzerohelampu1}\\
\vbar({\boldsymbol{\vec{p}}},\lambda)\,\Gamma\,u
(-{\boldsymbol{\vec{p}}},\lambda^\prime)&=&
\xi\ls{-\lambda^\prime}(\boldsymbol{\hat p})\biggl\{
\BDpos \half\delta_{\lambda^\prime\lambda}\,\Tr[\Gamma\gamma^0
(1-2\lambda\gamma_5)\slashchar{p}]-\lambda^\prime\,\Tr[\Gamma\gamma^0
(\slashchar{S}^1\tau^1_{\lambda^\prime\lambda}
+\slashchar{S}^2\tau^2_{\lambda^\prime\lambda})\,\slashchar{p}]\biggr\}.
\nonumber \\
&&\phantom{line}
\label{mzerohelampu2}
\eeqa

As an example, we consider once again the decay $Z^0\to f\fbar$.
The decay amplitude is equal to \eq{helampu1}, where $\Gamma$ is a
linear combination of $\half\gamma^\mu(1-\gamma\ls{5})$ and
$\half\gamma^\mu(1+\gamma\ls{5})$.  Evaluating the corresponding traces
yields:
\beqa
\ubar({\boldsymbol{\vec{p}}},\lambda)\,\half\gamma^\mu(1-\gamma\ls{5})\,v
(-{\boldsymbol{\vec{p}}},\lambda^\prime)&=&
2\lambda^\prime\,\xi\ls{\lambda^\prime}(\boldsymbol{\hat p})\,
\left[\BDpos mg^{\mu 0}\delta_{\lambda\lambda^\prime}
+ p^\mu \mathscr{S}^0_{\lambda\lambda^\prime}
- p^0 \mathscr{S}^\mu_{\lambda\lambda^\prime}
+ i\epsilon^{0\mu\nu\rho}(\mathscr{S}_{\lambda\lambda^\prime})_\nu
p_\rho\right]\,,
\nonumber \\
&&\phantom{line}\label{zff4a} \\
\ubar({\boldsymbol{\vec{p}}},\lambda)\,\half\gamma^\mu(1+\gamma\ls{5})\,v
(-{\boldsymbol{\vec{p}}},\lambda^\prime)&=&
2\lambda^\prime\,\xi\ls{\lambda^\prime}(\boldsymbol{\hat p})\,
\left[\BDneg mg^{\mu 0}\delta_{\lambda\lambda^\prime}
+ p^\mu \mathscr{S}^0_{\lambda\lambda^\prime}
- p^0 \mathscr{S}^\mu_{\lambda\lambda^\prime}
- i\epsilon^{0\mu\nu\rho}(\mathscr{S}_{\lambda\lambda^\prime})_\nu
p_\rho\right]\,.
\nonumber \\
&&\phantom{line}\label{zff4b}
\eeqa
Using \eq{scomm}, we see that \eqs{zff4a}{zff4b} reproduce exactly the
results of \eqs{zffhel1}{zffhel2}, respectively.

Finally, we note that if the two fermions do not
have the same mass, then the method presented above is not applicable.
However, generalizations of the above method exist in the literature
that can be employed to evaluate helicity amplitudes of the form of
\eq{wform} for unequal mass fermions; see, e.g., \Ref{vega}.
An alternative approach due to Hagiwara and
Zeppenfeld~\cite{Hagiwara:1985yu} is reviewed in \app{I.1}.

\section{\texorpdfstring{Helicity amplitudes and the spinor helicity method}{Helicity amplitudes and the spinor helicity method}}
\renewcommand{\theequation}{I.\arabic{equation}}
\renewcommand{\thefigure}{I.\arabic{figure}}
\renewcommand{\thetable}{I.\arabic{table}}
\setcounter{equation}{0}
\setcounter{figure}{0}
\setcounter{table}{0}

In \app{H}, we showed how to use the Bouchiat-Michel formulae (with
versions applicable to both two-component and four-component spinor
wave functions) to construct helicity amplitudes for processes with
two initial state and two final state equal mass fermions (or a
fermion-antifermion pair) in the center-of-mass frame of the two
fermions.  For practical applications, it is important to extend these
techniques to allow for final states with an arbitrary number of
particles.  The techniques should be powerful enough to allow for
pairs of fermions of unequal mass, and both massless and massive
spin-1 particles.  Ideally, these techniques should produce simple
analytic results (when possible) and yield efficient numerical
algorithms for the evaluation of the helicity amplitudes.

\subsection{The helicity amplitude technique of Hagiwara and Zeppenfeld}
\renewcommand{\theequation}{I.1.\arabic{equation}}
\renewcommand{\thefigure}{I.1.\arabic{figure}}
\renewcommand{\thetable}{I.1.\arabic{table}}
\setcounter{equation}{0}
\setcounter{figure}{0}
\setcounter{table}{0}

\indent

One method for computing helicity amplitudes for multi-particle final
states that is easily amenable to numerical analysis
was developed by Hagiwara and Zeppenfeld (HZ) \cite{Hagiwara:1985yu}.
The HZ formalism was subsequently employed in \refs{Hagiwara:1988pp}{helas}
in developing a fast numerical algorithm for the
computation of multi-parton processes.
In this section, we demonstrate how our two-component formalism
(denoted by DHM)
can be connected to theirs. In particular, we present a translation
between the two formalisms in Table~\ref{DHM-HZ}.

\begin{table}[t!]
\begin{center}
\begin{tabular}{|c|c|}\hline
DHM Formalism & HZ Formalism  \\ \hline
$x_\alpha(p,\lam)$ & $u(p,\lam)_-$ \\
$x^\alpha(p,\lam)$ & $v^\dagger(p,\lam)_+$ \\
$ x^{\dagger\dot{\alpha}}(p,\lam)$ & $v(p,\lam)_+$ \\
$ x^\dagger_{\dot{\alpha}}(p,\lam)$ & $u^\dagger(p,\lam)_-$ \\  \hline
$y_\alpha(p,\lam)$ & $v(p,\lam)_-$ \\
$y^\alpha(p,\lam)$ & $u^\dagger(p,\lam)_+$ \\
$ y^{\dagger\dot{\alpha}}(p,\lam)$ & $u(p,\lam)_+$ \\
$ y^\dagger_{\dot{\alpha}}(p,\lam)$ & $v^\dagger(p,\lam)_-$ \\ \hline
$p\newcdot\sigma$ & $(\slashchar{p})_+$ \\
$p\newcdot\sigmabar$ & $(
\slashchar{p})_-$ \\
$\sigma^\mu$ & $\sigma^\mu_+$ \\
$\sigmabar^\mu$ & $\sigma^\mu_-$ \\
\hline
$P_R$ & $P_+$ \\
$P_L$ & $P_-$ \\
$\lambda=\pm\half$ & $\lambda=\pm 1$ \\
$\chi\ls{\lambda}(-{\boldsymbol{\hat z}})$ &
$-\chi\ls{\lambda}(-{\boldsymbol{\hat z}})$ \\
\hline
\end{tabular}
\end{center}
\caption{Translation between our notation (denoted by DHM)
and the notation of Hagiwara and Zeppenfeld
(HZ) \cite{Hagiwara:1985yu}. The sign convention governing the definition
of $v(p,\lambda)_{\pm}$ is opposite to that of HZ (cf.~footnote~\ref{fnc}).}
\label{DHM-HZ}
\end{table}

After removing the propagator factors, an arbitrary tree amplitude with
external fermions can be expressed in terms of a four-component fermion string
\begin{equation}
\Psibar_i P_\tau \slashchar{a}_1\slashchar{a}_2\ldots
\slashchar{a}_n \Psi_j\,,\qquad\qquad \tau=\pm 1\,,
\label{HZ1}
\end{equation}
where $\Psi_j$ is a four-component spinor wave
function $u(p_j,\lam_j)$ or $v(p_j,-\lam_j)$,\footnote{\label{fnc}%
HZ defines $v(p,\lambda)=C\ubar^{\T}(p,\lambda)$, where
$C= i\gamma^2\gamma^0$,
which differs by an
overall minus sign from
the conventions employed in this review [cf.~\eq{acmatrix}].  In this
section, we will modify the HZ results in order to be consistent with
our sign convention.}
and $P_{\pm}=\half (1\pm\gamma_5)$ are the chiral projection operators.
Furthermore, $\slashchar{a}_k\equiv \gamma_\mu a^\mu_k$ where
$a_k$ represents an arbitrary Lorentz four-vector, which
can be a four-momentum $p_k ^\mu$, a vector boson wave function
$\epsilon^\mu(p_k,\lam_k)$, an axial vector (e.g., $\epsilon^{\mu\nu
  \rho\sigma}p_{k\nu}p_{m\rho}p_{n\sigma}$) or another fermion string
with uncontracted Lorentz indices (e.g.,
$\Psibar_m\gamma^\mu \Psi_n$).

In order to rewrite the fermion string, eq.~(\ref{HZ1}), in terms of
two-component spinors, HZ decomposes the four-component spinors as
follows:
\begin{equation}
\Psi_{j}\equiv\left(\begin{array}{c}(\psi_j)_-
\\ (\psi_j)_+ \end{array}\right)\,,\qquad u(p_j,\lam_j)\equiv
\left(\begin{array}{c}u(p_j,\lam_j)_-
\\ u(p_j,\lam_j)_+ \end{array}\right)\,, \qquad
v(p_j,\lam_j)\equiv\left(\begin{array}{c}
v(p_j,\lam_j)_- \\  v(p_j,\lam_j)_+\end{array}\right)\,.
\label{HZ3}
\end{equation}
Comparing with \eqs{uspin4}{vspin4},
the corresponding expressions in our notation are given in
Table~\ref{DHM-HZ}.   Note that $\lam=\pm1$ in the notation of HZ,
whereas in our notation (which we follow below) $\lam=
\pm\frac{1}{2}$.

The four-component fermion string is then replaced by the two-component
fermion string:
\beq \label{HZ4}
\Psibar_i P_\tau \slashchar{a}_1\slashchar{a}_2\ldots
\slashchar{a}_n
\Psi_j=(\psi_i)^\dagger_\tau
(\slashchar{a}_1)_\tau
(\slashchar{a}_2)_{-\tau} \ldots (\slashchar{a}_n)_{-\delta_n\tau}
(\psi_j)_{-\delta_n\tau}\,,\qquad \tau=\pm 1\,,\quad 
\text{where}~~\delta_n\equiv (-1)^n\,.
\eeq
In the notation of HZ,
\begin{equation}
(\slashchar{a})_\pm=a_\mu\sigma^\mu_\pm\,,
\label{HZ6}
\end{equation}
where $\sigma^\mu_+\equiv\sigma^\mu$ and
$\sigma^\mu_-\equiv\sigmabar^\mu$.  In \eq{HZ4}, the helicity labels
are suppressed; more explicitly,
\beq \label{hzconvention}
(\psi_k)_\tau\equiv \psi_k(p_k\,,\,\lambda_k)_\tau=
u(p_k\,,\,\lambda_k)_\tau\quad {\rm or} \quad v(p_k\,,\,-\lambda_k)_\tau\,.
\eeq
This convention of HZ (note the $-\lambda_k$ argument of $v$)
allows one to write simple generic formulae
in terms of $(\psi)_\pm$ that are
applicable to both $u_\pm$ and $v_\pm$.

Using the results of Table~\ref{DHM-HZ}, one can verify that
\eq{HZ4} is covariant with respect to dotted and undotted indices.
That is, the sign $\tau$ of $\psi_i^\dagger$ must match the sign of
the first $\sigma$-matrix in the string $(\slashchar{a}_1)_\tau
(\slashchar{a}_2)_{-\tau} \ldots (\slashchar{a}_n)_{-\delta_n\tau}$.
The signs of the sigma matrices within this string
alternate (either $+-+-\ldots$ or $-+-+\ldots$ in the case of
$\tau=+1$ or $-1$, respectively).  Finally, the sign of
the last $\sigma$-matrix in the string [which must be equal to
$-\delta_n\tau$ in light of the previous statement]
must match the sign of $\psi_j$ as indicated.

As noted above, it is possible that one of the $(\slashchar{a}_i)_\tau$
could be of the form $\sigma^\mu_\tau$ multiplied by another fermion
string with a free $\mu$-index.  One can uncouple the two fermion
strings by employing the Fierz identities given by
\eqst{twocompfierza}{twocompfierzc}.  For example,
\beqa \label{fierzex}
&&[(u_1)^\dagger_-(\slashchar{a}_1)_-\sigma^\mu_+(\slashchar{a}_2)_-
(u_2)_-]
[(u_3)^\dagger_-(\slashchar{a}_3)_-\sigma_{+\mu}(\slashchar{a}_4)_-
(u_4)_-] \nonumber \\[8pt]
&&\hspace{1in} =
\BDpos
[(u_1)^\dagger_-(\slashchar{a}_1)_-(\slashchar{a}_3)_+
(v_3)_+]
[(v_4)^\dagger_+(\slashchar{a}_4)_+(\slashchar{a}_2)_-
(u_2)_-]\,,
\eeqa
which is easily derived after translating to the DHM
notation.\footnote{Here, we
differ from HZ, who employ a Fierz identity that is not covariant with
respect to the dotted and undotted indices
[note the comment below \eq{eq:simplifysbarssbar}]. 
Thus, \eq{fierzex} differs from the result obtained in eq.~(3.17b) of HZ.}
As a result, the helicity tree amplitude for any process can be expressed
as the product of \textit{uncoupled} strings of two-component fermion spinors
\begin{equation}
\mathrm{FS}=(\psi_i)^\dagger_\tau
(\slashchar{a}_1)_\tau
(\slashchar{a}_2)_{-\tau} \ldots (\slashchar{a}_n)_{-\delta_n\tau}
(\psi_j)_{-\delta_n \tau}\,.
\label{HZ7}
\end{equation}

To evaluate the fermion string FS, we employ the explicit forms for
the two-component helicity spinor wave functions given in
\eqst{explicithelxa}{explicithelyb}, which can be rewritten as
\beq
\psi(p_k,\lambda_k)_\tau=C_k\,\omega\ls{\tau\lambda_k}
(\boldsymbol{\vec p})\,\chi\lsub{\lambda_k}(\boldsymbol{\hat p\ls{k}})\,,
\eeq
where, following the convention of \eq{hzconvention},
\begin{equation}
C_k=\left\{\begin{array}{ll}
1 & \mathrm{for}\quad (\psi_k)_\tau=u(p_k,\lam_k)_\tau\,, \\[2mm]
2\lam_k \tau & \mathrm{for}\quad (\psi_k)_\tau=v(p_k,-\lam_k)_\tau\,,
\end{array}\qquad\quad \tau=\pm1, \qquad \lam_k=\pm 1/2\,,
\right.
\end{equation}
and $\omega\ls\lambda(\boldsymbol{\vec p})\equiv
(E+2\lambda|{\boldsymbol{\vec p}}|)^{1/2}$ for $\lambda=\pm 1/2$.
Hence, the fermion string [\eq{HZ7}] is given by\cite{Hagiwara:1985yu}
\begin{equation} \label{HZ8}
\mathrm{FS}=C_i\, C_j\,\omega_{\tau\lam_i}(\boldsymbol{\vec p_i})\,\omega_
{-\delta_n\tau\lam_j}(\boldsymbol{\vec p_j})\,
S(p_i,a_1,a_2,\ldots,a_n,p_j)_{\lam_i\lam_j}^\tau\,,
\end{equation}
where the function $S$ is defined as
\beq \label{Sdef}
S(p_i,a_1,a_2,\ldots,a_n,p_j)_{\lam_i\lam_j}^\tau\equiv
\chi^\dagger\lsub{\lambda_i}(\boldsymbol{\hat p\ls{i}})
\left[\prod_{k=1}^{n}(\slashchar{a}_k)_{-\delta_k\tau}
\right]\chi\lsub{\lambda_j}(\boldsymbol{\hat p\ls{j}})\,,
\eeq
where $\delta_k\equiv (-1)^k$.
In the absence of the $\slashchar{a}_{\pm\tau}$ factors, we define
\beq \label{Tdef}
S(p_i\,,\,p_j)_{\lam_i\lam_j}\equiv
T({\boldsymbol{\hat p_i}}\,,\,{\boldsymbol{\hat p_j}})_{2\lam_i,2\lam_j}
= \chi^\dagger\lsub{\lambda_i}(\boldsymbol{\hat p\ls{i}})
\chi\lsub{\lambda_j}(\boldsymbol{\hat p\ls{j}})\,,
\eeq
where the
$T({\boldsymbol{\hat p_i}}\,,\,{\boldsymbol{\hat p_j}})_{2\lam_i,2\lam_j}$
are proportional to the (massless) spinorial products introduced
by Kleiss~\cite{Kleiss:1984dp} [cf.~\eqs{vevpq}{bracketspq}].

To evaluate $S$, we assume that the four-vectors $a_k^\mu$ are
real.\footnote{In the case of complex $a^\mu$, one should decompose
$a^\mu$ into its real and imaginary parts and evaluate separately the
real and imaginary parts of $S$.}
Then, we may employ the following identity:\footnote{To obtain
\eq{shashident}, we make use of \eq{chischis} applied to helicity
spinors: $\chi\ls\lambda\chi^\dagger\ls\lambda=\half(1+2\lambda
{\boldsymbol{\vec\sigma\newcdot\hat p}})$.}
\beq \label{shashident}
(\slashchar{a})_\tau=\sum_{\tau'=\pm}
\left[\BDpos a^0 \BDminus \tau^\prime\tau|{\boldsymbol{\vec a}}|\right]\chi\ls{\tau^\prime/2}
({\boldsymbol{\hat a}})\chi^\dagger\ls{\tau^\prime/2}
({\boldsymbol{\hat a}})\,,
\eeq
where $\chi\ls{\tau^\prime/2}({\boldsymbol{\hat a}})$ is a two-component
helicity spinor with three-momentum ${\boldsymbol{\vec a}}$.
Using \eq{shashident} in \eq{Sdef}, we end up with the desired
expression:
\begin{eqnarray}
S(p_i,a_1,a_2,\ldots,a_n,p_j)_{\lam_i\lam_j}^\tau
&=&\left[\prod_{k=1}^n\sum_{\tau_k=\pm}\bigl[
\BDpos a^0_k \BDplus {\tau_k
\delta_k\tau}|{\boldsymbol{\vec a_k}}|\bigr]\right]
T({\boldsymbol{\hat p}_i},{\boldsymbol{\hat a}_1})_{2\lam_i,\tau_1}
T({\boldsymbol{\hat a}_1},{\boldsymbol{\hat a}_2})_{\tau_1\tau_2}\qquad\;
\nonumber\\[2mm]
&&\;\; \hspace{0.5in}\times\,\cdots
T({\boldsymbol{\hat a}_{n-1}},{\boldsymbol{\hat a}_n})_{\tau_{n-1}\tau_n}
T({\boldsymbol{\hat a}_n},{\boldsymbol{\hat p}_j})_{\tau_n,2\lam_j}\,.
\label{masters}
\end{eqnarray}

All that remains is to evaluate the spinorial product
$T({\boldsymbol{\hat a}},{\boldsymbol{\hat b}})_{\tau_a\,\tau_b}$
($\tau_a$, $\tau_b=\pm 1$)
for arbitrary unit three-vectors ${\boldsymbol{\hat a}}$ and
${\boldsymbol{\hat b}}$.  Two properties of the spinorial product
$T({\boldsymbol{\hat a}}\,,\,{\boldsymbol{\hat b}})_{\tau_a,\tau_b}$
are noteworthy.  First, as this product is a scalar, it follows that
$T({\boldsymbol{\hat a}}\,,\,{\boldsymbol{\hat b}})_{\tau_a,\tau_b}^*=
T({\boldsymbol{\hat a}}\,,\,{\boldsymbol{\hat b}})_{\tau_a,\tau_b}^\dagger$.
Hence, \eq{Tdef} implies that
\beq \label{tproperties1}
T({\boldsymbol{\hat a}},{\boldsymbol{\hat b}})_{\tau_a\,\tau_b}
=T({\boldsymbol{\hat b}},{\boldsymbol{\hat a}})^*_{\tau_b\,\tau_a}\,.
\eeq
Second, we use \eq{chidefinition} to write:
\beqa \label{tabexplicit}
T({\boldsymbol{\hat a}},{\boldsymbol{\hat b}})_{\tau_a\,\tau_b}&=&
\chi^\dagger\ls{\tau_a/2}(\boldsymbol{\hat z})\,\exp(i\gamma_a
\sigma^{3}/2)
\,\exp(i\theta_a\sigma^{2}/2)\,\exp(i\phi_a
\sigma^{3}/2) \nonumber
\\
&&\qquad\quad \times
\exp(-i\phi_b\sigma^{3}/2)\,
\exp(-i\theta_b\sigma^{2}/2)\,\exp(-i\gamma_b\sigma^{3}/2)\,
\chi\ls{\tau_b/2}(\boldsymbol{\hat z})\,.
\eeqa
Complex conjugating this result, and using the fact that
$\chi(\boldsymbol{\hat z})$, $\sigma^{1}$ and $\sigma^{3}$
are real
and $\sigma^{2}$ is pure imaginary,
\beqa
\hspace{-0.3in}
T({\boldsymbol{\hat a}},{\boldsymbol{\hat b}})^\ast_{\tau_a\,\tau_b}&=&
\tau_a\tau_b\,
\chi^\dagger\ls{-\tau_a/2}(\boldsymbol{\hat z})
\,\sigma^{2}\,\exp(-i\gamma_a\sigma^{3}/2)
\,\exp(i\theta_a\sigma^{2}/2)\,\exp(-i\phi_a\sigma^{3}/2)
\nonumber \\
&&\qquad\quad \times
\exp(i\phi_b\sigma^{3}/2)\,
\exp(-i\theta_b\sigma^{2}/2)\,\exp(i\gamma_b\sigma^{3}/2)\,
\sigma^{2}\,\chi\ls{-\tau_b/2}(\boldsymbol{\hat z})\,,
\eeqa
after using \eq{twocomppropa}.  Since $\sigma^{2}$ anticommutes
with $\sigma^{3}$,
we end up with:
\beqa
\hspace{-0.3in}
T({\boldsymbol{\hat a}},{\boldsymbol{\hat b}})^\ast_{\tau_a\,\tau_b}&=&
\tau_a\tau_b\,
\chi^\dagger\ls{-\tau_a/2}(\boldsymbol{\hat
z})\,\exp(i\gamma_a\sigma^{3}/2)
\,\exp(i\theta_a\sigma^{2}/2)\,\exp(i\phi_a\sigma^{3}/2)
\nonumber
\\
&&\qquad\quad \times
\exp(-i\phi_b\sigma^{3}/2)\,
\exp(-i\theta_b\sigma^{2}/2)\,\exp(-i\gamma_b\sigma^{3}/2)\,
\chi\ls{-\tau_b/2}(\boldsymbol{\hat z})\nonumber \\
&=& \tau_a\tau_b\,
T({\boldsymbol{\hat a}},{\boldsymbol{\hat b}})_{-\tau_a\,,\,-\tau_b}\,.
\eeqa
Since $\tau_a$, $\tau_b=\pm 1$, it follows that
\beq \label{tproperties2}
T({\boldsymbol{\hat a}},{\boldsymbol{\hat b}})_{-\tau_a,-\tau_b}
= \tau_a\,\tau_b\,
T({\boldsymbol{\hat a}},{\boldsymbol{\hat b}})^*_{\tau_a\,\tau_b}\,.
\eeq
Using \eqs{tproperties1}{tproperties2},
it is sufficient to give explicit forms for only two of the
spinorial products~\cite{Kleiss:1985yh,Hagiwara:1985yu}.
\Eq{tabexplicit} yields:
\beqa
T({\boldsymbol{\hat a}},{\boldsymbol{\hat b}})_{++}
&=& e^{i(\phi_a-\phi_b+\gamma_a-\gamma_b)/2}\,
\cos\frac{\theta_a}{2}\cos\frac{\theta_b}{2}
+e^{-i(\phi_a-\phi_b-\gamma_a+\gamma_b)/2}\,
\sin\frac{\theta_a}{2}\sin\frac{\theta_b}{2}\,,\label{tplusplus}
\\[6pt]
T({\boldsymbol{\hat a}},{\boldsymbol{\hat b}})_{-+}
&=&e^{-i(\phi_a-\phi_b+\gamma_a+\gamma_b)/2}\,
\cos\frac{\theta_a}{2}\sin\frac{\theta_b}{2}
-e^{i(\phi_a-\phi_b-\gamma_a-\gamma_b)/2}\,
\sin\frac{\theta_a}{2}\cos\frac{\theta_b}{2}\,,\label{tminusplus}
\eeqa
where $(\theta_p,\phi_p)$ are the polar and azimuthal angles of
${\boldsymbol{\hat p}}$ (for ${\boldsymbol{\hat p}}={\boldsymbol{\hat a}}$
and ${\boldsymbol{\hat b}}$, respectively).  In the case where
${\boldsymbol{\hat a}}$ and/or ${\boldsymbol{\hat b}}$ are parallel
to the negative $z$-axis, we employ the convention of \eq{negativez} and
choose the corresponding azimuthal angle equal to $\pi$.\footnote{This
convention yields a value of $\chi\ls\lambda({\boldsymbol{-\hat z}})$ that
is opposite in sign to the convention adopted by HZ.}
Note that HZ employ a convention for their spinor wave functions
[cf.~\eq{chidefinition}] in which
$\gamma=-\phi$,  although the convention in which
$\gamma=0$ yields a slightly more symmetrical form for the spinorial products.

\Eqss{HZ7}{HZ8}{Sdef}
can be written in a form that is reminiscent of the results obtained in
\app{H.3}.   For example, using \eqst{explicithelxa}{explicithelyb},
\beqa
{x}^\dagger_{\dot{\alpha}}({\boldsymbol{\vec p}},\lambda)
\,\Gamma^{\dot{\alpha}\beta}\,
y_\beta({\boldsymbol{\vec p}^{\,\prime}},-\lambda^\prime)
&=&-2\lambda^\prime\omega_{-\lambda}({\boldsymbol{\vec p}})
\omega_{-\lambda^\prime}({\boldsymbol{\vec p}^{\,\prime}})
\chi^\dagger\ls\lambda
({\boldsymbol{\hat p}})
\Gamma
\chi\ls{\lambda^\prime}
({\boldsymbol{\hat p}^{\,\prime}})\,,
\label{hzhelamps1}
\\
y^\alpha({\boldsymbol{\vec p}},\lambda)
\,\Gamma_{\alpha\dot{\beta}}\,
{x}^{\dagger\dot{\beta}}({\boldsymbol{\vec p}^{\,\prime}},-\lambda^\prime)
&=&\phm 2\lambda^\prime\omega_{\lambda}({\boldsymbol{\vec p}})
\omega_{\lambda^\prime}({\boldsymbol{\vec p}^{\,\prime}})\chi^\dagger\ls\lambda
({\boldsymbol{\hat p}})
\Gamma
\chi\ls{\lambda^\prime}
({\boldsymbol{\hat p}^{\,\prime}})\,,
\label{hzhelamps2}
\\
y^\alpha({\boldsymbol{\vec p}},\lambda)
\,\Gamma_{\alpha}{}^\beta\,
y_\beta({\boldsymbol{\vec p}^{\,\prime}},-\lambda^\prime)
&=&-2\lambda^\prime\omega_{\lambda}({\boldsymbol{\vec p}})
\omega_{-\lambda^\prime}({\boldsymbol{\vec p}^{\,\prime}})\chi^\dagger\ls\lambda
({\boldsymbol{\hat p}})
\Gamma
\chi\ls{\lambda^{\,\prime}}
({\boldsymbol{\hat p}^\prime})\,,
\label{hzhelamps3}
\\
{x}^\dagger_{\dot{\alpha}}({\boldsymbol{\vec p}},\lambda)
\,\Gamma^{\dot{\alpha}}{}_{\dot{\beta}}\,
{x}^{\dagger\dot{\beta}}({\boldsymbol{\vec p}^{\,\prime}},-\lambda^\prime)
&=&\phm 2\lambda^\prime\omega_{-\lambda}({\boldsymbol{\vec p}})
\omega_{\lambda^\prime}({\boldsymbol{\vec p}^{\,\prime}})\chi^\dagger\ls\lambda
({\boldsymbol{\hat p}})
\Gamma
\chi\ls{\lambda^\prime}
({\boldsymbol{\hat p}^{\,\prime}})\,,\label{hzhelamps4}
\eeqa
where $\Gamma$ is a product of alternating $\sigma$ and $\sigmabar$
matrices.  The spinor index structure determines the identity of
the first and last matrix (e.g.,
$\Gamma^{\dot{\alpha}}{}_{\dot{\beta}}$ indicates a string of matrices
that begins with a $\sigmabar$ and ends with a $\sigma$, etc.).  By
suitable interchanges of $x$ and $y$, twelve additional equations of
similar type may be written.  Note that
$\chi\ls{\lambda}^\dagger\Gamma\chi\ls{\lambda^\prime}$ [appearing on
the right-hand side of \eqst{hzhelamps1}{hzhelamps4}] corresponds
precisely to the
$S(p,a_1,a_2,\ldots,a_n,p^\prime)^\tau_{\lambda\lambda^\prime}$ of
\eq{Sdef}, where the four possible $(\tau\,,\,\delta_n)$ combinations
are in one-to-one correspondence with the four possible spinor index
structures of $\Gamma$.  If ${\boldsymbol{\vec
    p}^{\,\prime}}=-{\boldsymbol{\vec p}}$, then one should recover
\eqst{helamps1}{helamps4}.  Thus, the HZ method provides a powerful
generalization of the helicity amplitude methods derived in
\app{H.3}.

\subsection{The spinor helicity method}
\renewcommand{\theequation}{I.2.\arabic{equation}}
\renewcommand{\thefigure}{I.2.\arabic{figure}}
\renewcommand{\thetable}{I.2.\arabic{table}}
\setcounter{equation}{0}
\setcounter{figure}{0}
\setcounter{table}{0}

\indent

In many practical calculations, the masses of the fermions can be
neglected.  In this case the
computation of multi-particle helicity amplitudes simplifies considerably.
In this section, we give a brief introduction to the
spinor helicity method; for a review, see \refs{heltechs}{helicityreviews}.
The spinor helicity method
is a powerful technique for computing helicity
amplitudes for multi-particle processes involving massless spin-1/2 and
spin-1 particles.  Although initially applied to tree-level
processes, more general techniques have also been developed that are
applicable to one-loop (and multiloop) diagrams~\cite{zbern}.
Rules for computing dimensionally regularized amplitudes within the
framework of the spinor helicity method have been given by \Ref{Kosower:1990ax}.
The spinor helicity techniques are ideal for QCD where
light quark masses can almost always be neglected.  Generalizations of
these methods that incorporate massive spin-1/2 and spin-1
particles exist, although they tend to be quite
cumbersome~\cite{massivemethods,massivemethods2}.
A Mathematica implementation of the spinor helicity formalism can be
found in \Ref{sandm}.  In this section, we
restrict the discussion to the massless case.

The spinor helicity technique described below is based
on a formalism developed by Xu, Zhang and
Chang~\cite{xzc} (denoted henceforth by XZC),
which is a modification of techniques established by the CALKUL
collaboration~\cite{calkul}.
The XZC formalism (which was also independently developed in
\refs{Kleiss:1985yh}{Gunion:1985vca}) is based on
the four-component spinor formalism.  Using \eq{vequalu},
XZC introduce a very useful notation for massless spinors
\beqa
\ket{p\pm}\equiv u(p,\pm\half) &=& v(p, \mp\half)\,,\label{brackets4a}\\
\bra{p\pm}\equiv \ubar(p,\pm\half) &=& \vbar(p, \mp\half)\,.\label{brackets4b}
\eeqa
Using these spinor wave functions, they define two non-trivial (massless) spinor
products (which are equivalent to the spinorial products introduced by
Kleiss~\cite{Kleiss:1984dp}):\footnote{Note that $\vev{p-|q-}=\vev{p+|q+}=0$
due to $P_L P_R=P_R P_L=0$.}
\beqa \label{bracketnotation}
\vev{p\,q}\equiv\vev{p-|q+}&=&\ubar(p,-\half)\,u(q,+\half)\,,\\
\phantom{}[p\,q]\equiv\vev{p+|q-}&=&\ubar(p,+\half)\,u(q,-\half)\,.
\eeqa
The $\pm$ notation specified by the bra and ket indicates the
chirality (i.e. the eigenvalue of $\gamma\ls{5}$)
of the corresponding four-component
spinor [cf.~\eq{mzerospinors}].

However, the two-component spinor formalism is especially economical
in the case of massless spin-1/2 fermions.  Hence, we shall reformulate
the XZC approach using two-component spinor notation.  First, we
consider the explicit forms for the two-component helicity spinor wave
functions [given by \eqst{helexplicitxa}{helexplicityb}]
in the massless limit:
\beqa
\hspace{-0.4in}
x_\alpha(\boldsymbol{\vec p},-\half)
&=&y_\alpha(\boldsymbol{\vec p},\half)=(2E)^{1/2}\,
\chi\ls{-1/2}(\boldsymbol{\hat p})\,,
\qquad\!\!
x^\alpha(\boldsymbol{\vec p},-\half)
=y^\alpha(\boldsymbol{\vec p},\half)=(2E)^{1/2}\,
\chi^\dagger\ls{1/2}(\boldsymbol{\hat p})\,,
\label{explicithelm0} \\
\hspace{-0.4in}
 x^{\dagger\dot{\alpha}}(\boldsymbol{\vec p},-\half)&=&
 y^{\dagger\dot{\alpha}}(\boldsymbol{\vec p},\half)=(2E)^{1/2}\,
\chi\ls{1/2}(\boldsymbol{\hat p})\,,
\qquad\,
 x^\dagger_{\dot{\alpha}}(\boldsymbol{\vec p},-\half)=
 y^\dagger_{\dot{\alpha}}(\boldsymbol{\vec p},\half)=(2E)^{1/2}\,
\chi^\dagger\ls{-1/2}(\boldsymbol{\hat p})\,,
\label{explicithelm0bar}
\eeqa
where $E=|\boldsymbol{\vec p}|$.  For all other choices of helicities,
the corresponding helicity spinor wave functions vanish.  Hence, we
define:\footnote{\label{fnwarned}%
The association of undotted and dotted indices in
\eqs{pplus}{pminus} is a consequence of our convention for the Dirac gamma
matrices given in \app{G} [cf.~\eq{gamma4}].  Note that in this convention,
the left-handed [right-handed] projection operator
$P_L$ [$P_R$] projects out the
lowered undotted [raised dotted] index components
of the four-component spinor [cf.~\eq{general4comp}].  However,
the reader is warned that
in the literature on the spinor helicity method, one often
finds $\ket{p+}$ associated with a lowered undotted index and
$\ket{p-}$ associated with an upper dotted index.  This is
due to a different convention for the sigma matrices,
such as the Wess and Bagger definition given in \eqs{sigswap1}{sigswap2}.
Numerically, this is equivalent to a convention for
the Dirac gamma matrices in which
$\sigma^\mu$ and $\sigmabar^\mu$ are interchanged
in \eq{gamma4}, resulting in an overall change of sign in the
matrix representation of $\gamma\ls{5}$.  As a result, in this latter
convention the lowered undotted [raised dotted] index components
are associated with positive [negative]
chirality. For an historical perspective, see the discussion
following \eq{sigswap2}.}
\beqa
\ket{p+} &=&
{y}^{\dagger\dot{\alpha}}(\boldsymbol{\vec p},\half)
={x}^{\dagger\dot{\alpha}}(\boldsymbol{\vec
  p},-\half)\,,\qquad\qquad
\bra{p+} = y^{\alpha}(\boldsymbol{\vec p},\half)
=x^{\alpha}(\boldsymbol{\vec p},-\half)\,,\label{pplus}\\
\ket{p-} &=& x_{\alpha}(\boldsymbol{\vec p},-\half)
=y_{\alpha}(\boldsymbol{\vec p},\half)\,,
\qquad\qquad
\bra{p-} = {x}^\dagger_{\dot{\alpha}}(\boldsymbol{\vec p},-\half)
={y}^\dagger_{\dot{\alpha}}(\boldsymbol{\vec p},\half)\,.\label{pminus}
\eeqa

The $\ket{p\pm}$ and $\bra{p\pm}$ satisfy the massless Dirac equation
[cf.~\eqst{onshellone}{onshellfour}]:
\beq \label{xzcdirac}
p\newcdot\sigma_\pm\ket{p\pm}=0,\qquad\qquad
\bra{p\pm} p\newcdot\sigma_\pm=0\,,
\eeq
where
$\sigma_+\equiv\sigma$ and $\sigma_-\equiv\sigmabar$ as indicated in
Table~\ref{DHM-HZ}.
The above and the following equations should each be read as two separate
equations corresponding to the upper and lower set of signs, respectively.
The following properties are also noteworthy:
\beqa
\ket{p\pm}\bra{p\pm}&=&
  \BDpos p\newcdot\sigma_\mp\,,\label{xzcprop3} \\
\bra{p\pm}\sigma^\mu_\pm\ket{p\pm}&=&
2p^\mu\,,\label{xzcprop1} \\
\vev{p\pm|q\mp}&=&-\vev{q\pm|p\mp}\,,\label{xzcprop5} \\
\bra{p+}\sigma^\mu_+\ket{q+}&=&
\bra{q-}\sigma^\mu_-\ket{p-}\,,\label{xzcprop2} \\
\bra{p\pm}\sigma^\mu_\pm\sigma^\nu_\mp\ket{q\mp} &=&
-\bra{q\pm}\sigma^\nu_\pm\sigma^\mu_\mp\ket{p\mp}\,. \label{xzcprop4}
\eeqa
Note that \eqst{xzcdirac}{xzcprop4} are covariant with respect to the
undotted and dotted spinor indices.  \Eqs{xzcprop3}{xzcprop1} follow from
\eqs{xxdagmassless}{yydagmassless}.  For example,
\beq
\bra{p+}\sigma_+^\mu\ket{p+}=y^\alpha(\boldsymbol{\vec p},\half)
\sigma^\mu_{\alpha\dot{\beta}}\,
{y}^{\dagger\dot{\beta}}(\boldsymbol{\vec p},\half)=
\sigma^\mu_{\alpha\dot{\beta}}\,
{y}^{\dagger\dot{\beta}}(\boldsymbol{\vec p},\half)
y^\alpha(\boldsymbol{\vec p},\half)
=\,\BDpos \Tr\,(\sigma^\mu p\newcdot\sigmabar)=
2p^\mu\,,
\eeq
and similarly for $\bra{p-}\sigma_-^\mu\ket{p-}$.  \Eqst{xzcprop5}{xzcprop4}
follow immediately from \eqst{zonetwo}{zsbarmunuz}.
\Eqs{xzcprop2}{xzcprop4} generalize easily to the case of a product of
an even and odd number of $\sigma$/$\sigmabar$ matrices.  For any
positive integer $n$,
\beqa
\bra{p+}\sigma^{\mu_1}_{+}\sigma^{\mu_2}_{-}\cdots
\sigma^{\mu_{2n-1}}_{+}\ket{q+}&=&
\bra{q-}\sigma^{\mu_{2n-1}}_{-}\cdots\sigma^{\mu_2}_{+}
\sigma^{\mu_1}_{-}\ket{p-}\,,\label{xzcpropodd} \\
\bra{p\pm}\sigma^{\mu_1}_\pm\sigma^{\mu_2}_\mp\cdots\sigma^{\mu_{2n}}_\mp\
\ket{q\mp}
&=&
-\bra{q\pm}\sigma^{\mu_{2n}}_\pm\cdots
\sigma^{\mu_2}_\pm\sigma^{\mu_1}_\mp\ket{p\mp}\,. \label{xzcpropeven}
\eeqa

Spinor products can be formed from the bras and kets in the usual
way and satisfy:
\beqa
\vev{p\pm|q\mp}^*&=&\vev{q\mp|p\pm}\,,\label{spinorprodast1} \\
\vev{p\pm|\sigma^\mu_\pm|q\pm}^*&=& \vev{q\pm|\sigma^\mu_\pm|p\pm}\,,
\label{spinorprodast2}
\eeqa
where we have used the fact that the $\sigma_\pm^\mu$ are hermitian.
Covariance with respect to the undotted and dotted spinors
allows only two possible spinor products:\footnote{Since we wish
to preserve the definition of the spinor products given in
\eq{bracketnotation}, $\vev{p\,q}$ is a sum over dotted indices
and $[p\,q]$ is a sum over undotted indices
in our two-component spinor conventions.  This is to be contrasted
with most of the literature on the spinor helicity method,
in which  $\vev{p\,q}$ is written as a sum over undotted indices
and $[p\,q]$ as a sum over dotted indices.  The origin of this
difference is explained in footnote~\ref{fnwarned}.}
\beqa
\vev{p\,q}\equiv\vev{p-|q+}& =&{x}^\dagger(\boldsymbol{\vec p},-\half)\,
{y}^\dagger(\boldsymbol{\vec q},\half)\,,\\
\phantom{}[p\,q]
\equiv\vev{p+|q-} & =& y(\boldsymbol{\vec p},\half)\,
x(\boldsymbol{\vec q},-\half)\,.
\eeqa
In particular, the products $\vev{p+|q+}$ and $\vev{p-|q-}$ never
arise in a computation using two-component spinor notation.
In terms of the spinorial products defined in \eq{Tdef},
\beqa
\vev{p\,q}\equiv\vev{p-|q+}& =& (2E_p)^{1/2} (2E_q)^{1/2}
T(\boldsymbol{\hat p}\,,\,\boldsymbol{\hat q})_{-+}\,,\label{vevpq}\\
\phantom{}[p\,q]
\equiv\vev{p+|q-}& =& (2E_p)^{1/2} (2E_q)^{1/2}
T(\boldsymbol{\hat p}\,,\,\boldsymbol{\hat q})_{+-}\,,\label{bracketspq}
\eeqa
where $E_p=|\boldsymbol{\vec p}|$ and $E_q=|\boldsymbol{\vec q}|$.
Explicit forms for $T(\boldsymbol{\hat p}\,,\,\boldsymbol{\hat q})_{-+}$
and  $T(\boldsymbol{\hat p}\,,\,\boldsymbol{\hat q})_{+-}=
-T(\boldsymbol{\hat p}\,,\,\boldsymbol{\hat q})^*_{-+}$ can be
obtained from \eq{tminusplus}.
Using \eqs{tproperties1}{tproperties2} [or equivalently, using
\eqs{xzcprop5}{spinorprodast1}],
the spinor products satisfy the following relations:
\beqa
\vev{p\,q} &=& -\vev{q\,p}\,, \label{spinprodrela}\\
\phantom{}[p\,q] &=& -[q\,p]\,, \label{spinprodrelb}\\
\vev{p\,q}^* &=& -
\phantom{}[p\,q]\,. \label{spinprodrelc}
\eeqa
 One immediate consequence of the
above results is:
\beqa
\vev{p\,p}&=&\vev{p-|p+}=0\,,\label{vevpp}\\
\phantom{}[p\,p] &=& \vev{p+|p-}=0\,.\label{bracketpp}
\eeqa
We next compute the absolute square of the spinor product:
\beqa
|\vev{p\,q}|^2 &=& {x}^\dagger_{\dot{\alpha}}(\boldsymbol{\vec p},-\half)\,
\,{y}^{\dagger\dot{\alpha}}(\boldsymbol{\vec q},\half)
\,x_\alpha(\boldsymbol{\vec p},-\half)\,
y^\alpha(\boldsymbol{\vec q},\half) =
x_\alpha(\boldsymbol{\vec p},-\half)\,
{x}^\dagger_{\dot{\alpha}}(\boldsymbol{\vec p},-\half)\,
{y}^{\dagger\dot{\alpha}}(\boldsymbol{\vec q},\half)\,
y^\alpha(\boldsymbol{\vec q},\half) \nonumber \\
&=& p\newcdot\sigma_{\alpha\dot{\alpha}}\,
q\newcdot\sigmabar^{\dot{\alpha}\alpha}=
p_\mu q_\nu\,
\Tr(\sigma^\mu\sigmabar^\nu)= \BDpos 2p\newcdot q\,.
\eeqa
Using this result and \eq{spinprodrelc} yields
\beq \label{spinsquared}
|\vev{p\,q}|^2 =
|[p\,q]|^2
= \BDpos 2p\newcdot q\,,
\eeq
which indicates that the spinor products are roughly the square roots
of the corresponding dot products.
One other noteworthy relation is:
\beqa \label{tridentity}
\vev{p_1\,p_2}[p_2\,p_3]\vev{p_3\,p_4}[p_4\,p_1] &=&
\Tr\left(\sigma\newcdot p_1\,\sigmabar\newcdot p_2\,\sigma\newcdot p_3\,
\sigmabar\newcdot p_4\right) \nonumber \\
&=& 2(g_{\mu\nu} g_{\rho\kappa}-g_{\mu\rho} g_{\nu\kappa}+
g_{\mu\kappa} g_{\nu\rho}+i\epsilon_{\mu\nu\rho\kappa})
p_1^\mu p_2^\nu p_3^\rho p_4^\kappa\,,
\eeqa
where the trace has been evaluated using \eq{APPtrssbarssbar}.
Note that the first line of
\eq{tridentity} immediately follows from \eqs{xxdagmassless}{yydagmassless}
after plugging in the definition of the spinor products.

In \app{I.1}, we showed that a
fermion string can be expressed in terms of products of the spinorial
products $T$ [cf.~\eq{masters}].  When applied to massless spinors,
\eq{spinsquared} indicates that the square of the helicity amplitude
of a multi-fermion scattering process can be expressed in terms of
products of dot products of pairs of fermion momenta.  If more than
one diagram contributes to a helicity amplitude, then it is often
possible to combine the contributions after a rearrangement of momenta
via the Fierz identities.  Using \eqst{eq:twocompfierzone}{twocompfierzc},
it follows that:
\beqa
\label{Schouten1}
&& \vev{p_1\,p_2}\vev{p_3\,p_4}=
\vev{p_1\,p_3}\vev{p_2\,p_4}
+
\vev{p_1\,p_4}\vev{p_3\,p_2}
\,,\\
\label{Schouten2}
&&\,\,\phantom{}[p_1\,p_2]\,[p_3\,p_4]\,=\,
\phantom{}
[p_1\,p_3]\,[p_2\,p_4]
\,+\,
[p_1\,p_4]\,[p_3\,p_2]
\,,\\
&&\vev{p_1+|\sigma^\mu_+|p_2+}
\vev{p_3+|\sigma_{+\mu}|p_4+}=
\BDpos 2\,[p_1\,p_3]\,\vev{p_4\,p_2}\,,\\
&&\vev{p_1-|\sigma^\mu_-|\,p_2-}
\vev{p_3-|\sigma_{-\mu}|p_4-}=
\BDpos 2\vev{p_1\,p_3}\,[p_4\,p_2]\,,\\
&&\vev{p_1+|\sigma^\mu_+|p_2+}
\vev{p_3-|\sigma_{-\mu}|p_4-}=
\BDpos 2\,[p_1\,p_4]\,\vev{p_3\,p_2}\,.
\eeqa
\Eqs{Schouten1}{Schouten2} are often called
the Schouten identities, as they follow from \eq{schouten}.

It is desirable to extend the spinor helicity formalism to
multi-particle processes
involving massless fermions and massless spin-1 bosons.  In
particular, XZC developed a simple technique for expressing
the squares of the corresponding helicity amplitudes
in terms of ratios of products of dot
products.  Their trick was to introduce
a convenient expression for the massless spin-1
polarization vector in terms of products of massless spin-1/2
spinor wave functions.
Before exhibiting their result,  we provide a brief review of spin-1
polarization vectors in the helicity basis.

We first consider
a massless spin-1 particle moving in the $z$-direction with
four-momentum $k^\mu=E(1\,;\,0\,,\,0\,,\,1)$.
The textbook expression for the helicity
$\pm 1$ polarization vectors of a
massless spin-1 boson is given by~\cite{auvil,Carruthers,leader,pilkuhn}:
\beq \label{varepsz}
\varepsilon^\mu(\boldsymbol{\hat z},\pm 1)=\frac{1}{\sqrt{2}}
\left(0\,;\,\mp 1\,,\,-i\,,\,0\right)\,.
\eeq
Note that the $\varepsilon^\mu(\boldsymbol{\hat z},\lambda)$ are normalized
eigenvectors of the spin-1 operator $\boldsymbol{\vec{\mathcal{S}}\newcdot\hat z}$,
\beq
(\boldsymbol{\vec{\mathcal{S}}\newcdot\hat z})^\mu{}_\nu\,
\varepsilon^\nu(\boldsymbol{\hat z},\lambda)=\lambda\,
\varepsilon^\mu(\boldsymbol{\hat z},\lambda)\,,
\eeq
where $\mathcal{S}^i\equiv \half\epsilon^{ijk}\mathcal{S}_{jk}$,
and the matrix elements of the $4\times 4$
matrices $\mathcal{S}_{jk}$ are given by \eq{explicitsmunu}.

If we transform $\varepsilon^\mu(\boldsymbol{\hat z},\lambda)$ by
employing a three-dimensional rotation $\mathcal{R}$ such that
$\boldsymbol{\hat k}=\mathcal{R}\,\boldsymbol{\hat z}$, then we can obtain
the polarization vector for
a massless spin-1 boson of energy $E$ moving in the direction
$\boldsymbol{\hat k}=(\sin\theta\cos\phi\,,\,\sin\theta\sin\phi\,,\,\cos\theta)$.
That is,
\beq \label{rotateone}
\varepsilon^\mu(\boldsymbol{\hat k},\lambda)=\Lambda^\mu{}_\nu(\phi\,,\theta\,,\,\gamma)\,
\varepsilon^\nu(\boldsymbol{\hat z},\lambda)\,,
\eeq
where
\beq \label{lamrij}
\Lambda^0{}_0=1\,,\quad\quad \Lambda^i{}_0=\Lambda^0{}_i=0\,,\quad {\rm and} \quad
\Lambda^i{}_j=\mathcal{R}^{ij}(\phi\,,\theta\,,\,\gamma)\,,
\eeq
and $\mathcal{R}(\phi\,,\theta\,,\,\gamma)$ is the rotation matrix introduced
in \eq{calrdef}.  A simple computation yields:
\beq \label{varepsk}
\varepsilon^\mu(\boldsymbol{\hat k},\pm 1)=\frac{1}{\sqrt{2}}
\,e^{\mp i\gamma}\left(0\,;\,\mp\cos\theta\cos\phi+i\sin\phi\,,\,
\mp\cos\theta\sin\phi-i\cos\phi\,,\,\pm\sin\theta\right)\,.
\eeq
Note that $\varepsilon^\mu(\boldsymbol{\hat k},\pm 1)$
depends only on the direction of $\boldsymbol{\vec k}$ and not on its
magnitude $E=|\boldsymbol{\vec k}|$.  One can easily check that
the $\varepsilon^\mu(\boldsymbol{\hat k},\pm 1)$ are normalized
eigenstates of $\boldsymbol{\vec{\mathcal{S}}\newcdot\hat k}$ with corresponding
eigenvalues $\pm 1$.

Similar to the corresponding discussion in \app{C} for the spin-1/2
spinor wave functions, the Euler angle $\gamma$ is arbitrary.
In the literature, one typically finds conventions where
$\gamma=-\phi$~\cite{Cbook,auvil,Carruthers} or $\gamma=0$~\cite{leader}, and
we will consider both possibilities below.

Although we will not need it here,
the expressions given by \eqs{varepsz}{varepsk}
also apply in the case of a massive spin-1 particle.  In
addition, there is a helicity $\lambda=0$ polarization vector
which depends on the magnitude of the momentum as well as its direction:
\beq
\varepsilon^\mu(\boldsymbol{|\boldsymbol{\vec{k}}|\boldsymbol{\hat z}}\,,\, 0)=
(|\boldsymbol{\vec{k}}|/m\,;\,0\,,\,0\,,\,E/m)\,,
\eeq
where
$E=(|\boldsymbol{\vec{k}}|^2+m^2)^{1/2}$.  One can
use \eq{rotateone} to obtain the helicity zero polarization vector
for a massive spin-1 particle moving in an arbitrary direction
\beq \label{massivespinone}
\varepsilon^\mu(\boldsymbol{\vec{k}}\,,\, 0)=
\frac{1}{m}\left(|\boldsymbol{\vec k}|\,;\,E\sin\theta\cos\phi\,,\,
E\sin\theta\sin\phi\,,\,E\cos\theta\right)\,.
\eeq
Note that both the massless and massive spin-1 polarization vectors
satisfy:\footnote{\label{fnylm}%
Some authors introduce polarization vectors where the
sign factor $(-1)^\lambda$ in \eq{epsilonphase} is omitted.
One motivation for \eq{epsilonphase} is to maintain
consistency with the Condon-Shortley phase conventions~\cite{edmonds}
for the eigenfunctions of the spin angular momentum operators
$\boldsymbol{\vec S^2}$ and $S_z$ (for spin-1 particles).
In particular, note the relation
$\boldsymbol{\hat r\newcdot\hat\varepsilon}^\mu
(\boldsymbol{\hat z},\pm 1)=(4\pi/3)^{1/2}Y_{1,\pm 1}(\theta,\phi)$
between the polarization three-vector and
the $\ell=1$ spherical harmonics without any additional sign factors.}
\beq \label{epsilonphase}
\varepsilon^\mu(\boldsymbol{\vec{k}}\,,\,\lambda)^* = (-1)^\lambda
\varepsilon^\mu(\boldsymbol{\vec{k}}\,,\,-\lambda)\,.
\eeq
One can check that the $\varepsilon^\mu(\boldsymbol{\vec{k}},\lambda)$
also satisfies the standard conditions for a valid
polarization four-vector:
\beq \label{spinoneprops}
k\newcdot\varepsilon(\boldsymbol{\vec{k}}\,,\,\lambda)=0\,,\,\qquad\qquad
\varepsilon(\boldsymbol{\vec{k}}\,,\,\lambda)\newcdot\varepsilon(\boldsymbol{\vec{k}}\,,\,\lambda^\prime)^*
= \BDneg \delta_{\lambda\lambda^\prime}\,.
\eeq

If the spin-1 boson three-momentum is $-\boldsymbol{\vec k}$, then its
polarization vector can be obtained from \eqs{varepsk}{massivespinone}
by taking $\theta\to\pi-\theta$ and $\phi\to\phi+\pi$.   It can also
be derived from \eqs{rotateone}{lamrij} by making use of the spin-1 analogue
of \eq{dgamgam},
\beq
\mathcal{R}(\phi+\pi\,,\,\pi-\theta\,,\,\gamma(-\boldsymbol{\hat k}))=
\mathcal{R}(\phi\,,\,\theta\,,\,\gamma(\boldsymbol{\hat k}))
\,R(\boldsymbol{\hat z},-\gamma(\boldsymbol{\hat k})
-\gamma(-\boldsymbol{\hat k}))\,R(\boldsymbol{\hat x},\pi)\,,
\eeq
where we have exhibited the possible dependence of $\gamma$ on the
direction of $\boldsymbol{\hat k}$, and $R$ is the rotation matrix
given by \eq{rij}.  Introducing the notation
$\varepsilon^\mu\equiv(\varepsilon^0\,;\,\boldsymbol{\hat\varepsilon})$,
and noting the relations:
\beqa
R(\boldsymbol{\hat x},\pi)\boldsymbol{\hat\varepsilon}(\boldsymbol{\hat z},
\lambda)&=&-\boldsymbol{\hat\varepsilon}(\boldsymbol{\hat z},
-\lambda)\,,\\
R(\boldsymbol{\hat z},\beta)\boldsymbol{\hat\varepsilon}(\boldsymbol{\hat z},
\lambda)&=&e^{-i\lambda\beta}\,\boldsymbol{\hat\varepsilon}(\boldsymbol{\hat z},
\lambda)\,,
\eeqa
it follows that:
\beq \label{minuskspinone}
\varepsilon^\mu(\boldsymbol{-\vec{k}}\,,\, \lambda)=
\BDneg g_{\mu\mu}
\,\xi\ls{-\lambda}(\boldsymbol{\hat k})\,
\varepsilon^\mu(\boldsymbol{\vec{k}}\,,\, -\lambda)\,,
\qquad \lambda=0\,,\pm1\,,
\eeq
where there is no sum over the repeated index $\mu$, and
\beq \label{xipspinone}
\xi_\lambda(\boldsymbol{\hat k}) = -e^{i\lambda[\gamma(\boldsymbol{\hat k})
+\gamma(-\boldsymbol{\hat k})]}\,,\qquad \lambda=0,\pm 1\,.
\eeq
Note that for
$\lambda=\pm 1$, the phase factor $\xi\ls{\lambda}(\boldsymbol{\hat k})$
depends on the convention for the definition of the Euler angle $\gamma$
used to define the spin-1 polarization vector.
As an example, corresponding to the
two conventional choices for $\gamma$,
\beq \label{xispinone}
\xi\ls{\lambda}(\boldsymbol{\hat k})=\begin{cases}
(-1)^{1-\lambda} e^{-2i\lambda\phi} & \textrm{for}~~
\gamma(\boldsymbol{\hat k})=-\phi\,,\quad
\gamma(-\boldsymbol{\hat k})=-\phi-\pi\,,\\
\qquad -1 & \textrm{for}~~\gamma(\boldsymbol{\hat k})=
\gamma(-\boldsymbol{\hat k})=0\,.\end{cases}
\eeq

To motivate the XZC form for the massless spin-1 polarization vectors,
we first introduce a four-vector
\beq
\widetilde k^\mu \equiv  E(1\,;\,-\boldsymbol{\hat k})\,,
\eeq
corresponding to the four-momentum $k^\mu= E(1\,;\,\boldsymbol{\hat k})$
of the massless spin-1 boson.  A straightforward calculation then shows that
\beq \label{kbarspinone}
\varepsilon^\mu(k\,,\pm 1) = \frac{1}{\sqrt{2}}\,
 \frac{\bigl\langle k\mp\left|\sigma^\mu_\mp\right|\widetilde k\mp\bigr\rangle}
       {\bigl\langle k\pm\bigl|\widetilde k\mp\bigr\rangle}
\eeq
precisely reproduces the result of \eq{varepsk}, where the massless
spinor wave functions are defined according to \eq{chidefinition}.
\Eq{kbarspinone} is somewhat inconvenient because the four-vector $\widetilde k$
cannot be covariantly defined in terms of $k$.  XZC finessed this
problem by introducing a ``reference'' four-vector $p$ (in practical
computations, $p$ is
taken to be another four-momentum vector in the scattering process of
interest), with the properties that $p^2=0$ and $p\newcdot k\neq 0$.
The XZC spin-1 polarization vectors are given
by [cf.~\eq{spinorvector}]~\footnote{In the literature on the
spinor helicity method,
the spin-1 polarization vector $\varepsilon$ is employed
in Feynman diagram computations for
an \textit{outgoing} final state boson, in contrast to
the standard conventions of most quantum field theory textbooks.
In this review, we subscribe to the latter [as indicated at
the end of \sec{externalrules}].  Hence,
to be consistent with our conventions above for the spin-1
polarization vector, we have taken the complex conjugate of
the original definition of the XZC spin-1 polarization vectors.
In addition, we have removed an overall $\pm$ sign in order to conform
to \eq{epsilonphase} [cf.~footnote \ref{fnylm}].}
\beq \label{xzcspinone}
\hspace{0.5in}
\varepsilon^\mu(k\,,\pm1) =  \frac{1}{\sqrt{2}}\,
     \frac{\bra{k\mp}\sigma^\mu_\mp \ket{p\mp}}
       {\vev{k\pm|p\mp} }\,.
\eeq
One can immediately check that $\varepsilon^\mu(k,\lambda)$
so defined
satisfy the standard conditions for a valid
polarization four-vector given in \eq{spinoneprops} and
the phase convention of \eq{epsilonphase}.
The representation of the massless spin-1 polarization vector
in terms of spinor products [\eqs{kbarspinone}{xzcspinone}] is an
application of the spinor calculus that was first developed by
van der Waerden~\cite{vdWaerden1}.

The significance of the reference four-vector $p$ can be discerned
from the property that if a different reference momentum is chosen then
$\varepsilon^\mu$ is shifted by a factor proportional to $k^\mu$.
Explicitly, if $\varepsilon^\mu(k\,,\,p\,,\,\lambda)$ is a polarization vector
with reference momentum $p$, then\footnote{To derive \eq{ekq},
evaluate $\varepsilon^\mu(k\,,\,q\,,\,\lambda)-
\varepsilon^\mu(k\,,\,p\,,\,\lambda)$, and simplify the
resulting expression using \eqss{xzcprop3}{xzcprop2}{xzcprop4}.}
\beq \label{ekq}
\varepsilon^\mu(k\,,\,q\,,\,\pm 1)=\varepsilon^\mu(k\,,\,p\,,\,\pm 1)+
\frac{\sqrt{2}\vev{q\pm|p\mp}}{\vev{k\pm|q\mp}\vev{k\pm|p\mp}}\,k^\mu\,.
\eeq
In particular, if we choose $q=\widetilde k$, we see that the
difference of the XZC
spin-1 polarization vector and the polarization vector
given by \eq{varepsk} is proportional to $k^\mu$.
This shift of the reference momentum from $p$ to $q$
in the XZC definition of the polarization vector does
not affect \eq{spinoneprops} since $k^2=0$ for massless
spin-1 particles.  Moreover, this shift does not affect
the final result for any observable (in particular the sum of
amplitudes of any gauge invariant set of Feynman diagrams
remains unchanged).  Thus, the
presence of the arbitrary four-vector $p$ just reflects the gauge
invariance of the theory of massless spin-1 particles.

We can also verify that $\varepsilon^\mu(k\,,\,p\,,\,\lambda)$
defined in \eq{xzcspinone} behaves as expected under rotations.
Using \eq{chidefinition}, massless spinors transform as:
\beq
\ket{k\pm}\longrightarrow \mathcal{D}(\phi,\theta,\gamma)
\ket{k\pm}\,,\qquad\quad
\bra{k\pm}\longrightarrow \bra{k\pm}[\mathcal{D}(\phi,\theta,\gamma)]^{-1}\,,
\eeq
under a rotation specified by the Euler angles $\phi$, $\theta$ and $\gamma$.
We shall rotate the spin-1 polarization vectors by rotating both
$\boldsymbol{\vec k}$ and the reference momentum $\boldsymbol{\vec p}$
simultaneously (since one is always free to shift the reference vector
with no physical consequence).  Using \eq{dsigd}, it follows that:
\beq
[\mathcal{D}(\phi,\theta,\gamma)]^{-1}\,\sigma_\pm^\mu
\,\mathcal{D}(\phi,\theta,\gamma)
=\Lambda^\mu{}_\nu\sigma^\nu_\pm\,,
\eeq
where $\Lambda^\mu{}_\nu$ is specified by \eq{lamrij}.  Indeed,
if we simultaneously rotate both $k$ and $p$ via $k^\mu\to
\Lambda^\mu{}_\nu k^\nu$ and $p^\mu\to
\Lambda^\mu{}_\nu p^\nu$, then
\beq
\varepsilon^\mu(k\,,\,p\,,\,\lambda)\longrightarrow \Lambda^\mu{}_\nu\,
\varepsilon^\nu(k\,,\,p\,,\,\lambda)\,,
\eeq
as expected.  By a similar computation, one can check that
under $\boldsymbol{\vec k}\to -\boldsymbol{\vec k}$ and
$\boldsymbol{\vec p}\to -\boldsymbol{\vec p}$, \eq{minuskspinone}
is satisfied.\footnote{Here, we have used \eqs{sig3a}{sig3b} to write
$\sigma^0_\pm\sigma^\mu_\mp\sigma^0_\pm
=
-\sigma^\mu_\pm \BDplus 2g^{\mu 0}\sigma^0_\pm
= \BDpos g^{\mu\mu}\sigma^\mu_\pm$ (no sum over~$\mu$).}
In terms of the $\xi_{\pm 1/2}$ defined in \eq{xip},
we find
\beq
\xi_\lambda(\boldsymbol{\vec k})=-\left(\frac{\xi_{-\lambda/2}(\boldsymbol{\vec k})}
{\xi_{\lambda/2}(\boldsymbol{\vec k})}\right)^* =
\left[\xi\ls{\lambda/2}(\boldsymbol{\vec k})\right]^2
\,,\qquad \lambda=\pm 1\,,
\eeq
which agrees with \eq{xipspinone}.

The following additional properties of
$\varepsilon^\mu(k\,,\,p\,,\,\lambda)$ defined in \eq{xzcspinone} are noteworthy:
\beqa
p\newcdot \varepsilon(k\,,\,p\,,\,\lambda) &=& 0\,, \label{epstwo} \\
\sum_{\lambda=\pm 1} \varepsilon_\mu(k\,,\,p\,,\,\lambda)
\varepsilon_\nu(k\,,\,p\,,\,\lambda)^\ast
     & =&
\BDneg g_{\mu \nu} \BDplus \frac{p_\mu k_\nu + p_\nu k_\mu}{p\newcdot k}
\,.\label{epsthree}
\eeqa
For example, to prove \eq{epsthree}, we use
\eqs{spinorprodast1}{spinorprodast2}, and simplify the resulting
expression with the help of \eqs{xzcprop3}{xzcpropodd}, which yields:
\beq
\sum_{\lambda=\pm 1} \varepsilon_\mu(k\,,\,p\,,\,\lambda)
\varepsilon_\nu(k\,,\,p\,,\,\lambda)^\ast
     =\frac{\bra{k+}(\sigma_\mu p\newcdot\sigmabar\sigma_\nu+
\sigma_\nu p\newcdot\sigmabar\sigma_\mu)\ket{k+}}{2\bra{k+}p\newcdot\sigma
\ket{k+}}\,.
\eeq
Using \eq{sig3b} to simplify the product of three
$\sigma/\sigmabar$ matrices, and employing \eq{xzcprop1} then
yields \eq{epsthree}.

Finally, using the Fierz identities given in \eqst{Fierz1}{Fierz3},
one derives from \eq{xzcspinone} that
\beqa
\sigma_\pm\newcdot{\varepsilon}(k\,,\,\pm 1) = \BDpos \frac{\sqrt{2}\ket{p\mp}
\bra{k\,\mp}}{\vev{k\pm|p\mp}}\,,
 \qquad && \qquad
\sigma_\pm\newcdot{\varepsilon}(k\,,\,\pm 1)^\ast = \BDpos \frac{\sqrt{2}\ket{k\mp}
\bra{p\mp}}{\vev{p\mp|k\pm}}\,,
\label{epsfour} \\
\sigma_\mp\newcdot{\varepsilon}(k\,,\,\pm 1) = \BDpos
\frac{\sqrt{2}\ket{k\pm}\bra{p\pm}}{\vev{k\pm|p\mp}}\,,
\qquad && \qquad
\sigma_\mp\newcdot{\varepsilon}(k\,,\,\pm 1)^\ast = \BDpos
\frac{\sqrt{2}\ket{p\pm}\bra{k\pm}}{\vev{p\mp|k\pm}}
\,. \label{epsfive}
\eeqa
Note that each equation in \eqs{epsfour}{epsfive}
represents two separate equations, corresponding
to the upper and lower signs in each equation, respectively.

It should now be clear how to convert the square of a helicity amplitude
for a multi-particle process involving massless spin-1/2 and massless
spin-1 particles into a ratio of products of dot products of momenta.  By
writing all massless spin-1 polarization vectors in the form of
\eq{xzcspinone} and using the properties given above, the helicity
amplitudes can easily be expressed as a ratio of two quantities, each
of which is a product of spinor products.  Squaring the corresponding
amplitude then yields a ratio of products of dot products of
four-momenta.  A following simple example will demonstrate the technique.

Consider Compton scattering in QED, $e^-(\boldsymbol{\vec p}_1,\lambda_1)
\gamma(\boldsymbol{\vec k}_1,\lambda_1^\prime)\rightarrow
e^-(\boldsymbol{\vec p}_2,\lambda_2)
\gamma(\boldsymbol{\vec k}_2,\lambda_2^\prime)$, in the limit
of massless electrons.  The amplitude for this process is given
by \eq{gfscatter} with $m=0$ and $G_L=G_R=-e$.  Writing out the ``crossed''
term explicitly, and noting that for massless particles,
$s\equiv \BDpos (p_1+k_1)^2 = \BDpos 2p_1\newcdot k_1$ and
$u\equiv
\BDpos (p_1-k_2)^2
=
\BDneg 2p_1\newcdot k_2$,
\beqa
i\mathcal{M} &=& \frac{- ie^2}{2p_1\newcdot k_1}\biggl\{
 x^\dagger(
\boldsymbol{\vec p}_2,\lambda_2)\,
\sigmabar\newcdot\varepsilon\ls{2}^*\,\sigma\newcdot (p_1+k_1)\,
\sigmabar\newcdot\varepsilon\ls{1}\, x(\boldsymbol{\vec p}_1,\lambda_1)
+y(\boldsymbol{\vec p}_2,\lambda_2)\,\sigma\newcdot\varepsilon\ls{2}^*\,
\sigmabar\newcdot (p_1+k_1)
\,\sigma\newcdot\varepsilon\ls{1} y^\dagger(\boldsymbol{\vec p}_1,\lambda_1)
\biggr\}\nonumber \\
&& \hspace{-0.1in}+ \frac{ie^2}{2p_1\newcdot k_2}\biggl\{
 x^\dagger(
\boldsymbol{\vec p}_2,\lambda_2)\,
\sigmabar\newcdot\varepsilon\ls{1}\,\sigma\newcdot (p_1-k_2)\,
\sigmabar\newcdot\varepsilon\ls{2}^*\, x(\boldsymbol{\vec p}_1,\lambda_1)
+y(\boldsymbol{\vec p}_2,\lambda_2)\,\sigma\newcdot\varepsilon\ls{1}\,
\sigmabar\newcdot (p_1-k_2)
\,\sigma\newcdot\varepsilon\ls{2}^* y^\dagger(\boldsymbol{\vec p}_1,\lambda_1)
\biggr\}\,.\nonumber \\
&&\phantom{line}
\eeqa
The results of \eqst{helexplicitxa}{helexplicityb} imply that
the helicity amplitudes with $\lambda_1\neq\lambda_2$ vanish.
Using \eqs{pplus}{pminus}, we identify:
\beqa
i\mathcal{M}(\lambda_1=\lambda_2=\half) &=&
\frac{-ie^2}{2p_1\newcdot k_1}
\vev{p_2+|\sigma_+\newcdot\varepsilon_2^*\,\sigma_-\newcdot(p_1+k_1)\,\sigma_+
\newcdot\varepsilon_1|p_1+}\nonumber \\
&&\qquad +\frac{ie^2}{2p_1\newcdot k_2}
\vev{p_2+|\sigma_+\newcdot\varepsilon_1\,\sigma_-\newcdot(p_1-k_2)\,\sigma_+
\newcdot\varepsilon_2^*|p_1+}\,,\label{comptons}
\eeqa
\beqa
i\mathcal{M}(\lambda_1=\lambda_2=-\half) &=&
\frac{-ie^2}{2p_1\newcdot k_1}
\vev{p_2-|\sigma_-\newcdot\varepsilon_2^*\,\sigma_+\newcdot(p_1+k_1)\,\sigma_-
\newcdot\varepsilon_1|p_1-}\nonumber \\
&&\qquad +\frac{ie^2}{2p_1\newcdot k_2}
\vev{p_2-|\sigma_-\newcdot\varepsilon_1\,\sigma_+\newcdot(p_1-k_2)\,\sigma_-
\newcdot\varepsilon_2^*|p_1-}\,.\label{comptonu}
\eeqa
Further simplification ensues when we apply the results of
\eqs{epsfour}{epsfive}.  To use these results, we must select a
reference momentum $p$, which can be any lightlike four-vector that is
not parallel to the corresponding photon polarization vector.   One is free
to choose a different reference momentum for each photon polarization
vector.  Moreover, when computing two different helicity amplitudes
(each of which are gauge invariant quantities), one may select a
different reference momentum for the \textit{same} photon polarization
vector in the two computations.  The decision on which reference
momenta to choose is somewhat of an art; experience will teach you
which choices lead to the most simplification in a given calculation.

We shall consider two possible choices for the reference momenta for
$\varepsilon_1$ and $\varepsilon_2$, which we denote as
$p^{(1)}$ and $p^{(2)}$, respectively:
\begin{enumerate}
\item
$p^{(1)}=p_1$ and $p^{(2)}=p_2$\,,
\item
$p^{(1)}=p_2$ and $p^{(2)}=p_1$\,.
\end{enumerate}
With either choice, it is straightforward to show that
$\mathcal{M}(\lambda_1=\lambda_2=\pm\half)$ vanish unless the photon
helicities are equal, i.e. $\lambda_1^\prime=\lambda_2^\prime$.  This
leaves only four possible non-vanishing helicity amplitudes.

For the case of $\lambda_1=\lambda_2=\pm\half$ and
$\lambda^\prime_1=\lambda^\prime_2=\pm 1$
(i.e., $\lambda_1\lambda^\prime_1>0$),
we choose reference momenta $p^{(1)}=p_2$ and $p^{(2)}=p_1$.
Then, the second term vanishes
on the right-hand side of \eqs{comptons}{comptonu}, respectively.
Making use of \eqss{xzcprop3}{epsfour}{epsfive}, we find
\beqa
 i\mathcal{M}(\lambda_1=\lambda_2
=\half\,,\,\lambda^\prime_1=\lambda^\prime_2=1)
 &=& \frac{\BDneg ie^2}{p_1\newcdot k_1}
\frac{\vev{p_2+|k_2-}\,\vev{p_1-|k_1+}\,\vev{k_1+|p_2-}\,\vev{k_1-|p_1+}}
{\vev{p_1-|k_2+}\,\vev{k_1+|p_2-}} \nonumber \\
&=& \frac{\BDneg ie^2}{p_1\newcdot k_1}
\frac{\vev{p_1\,k_1}\vev{k_1\,p_1}\,[p_2\,k_2]}
{\vev{p_1\,k_2}}\,.
\eeqa
Using \eqs{spinprodrela}{spinsquared} to write the dot product in terms of
spinor products, we obtain:
\beq \label{compton1}
i\mathcal{M}(\lambda_1=
\lambda_2=\half\,,\,\lambda^\prime_1=\lambda^\prime_2=1)
 = 2i e^2\frac{\vev{p_1\,k_1}}{\vev{p_1\,k_1}^*}\,
\frac{[p_2\,k_2]}{\vev{p_1\,k_2}}\,.
\eeq
A similar computation yields
\beq \label{compton2}
 i\mathcal{M}(\lambda_1=\lambda_2=
-\half\,,\,\lambda^\prime_1=\lambda^\prime_2=-1)
 = 2i e^2\frac{[p_1\,k_1]}{[p_1\,k_1]^*}\,
\frac{\vev{p_2\,k_2}}{[p_1\,k_2]}\,.
 \eeq

For the case of $\lambda_1=\lambda_2=\pm\half$ and
$\lambda^\prime_1=\lambda^\prime_2=\mp 1$ (i.e.,
$\lambda_1\lambda^\prime_1<0$), we choose reference momenta
$p^{(1)}=p_1$ and $p^{(2)}=p_2$.  Then, the first term vanishes
on the right-hand side of \eqs{comptons}{comptonu}, respectively.
A similar calculation to the one given above yields:
\beqa
i\mathcal{M}(\lambda_1=\lambda_2
=-\half\,,\,\lambda^\prime_1=\lambda^\prime_2=1)
&=& 2i
e^2\frac{[p_1\,k_2]}{[p_1\,k_2]^*}\,\frac{\vev{p_2\,k_1}}{[p_1\,k_1]}\,,
 \label{compton3} \\[8pt]
i\mathcal{M}(\lambda_1=
\lambda_2=\half\,,\,\lambda^\prime_1=\lambda^\prime_2=-1)
&=& 2i e^2\frac{\vev{p_1\,k_2}}{\vev{p_1\,k_2}^*}\,
\frac{[p_2\,k_1]}{\vev{p_1\,k_1}}\,. \label{compton4}
\eeqa

Note that each pair of helicity amplitudes above is simply related:
\beq \label{parityinv}
\mathcal{M}_{\lambda_1\,,\,\lambda_1^\prime\,;\,\lambda_2\,,\,
 \lambda^\prime_2}(s,\theta,\phi)^\ast
=\mathcal{M}_{-\lambda_1\,,\,-\lambda_1^\prime\,;\,-\lambda_2\,,\,
 -\lambda^\prime_2}(s,\theta,\phi)\,,
 \eeq
which is a consequence of rotational and parity invariance
(as shown below).   Thus in this example,
we only need to evaluate two non-zero helicity amplitudes.
It is clear that we have simplified the computation enormously by our
choice of reference momenta.  With a less judicious choice, the
calculation is significantly more tedious, although gauge invariance
guarantees that one must arrive at the same result for the helicity
amplitudes quoted above.

One can easily evaluate the spinor products
above in the center-of-mass system.  Writing $p_1^\mu =
E(1\,;\,\boldsymbol{\hat z})$,  $k_1^\mu =
E(1\,;\,-\boldsymbol{\hat z})$, $p_2^\mu=
E(1\,;\,\boldsymbol{\hat p}_{\rm CM})$ and $k_2^\mu=
E(1\,;\,-\boldsymbol{\hat p}_{\rm CM})$, and using
the results of \eqs{tminusplus}{vevpq}, we obtain:
\beqa
\hspace{-0.1in}
\vev{p_1\,k_1}&=&2E\xi_{-1/2}(\boldsymbol{\hat z})\,,\qquad\qquad\quad
 \vev{p_1\,k_2}=2E e^{i[\phi+\gamma(\boldsymbol{\hat p}_{\rm CM})]/2}
\xi_{-1/2}(\boldsymbol{\hat p}_{\rm CM})\cos(\theta/2)\,,
\label{sprods1}\\
\hspace{-0.1in}
\vev{p_2\,k_2}&=&2E\xi_{-1/2}(\boldsymbol{\hat p}_{\rm CM})\,,\qquad\quad\,\,\,\,
\vev{p_2\,k_1}=2E e^{-i[\phi+\gamma(\boldsymbol{\hat p}_{\rm CM})]/2}
\xi_{-1/2}(\boldsymbol{\hat z})\cos(\theta/2)\,,\label{sprods2}
\eeqa
where $\theta$ and $\phi$ are the polar and azimuthal angles of
$\boldsymbol{\hat p}_{\rm CM}$.  Phase factors involving
$\xi_{-1/2}$ arise from the use of \eqs{cphase}{xip}.  For example,
corresponding to the two conventional choices for $\gamma$,
we use \eq{xphase} to obtain
\beqa
\xi_{-1/2}(\boldsymbol{\hat z})&=&\begin{cases} -1 & \quad\textrm{for}
~\gamma(\boldsymbol{\hat z})=0\,,\quad
\gamma(-\boldsymbol{\hat z})=-\pi\,,\\
\phm i& \quad\textrm{for}~\gamma(\boldsymbol{\hat z})=
\gamma(-\boldsymbol{\hat z})=0\,,\end{cases}
\\[8pt]
\xi_{-1/2}(\boldsymbol{\hat p}_{\rm CM})&=&\begin{cases} -e^{i\phi} &
\,\,\,\textrm{for}~\gamma(\boldsymbol{\hat p}_{\rm CM})=-\phi\,,
\quad ~\gamma(-\boldsymbol{\hat p}_{\rm CM})=-\phi-\pi\\
\phm i& \,\,\,\textrm{for}~\gamma(\boldsymbol{\hat p}_{\rm CM})=
\gamma(-\boldsymbol{\hat p}_{\rm CM})=0\,.\end{cases}
\eeqa
All other relevant spinor products can be found using
\eqst{spinprodrela}{spinprodrelc}.

It is always possible
to define the plane of the scattering process to be the $x$--$z$ plane,
in which case $\phi=0$ and all the spinor products in \eqs{sprods1}{sprods2}
are manifestly real.  Nevertheless, by keeping the explicit $\phi$-dependence,
one maintains a useful check of the calculation.
Inserting the explicit forms for the spinor products into
\eqst{compton1}{compton4}, we confirm that the $\phi$-dependence
of the helicity amplitudes is given by~\cite{barut,leader}:
\beq \label{phiphase}
\mathcal{M}_{\lambda_1\,,\,\lambda_1^\prime\,;\,
\lambda_2\,,\,\lambda_2^\prime}(s,\theta,\phi)
=\begin{cases}
e^{i(\lambda_1-\lambda_1^\prime-\lambda_2-\lambda_2^\prime)\phi}
\mathcal{M}_{\lambda_1\,,\,\lambda_1^\prime\,;\,\lambda_2\,,\,
\lambda_2^\prime}(s,\theta)\,, & \textrm{for}~
\gamma(\boldsymbol{\hat p}_{\rm CM})=-\phi\,,{\rm and}\\[-2pt]
\phantom{e^{i(\lambda_1-\lambda_1^\prime-\lambda_2-\lambda_2^\prime)\phi}
\mathcal{M}_{\lambda_1\,,\,\lambda_1^\prime\,;\,\lambda_2\,,\,
\lambda_2^\prime}(s,\theta)\,,} & \phantom{\textrm{for}}~\gamma
(-\boldsymbol{\hat p}_{\rm CM})=-\phi-\pi\,,\\[6pt]
e^{i(\lambda_1-\lambda_1^\prime)\phi}
\mathcal{M}_{\lambda_1\,,\,\lambda_1^\prime\,;\,\lambda_2\,,\,
\lambda_2^\prime}(s,\theta)\,, & \textrm{for}~
\gamma(\boldsymbol{\hat p}_{\rm CM})=\gamma(-\boldsymbol{\hat p}_{\rm CM})=0
\,,\end{cases}
\eeq
as a consequence of rotational invariance~\cite{jacobwick}.\footnote{In the
first case, where $\gamma(\boldsymbol{\hat p}_{\rm CM})=-\phi$
and $\gamma(-\boldsymbol{\hat p}_{\rm CM})=-\phi-\pi$,
the sign of $\lambda_2^\prime$ in the $\phi$-dependent
phase factor of \eq{phiphase} is opposite to the one given in
\Ref{jacobwick}, due to the Jacob-Wick second-particle
convention, which we do not employ here.  Since $\lambda_1=\lambda_2$ and
$\lambda_1^\prime=\lambda_2^\prime$, the latter would imply that the $\phi$-dependent
phase cancels exactly if the Jacob-Wick second-particle convention is used.
This is easily checked by putting $\gamma(\boldsymbol{\hat p}_{\rm CM})=-\phi$
and $\xi_\lambda=1$ in \eqs{sprods1}{sprods2}, in which case
all the spinor products are real.}
The remaining $\theta$-dependent amplitudes are easily evaluated and
are in agreement with the results
of \refs{leader}{anselmino}.
Note that parity invariance implies that \eqst{compton1}{compton4}
must satisfy~\cite{jacobwick,leader,anselmino}
\beq \label{mparity}
 \mathcal{M}_{\lambda_1\,,\,\lambda_1^\prime\,;\,\lambda_2\,,\,
 \lambda^\prime_2}(s,\theta)=
\mathcal{M}_{-\lambda_1\,,\,-\lambda_1^\prime\,;\,-\lambda_2\,,\,
 -\lambda^\prime_2}(s,\theta)\,.
\eeq
Indeed, in our computation above, \eq{parityinv} is satisfied,
which is consistent with
\eq{mparity} in light of \eq{phiphase}.

To compute the unpolarized cross-section for Compton scattering, one
must sum the absolute squares of the helicity amplitudes and divide by
4 to average over the initial helicities.  Since quantities such as
$\vev{p_1\,k_1}/\vev{p_1\,k_1}^*$ are pure phases, one immediately obtains:
\beqa
\hspace{-0.3in} |\mathcal{M}(\lambda_1=\lambda_2
=\half\,,\,\lambda^\prime_1=\lambda^\prime_2=1)|^2
 =|\mathcal{M}(\lambda_1=\lambda_2=
-\half\,,\,\lambda^\prime_1=\lambda^\prime_2=-1)|^2
 &=& 4e^4\frac{p_1\newcdot k_1}{p_1\newcdot k_2}\,,\\[8pt]
\hspace{-0.3in}
|\mathcal{M}(\lambda_1=\lambda_2=-\half\,,\,\lambda^\prime_1
=\lambda^\prime_2=1)|^2
=|\mathcal{M}(\lambda_1=\lambda_2=\half\,,\,\lambda^\prime_1
=\lambda^\prime_2=-1)|^2
&=& 4e^4\frac{p_1\newcdot k_2}{p_1\newcdot k_1}\,, \eeqa after
employing \eq{spinsquared} and noting that $p_1\newcdot
k_1=p_2\newcdot k_2$ and $p_1\newcdot k_2=p_2\newcdot k_1$
(which follow from four-momentum conservation,
$p_1+k_1=p_2+k_2$, for the scattering of massless particles).  Thus,
\beq
\quarter\sum_{\rm spins}\,|\mathcal{M}|^2
=2e^4\left(\frac{p_1\newcdot k_1}{p_1\newcdot k_2}
+\frac{p_1\newcdot k_2}{p_1\newcdot k_1}\right)\,,
\eeq
which coincides with the well-known result quoted in \Ref{Peskin:1995ev}.

\section{\texorpdfstring{The Standard Model and its seesaw extension}{The Standard Model and its seesaw extension}}
\label{app:J}
\renewcommand{\theequation}{J.\arabic{equation}}
\renewcommand{\thefigure}{J.\arabic{figure}}
\renewcommand{\thetable}{J.\arabic{table}}
\setcounter{equation}{0}
\setcounter{figure}{0}
\setcounter{table}{0}

In the Standard Model, three generations of quarks and leptons
are described by the
two-component fermion fields listed in Table \ref{SMfermions},
where $Y$ is the weak hypercharge, $T_3$ is the third component of the
weak isospin, and $Q = T_3 + Y$ is the electric charge.
After SU(2)$_L\times$U(1)$_Y$ breaking, the quark and lepton
fields gain mass in such a way that the above two-component fields
combine to make up four-component Dirac fermions:
\beq \label{quarks4}
U_i = \begin{pmatrix} u_i\\[4pt] {\bar u}{}^{\dagger i}\end{pmatrix}\,,
\qquad\qquad
D_i = \begin{pmatrix} d_i\\[4pt] {\bar d}{}^{\dagger i}\end{pmatrix}\,,
\qquad\qquad
L_i = \begin{pmatrix} \ell_i\\[4pt]
{\bar\ell}{}^{\dagger i}\end{pmatrix}\,,
\eeq
while the neutrinos $\nu_i$ remain massless.
The extension of the Standard
Model to include neutrino masses will be treated in \app{J.2}.

\begin{table}[t!]
\begin{center}
\renewcommand{\arraystretch}{1.5}
\begin{tabular*}{0.95\textwidth}%
     {@{\extracolsep{\fill}}cccccc}
Two-component&&&& \\ [-4pt]
fermion fields&   SU(3) &
 SU(2)$_L$&  $Y$& $T_3$& $Q=T_3+Y$ \\
\hline \\
$\boldsymbol{Q}_i\equiv\begin{pmatrix} u_i\\[4pt] d_i\end{pmatrix}$ &
$\begin{array}{c} {\rm triplet}\\[4pt] {\rm triplet}\end{array}$
& doublet& $\begin{matrix} \phm\tf16\\[3pt]\phm\tf16\end{matrix}$&
    $\begin{matrix} \phm\tf12\\[3pt]-\tf12\end{matrix}$&
$\begin{matrix} \phm\tf23\\[3pt]
-\tf13\end{matrix}$ \\[18pt]
$\phantom{U^i\equiv}
{\bar u}^{i}$ & anti-triplet &   singlet&   $-\tf23$&  $\phm 0$&  $-\tf23$ \\[8pt]
$\phantom{D^i\equiv}{\bar d}^{i}$&
anti-triplet & singlet&  $\phm\tf13$&  $\phm 0$&   $\phm\tf13$ \\[18pt]
$\boldsymbol{L}_i\equiv\begin{pmatrix} \nu_i\\[4pt] \ell_i\end{pmatrix}$ &
$\begin{array}{c} {\rm singlet} \\[4pt] {\rm singlet}\end{array}$
& doublet& $\begin{matrix} -\tf12\\[3pt] -\tf12 \end{matrix}$&
     $\begin{matrix} \phm\tf12\\[3pt]-\tf12\end{matrix}$&
$\begin{matrix}\phm 0 \\[3pt] -1\end{matrix}$ \\[18pt]
$\phantom{E^i\equiv}
{\bar\ell}^{i}$& singlet &  singlet&   $\phm 1$&  $\phm 0$&  $\phm 1$\\[8pt]
\hline
\end{tabular*}
\end{center}
\caption{\label{SMfermions}
Fermions of the Standard Model (following the naming conventions of
Table~\ref{tab:nomenclature}) and their
SU(3)$\times$SU(2)$_L\times$U(1)$_Y$ quantum numbers.
The generation indices run over $i=1,2,3$.  Color indices for
the quarks are suppressed.
The bars on the two-component antifermion
fields are part of their names, and do not
denote some form of complex conjugation.}
\end{table}

Here, we follow the
convention for particle symbols established in Table~\ref{tab:nomenclature}.
Note that $u$, $\bar u$, $d$, $\bar d$, $\ell$ and $\bar\ell$
are two-component fields, whereas the usual four-component
quark and charged lepton fields are denoted by capital letters $U$,
$D$ and $E$.  Consider a generic
four-component field expressed in terms of the corresponding
two-component fields:
\beq
F = \begin{pmatrix} f \\[4pt] {\bar f}^\dagger
\end{pmatrix}\,.
\eeq
The electroweak quantum numbers of $f$ are denoted by $T_3^f$, $Y_f$ and $Q_f$,
whereas the corresponding quantum numbers
for $\bar f$ are $T_3^{\bar f}=0$ and $Q_{\bar f}= Y_{\bar f}=-Q_f$.
Thus we have the correspondence to our general notation
[\eq{general4comp}]
\beq
f \longleftrightarrow \chi,\qquad \bar f \longleftrightarrow \eta\,.
\eeq
We can then immediately translate the couplings given in the
general case in \fig{fig:GaugevertexDirac} to the Standard Model.

\subsection{Standard Model fermion interaction vertices}
\label{app:J1}
\renewcommand{\theequation}{J.1.\arabic{equation}}
\renewcommand{\thefigure}{J.1.\arabic{figure}}
\renewcommand{\thetable}{J.1.\arabic{table}}
\setcounter{equation}{0}
\setcounter{figure}{0}
\setcounter{table}{0}

The QCD color interactions of the quarks are governed by the following
interaction Lagrangian:
\beq
\mathscr{L}_{\rm int} =
\BDneg g_s A_a^{\mu} {q}^{\dagger mi}\,
\sigmabar_\mu ({\boldsymbol T}^a)_m{}^n q_{ni}
\BDplus g_s A_a^{\mu} {\bar q}^{\dagger}_{ni}\,
\sigmabar_\mu ({\boldsymbol T}^a)_m{}^n {\bar q}^{mi} \,,
\label{appeq:lintG}
\eeq
summed over the generations $i$,
where $q$ is a (mass eigenstate) quark field, $m$ and $n$ are
SU(3) color triplet indices, $A^\mu_a$ is the gluon field (with the
corresponding gluons denoted by $g_a$),
and ${\boldsymbol T}^a$ are the color generators in
the triplet representation of SU(3).  The corresponding Feynman rules
are given in \fig{fig:QCDgluonrules}.

\begin{figure}[t!]
\begin{flushleft}
\begin{picture}(200,72)(0,5)
\Gluon(10,40)(60,40){5}{4}
\ArrowLine(60,40)(100,70)
\ArrowLine(100,10)(60,40)
\Text(25,53)[c]{$\mu,a$}
\Text(70,20)[]{$q_{nj}$}
\Text(70,63)[]{$q_{mi}$}
\Text(130,40)[l]{$\BDneg ig_s \delta^j_i
(\boldsymbol{T^a})_m{}^{n}\,\sigmabar^{\dot{\alpha}\beta}_\mu$}
\Text(110,70)[]{$\dot{\alpha}$}
\Text(110,10)[]{$\beta$}
\end{picture}
\hspace{1.2cm}
\begin{picture}(200,72)(0,5)
\Gluon(10,40)(60,40){5}{4}
\ArrowLine(60,40)(100,70)
\ArrowLine(100,10)(60,40)
\Text(25,53)[c]{$\mu,a$}
\Text(70,15)[]{${\bar q}^{mi}$}
\Text(70,66)[]{${\bar q}^{nj}$}
\Text(130,40)[l]{$\BDpos ig_s \delta^j_i
(\boldsymbol{T^a})_m{}^{n}\,\sigmabar^{\dot{\alpha}\beta}_\mu$}
\Text(110,70)[]{$\dot{\alpha}$}
\Text(110,10)[]{$\beta$}
\end{picture}
\end{flushleft}
\caption{Fermionic Feynman rules for QCD that involve the
gluon, with $q = u,d,c,s,t,b$.
Lowered (raised) indices $m,n$ correspond to
the fundamental (anti-fundamental) representation of $SU(3)_c$.
The gluon interactions are flavor-diagonal (where $i$, $j$ are
flavor indices).
For each rule, a corresponding one with lowered spinor
indices is obtained by $\sigmabar_\mu^{\dot\alpha\beta} \rightarrow
-\sigma_{\mu\beta\dot\alpha}$.
\label{fig:QCDgluonrules}}
\end{figure}
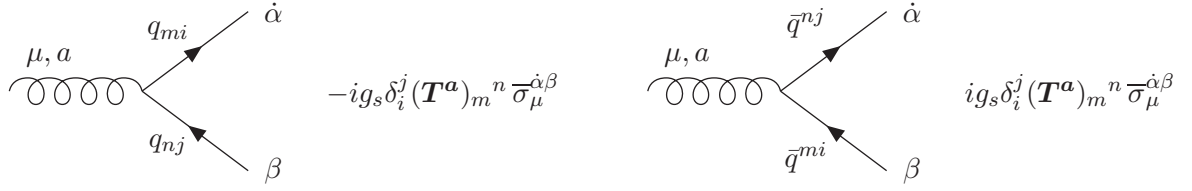

Next, we write out the Feynman rules for the electroweak interactions
of quarks and leptons.  Using \eqs{eq:lintG}{fnew}, the
interactions of the gauge bosons and quarks are given by:
\beqa \label{wqqint}
\mathscr{L}_{\rm int}&=&
\BDneg \frac{g}{\sqrt{2}}\left[
({\hat u}^{\dagger i} \sigmabar^\mu\hat d_i +
{\hat \nu}^{\dagger i} \sigmabar^\mu\hat \ell_i) W_\mu^+
+({\hat d}^{\dagger i}\sigmabar^\mu \hat u_i
+{\hat \ell}^{\dagger i}\sigmabar^\mu \hat \nu_i)W_\mu^-\right]
\nonumber \\
&& \BDminus \frac{g}{c_W}\sum_{f=u,d,\nu,\ell}
\left\{(T_3^f-s_W^2 Q_f)
{\hat f}^{\dagger i}\sigmabar^\mu\hat f_i
+s_W^2 Q_f \hat{\bar f}^{\dagger i}
\sigmabar^\mu\hat{\bar f}_i\right\}Z_\mu \nonumber \\
&&
\BDminus e\sum_{f=u,d,\ell}Q^f ({\hat f}^{\dagger i}
\sigmabar^\mu\hat f_i- \hat{\bar f}^{\dagger i}
\sigmabar^\mu\hat {\bar f}_i)A_\mu\,,
\eeqa
where $s_W\equiv\sin\theta_W$, $c_W\equiv\cos\theta_W$,
the hatted symbols indicate fermion interaction eigenstates
and $i$ labels the generations.  Following
the discussion of \sec{subsec:generalmass},
we must convert from fermion interaction eigenstates
to mass eigenstates.
In order to accomplish this step, we must first identify the
quark and lepton mass matrices.  In the electroweak theory, the
fermion mass matrices
originate from the fermion-Higgs Yukawa interactions.

The Higgs field of the Standard Model is a complex SU(2)$_L$ doublet
of hypercharge $Y=\half$,
\beq
\boldsymbol{\Phi}_a\equiv \begin{pmatrix} \Phi^+ \\ \Phi^0
\end{pmatrix}\,,
\eeq
where the SU(2)$_L$ index $a=1,2$ is defined such that
$\boldsymbol{\Phi}_1\equiv\Phi^+$ and
$\boldsymbol{\Phi}_2\equiv\Phi^0$.  Here, the superscripts $+$ and $0$
refer to the electric charge of the Higgs field, $Q=T_3+Y$, with
$Y=\half$ and $T_3=\pm\half$.  Since $\boldsymbol{\Phi_a}$ is
complex, we can also introduce the complex conjugate Higgs doublet field
with hypercharge $Y=-\half$,
\beq
\boldsymbol{\Phi}^{\dagger\,a} \equiv
\left(\Phi^-\,,\,\,\,(\Phi^0)^\dagger\right)\,,
\eeq
where $\Phi^-\equiv(\Phi^+)^\dagger$.
The SU(2)$_L\times$U(1)$_Y$ gauge invariant
Yukawa interactions of the quarks and leptons with the Higgs field
are then given by:
\beq \label{hsmyukawa}
\mathscr{L}_{\rm Y}=\epsilon^{ab}(\boldsymbol{Y}_u)^i{}_j
\boldsymbol{\Phi}_a\boldsymbol{\hat Q}_{bi} {\bar u}^{j}
-(\boldsymbol{Y}_d)^i{}_j \boldsymbol{\Phi}^{\dagger\,a}
\boldsymbol{\hat Q}_{ai}\hat{\bar d}^{j}
-(\boldsymbol{Y}_\ell)^i{}_j \boldsymbol{\Phi}^{\dagger\,a}
\boldsymbol{\hat L}_{ai}\hat{\bar\ell}^{j}+{\rm h.c.}
\eeq
where $\epsilon^{ab}$ is the antisymmetric invariant tensor of
SU(2)$_L$, defined such that $\epsilon^{12}=-\epsilon^{21}=+1$.
Using the definitions of the SU(2)$_L$ doublet quark and lepton
fields given in Table \ref{SMfermions},
one can rewrite \eq{hsmyukawa} more explicitly as:
\beq \label{hsmyuk}
-\mathscr{L}_{\rm Y}=(\boldsymbol{Y}_u)^i{}_j
\left[\Phi^0 \hat u_i \hat{\bar u}^{j} -\Phi^+
  \hat d_i \hat{\bar u}^{j} \right]
+ (\boldsymbol{Y}_d)^i{}_j
\left[\Phi^- \hat u_i \hat{\bar d}^{j}
+\Phi^{0\ast}\hat d_i \hat{\bar d}^{j}\right]
+(\boldsymbol{Y}_\ell)^i{}_j\left[\Phi^- \hat \nu_i \hat {\bar\ell}^{j}
+\Phi^{0\ast}\hat \ell_i \hat{\bar\ell}^{j}\right]+ {\rm h.c.}
\eeq
The Higgs fields can be written in terms of the physical Higgs scalar
$h_{\rm SM}$ and Nambu-Goldstone bosons $G^0, G^\pm$ as
\beqa
&&\Phi^0 = v + \frac{1}{\sqrt{2}} (h_{\rm SM} + i G^0)\,,
\label{goldn} \\
&&\Phi^+ = G^+ = (\Phi^-)^\dagger = (G^-)^\dagger.\label{goldc}
\eeqa
where $v = \sqrt{2} m_W/g \simeq 174$ GeV.
In the unitary gauge appropriate for tree-level calculations,
the Nambu-Goldstone bosons become infinitely heavy
and decouple.  We identify the quark and lepton mass matrices by
setting $\Phi^0=v$ and $\Phi^+=\Phi^-=0$ in \eq{hsmyuk}:
\beq \label{mvy}
(\boldsymbol{M}_u)^i{}_j = v (\boldsymbol{Y}_u)^i{}_j\,,\qquad
(\boldsymbol{M}_d)^i{}_j = v (\boldsymbol{Y}_d)^i{}_j\,,\qquad
(\boldsymbol{M}_\ell)^i{}_j = v (\boldsymbol{Y}_\ell)^i{}_j\,.
\eeq
The neutrinos remain massless.  An extension of the Standard Model
that incorporates massive neutrinos is treated in \app{J.2}.

To diagonalize the quark and lepton mass matrices, we introduce
four unitary matrices for the quark mass diagonalization,
$L_u$, $L_d$, $R_u$ and $R_d$, and two unitary matrices for the
lepton mass diagonalization, $L_\ell$ and $R_\ell$
[cf.~\eq{lrdef}] such that
\beqa
\hat u_i &=& (L_u)_i{}^j u_j\,,\qquad \hat d_i=(L_d)_i{}^j d_j\,,
\qquad \hat
{\bar u}^{i}= (R_u)^i{}_j {\bar u}^{j}\,,
\qquad
\hat{\bar d}^{i} = (R_d)^i{}_j {\bar d}^{j}\,, \label{quarkmasseigenstates}\\
\hat \ell_i&=&(L_\ell)_i{}^j \ell_j\,,
\qquad
\hat{\bar\ell}^{i} = (R_\ell)^i{}_j {\bar\ell}^{j}\,,\label{leptonmasseigenstates}
\eeqa
where the unhatted fields
$u$, $d$, $\bar u$ and $\bar d$ are the corresponding quark
mass eigenstates and $\nu$, $\ell$ and $\bar\ell$ are the corresponding
lepton mass eigenstates.
The fermion mass diagonalization procedure
consists of the singular value decomposition of the quark and lepton
mass matrices:
\beqa
L_u^{\T} \boldsymbol{M}_u R_u
&=&{\rm diag}(m_u\,,\,m_c\,,\,m_t)\,,\label{luru}\\
L_d^{\T} \boldsymbol{M}_d R_d
&=&{\rm diag}(m_d\,,\,m_s\,,\,m_b)\,,\label{kdrd}\\
L_\ell^{\T} \boldsymbol{M}_\ell R_\ell
&=&{\rm diag}(m_e\,,\,m_\mu\,,\,m_\tau)\,,
\label{lere}
\eeqa
where the diagonalized masses are real and non-negative (cf.~\app{D.1}). Since
the neutrinos are massless, we are free to define the physical neutrino
fields, $\nu_i$, as the weak SU(2)
partners of the corresponding charged lepton mass eigenstate fields.
That is,
\beq \label{neutrinodef}
\hat \nu_i = (L_\ell)_i{}^j \nu_j\,.
\eeq

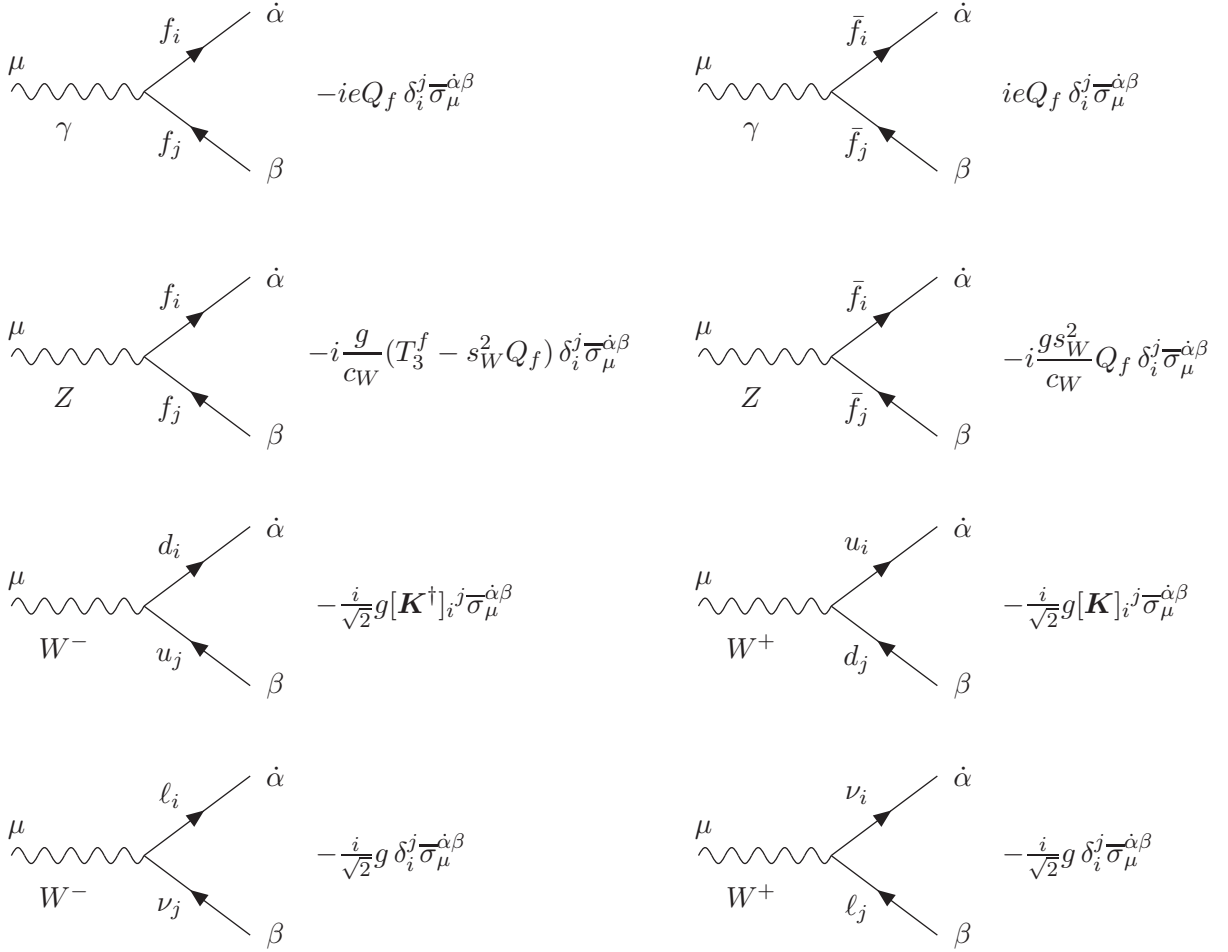
\begin{figure}[t]
\begin{flushleft}
\begin{picture}(190,68)(0,0)
\Photon(60,40)(10,40){3}{5}
\ArrowLine(60,40)(100,70)
\ArrowLine(100,10)(60,40)
\Text(30,25)[]{$\gamma$}
\Text(70,20)[]{$f_j$}
\Text(70,63)[]{$f_i$}
\Text(125,40)[l]{$\BDneg ie Q_f\,\delta_i^j\sigmabar_\mu^{\dot{\alpha}\beta}$}
\Text(110,70)[]{$\dot{\alpha}$}
\Text(110,10)[]{$\beta$}
\Text(12,50)[]{$\mu$}
\end{picture}
\hspace{2.2cm}
\begin{picture}(190,68)(0,0)
\Photon(60,40)(10,40){3}{5}
\ArrowLine(60,40)(100,70)
\ArrowLine(100,10)(60,40)
\Text(30,25)[]{$\gamma$}
\Text(70,20)[]{$\bar f_j$}
\Text(70,63)[]{$\bar f_i$}
\Text(125,40)[l]{$\BDpos ie Q_f\,\delta_i^j\sigmabar_\mu^{\dot{\alpha}\beta}$}
\Text(110,70)[]{$\dot{\alpha}$}
\Text(110,10)[]{$\beta$}
\Text(12,50)[]{$\mu$}
\end{picture}
\end{flushleft}
\vspace{0.05cm}
\begin{flushleft}
\begin{picture}(190,68)(0,0)
\Photon(60,40)(10,40){3}{5}
\ArrowLine(60,40)(100,70)
\ArrowLine(100,10)(60,40)
\Text(30,25)[]{$Z$}
\Text(70,20)[]{$f_j$}
\Text(70,63)[]{$f_i$}
\Text(122,40)[l]{$\BDneg i\displaystyle\frac{g}{c_W}(T_3^f-s_W^2Q_f)
\,\delta_i^j\sigmabar_\mu^{\dot{\alpha}\beta}$}
\Text(110,70)[]{$\dot{\alpha}$}
\Text(110,10)[]{$\beta$}
\Text(12,50)[]{$\mu$}
\end{picture}
\hspace{2.2cm}
\begin{picture}(190,74)(0,0)
\Photon(10,40)(60,40){3}{5}
\ArrowLine(60,40)(100,70)
\ArrowLine(100,10)(60,40)
\Text(30,25)[]{$Z$}
\Text(70,20)[]{$\bar f_j$}
\Text(70,63)[]{$\bar f_i$}
\Text(125,40)[l]{$\BDneg i\displaystyle\frac{gs_W^2}{c_W}
Q_f\,\delta_i^j\sigmabar_\mu^{\dot{\alpha}\beta}$}
\Text(110,70)[]{$\dot{\alpha}$}
\Text(110,10)[]{$\beta$}
\Text(12,50)[]{$\mu$}
\end{picture}
\end{flushleft}
\vspace{0.05cm}
\begin{flushleft}
\begin{picture}(190,68)(0,0)
\Photon(60,40)(10,40){3}{5}
\ArrowLine(60,40)(100,70)
\ArrowLine(100,10)(60,40)
\Text(30,25)[]{$W^-$}
\Text(70,20)[]{$u_j$}
\Text(70,63)[]{$d_i$}
\Text(125,40)[l]{$\BDneg \nicefrac{i}{\sqrt{2}}
g [\boldsymbol{K}^\dagger]_i{}^j\sigmabar_\mu^{\dot{\alpha}\beta}$}
\Text(110,70)[]{$\dot{\alpha}$}
\Text(110,10)[]{$\beta$}
\Text(12,50)[]{$\mu$}
\end{picture}
\hspace{2.2cm}
\begin{picture}(190,68)(0,0)
\Photon(60,40)(10,40){3}{5}
\ArrowLine(60,40)(100,70)
\ArrowLine(100,10)(60,40)
\Text(30,25)[]{$W^+$}
\Text(70,20)[]{$d_j$}
\Text(70,63)[]{$u_i$}
\Text(125,40)[l]{$\BDneg \nicefrac{i}{\sqrt{2}}
g[\boldsymbol{K}]_i{}^j\sigmabar_\mu^{\dot{\alpha}\beta}$}
\Text(110,70)[]{$\dot{\alpha}$}
\Text(110,10)[]{$\beta$}
\Text(12,50)[]{$\mu$}
\end{picture}
\end{flushleft}
%
\vspace{0.05cm}
\begin{flushleft}
\begin{picture}(190,50)(0,18)
\Photon(60,40)(10,40){3}{5}
\ArrowLine(60,40)(100,70)
\ArrowLine(100,10)(60,40)
\Text(30,25)[]{$W^-$}
\Text(70,20)[]{$\nu_j$}
\Text(70,63)[]{$\ell_i$}
\Text(125,40)[l]{$\BDneg \nicefrac{i}{\sqrt{2}}
            g\,\delta_i^j \sigmabar_\mu^{\dot{\alpha}\beta}$}
\Text(110,70)[]{$\dot{\alpha}$}
\Text(110,10)[]{$\beta$}
\Text(12,50)[]{$\mu$}
\end{picture}
\hspace{2.2cm}
\begin{picture}(190,50)(0,18)
\Photon(60,40)(10,40){3}{5}
\ArrowLine(60,40)(100,70)
\ArrowLine(100,10)(60,40)
\Text(30,25)[]{$W^+$}
\Text(70,20)[]{$\ell_j$}
\Text(70,63)[]{$\nu_i$}
\Text(125,40)[l]{$\BDneg \nicefrac{i}{\sqrt{2}}
            g\,\delta_i^j\sigmabar_\mu^{\dot{\alpha}\beta}$}
\Text(110,70)[]{$\dot{\alpha}$}
\Text(110,10)[]{$\beta$}
\Text(12,50)[]{$\mu$}
\end{picture}
\end{flushleft}
%
\caption{Feynman rules for the two-component fermion interactions with
electroweak gauge bosons in the Standard Model.  The couplings of
the fermions to $\gamma$ and $Z$ are flavor-diagonal.  
In all couplings, $i$ and $j$ label the fermion generations;
an upper [lowered] flavor index in the corresponding Feynman rule
is associated with a fermion line that points into [out from] the vertex. 
For the $W^\pm$ bosons, the
charge indicated is flowing into the vertex.
The electric charge is denoted by $Q_f$ (in units of $e>0$),
with $Q_e = -1$ for the electron.  $T_3^f=1/2$
for $f=u$, $\nu$, and $T_3^f=-1/2$ for $f=d$, $\ell$.
The CKM mixing matrix is denoted by $\boldsymbol{K}$,
and $s_W\equiv\sin\theta_W$, $c_W\equiv\cos\theta_W$ and $e\equiv
g\sin\theta_W$.
For each rule, a corresponding one with lowered spinor
indices is obtained by $\sigmabar_\mu^{\dot\alpha\beta} \rightarrow
-\sigma_{\mu\beta\dot\alpha}$.
\label{SMintvertices}}
\end{figure}

We can now write out the couplings of the mass eigenstate quarks and
leptons to the gauge bosons and Higgs bosons.
Consider first the charged current interactions of the quarks and
leptons.  Using \eq{quarkmasseigenstates}, it follows that
${\hat u}^{\dagger i}\sigmabar^\mu\hat d_i
=\boldsymbol{K}_i{}^j u^{\dagger i}\sigmabar^\mu d_j$, where
\beq \label{ckm-matrix}
\boldsymbol{K}=L^\dagger_u L_d
\eeq
is the unitary Cabibbo-Kobayashi-Maskawa (CKM) matrix~\cite{CKM}.\footnote{The
CKM matrix elements $V_{ij}$ as defined in ref.~\cite{RPP} are related
by, for example, $V_{tb} = {\boldsymbol{K}}_3{}^3$ and $V_{us} =
{\boldsymbol{K}}_1{}^2$.}
Due to \eq{neutrinodef}, the corresponding leptonic CKM matrix is
the unit matrix.  Hence, the
charged current interactions take the
form
\beq \label{wqqintm}
\mathscr{L}_{\rm int}= \BDneg
\frac{g}{\sqrt{2}}\left[\boldsymbol{K}_i{}^j
 u^{\dagger i}\sigmabar^\mu  d_j W_\mu^+ + (\boldsymbol{K}^\dagger)_i{}^j
 d^{\dagger i}\sigmabar^\mu u_j W_\mu^-
+  \nu^{\dagger i} \sigmabar^\mu \ell_i W_\mu^+
+  \ell^{\dagger i} \sigmabar^\mu \nu_i W_\mu^-
\right ]\,,
\eeq
where
$[\boldsymbol{K}^\dagger]_i{}^j
\equiv [\boldsymbol{K}_j{}^i]^*$.
Note that in the Standard Model, $\bar u$, $\bar d$ and $\bar\ell$
do not couple to the $W^\pm$.

To obtain the neutral current interactions, we insert
\eqst{quarkmasseigenstates}{neutrinodef} into \eq{wqqint}.
All factors of the \textit{unitary} matrices $L_f$ and $R_f$
($f=u,d,\ell$) cancel out,
and the resulting interactions are flavor-diagonal.  That is, we may
simply remove the hats from the quark and lepton fields that
couple to the $Z$ and photon fields in \eq{wqqint}.
This is the well-known Glashow-Iliopoulos-Maiani
(GIM) mechanism for the flavor-conserving neutral
currents~\cite{GIM}.\footnote{This
also provides the justification for employing mass eigenstate quark fields in
the QCD interaction Lagrangian in \eq{appeq:lintG}.}

The Feynman rules for the interactions of the quarks and leptons with
the charged and neutral gauge bosons are exhibited in \fig{SMintvertices}.
For each of the rules of \fig{SMintvertices},
we have chosen to employ
$\sigmabar_\mu^{\dot{\alpha}\beta}$. If the indices
are lowered one should take $\sigmabar_\mu^{\dot{\alpha}\beta}
\ra -\sigma_{\mu\beta\dot{\alpha}}$.

Finally, we exhibit the
interactions of the quark and lepton mass eigenstates with the
Higgs fields.  The diagonalization of the fermion mass matrices
is equivalent to the diagonalization of the Yukawa couplings
[cf.~eqs.~(\ref{mvy}) and (\ref{luru})--(\ref{lere})].  Thus, we
define\footnote{Boldfaced symbols are used for the
non-diagonal Yukawa matrices, while non-boldfaced symbols are used
for the diagonalized Yukawa couplings.}
\beq
Y_{fi}=m_{fi}/v\,,\qquad  f=u,d,\ell\,,
\eeq
where $i$ labels the fermion generation.  It is convenient to rewrite
\eqst{luru}{lere} as follows:
\beq \label{lfrf}
(L_f)_k{}^j (\boldsymbol{Y}_f)^k{}_m (R_f)^m{}_i = Y_{fi}\delta^j_i\,,
\qquad\quad f=u,d,\ell\,,
\eeq
with no sum over the repeated index $i$.  Using
the unitarity of $L_f$ ($f=u$, $d$), \eq{lfrf} is equivalent to the
following convenient form:
\beq \label{lfrf2}
(\boldsymbol{Y}_f R_f)^k{}_i=Y_{fi}(L_f^\dagger)_i{}^k\,.
\eeq
Inserting \eqss{quarkmasseigenstates}{neutrinodef}{lfrf} into
\eq{hsmyuk}, the resulting Higgs-fermion Lagrangian is
flavor-diagonal:
\beq \label{SMhbosonrules}
\mathscr{L}_{\rm int}= -\frac{1}{\sqrt{2}}
h_{\rm SM}
\left[
Y_{ui} u_i {\bar u}^{i} +
Y_{di} d_i {\bar d}^{i} +
Y_{\ell i} \ell_i {\bar\ell}^{i}
\right]
+{\rm h.c.}
\eeq
The corresponding Feynman rules for the Higgs-fermion interaction are
shown in \fig{qqhiggs}.
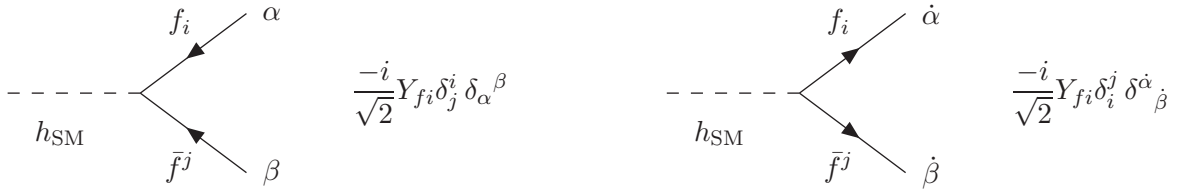
\begin{figure}[ht!]
\begin{center}
\begin{picture}(200,63)(5,15)
\DashLine(60,40)(10,40)5
\ArrowLine(100,70)(60,40)
\ArrowLine(100,10)(60,40)
\Text(30,25)[]{$\hsm$}
\Text(75,12)[]{${\bar f}^{j}$}
\Text(75,67)[]{$f_i$}
\Text(140,40)[l]{$\displaystyle{\frac{-i}{\sqrt{2}} Y_{fi} \delta^i_j
\,\delta_\alpha{}^\beta}$}
\Text(110,70)[]{$\alpha$}
\Text(110,10)[]{$\beta$}
\end{picture}
\hspace{1.3cm}
\begin{picture}(200,53)(0,15)
\DashLine(60,40)(10,40)5
\ArrowLine(60,40)(100,70)
\ArrowLine(60,40)(100,10)
\Text(30,25)[]{$\hsm$}
\Text(75,12)[]{${\bar f}^{j}$}
\Text(75,67)[]{$f_i$}
\Text(140,40)[l]{$\displaystyle{\frac{-i}{\sqrt{2}} Y_{fi} \delta^j_i\,
\delta^{\dot{\alpha}}{}_{\dot{\beta}}}$}
\Text(110,70)[]{$\dot{\alpha}$}
\Text(110,10)[]{$\dot{\beta}$}
\end{picture}
\end{center}
\caption{\label{qqhiggs} Feynman rules for the Standard Model
Higgs boson interactions with fermions, where $Y_{fi}\equiv m_{fi}/v$,
and $i$, $j$ label the generations.}
\end{figure}

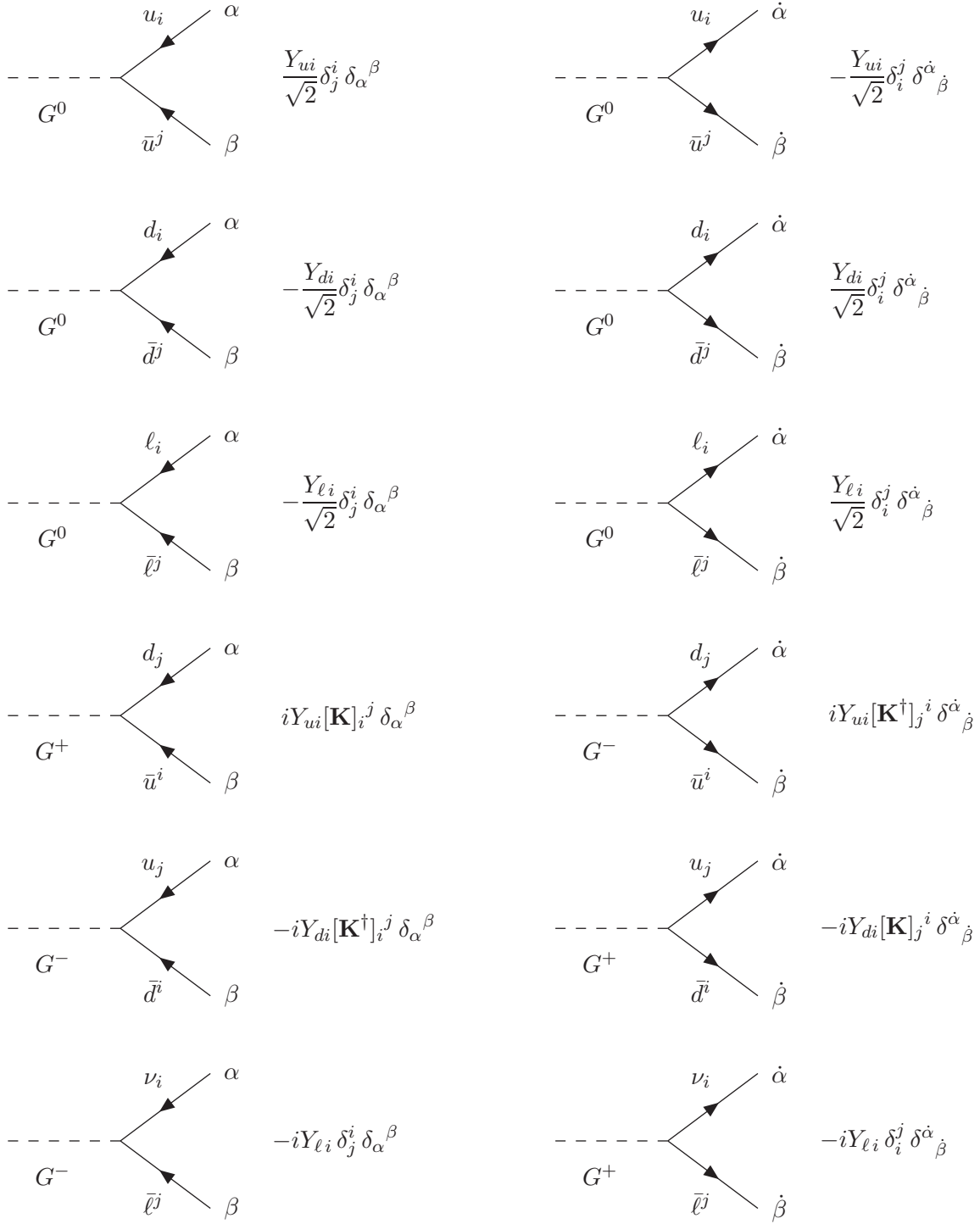
\begin{figure}[t!]
\begin{flushleft}
\begin{picture}(200,60)(0,10)
\DashLine(60,40)(10,40)5
\ArrowLine(100,70)(60,40)
\ArrowLine(100,10)(60,40)
\Text(30,25)[]{$G^0$}
\Text(75,12)[]{${\bar u}^{j}$}
\Text(75,67)[]{$u_i$}
\Text(132,40)[l]{$\displaystyle{\frac{Y_{ui}}{\sqrt{2}}
\delta_j^i \,\delta_\alpha{}^\beta}$}
\Text(110,70)[]{$\alpha$}
\Text(110,10)[]{$\beta$}
\end{picture}
\hspace{1.3cm}
\begin{picture}(200,60)(0,10)
\DashLine(60,40)(10,40)5
\ArrowLine(60,40)(100,70)
\ArrowLine(60,40)(100,10)
\Text(30,25)[]{$G^0$}
\Text(75,12)[]{${\bar u}^{j}$}
\Text(75,67)[]{$u_i$}
\Text(132,40)[l]{$\displaystyle{-\frac{Y_{ui}}{\sqrt{2}}
\delta_i^j \,\delta^{\dot\alpha}{}_{\dot\beta}}$}
\Text(110,70)[]{$\dot\alpha$}
\Text(110,10)[]{$\dot\beta$}
\end{picture}
\end{flushleft}
\vspace{0.35cm}
\begin{flushleft}
\begin{picture}(200,60)(0,10)
\DashLine(60,40)(10,40)5
\ArrowLine(100,70)(60,40)
\ArrowLine(100,10)(60,40)
\Text(30,25)[]{$G^0$}
\Text(75,12)[]{${\bar d}^{j}$}
\Text(75,67)[]{$d_i$}
\Text(132,40)[l]{$\displaystyle{-\frac{Y_{di}}{\sqrt{2}}
\delta_j^i \,\delta_\alpha{}^\beta}$}
\Text(110,70)[]{$\alpha$}
\Text(110,10)[]{$\beta$}
\end{picture}
\hspace{1.3cm}
\begin{picture}(200,60)(0,10)
\DashLine(60,40)(10,40)5
\ArrowLine(60,40)(100,70)
\ArrowLine(60,40)(100,10)
\Text(30,25)[]{$G^0$}
\Text(75,12)[]{${\bar d}^{j}$}
\Text(75,67)[]{$d_i$}
\Text(132,40)[l]{$\displaystyle{\frac{Y_{di}}{\sqrt{2}}}
\delta_i^j \,\delta^{\dot{\alpha}}{}_{\dot{\beta}}$}
\Text(110,70)[]{$\dot\alpha$}
\Text(110,10)[]{$\dot\beta$}
\end{picture}
\end{flushleft}
\vspace{0.35cm}
\begin{flushleft}
\begin{picture}(200,60)(0,10)
\DashLine(60,40)(10,40)5
\ArrowLine(100,70)(60,40)
\ArrowLine(100,10)(60,40)
\Text(30,25)[]{$G^0$}
\Text(75,12)[]{${\bar\ell}^{j}$}
\Text(75,67)[]{$\ell_i$}
\Text(132,40)[l]{$\displaystyle{-\frac{Y_{\ell\,i}}{\sqrt{2}}
\delta^i_j\,\delta_\alpha{}^\beta}$}
\Text(110,70)[]{$\alpha$}
\Text(110,10)[]{$\beta$}
\end{picture}
\hspace{1.3cm}
\begin{picture}(200,60)(0,10)
\DashLine(60,40)(10,40)5
\ArrowLine(60,40)(100,70)
\ArrowLine(60,40)(100,10)
\Text(30,25)[]{$G^0$}
\Text(75,12)[]{${\bar\ell}^{j}$}
\Text(75,67)[]{$\ell_i$}
\Text(132,40)[l]{$\displaystyle{{\frac{Y_{\ell\,i}}{\sqrt{2}}}
\,\delta_i^j\,\delta^{\dot{\alpha}}{}_{\dot{\beta}}}$}
\Text(110,70)[]{$\dot\alpha$}
\Text(110,10)[]{$\dot\beta$}
\end{picture}
\end{flushleft}
\vspace{0.35cm}
\begin{flushleft}
\begin{picture}(200,60)(0,10)
\DashLine(60,40)(10,40)5
\ArrowLine(100,70)(60,40)
\ArrowLine(100,10)(60,40)
\Text(30,25)[]{$G^+$}
\Text(75,12)[]{${\bar u}^{i}$}
\Text(75,67)[]{$d_j$}
\Text(132,40)[l]{$\displaystyle{i Y_{ui} [{\bf K}]_i{}^j
\,\delta_\alpha{}^\beta}$}
\Text(110,70)[]{$\alpha$}
\Text(110,10)[]{$\beta$}
\end{picture}
\hspace{1.3cm}
\begin{picture}(200,60)(0,10)
\DashLine(60,40)(10,40)5
\ArrowLine(60,40)(100,70)
\ArrowLine(60,40)(100,10)
\Text(30,25)[]{$G^-$}
\Text(75,12)[]{${\bar u}^{i}$}
\Text(75,67)[]{$d_j$}
\Text(132,40)[l]{$\displaystyle{i Y_{ui} [{\bf K}^\dagger]_j{}^i
\,\delta^{\dot\alpha}{}_{\dot\beta}}$}
\Text(110,70)[]{$\dot\alpha$}
\Text(110,10)[]{$\dot\beta$}
\end{picture}
\end{flushleft}
\vspace{0.35cm}
\begin{flushleft}
\begin{picture}(200,60)(0,10)
\DashLine(60,40)(10,40)5
\ArrowLine(100,70)(60,40)
\ArrowLine(100,10)(60,40)
\Text(30,25)[]{$G^-$}
\Text(75,12)[]{${\bar d}^{i}$}
\Text(75,67)[]{$u_j$}
\Text(128,40)[l]{$\displaystyle{-i Y_{di} [{\bf K}^\dagger]_i{}^j
\,\delta_\alpha{}^\beta}$}
\Text(110,70)[]{$\alpha$}
\Text(110,10)[]{$\beta$}
\end{picture}
\hspace{1.3cm}
\begin{picture}(200,60)(0,10)
\DashLine(60,40)(10,40)5
\ArrowLine(60,40)(100,70)
\ArrowLine(60,40)(100,10)
\Text(30,25)[]{$G^+$}
\Text(75,12)[]{${\bar d}^{i}$}
\Text(75,67)[]{$u_j$}
\Text(128,40)[l]{$\displaystyle{-i Y_{di} [{\bf K}]_j{}^i
\,\delta^{\dot\alpha}{}_{\dot\beta}}$}
\Text(110,70)[]{$\dot\alpha$}
\Text(110,10)[]{$\dot\beta$}
\end{picture}
\end{flushleft}
\vspace{0.35cm}
\begin{flushleft}
\begin{picture}(200,60)(0,10)
\DashLine(60,40)(10,40)5
\ArrowLine(100,70)(60,40)
\ArrowLine(100,10)(60,40)
\Text(30,25)[]{$G^-$}
\Text(75,12)[]{${\bar\ell}^{j}$}
\Text(75,67)[]{$\nu_i$}
\Text(128,40)[l]{$\displaystyle{-i Y_{\ell\,i}
\,\delta^i_j\,\delta_\alpha{}^\beta}$}
\Text(110,70)[]{$\alpha$}
\Text(110,10)[]{$\beta$}
\end{picture}
\hspace{1.3cm}
\begin{picture}(200,60)(0,10)
\DashLine(60,40)(10,40)5
\ArrowLine(60,40)(100,70)
\ArrowLine(60,40)(100,10)
\Text(30,25)[]{$G^+$}
\Text(75,12)[]{${\bar\ell}^{j}$}
\Text(75,67)[]{$\nu_i$}
\Text(128,40)[l]{$\displaystyle{-i Y_{\ell\,i}
\,\delta_i^j\,\delta^{\dot\alpha}{}_{\dot\beta}}$}
\Text(110,70)[]{$\dot\alpha$}
\Text(110,10)[]{$\dot\beta$}
\end{picture}
\end{flushleft}
%
\caption{\label{fig:SMNGbosons} Feynman rules for the
Standard Model Nambu-Goldstone boson
interactions with quarks and leptons, where $Y_{fi}\equiv m_{fi}/v$,
and $i$, $j$ label the generations.}
\vskip 0.5in
\end{figure}

In the case of more general covariant gauge-fixing (e.g., the 't Hooft-Feynman
gauge or Landau gauge), the Goldstone bosons appear explicitly in
internal lines of Feynman diagrams.  The Feynman rules for
$G^0$-fermion interactions are flavor-diagonal, whereas the
corresponding rules for $G^\pm$ exhibit flavor-changing
interactions that depend on the CKM matrix elements, as shown in
\fig{fig:SMNGbosons}.  In the derivation of the couplings
of the Nambu-Goldstone bosons to the fermion mass eigenstates
[cf.~\eqst{hsmyuk}{goldc}],
the following quantities appear:
\beqa
(L_d)_k{}^j (\boldsymbol{Y}_u)^{k}{}_m (R_u)^m{}_i &=& Y_{u i} (L_d)_k{}^j (L_u^\dagger)_i{}^k
=Y_{u i} (L^\dagger_u L_d)_i{}^j=[\boldsymbol{K}]_i{}^j Y_{ui}\,,
\label{lyr1}
\\
(L_u)_k{}^j (\boldsymbol{Y}_d)^{k}{}_m (R_d)^m{}_i &=& Y_{d i} (L_u)_k{}^j (L_d^\dagger)_i{}^k
=Y_{d i} (L^\dagger_d L_u)_i{}^j=[\boldsymbol{K}^\dagger]_i{}^j Y_{di}\,,
\label{lyr2}
\eeqa
with no sum over the repeated index $i$.  The CKM matrix, $\boldsymbol{K}$, appears by virtue
of \eqs{ckm-matrix}{lfrf2}.
Hence, the interaction
Lagrangian for the coupling of the Nambu-Goldstone bosons to the
fermion mass eigenstates is given by:
\beq \label{gbosonrules}
\mathscr{L}_{\rm int}=
Y_{ui}[\boldsymbol{K}]_i{}^j d_j {\bar u}^{i} G^+
-Y_{di}[\boldsymbol{K}^\dagger]_i{}^j u_j {\bar d}^{i} G^-
-Y_{\ell i} \nu_i {\bar\ell}^{i} G^-
+\frac{i}{\sqrt{2}}\left[Y_{di} d_i {\bar d}^{i} - Y_{ui} u_i {\bar u}^{i}
+Y_{\ell\, i} \ell_i {\bar\ell}^{i} \right]G^0+{\rm h.c.},
\eeq
which yields the diagrammatic Feynman rules shown in
\fig{fig:SMNGbosons}.

\subsection{Incorporating massive neutrinos into
the Standard Model}
\label{app:J2}
\renewcommand{\theequation}{J.2.\arabic{equation}}
\renewcommand{\thefigure}{J.2.\arabic{figure}}
\renewcommand{\thetable}{J.2.\arabic{table}}
\setcounter{equation}{0}
\setcounter{figure}{0}
\setcounter{table}{0}

To accommodate massive neutrinos,
we must slightly extend the
Standard Model~\cite{nureviews}.  The simplest approach is
to introduce an SU(2)$\times$U(1) gauge invariant
dimension-five operator~\cite{refdim5},
\beqa \label{dim5}
\mathscr{L}_{5}&= & -\frac{\boldsymbol{\hat F}^{ij}}{2\Lambda}
(\epsilon^{ab}\boldsymbol{\Phi}_a\boldsymbol{\hat L}_{bi})
(\epsilon^{cd}\boldsymbol{\Phi}_c\boldsymbol{\hat L}_{dj})+{\rm h.c.}
\nonumber \\
&=&-\frac{\boldsymbol{\hat F}^{ij}}{2\Lambda}
(\Phi^0\hat\nu_i-\Phi^+\hat \ell_i)
(\Phi^0\hat\nu_j-\Phi^+\hat \ell_j)+{\rm h.c.}\,,
\eeqa
where $\boldsymbol{\hat F}^{ij}$ are generalized Yukawa couplings,
the hatted fields indicate two-component fermion
interaction eigenstates (with spinor indices suppressed), and $i$, $j$
label the three generations.  After electroweak symmetry breaking,
the neutral component of the
doublet Higgs field acquires a vacuum expectation value, and a
Majorana mass matrix for the neutrinos is generated.

The diagonalization of the charged lepton mass matrix is unmodified
from the treatment given in \app{J.1}, where the
unhatted mass eigenstate charged lepton fields
are given by \eq{leptonmasseigenstates},
and $L_\ell$ and $R_\ell$ satisfy \eq{lere}.
However, the unhatted neutrino field introduced in \eq{neutrinodef}
is \textit{not} a neutrino mass eigenstate field when
the effect of the dimension-five Lagrangian, \eq{dim5}, is taken into account.
To avoid confusion, we replace the
unhatted neutrino fields of \eq{neutrinodef}
with new neutrino fields $\breve\nu_j$.
That is, we define
\beq \label{brevenu}
\hat\nu_i=(L_\ell)_i{}^j\breve{\nu}_j\,.
\eeq
We then rewrite \eq{dim5} in terms of the charged lepton mass eigenstate
field and the new neutrino field defined by \eq{brevenu}:
\beq \label{dimen5}
\mathscr{L}_{5}
=-\frac{\boldsymbol{F}^{ij}}{2\Lambda}
(\Phi^0\breve\nu_i-\Phi^+ \ell_i)
(\Phi^0\breve\nu_j-\Phi^+ \ell_j)+{\rm h.c.}\,,
\eeq
where $\boldsymbol{F}\equiv L_\ell^{\T}\boldsymbol{\hat F}L_\ell$.
Setting $\Phi^0=v$ and
$\Phi^+=\Phi^-=0$, we identify the $3\times 3$
complex symmetric effective light neutrino mass
matrix, $\boldsymbol{M_{\nu_\ell}}$, by
\beq
-\mathscr{L}_{m_\nu}=\half (\boldsymbol{M_{\nu_\ell}})^{ij}\breve\nu_i
\breve\nu_j+{\rm h.c.}\,,
\eeq
where
\beq \label{mnueff}
\boldsymbol{M_{\nu_\ell}}=\frac{v^2}{\Lambda}\boldsymbol{F}\,.
\eeq
Current bounds on light
neutrino masses suggest that $v^2/\Lambda\lsim 1~{\rm
eV}$, or $\Lambda\gsim 10^{13}~{\rm GeV}$~\cite{RPP,numass}.

The physical neutrino mass eigenstate fields
can be identified by
introducing the unitary Maki-Nakagawa-Sakata (MNS)
matrix, $\boldsymbol{U}_{\rm MNS}$,
such that~\cite{PMNSref},\footnote{In the literature, the MNS matrix is
often defined such that
$\boldsymbol{U}^*_{\rm MNS}$ (and \textit{not}
$\boldsymbol{U}_{\rm MNS}$) appears in \eq{PMNS}.}
\beq \label{PMNS}
(\breve\nu_{\ell})_i=(\boldsymbol{U}_{\rm MNS})_i{}^j (\nu_\ell)_j\,,
\eeq
where the unhatted $(\nu_\ell)_j$ fields [$j=1,2,3$] denote the physical
(mass eigenstate) Majorana neutrino fields.
$\boldsymbol{U}_{\rm MNS}$ is determined by the
Takagi diagonalization of $\boldsymbol{M_{\nu_\ell}}$ [cf.~\app{D.2}]:
\beq
\boldsymbol{U}^{\T}_{\rm MNS} \boldsymbol{M_{\nu_\ell}}
\boldsymbol{U}_{\rm MNS} = {\rm
diag}(m_{\nu_{\ell}1}\,,\,m_{\nu_{\ell}2}\,,\,m_{\nu_{\ell}3})\,,
\label{eq:vmphys}
\eeq
where the $m_{\nu_{\ell}j}$ 
are the (real non-negative) masses of the physical neutrinos.

The interaction Lagrangian of the neutrino mass eigenstates can
now be determined.
The charged current neutrino
interactions are given by [cf.~\eq{wqqintm}]:
\beqa \label{wqqintmnu}
\mathscr{L}_{\rm int}&=& \BDneg \frac{g}{\sqrt{2}}\left[
{\breve\nu}^{\dagger i} \sigmabar^\mu \ell_i W_\mu^+
+  \ell^{\dagger i} \sigmabar^\mu \breve\nu_i W_\mu^-
\right] \nonumber \\
&=& \BDneg
\frac{g}{\sqrt{2}}\left[(\boldsymbol{U}^\dagger_{\rm MNS})_j{}^i
{\nu}_\ell^{\dagger j} \sigmabar^\mu  \ell_i W_\mu^+ +
(\boldsymbol{U}_{\rm MNS})_i{}^j
 \ell^{\dagger i}\sigmabar^\mu \nu_{\ell j} W_\mu^-\right]
\,,
\eeqa
where we have used \eq{PMNS} to express the interaction Lagrangian
in terms of the neutrino mass eigenstate fields.  The
neutral current neutrino interactions are flavor-diagonal
(which follows from the unitarity of $U_{\rm MNS}$),
and are thus
equivalent to those of the Standard Model.  Finally, the
couplings of the neutrinos to the Higgs and Nambu-Goldstone
fields arise from \eq{dimen5} and from the term in \eq{hsmyuk} proportional
to $\boldsymbol{Y}_{\!\ell}$.  Neglecting terms
of $\mathcal{O}(m_\nu^2/v^2)$, one obtains:
\beqa \label{nunuhiggs}
\mathscr{L}_{\rm int}&=&
\frac{1}{v}\sum_{i,j} \left[(m_{\nu_\ell})_j
(\boldsymbol{U}^\dagger_{\rm MNS})_j{}^i
(\nu_\ell)_j\,\ell_i\, G^+
-(m_\ell)_i (\boldsymbol{U}_{\rm MNS})_i{}^j
(\nu_\ell)_j\,{\bar\ell}^{i}\, G^- +{\rm h.c.}\right]  \nonumber \\
&&-\frac{1}{\sqrt{2} v}\sum_j (m_{\nu_\ell})_j\left[(\nu_\ell)_j
(\nu_\ell)_j (h_{\rm SM}+iG^0) +{\rm h.c.}\right]\,.
\eeqa
The Feynman rules for the interactions of the neutrino
with the electroweak gauge bosons, the Higgs boson and
the Nambu-Goldstone bosons
are exhibited in \fig{SMnuintvertices}.
\begin{figure}[ht!]
\begin{flushleft}
\begin{picture}(190,50)(10,18)
\Photon(60,40)(10,40){3}{5}
\ArrowLine(60,40)(100,70)
\ArrowLine(100,10)(60,40)
\Text(30,25)[]{$W^-$}
\Text(70,20)[]{$(\nu_\ell)_j$}
\Text(70,63)[]{$\ell_i$}
\Text(125,40)[l]{$\BDneg \displaystyle\frac{ig}{\sqrt{2}}
[\boldsymbol{U}_{\rm MNS}]_i{}^j \sigmabar_\mu^{\dot{\alpha}\beta}$}
\Text(110,70)[]{$\dot{\alpha}$}
\Text(110,10)[]{$\beta$}
\Text(12,50)[]{$\mu$}
\end{picture}
\hspace{2.3cm}
\begin{picture}(190,50)(30,18)
\Photon(60,40)(10,40){3}{5}
\ArrowLine(60,40)(100,70)
\ArrowLine(100,10)(60,40)
\Text(30,25)[]{$W^+$}
\Text(70,20)[]{$\ell_j$}
\Text(70,63)[]{$(\nu_\ell)_i$}
\Text(125,40)[l]{$\BDneg \displaystyle\frac{ig}{\sqrt{2}}
            [\boldsymbol{U}^\dagger_{\rm MNS}]_i{}^j
\sigmabar_\mu^{\dot{\alpha}\beta}$}
\Text(110,70)[]{$\dot{\alpha}$}
\Text(110,10)[]{$\beta$}
\Text(12,50)[]{$\mu$}
\end{picture}
\end{flushleft}
\begin{flushleft}
\begin{picture}(190,78)(10,20)
\Photon(60,40)(10,40){3}{5}
\ArrowLine(60,40)(100,70)
\ArrowLine(100,10)(60,40)
\Text(30,25)[]{$Z$}
\Text(70,20)[]{$(\nu_{\ell})_j$}
\Text(70,63)[]{$(\nu_{\ell})_i$}
\Text(122,40)[l]{$\BDneg i\displaystyle\frac{g}{2\cos\theta_W}
\sigmabar_\mu^{\dot{\alpha}\beta}\delta_i^j$}
\Text(110,70)[]{$\dot{\alpha}$}
\Text(110,10)[]{$\beta$}
\Text(12,50)[]{$\mu$}
\end{picture}
%
\hspace{2.3cm}
\begin{picture}(190,50)(30,18)
\Photon(60,40)(10,40){3}{5}
\ArrowLine(60,40)(100,70)
\ArrowLine(100,10)(60,40)
\Text(30,25)[]{$\gamma$}
\Text(70,20)[]{$(\nu_\ell)_j$}
\Text(70,63)[]{$(\nu_\ell)_i$}
\Text(145,40)[l]{0}
\Text(12,50)[]{$\mu$}
\Text(110,70)[]{$\dot{\alpha}$}
\Text(110,10)[]{$\beta$}
\end{picture}
\end{flushleft}
\vspace{0.35cm}
\begin{flushleft}
\begin{picture}(200,60)(10,10)
\DashLine(60,40)(10,40)5
\ArrowLine(100,70)(60,40)
\ArrowLine(100,10)(60,40)
\Text(30,25)[]{$h_{\rm SM}^0$}
\Text(75,12)[]{$(\nu_\ell)_j$}
\Text(75,67)[]{$(\nu_\ell)_i$}
\Text(132,40)[l]{$\displaystyle{-\frac{i\sqrt{2}}{v}(m_{\nu_\ell})_i
\,\delta^{ij} \,\delta_\alpha{}^\beta}$}
\Text(110,70)[]{$\alpha$}
\Text(110,10)[]{$\beta$}
\end{picture}
\hspace{1.3cm}
\begin{picture}(200,60)(10,10)
\DashLine(60,40)(10,40)5
\ArrowLine(60,40)(100,70)
\ArrowLine(60,40)(100,10)
\Text(30,25)[]{$h_{\rm SM}^0$}
\Text(75,12)[]{$(\nu_\ell)_j$}
\Text(75,67)[]{$(\nu_\ell)_i$}
\Text(132,40)[l]{$\displaystyle{-\frac{i\sqrt{2}}{v} (m_{\nu_\ell})_i
\,\delta_{ij} \,\delta^{\dot\alpha}{}_{\dot\beta}}$}
\Text(110,70)[]{$\dot\alpha$}
\Text(110,10)[]{$\dot\beta$}
\end{picture}
\end{flushleft}
\vspace{0.35cm}
\begin{flushleft}
\begin{picture}(200,60)(10,10)
\DashLine(60,40)(10,40)5
\ArrowLine(100,70)(60,40)
\ArrowLine(100,10)(60,40)
\Text(30,25)[]{$G^0$}
\Text(75,12)[]{$(\nu_\ell)_j$}
\Text(75,67)[]{$(\nu_\ell)_i$}
\Text(132,40)[l]{$\displaystyle{\frac{\sqrt{2}}{v}(m_{\nu_\ell})_i
\,\delta^{ij} \,\delta_\alpha{}^\beta}$}
\Text(110,70)[]{$\alpha$}
\Text(110,10)[]{$\beta$}
\end{picture}
\hspace{1.3cm}
\begin{picture}(200,60)(10,10)
\DashLine(60,40)(10,40)5
\ArrowLine(60,40)(100,70)
\ArrowLine(60,40)(100,10)
\Text(30,25)[]{$G^0$}
\Text(75,12)[]{$(\nu_\ell)_j$}
\Text(75,67)[]{$(\nu_\ell)_i$}
\Text(132,40)[l]{$\displaystyle{-\frac{\sqrt{2}}{v}(m_{\nu_\ell})_i
\,\delta_{ij} \,\delta^{\dot\alpha}{}_{\dot\beta}}$}
\Text(110,70)[]{$\dot\alpha$}
\Text(110,10)[]{$\dot\beta$}
\end{picture}
\end{flushleft}
\vspace{0.35cm}
\begin{flushleft}
\begin{picture}(200,60)(10,10)
\DashLine(60,40)(10,40)5
\ArrowLine(100,70)(60,40)
\ArrowLine(100,10)(60,40)
\Text(30,25)[]{$G^+$}
\Text(75,12)[]{$(\nu_\ell)_j$}
\Text(75,67)[]{$\ell_i$}
\Text(127,40)[l]{$\displaystyle{\frac{i}{v} (m_{\nu_\ell})_j
[\boldsymbol{U}^\dagger_{\rm MNS}]_j{}^i
\,\delta_\alpha{}^\beta}$}
\Text(110,70)[]{$\alpha$}
\Text(110,10)[]{$\beta$}
\end{picture}
\hspace{1.3cm}
\begin{picture}(200,60)(10,10)
\DashLine(60,40)(10,40)5
\ArrowLine(60,40)(100,70)
\ArrowLine(60,40)(100,10)
\Text(30,25)[]{$G^-$}
\Text(75,12)[]{$(\nu_{\ell})_j$}
\Text(75,67)[]{$\ell_i$}
\Text(132,40)[l]{$\displaystyle{\frac{i}{v} (m_{\nu_\ell})_j
[\boldsymbol{U}_{\rm MNS}]_i{}^j
\,\delta^{\dot\alpha}{}_{\dot\beta}}$}
\Text(110,70)[]{$\dot\alpha$}
\Text(110,10)[]{$\dot\beta$}
\end{picture}
\end{flushleft}
\vspace{0.35cm}
\begin{flushleft}
\begin{picture}(200,60)(10,10)
\DashLine(60,40)(10,40)5
\ArrowLine(100,70)(60,40)
\ArrowLine(100,10)(60,40)
\Text(30,25)[]{$G^-$}
\Text(75,12)[]{${\bar\ell}^{j}$}
\Text(75,67)[]{$(\nu_\ell)_i$}
\Text(128,40)[l]{$\displaystyle{-\frac{i}{v}
    (m_\ell)_j[\boldsymbol{U}_{\rm MNS}]_j{}^i
\,\delta_\alpha{}^\beta}$}
\Text(110,70)[]{$\alpha$}
\Text(110,10)[]{$\beta$}
\end{picture}
\hspace{1.3cm}
\begin{picture}(200,60)(10,10)
\DashLine(60,40)(10,40)5
\ArrowLine(60,40)(100,70)
\ArrowLine(60,40)(100,10)
\Text(30,25)[]{$G^+$}
\Text(75,12)[]{${\bar\ell}^{j}$}
\Text(75,67)[]{$(\nu_\ell)_i$}
\Text(128,40)[l]{$\displaystyle{-\frac{i}{v}
(m_\ell)_j[\boldsymbol{U}^\dagger_{\rm MNS}]_i{}^j
\,\delta^{\dot\alpha}{}_{\dot\beta}}$}
\Text(110,70)[]{$\dot\alpha$}
\Text(110,10)[]{$\dot\beta$}
\end{picture}
\end{flushleft}
%
\caption{Feynman rules for the interactions of the
two-component light neutrino ($\nu_\ell$) with
electroweak gauge bosons, the Standard Model Higgs boson
and the Nambu-Goldstone bosons, where $i$, $j$ label the generation.
For the $W^\pm$ bosons and $G^\pm$ scalars, the
charge indicated is flowing into the vertex.
The MNS mixing matrix is denoted by $\boldsymbol{U}_{\rm MNS}$.
For the rules involving $W^\pm$ and $Z$ bosons,
a corresponding one with lowered spinor
indices is obtained by $\sigmabar_\mu^{\dot\alpha\beta} \rightarrow
-\sigma_{\mu\beta\dot\alpha}$.  
In the $h^0_{\rm SM}$ and $G^0$
interactions, a factor of 2 is included to account for
the identical neutrinos.
\label{SMnuintvertices}}
\end{figure}
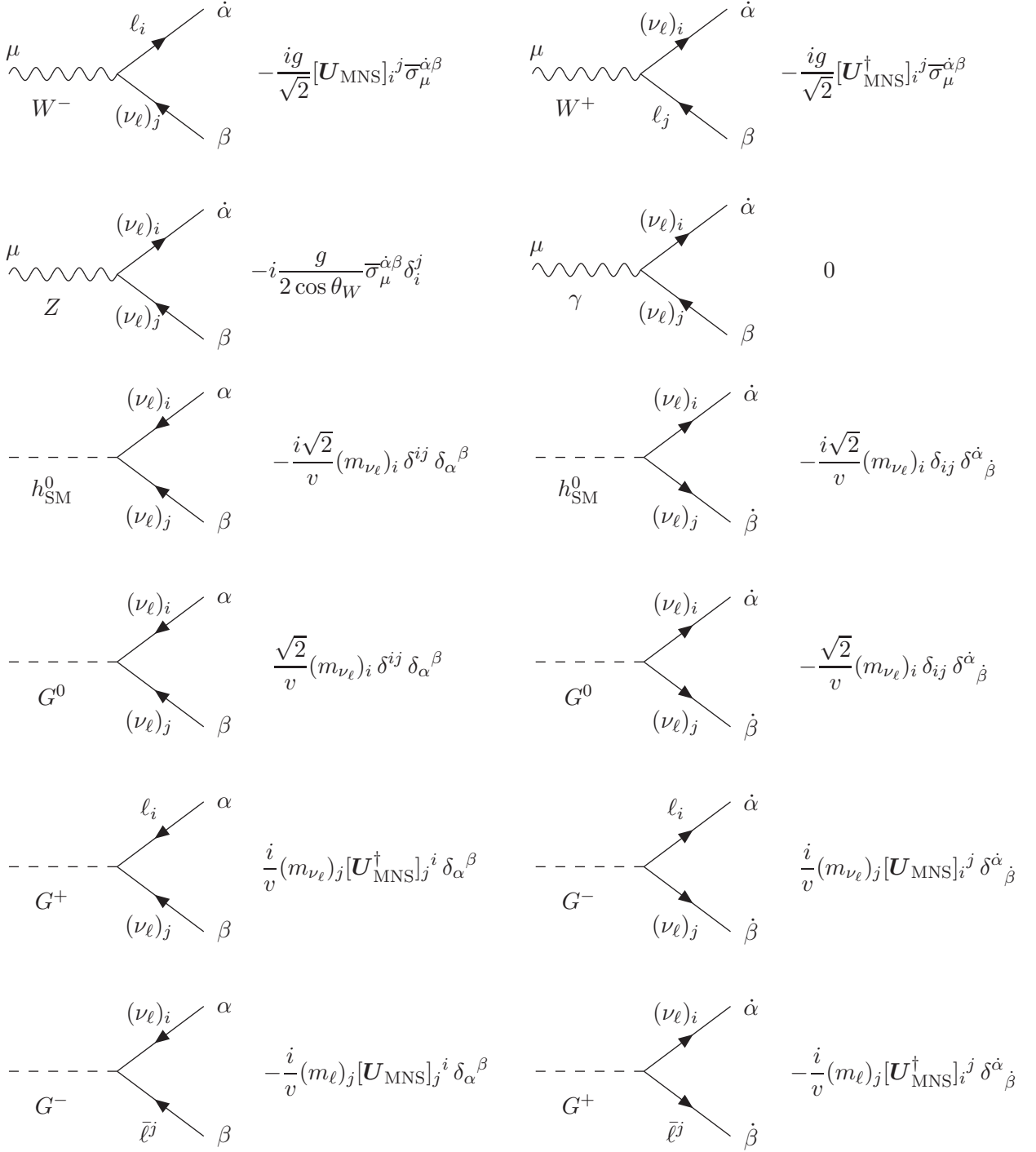

The dimension-five Lagrangian, \eq{dim5}, is generated by new physics beyond the
Standard Model at the scale $\Lambda$.
A possible realization of \eq{dim5} is the seesaw mechanism,
which was independently discovered on a number of
occasions~\cite{seesaw}.  In the seesaw extension of the
Standard Model~\cite{Valle-Schechter1}, one introduces the
SU(3)$\times$SU(2)$\times$U(1) gauge singlet two-component
neutrino fields ${\bar\nu}^{I}$ ($I=1,2,\ldots,n$)
and writes down the most
general renormalizable couplings of the ${\bar\nu}^{I}$ to the Standard Model
fields:
\beq \label{seesawlagrange}
\mathscr{L}_{\rm seesaw}= \epsilon^{ab}(\boldsymbol{\hat{Y}}_\nu)^i{}_J
\boldsymbol{\Phi}_a \boldsymbol{\hat L}_{bi}\hat{\bar \nu}^{J}-\half
\boldsymbol{\hat M}_{IJ}
\hat{\bar\nu}^{I}\hat{\bar\nu}^{J}+{\rm h.c.}\,,
\eeq
where the Yukawa coupling proportional to $\boldsymbol{\hat{Y}}_\nu$
is the leptonic analogue of the Higgs-quark Yukawa coupling proportional
to $\boldsymbol{\hat{Y}}_u$ [cf.~\eq{hsmyukawa}].
In \eq{seesawlagrange}, we have distinguished the flavor labels of
three generations of Standard Model neutrino and
charged lepton fields (denoted
by lower case Roman letters $i$, $j,\ldots$)
and the flavor labels of singlet
neutrino fields (denoted by upper case Roman letters $I$, $J,\ldots$).
Note that $\boldsymbol{\hat Y}_\nu$ is a $3\times n$ matrix
and $\boldsymbol{\hat M}$ is an $n\times n$ matrix, where $n$ is
the number of singlet neutrino flavors.  In general, we shall
not specify the value of $n$, which may differ
from the number of Standard Model lepton generations.

If $\Lambda\equiv\|\boldsymbol{\hat M}\|\gg v$,\footnote{The
Euclidean matrix norm is defined by
$\|A\|\equiv\left[\Tr(A^\dagger A)\right]\llsup{1/2}=\left[\sum_{i,j}
|a_{ij}|^2\right]\llsup{1/2}$, for a matrix $A$ whose matrix elements
are given by $a_{ij}$.}
then a dimension-five
operator of the form given by \eq{dim5} is generated
in the effective theory at energy scales below $\Lambda$.
In this limit, we may neglect the kinetic
energy term of the gauge singlet neutrino fields.  Using the Lagrange
field equations, we may solve for $\hat{\bar\nu}^{I}$.  Inserting the solution
back into \eq{seesawlagrange} then yields \eq{dim5}, with
$\boldsymbol{\hat F}/\Lambda$ given by
\beq \label{hatFoverL}
\boldsymbol{\hat F}^{ij}/\Lambda=
-(\boldsymbol{\hat Y}_\nu)^i{}_K (\boldsymbol{\hat Y}_\nu)^j{}_N
(\boldsymbol{\hat M}^{-1})^{KN}\,.
\eeq

Using the definition of the SU(2)$_L$ doublet lepton
field given in Table \ref{SMfermions},
one can rewrite \eq{seesawlagrange} more explicitly as:
\beq \label{seesawlag}
\mathscr{L}_{\rm seesaw}= -(\boldsymbol{\hat{Y}}_\nu)^i{}_J
\left[\Phi^0\hat\nu_i
\hat{\bar\nu}^{J}-\Phi^+\hat\ell_i\hat{\bar\nu}^{J}\right]-\half
\boldsymbol{\hat M}_{IJ}
\hat{\bar\nu}^{I}\hat{\bar\nu}^{J}+{\rm h.c.}
\eeq
To analyze the physical consequences of the seesaw Lagrangian,
we first express \eq{seesawlag} in terms of the
unhatted mass eigenstate charged lepton fields
[cf.~\eq{leptonmasseigenstates}], and the light neutrino fields
$\breve\nu_i$ introduced in \eq{brevenu}.
It is also convenient to introduce new gauge singlet neutrino fields
$\breve{\bar\nu}^{J}$ by defining
\beq
\hat{\bar\nu}^{I}=N^I{}_J\breve{\bar\nu}^{J}\,,
\eeq
where $N$ is the unitary matrix that Takagi-diagonalizes the
complex symmetric matrix $\boldsymbol{\hat M}$.  That~is,
\beq \label{rhnumassmatrix}
\boldsymbol{M}\equiv N^{\T}\boldsymbol{\hat M} N=
{\rm diag}(M_1\,,\,M_2\,,\ldots\,,M_n)\,,
\eeq
where the $M_I$ are the singular values of $\boldsymbol{\hat M}$ (i.e., the
non-negative square roots of the eigenvalues
of $\boldsymbol{\hat M}^\dagger\boldsymbol{\hat M}$).
In terms of the mass eigenstate charged lepton fields $\ell_i$ and the neutrino
fields $\breve{\nu}_i$ and $\breve{\bar\nu}^{I}$, the seesaw Lagrangian
[\eq{seesawlag}] is then given by:
\beq \label{seesawlag2}
\mathscr{L}_{\rm seesaw}= -(\boldsymbol{Y}_\nu)^i{}_J
\left[\Phi^0\breve\nu_i
\breve{\bar\nu}^{J}-\Phi^+\ell_i\breve{\bar\nu}^{J}\right]-\half
\boldsymbol{M}_{IJ}
\breve{\bar\nu}^{I}\breve{\bar\nu}^{J}+{\rm h.c.}\,,
\eeq
where
\beq \label{boldY}
\boldsymbol{Y}_\nu\equiv L_\ell^{\T}\boldsymbol{\hat Y}_\nu N\,.
\eeq

As above, in the limit of $\Lambda\equiv\|\boldsymbol{\hat M}\|=
\|\boldsymbol{M}\|\gg v$,
it is also possible to directly generate the effective dimension-five
operator [\eq{dimen5}] in terms of the mass eigenstate charged lepton fields
and the new neutrino fields $\breve\nu_j$.  We then identify the
corresponding coefficient, $\boldsymbol{F}/\Lambda$, as
\beq \label{FoverL}
\boldsymbol{F}^{ij}/\Lambda=
- (\boldsymbol{Y}_\nu)^i{}_K (\boldsymbol{Y}_\nu)^j{}_N
(\boldsymbol{M}^{-1})^{KN}\,.
\eeq
Recalling that $\boldsymbol{F}=L_\ell^{\T}\boldsymbol{\hat F}L_\ell$,
one can check that \eq{FoverL} indeed follows from
\eqss{hatFoverL}{rhnumassmatrix}{boldY}.

To identify the neutrino mass matrix, we set $\Phi^0=v$ and $\Phi^+=\Phi^-=0$
in \eq{seesawlag2}:
\beq
-\mathscr{L}_{m_\nu} =
\half\, (\breve\nu_i \quad \breve{\bar\nu}^{J})\,\mathcal{M}_\nu
\left(\begin{array}{c}\breve\nu_k \\ \breve{\bar\nu}^{M} \\
\end{array}\right)+ {\rm h.c.}
\label{eq:mterms}
\eeq
The neutrino mass matrix $\mathcal{M}_\nu$ is a $(3+n)\times (3+n)$ complex
symmetric matrix given in block form by:
\beq \label{eq:vmass}
\mathcal{M}_\nu\equiv\left(\begin{array}{cc}
\mathds{O} & \,\,\, \boldsymbol{M_D} \\
\boldsymbol{M}^{\T}_{\boldsymbol{D}} & \,\,\, \boldsymbol{M}
\end{array}\right)\,,
\eeq
where $\mathds{O}$ is the $3\times 3$ zero matrix,
$\boldsymbol{M}$ is the diagonal matrix defined in
\eq{rhnumassmatrix} and
$\boldsymbol{M_D}$ is a $3\times n$ complex matrix (called the
Dirac neutrino mass matrix),
\beq \label{diracmm}
(\boldsymbol{M_D})^i{}_J\equiv v(\boldsymbol{Y}_\nu)^i{}_J\,.
\eeq
Note that if $n=3$ and
$\boldsymbol{M}=\mathds{O}$, then $\boldsymbol{M_D}$ is
a $3\times 3$ matrix that is simply
the leptonic analogue of the up-type quark mass matrix $\boldsymbol{M}_u$.
In this case, we would perform a singular value decomposition of
$\boldsymbol{M_D}$ and identify the unhatted neutrino mass eigenstate
fields, which can be assembled into three generations of
four-component Dirac neutrinos,
\beq
N_i= \begin{pmatrix} \nu_i\\[4pt] {\bar\nu}^{\dagger i}
\end{pmatrix}\,,\qquad i=1,2,3\,.
\eeq

In the seesaw model (with $n$ not specified), we assume that
$\|\boldsymbol{M}\|\gg \|\boldsymbol{M_D}\|$.  In this case, the neutrino
mass matrix can be perturbatively Takagi-block-diagonalized
as follows~\cite{dedes2,herrero,dhr}.
Introduce the $(3+n)\times (3+n)$ (approximate) unitary matrix:
\beq \label{udef}
\mathcal{U}=\left(\begin{array}{cc} \mathds{1}_{3\times 3} -\half
\boldsymbol{M}^*_{\boldsymbol{D}}\boldsymbol{M}^{-2}
\boldsymbol{M}^{\T}_{\boldsymbol{D}} &
\quad  \boldsymbol{M}^*_{\boldsymbol{D}}\boldsymbol{M}^{-1}
\\ -\boldsymbol{M}^{-1}\boldsymbol{M}^{\T}_{\boldsymbol{D}}
& \quad \mathds{1}_{n\times n} -\half
\boldsymbol{M}^{-1}\boldsymbol{M}^{\T}_{\boldsymbol{D}}
\boldsymbol{M}^*_{\boldsymbol{D}}\boldsymbol{M}^{-1}
\end{array}\right)\,,
\eeq
where $\mathds{1}$ is the identity matrix (whose dimension is explicitly
specified above).
We define transformed [light ($\ell$) and heavy ($h$)] neutrino states
$(\breve\nu_\ell)_i$ and $(\breve{\bar\nu}_h)^j$ by:
\beq \label{nutransform}
\begin{pmatrix} \breve\nu_i \\ \breve{\bar\nu}^{J} \end{pmatrix}=
\mathcal{U}\begin{pmatrix}(\breve\nu_\ell)_k \\
(\breve{\bar\nu}_h)^M \end{pmatrix}\,.
\eeq
By straightforward matrix multiplication, one can verify that
to second order accuracy in perturbation theory,
\beq \label{nublockdiag}
\mathcal{U}^{\T}\mathcal{M}_\nu \,\mathcal{U}\simeq
\begin{pmatrix} -\boldsymbol{M_D} \boldsymbol{M}^{-1}
\boldsymbol{M}^{\T}_{\boldsymbol{D}} & \mathds{O} \\
\mathds{O}^{\T} & \boldsymbol{M}+\half(\boldsymbol{M}^{-1}
\boldsymbol{M}^\dagger_{\boldsymbol{D}} \boldsymbol{M_D}+
\boldsymbol{M}^{\T}_{\boldsymbol{D}} \boldsymbol{M}^{*}_{\boldsymbol{D}}
\boldsymbol{M}^{-1})\end{pmatrix}\,,
\eeq
where $\mathds{O}$ is the $3\times n$ zero matrix.

We now can identify an effective $3\times 3$ complex symmetric mass
matrix $\boldsymbol{M_{\nu_\ell}}$ for the three light neutrinos
as the upper left-hand block of \eq{nublockdiag},
\beq
\boldsymbol{M_{\nu_\ell}} \simeq  -\boldsymbol{M\ls{D}}
\boldsymbol{M}^{-1} \boldsymbol{M}^{\T}_{\boldsymbol{D}} \,,
\label{eq:mnu}
\eeq
where corrections of $\mathcal{O}(v^4/\Lambda^3)$ have been neglected.
Using \eqs{FoverL}{diracmm}, we see that the light neutrino mass
matrix obtained in \eq{mnueff} has been correctly reproduced to leading
order in $v^2/\Lambda^2$.

The physical light neutrino mass eigenstate fields and their masses
are identified by \eqs{PMNS}{eq:vmphys}.
At energy scales below the heavy neutrino mass
scale, $\Lambda\equiv\|\boldsymbol{M}\|$, and
we can set $\breve{\bar\nu}_h=0$.  Neglecting corrections of
$\mathcal{O}(v^2/\Lambda^2)$, \eqst{diracmm}{eq:mnu}
imply that\footnote{Strictly speaking,
\eq{heavynu} should be written as:
\beq
(\boldsymbol{Y}_\nu)^i{}_J\breve{\bar\nu}^{J}\simeq\frac{1}{v}
\sum_{k,n} (\boldsymbol{U}^\dagger_{\rm{MNS}})_n{}^i\,
\delta^{nk}(m_{\nu_\ell})_k(\nu_{\ell})_k\,,\nonumber
\eeq
to maintain covariance in the flavor indices.}
\beqa
\breve\nu_i&\simeq &(\boldsymbol{U}_{\rm MNS})_i{}^j(\nu_\ell)_j\,,
\label{lightnu}\\[6pt]
(\boldsymbol{Y}_\nu)^i{}_J\breve{\bar\nu}^{J}&\simeq &
\frac{1}{v}(\boldsymbol{M_{\nu_\ell}U}_{\rm MNS})^{ik}(\nu_\ell)_k
=\frac{1}{v}
\sum_k (\boldsymbol{U}^\dagger_{\rm{MNS}})_k{}^i\,(m_{\nu_\ell})_k
(\nu_{\ell})_k\,,
\label{heavynu}
\eeqa
where in the last step above we have used \eq{eq:vmphys}
and $(\boldsymbol{U}^\dagger_{\rm MNS})_j{}^i \equiv
[(\boldsymbol{U}_{\rm MNS})_i{}^j]^*$.
Using \eqs{lightnu}{heavynu} to express the seesaw Lagrangian
in terms of the light neutrino mass eigenstate fields, one
can verify that the resulting interactions of the light neutrinos (and charged
leptons) to gauge bosons, the Higgs boson and the Nambu-Goldstone bosons
reproduce the results
of \eqs{wqqintmnu}{nunuhiggs} at leading order in $v^2/\Lambda^2$.

For completeness, we examine the effective $n\times n$
complex symmetric mass matrix of the heavy
neutrino states, $\boldsymbol{M_{\nu_h}}$, which is identified as the lower
right-hand block in \eq{nublockdiag},
\beq \label{effheavy}
\boldsymbol{M_{\nu_h}}\simeq\boldsymbol{M}+\half(\boldsymbol{M}^{-1}
\boldsymbol{M}^\dagger_{\boldsymbol{D}} \boldsymbol{M_D}+
\boldsymbol{M}^{\T}_{\boldsymbol{D}} \boldsymbol{M}^{*}_{\boldsymbol{D}}
\boldsymbol{M}^{-1})\,.
\eeq
Although $\boldsymbol{M}$ is diagonal by definition
[cf.~\eq{rhnumassmatrix}], the right-hand side of
\eq{effheavy} is no longer diagonal due to
the second order perturbative correction.  However, we do not have to
perform another Takagi diagonalization, since the off-diagonal
elements of the lower right-hand block only affect the
physical (diagonal) masses at higher order in perturbation theory.
Thus, we
identify the physical heavy neutrino mass eigenstates to leading order
by the unhatted fields,
\beq
{\bar\nu}_h^{J}\simeq \breve{\bar\nu}^{J}_h\,,
\eeq
with masses
\beq \label{heavynumass}
m_{\nu_{hJ}}\simeq M_J\left(1+\frac{1}{M_J^2}\sum_i |
(\boldsymbol{M_D})^i{}_J|^2\right)\,,
\eeq
where the $M_J$ are the diagonal elements of $\boldsymbol{M}$
(and no sum over the repeated index $J$ is implied).  That is, the masses
of the heavy neutrinos are simply given by $m_{\nu_{hJ}}\simeq M_J$,
up to corrections that are of the same order as the light neutrino masses.

The interactions of the heavy neutrinos can be likewise obtained.
The only unsuppressed interactions are heavy neutrino couplings to
the Higgs boson and Nambu-Goldstone bosons that are
proportional to the Dirac neutrino mass matrix,
\beq
\mathscr{L}_{\rm int}=-\frac{1}{\sqrt{2}v}(\boldsymbol{U}^{\T}_{\rm MNS}
\boldsymbol{M_D})^k{}_J{\bar\nu}_h^{J}(\nu_\ell)_k (h^0_{\rm SM}+iG^0)
+\frac{1}{v}(\boldsymbol{M_D})^i{}_J\ell_i{\bar\nu}_h^{J}G^+ + {\rm h.c.}
\eeq
All other couplings of the heavy neutrinos to the
$W^\pm$ and $Z$ bosons
(and additional contributions to the couplings of the heavy neutrinos to
the Higgs boson and Nambu-Goldstone bosons) are suppressed
by (at least) a factor of $\mathcal{O}(v/\Lambda)$.

\section{\texorpdfstring{MSSM fermion interaction vertices}{MSSM fermion interaction vertices}} 
\renewcommand{\theequation}{K.\arabic{equation}}
\renewcommand{\thefigure}{K.\arabic{figure}}
\renewcommand{\thetable}{K.\arabic{table}}
\setcounter{equation}{0}
\setcounter{figure}{0}
\setcounter{table}{0}

In this section, we provide the Feynman rules for the MSSM interaction
vertices.  To complete the tabulation of all MSSM Feynman rules, one
requires the rules for the purely bosonic interactions of the MSSM.
These can be found in \refs{rosiek}{mssmrules}.

\subsection{Higgs-fermion interaction vertices in the MSSM}
\label{subsec:Higgsfermionrules}
\renewcommand{\theequation}{K.1.\arabic{equation}}
\renewcommand{\thefigure}{K.1.\arabic{figure}}
\renewcommand{\thetable}{K.1.\arabic{table}}
\setcounter{equation}{0}
\setcounter{figure}{0}
\setcounter{table}{0}

The MSSM Higgs sector is a
two Higgs doublet model containing eight real scalar
degrees of freedom:
one complex $Y=-\half$ doublet, $\boldsymbol{H_d}=(H_d^0\,,\,H_d^-)$
and one complex $Y=+\half$ doublet,
$\boldsymbol{H_u}=(H_u^+\,,\, H_u^0)$.  The notation reflects the
form of the MSSM Higgs sector coupling to fermions:
\beq \label{hmssmyukawa}
\mathscr{L}_{\rm Y}=\epsilon^{ab}\left[(\boldsymbol{Y}_u)^i{}_j
(\boldsymbol{H_u})_a\boldsymbol{\hat Q}_{bi} \hat{\bar u}^{j}
-(\boldsymbol{Y}_d)^i{}_j (\boldsymbol{H_d})_a
\boldsymbol{\hat Q}_{bi}\hat{\bar d}^{j}
-(\boldsymbol{Y}_\ell)^i{}_j (\boldsymbol{H_d})_a
\boldsymbol{\hat L}_{bi}\hat{\bar\ell}^{j}\right]+{\rm h.c.}\,,
\eeq
where the hatted fields are interaction eigenstate quark and lepton
fields (with generation labels $i$ and $j$),
$a$ and $b$ are SU(2)$_L$ indices and the invariant SU(2)$_L$
tensor $\epsilon^{ab}$ is defined below \eq{hsmyukawa}.
That is, the neutral Higgs fields $H_d^0$
[$H_u^0$] couple exclusively to down-type [up-type] fermion pairs,
respectively.  In the supersymmetric model,
both hypercharge $Y=-\half$ and $Y=+\half$ complex Higgs doublets are
required in order that the theory (which now contains the corresponding
higgsino superpartners) remain anomaly free.
The supersymmetric structure of the theory forbids the coupling
of $\boldsymbol{H_u^\dagger}$ to $\hat{\bar d}^{j}$ and $\hat{\bar\ell}^{j}$
or the coupling of $\boldsymbol{H_d^\dagger}$ to $\hat{\bar u}^{j}$,
as such couplings would not be holomorphic.  Consequently, (at least)
two Higgs doublets are required in the MSSM
to generate mass for both ``up''-type and ``down''-type
quarks and charged leptons~\cite{fayethiggs,susyhiggs,gunhab}.

To find the couplings of the Higgs fields, we expand them around
the neutral Higgs field vacuum
expectation values $v_d\equiv \vev{H_d^0}$ and
$v_u\equiv\vev{H_u^0}$. Depending on the application, these
may be chosen to be the minimum of the tree-level scalar potential, or of the
full loop-corrected effective potential, or just left arbitrary.
It is always possible to choose the phases of the Higgs fields such
that $v_u$ and $v_d$ are real and positive.   We then define
\beq \label{tanbetadef}
\beta \equiv \tan^{-1}\left(\frac{v_u}{v_d}\right)\,,\qquad\qquad
0\leq\beta\leq\frac{\pi}{2}\,.
\eeq

The one potentially complex squared-mass
parameter that appears in the tree-level MSSM Higgs scalar potential is
necessarily real in the convention where the vacuum expectation values
of the neutral Higgs fields are real and positive.\footnote{The
coefficients of the quartic terms of the tree-level MSSM Higgs potential
are related to the electroweak gauge couplings and are manifestly real,
independently of the convention for the phases of the Higgs fields.}
Consequently, the tree-level MSSM
Higgs sector conserves CP, which implies that the neutral Higgs
mass eigenstates possess definite CP quantum numbers.\footnote{When
one-loop corrections are taken into account, new MSSM phases can enter
in the loops that cannot be removed.  In this case, the physical
neutral Higgs states can be mixtures of
CP-even and CP-odd scalar states~\cite{cpreview}.}
Spontaneous electroweak
symmetry breaking results in three Goldstone bosons $G^\pm$, $G^0$
(the neutral Goldstone boson is a CP-odd scalar field), which
are absorbed and become the longitudinal components of the $W^\pm$ and
$Z$.
The remaining five physical Higgs particles consist of a charged
Higgs pair $H^\pm$, one CP-odd scalar $A^0$, and two CP-even scalars
$h^0$ and $H^0$.

It is convenient to define $H_u^-\equiv (H_u^+)^\dagger$ and $H_d^+\equiv
(H_d^-)^\dagger$.
One can then parameterize the mixing angles between Higgs
gauge eigenstates and mass eigenstates by writing:
\beqa
H_u^0 &=& v_u + \frac{1}{\sqrt{2}} \sum_{\phi^0} k_{u\phi^0} \phi^0,\qquad
\qquad H_u^{\pm} = \sum_{\phi^{\pm}} k_{u\phi^{\pm}} \phi^{\pm}\,,
\label{eq:Huneutmasseigenstates}
\\
H_d^0 &=& v_d + \frac{1}{\sqrt{2}} \sum_{\phi^0} k_{d\phi^0} \phi^0,
\qquad\qquad H_d^{\pm}= \sum_{\phi^{\pm}} k_{d\phi^{\pm}} \phi^{\pm}\,.
\label{eq:Hdminusmasseigenstates}
\eeqa
For $\phi^\pm = (H^\pm,\> G^\pm)$,\footnote{Note that
$\phi^-\equiv(\phi^+)^\dagger$.  Since the
$k_{f\phi^{\pm}}$ (for $f=u,d$) are real quantities,
we adopt the notation in which
$k_{f\phi^+}=k_{f\phi^-}\equiv k_{\phi^\pm}$ and
$\beta_+=\beta_{-}\equiv\beta_\pm$.}
\beqa
k_{u\phi^\pm} &=& (\cos\beta_\pm, \>\, \sin\beta_{\pm})\,,
\label{kdef1}\\
k_{d\phi^\pm} &=& (\sin\beta_\pm, \>\, -\cos\beta_{\pm})\,,
\label{kdef2}
\eeqa
and for $\phi^0 = (h^0,\, H^0,\, A^0,\, G^0)$,
\beqa
k_{u\phi^0} &=&
(\cos\alpha,\>\,\sin\alpha,\>\,i\cos\beta_0,\>\,i\sin\beta_0)\,,
\label{eq:defkuphi0}
\\
k_{d\phi^0} &=&
(-\sin\alpha,\>\,\cos\alpha,\>\,i\sin\beta_0,\>\,-i\cos\beta_0)\,,
\label{eq:defkdphi0}
\eeqa
where the mixing angle $\alpha$ parameterizes the orthogonal
matrix that diagonalizes the $2\times 2$ CP-even Higgs squared-mass
matrix $\mathcal{M}_0^2$ [defined in \eq{cpevenmatrix} below].  

In \eqs{eq:Huneutmasseigenstates}{eq:Hdminusmasseigenstates}, the
normalization of the vacuum expectation values is
\beq \label{vuvddef}
v_d^2+v_u^2={2m_W^2/g^2} \simeq (174~{\rm GeV})^2\,,
\eeq
if one chooses
$v_u$, $v_d$ to be near the true minimum
of the Higgs effective potential.
Note that in the special case that $v_u$ and
$v_d$ are at the
minimum of the {\em tree-level} potential, the mixing angles
$\beta_\pm$
in the charged Higgs sector and
$\beta_0$ in the CP-odd neutral Higgs sectors coincide such that
$\beta_\pm=\beta_0=\beta$, where $\beta$ is defined in \eq{tanbetadef}.
However, if one expands around a more general choice of $v_u,v_d$,
including for example the minimum of the full effective potential,
then the tree-level mixing angles $\beta_0$ and $\beta_\pm$ are
distinct from each other and from $\beta$. (Depending on the choice of
renormalization scale for a particular calculation, the tree-level
potential in the MSSM may have a very different minimum from the true
minimum of the full effective potential, or may not have a proper
minimum at all.) Therefore, we do not assume anything specific about
$v_u$ and $v_d$ except that they are real and positive by convention.

All MSSM Higgs boson masses and the mixing angle
$\alpha$ are determined at tree level by
two Higgs sector parameters, usually taken to be 
the ratio of the tree-level vacuum expectation values, 
$\tan\beta=v_u/v_d$, and
the mass of the CP-odd Higgs scalar, $m_A$~\cite{susyhiggs,gunhab}.
The tree-level value of the squared mass of the charged Higgs boson is given by
\beq
m_{H^\pm}^2 =m_{A}^2+m_W^2\,.
\eeq
The CP-even Higgs bosons $h^0$ and $H^0$ are eigenstates of the
tree-level squared-mass matrix,
\beq \label{cpevenmatrix}
\mathcal{M}_0^2 =
\begin{pmatrix}m_A^2 \sin^2\beta + m^2_Z \cos^2\beta &\quad
           -(m_A^2+m^2_Z)\sin\beta\cos\beta \\
  -(m_A^2+m^2_Z)\sin\beta\cos\beta&\quad
  m_A^2\cos^2\beta+ m^2_Z \sin^2\beta \end{pmatrix}\,.
\eeq
The eigenvalues of $\mathcal{M}_0^2$ are
the tree-level squared masses of the two CP-even Higgs scalars,
\beq \label{h0mass}
  m^2_{H,h} = \half \left( m_A^2 + m^2_Z \pm
                  \sqrt{(m_A^2+m^2_Z)^2 - 4m^2_Z m_A^2 \cos^2 2\beta}
                  \; \right)\,,
\eeq
with $m_h\leq m_H$.
The angle $\alpha$ of the orthogonal matrix
that diagonalizes $\mathcal{M}_0^2$ is given
by~\cite{gunhab2}:
\beq
\sin 2\alpha=-\sin 2\beta\left(\frac{m_A^2+m_Z^2}{m_H^2-m_h^2}\right)\,,
\qquad\quad
\cos 2\alpha=-\cos 2\beta\left(\frac{m_A^2-m_Z^2}{m_H^2-m_h^2}\right)\,.
\eeq
Since $\sin 2\alpha\leq 0$, the tree-level value of $\alpha$ is
restricted to lie in the range $-\pi/2\leq\alpha\leq 0$.

Radiative corrections can have a significant impact on the 
tree-level Higgs masses and mixing angle
$\alpha$~\cite{cpreview,hempfling}.  For example, the tree-level bound
$m_h\leq m_Z|\cos 2\beta|\leq m_Z$ [which follows from \eq{h0mass}]
is significantly modified by an incomplete cancellation of top
quark and top squark loop corrections.  Including the latter implies
that $m_h\lsim 135$~GeV~\cite{higgsradcor}, which (in contrast to 
the tree-level prediction) is not experimentally excluded.

The Higgs-fermion Yukawa couplings in the gauge-interaction basis are
given by \eq{hmssmyukawa}.
Explicitly,
\beqa \label{lintyuk}
-\mathscr{L}_{\rm Y} &=& (\mathbold{Y}_u)^i{}_j
\left[\hat u_i\hat{\bar u}^{j} H_u^0
-
\hat d_i\hat{\bar u}^{j} H_u^+\right]
+(\mathbold{Y}_d)^i{}_j
\left[\hat d_i\hat{\bar d}^{j} H_d^0-\hat u_i\hat{\bar d}^{j} H_d^-\right]
\nonumber \\
&&\qquad\qquad\qquad
+(\mathbold{Y}_\ell)^i{}_j
\left[\hat \ell_i\hat{\bar\ell}^{j} H_d^0-\hat \nu_i\hat{\bar\ell}^{j}
H_d^-\right]+ {\rm h.c.}
\eeqa
We use
\eqs{eq:Huneutmasseigenstates}{eq:Hdminusmasseigenstates} to express
the interaction-eigenstate Higgs fields in terms of the physical
Higgs fields and Goldstone fields.  We can identify the quark
and lepton mass matrices simply by setting $H_u^0=v_u$, $H_d^0=v_d$
and $H_u^+=H_d^-=0$ in \eq{lintyuk}\,,
\beq
(\boldsymbol{M}_u)^i{}_j = v_u (\mathbold{Y}_u)^i{}_j\,,\qquad
(\boldsymbol{M}_d)^i{}_j = v_d (\mathbold{Y}_d)^i{}_j\,,\qquad
(\boldsymbol{M}_\ell)^i{}_j = v_d (\mathbold{Y}_\ell)^i{}_j\,.
\eeq
We then use \eqs{quarkmasseigenstates}{leptonmasseigenstates} to express
the interaction-eigenstate quark and lepton fields in terms
of the corresponding mass eigenstate fields.  \Eqs{luru}{lere} ensure that
the fermion mass matrices are diagonal (with real non-negative elements)
in the fermion mass eigenstate basis.  In this basis, the
resulting neutral Higgs-fermion interactions are diagonal.
Here, the diagonalized Higgs-fermion Yukawa coupling matrices
appear:
\beqa
{\rm diag}(Y_{u1}, Y_{u2},Y_{u3})
\equiv
{\rm diag}(Y_{u}, Y_{c},Y_{t})
&=& L_u^{\T} \mathbold{Y}_u R_u\,,
\label{Yudef}
\\
{\rm diag}(Y_{d1}, Y_{d2},Y_{d3})
\equiv
{\rm diag}(Y_{d}, Y_{s},Y_{b})
&=& L_d^{\T} \mathbold{Y}_d R_d\,,
\label{Yddef}
\\
{\rm diag}(Y_{\ell 1}, Y_{\ell 2},Y_{\ell 3})
\equiv
{\rm diag}(Y_{e}, Y_{\mu},Y_{\tau})
&=& L_\ell^{\T} \mathbold{Y}_\ell R_\ell\,.
\label{Yelldef}
\eeqa
The diagonalized Yukawa couplings
are related to the corresponding
fermion masses by
\beq
Y_{ui}={m_{u_i}/v_u}\,,\qquad\qquad
Y_{di}={m_{d_i}/v_d}\,,\qquad\qquad
Y_{\ell i}={m_{\ell _i}/v_d}\,.
\eeq

We have used the same symbol for the Yukawa couplings in
the MSSM as we did for the Standard
Model Yukawa couplings in \app{J.1}. However, it is important
to note that the MSSM Yukawa couplings are normalized differently because of
the presence of two neutral Higgs field vacuum expectation values.
Using a superscript SM to denote the Standard Model Yukawa
couplings of \app{J.1}, the MSSM Yukawa couplings defined here
are related by:
\beq \label{yukcompare}
Y_{ui} = Y^{\rm SM}_{ui}/\sin\beta\,,\qquad\quad
Y_{di} = Y^{\rm SM}_{di}/\cos\beta\,,\qquad\quad
Y_{\ell i} = Y^{\rm SM}_{\ell i}/\cos\beta\,.
\eeq

\begin{figure}[t!]
\begin{center}
\begin{picture}(200,57)(7,15)
\DashLine(60,40)(10,40)5
\ArrowLine(100,70)(60,40)
\ArrowLine(100,10)(60,40)
\Text(30,29)[]{$\phi^0$}
\Text(70,20)[]{${\bar u}^{j}$}
\Text(70,62)[]{$u_i$}
\Text(165,40)[]{$-\nicefrac{i}{\sqrt{2}} Y_{ui} k_{u\phi^0}
\delta_j^i \, \delta_{\alpha}{}^{\beta} $}
\Text(110,70)[]{$\alpha$}
\Text(110,10)[]{$\beta$}
\end{picture}
\hspace{1.4cm}
\begin{picture}(200,57)(7,15)
\DashLine(10,40)(60,40)5
\ArrowLine(60,40)(100,70)
\ArrowLine(60,40)(100,10)
\Text(30,29)[]{$\phi^0$}
\Text(70,20)[]{${\bar u}^{j}$}
\Text(70,62)[]{$u_i$}
\Text(165,40)[]{$-\nicefrac{i}{\sqrt{2}} Y_{ui} k_{u\phi^0}^*
\delta_i^j
\, \delta^{\dot{\alpha}}{}_{\dot{\beta}} $}
\Text(110,70)[]{$\dot{\alpha}$}
\Text(110,10)[]{$\dot{\beta}$}
\end{picture}
\end{center}
\vspace{0.35cm}
\begin{center}
\begin{picture}(200,53)(7,16)
\DashLine(60,40)(10,40)5
\ArrowLine(100,70)(60,40)
\ArrowLine(100,10)(60,40)
\Text(30,29)[]{$\phi^0$}
\Text(70,20)[]{${\bar d}^{j}$}
\Text(70,62)[]{$d_i$}
\Text(165,40)[]{$-\nicefrac{i}{\sqrt{2}} Y_{di} k_{d\phi^0}
\delta_j^i
\, \delta_{\alpha}{}^{\beta} $}
\Text(110,70)[]{$\alpha$}
\Text(110,10)[]{$\beta$}
\end{picture}
\hspace{1.4cm}
\begin{picture}(200,53)(7,16)
\DashLine(10,40)(60,40)5
\ArrowLine(60,40)(100,70)
\ArrowLine(60,40)(100,10)
\Text(30,29)[]{$\phi^0$}
\Text(70,20)[]{${\bar d}^{j}$}
\Text(70,62)[]{$d_i$}
\Text(165,40)[]{$-\nicefrac{i}{\sqrt{2}} Y_{di} k_{d\phi^0}^*
\delta_i^j
\, \delta^{\dot{\alpha}}{}_{\dot{\beta}} $}
\Text(110,70)[]{$\dot{\alpha}$}
\Text(110,10)[]{$\dot{\beta}$}
\end{picture}
\end{center}
\vspace{0.35cm}
\begin{center}
\begin{picture}(200,53)(7,16)
\DashLine(60,40)(10,40)5
\ArrowLine(100,70)(60,40)
\ArrowLine(100,10)(60,40)
\Text(30,29)[]{$\phi^0$}
\Text(70,20)[]{$\bar\ell$}
\Text(70,62)[]{$\ell$}
\Text(165,40)[]{$-\nicefrac{i}{\sqrt{2}} Y_{\ell} k_{d\phi^0}
\, \delta_{\alpha}{}^{\beta} $}
\Text(110,70)[]{$\alpha$}
\Text(110,10)[]{$\beta$}
\end{picture}
\hspace{1.4cm}
\begin{picture}(200,53)(7,16)
\DashLine(10,40)(60,40)5
\ArrowLine(60,40)(100,70)
\ArrowLine(60,40)(100,10)
\Text(30,29)[]{$\phi^0$}
\Text(70,20)[]{$\bar\ell$}
\Text(70,62)[]{$\ell$}
\Text(165,40)[]{$-\nicefrac{i}{\sqrt{2}} Y_{\ell} k_{d\phi^0}^*
\, \delta^{\dot{\alpha}}{}_{\dot{\beta}} $}
\Text(110,70)[]{$\dot{\alpha}$}
\Text(110,10)[]{$\dot{\beta}$}
\end{picture}
\end{center}
\caption{\label{nehiggsqq}Feynman rules for the interactions of neutral
Higgs bosons $\phi^0 = (h^0, H^0, A^0, G^0)$ with fermion-antifermion
pairs in the MSSM. The repeated index $i$ is not summed.}
\end{figure}
The interactions of the neutral Higgs and Goldstone scalars
$\phi^0 = (h^0, H^0, A^0, G^0)$
with Standard Model fermions are given in
\fig{nehiggsqq}.
Note that the rules involving undotted spinor indices are proportional to either
couplings $k_{d\phi^0}$ and
$k_{u\phi^0}$, whereas the rules involving dotted spinor indices
are proportional to the corresponding complex
conjugated couplings.  For the CP-even scalars, $h^0$ and $H^0$, the
corresponding couplings are real.  Hence,
starting with the rule for the coupling of the CP-even neutral scalars to
fermions with undotted indices, one obtains the
corresponding rule for the coupling to
fermions with dotted indices (with the direction of the
arrows reversed) by taking $\delta_{\alpha}{}^{\beta}\rightarrow
\delta^{\dot{\alpha}}{}_{\dot{\beta}}$.
In contrast, for the CP-odd scalars, $A^0$ and $G^0$,
the corresponding couplings $k_{d\phi^0}$ and
$k_{u\phi^0}$ are purely imaginary.
Therefore, starting with the rule for the coupling of the CP-odd neutral scalars to
fermions with undotted indices, one obtains the
corresponding rule for the coupling to
fermions with dotted indices (with the direction of the
arrows reversed) by taking $\delta_{\alpha}{}^{\beta}\rightarrow
-\delta^{\dot{\alpha}}{}_{\dot{\beta}}$.
The latter minus sign is a signal that
$\ha$ and $G^0$ are CP-odd scalars.
In particular, due to the fact that
the Feynman rules for
$\ha$ and $G^0$ arise from a term in $\mathscr{L}_{\rm int}$
proportional to $i~{\rm Im}~H^0$, the latter $i$ flips sign when the
rule is conjugated resulting in the extra minus sign noted above.
As an additional consequence, since the Feynman rules
are obtained from $i\mathscr{L}_{\rm int}$, the overall $\ha$ and
$G^0$ rules are real.
\begin{figure}[t!]
\begin{center}
\begin{picture}(200,63)(0,16)
\DashArrowLine(10,40)(62,40)5
\ArrowLine(100,70)(60,40)
\ArrowLine(100,10)(60,40)
\Text(30,30)[]{$\phi^+$}
\Text(70,20)[]{$d_i$}
\Text(70,62)[]{${\bar u}^{j}$}
\Text(165,40)[]{$i Y_{uj} [\mathbold{K}]_j{}^i
k_{u\phi^\pm}
\, \delta_{\alpha}{}^{\beta} $}
\Text(110,70)[]{$\alpha$}
\Text(110,10)[]{$\beta$}
\end{picture}
\hspace{1.4cm}
\begin{picture}(200,63)(0,16)
\DashArrowLine(10,40)(60,40)5
\ArrowLine(60,40)(100,70)
\ArrowLine(60,40)(100,10)
\Text(30,30)[]{$\phi^-$}
\Text(70,20)[]{$d_i$}
\Text(70,62)[]{${\bar u}^{j}$}
\Text(165,40)[]{$iY_{uj}[\mathbold{K}^\dagger]_i{}^j
k_{u\phi^\pm}
\, \delta^{\dot{\alpha}}{}_{\dot{\beta}}$}
\Text(110,70)[]{$\dot{\alpha}$}
\Text(110,10)[]{$\dot{\beta}$}
\end{picture}
\end{center}
\vspace{0.35cm}
\begin{center}
\begin{picture}(200,55)(0,15)
\DashArrowLine(10,40)(60,40)5
\ArrowLine(100,70)(60,40)
\ArrowLine(100,10)(60,40)
\Text(30,30)[]{$\phi^-$}
\Text(70,20)[]{${\bar d}^{j}$}
\Text(70,62)[]{$u_i$}
\Text(165,40)[]{$iY_{dj}[\mathbold{K}^\dagger]_j{}^i k_{d\phi^\pm}
\,\delta_{\alpha}{}^{\beta} $}
\Text(110,70)[]{$\alpha$}
\Text(110,10)[]{$\beta$}
\end{picture}
\hspace{1.4cm}
\begin{picture}(200,55)(0,15)
\DashArrowLine(10,40)(60,40)5
\ArrowLine(60,40)(100,70)
\ArrowLine(60,40)(100,10)
\Text(30,30)[]{$\phi^+$}
\Text(70,20)[]{${\bar d}^{j}$}
\Text(70,62)[]{$u_i$}
\Text(165,40)[]{$iY_{dj}[\mathbold{K}]_i{}^j k_{d\phi^\pm}
\, \delta^{\dot{\alpha}}{}_{\dot{\beta}}$}
\Text(110,70)[]{$\dot{\alpha}$}
\Text(110,10)[]{$\dot{\beta}$}
\end{picture}
\end{center}
\vspace{0.35cm}
\begin{center}
\begin{picture}(200,55)(0,15)
\DashArrowLine(10,40)(60,40)5
\ArrowLine(100,70)(60,40)
\ArrowLine(100,10)(60,40)
\Text(30,30)[]{$\phi^-$}
\Text(70,20)[]{$\bar\ell$}
\Text(70,62)[]{$\nu_\ell$}
\Text(165,40)[]{$iY_{\ell} k_{d\phi^\pm}
\,\delta_{\alpha}{}^{\beta} $}
\Text(110,70)[]{$\alpha$}
\Text(110,10)[]{$\beta$}
\end{picture}
\hspace{1.4cm}
\begin{picture}(200,55)(0,15)
\DashArrowLine(10,40)(60,40)5
\ArrowLine(60,40)(100,70)
\ArrowLine(60,40)(100,10)
\Text(30,30)[]{$\phi^+$}
\Text(70,20)[]{$\bar\ell$}
\Text(70,62)[]{$\nu_\ell$}
\Text(165,40)[]{$iY_{\ell} k_{d\phi^\pm}
\, \delta^{\dot{\alpha}}{}_{\dot{\beta}}$}
\Text(110,70)[]{$\dot{\alpha}$}
\Text(110,10)[]{$\dot{\beta}$}
\end{picture}
\end{center}
\caption{Feynman rules for the interactions of charged Higgs bosons
$\phi^\pm = (H^\pm,G^\pm)$ with
fermion-antifermion pairs in the MSSM. The repeated index $j$ is not summed.}
\label{chiggsqq}
\end{figure}

The couplings of the charged Higgs and
Goldstone bosons to quark-antiquark pairs
are not flavor-diagonal and involve the CKM
matrix $\mathbold{K}$.  Starting with \eq{lintyuk},
and changing to the mass eigenstate basis as before,
we make use of \eqs{lyr1}{lyr2} to obtain
\beqa
\mathscr{L}_{\rm int}&=& Y_{ui}[\boldsymbol{K}]_i{}^j  d_j {\bar u}^{i}
H^+\cos\beta_{\pm} +
Y_{di}[\boldsymbol{K}^\dagger]_i{}^j u_j {\bar d}^{i} H^-\sin\beta_{\pm}+
Y_{\ell i}\nu_i {\bar\ell}^{i} H^-\sin\beta_{\pm}
\phantom{aaaaaaaaaaaaa}\nonumber \\
&&
+Y_{ui}[\boldsymbol{K}]_i{}^j  d_j {\bar u}^{i}G^+\sin\beta_{\pm} -
Y_{di}[\boldsymbol{K}^\dagger]_i{}^j u_j {\bar d}^{i} G^-\cos\beta_{\pm} -
Y_{\ell i}\nu_i {\bar\ell}^{i} G^-\cos\beta_{\pm}
+{\rm h.c.}
\eeqa
The resulting charged scalar Feynman
rules of the MSSM are given in \fig{chiggsqq}.  Note that
when \eq{yukcompare} is taken into account, 
the fermion couplings to the neutral and charged
Goldstone bosons are equivalent
to those of the Standard Model [cf.~\eq{gbosonrules}] if we choose
$\beta_0=\beta_{\pm}=\beta$.

\subsection{Gauge interaction vertices for neutralinos and charginos}
\renewcommand{\theequation}{K.2.\arabic{equation}}
\renewcommand{\thefigure}{K.2.\arabic{figure}}
\renewcommand{\thetable}{K.2.\arabic{table}}
\setcounter{equation}{0}
\setcounter{figure}{0}
\setcounter{table}{0}

Following eqs.~(C83) and (C88)
of \Ref{HaberKane}, we define:
\beqa
O^L_{ij}&=&-\nicefrac{1}{\sqrt{2}}N_{i4}V_{j2}^*+N_{i2}V_{j1}^*\,,
\label{eq:defOL}
\\
O^R_{ij}&=&\phm\nicefrac{1}{\sqrt{2}}N_{i3}^\ast U_{j2}+N_{i2}^\ast U_{j1}\,,
\label{eq:defOR}
\\
O^{\prime L}_{ij}&=&-V_{i1}V_{j1}^*-\half V_{i2}V_{j2}^*+\delta_{ij}s_W^2\,,
\label{eq:defOLp}
\\
O^{\prime R}_{ij}&=&-U_{i1}^\ast U_{j1}-\half U_{i2}^\ast U_{j2}+\delta_{ij}s_W^2\,,
\label{eq:defORp}
\\
O^{\prime\prime L}_{ij}&=&-O^{\prime\prime R}_{ji}=
\half(N_{i4}N_{j4}^*-N_{i3}N_{j3}^*)\,,
\label{eq:defOLpp}
\eeqa
where $s_W\equiv\sin\theta_W$.
Here $U$ and $V$ are the unitary matrices that diagonalize the chargino
mass matrix via the singular value decomposition:
\beq
U^* M_{\psi^\pm}
V^{-1}={\rm diag}(m_{{\widetilde C}_1},m_{{\widetilde C}_2})\,,
\label{u-and-v}
\eeq
with
\beq \label{Charginomassmatrix}
M_{\psi^\pm} =\begin{pmatrix} M_2 & g v_u \\[4pt] g v_d & \mu\end{pmatrix}\,.
\eeq
Similarly, $N$ is a unitary matrix that Takagi-diagonalizes the neutralino mass
matrix,
\beq
N^* M_{\psi^0} N^{-1}
={\rm diag}(m_{{\widetilde N}_1},m_{{\widetilde N}_2},
m_{{\widetilde N}_3},m_{{\widetilde N}_4})\,,
\label{eq:neutmix}
\eeq
with
\beq \label{Neutralinomassmatrix}
M_{\psi^0} =\begin{pmatrix}
M_1 & 0   & -g' v_d/\sqrt{2} & g' v_u/\sqrt{2}  \\[4pt]
0   & M_2 &  g v_d/\sqrt{2}  &  -g v_u/\sqrt{2} \\[4pt]
-g' v_d/\sqrt{2} & g v_d/\sqrt{2} & 0  & -\mu \\[4pt]
g' v_u/\sqrt{2} & -g v_u/\sqrt{2} & -\mu & 0\end{pmatrix}\,.
\eeq
As noted above \eq{tanbetadef},
we work in a convention in which $v_u$ and $v_d$
are real and positive.  The gaugino mass parameters $M_1$, $M_2$
and the higgsino mass parameter $\mu$ are potentially complex.

We now list the gauge boson interactions with the neutralinos and
charginos in the form of Feynman rules.  Here, we make use of the results
presented in \figst{fig:Gaugevertexrules}{fig:GaugevertexMajDirac}.
The Feynman rules for $Z$ and $\gamma$
interactions with charginos and neutralinos are given in \fig{nnboson}
and the corresponding rules for $W^\pm$ interactions are given in
\fig{ccboson}.
For each of these rules, one has a version with lowered
spinor indices by replacing $\sigmabar_\mu^{\dot{\alpha}\beta} \ra
-\sigma_{\mu\beta\dot{\alpha}}$.  We label fermion lines with the symbols
of the two-component fermion fields as given in
Table~\ref{tab:nomenclature}.
\begin{figure}[b!]
\begin{flushleft}
\begin{picture}(200,68)(0,0)
\Photon(60,40)(10,40){3}{5}
\ArrowLine(60,40)(100,70)
\ArrowLine(100,10)(60,40)
\Text(30,25)[]{$\gamma$}
\Text(70,20)[]{$\chi_j^+$}
\Text(70,67)[]{$\chi_i^+$}
\Text(135,40)[l]{$\BDneg ie\,
\delta_{ij}\sigmabar_\mu^{\dot{\alpha}\beta}$}
\Text(110,70)[]{$\dot{\alpha}$}
\Text(110,10)[]{$\beta$}
\Text(12,50)[]{$\mu$}
\end{picture}
\hspace{1.4cm}
\begin{picture}(200,68)(0,0)
\Photon(60,40)(10,40){3}{5}
\ArrowLine(60,40)(100,70)
\ArrowLine(100,10)(60,40)
\Text(30,25)[]{$\gamma$}
\Text(70,20)[]{$\chi_j^-$}
\Text(70,67)[]{$\chi_i^-$}
\Text(135,40)[l]{$\BDpos
ie\,\delta_{ij}\sigmabar_\mu^{\dot{\alpha}\beta}$}
\Text(110,70)[]{$\dot{\alpha}$}
\Text(110,10)[]{$\beta$}
\Text(12,50)[]{$\mu$}
\end{picture}
\end{flushleft}
\vspace{0.1cm}
\begin{flushleft}
\begin{picture}(200,68)(0,0)
\Photon(60,40)(10,40){3}{5}
\ArrowLine(60,40)(100,70)
\ArrowLine(100,10)(60,40)
\Text(30,25)[]{$Z$}
\Text(70,20)[]{$\chi_j^+$}
\Text(70,67)[]{$\chi_i^+$}
\Text(130,40)[l]{$\BDpos i{\displaystyle\frac{g}{\cw}}O^{'L}_{ij}
\sigmabar^{\dot{\alpha}\beta}_\mu
$}
\Text(110,70)[]{$\dot{\alpha}$}
\Text(110,10)[]{$\beta$}
\Text(12,50)[]{$\mu$}
\end{picture}
\hspace{1.4cm}
\begin{picture}(200,58)(0,10)
\Photon(60,40)(10,40){3}{5}
\ArrowLine(60,40)(100,70)
\ArrowLine(100,10)(60,40)
\Text(30,25)[]{$Z$}
\Text(70,20)[]{$\chi_i^-$}
\Text(70,67)[]{$\chi_j^-$}
\Text(132,40)[l]{$\BDneg i{\displaystyle\frac{g}{\cw}}O^{'R}_{ij}
\sigmabar^{\dot{\alpha}\beta}_\mu$}
\Text(110,70)[]{$\dot{\alpha}$}
\Text(110,10)[]{$\beta$}
\Text(12,50)[]{$\mu$}
\end{picture}
\end{flushleft}
\vspace{0.1cm}
\begin{flushleft}
\begin{picture}(200,58)(0,10)
\Photon(60,40)(10,40){3}{5}
\ArrowLine(60,40)(100,70)
\ArrowLine(100,10)(60,40)
\Text(30,25)[]{$Z$}
\Text(70,20)[]{$\chi^0_j$}
\Text(70,67)[]{$\chi^0_i$}
\Text(130,40)[l]{$\BDpos i{\displaystyle\frac{g}{c_W}}O^{''L}_{ij}
\sigmabar^{\dot{\alpha}\beta}_\mu$}
\Text(110,70)[]{$\dot{\alpha}$}
\Text(110,10)[]{$\beta$}
\Text(12,50)[]{$\mu$}
\end{picture}
\end{flushleft}
\caption{Feynman rules for the
chargino and neutralino interactions with neutral gauge bosons.
The coupling matrices are defined in
\eqst{eq:defOLp}{eq:defOLpp} and $\cw\equiv\cos\theta_W$.
For each rule, a corresponding one with lowered spinor
indices is obtained by $\sigmabar_\mu^{\dot\alpha\beta} \rightarrow
-\sigma_{\mu\beta\dot\alpha}$.
}
\label{nnboson}
\end{figure}
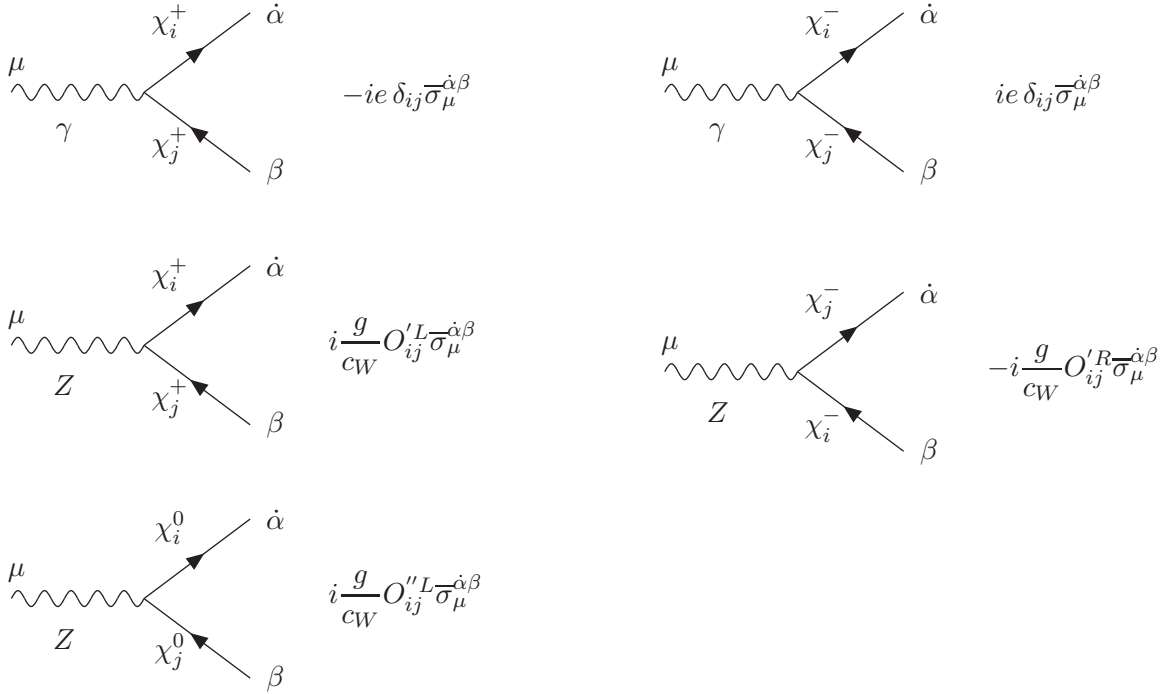
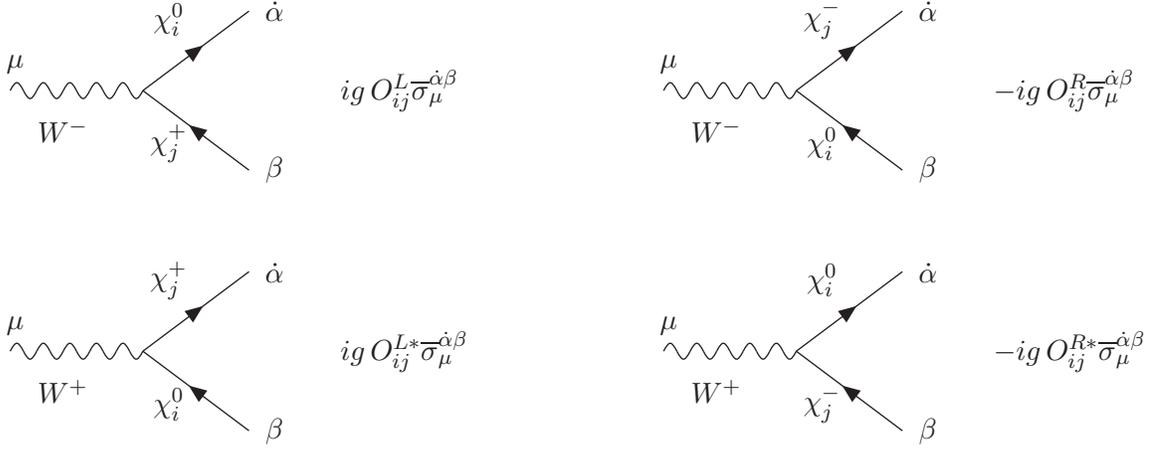
\begin{figure}[tpb]
\begin{center}
\begin{picture}(200,68)(0,0)
\Photon(60,40)(10,40){3}{5}
\ArrowLine(60,40)(100,70)
\ArrowLine(100,10)(60,40)
\Text(30,25)[]{$W^-$}
\Text(70,20)[]{$\chi_j^+$}
\Text(70,67)[]{$\chi_i^0$}
\Text(135,40)[l]{$\BDpos ig\,O_{ij}^L\sigmabar_\mu^{\dot{\alpha}\beta}$}
\Text(110,70)[]{$\dot{\alpha}$}
\Text(110,10)[]{$\beta$}
\Text(12,50)[]{$\mu$}
\end{picture}
\hspace{1.4cm}
\begin{picture}(200,68)(0,0)
\Photon(60,40)(10,40){3}{5}
\ArrowLine(60,40)(100,70)
\ArrowLine(100,10)(60,40)
\Text(30,25)[]{$W^-$}
\Text(70,20)[]{$\chi^0_i$}
\Text(70,67)[]{$\chi_j^-$}
\Text(135,40)[l]{$\BDneg ig\,O_{ij}^R\sigmabar_\mu^{\dot{\alpha}\beta}$}
\Text(110,70)[]{$\dot{\alpha}$}
\Text(110,10)[]{$\beta$}
\Text(12,50)[]{$\mu$}
\end{picture}
\end{center}
\vspace{0.2cm}
\begin{center}
\begin{picture}(200,58)(0,10)
\Photon(60,40)(10,40){3}{5}
\ArrowLine(60,40)(100,70)
\ArrowLine(100,10)(60,40)
\Text(30,25)[]{$W^+$}
\Text(70,20)[]{$\chi^0_i$}
\Text(70,67)[]{$\chi_j^+$}
\Text(135,40)[l]{$\BDpos
ig\,O_{ij}^{L*}\sigmabar_\mu^{\dot{\alpha}\beta}$}
\Text(110,70)[]{$\dot{\alpha}$}
\Text(110,10)[]{$\beta$}
\Text(12,50)[]{$\mu$}
\end{picture}
\hspace{1.4cm}
\begin{picture}(200,58)(0,10)
\Photon(60,40)(10,40){3}{5}
\ArrowLine(60,40)(100,70)
\ArrowLine(100,10)(60,40)
\Text(30,25)[]{$W^+$}
\Text(70,20)[]{$\chi_j^-$}
\Text(70,67)[]{$\chi^0_i$}
\Text(135,40)[l]{$\BDneg
ig\,O_{ij}^{R*}\sigmabar_\mu^{\dot{\alpha}\beta}$}
\Text(110,70)[]{$\dot{\alpha}$}
\Text(110,10)[]{$\beta$}
\Text(12,50)[]{$\mu$}
\end{picture}
\end{center}
\caption{Feynman rules for the chargino and neutralino interactions with
$W^\pm$ gauge bosons. The charge indicated on the $W$ boson is flowing
into the vertex in each case. The coupling matrices are defined in
eqs.~(\ref{eq:defOL}) and (\ref{eq:defOR}).
For each rule, a corresponding one with lowered spinor
indices is obtained by $\sigmabar_\mu^{\dot\alpha\beta} \rightarrow
-\sigma_{\mu\beta\dot\alpha}$.}
\label{ccboson}
\end{figure}
The $Z\Ni\Nj$ interaction vertex also subsumes the
$O_{ij}^{\prime\prime R}$ interaction found in four-component Majorana
Feynman rules as in ref.~\cite{HaberKane}, due to the result of
\eq{majidentitya} and the relation $O_{ij}^{\prime\prime R} =
-O_{ji}^{\prime\prime L}$ of \eq{eq:defOLpp}.

The chargino sector is CP-conserving if
$\Im(M_2\mu^*)=0$.  In this case,
the chargino fields can be rephased such that $M_2$ and $\mu$ are
real, and the chargino mixing matrices $U$ and $V$ can be
chosen to be real orthogonal.  In particular, the couplings
$O^{\prime\,L}$ and $O^{\prime\,R}$ are manifestly real.
Likewise, the neutralino sector is
CP-conserving if $\Im(M_1\mu^*)\!=\!
\Im(M_2\mu^*)\!=\!\Im(M_1 M_2^*)\!=\!0$.\footnote{If all three
of the potentially complex parameters
$M_1$, $M_2$ and $\mu$ are non-zero, then only two of the three
conditions for a CP-conserving neutralino sector
are independent, since the third condition follows
automatically from the first two conditions.}
In this case, the neutralino fields can be rephased
such that $M_1$, $M_2$ and $\mu$ are all real, and
the neutralino mixing matrix can be chosen
[cf.~\eqs{zmz}{oz}] such that~\cite{gunhab}:
\beq \label{Nz}
N_{ij}=\varepsilon_i^{1/2} Z_{ij}\,,\qquad\quad \text{no sum over $i$}\,,
\eeq
where $Z$ is a real orthogonal matrix, and $\varepsilon_i$
is the sign (either $\pm 1$) of the $i$th eigenvalue of
the real symmetric neutralino mass matrix, $M_{\psi^0}$.  That is,
the $i$th row of $N$ is purely real [imaginary] if $\varepsilon_i=+1$
[$-1$].
In particular, the matrix element $O^{\prime\prime\,L}_{ij}$
is purely real [imaginary] if $\varepsilon_i\varepsilon_j=+1$ [$-1$].
More generally, the neutralino and chargino interactions with the
electroweak gauge bosons are CP-conserving if the
corresponding Feynman rules for the interaction vertices
are either purely real or purely imaginary.

In the CP-violating case, the matrices $U$ and $V$ cannot be chosen
to be real orthogonal,
and $N$ cannot be written in the form of \eq{Nz}.\footnote{Since
$M_{\psi^0}$ is in general a \textit{complex}
symmetric matrix, its eigenvalues are not necessarily all
real.  In particular, if the $i$th eigenvalue is not real, then there is no
longer any meaning to the sign $\varepsilon_i$.}  Nevertheless, the
diagonal couplings $O^{\prime\,L}_{ii}$, $O^{\prime\,R}_{ii}$
and $O^{\prime\prime\,L}_{ii}$ are manifestly real.
This indicates that
the diagonal $Z^0\Ci^+\Ci^-$ and $Z^0\Ni\Ni$ couplings
are CP-conserving at tree level, even in the presence of a CP-violating
chargino and neutralino sector.  Similarly, the
diagonal $\gamma\Ci^+\Ci^-$ couplings
are CP-conserving, whereas the off-diagonal
$\gamma\Ci^\pm\Cj^\mp$ couplings ($i\neq j$) vanish at tree level, as
expected from gauge invariance.

\subsection{Higgs interactions with charginos and neutralinos}
\label{inohiggsrules}
\renewcommand{\theequation}{K.3.\arabic{equation}}
\renewcommand{\thefigure}{K.3.\arabic{figure}}
\renewcommand{\thetable}{K.3.\arabic{table}}
\setcounter{equation}{0}
\setcounter{figure}{0}
\setcounter{table}{0}

The couplings of chargino and neutralino mass eigenstates to the Higgs
mass eigenstates can be written, in terms of the Higgs mixing parameters
of eqs.~(\ref{eq:defkuphi0}) and (\ref{eq:defkdphi0}) and the
neutralino and chargino mixing matrices of \app{K.2}, as~\cite{gunhab}:
\beqa
Y^{\phi^0\chi^0_i \chi^0_j} &=&
\frac{1}{2} (k_{d\phi^0}^* N_{i3}^*  -  k_{u\phi^0}^* N_{i4}^*)
(g N_{j2}^* - g' N_{j1}^*) + (i \leftrightarrow j)\,,
\label{higgs-gauginos1}
\\
Y^{\phi^0\chi^-_i \chi^+_j} &=&
\frac{g}{\sqrt{2}} (
   k_{u\phi^0}^* U_{i1}^* V_{j2}^*
  +k_{d\phi^0}^* U_{i2}^* V_{j1}^*  )\,,\label{higgs-gauginos2}
\\
Y^{\phi^+\chi_i^0\chi^-_j} &=&
k_{d\phi^\pm} \bigl [
g (N_{i3}^* U_{j1}^*  - \frac{1}{\sqrt{2}} N_{i2}^* U_{j2}^* )
-\frac{g'}{\sqrt{2}} N_{i1}^* U_{j2}^* \bigr ]\,,\label{higgs-gauginos3}
\\
Y^{\phi^-\chi_i^0\chi^+_j} &=&
k_{u\phi^\pm} \bigl [
g (N_{i4}^* V_{j1}^*  + \frac{1}{\sqrt{2}} N_{i2}^* V_{j2}^* )
+\frac{g'}{\sqrt{2}} N_{i1}^* V_{j2}^*  \bigr ]\,,
\label{higgs-gauginos4}
\eeqa
for $\phi^0=h^0, H^0, A^0, G^0$ and $\phi^\pm=H^\pm, G^\pm$.
We exhibit the Higgs boson and Goldstone boson
interactions with the neutralinos and
charginos in \fig{inohiggsboson}.  For each of the Feynman rules in
\fig{inohiggsboson}, one can reverse all arrows by taking
$\delta_\alpha{}^\beta \rightarrow \delta^{\dot{\alpha}}{}_{\dot{\beta}}$
and complex conjugating the corresponding coupling (but not the overall
factor of $-i$).
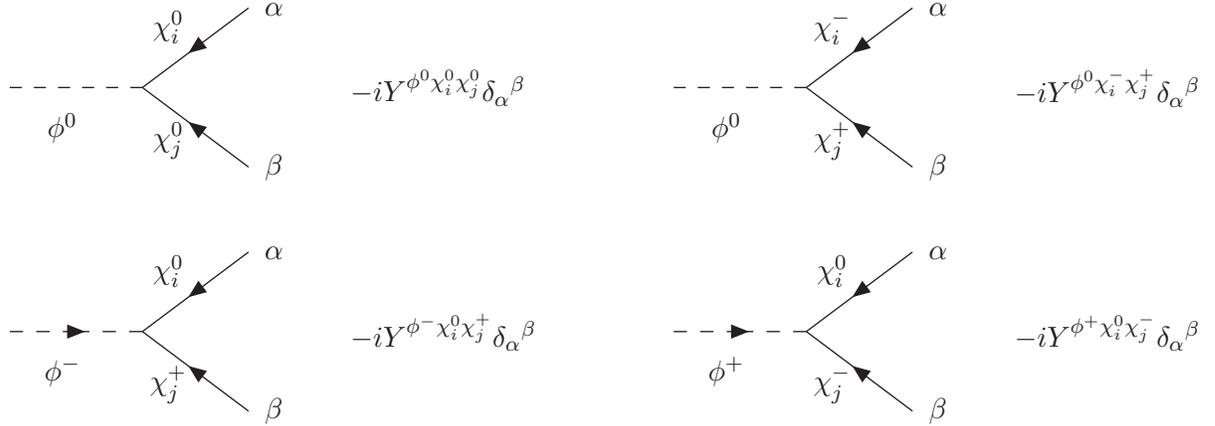
\begin{figure}[t!]
\begin{center}
\begin{picture}(195,80)(15,10)
\DashLine(60,40)(10,40){5} \ArrowLine(100,70)(60,40)
\ArrowLine(100,10)(60,40)
\Text(30,25)[]{$\phi^0$}
\Text(70,20)[]{$\chi^0_j$}
\Text(70,63)[]{$\chi^0_i$}
\Text(139,40)[l]{$-i Y^{\phi^0\chi_i^0\chi_j^0} \delta_\alpha{}^\beta$}
\Text(110,70)[]{$\alpha$} \Text(110,10)[]{$\beta$}
\end{picture}
\hspace{1.4cm}
\begin{picture}(195,53)(6,10)
\DashLine(60,40)(10,40){5}
\ArrowLine(100,70)(60,40)
\ArrowLine(100,10)(60,40)
\Text(30,25)[]{$\phi^0$}
\Text(70,20)[]{$\chi^+_j$}
\Text(70,63)[]{$\chi^-_i$}
\Text(139,40)[l]{$-i Y^{\phi^0\chi_i^-\chi_j^+} \delta_\alpha{}^\beta$}
\Text(110,70)[]{$\alpha$}
\Text(110,10)[]{$\beta$}
\end{picture}
\end{center}
\vspace{0.4cm}
\begin{center}
\begin{picture}(195,53)(15,13)
\DashArrowLine(10,40)(60,40){5}
\ArrowLine(100,70)(60,40)
\ArrowLine(100,10)(60,40)
\Text(30,25)[]{$\phi^-$}
\Text(70,20)[]{$\chi^+_j$}
\Text(70,63)[]{$\chi^0_i$}
\Text(139,40)[l]{$-iY^{\phi^-\chi^0_i\chi^+_j} \delta_\alpha{}^\beta$}
\Text(110,70)[]{$\alpha$}
\Text(110,10)[]{$\beta$}
\end{picture}
\hspace{1.4cm}
\begin{picture}(195,53)(6,13)
\DashArrowLine(10,40)(60,40){5}
\ArrowLine(100,70)(60,40)
\ArrowLine(100,10)(60,40)
\Text(30,25)[]{$\phi^+$}
\Text(70,20)[]{$\chi^-_j$}
\Text(70,63)[]{$\chi^0_i$}
\Text(139,40)[l]{$-iY^{\phi^+\chi^0_i\chi^-_j} \delta_\alpha{}^\beta$}
\Text(110,70)[]{$\alpha$}
\Text(110,10)[]{$\beta$}
\end{picture}
\end{center}
\caption{Feynman rules for the interactions of neutral Higgs bosons
$\phi^0 = (h^0,H^0,A^0,G^0)$ with neutralino pairs and chargino pairs,
respectively,
and the interaction of charged Higgs bosons $\phi^\pm = (H^\pm, G^\pm)$
with
chargino-neutralino pairs.
For each rule, there is a corresponding one
with all arrows reversed, undotted indices changed to dotted indices with
the opposite height, and the $Y$ coupling (without the explicit $-i$)
replaced by its complex conjugate.
}
\label{inohiggsboson}
\end{figure}

Goldstone bosons may appear as internal lines in of Feynman graphs
that are evaluated in the 't Hooft-Feynman gauge.  The
propagation of a Goldstone boson yields a result
that is identical to the
propagation of the corresponding longitudinal gauge boson
in the unitary gauge.  It is thus
convenient to express the Goldstone boson couplings
to the neutralinos and charginos in terms of the
corresponding gauge boson couplings.  To accomplish this, we first
record a number of identities among the neutralino and chargino mixing
matrices.  First, we use \eqs{u-and-v}{Charginomassmatrix} to derive:
\beqa
  M_2 U_{i1}^* + g v_d U_{i2}^* &=&m_{\widetilde C_i}V_{i1}\,,\qquad\qquad
  gv_u U_{i1}^*+\mu U_{i2}^* =m_{\widetilde C_i}V_{i2} \,,\\
  M_2 V_{i1}^* + g v_u V_{i2}^* &=&m_{\widetilde C_i}U_{i1}\,,\qquad\qquad
  gv_d V_{i1}^*+\mu V_{i2}^* =m_{\widetilde C_i}U_{i2}\,.
\eeqa
Next, we  use \eqs{eq:neutmix}{Neutralinomassmatrix} to derive:
\beqa
m_{\Ni}N_{i4}&=&\sum_{j=1}^4 N_{ij}^*(M_{\psi^0})_{j4}=\frac{v_u}{\sqrt{2}}
\left(g' N_{i1}^*-gN_{i2}^*\right)-\mu N_{i3}^* \,,\\
m_{\Ni}N_{i3}&=&\sum_{j=1}^4 N_{ij}^*(M_{\psi^0})_{j3}=-\frac{v_d}{\sqrt{2}}
\left(g' N_{i1}^*-gN_{i2}^*\right)-\mu N_{i4}^* \,,\\
m_{\Ni}N_{i2}&=&\sum_{j=1}^4 N_{ij}^*(M_{\psi^0})_{j2}=N_{i2}^*M_2
+\frac{g}{\sqrt{2}}\left(v_d N_{i3}^*-v_u N_{i4}^*\right)\,.
\eeqa
By a judicious combination of the above identities,
$\mu$ and $M_2$ can be eliminated.  One can then rewrite the Goldstone boson
couplings of \eqst{higgs-gauginos1}{higgs-gauginos4} in terms of
the gauge boson couplings
$O^{L,R}$, $O^{\prime\,L,R}$ and $O^{\prime\prime\,L,R}$
defined in \eqst{eq:defOL}{eq:defOLpp}.
It then follows that:
\beqa
iY^{G^0\chi^0_i\chi^0_j}&=&\frac{\sqrt{2}}{v}
\left(m_{\Ni}O^{\prime\prime\,L}_{ij}
-m_{\Nj}O^{\prime\prime\,R}_{ij}\right)
\,,\label{gxx1}\\
iY^{G^0\chi^-_i\chi^+_j}&=&\frac{\sqrt{2}}{v}
\left(m_{\Ci}O^{\prime\,L}_{ij}
-m_{\Cj}O^{\prime\,R}_{ij}\right)\,,\label{gxx2}\\
Y^{G^+\chi_i^0\chi_j^-}&=&\frac{\sqrt{2}}{v}
\left(m_{\Cj}O^{L\,*}_{ij}
-m_{\Ni}O^{R\,*}_{ij}\right)\,,\\
Y^{G^-\chi_i^0\chi_j^+}&=&-\frac{\sqrt{2}}{v}
\left(m_{\Ni}O^{L}_{ij}
-m_{\Cj}O^{R}_{ij}\right)\,.
\eeqa
Note that by using $O^{\prime\prime\,R}_{ij}=-O^{\prime\prime\,L}_{ji}$,
it follows from \eq{gxx1}
that $iY^{G^0\chi_i^0\chi_j^0}$ is symmetric under the
interchange of $i$ and $j$, as expected.

In general, for a CP-violating chargino and neutralino sector,
the couplings $Y^{\phi^0\chi^0_i\chi^0_i}$
and $Y^{\phi^0\chi_i^+\chi_i^-}$ for $\phi^0=h^0, H^0, A^0$
are neither purely real nor purely imaginary.  That is,
the diagonal neutralino and chargino couplings to the physical neutral
Higgs bosons are generically CP-violating.
However for $\phi^0=G^0$, the diagonal neutralino and chargino couplings to the
neutral Goldstone boson (when multiplied by~$i$) are manifestly real.
In particular,
\eqs{gxx1}{gxx2} yield:
\beqa
iY^{G^0\chi_i^0\chi_i^0}&=&\frac{2\sqrt{2}m_{\Ni}}{v}\,
O^{\prime\prime\,L}_{ii}=
\frac{\sqrt{2}\,m_{\Ni}}{v}\left[|N_{i4}|^2-
|N_{i3}|^2\right]
\,,\label{iy} \\
iY^{G^0\chi_i^-\chi_i^+}
&=& \frac{\sqrt{2}\,m_{\widetilde C_i}}{v}(O^{\prime\,L}_{ii}-
O^{\prime\,R}_{ii})=
\frac{m_{\widetilde C_i}}{\sqrt{2}\,v}
\left[|V_{i2}|^2-|U_{i2}|^2\right]\label{gchichi}\,,
\eeqa
where the unitarity of $U$ and $V$ has been used to obtain the
final expression in \eq{gchichi}.
It follows that the diagonal neutralino and chargino couplings to the
neutral Goldstone boson are CP-conserving.
This result
is not surprising, as the
corresponding diagonal tree-level couplings of the (longitudinal)
$Z^0$ boson are always CP-conserving as noted at the end of \app{K.2}.

\subsection{Chargino and neutralino interactions with
fermions and sfermions}
\label{inorules}
\renewcommand{\theequation}{K.4.\arabic{equation}}
\renewcommand{\thefigure}{K.4.\arabic{figure}}
\renewcommand{\thetable}{K.4.\arabic{table}}
\setcounter{equation}{0}
\setcounter{figure}{0}
\setcounter{table}{0}

In the MSSM, the scalar partners of the two-component fields $q$ and
${\bar q}^\dagger$
are the squarks,
denoted by $\widetilde q_L$ and $\widetilde q_R$, respectively.
In our notation, $\widetilde q^{\,*}_L$ and $\widetilde q^{\,*}_R$ denote
both the complex conjugate fields and the
names of the corresponding anti-squarks. Thus
$u$, $\widetilde u_L$ and $\widetilde u_R$ all have
electric charges  $+2/3$, whereas
$\bar u$, $\widetilde u^{\,*}_L$ and $\widetilde u^{\,*}_R$ all have
electric charges  $-2/3$.
Likewise, the scalar partners of the two-component fields $\ell$ and
${\bar\ell}^\dagger$ are the charged sleptons, denoted by
$\widetilde \ell_L$ and $\widetilde \ell_R$, respectively,
with $\ell = e,\mu,\tau$.
The sneutrino, $\widetilde\nu$ is the superpartner of the neutrino.
There is no $\widetilde\nu_R$, since there is no $\bar\nu$ in
the theory.\footnote{It is possible to construct a seesaw-extended
MSSM that would be the minimal supersymmetric extension of the
seesaw-extended Standard Model described in \app{J.2}.  In the
seesaw-extended MSSM, both $\bar\nu$ and its supersymmetric
partner $\widetilde\nu_R$ exist.  For further details on the
sneutrino sector of the seesaw-extended MSSM, see \Ref{dhr}.}

The Feynman rules for the chargino-quark-squark interactions
are given in \fig{cqsq}, and the rules for the neutralino-quark-squark
interactions are given in \fig{nqsq}.
Here we have taken the quark and lepton two-component fields to
be in a mass eigenstate basis, and the squark and
slepton field basis consists of the superpartners of these fields,
as described above.
Therefore, in practical applications,
one must include unitary rotation matrix elements relating the 
squarks and sleptons as given to the mass eigenstates, which can be
different.

In principle, all sfermions with a given electric charge can mix with
each other.  However, there is a popular, and perhaps
phenomenologically and theoretically favored, approximation in which
only the sfermions of the third family have significant mixing. For $f
= t,b,\tau$, one can then write the relationship between the gauge
eigenstates $\tilde f_L$, $\tilde f_R$ and the mass eigenstates
$\tilde f_1$, $\tilde f_2$ as~\cite{rudaz}
\beq
\begin{pmatrix}
\tilde f_R
\\
\tilde f_L
\end{pmatrix}
= X_{\tilde f}
\begin{pmatrix}
\tilde f_1
\\
\tilde f_2
\end{pmatrix}
\,,\qquad\quad
X_{\tilde f} \equiv
\begin{pmatrix}
R_{\tilde f_1} & R_{\tilde f_2}
\\
L_{\tilde f_1} & L_{\tilde f_2}
\end{pmatrix}\,,
\label{eq:sfermionmix}
\eeq
where $X$ is a $2\times 2$ unitary matrix. Then one can choose
$R_{\tilde f_1} = L^*_{\tilde f_2} = c_{\tilde f}$,
and
$L_{\tilde f_1} = -R^*_{\tilde f_2} = s_{\tilde f}$
with
\beqa
|c_{\tilde f}|^2 + |s_{\tilde f}|^2 = 1.
\label{eq:cfsfunitary}
\eeqa
If there is no CP violation, then
$c_{\tilde f}$ and $s_{\tilde f}$ can be taken real,
and they are the cosine
and sine of a sfermion mixing angle.\footnote{Our convention for
$c_{\tilde f}, s_{\tilde f}$ has the property that for zero mixing angle,
$\stilde f_1 = \stilde f_R$ and $\stilde f_2 = \stilde f_L$.
The conventions most commonly found in the literature
unfortunately do not
have this nice property.}
For the
other charged sfermions
($\widetilde{f}=
\widetilde{u},\widetilde{d},\widetilde{c},\widetilde{s},
\widetilde{e},\widetilde\mu$), one can use the same notation, and approximate
$L_{\tilde f_2} = R_{\tilde f_1} = 1$ and
$L_{\tilde f_1} = R_{\tilde f_2} = 0$.
The resulting Feynman rules for squarks and sleptons that mix
within each generation are shown in \figs{cqsqmixed}{nqsqmixed}.

For each Feynman rule in Figs.~\ref{cqsq}--\ref{nqsqmixed},
one can reverse all arrows by taking
$\delta _\alpha{}^\beta \rightarrow \delta^{\dot{\alpha}}{}_{\dot{\beta}}$
and complex conjugating the corresponding rule
(but leaving the explicit factor of $i$ intact).

\begin{figure}[tpb]
\begin{flushleft}
\begin{picture}(200,72)(4,7)
\DashArrowLine(60,40)(10,40)5
\ArrowLine(100,70)(60,40)
\ArrowLine(100,10)(60,40)
\Text(30,25)[]{$\widetilde d_{Lj}$}
\Text(70,20)[]{$u_k$}
\Text(70,67)[]{$\chi^-_i$}
\Text(125,40)[l]{$-i g U^*_{i1}
[\mathbold{K}^\dagger]_j{}^k
\, \delta_{\alpha}{}^{\beta}$}
\Text(110,70)[]{$\alpha$}
\Text(110,10)[]{$\beta$}
\end{picture}
\hspace{1.1cm}
\begin{picture}(180,72)(-4,7)
\DashArrowLine(60,40)(10,40)5
\ArrowLine(100,70)(60,40)
\ArrowLine(100,10)(60,40)
\Text(30,25)[]{$\widetilde u_{Lj}$}
\Text(70,20)[]{$d_k$}
\Text(70,67)[]{$\chi^+_i$}
\Text(125,40)[l]{$-i g V^*_{i1}[\mathbold{K}]_j{}^k\,
\delta_{\alpha}{}^{\beta}$}
\Text(110,70)[]{$\alpha$}
\Text(110,10)[]{$\beta$}
\end{picture}
\end{flushleft}
\begin{flushleft}
\begin{picture}(200,72)(4,7)
\DashArrowLine(60,40)(10,40)5
\ArrowLine(100,70)(60,40)
\ArrowLine(100,10)(60,40)
\Text(30,25)[]{$\widetilde\ell_{L}$}
\Text(70,20)[]{$\nu_\ell$}
\Text(70,67)[]{$\chi^-_i$}
\Text(125,40)[l]{$-i g U^*_{i1}
\, \delta_{\alpha}{}^{\beta}$}
\Text(110,70)[]{$\alpha$}
\Text(110,10)[]{$\beta$}
\end{picture}
\hspace{1.1cm}
\begin{picture}(180,72)(-4,7)
\DashArrowLine(60,40)(10,40)5
\ArrowLine(100,70)(60,40)
\ArrowLine(100,10)(60,40)
\Text(30,25)[]{$\widetilde\nu_\ell$}
\Text(70,20)[]{$\ell$}
\Text(70,67)[]{$\chi^+_i$}
\Text(125,40)[l]{$-i g V^*_{i1}\,
\delta_{\alpha}{}^{\beta}$}
\Text(110,70)[]{$\alpha$}
\Text(110,10)[]{$\beta$}
\end{picture}
\end{flushleft}
\begin{flushleft}
\begin{picture}(200,72)(4,7)
\DashArrowLine(10,40)(60,40)5
\ArrowLine(100,70)(60,40)
\ArrowLine(100,10)(60,40)
\Text(30,25)[]{$\widetilde d_{Lj}$}
\Text(70,20)[]{${\bar u}^{k}$}
\Text(70,67)[]{$\chi^+_i$}
\Text(125,40)[l]{$i
V_{i2}^*
[\mathbold{K}]_k{}^j Y_{uk}
\, \delta_{\alpha}{}^{\beta} $}
\Text(110,70)[]{$\alpha$}
\Text(110,10)[]{$\beta$}
\end{picture}
\hspace{1.1cm}
\begin{picture}(180,72)(-4,7)
\DashArrowLine(10,40)(60,40)5
\ArrowLine(100,70)(60,40)
\ArrowLine(100,10)(60,40)
\Text(30,25)[]{$\widetilde u_{Lj}$}
\Text(70,20)[]{${\bar d}^{k}$}
\Text(70,67)[]{$\chi^-_i$}
\Text(125,40)[l]{$i
U_{i2}^*
[\mathbold{K}^\dagger]_k{}^j
Y_{dj}
\, \delta_{\alpha}{}^{\beta} $}
\Text(110,70)[]{$\alpha$}
\Text(110,10)[]{$\beta$}
\end{picture}
\end{flushleft}
\begin{flushleft}
\begin{picture}(200,72)(4,7)
\DashArrowLine(60,40)(10,40)5
\ArrowLine(100,70)(60,40)
\ArrowLine(100,10)(60,40)
\Text(30,25)[]{$\widetilde d_{Rj}$}
\Text(70,20)[]{$u_k$}
\Text(70,67)[]{$\chi^-_i$}
\Text(125,40)[l]{$i
U_{i2}^*
[\mathbold{K}^\dagger]_j{}^k
Y_{dj}
\, \delta_{\alpha}{}^{\beta} $}
\Text(110,70)[]{$\alpha$}
\Text(110,10)[]{$\beta$}
\end{picture}
\hspace{1.1cm}
\begin{picture}(180,72)(-4,7)
\DashArrowLine(60,40)(10,40)5
\ArrowLine(100,70)(60,40)
\ArrowLine(100,10)(60,40)
\Text(30,25)[]{$\widetilde u_{Rj}$}
\Text(70,20)[]{$d_k$}
\Text(70,67)[]{$\chi^+_i$}
\Text(125,40)[l]{$i
V_{i2}^*
[\mathbold{K}]_j{}^k Y_{uj}
\, \delta_{\alpha}{}^{\beta} $}
\Text(110,70)[]{$\alpha$}
\Text(110,10)[]{$\beta$}
\end{picture}
\end{flushleft}
\begin{flushleft}
\begin{picture}(200,72)(4,7)
\DashArrowLine(60,40)(10,40)5
\ArrowLine(100,70)(60,40)
\ArrowLine(100,10)(60,40)
\Text(30,25)[]{$\widetilde \ell_R$}
\Text(70,20)[]{$\nu_\ell$}
\Text(70,67)[]{$\chi^-_i$}
\Text(125,40)[l]{$i
U_{i2}^* Y_\ell \, \delta_{\alpha}{}^{\beta} $}
\Text(110,70)[]{$\alpha$}
\Text(110,10)[]{$\beta$}
\end{picture}
\hspace{1.1cm}
\begin{picture}(180,72)(-4,7)
\DashArrowLine(10,40)(60,40)5
\ArrowLine(100,70)(60,40)
\ArrowLine(100,10)(60,40)
\Text(30,25)[]{$\widetilde \nu_\ell$}
\Text(70,20)[]{$\bar\ell$}
\Text(70,67)[]{$\chi^-_i$}
\Text(125,40)[l]{$i
U_{i2}^* Y_\ell \, \delta_{\alpha}{}^{\beta} $}
\Text(110,70)[]{$\alpha$}
\Text(110,10)[]{$\beta$}
\end{picture}
\end{flushleft}
\caption{Feynman rules for the interactions of charginos with
fermion/sfermion pairs in the MSSM.  The fermions are taken to be
in a mass eigenstate basis, and the sfermions are in a basis
whose elements are the supersymmetric partners of them.
For each rule, there is a corresponding one
with all arrows reversed, undotted indices changed to dotted indices with
the opposite height, and the coupling (without the explicit $i$)
replaced by its complex conjugate.
Note that chargino interaction vertices
involving $\bar u\widetilde d_R$ and $\bar d\widetilde u_R$ do not occur
in the MSSM.  An alternative version of these rules, for the case that mixing
is allowed only among third-family sfermions, is given in \fig{cqsqmixed}.
}
\label{cqsq}
\end{figure}
\begin{figure}[tbp]
\begin{flushleft}
\begin{picture}(200,68)(0,0)
\DashArrowLine(60,40)(10,40)5
\ArrowLine(100,70)(60,40)
\ArrowLine(100,10)(60,40)
\Text(30,25)[]{$\widetilde f_{Lj}$}
\Text(70,20)[]{$f_k$}
\Text(70,67)[]{$\chi^0_i$}
\Text(140,40)[l]{$-i \sqrt{2}
\left [ g T_3^f N_{i2}^* + g'
(Q_f - T_3^f) N_{i1}^* \right ]
\delta_j^k\, \delta_{\alpha}{}^{\beta}$}
\Text(110,70)[]{$\alpha$}
\Text(110,10)[]{$\beta$}
\end{picture}
\end{flushleft}
\begin{flushleft}
\begin{picture}(200,68)(0,0)
\DashArrowLine(10,40)(60,40)5
\ArrowLine(100,70)(60,40)
\ArrowLine(100,10)(60,40)
\Text(30,25)[]{$\widetilde f_{Rj}$}
\Text(70,20)[]{${\bar f}^{k}$}
\Text(70,67)[]{$\chi^0_i$}
\Text(140,40)[l]{$i \sqrt{2} g' Q_f N_{i1}^*
\,\delta_k^j \delta_{\alpha}{}^{\beta}$}
\Text(110,70)[]{$\alpha$}
\Text(110,10)[]{$\beta$}
\end{picture}
\phantom{xxxxxxx}
\end{flushleft}
\begin{center}
\begin{picture}(200,63)(10,10)
\DashArrowLine(10,40)(60,40)5
\ArrowLine(100,70)(60,40)
\ArrowLine(100,10)(60,40)
\Text(30,25)[]{$\widetilde u_{Lj}$}
\Text(70,20)[]{${\bar u}^{k}$}
\Text(70,67)[]{$\chi^0_i$}
\Text(127,40)[l]{$-i
N_{i4}^*
Y_{uj} \delta_k^j
\, \delta_{\alpha}{}^{\beta} $}
\Text(110,70)[]{$\alpha$}
\Text(110,10)[]{$\beta$}
\end{picture}
\hspace{1.1cm}
\begin{picture}(200,63)(10,10)
\DashArrowLine(60,40)(10,40)5
\ArrowLine(100,70)(60,40)
\ArrowLine(100,10)(60,40)
\Text(30,25)[]{$\widetilde u_{Rj}$}
\Text(70,20)[]{$u_k$}
\Text(70,67)[]{$\chi^0_i$}
\Text(127,40)[l]{$-i
N_{i4}^* Y_{uj} \delta_j^k
\, \delta_{\alpha}{}^{\beta} $}
\Text(110,70)[]{$\alpha$}
\Text(110,10)[]{$\beta$}
\end{picture}
\end{center}
\begin{center}
\begin{picture}(200,67)(10,18)
\DashArrowLine(10,40)(60,40)5
\ArrowLine(100,70)(60,40)
\ArrowLine(100,10)(60,40)
\Text(30,25)[]{$\widetilde d_{Lj}$}
\Text(70,20)[]{${\bar d}^{k}$}
\Text(70,67)[]{$\chi^0_i$}
\Text(127,40)[l]{$-i
N_{i3}^* Y_{dj} \delta_k^j
\, \delta_{\alpha}{}^{\beta} $}
\Text(110,70)[]{$\alpha$}
\Text(110,10)[]{$\beta$}
\end{picture}
\hspace{1.1cm}
\begin{picture}(200,67)(10,18)
\DashArrowLine(60,40)(10,40)5
\ArrowLine(100,70)(60,40)
\ArrowLine(100,10)(60,40)
\Text(30,25)[]{$\widetilde d_{Rj}$}
\Text(70,20)[]{$d_k$}
\Text(70,67)[]{$\chi^0_i$}
\Text(127,40)[l]{$-i
N_{i3}^* Y_{dj} \delta_j^k \, \delta_{\alpha}{}^{\beta} $}
\Text(110,70)[]{$\alpha$}
\Text(110,10)[]{$\beta$}
\end{picture}
\end{center}
\begin{center}
\begin{picture}(200,67)(10,18)
\DashArrowLine(10,40)(60,40)5
\ArrowLine(100,70)(60,40)
\ArrowLine(100,10)(60,40)
\Text(30,25)[]{$\widetilde \ell_{L}$}
\Text(70,20)[]{$\bar\ell$}
\Text(70,67)[]{$\chi^0_i$}
\Text(127,40)[l]{$-i
N_{i3}^* Y_{\ell}
\, \delta_{\alpha}{}^{\beta} $}
\Text(110,70)[]{$\alpha$}
\Text(110,10)[]{$\beta$}
\end{picture}
\hspace{1.1cm}
\begin{picture}(200,67)(10,18)
\DashArrowLine(60,40)(10,40)5
\ArrowLine(100,70)(60,40)
\ArrowLine(100,10)(60,40)
\Text(30,25)[]{$\widetilde \ell_{R}$}
\Text(70,20)[]{$\ell$}
\Text(70,67)[]{$\chi^0_i$}
\Text(127,40)[l]{$-i
N_{i3}^* Y_{\ell} \, \delta_{\alpha}{}^{\beta} $}
\Text(110,70)[]{$\alpha$}
\Text(110,10)[]{$\beta$}
\end{picture}
\end{center}
\caption{Feynman rules for the interactions of neutralinos with
fermion/sfermion pairs in the MSSM.
The fermions are taken to be
in a mass eigenstate basis, and the sfermions are in a basis
whose elements are the supersymmetric partners of them.
For each rule, there is a corresponding one
with all arrows reversed, undotted indices changed to dotted indices with
the opposite height, and the coupling (without the explicit $i$)
replaced by its complex conjugate.
An alternative version of these rules, for the case that mixing
is allowed only among third-family sfermions, is given in \fig{nqsqmixed}.
}
\label{nqsq}
\end{figure}

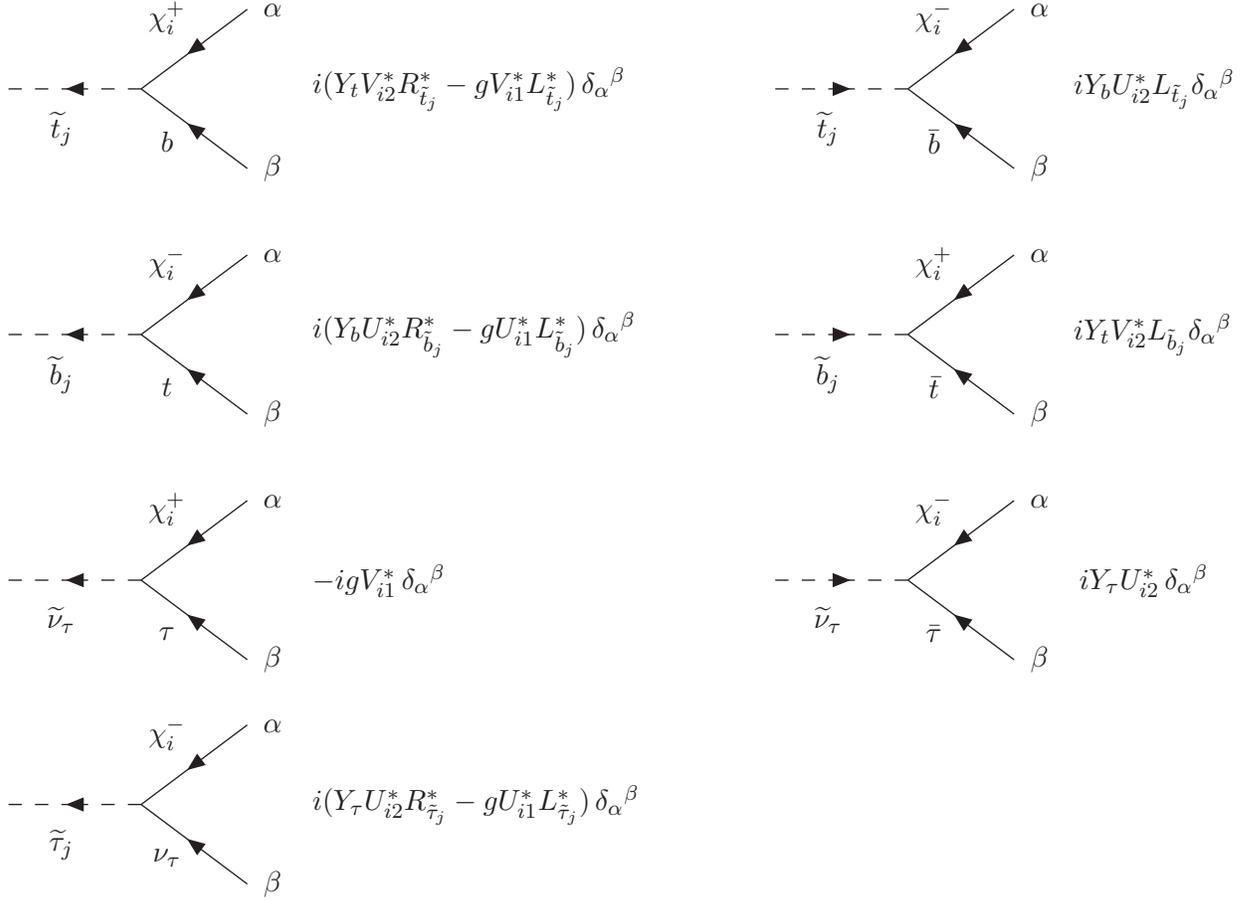
\begin{figure}[b!]
\begin{flushleft}
\begin{picture}(250,80)(9,7)
\DashArrowLine(60,40)(10,40)5
\ArrowLine(100,70)(60,40)
\ArrowLine(100,10)(60,40)
\Text(30,25)[]{$\widetilde t_{j}$}
\Text(70,20)[]{$b$}
\Text(70,67)[]{$\chi^+_i$}
\Text(125,40)[l]{$i (Y_t V_{i2}^* R_{\tilde t_j}^* -g V^*_{i1}
L_{\tilde t_j}^*)
\, \delta_{\alpha}{}^{\beta}$}
\Text(110,70)[]{$\alpha$}
\Text(110,10)[]{$\beta$}
\end{picture}
\hspace{1.15cm}
\begin{picture}(160,80)(9,7)
\DashArrowLine(10,40)(60,40)5
\ArrowLine(100,70)(60,40)
\ArrowLine(100,10)(60,40)
\Text(30,25)[]{$\widetilde t_{j}$}
\Text(70,20)[]{$\bar b$}
\Text(70,67)[]{$\chi^-_i$}
\Text(123,40)[l]{$i Y_b U_{i2}^* L_{\tilde t_j}
 \delta_{\alpha}{}^{\beta} $}
\Text(110,70)[]{$\alpha$}
\Text(110,10)[]{$\beta$}
\end{picture}
\end{flushleft}
\begin{flushleft}
\begin{picture}(250,80)(9,7)
\DashArrowLine(60,40)(10,40)5
\ArrowLine(100,70)(60,40)
\ArrowLine(100,10)(60,40)
\Text(30,25)[]{$\widetilde b_{j}$}
\Text(70,20)[]{$t$}
\Text(70,67)[]{$\chi^-_i$}
\Text(125,40)[l]{$i (Y_b U_{i2}^* R_{\tilde b_j}^*-g U^*_{i1} L_{\tilde b_j}^*)
\, \delta_{\alpha}{}^{\beta}$}
\Text(110,70)[]{$\alpha$}
\Text(110,10)[]{$\beta$}
\end{picture}
\hspace{1.15cm}
\begin{picture}(160,80)(9,7)
\DashArrowLine(10,40)(60,40)5
\ArrowLine(100,70)(60,40)
\ArrowLine(100,10)(60,40)
\Text(30,25)[]{$\widetilde b_{j}$}
\Text(70,20)[]{$\bar t$}
\Text(70,67)[]{$\chi^+_i$}
\Text(123,40)[l]{$i Y_t V_{i2}^* L_{\tilde b_j}
 \delta_{\alpha}{}^{\beta} $}
\Text(110,70)[]{$\alpha$}
\Text(110,10)[]{$\beta$}
\end{picture}
\end{flushleft}
\begin{flushleft}
\begin{picture}(250,80)(9,7)
\DashArrowLine(60,40)(10,40)5
\ArrowLine(100,70)(60,40)
\ArrowLine(100,10)(60,40)
\Text(30,25)[]{$\widetilde\nu_\tau$}
\Text(70,20)[]{$\tau$}
\Text(70,67)[]{$\chi^+_i$}
\Text(125,40)[l]{$-i g V^*_{i1}\,
\delta_{\alpha}{}^{\beta}$}
\Text(110,70)[]{$\alpha$}
\Text(110,10)[]{$\beta$}
\end{picture}
\hspace{1.15cm}
\begin{picture}(160,72)(9,7)
\DashArrowLine(10,40)(60,40)5
\ArrowLine(100,70)(60,40)
\ArrowLine(100,10)(60,40)
\Text(30,25)[]{$\widetilde \nu_\tau$}
\Text(70,20)[]{$\bar\tau$}
\Text(70,67)[]{$\chi^-_i$}
\Text(125,40)[l]{$i Y_\tau U_{i2}^* \, \delta_{\alpha}{}^{\beta} $}
\Text(110,70)[]{$\alpha$}
\Text(110,10)[]{$\beta$}
\end{picture}
\end{flushleft}
\begin{flushleft}
\begin{picture}(250,72)(9,7)
\DashArrowLine(60,40)(10,40)5
\ArrowLine(100,70)(60,40)
\ArrowLine(100,10)(60,40)
\Text(30,25)[]{$\widetilde\tau_j$}
\Text(70,20)[]{$\nu_\tau$}
\Text(70,67)[]{$\chi^-_i$}
\Text(125,40)[l]{$i(Y_\tau U_{i2}^*R_{\tilde \tau_j}^*
- g U^*_{i1}L_{\tilde \tau_j}^* )
\, \delta_{\alpha}{}^{\beta}$}
\Text(110,70)[]{$\alpha$}
\Text(110,10)[]{$\beta$}
\end{picture}
\hspace{1.15cm}
\end{flushleft}
\caption{Feynman rules for the interactions of charginos with
  third-family fermion/sfermion pairs in the MSSM.  The fermions are
  taken to be in a mass eigenstate basis. CKM mixing is neglected, and
  the sfermions are assumed to only mix within the third family. The
  corresponding rules for the first and second families with the
  approximation of no mixing and vanishing fermion masses can be
  obtained from these by setting $Y_f = 0$ and $L_{\tilde f_2} =
  R_{\tilde f_1} = 1$ and $L_{\tilde f_1} = R_{\tilde f_2} = 0$ (so
  that $\stilde f_1 = \stilde f_R$ and $\stilde f_2 = \stilde f_L$).
  For each rule, there is a corresponding one with all arrows
  reversed, undotted indices changed to dotted indices with the
  opposite height, and the coupling (without the explicit $i$)
  replaced by its complex conjugate.}
\label{cqsqmixed}
\end{figure}
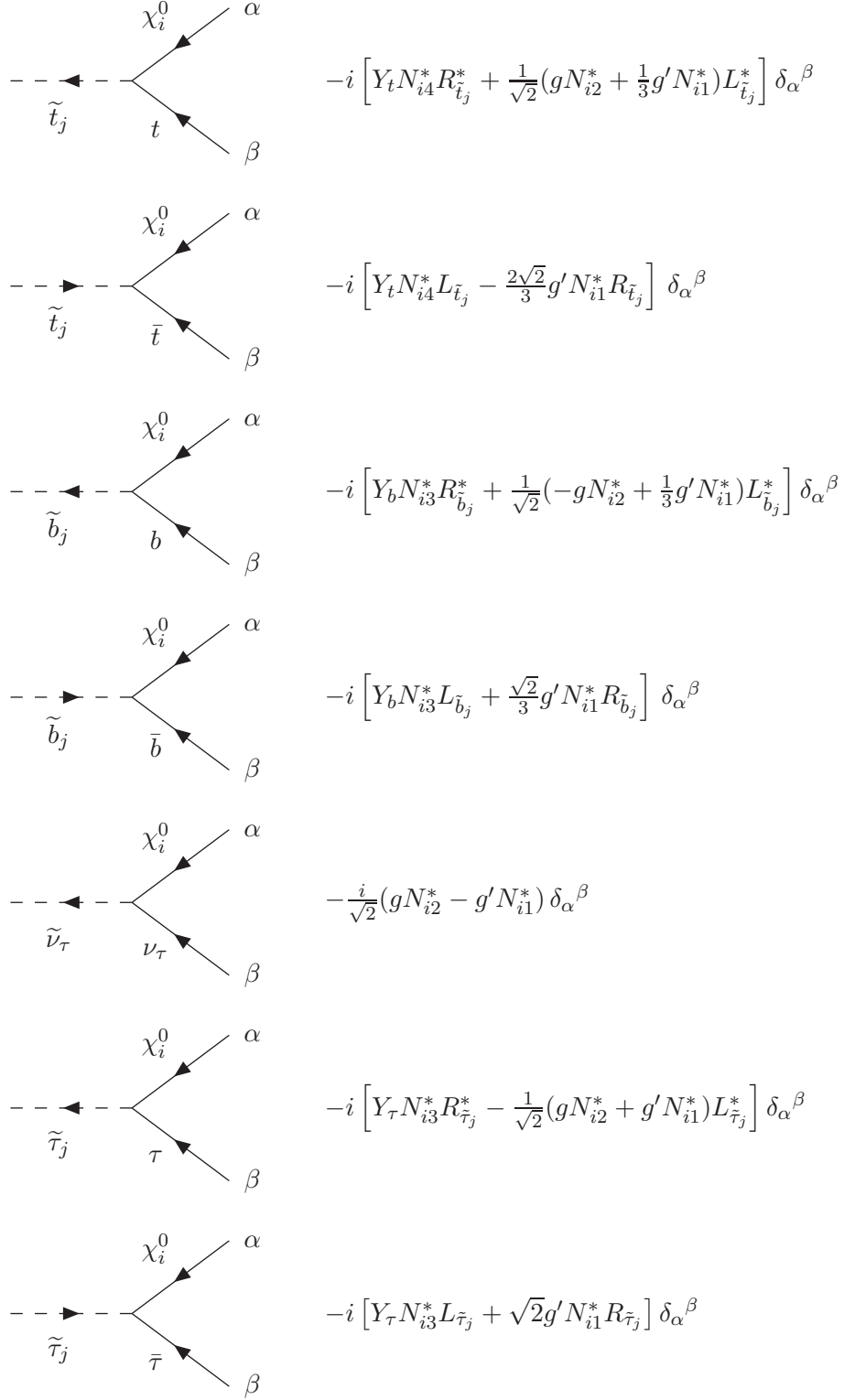
\begin{figure}[tbp]
\begin{flushleft}
\begin{picture}(200,68)(-50,0)
\DashArrowLine(60,40)(10,40)5
\ArrowLine(100,70)(60,40)
\ArrowLine(100,10)(60,40)
\Text(30,25)[]{$\widetilde t_{j}$}
\Text(70,20)[]{$t$}
\Text(70,67)[]{$\chi^0_i$}
\Text(140,40)[l]{$
-i \left [
Y_t N_{i4}^* R_{\tilde t_j}^*
+\frac{1}{\sqrt{2}} (g N_{i2}^* + \third g'
N_{i1}^*) L_{\tilde t_j}^* \right ]
 \delta_{\alpha}{}^{\beta}$}
\Text(110,70)[]{$\alpha$}
\Text(110,10)[]{$\beta$}
\end{picture}
\end{flushleft}
\begin{flushleft}
\begin{picture}(200,72)(-50,0)
\DashArrowLine(10,40)(60,40)5
\ArrowLine(100,70)(60,40)
\ArrowLine(100,10)(60,40)
\Text(30,25)[]{$\widetilde t_{j}$}
\Text(70,20)[]{$\bar t$}
\Text(70,67)[]{$\chi^0_i$}
\Text(140,40)[l]{$-i \left [Y_t N_{i4}^*L_{\tilde t_j}
- \frac{2\sqrt{2}}{3} g' N_{i1}^* R_{\tilde t_j}
\right ]
\, \delta_{\alpha}{}^{\beta}$}
\Text(110,70)[]{$\alpha$}
\Text(110,10)[]{$\beta$}
\end{picture}
\phantom{xxxxxxx}
\end{flushleft}
\begin{flushleft}
\begin{picture}(200,72)(-50,0)
\DashArrowLine(60,40)(10,40)5
\ArrowLine(100,70)(60,40)
\ArrowLine(100,10)(60,40)
\Text(30,25)[]{$\widetilde b_{j}$}
\Text(70,20)[]{$b$}
\Text(70,67)[]{$\chi^0_i$}
\Text(140,40)[l]{$
-i \left [
Y_b N_{i3}^* R_{\tilde b_j}^*
+\frac{1}{\sqrt{2}} (-g N_{i2}^* + \third g'
N_{i1}^*) L_{\tilde b_j}^* \right ]
 \delta_{\alpha}{}^{\beta}$}
\Text(110,70)[]{$\alpha$}
\Text(110,10)[]{$\beta$}
\end{picture}
\end{flushleft}
\begin{flushleft}
\begin{picture}(200,72)(-50,0)
\DashArrowLine(10,40)(60,40)5
\ArrowLine(100,70)(60,40)
\ArrowLine(100,10)(60,40)
\Text(30,25)[]{$\widetilde b_{j}$}
\Text(70,20)[]{$\bar b$}
\Text(70,67)[]{$\chi^0_i$}
\Text(140,40)[l]{$-i \left [Y_b N_{i3}^* L_{\tilde b_j}
 + \frac{\sqrt{2}}{3} g' N_{i1}^* R_{\tilde b_j}
\right ]
\, \delta_{\alpha}{}^{\beta}$}
\Text(110,70)[]{$\alpha$}
\Text(110,10)[]{$\beta$}
\end{picture}
\phantom{xxxxxxx}
\end{flushleft}
\begin{flushleft}
\begin{picture}(200,72)(-50,0)
\DashArrowLine(60,40)(10,40)5
\ArrowLine(100,70)(60,40)
\ArrowLine(100,10)(60,40)
\Text(30,25)[]{$\widetilde \nu_\tau$}
\Text(70,20)[]{$\nu_\tau$}
\Text(70,67)[]{$\chi^0_i$}
\Text(140,40)[l]{$
-\frac{i}{\sqrt{2}} (g N_{i2}^* - g'
N_{i1}^*)
\, \delta_{\alpha}{}^{\beta}$}
\Text(110,70)[]{$\alpha$}
\Text(110,10)[]{$\beta$}
\end{picture}
\end{flushleft}
\begin{flushleft}
\begin{picture}(200,72)(-50,0)
\DashArrowLine(60,40)(10,40)5
\ArrowLine(100,70)(60,40)
\ArrowLine(100,10)(60,40)
\Text(30,25)[]{$\widetilde \tau_{j}$}
\Text(70,20)[]{$\tau$}
\Text(70,67)[]{$\chi^0_i$}
\Text(140,40)[l]{$
-i \left [
Y_\tau N_{i3}^* R_{\tilde \tau_j}^*
-\frac{1}{\sqrt{2}} (g N_{i2}^* + g'
N_{i1}^*) L_{\tilde \tau_j}^* \right ]
 \delta_{\alpha}{}^{\beta}$}
\Text(110,70)[]{$\alpha$}
\Text(110,10)[]{$\beta$}
\end{picture}
\end{flushleft}
\begin{flushleft}
\begin{picture}(200,72)(-50,0)
\DashArrowLine(10,40)(60,40)5
\ArrowLine(100,70)(60,40)
\ArrowLine(100,10)(60,40)
\Text(30,25)[]{$\widetilde \tau_{j}$}
\Text(70,20)[]{$\bar\tau$}
\Text(70,67)[]{$\chi^0_i$}
\Text(140,40)[l]{$-i \left [Y_\tau N_{i3}^* L_{\tilde \tau_j}
+ \sqrt{2} g' N_{i1}^* R_{\tilde \tau_j}
\right ]
 \delta_{\alpha}{}^{\beta}$}
\Text(110,70)[]{$\alpha$}
\Text(110,10)[]{$\beta$}
\end{picture}
\phantom{xxxxxxx}
\end{flushleft}
\caption{Feynman rules for the interactions of neutralinos with
third-family fermion/sfermion pairs in the MSSM.
The comments of the caption of \fig{cqsqmixed} also apply here.}
\label{nqsqmixed}
\end{figure}

\clearpage

\subsection{SUSY-QCD Feynman rules}
\label{susyQCDrules}
\renewcommand{\theequation}{K.5.\arabic{equation}}
\renewcommand{\thefigure}{K.5.\arabic{figure}}
\renewcommand{\thetable}{K.5.\arabic{table}}
\setcounter{equation}{0}
\setcounter{figure}{0}
\setcounter{table}{0}

In supersymmetric (SUSY) QCD, the Lagrangian governing the gluon
interactions with colored fermions (gluinos and quarks)
in two-component spinor notation,
which derives from the covariant derivatives in the kinetic terms,
is given by
\beq
\mathscr{L}_{\rm int} \>=\>
\BDpos i g_s f^{abd}
\,({\tilde g}^\dagger_a\,\sigmabar_\mu\, {\stilde g}_b) A^\mu_d
\BDminus g_s T_j^{ak} \sum_q \left [{q}^{\dagger j}\sigmabar_\mu q_k
  - {\bar q}^\dagger_k \sigmabar_\mu {\bar q}^j \right ] A^\mu_a \, .
\eeq
Here $g_s$ is the strong coupling constant,
$a,b,d= 1,2,\ldots ,8$ are $SU(3)_c$
adjoint representation
indices, and $f^{abd}$ are the $SU(3)$ structure constants.
Raised (lowered) indices $j,k = 1,2,3$ are color indices in the fundamental
(anti-fundamental) representation. We
have denoted the two-component gluino field by ${\stilde g}_a$ as in
Table~\ref{tab:nomenclature} and the gluon field by $A^\mu_a$.
The sum $\sum_q$ is over the six flavors $q=u,d,s,c,b,t$ (in either the
mass eigenstate or electroweak gauge-eigenstate basis).
The
corresponding Feynman rules are shown in \fig{fig:SUSYQCDgluonrules}.
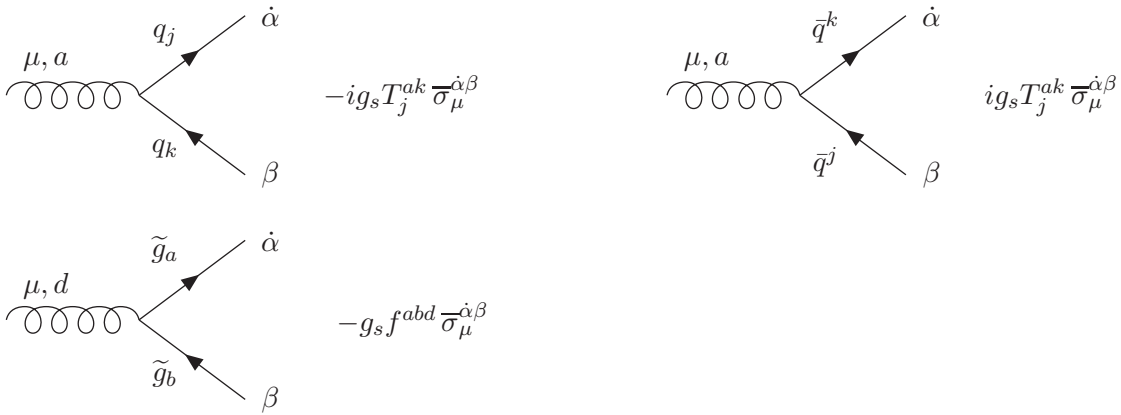
\begin{figure}[b!]
\begin{flushleft}
\begin{picture}(200,72)(0,5)
\Gluon(10,40)(60,40){5}{4}
\ArrowLine(60,40)(100,70)
\ArrowLine(100,10)(60,40)
\Text(25,53)[c]{$\mu,a$}
\Text(70,20)[]{$q_k$}
\Text(70,63)[]{$q_j$}
\Text(130,40)[l]{$\BDneg ig_s
T_j^{ak}\,\sigmabar^{\dot{\alpha}\beta}_\mu$}
\Text(110,70)[]{$\dot{\alpha}$}
\Text(110,10)[]{$\beta$}
\end{picture}
\hspace{1.5cm}
\begin{picture}(200,72)(0,5)
\Gluon(10,40)(60,40){5}{4}
\ArrowLine(60,40)(100,70)
\ArrowLine(100,10)(60,40)
\Text(25,53)[c]{$\mu, a$}
\Text(70,15)[]{${\bar q}^j$}
\Text(70,66)[]{${\bar q}^k$}
\Text(130,40)[l]{$\BDpos ig_s
T_j^{ak}\,\sigmabar^{\dot{\alpha}\beta}_\mu$}
\Text(110,70)[]{$\dot{\alpha}$}
\Text(110,10)[]{$\beta$}
\end{picture}
\end{flushleft}
\begin{flushleft}
\begin{picture}(200,72)(0,5)
\Gluon(10,40)(60,40){5}{4}
\ArrowLine(60,40)(100,70)
\ArrowLine(100,10)(60,40)
\Text(25,53)[c]{$\mu, d$}
\Text(70,20)[]{${\stilde g}_b$}
\Text(70,67)[]{${\stilde g}_a$}
\Text(135,40)[l]{$\BDneg g_s f^{abd}\,\sigmabar^{\dot{\alpha}\beta}_\mu$}
\Text(110,70)[]{$\dot{\alpha}$}
\Text(110,10)[]{$\beta$}
\end{picture}
\end{flushleft}
\caption{Fermionic Feynman rules for SUSY-QCD that involve the
gluon, with $q = u,d,c,s,t,b$.
Lowered (raised) indices $j,k$ correspond to
the fundamental (anti-fundamental) representation of $SU(3)_c$.
For each rule, a corresponding one with lowered spinor
indices is obtained by $\sigmabar_\mu^{\dot\alpha\beta} \rightarrow
-\sigma_{\mu\beta\dot\alpha}$.
\label{fig:SUSYQCDgluonrules}}
\end{figure}
The gluino-squark-quark Lagrangian is given by:
\beq
\mathscr{L}_{\rm int}  =-\sqrt{2}g_sT_{j}^{ak}
\sum_q \left[ {\tilde g}_a q_k\, {\tilde q}_{L}^{\ast j}
+  {\tilde g}^\dagger_a {q}^{\dagger j} \, {\tilde q}_{Lk}
- {\tilde g}_a {\bar q}^j \, {\tilde q}_{Rk}
-   {\tilde g}^\dagger_a {\bar q}^\dagger_k\, {\tilde q}_{R}^{\ast j}
\right] ,
\eeq
where the squark fields are taken to be in the same basis as the quarks.
The Feynman rules resulting from these Lagrangian terms are shown in
\fig{fig:SUSYQCDsquarkrules}.
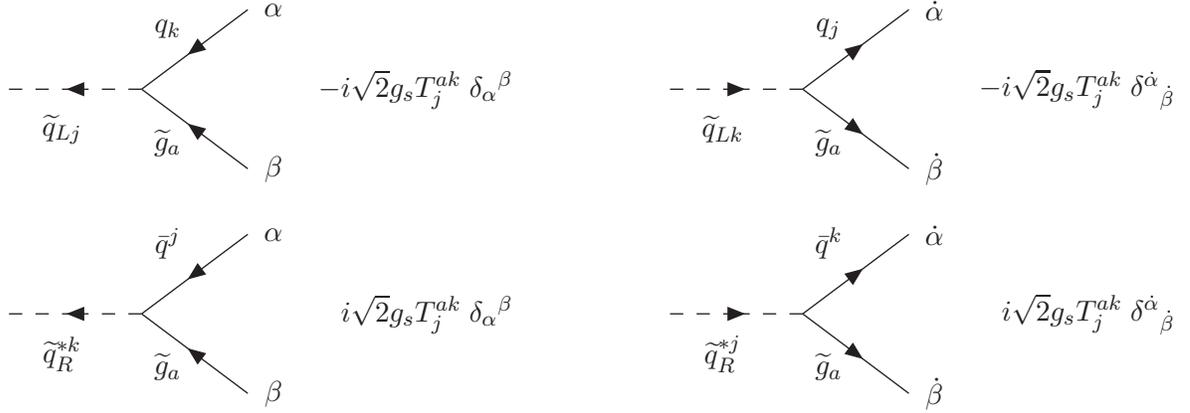
\begin{figure}[tbp]
\begin{flushleft}
\begin{picture}(200,72)(5,5)
\DashArrowLine(60,40)(10,40){5}
\ArrowLine(100,70)(60,40)
\ArrowLine(100,10)(60,40)
\Text(30,25)[]{${\stilde q}_{Lj}$}
\Text(70,20)[]{${\stilde g}_a$}
\Text(70,63)[]{$q_k$}
\Text(127,40)[l]{$-i\sqrt{2}g_s T_{j}^{ak}\;\delta_{\alpha}{}^{\beta}$}
\Text(110,70)[]{$\alpha$}
\Text(110,10)[]{$\beta$}
\end{picture}
\hspace{1.5cm}
\begin{picture}(200,72)(5,5)
\DashArrowLine(10,40)(60,40){5}
\ArrowLine(60,40)(100,70)
\ArrowLine(60,40)(100,10)
\Text(30,25)[]{${\stilde q}_{Lk}$}
\Text(70,20)[]{${\stilde g}_a$}
\Text(70,63)[]{$q_j$}
\Text(127,40)[l]{$-i\sqrt{2}g_s T_{j}^{ak}\;\delta^{\dot\alpha}{}_{\dot\beta}$}
\Text(110,70)[]{$\dot\alpha$}
\Text(110,10)[]{$\dot\beta$}
\end{picture}
\end{flushleft}
\begin{flushleft}
\begin{picture}(200,72)(5,5)
\DashArrowLine(60,40)(10,40){5}
\ArrowLine(100,70)(60,40)
\ArrowLine(100,10)(60,40)
\Text(30,25)[]{${\stilde q}_R^{*k}$}
\Text(70,20)[]{${\stilde g}_a$}
\Text(70,67)[]{${\bar q}^j$}
\Text(127,40)[l]{$\phm i\sqrt{2}g_s T_{j}^{ak}\;\delta_{\alpha}{}^{\beta}$}
\Text(110,70)[]{$\alpha$}
\Text(110,10)[]{$\beta$}
\end{picture}
\hspace{1.5cm}
\begin{picture}(200,72)(5,5)
\DashArrowLine(10,40)(60,40){5}
\ArrowLine(60,40)(100,70)
\ArrowLine(60,40)(100,10)
\Text(30,25)[]{${\stilde q}_R^{*j}$}
\Text(70,20)[]{${\stilde g}_a$}
\Text(70,67)[]{${\bar q}^k$}
\Text(127,40)[l]{$\phm i\sqrt{2}g_s T_{j}^{ak}\;
\delta^{\dot\alpha}{}_{\dot\beta}$}
\Text(110,70)[]{$\dot\alpha$}
\Text(110,10)[]{$\dot\beta$}
\end{picture}
\end{flushleft}
\caption{Fermionic Feynman rules for SUSY-QCD that involve the
squarks, in a basis corresponding to the quark mass eigenstates
$q = u,d,c,s,t,b$.
Lowered (raised) indices $j,k$ correspond to
the fundamental (anti-fundamental) representation of $SU(3)_c$,
and the index $a$ labels the adjoint representation carried by
the gluino.
The spinor index heights can be exchanged in each case, by
replacing
$\delta_\alpha{}^\beta \rightarrow \delta_\beta{}^\alpha$
or
$\delta^{\dot\alpha}{}_{\dot\beta} \rightarrow
\delta^{\dot\beta}{}_{\dot\alpha}$.
For an alternative set of rules, incorporating
$\widetilde q_L$--$\widetilde q_R$ mixing, see
\fig{fig:SUSYQCDsquarkmasseigrules}.
\label{fig:SUSYQCDsquarkrules}}
\end{figure}

For practical applications, one typically takes the
quark fields as the familiar mass eigenstates, and then performs a unitary
rotation
on the squarks in the corresponding basis to obtain their mass eigenstate
basis.
In the approximation described at the end of \app{K.4}
[cf \eqs{eq:sfermionmix}{eq:cfsfunitary} and the accompanying text], one
obtains the Feynman rules
of \fig{fig:SUSYQCDsquarkmasseigrules},
as an alternative to those of \fig{fig:SUSYQCDsquarkrules}.
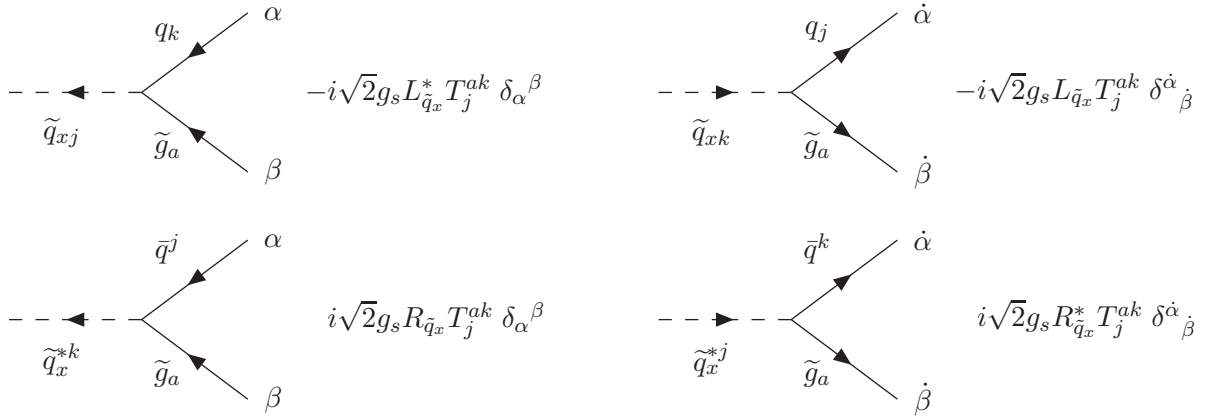
\begin{figure}[tbp]
\begin{flushleft}
\begin{picture}(200,68)(5,10)
\DashArrowLine(60,40)(10,40){5}
\ArrowLine(100,70)(60,40)
\ArrowLine(100,10)(60,40)
\Text(30,25)[]{${\stilde q}_{xj}$}
\Text(70,20)[]{${\stilde g}_a$}
\Text(70,63)[]{$q_k$}
\Text(122,40)[l]{$-i\sqrt{2}g_s L_{\tilde q_x}^* T_{j}^{ak}\;
\delta_{\alpha}{}^{\beta}$}
\Text(110,70)[]{$\alpha$}
\Text(110,10)[]{$\beta$}
\end{picture}
\hspace{1.35cm}
\begin{picture}(200,68)(5,10)
\DashArrowLine(10,40)(60,40){5}
\ArrowLine(60,40)(100,70)
\ArrowLine(60,40)(100,10)
\Text(30,25)[]{${\stilde q}_{xk}$}
\Text(70,20)[]{${\stilde g}_a$}
\Text(70,63)[]{$q_j$}
\Text(122,40)[l]{$-i\sqrt{2}g_s L_{\tilde q_x}
T_{j}^{ak}\;\delta^{\dot\alpha}{}_{\dot\beta}$}
\Text(110,70)[]{$\dot\alpha$}
\Text(110,10)[]{$\dot\beta$}
\end{picture}
\end{flushleft}
\begin{flushleft}
\begin{picture}(200,68)(5,10)
\DashArrowLine(60,40)(10,40){5}
\ArrowLine(100,70)(60,40)
\ArrowLine(100,10)(60,40)
\Text(30,25)[]{${\stilde q}_x^{*k}$}
\Text(70,20)[]{${\stilde g}_a$}
\Text(70,67)[]{${\bar q}^j$}
\Text(122,40)[l]{$\phm i\sqrt{2}g_s R_{\tilde q_x}
T_{j}^{ak}\;\delta_{\alpha}{}^{\beta}$}
\Text(110,70)[]{$\alpha$}
\Text(110,10)[]{$\beta$}
\end{picture}
\hspace{1.35cm}
\begin{picture}(200,73)(5,10)
\DashArrowLine(10,40)(60,40){5}
\ArrowLine(60,40)(100,70)
\ArrowLine(60,40)(100,10)
\Text(30,25)[]{${\stilde q}_x^{*j}$}
\Text(70,20)[]{${\stilde g}_a$}
\Text(70,67)[]{${\bar q}^k$}
\Text(122,40)[l]{$\phm i\sqrt{2}g_s R_{\tilde q_x}^*
T_{j}^{ak}\; \delta^{\dot\alpha}{}_{\dot\beta}$}
\Text(110,70)[]{$\dot\alpha$}
\Text(110,10)[]{$\dot\beta$}
\end{picture}
\end{flushleft}
\caption{Fermionic Feynman rules for SUSY-QCD that involve the squarks in
the mass eigenstate basis labeled by $x=1,2$ and $q = u,d,c,s,t,b$, in the
approximation where mixing is allowed only within a given flavor
(typically, for the third family only), as in eq.~(\ref{eq:sfermionmix}).
Lowered (raised) indices $j,k$ correspond to the fundamental
(anti-fundamental) representation of $SU(3)_c$, and the index $a$ labels
the adjoint representation carried by the gluino. The spinor index heights
can be exchanged in each case, by replacing $\delta_\alpha{}^\beta
\rightarrow \delta_\beta{}^\alpha$ or $\delta^{\dot\alpha}{}_{\dot\beta}
\rightarrow \delta^{\dot\beta}{}_{\dot\alpha}$.
\label{fig:SUSYQCDsquarkmasseigrules}}
\end{figure}

\clearpage

\section{\texorpdfstring{Trilinear R-parity-violating Yukawa interactions}{Trilinear R-parity-violating Yukawa interactions}}
\renewcommand{\theequation}{L.\arabic{equation}}
\renewcommand{\thefigure}{L.\arabic{figure}}
\renewcommand{\thetable}{L.\arabic{table}}
\setcounter{equation}{0}
\setcounter{figure}{0}
\setcounter{table}{0}

In the MSSM, a multiplicative R-parity invariance is imposed, where
${R=(-1)^{3(B\!-\!L)+2S}}$
for a particle of baryon number $B$, lepton number $L$ and
spin $S$\cite{Fayet77}. Equivalently, R-parity can be defined
to be an additive quantum number modulo 2, where $R=+1$ corresponds
to an even R-parity and $R=-1$ corresponds to an odd R-parity.
In particular, all the ordinary Standard Model
particles are R~parity even, whereas the corresponding
supersymmetric partners are R~parity odd.
In the R-parity-violating extension of the MSSM (denoted below
as RPV-MSSM), new
interactions are allowed that violate R-parity.  Such interactions
necessarily violate the $B-L$ global symmetry.
R-parity-violating interactions can significantly
alter the phenomenology at colliders (see for example
\cite{Dimopoulos:1988fr,Dreiner:1991pe}), especially as the lightest
supersymmetric particle (LSP)
is no longer stable \cite{Hall:1983id,Dawson:1985vr}.
Moreover, the LSP need not be restricted to the lightest neutralino
(or perhaps the sneutrino)
as in the MSSM, but can be any supersymmetric particle
\cite{Dreiner:2008ca}.

In this appendix, we focus
on new trilinear supersymmetric Yukawa interactions that can appear in an
RPV-MSSM~\cite{rpv-superpot,Hall:1983id,Ellis:1984gi,rpvreviews}:
\beqa
\mathscr{L}_{LL\bar{e}}
&=&
-\half\lam_{ijk}\left(
{\widetilde \ell}^*_{Rk}\nu_i\ell_j
+ {\widetilde\nu}_{i} \ell_j \bar\ell_k
+{\widetilde \ell}_{Lj} \bar\ell_k\nu_i
-{\widetilde \ell}^*_{Rk}\ell_i\nu_j
- {\widetilde\nu}_{j}  \bar\ell_k\ell_i
-{\widetilde \ell}_{Li} \nu_j\bar\ell_k
\right)+{\rm h.c.}\,,
\label{rpvyuk1} \\
\mathscr{L}_{LQ\bar{d}}
&=&
-\lam'_{ijk} \left(
{\widetilde d}^*_{Rk}\nu_i d_j
+ {\widetilde\nu}_{i} d_j \bar d_k
+ {\widetilde d}_{Lj} \bar d_k \nu_i
- {\widetilde d}^*_{Rk}\ell_i u_j
- {\widetilde u}_{Lj} \bar d_k \ell_i
- {\widetilde \ell}_{Li} u_j \bar d_k \right)+{\rm h.c.} \,,
\phantom{xxx}\label{rpvyuk2}
\\
\mathscr{L}_{\bar{u}\bar{d}\bar{d}} &=&
-\half \lam''_{ijk}\eps_{pqr}\left[
{\widetilde u}_{Ri}^{p*} {\bar d}_j^{q} {\bar d}_k^{r}
+ {\widetilde d}_{Rj}^{q*} {\bar u}_i^{p} {\bar d}_k^{r}
+ {\widetilde d}_{Rk}^{r*} {\bar u}_i^{p} {\bar d}_j^{q}
 \right]+{\rm h.c.}
\,,\label{rpvyuk3}
\eeqa
where repeated indices are summed over.\footnote{The
extra factors of $\half$ in \eqs{rpvyuk1}{rpvyuk3} have been chosen
for convenience, due to the
antisymmetry properties of the corresponding couplings:
$\lam_{ijk} =-\lam_{jik}$, $\lam''_{ijk}=-\lam''_{ikj}$.}  
In \eqst{rpvyuk1}{rpvyuk3},
$\lam_{ijk},\,\lam'_{ijk} ,\,\lam''_{ijk}$ are dimensionless
coupling constants, $i,j,k$ are generation indices, and
$p,q,r=1,2,3$ are color $SU(3)$ indices, respectively.
The couplings proportional to $\lam$ and $\lam'$ violate $L$
and conserve $B$, whereas the couplings proportional to $\lam''$ violate
$B$ and conserve $L$.  Various phenomenological constraints on these couplings
are summarized in refs.~\cite{rpvreviews}.

In addition to $\lam_{ijk},\,\lam'_{ijk} ,\,\lam''_{ijk}$, the
Lagrangian of the RPV-MSSM contains one additional
supersymmetric $L$-violating mass
parameter, $\kap_i$, which leads to slepton--Higgs mixing and
lepton--higgsino mixing.
Finally, supersymmetry-breaking R-parity-violating parameters would
also contribute to slepton--Higgs mixing and
yields new trilinear scalar interactions.
These effects modify the Feynman rules of
\app{K} through additional mixing matrices, which we do not include
here (for further details, see e.g. ref.~\cite{Allanach:2003eb}).

Recently, the two-component fermion
Feynman rules for the neutral fermions have been given in
refs.~\cite{dedes,dedes2}.  Using
\eq{eq:lintY} and \fig{fig:Yukawavertexrules} we can now directly
determine the corresponding Feynman rules. These are given in
Figs.~\ref{LLE-rules}, \ref{LQD-rules}, and \ref{UDD-rules}. The same
Lagrangian for the Yukawa interactions is given in terms of four-component
fermions in refs.~\cite{Dreiner:1999qz,Richardson:2000nt}.
Two sample computations that make use of these rules are presented in
\secs{rpv-decay1}{rpv-decay2}.

\begin{figure}[tbp!]
\begin{flushleft}
\begin{picture}(200,68)(-10,8)
\DashArrowLine(60,40)(10,40)5
\ArrowLine(100,70)(60,40)
\ArrowLine(100,10)(60,40)
\Text(30,30)[]{$\widetilde \ell_{Rk}$}
\Text(70,20)[]{$\nu_i$}
\Text(70,66)[]{$\ell_j$}
\Text(140,40)[l]{$-i\lam_{ijk}$}
\end{picture}
\hspace{1.5cm}
\begin{picture}(200,68)(-10,8)
\DashArrowLine(10,40)(60,40)5
\ArrowLine(100,70)(60,40)
\ArrowLine(100,10)(60,40)
\Text(30,30)[]{$\widetilde \nu_{i}$}
\Text(70,17)[]{$\bar\ell_k$}
\Text(70,66)[]{$\ell_j$}
\Text(140,40)[l]{$-i\lam_{ijk}$}
\end{picture}
\end{flushleft}
\begin{flushleft}
\begin{picture}(200,53)(-10,20)
\DashArrowLine(10,40)(60,40)5
\ArrowLine(100,70)(60,40)
\ArrowLine(100,10)(60,40)
\Text(30,30)[]{$\widetilde \ell_{Lj}$}
\Text(70,17)[]{$\bar\ell_k$}
\Text(70,66)[]{$\nu_i$}
\Text(140,40)[l]{$-i\lam_{ijk}$}
\end{picture}
\end{flushleft}
\caption{\label{LLE-rules}
Feynman rules for the Yukawa couplings of two-component fermions due
to the supersymmetric, R-parity-violating Yukawa Lagrangian
$\mathscr{L}_{LL\bar e}$ [cf.~\eq{rpvyuk1}].
For each diagram, there is another with all arrows
reversed and $\lambda_{ijk} \rightarrow \lambda_{ijk}^*$.  }
\end{figure}
\begin{figure}[tbp!]
\begin{flushleft}
\begin{picture}(200,70)(0,14)
\DashArrowLine(60,40)(10,40)5
\ArrowLine(100,70)(60,40)
\ArrowLine(100,10)(60,40)
\Text(30,30)[]{$\widetilde d_{Rk}$}
\Text(70,20)[]{$\nu_i$}
\Text(70,66)[]{$d_j$}
\Text(140,40)[l]{$-i\lam'_{ijk}$}
\end{picture}
\hspace{1.4cm}
\begin{picture}(200,70)(0,14)
\DashArrowLine(10,40)(60,40)5
\ArrowLine(100,70)(60,40)
\ArrowLine(100,10)(60,40)
\Text(30,30)[]{$\widetilde \nu_i$}
\Text(70,17)[]{$\bar d_k$}
\Text(70,66)[]{$d_j$}
\Text(140,40)[l]{$-i\lam'_{ijk}$}
\end{picture}
\end{flushleft}
\begin{flushleft}
\begin{picture}(200,70)(0,14)
\DashArrowLine(10,40)(60,40)5
\ArrowLine(100,70)(60,40)
\ArrowLine(100,10)(60,40)
\Text(30,30)[]{$\widetilde d_{Lj}$}
\Text(70,17)[]{$\bar d_k$}
\Text(70,66)[]{$\nu_i$}
\Text(140,40)[l]{$-i\lam'_{ijk}$}
\end{picture}
\hspace{1.4cm}
\begin{picture}(200,70)(0,14)
\DashArrowLine(60,40)(10,40)5
\ArrowLine(100,70)(60,40)
\ArrowLine(100,10)(60,40)
\Text(30,30)[]{$\widetilde d_{Rk}$}
\Text(70,20)[]{$\ell_i$}
\Text(70,66)[]{$u_j$}
\Text(140,40)[l]{$\phm i\lam'_{ijk}$}
\end{picture}
\end{flushleft}
\begin{flushleft}
\begin{picture}(200,66)(0,18)
\DashArrowLine(10,40)(60,40)5
\ArrowLine(100,70)(60,40)
\ArrowLine(100,10)(60,40)
\Text(30,30)[]{$\widetilde u_{Lj}$}
\Text(70,17)[]{$\bar d_k$}
\Text(70,66)[]{$\ell_i$}
\Text(140,40)[l]{$\phm i\lam'_{ijk}$}
\end{picture}
\hspace{1.4cm}
\begin{picture}(200,66)(0,18)
\DashArrowLine(10,40)(60,40)5
\ArrowLine(100,70)(60,40)
\ArrowLine(100,10)(60,40)
\Text(30,30)[]{$\widetilde\ell_{Li}$}
\Text(70,17)[]{$\bar d_k$}
\Text(70,66)[]{$u_j$}
\Text(140,40)[l]{$\phm i\lam'_{ijk}$}
\end{picture}
\end{flushleft}
\caption{\label{LQD-rules}
Feynman rules for the Yukawa couplings of two-component fermions for
the supersymmetric, R-parity-violating Yukawa Lagrangian
$\mathscr{L}_{LQ\bar d}$ [cf.~\eq{rpvyuk2}].
For each diagram, there is another with all arrows reversed and
$\lambda_{ijk}^{\prime} \rightarrow \lambda_{ijk}^{\prime *} $.}
\end{figure}
\begin{figure}[tbp]
\begin{center}
\begin{picture}(200,54)(0,22)
\DashArrowLine(60,40)(10,40)5
\ArrowLine(100,70)(60,40)
\ArrowLine(100,10)(60,40)
\Text(30,30)[]{$\widetilde u_{Ri}^{p}$}
\Text(70,20)[]{${\bar d}_j^{q}$}
\Text(70,66)[]{${\bar d}_k^{r}$}
\Text(140,40)[l]{$-i\eps_{pqr}\lam''_{ijk}$}
\end{picture}
\hspace{1.4cm}
\begin{picture}(200,54)(0,22)
\DashArrowLine(60,40)(10,40)5
\ArrowLine(100,70)(60,40)
\ArrowLine(100,10)(60,40)
\Text(30,30)[]{$\widetilde d_{Rk}^{r}$}
\Text(70,17)[]{${\bar u}_i^{p}$}
\Text(70,66)[]{${\bar d}_j^{q}$}
\Text(140,40)[l]{$-i\eps_{pqr}\lam''_{ijk}$}
\end{picture}
\end{center}
\caption{\label{UDD-rules}
Feynman rules for the Yukawa couplings of two-component fermions due
to the supersymmetric,  R-parity-violating Yukawa Lagrangian
$\mathscr{L}_{\bar u\bar d\bar d}$ [cf.~\eq{rpvyuk3}].
For each diagram, there
is another with all arrows reversed and $\lambda_{ijk}^{\prime\prime}
\rightarrow \lambda_{ijk}^{\prime\prime *}$.  }
\end{figure}

\end{appendices}

\clearpage


\end{document}